\documentclass[10pt]{report} 

\usepackage[utf8]{inputenc} 
\usepackage[backend=biber,style=numeric,sorting=nyt]{biblatex} 
\pdfoutput=1

\usepackage{geometry} 
\usepackage{blindtext}
\usepackage{rotating}
\usepackage{graphicx} 
\usepackage{subfigure} 
\usepackage{amsmath, amssymb} %
\usepackage{IEEEtrantools}
\usepackage{xparse}
\usepackage{physics}
\usepackage[font=footnotesize,labelfont=bf]{caption}

\usepackage{booktabs} 
\usepackage{array} 
\usepackage{paralist} 
\usepackage{verbatim} 
\usepackage[official]{eurosym}

\usepackage[colorlinks]{hyperref} 
\usepackage{fancyhdr} 
\usepackage{sectsty}
\allsectionsfont{\sffamily\mdseries\upshape} 

\usepackage[nottoc]{tocbibind} 
\usepackage[titles,subfigure]{tocloft} 

\begin{document}
\begin{titlepage}
   \begin{center}
       \vspace*{1cm}
       \Large
       \textbf{Separating the Sources of Chaos in the Relativistic 3 Body Problem }\\
       \vspace{1.5cm}
       \Large
       \textbf{Patric de Gentile-Williams}\\
       \vspace{0.5cm}
       \normalsize
       \textbf{Student  Number 21178214}\\
       \vspace{1.5cm}
       \Large
       \textbf{MSc Physics}\\
       \vspace{1.5cm}
       \textbf{PHAS0062 -MSc Research Project (21/22)}\\
       \vspace{2.5cm}
       \textbf{Supervisor: Dr Betti Hartmann}\\
       \vspace{0.5cm}
       \textbf{ Dr Christian Böhmer}\\
       \vspace{1.5cm}
       \textbf{ Updated November 5th, 2023}
       \vfill
       \vspace{0.8cm}
       \includegraphics[width=0.4\textwidth]{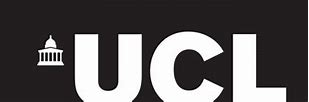}
   \end{center}
\end{titlepage}
\begin{abstract}
The dynamics of the Restricted 3 Body Problem in the Post Newtonian context have been, and continue to be, studied extensively and a number of characteristics such as ejections of bodies from the system, precession of orbits, chaotic trajectories and collisions are investigated and classified. In this paper, I examine the extent to which these characteristics are attributable to Relativistic causes (more correctly Post Newtonian approximations of Relativistic causes) or more prosaically whether these characteristics already exist in the Newtonian case and whether these or other effects would also be caused by the time of propagation of the field, the impact of which is less well covered in the literature.  In this project I have written bespoke code for the simulations and present the results in a methodical way showing many of the steps in the process.  In the Appendices, I discuss the technical challenges of constructing, testing and calibrating the code, so that other non-experts in the field can reproduce the code and results and build on the work that I have done. Although the work done here must be considered preliminary, results suggest that the impact of the field propagation time on the behaviour of the test mass in the restricted 3 body problem outweighs the impact of the Post Newtonian corrections, in particular due to a transfer of angular momentum from the binary system to the test mass, and this effect may therefore be of interests to those studying accretion discs of binary systems.
\end{abstract}
\tableofcontents

\chapter{Introduction}
\begin{figure}[!ht]
\includegraphics[width=0.475\textwidth]{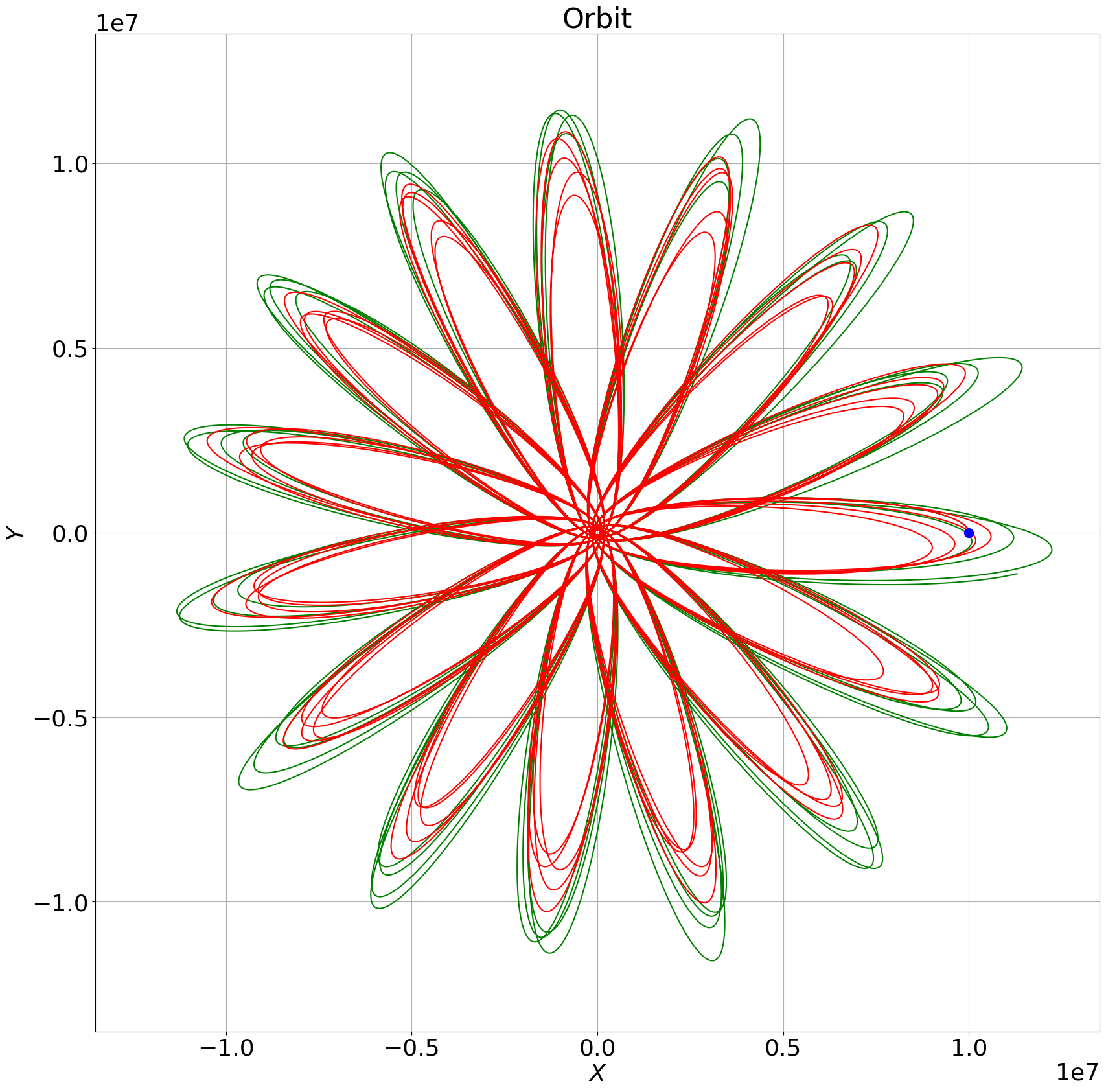}
\includegraphics[width=0.508\textwidth]{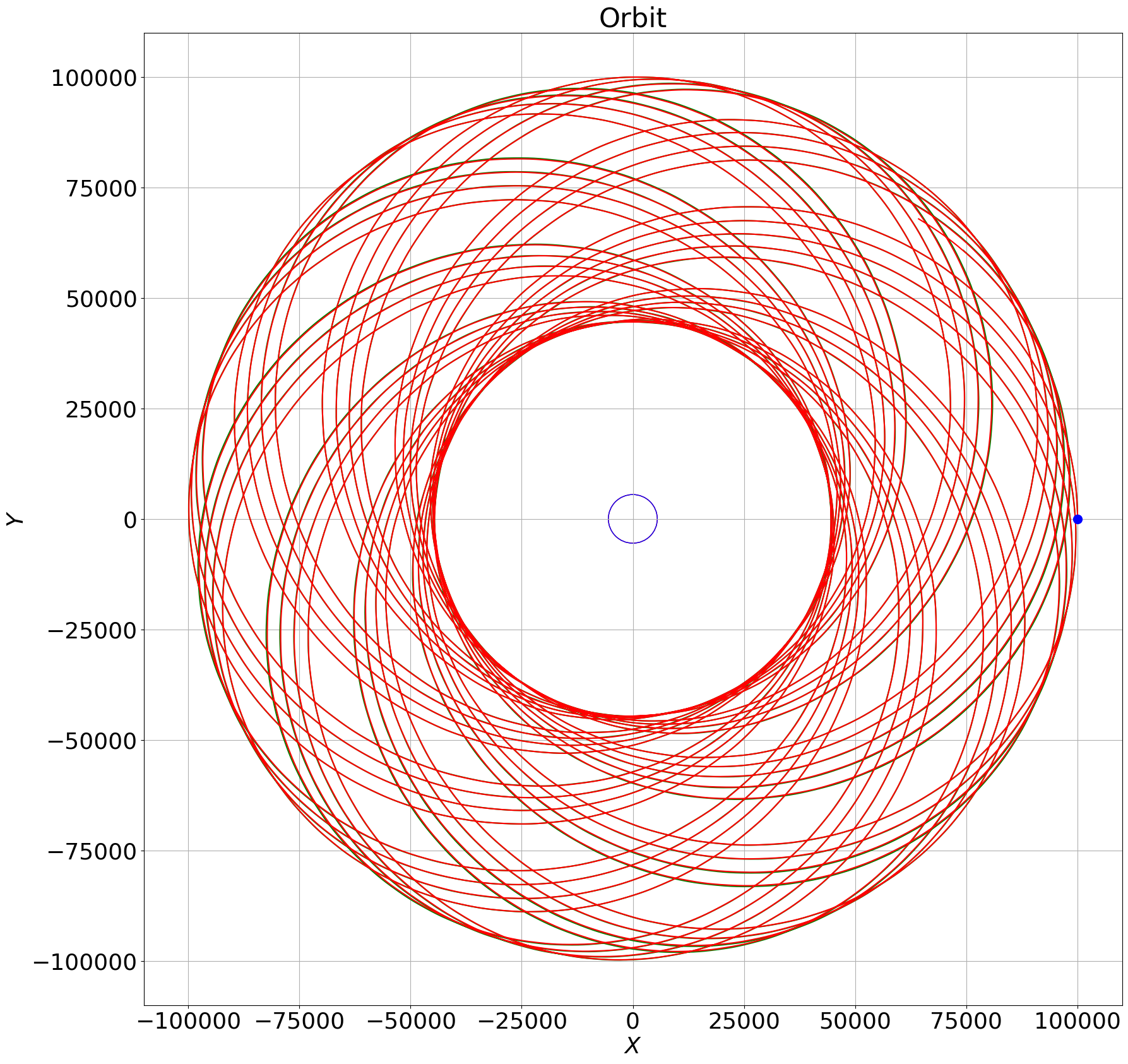}
\caption{Examples of Precession}
\end{figure}
\noindent \\
As discussed in my Research Essay \cite{PDGW}, unlike the case of the Newtonian two body problem, there is no general analytical solution to the geodesic paths followed by test masses in the spacetime of two large gravitational masses (although some special cases have been solved). In the General Relativity setting, the problem is only exacerbated by the form of the forces / accelerations felt by a test mass which require more sophisticated mathematical expressions to model them.\\ \\
Researchers have developed increasingly sophisticated numerical techniques, culminating in the 2005 breakthroughs when Baker et al. of the Goddard Group \cite{Baker2005}, Pretorius \cite{Pretorius2005} and  Campanelli et al. of the Brownsville Group \cite{Campanelli2005} independently developed their moving puncture technique which allowed the accurate simulation of binary black hole systems and these techniques underlie the models used by the gravitational observatories Ligo, Virgo, etc. (the relativistic two body problem). \\ \\
However, to study the Relativistic Three Body Problem (even the restricted -planar- version of the problem) researchers rely on what are known as 1 PN, 2 PN and 2.5 PN techniques developed by numerous parties during the 20\textsuperscript{th} century (as discussed in \cite{PDGW}), where the number in the $\#$ PN name refers to the power to which the expansion term $\frac{1}{c^2}$ is raised.  \\ \\
In its most general form, the relativistic three body problem has 33 degrees of freedom. Each mass has 3 position coordinates, 3 velocity coordinates, 3 coordinates for its spin, an electric charge and a mass. Symmetries and a judicious choice of coordinates will reduce this number somewhat but still leave the investigator with an unmanageable search space. To this complexity we must add a variety of technical choices: the mathematical models of Gravity and the numerical techniques used. The investigator is therefore obliged to search only highly constrained parts of the available space.\\ \\
Two topics that appear less well covered in the literature are: 
\begin{itemize}
\item the extent to which the corrections obtained from the Schwarzschild metric can be used to model the 3 body problem 
\item whether allowing for the propagation time of the gravitational field has much impact on the simulations (what I refer to as the propagation time of the gravitational field is typically referred to in the literature as `retarded gravitational potential') 
\end{itemize}
In studying these questions, I have chosen to write my own code (rather than use pre-existing ones) as I feel the effort of doing so will be amply justified by the deeper insight that I will gain. Writing my own code will allow me to explore new questions such as whether using the Schwarzschild corrections as a low 'cost' version of the more traditional Post Newtonian correction models produces usable results; it will also allow me to study the impact of including field propagation time in the models. Writing my own code will also allow me to develop bespoke metrics that I may wish to include, as well as allow complete freedom in the choice of initial conditions; for instance, whilst I have largely studied the Restricted Three Body Relativistic System, I have in fact built the code as fully three dimensional and could therefore also use the code to study a more general set of conditions.\\ \\
Importantly, writing my own code is a powerful learning tool and I believe will teach me more about the problem, the process of research and meaningfully improve my programming abilities when contrasted with the impact of simply using existing tools.  \\ \\
The code and results data is available on request from the author.\\

\chapter{Physical Model: Mathematical Considerations}
\section{Field Propagation Time}
\begin{figure}[!ht]
\includegraphics[width=0.99\textwidth]{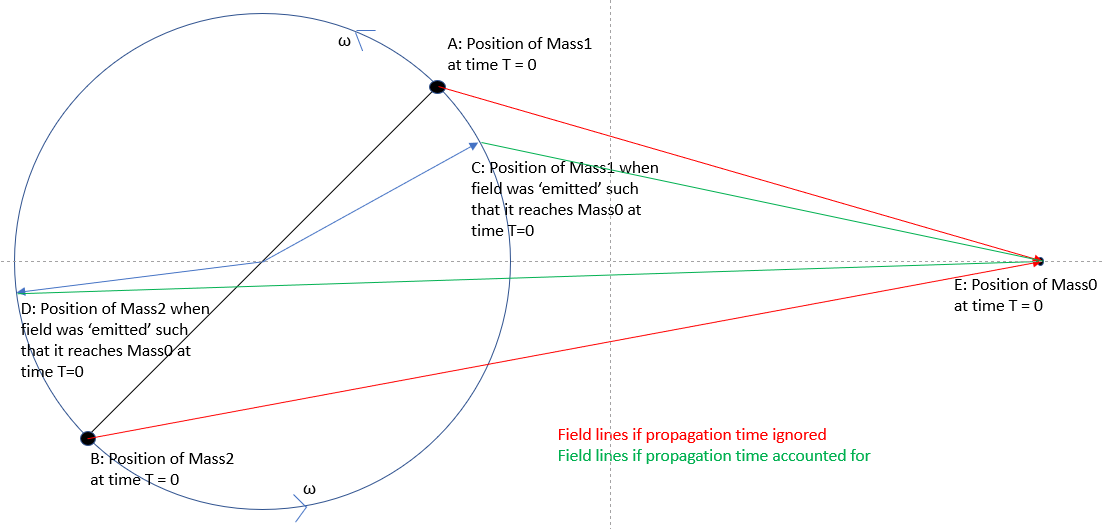}
\caption{Field Propagation Dynamics}
\label{fieldfig}
\end{figure}
\begin{figure}[!ht]
\includegraphics[width=0.99\textwidth]{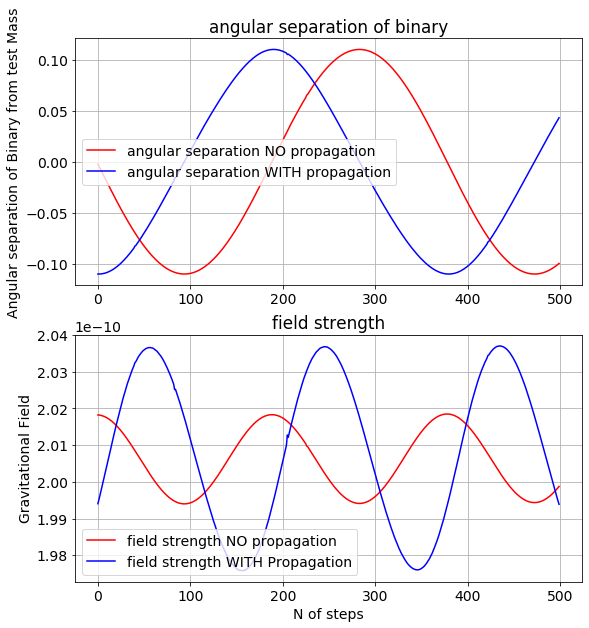}
\caption{Newtonian Gravity with and without the Field Propagation Effect:\\  Angular Separation and Field Strength Dynamics}
\label{fieldprop1}
\end{figure}
\begin{figure}[!ht]
\includegraphics[width=0.99\textwidth]{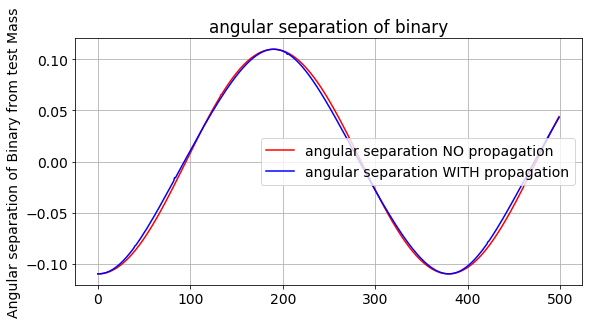}
\caption{Newtonian Gravity with and without the Field Propagation Effect:\\ Angular Separation with Phase Manually Adjusted to Line-up the Two Graphs}
\label{fieldprop2}
\end{figure}
\begin{figure}[!ht]
\centering
\includegraphics[width=0.9\textwidth]{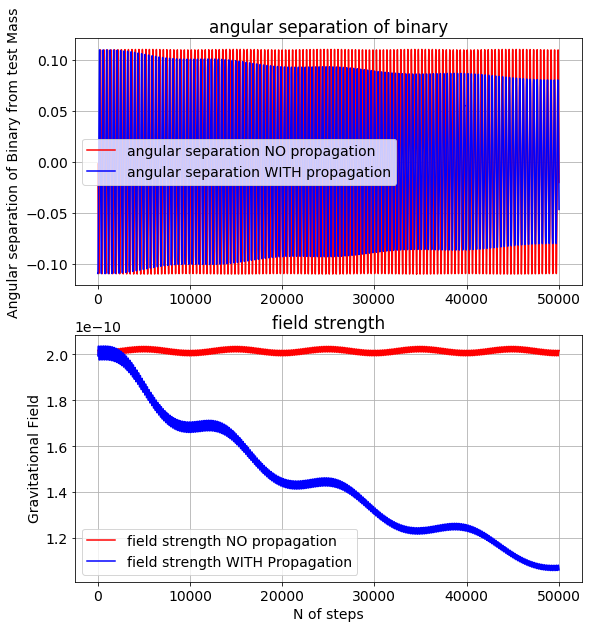}
\caption{Newtonian vs Field Propagation: Field Strength Oscillation (Prograde)}
\label{fieldprop3}
\end{figure}
\begin{figure}[!ht]
\centering
\includegraphics[width=0.9\textwidth]{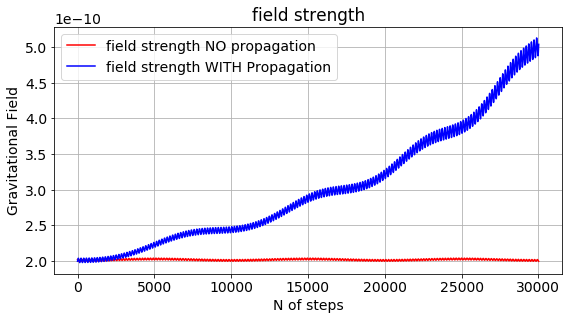}
\caption{Newtonian vs Field Propagation: Field Strength Oscillation (Retrograde)}
\label{fieldprop4}
\end{figure}
\begin{figure}[!h]
\centering
\includegraphics[width=0.7\textwidth]{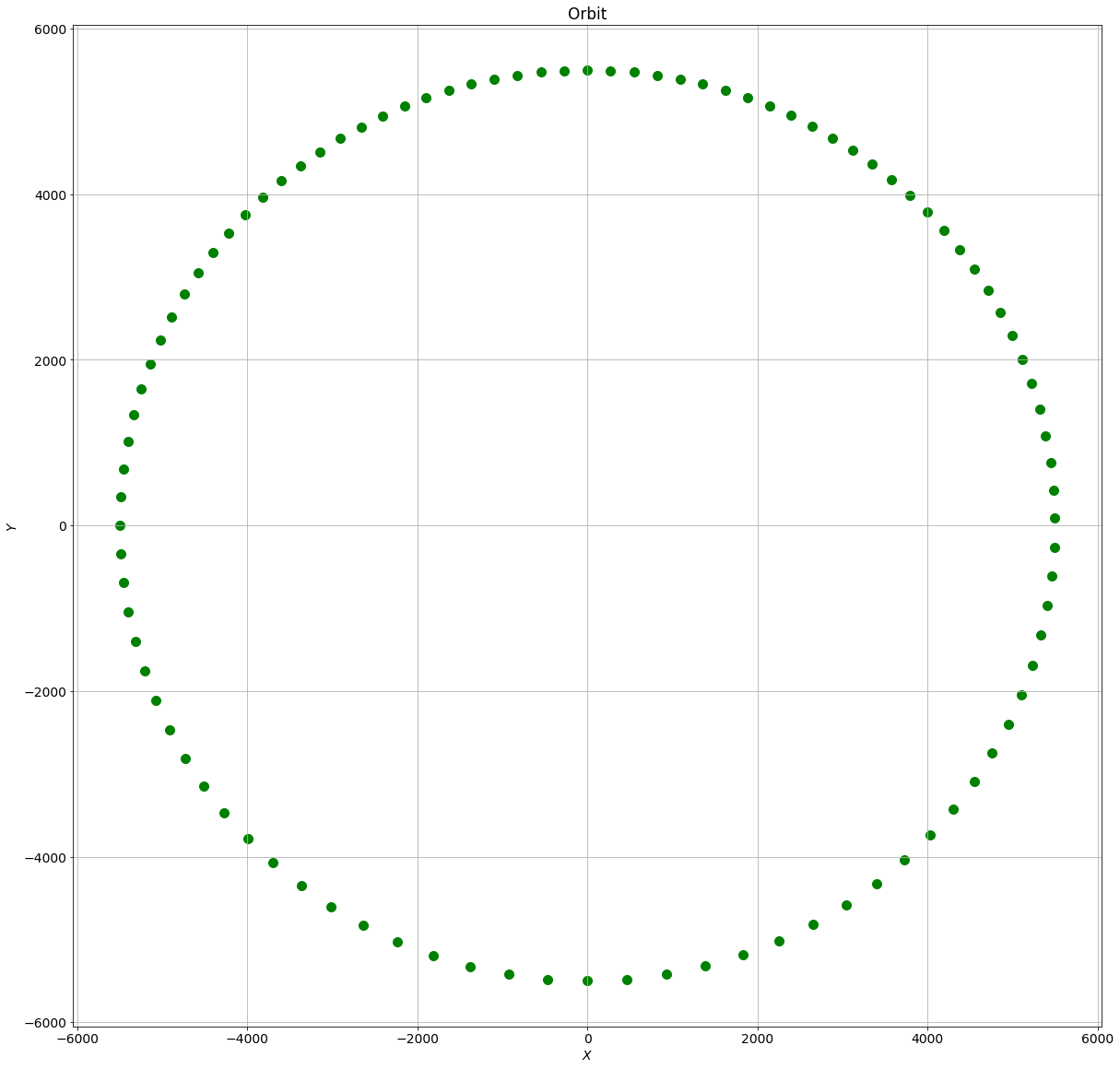}
\caption[Model of perceived position of primary as seen by test mass - no extrapolation]{In this picture, we see position of one of the primaries in the X / Y plane, with an orbital radius of 5,500$m$. Each dot is the position of the primary as seen by the test mass, using the test mass' clock to sample the orbit with a frequency of 100 snapshots per orbit of the binary. The test mass knows the period of the binary simply by timing the primary's return to its position of maximum angle}
\label{doppler}
\end{figure}
\begin{figure}[!h]
\centering
 \subfigure[Blue Frame]{\includegraphics[width=0.49\textwidth]{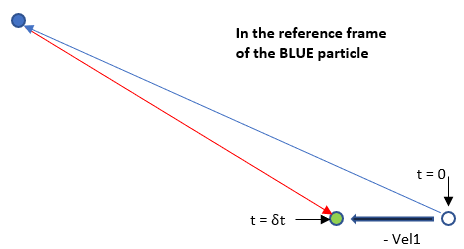}}
 \subfigure[Green Frame]{\includegraphics[width=0.49\textwidth]{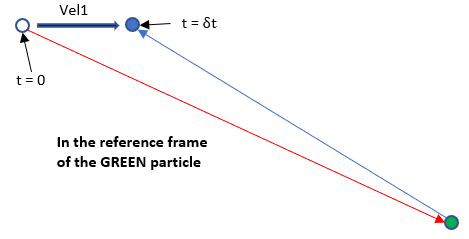}}
\caption{Field forces for moving particles \\}
\label{spec_rel1}
\end{figure}
\begin{figure}[!h]
\centering
\includegraphics[width=0.50\textwidth]{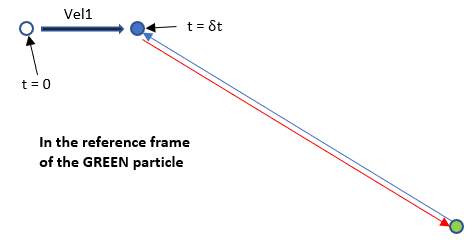}
\caption{Corrected field forces for moving particles \\}
\label{spec_rel3}
\end{figure}
\begin{figure}[!h]
\centering
\includegraphics[width=0.75\textwidth]{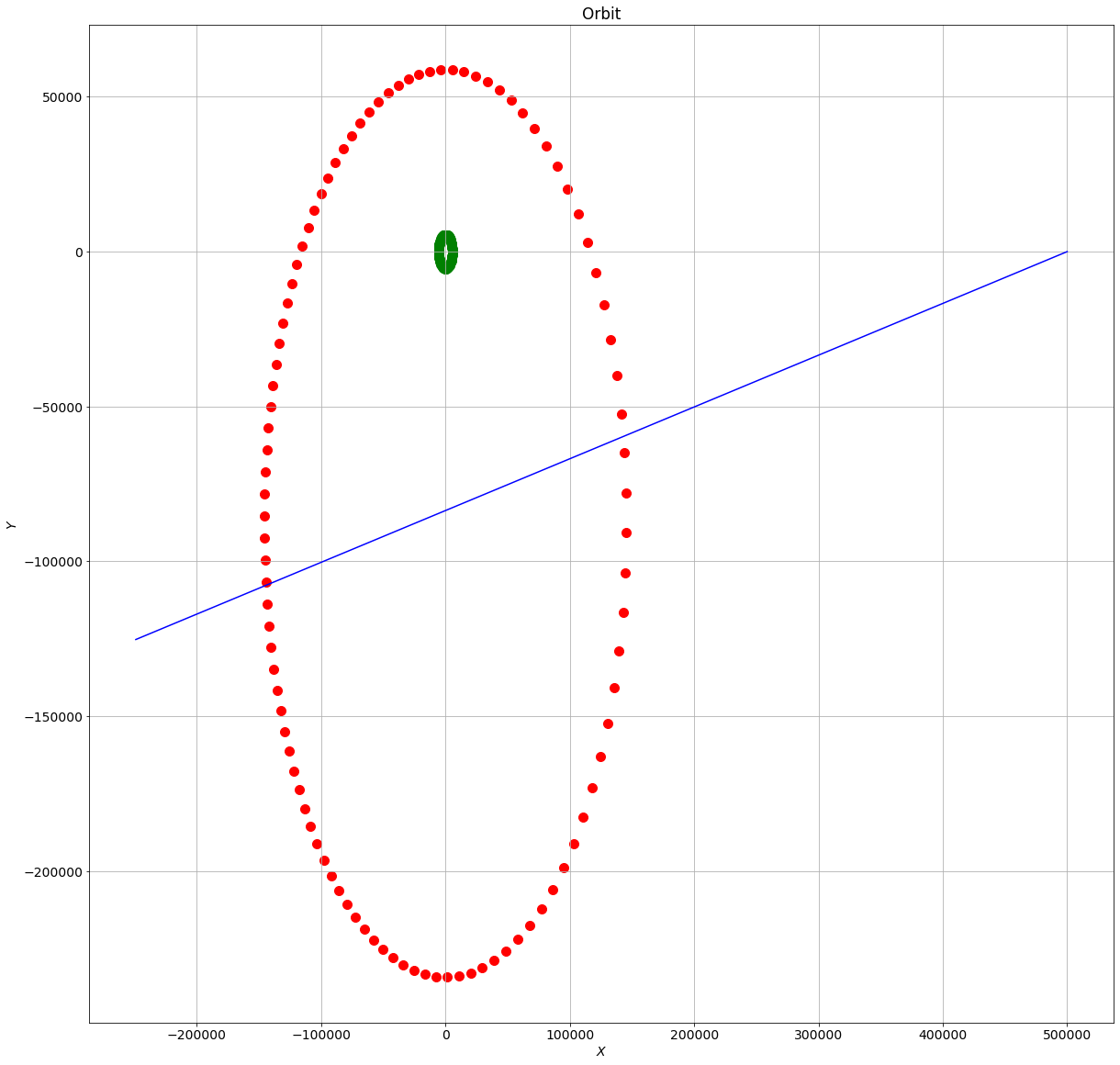}
\caption[Model of perceived position of primary as seen by test mass - with extrapolation]{Actual positions of $P1$ binary in green and perceived position of $P1$ as felt by the test mass due to special relativistic effects. }
\label{projected}
\end{figure}
\begin{figure}[!h]
\centering
\includegraphics[width=0.75\textwidth]{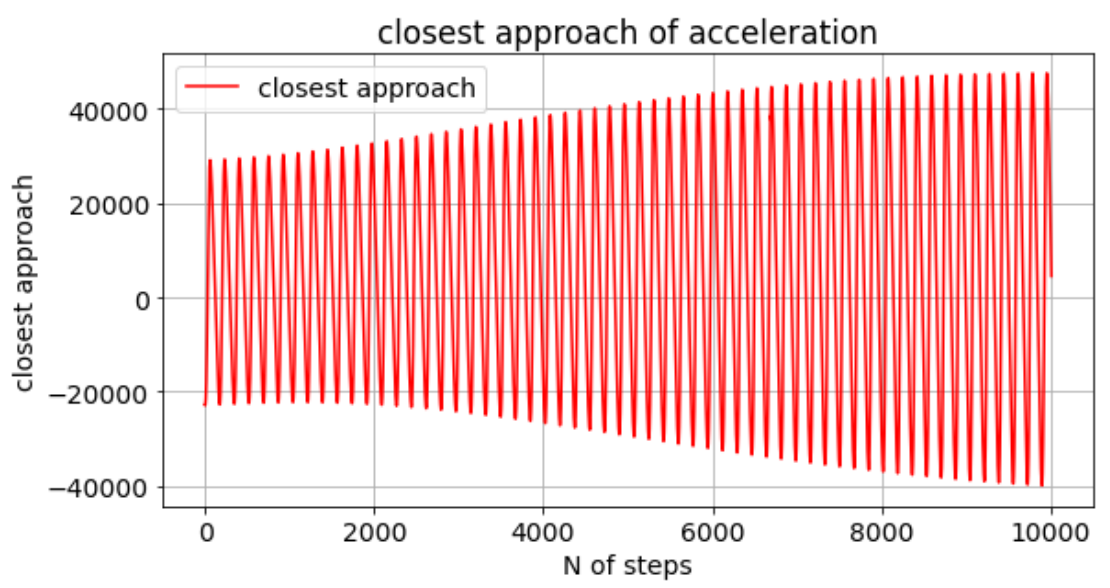}
\caption{Distance of closest approach to the origin of the acceleration felt by the test mass. }
\label{clapp}
\end{figure}
\begin{figure}[!h]
\centering
\includegraphics[width=0.75\textwidth]{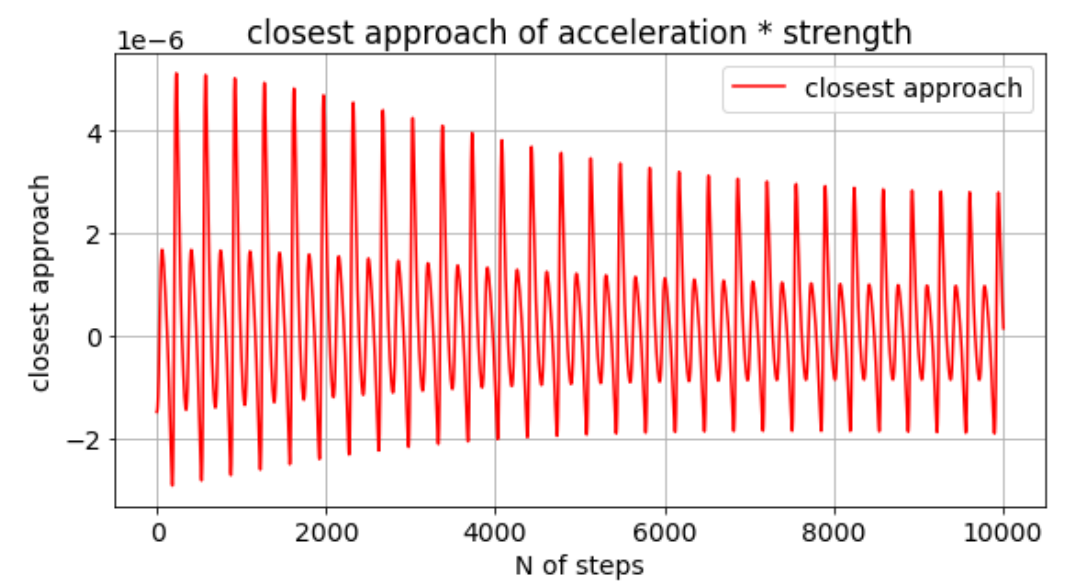}
\caption{Distance of closest approach of the acceleration of the test mass scaled by amplitude.}
\label{clapp-strength}
\end{figure}

\noindent \\
In gravity we would a expect to observe the equivalent phenomenon to retarded potentials in electro-magnetism. However, the standard equations for Post-Newtonian corrections for the gravitational field such as \eqref{1PN} and \eqref{2PN} whilst containing relativistic corrections, nonetheless in some sense do not allow for the propagation time of the field. We can see this in \eqref{1PN} in as much as there is no explicit time coordinate in the expression and more importantly the velocity and direction vectors are not time dependent.  This means that these equations do not take account of the fact that the gravitational sources move between the time the field is `generated' and the time the field is `felt' by the test mass and consequently these expressions are only good approximations when the velocity of bodies is slow and the likely position uncertainty is small. If one considers the system in a rotating frame (where the axis between the two primaries is taken say as the X axis) then these equations are `correct' but in a fixed frame of reference, or indeed any frame other that one rotating with the binary, then the gravitational sources are not static and hence these equations miss the physical phenomenon: that the sources move between the time of field emission and the time the field is felt by other bodies. \\ \\
The literature is relatively sparse on the effects of gravitational field propagation time, although Eddington in 1920 had argued that two-body orbits would be unstable in a `retarded gravitational field' \cite{eddington1920} and Kennedy developed a theory of planetary motion in a retarded Newtonian potential \cite{kennedy1929} in 1929, which he himself dismisses on the ground of the theory not being covariant.  More recently Raju has published extensively on retarded gravitation theory in a Newtonian context, \cite{Raju2015} but does not extend his treatment to the three body problem.   \\ \\
In order to properly take account of the evolution of the field, any set of initial conditions would have to specify a complete path history for all bodies in the system, which is in general not feasible and indeed is a circular problem (since we are trying to determine the path, so requiring a specification of the path history in order to determine the path is circular). There are some special circumstances where specifying the initial conditions is possible, for instance: 
\begin{itemize}
\item all bodies start infinitely far apart spatially and infinitely far in the past and the bodies are all on a trajectory towards a roughly common point
\item in the restricted 3 body problem, two of the bodies are in a predictable orbit around each other and the third body is of negligible mass. This is the case I am studying
\end{itemize}
I assume a finite speed of light and that the gravitational field also propagates at the speed of light, then in the time frame of the test mass, as we see in Figure \ref{fieldfig}, in order to compute the effects of the fields generated by $Mass1$ and $Mass2$ on $Mass0$ at time $t=0$, I need to search back in time for the positions C and D of $Mass1$ and $Mass2$ respectively at times $-t_1$ and $-t_2$ (Noting that the test mass knows the period of the binary system from measuring the time taken for complete orbits and hence `knows' the angular velocity).  I obtain expressions for these positions by equating the time taken for light to travel from position C to position E (or position D to position E) with the time taken for $Mass1$ to move at speed $\omega$ from position C to position A (or $Mass2$ from position D to position B). So, with the distance between $Mass1$ and $Mass2$ being $\ R \ $ and the distance from the Centre of Mass (CoM) of the binary to $Mass0$ being $\ r \ $ and for simplicity of expression (without loss of generality) I assume that at time $t = 0$, the three masses lie on the X axis, (from the perspective of a distant observer on the z axis) then for $Mass1$ we have:
\begin{align}
c\ t_1 &= \sqrt{\big(\frac{R}{2} \ \cos(\omega \ t_1) - r \big)^2 + \big( \frac{R}{2} \ \sin(\omega \ t_1) \big)^2}
\intertext{where the coordinates of position C are: $(\frac{R}{2} \cos(\omega \ t_1) , \frac{R}{2} \sin(\omega \ t_1))$ then}
t_1 &= \frac{1}{c}\ \sqrt{\big(\frac{R}{2} \ \cos(\omega \ t_1) - r \big)^2 + \big( \frac{R}{2} \ \sin(\omega \ t_1) \big)^2}
\intertext{and for $Mass2$ }
c\ t_2 &= \sqrt{\big(\frac{R}{2} \ \cos(\omega \ t_2 + \pi) - r \big)^2 + \big( \frac{R}{2} \ \sin(\omega \ t_2 +\pi) \big)^2} 
\intertext{where the coordinates of position D are: $(- \frac{R}{2}\cos(\omega \ t_2) , -\frac{R}{2} \sin(\omega \ t_2))$ then}
t_2 &= \frac{1}{c}\ \sqrt{\big(\frac{R}{2} \ \cos(\omega \ t_2 + \pi) - r \big)^2 + \big( \frac{R}{2} \ \sin(\omega \ t_2 +\pi) \big)^2}
\end{align}
noting that $\omega$ the angular velocity of the binary is a function of the separation of the binaries (and their masses) I can therefore solve for  $r$  as a function of $t_1$ and separately as a function of $t_2$:
\begin{align}
c^2 t_1^2 \ -  \ &\big( \frac{R}{2} \ \sin(\omega \ t_1) \big)^2 = \big(\frac{R}{2} \ \cos(\omega \ t_1) - r \big)^2 \\ \nonumber \\
r &= \frac{R}{2} \ \cos(\omega \ t_1) \pm \sqrt{c^2 t_1^2 -  \big( \frac{R}{2} \ \sin(\omega \ t_1) \big)^2}
\intertext{and for $Mass2$}
r &= \frac{R}{2} \ \cos(\omega \ t_2 + \pi) \pm \sqrt{c^2 t_1^2 -  \big( \frac{R}{2} \ \sin(\omega \ t_2 + \pi) \big)^2}
\end{align}
To get an idea of the scale of this effect, I consider the case where the angular velocity of the binary system is high and the distance from the binary to the test mass is relatively low in order  to check whether the effects are of measurable amplitude.\\ \\
For example, if we have two solar mass black holes (which, for reference, have Schwarzschild radii of approximately 3,000m) orbiting each other at a distance of 50,000m, then their angular velocity is (using the model above) 1,495 radians per second which corresponds to a velocity of 37,377,304 metres per second or 0.1247$c$. In this scenario, with the binaries and the test mass all aligned along the X axis, and the test mass at a distance of 250,000 metres from the binary CoM,  then the test mass would feel the field from $Mass1$ when it was at an angle 1.2082 radians to the X axis and $Mass2$ would be at an angle 1.2872 radians (these angles are measured at the origin). However, what is important here is not the absolute angles but the difference between them, which in the strong field, high velocity limit are of sufficient scale to affect trajectories. Looking at the Newtonian gravitational forces / accelerations, the acceleration of the test mass is $4.4017 \cdot 10^9 \  m  s^{-2}$ whereas with the field propagation time effect, as described above, the acceleration is $4.2779 \cdot 10^9 \ m  s^{-2}$, which is a $2.8\%$ reduction in the field strength; what is more, this difference in field strength oscillates as the system evolves.\\ \\
In contrast, in the weaker field, slower moving case, if the separation between the primaries is $10^{10}$ metres and the test mass is at a distance $10^{12}$ from the binary system, then the angular velocity of the binary is $1.63 10^{-5}$ radians per second, the angular difference between the angles of emission points with the X axis is 0.0005442 radians and the change in the field strength due to the field propagation time effect is of order $10^{-10}$ and this effect will become progressively less measurable as the distances increase.\\ 
\noindent \\
In Figure \ref{fieldprop1}, in order to show the effect where it is highly visible (in the strong field limit) I have set the distance between the primaries at 11,000$m$ and the test mass at 100,000$m$ from the origin and plotted the angular separation of the binaries (upper pane) as well as the field strength (lower pane) for the Newtonian case both with and without the field propagation time effect. Note that in this case the angles are measured not at the origin as I had done in the previous paragraph but are measured from the position of the test mass. In Figures \ref{fieldprop1}, \ref{fieldprop2}, \ref{fieldprop3} and \ref{fieldprop4}, the X axis is a time coordinate, here shown as the number of computational steps of the model.\\ \\
For ease, and for the only time in this paper, I have set the mass of the binary and $G$ equal to 1. In the Newtonian case, we would expect to see the angle between the binaries oscillating between plus or minus  $ \arctan( 5500/100000) $  or plus or minus 0.1099 radians; and we would expect the field intensity to vary between:
\begin{align}
maxfield &= \frac{1}{105500^2} + \frac{1}{94500^2} &= 2.01824 \cdot 10^{-10} \\
minfield &= 2 \cdot \frac{1}{100000^2 + 5500^2} &= 1.99397 \cdot 10^{-10} 
\end{align}
which is exactly what we see in the red lines of the upper and lower panes of Figure \ref{fieldprop1},
\noindent
and where the blue lines are the equivalent data set for the field propagation mode. As expected, we observe a phase shift between the two modes, but in a rotating system, such a phase shift contains little physics.\\ \\
However, what is also obvious from the field strength picture is that the amplitude of oscillation of the field strength is about 3 times greater in the field propagation mode than in the Newtonian mode.  Manually removing the phase difference in the angular separation picture we see in Figure \ref{fieldprop2} a changed version of the upper pane of Figure \ref{fieldprop1} where a subtle but clear effect on the angular separation in the field propagation mode which, whilst having the same maxima as the Newtonian mode, we see that the angular separation in the field propagation mode is consistently smaller than it is in the Newtonian mode (the blue line is ``inside'' the red line).\\ 
\noindent \\
The effect of this increase in amplitude of oscillation and change of shape of the angular separation curve (which translates into variations in distances between the test mass and the primaries) is surprisingly strong as we see in Figure \ref{fieldprop3}.  We see that for the field propagation mode, there is a very strong transfer of angular momentum and energy from the binary to the test mass, and that the orbit of test mass increases rapidly as a result, whereas without the field propagation time effect (in Red) there is no increase in orbital radius. (Note that the oscillations with a period of 10,000 time steps - equivalent to one orbit - are just a reflection of the fact that is almost impossible to start the system in a perfectly circular orbit and hence the orbit is elliptical, I will therefore not comment on this frequency of oscillation any further).  \\ \\
Figure \ref{fieldprop4} is the identical scenario to that of Figure \ref{fieldprop3}, but set off in the retrograde direction, and consequently the transfer of angular momentum is in the opposite direction and we see the test mass spiral into the binary (the scenario is also run over fewer orbits).  \\ 
\noindent \\
A simple model for the propagation time of the field discussed above would lead us to observe a Doppler like effect where, as the test mass orbits the binary, it perceives/'feels' the primaries receding from it as moving more slowly than the primaries that are moving towards it, which it `feels' as moving faster.  Consequently, the test mass would feel the primaries spending more time on the receding part of the orbit and less time on the approaching part of the orbit and hence the symmetry of forces would be broken and the test mass would `feel' a stronger force coming from the receding side. This is best illustrated in Figure \ref{doppler} where I have plotted the perceived position of the one of the primaries from the perspective of the test mass at equal time intervals (again from the perspective of the test mass); the primaries are orbiting counter-clockwise. There is a clear excess `density' of presence of the primary at the top of the orbit (receding part) compared to the bottom (approaching) part of the orbit. Also, the points of highest and lowest `density' in this figure are not at the top and bottom of the binary orbit but at the points of the orbits where the tangents to the orbit pass through the position of the test mass. This Figure would appear to explain why the test mass will acquire angular momentum since the forces due to the binary are not symmetric about the binary's centre of mass, and the test mass will feel a greater force coming from the receding primaries, thus imparting angular momentum to the test mass relative to the binary's CoM.\\ 
\noindent \\
However, as Poincaré observed \cite{Poincaré1905} velocity dependent effects must be accounted for. In Figures \ref{spec_rel1} and \ref{spec_rel3} we see two charged particles passing each other in Minkowski spacetime, it does not matter whether we consider the charge to be electric charge or gravitational mass. In Figures \ref{spec_rel1} a) $\&$ b) we see the same physics in the frames of reference of each of the particles respectively, and in both cases the field `emitted' by the static particle is itself static whereas the field `emitted' by the moving particle is time dependent. Consequently, the moving particle emits its field at time $t = 0$ and it is `felt' by the static particle at time $t = \delta t$. If we measure the field `felt' by the moving particle at $t = \delta t$ we see that it does not `point' in the same direction nor is it of the same magnitude as the force felt by the static particle, thus causing a dipole which would contradict observation. What in fact happens, as shown in Figure \ref{spec_rel3} is that there is a relativistic effect (correction) to the interactions which causes the static particle to feel a force `emitted' from a point extrapolated for the moving particle's velocity times (multiplied by) the time interval $ \delta t$ thus removing the dipole and restoring the correspondence between theory and observation. In electromagnetism these corrections are the Liénard-Wiechert potential. \\
\noindent \\
In Figure \ref{projected} I have modified the model used to generate the same effect as in Figure \ref{doppler} where I include the special relativistic effects of Liénard Wiechert potentials. This simply corresponds to interpolating the position `felt' by the test mass of the primary $P1$ for the same 100 time steps shown in Figure \ref{doppler}. The positions are therefore given by:
\begin{align}
\vb{NP1} = \vb{P1} + (\vb{v_{P1}} -  \vb{v_{TM}}) * \delta t
\end{align}
\noindent \\
where $\vb{NP1}$ is the projected position of $P1$ as `felt' by the test mass, $\vb{P1}$ is the actual position of $P1$ when the field was emitted, $\vb{v_{P1}}$ is the velocity of $P1$ at the time field was emitted,   $\vb{v_{TM}}$ is the velocity of the test mass at the time the field was emitted and $\delta t$ is the time the field takes to propagate from $\vb{P1}$ to the test mass.\\
\noindent \\
We see in this figure\ref{projected} the same positions of the primary $(Mass1)$ that we saw in Figure \ref{doppler} (in green) and we observe that the perceived positions of $P1$ (in red) now describe a much bigger ellipse than their actual orbit. This may seem surprising but is in fact quite straightforward. As with the previous examples, I have set binaries total mass at $4 . 10^{30}kg$, binary separation at $11,000m$, the binaries are in a stable orbit, test mass is at position $X = 500,000m, \ \ Y = 0m$. Using simple Newtonian mechanics gives us a linear velocity for the binaries of $\approx 0.26 c$ and a linear velocity for the test mass of $ \approx 0.17 c$. The shortest time the field can take to propagate from the primaries to the test mass is $ \frac{500000 - 5500}{c} \approx 0.00165 seconds$. In this time, the interpolated distance assumed to be travelled by $Mass1$ is $0.26 * c * 0.00165 = 1.28 \cdot  10^5m$ or approximately eleven times the separation of the primaries. In blue, I have drawn the mean direction of the gravitational field felt by the test mass taken over these one hundred points.\\ \\
Several comments should be made here:
\begin{itemize}
\item  we see that the spacings between points appear to have rotated by $\frac{\pi}{2}$ which is because the dominant part of these positions are driven by velocities and these are given by the time derivative of the position vectors and so sine(s) go to cosine(s) etc.  
\item the red ellipse of `interpolated' trajectories is offset relative to the original green trajectories. This is due to the linear velocity of the test mass which in this model is simply in the $Y$ direction. When the test mass is at large distances from the binary, the linear velocity of the test mass is small, and the mean gravitational force exerted by the binary on the test mass is pointing almost exactly at the binary's CoM
\item that as the test mass moves further away, the trajectory described by the primaries tends to a circle centred on the origin with radius $\vb{v_{PR}} * X$ where $\vb{v_{PR}}$ is the linear velocity of the primaries expressed in units of $c$ and $X$ is the distance between the test mass and CoM of the binary. So as $X$ increases, so does the perceived orbit of the primaries but as the separation of the primaries $R$ increases, the linear velocity of the primaries falls as $\frac{1}{R^{0.5}}$ and hence so does the perceived radius of the primaries
\item the blue line is pointing \emph{down} and as such would tend to reduce the angular momentum of the test mass, which is the reverse of the phenomenon I observed in the PN model, however as mentioned above, this slope is entirely due to the velocity of the test mass, if the test mass were stationary then the blue line would point at the CoM of the binary
\end{itemize}
\noindent \\
It is at this point that I have to recognise the limitations of special relativistic interpolations and the model I used to generate Figures \ref{doppler} and \ref{projected}, although some features are still useful, for instance the general picture of interpolating a much wider trajectory of the primaries still has merit, it is simply that we cannot use this model to compute time integrated forces. \\ \\
Using the full simulation model with PN corrections, in Figure \ref{clapp} I have plotted the distance of closest approach to the origin of the acceleration felt by the test mass. A distance of closest approach of zero would mean that the acceleration is pointed directly at the CoM of the binary, a positive value of the distance of closest approach represents a force that would tend to increase the angular momentum of the test mass and conversely for negative values. Here we observe that:
\begin{itemize}
\item the direction of the acceleration asymmetrically oscillates about the origin with a clear positive bias. This positive bias means that the acceleration felt by the test mass is not (when averaged over an orbit of the primaries) oriented towards the CoM of the binary but slightly \emph{above} the CoM, this means that at least some of the acceleration is along the tangent of the test mass' velocity vector and hence the absolute velocity of the test mass will increase thereby imparting angular momentum and energy to the test mass
\item this acceleration along the tangent of velocity the test mass can be understood by looking at the last term of \eqref{1PN} which is a (positive) term proportional to $\vb{v}_{ij}$  the relative velocity of the test mass and the primaries ($\vb{v}_{ij} = \vb{v}_{i}  - \vb{v}_{j}$). However, averaged over a primary orbit the mean velocity of the primaries is zero (the primaries are going around in a circle) and hence averaged over time the last term in \eqref{1PN} is in fact tangent to $\vb{v}_i$. Similar comments can be made about the $\vb{v}_{ij}$  terms in \eqref{2PN} and \eqref{2.5PN} and the associated notes
\item in Figures \ref{clapp} and \ref{clapp-strength} we see that the oscillatory part of the plots correspond to the $\vb{v}_{j}$ terms described above (the velocities of the primaries) and that the asymmetry of the oscillations corresponds to the $ \vb{v}_{i}$ term (the velocity of the test mass)
\item as the system evolves the amplitude of oscillation becomes larger and more symmetrical. This is because, as the system evolves, the test mass is moving away from the binary and hence the time taken for the field to reach the test mass increases and therefore the interpolated position of the primaries describes a bigger ellipse.
\item at the same time, the linear velocity of the test mass falls and hence the asymmetry in the perceived orbit of the primaries is reduced
\end{itemize}
\noindent \\
In Figure \ref{clapp-strength}, I have scaled (multiplied) the acceleration / direction (unit) vectors by the magnitude of the acceleration felt by the test mass. Here we observe:
\begin{itemize}
\item that half the cycles have smaller amplitude, this simply reflects the fact that the force is much greater when the primary is close to the test mass and this effect is amplified by the interpolated position of the primary
\item that in both half cycles (lower amplitude and higher amplitude) are asymmetrically positive and hence as above cause a transfer of angular momentum to the test mass
\item that as the test mass is driven away from the binary the amplitudes reduce
\item that as the test mass is driven away from the binary the asymmetry reduces
\end{itemize}
Importantly, the phenomenon will not exist in the Newtonian case as it only exists with finite speeds of light. Equally importantly, this phenomenon does not come to light if the system is studied in a rotating frame, since in that frame the spacetime of the binary is static. This phenomenon has a strong impact as will be described in the results section: Chapter \ref{chapresults}.\\ 
\noindent \\  

In Chapter \ref{chapresults}, I also briefly discuss the impact of this increase of orbit on cosmological timescales.

\clearpage

\section{Corrections from the Schwarzschild metric} \label{chapschwar}
The Schwarzschild metric is an exact solution to Einstein's field equations for the vacuum outside a static, spherically symmetric massive body, assuming no charge, angular momentum or cosmological constant. It is not therefore a solution to the spacetime around two large masses orbiting each other. Nonetheless I will use the correction terms from this metric as a lowest order Post Newtonian correction (which I will refer to as 0.5PN despite this not respecting the naming convention of referring to powers of $\frac{1}{c^2}$, and I will also at times refer to the Newtonian model as the 0 PN model) and discuss the applicability of this correction in Chapter \ref{chapresults}.\\ \\
The Schwarzschild metric as given by \cite{hartle:2003} is:
\begin{align}
ds &= - \bigg(1 - \frac{2GM}{c^2r} \bigg) (cdt)^2 + \frac{1}{\bigg(1 - \frac{2GM}{c^2r} \bigg) } dr^2 + r^2 ( d\theta^2 + \sin^2(\theta) ) \ d\phi^2
\intertext{if we set $\theta = \frac{\pi}{2}$,  which we can do as the system in symmetrical, we can extract an expression for the effective potential for a particle of mass $m$}
V_{eff}(r) &= -\frac{GMm}{r}  + \frac{L^2}{2 \ r^2 \ m} - \frac{GM\ L^2}{c^2 \ r^3 \ m}
\end{align}
where $L$ is the total angular momentum of the system. Differentiating $V_{eff}(r)$ with respect to $r$ in order to get an expression for the force acting on the particle, noting that when restricted to the plane and for a circular orbit we have $L = m \cdot \ v_{tangential} \cdot r \ \ $   and therefore $ \frac{L^2}{2r^2m} = \frac{v_{tan}^2 \ m }{2 }$
\begin{align}
F &= - \pdv{r} \bigg( - \frac{G M m}{r} + \frac{v_{tan}^2 \ m}{2 } - \frac{ G M  L^2}{c^2 \ m \ r^3} \bigg) \\  \nonumber \\
&= - \frac{GMm}{r^2}\ \big( 1 + \frac{3L^2}{c^2 \ r^2 \ m^2} \big)
\end{align}
the term in the bracket above is the correction I will use for 0.5 PN corrections. In addition, I have used this correction to change the energy and thus the gravitational mass of the binary system as well as the basis for computing the orbital velocity of the binary in all non-Newtonian computations in the rest of this paper.\\ \\
In a binary system where $M$ is the total mass of the two bodies and $\mu$ is the ratio of the mass of the first body to the total mass and $R_1$ and $R_2$ are the distances of the two bodies to the CoM of the system we have:
\begin{align}
M_1 &= M \mu \\
M_2 &= M (1 - \mu) \\
R_1 &= r (1 - \mu) \\
R_2 &= r \mu  \\
L &=  \omega \ ( R_2^2  M_2 + R^2_1  M_1) \nonumber
\intertext{where $\omega$ is the angular velocity of the binary system} 
&= \frac{V_2}{R_2 }\  ( r^2 \mu^2 M (1-\mu) + r^2 (1-\mu )^2 M \mu ) \nonumber \\
  &= \frac{V_2}{r \mu} \  r^2 \ M \ \mu \bigg( \big(1 - \mu) ( \mu + (1-\mu) \big) \bigg) \nonumber \\
    & =    V_2 \ r \ M \ (1 - \mu)      =         V_2 \ r \ M_2                \label{angmom}               
\end{align}
and we note that in the planar restricted system, the angular momentum vectors of the test mass relative to each primary are parallel and therefore in is appropriate to sum them as I have done in \eqref{angmom}.\\ \\ 
Therefore, substituting into the above expression for the force of $M_1$ acting on $M_2$\\
\begin{align}
F &= - \frac{G M_1 M_2}{r^2} \ \Big( 1 + \frac{3\ ( V_2 \ r \ M_2)^2}{c^2 \ r^2 \ M_2^2} \Big)   \\ \nonumber \\
  &= - \frac{GM^2 \ \mu (1-\mu)}{r^2} \ \Big( 1 + \frac{3\ ( V_2 \ r    \ M \  (1-\mu))^2}{c^2 \ r^2 \ M^2 \ (1-\mu)^2} \Big)
\intertext{and this must be equal to the centrifugal force acting $M_2$, where again $V_2$ is the component of the velocity orthogonal to $\vb{r}$}
F &= \frac{V_2^2 \ M_2}{R_2} = \frac{ V_2^2 M\ (1-\mu)}{r \ \mu}
\end{align}
So we have:\\
\begin{align}
\frac{GM^2\  \mu (1-\mu)}{r^2} \ \Big( 1 + \frac{3\ ( V_2 \ r    \ M \  (1-\mu))^2}{c^2 \ r^2 \ M^2 \ (1-\mu)^2} \Big) &= \frac{ V_2^2 M\ (1-\mu)}{r\ \mu}\\ \nonumber\\
\frac{GM \mu }{r} \ \Big( 1 + \frac{3\ ( V_2 \ r     \ M  \ (1-\mu))^2}{c^2 \ r^2 \ M^2 \ (1-\mu)^2} \Big) &= \frac{ V_2^2 }{\mu}\\ \nonumber\\
\frac{GM \mu^2 }{r} \ \Big( 1 + \frac{3\  V_2^2 }{c^2  } \Big) &=V_2^2
\end{align}
which gives
\begin{align}
\frac{GM \mu^2 }{r} &= V_2^2 \ \Big( 1 - 3\ \frac{ G M \mu^2}{r c^2} \Big) \\ \nonumber\\
G M \mu^2  c^2 &= V_2^2 \ ( rc^2 - 3\ GM\mu^2) 
\intertext{so}
V_2^2 &= \frac{GM\mu^2c^2}{r c^2 - 3\ GM\mu^2}
\intertext{and therefore}
\omega_2^2 = \omega_1^2 = \frac{V_2^2}{R_2^2} =   \frac{V_2^2}{(r \mu)^2}& = \frac{1}{r^2} \ \frac{GM c^2}{rc^2 - 3\ G M \mu^2} \\ \nonumber\\
\omega_2 = \omega_1& = \frac{1}{r} \ \sqrt{\frac{GM c^2}{r c^2 - 3\  G M \mu^2}}
\end{align}
From this expression I construct the mass equivalent of the primaries' kinetic energy and use this to restate the effective mass of the binary system\\
\begin{align}
M  &= M_{Cl} \ \Big( 1 +  \frac{0.5}{ c^2} \ \big( \mu \ V_1^2                     + (1 - \mu) \ V_2^2                  \big) \Big) \\ 
	   &= M_{Cl} \ \Big( 1 +  \frac{0.5}{ c^2} \ \big( \mu \ ( \omega_1 \ R_1)^2 + (1 - \mu) \ (\omega_2 R_2)^2 \big) \Big)\\ 
            &= M_{Cl}  \ \Big( 1 +  \frac{0.5}{ c^2} \ \omega_1^2 \ r^2 \ \big( \mu \  (1-\mu)^2 + (1 - \mu) \ (\mu)^2  \big) \Big) \label{schwar}
\end{align} \\ \\
Clearly, this approximation become less and less valid the more the test mass can resolve the positions of the primaries. \\

\section{Post Newtonian Approximations} \label{PNapprox}
As discussed above, Post Newtonian approximations are named / numbered by their expansions in power of $\frac{1}{c^2}$. 
\subsection{ 1 PN Approximation}
For the 1 PN, 2 PN and 2.5 PN  approximations, I use the expression given in Boekolt, Moerman and Zwart 2021 \cite{Boekholt_2021}, from which we have an expression for the acceleration of body $\boldsymbol{i}$ under the influence of bodies $\boldsymbol{j}$ and $\boldsymbol{k}$. Starting with the 1 PN terms only, we have:
\begin{align}
\vb{a}_i = &- \sum_{j \neq i} \frac{ m_j \vb{n}_{ij}}{r_{ij}^2} \qquad \qquad \qquad \text{(the standard Newtonian term)} \label{1PN} \\ \nonumber 
&+ \frac{1}{c^2} \sum_{j \neq i} \frac{ m_j \vb{n}_{ij}}{r_{ij}^2}  \Bigg[ 4\frac{m_j}{r_{ij}} + 5\frac{m_i}{r_{ij}} + \sum_{k \neq i,j} \bigg( \frac{m_j}{r_{jk}} + 4 \frac{m_k}{r_{ik}} - \frac{m_k r_{ij}}{2r_{jk}^2} (\vb{n}_{ij}\cdot \vb{n}_{jk}) \bigg) - v_i^2 \\ \nonumber
&+ 4 \vb{v}_i \cdot \vb{v}_j - 2 \vb{v}_j^2 + \frac{3}{2} ( \vb{v}_j \cdot \vb{n}_{ij} )^2 \Bigg] \\ \nonumber
& - \frac{7}{2c^2} \sum_{ j \neq i} \frac{m_j}{r_{ij}} \sum_{k \neq i,j} \frac{m_k \vb{n}_{jk}}{r_{jk}^2}\\ \nonumber
&+ \frac{1}{c^2} \sum_{j \neq i} \frac{m_j}{r_{ij}^2} \vb{n}_{ij} \cdot (4\vb{v}_i -3\vb{v}_j)(\vb{v}_i - \vb{v}_j)  
\end{align}
where we have
\begin{align}
r_{ij} &= \abs{ \vb{r}_i -  \vb{r}_j } \nonumber \\
\vb{n}_{ij} & = \frac{ \vb{r}_i -  \vb{r}_j }{r_{ij}} \nonumber \\
\vb{v}_{ij} &= \vb{v}_{i} - \vb{v}_{j}
\end{align}
and $\vb{r}$, $\vb{v}$, and $\vb{a}$ represent the Cartesian position, velocity and accelerations vectors respectively. In this expression the authors had explicitly set $G$ equal to 1 as is common practice in many GR texts.\\ \\
Here we recognise the Newtonian term followed by four lines of Post Newtonian corrections. We note that in the second and fourth lines of the corrections we have terms in the square of the velocities of the primary masses: $- 2 v_j^2 + \frac{3}{2}(\vb{v}_j \cdot \vb{n}_{ij} )^2$ and $(4v_i -3v_j)(v_i-v_j)$, so when the primaries are close together and the test mass is far from the primaries, the corrections are likely to be dominated by these terms and hence the corrections will simplify to:
\begin{align}
+2.5 \ \frac{v_j^2}{c^2}
\end{align}
multiplied by the standard Newtonian acceleration term. Crucially I note that this term is repulsive and that as the orbital velocities of the primary bodies are inversely proportional to the square root of their separation, that for binary systems separated by a distance of order 100 times the Schwarzschild radius or less these terms become important, and indeed very large as the separation of the primaries decreases. \\ \\
The other terms which may become large are in the first and third lines of the corrections and these are products of the primary masses and are only divided by one power of the distance to the test mass and two powers of the distance between the primaries (which can be very small) and all divided by $c^2$:  
\begin{align}
\frac{1}{c^2} \frac{m_j m_k}{r_{ij}r_{jk}^2} 
\end{align}
In the simple case where I have two equal mass primaries (the scenario in which this term is largest) and the ratio of the distance from the test mass to the CoM of the primaries \emph{vs} the distance between the primaries: $\frac{r_{ij}}{r_{jk}}$, is of order $10^6$ we observe that:
\begin{align}
\frac{1}{c^2} \frac{m_j m_k}{r_{ij}r_{jk}^2}  = \frac{1}{4c^2} \frac{M^2}{r_{ij}^3} 10^{12} = \frac{10^{12} \ M}{4c^2 \ r_{ij} } \ \ a_{Newton}
\end{align}
which reinstating units gives us
\begin{align}
 = \frac{10^{12} \ G M}{4c^2 \ r_{ij} } \ \ a_{Newton}
\end{align}
and, as I am discussing situations where $r_{ij}$ is very large compared to the Schwarzschild radius, this term is small.\\ \\
\subsection{ 2 PN and 2.5 PN Approximations}
For the 2 PN approximation we add further terms to the acceleration, this time with coefficient $ \frac{1}{c^4}$ and $ \frac{1}{c^5}$ for the 2.5 PN terms.
\begin{align}
\vb{a}_i =  \vb{a}_{i (for 1 PN)} \label{2PN}  \nonumber \\
        + \frac{1}{c^4} &\sum_{j \neq i} \frac{m_j \vb{n}_{ij}}{r^2_{ij}} \bigg[ -2 \vb{v}_j^4 + 4\vb{v}^2_j (\vb{v}_i \cdot \vb{v}_j) -2 (\vb{v}_i \cdot \vb{v}_j)^2 + \frac{3}{2} \vb{v}_i^2(\vb{n}_{ij} \cdot \vb{v}_j )^2 + \frac{9}{2} \vb{v}^2_j (\vb{n}_{ij} \cdot \vb{v}_j )^2  \nonumber \\
        & -6  (\vb{v}_i \cdot \vb{v}_j) (\vb{n}_{ij} \cdot \vb{v}_j )^2  - \frac{15}{8} (\vb{n}_{ij} \cdot \vb{v}_j )^4 - \frac{57}{4} \frac{m_i^2}{r^2_{ij}} - 9 \frac{m_j^2}{r^2_{ij}} -\frac{69}{2} \frac{m_im_j}{r^2_{ij}}  \nonumber \\
        & +\frac{m_i}{r_{ij}} \bigg(-\frac{15}{4} \vb{v}_i^2 + \frac{5}{4} \vb{v}_j^2 -\frac{5}{2} (\vb{v}_i \cdot \vb{v}_j)  + \frac{39}{2} (\vb{n}_{ij} \cdot \vb{v}_i)^2 - 39(\vb{n}_{ij} \cdot \vb{v}_i )(\vb{n}_{ij} \cdot \vb{v}_j ) \nonumber \\
        &+\frac{17}{2}(\vb{n}_{ij} \cdot \vb{v}_j )^2 \bigg)  \nonumber \\
        & +\frac{m_j}{r_{ij}} \big (4 \vb{v}_j^2 - 8 (\vb{v}_i \cdot \vb{v}_j) + 2(\vb{n}_{ij} \cdot \vb{v}_i)^2 -4 (\vb{n}_{ij} \cdot \vb{v}_i)(\vb{n}_{ij} \cdot \vb{v}_j) -6(\vb{n}_{ij} \cdot \vb{v}_j)^2 \big) \bigg]  \nonumber \\
        + \frac{1}{c^4} &\sum_{j \neq i} \frac{m_j \vb{v}_{ij}}{r^2_{ij}} \bigg[ \frac{m_i}{r_{ij}} \bigg( \frac{55}{4} (\vb{n}_{ij} \cdot \vb{v}_j) -\frac{63}{4}(\vb{n}_{ij} \cdot \vb{v}_i) \bigg) -2 \frac{m_j}{r_{ij}}((\vb{n}_{ij} \cdot \vb{v}_i)+(\vb{n}_{ij} \cdot \vb{v}_j)) \nonumber \\
        &+ \vb{v}_i^2 (\vb{n}_{ij} \cdot \vb{v}_j) + 4 \vb{v}_j^2(\vb{n}_{ij} \cdot \vb{v}_i) - 5 \vb{v}_j^2 (\vb{n}_{ij} \cdot \vb{v}_j) - 4(\vb{v}_i \cdot \vb{v}_j)(\vb{n}_{ij} \cdot \vb{v}_{ij}) \nonumber \\
        &-6 (\vb{n}_{ij} \cdot \vb{v}_i)(\vb{n}_{ij} \cdot \vb{v}_j)^2 + \frac{9}{2} (\vb{n}_{ij} \cdot \vb{v}_j)^3 \bigg] \\
\intertext{ and we then add the 2.5 PN terms}
       + \frac{1}{c^5} &\sum_{j \neq i} \frac{4 m_i m_j}{5r^3_{ij}} \bigg[ \vb{n}_{ij}    (\vb{n}_{ij} \cdot \vb{v}_{ij})    \bigg(\frac{52}{3} \frac{m_j}{r_{ij}} - 6\frac{m_i}{r_{ij}} +3 \vb{v}^2_{ij} \bigg) + \vb{v}_{ij} \bigg( 2  \frac{m_i}{r_{ij}} - 8\frac{m_j}{r_{ij}} - \vb{v}^2_{ij} \bigg) \bigg] \nonumber \\
        &+ \mathcal{O} \bigg(\frac{1}{c^6} \bigg)
\end{align}
and again we have
\begin{align}
r_{ij} &= \abs{ \vb{r}_i -  \vb{r}_j } \label{2.5PN}\nonumber \\
\vb{n}_{ij} & = \frac{ \vb{r}_i -  \vb{r}_j }{r_{ij}} \nonumber \\
\vb{v}_{ij} &= \vb{v}_{i} - \vb{v}_{j}
\end{align}
and $\vb{r}$, $\vb{v}$, and $\vb{a}$ represent the Cartesian position, velocity and accelerations vectors respectively.
We observe that the 2.5 PN terms will be very small since the 2 PN terms contain expressions in $\vb{v}_j^4$ times $\frac{1}{c^4}$ times $\frac{1}{r_{ij}^2}$ whereas the 2.5 PN terms only contain expressions in $\vb{v}_{ij}^2$ times $\frac{1}{c^5}$ times $\frac{1}{r_{ij}^3}$ and as we expect $\abs{\vb{v}_{j}} >= \abs{\vb{v}_{ij}}$  then as an order of magnitude approximation, we would expect the 2.5 PN terms to be of order $\frac{1}{\vb{v}_j^2 \cdot c  \cdot  r_{ij}}$ smaller than the 2 PN terms. So if $\vb{v}_{j}$ is within a few orders of magnitude of $c$ and $r_{ij}$ is within a few order of magnitude of the Schwarzschild radii, then 2 PN terms could be 20 to 30 orders of magnitude larger than 2.5 PN terms and I will therefore observe virtually no effects of 2.5 PN relative to 2 PN in the simulations except in extremely long runs, not least because these simulations have been run with machine precision of circa 17 significant figures.

\chapter{Physical Model: Numerical Implementation}

\section{Practical Hurdles in Exploring the 3 Body Problem}	
As there are no analytical solutions to the differential equations governing the system, the investigator must construct an `Integration' scheme in order to follow the evolution of the system. Integration schemes are stepwise procedures whereby, knowing the initial conditions and the forces at some time $t$, it is possible to generate a new set of positions and velocities for all the bodies of the system after some short time interval $\delta t$. \\ \\    
The first challenge is to develop the integrator and the simplest and most intuitive procedure is simply to apply the velocities to the respective bodies for the chosen time interval $\delta t$ generating new positions and to apply the accelerations to the velocities for the same interval $\delta t$ generating new velocities. At this point the forces are recomputed from the differential equations, and the procedure is repeated for the next time interval $\delta t$ using the new positions, velocities and accelerations. This method was first proposed by Euler (and is known as Euler's method \cite{Euler1768}).\\ \\
Intuitively one might suppose that by choosing small enough time intervals this process would be quite accurate. However, simple experimentation shows that in fact the size of time interval required for a given required accuracy is very small indeed, leading to a very high computational burden if the investigator wishes to follow the systems for extended periods (see Appendix \ref{accuracy} for further discussion on this point).\\ \\
A simple way to test the accuracy of the integration scheme is to compare its predictions in the weak field limit with the predictions of an analytical Newtonian model for a stable circular orbit. \\ \\
I explore the behaviour of a binary system composed of two heavy bodies orbiting each other with some distance $R$ between their centres of mass, and a third test body (of negligible (unit) mass) orbiting the binary system at a much greater distance, say $10^6 \cdot R$. From the test body's perspective, the two bodies in the binary system will be indistinguishable from a single (larger) body, whose mass is the sum of the masses of the two bodies making up the binary, in other words the system is reduced to a two body problem for which we can use Newton's expression for the velocity of a stable circular orbit:
\begin{align}
V = \sqrt{ \frac{G \cdot (M_1 + M_2)}{10^6 \cdot R} }
\end{align}
where $G$ is Newton's gravitational constant, $M_1$ and $M_2$ are the masses of the two bodies making up the binary and $10^6 \cdot R$ is the distance from the test body to the centre of mass (CoM) of the binary system (all expressed in units of kg, m, s).\\ \\
By setting up an integration scheme with the binary system's CoM located at the origin and giving the test body an initial position of $( 10^6 \cdot R, 0, 0)$ and initial velocity of $(0, V, 0)$, one would expect the system to describe circular orbits of radius $10^6 \cdot R$ with period $T$ given by:
\begin{align}
T = \sqrt{  \frac{4 \cdot \pi \cdot (10^6 \cdot R)^3}{G \cdot (M_1 + M_2)}}
\end{align}
There are many tests that could be performed to test the precision of the Integration scheme, but one of the simplest is to measure the variation in the orbital radius over one orbit. In other words, define an accuracy measure as:
\begin{align}
accuracy  = \frac{R_1 - R_0}{ R_0}
\end{align}
where $R_0$ and $R_1$ are the distances of the test body to the origin at time zero and after one orbit respectively and indeed at any point in that orbit. \\ \\
I choose a desired target value for $accuracy$ and then vary the time interval $\delta t$ until I find a value of $\delta t$ such that $accuracy$ is less than the desired target. The Euler method of integration turns out to be very inefficient and whilst Heun's Method \cite{Heun} is a more efficient integration scheme again the accuracy of this method is insufficient for precision projects.\\ \\
Many investigators now use Runge–Kutta methods with an adaptive step such as the Dormand Prince method \cite{DORMAND1980}. As with the Euler and Heun methods described above, the Runge-Kutta and Dormand Prince tools only provide approximations for first order differential equations, so it is necessary to split the second order equations (for force or acceleration) into systems of two first order equations: velocities in terms of accelerations and positions in terms of velocities and in principle this needs to be done for each body and each spatial dimension giving us a total of eighteen equations.  Please see Appendix A for a discussion of the accuracies achieved used the Euler and Dormand Prince methods as well as their CPU run times. \\ \\
As discussed above, I will dramatically simplify the problem by assuming that the 3\textsuperscript{rd} body which I call the `test mass' is very light (in fact I assume it has unit mass and therefore does not travel at the speed of light). This means that the binary system is not affected by the presence and behaviour of the test mass and can therefore be treated as a Two Body system (although in General Relativity the two body system is still complex when the two bodies are in close proximity). I can therefore make increasingly sophisticated levels of approximation for the dynamics (positions and velocities) of the binary system and then use these approximations to explore the behaviour of the test mass in the spacetime of the binary system (as modelled).
One important choice which needs to be made in the construction of the simulation is the choice of frame of reference. In all that follows I have chosen a fixed frame of reference with the CoM of the binary system at the origin and the orbits of the Primaries constrained to the $X - Y $ plane (the Primaries are in practice black holes, and therefore when studying a binary system of two equal mass black holes, each of one Solar mass, I have set the capture / collision distance at 3,000 metres, i.e. the Schwarzschild radius of a Solar Mass body). Many authors choose frames of reference which rotate with the binary system but as one of the effects which I expect to be significant is the transfer of angular momentum from the binaries to the test mass, I chose a fixed frame.\\ 
	
\section{Initial Choices of Model and Units}	\label{initchoice}
\begin{figure}[!ht]
    \subfigure[]{\includegraphics[width=0.495\textwidth]{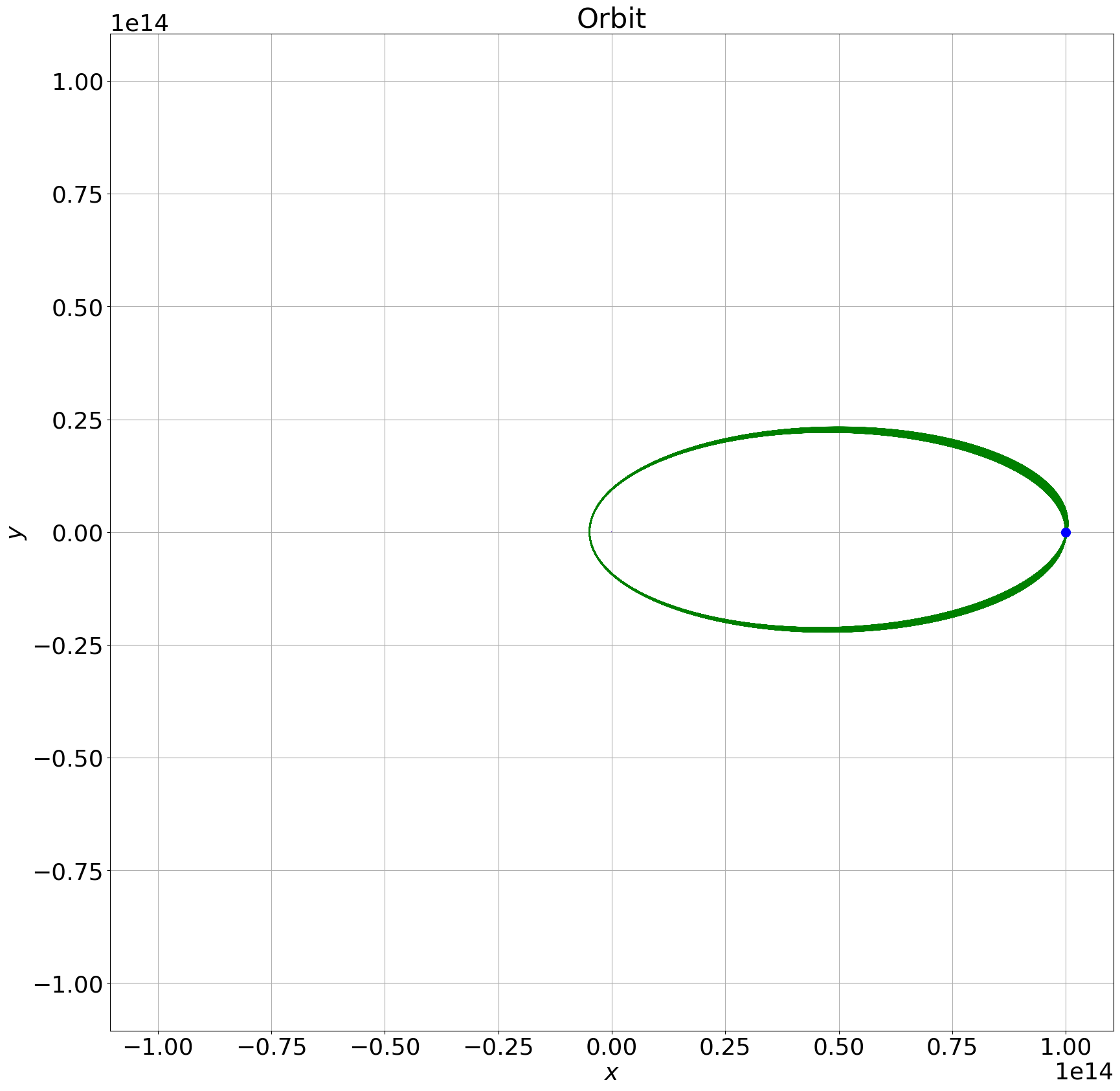}}
    \subfigure[]{\includegraphics[width=0.48\textwidth]{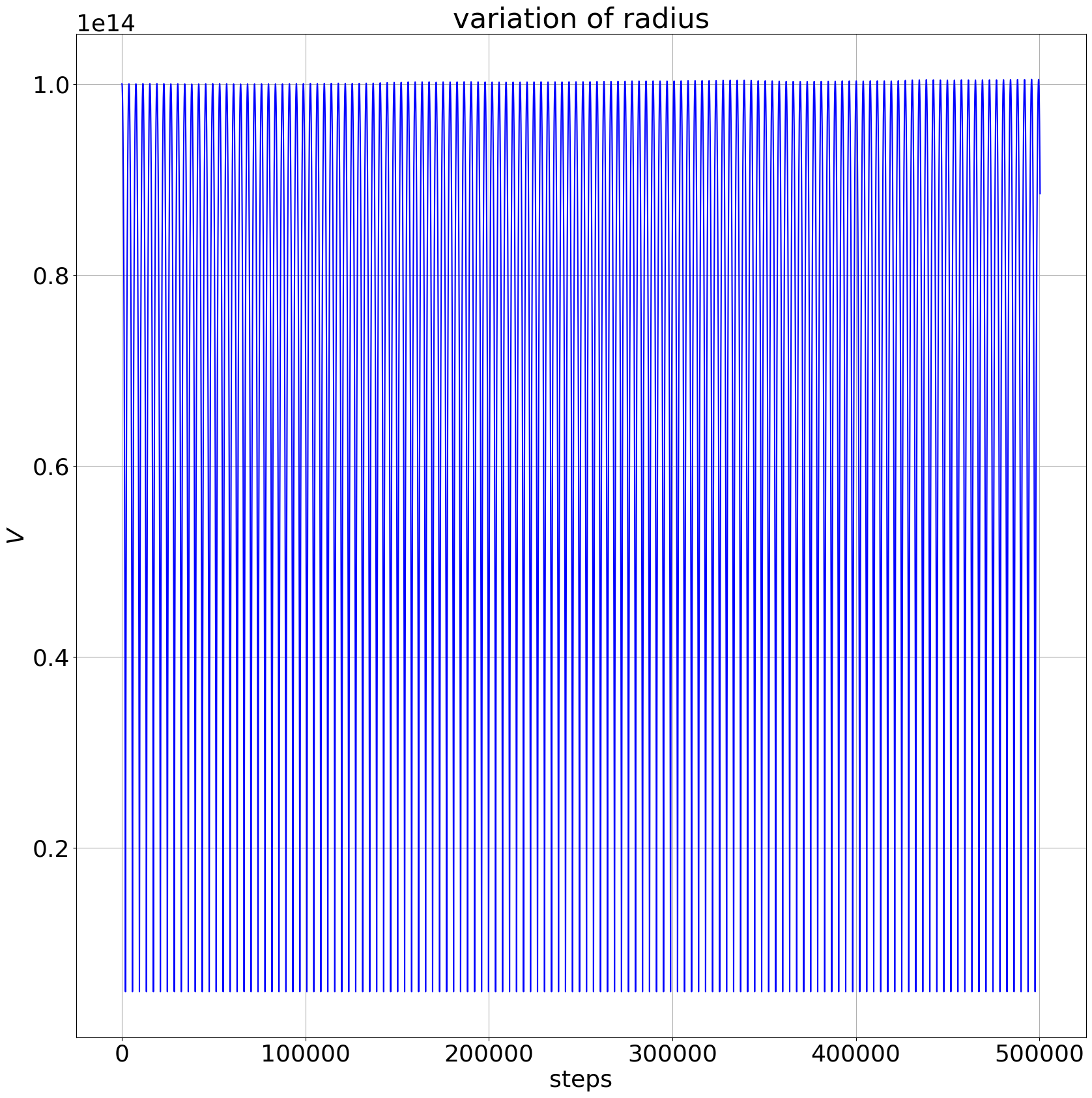}}
 \caption{Initial Velocity: 500$ms^{-1}$ (a) Position of the Test Mass in the X-Y plane (b) Variation of Distance from Test Mass to Binary CoM with time\\ Separation of primaries:  $1.1 \cdot 10^{11} m$, Test Mass at X =  $10^{14} m$, 500,000 time steps}
 \label{elip-500}
\end{figure}

\begin{figure}[!ht]
\includegraphics[width=145mm]{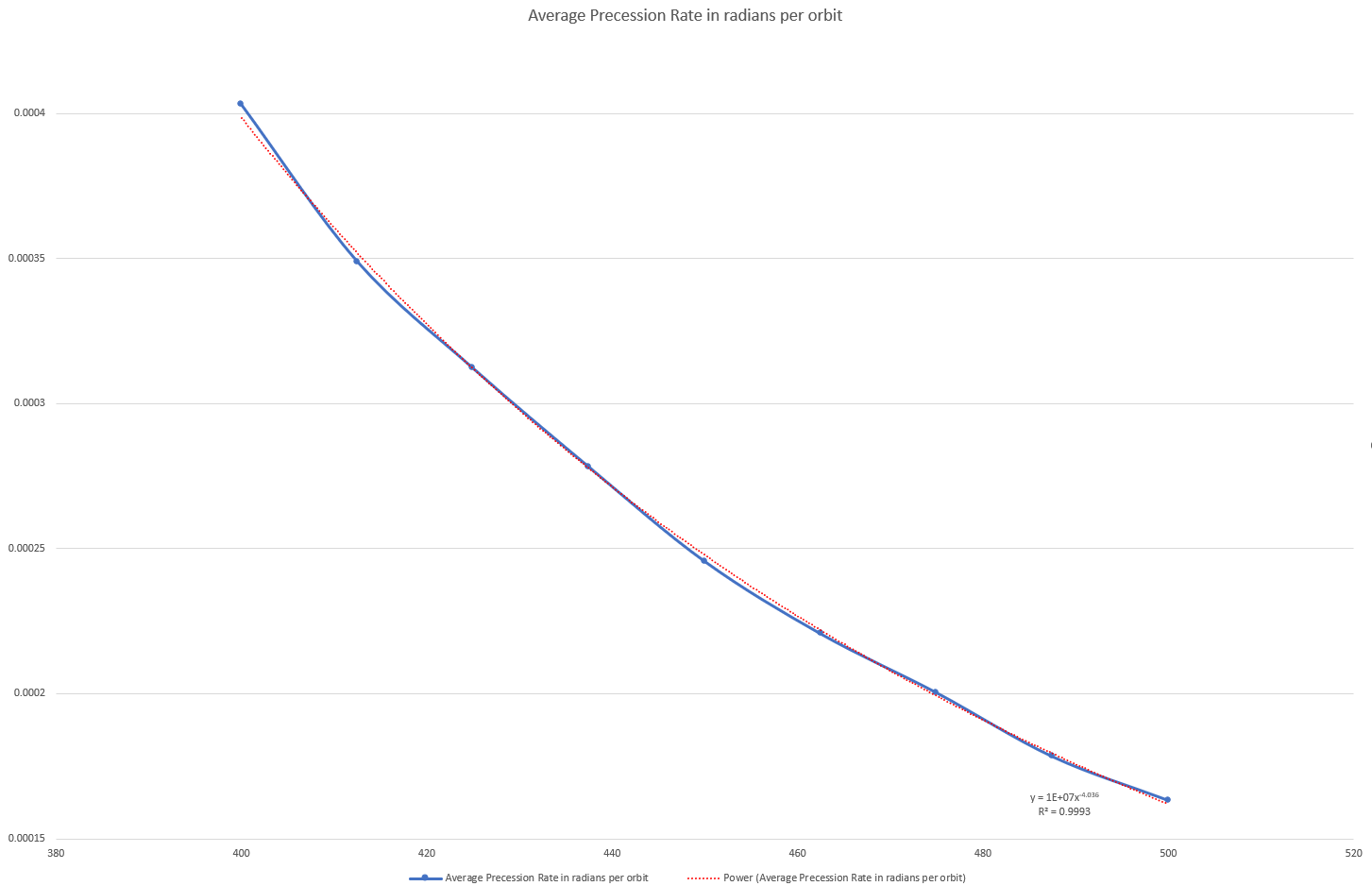}
\caption{Rate of Precession of Test Mass Orbit with Change of Initial Velocity and Regression Line in Red}
\label{prec-rate-fig}
\end{figure}

\begin{figure}[!ht]
\subfigure[]{\includegraphics[width=0.49\textwidth]{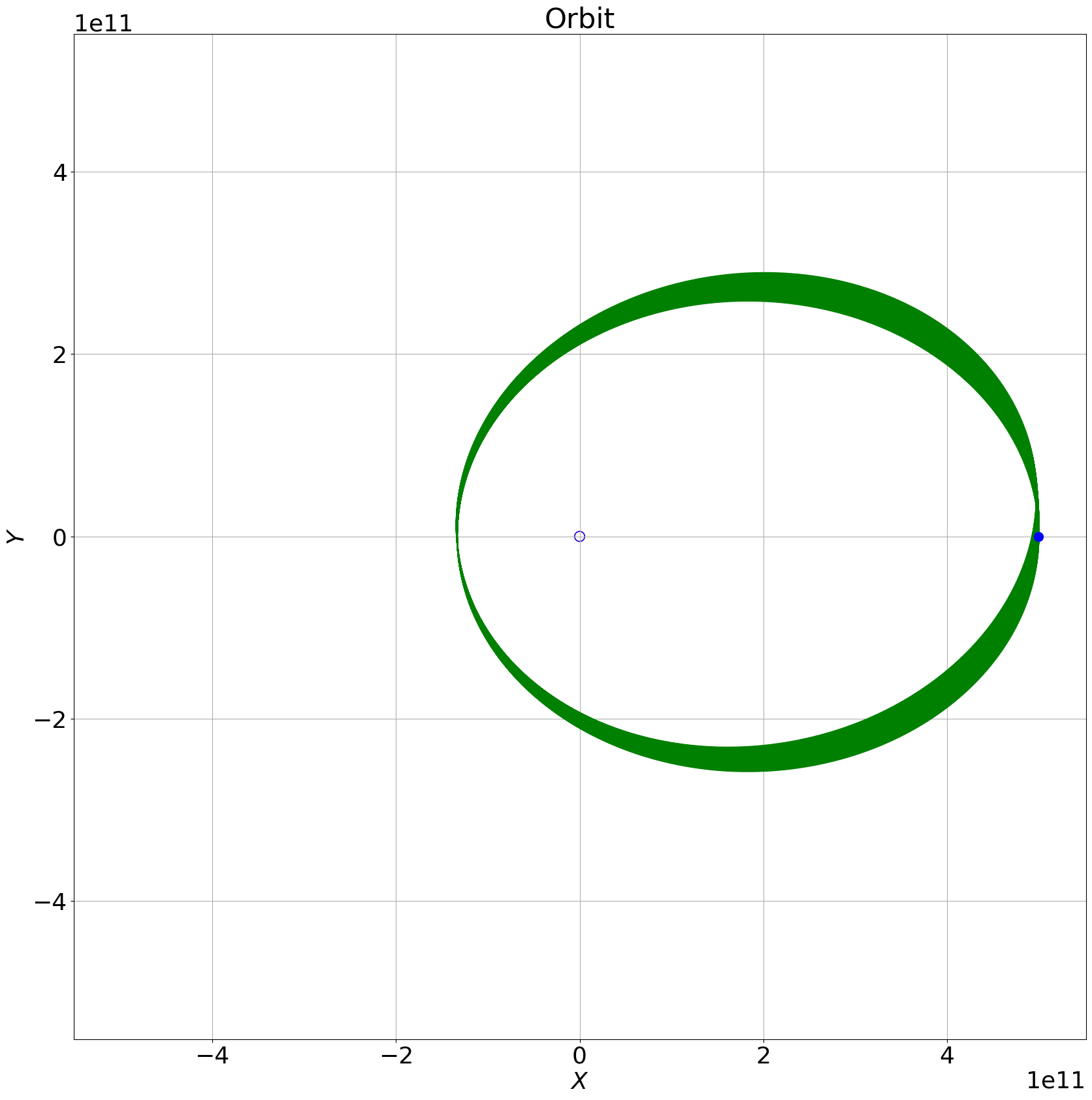}}
\subfigure[]{\includegraphics[width=0.49\textwidth]{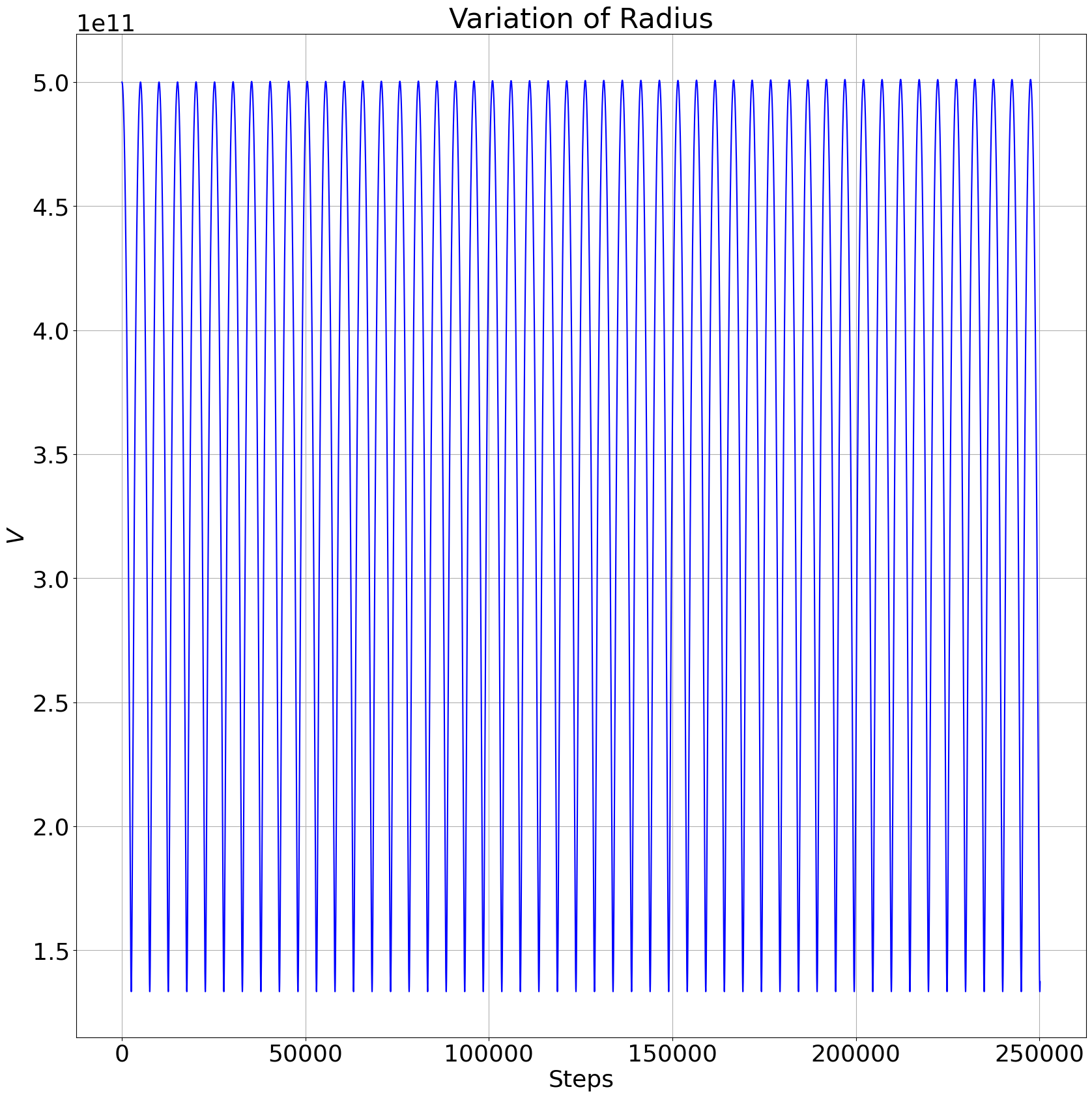}}  
\caption{Velocity 15,000 $ms^{-1}$ (a) Position Plot  (b) Distance to Origin \\ Separation of Binary: $1.1 \cdot 10^{10}m$, Test Mass at X = $5 \cdot 10^{11}$, 25 Orbits}
\label{cm-15000} 
\subfigure[]{\includegraphics[width=0.50\textwidth]{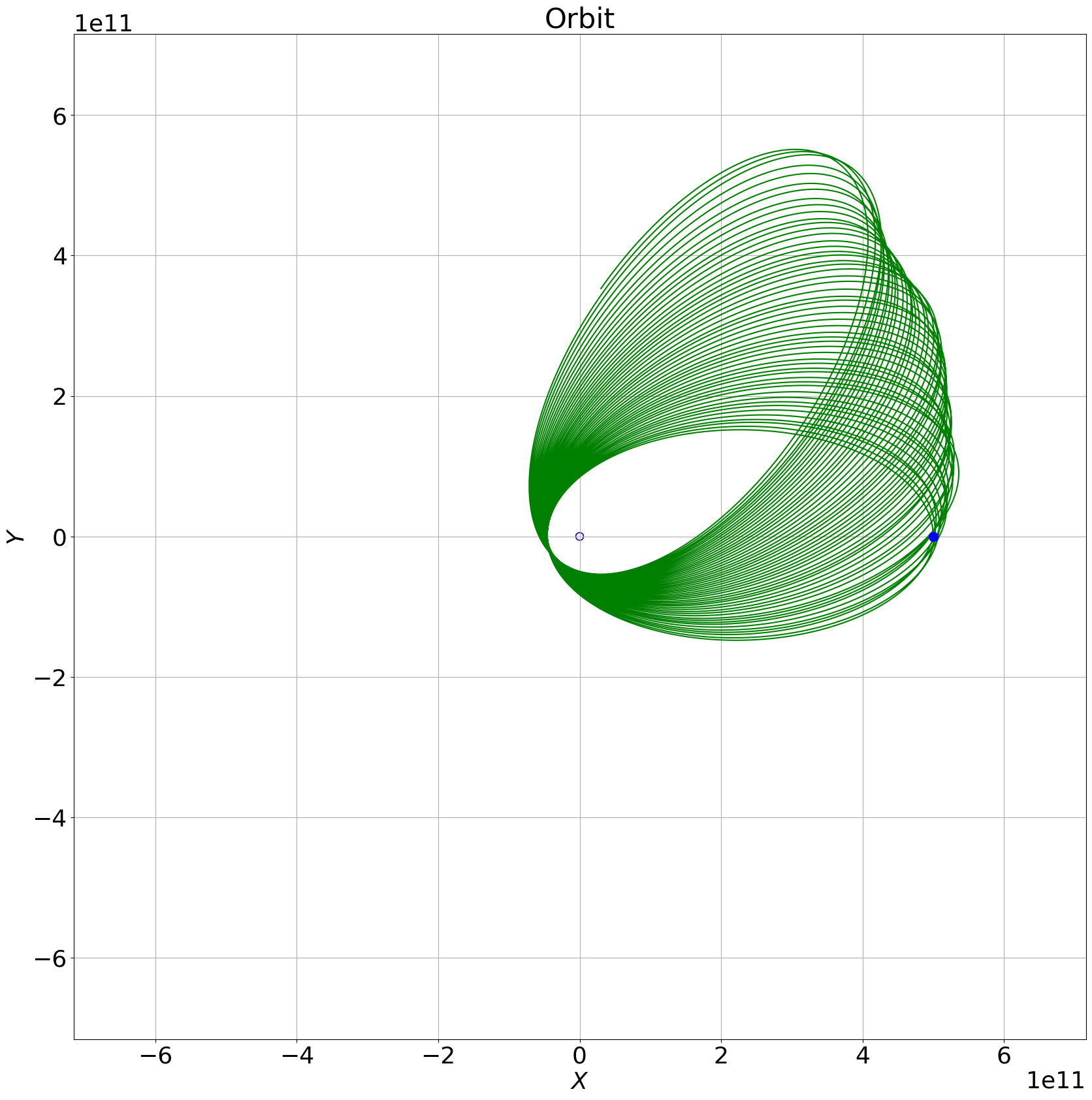}}
\subfigure[]{\includegraphics[width=0.49\textwidth]{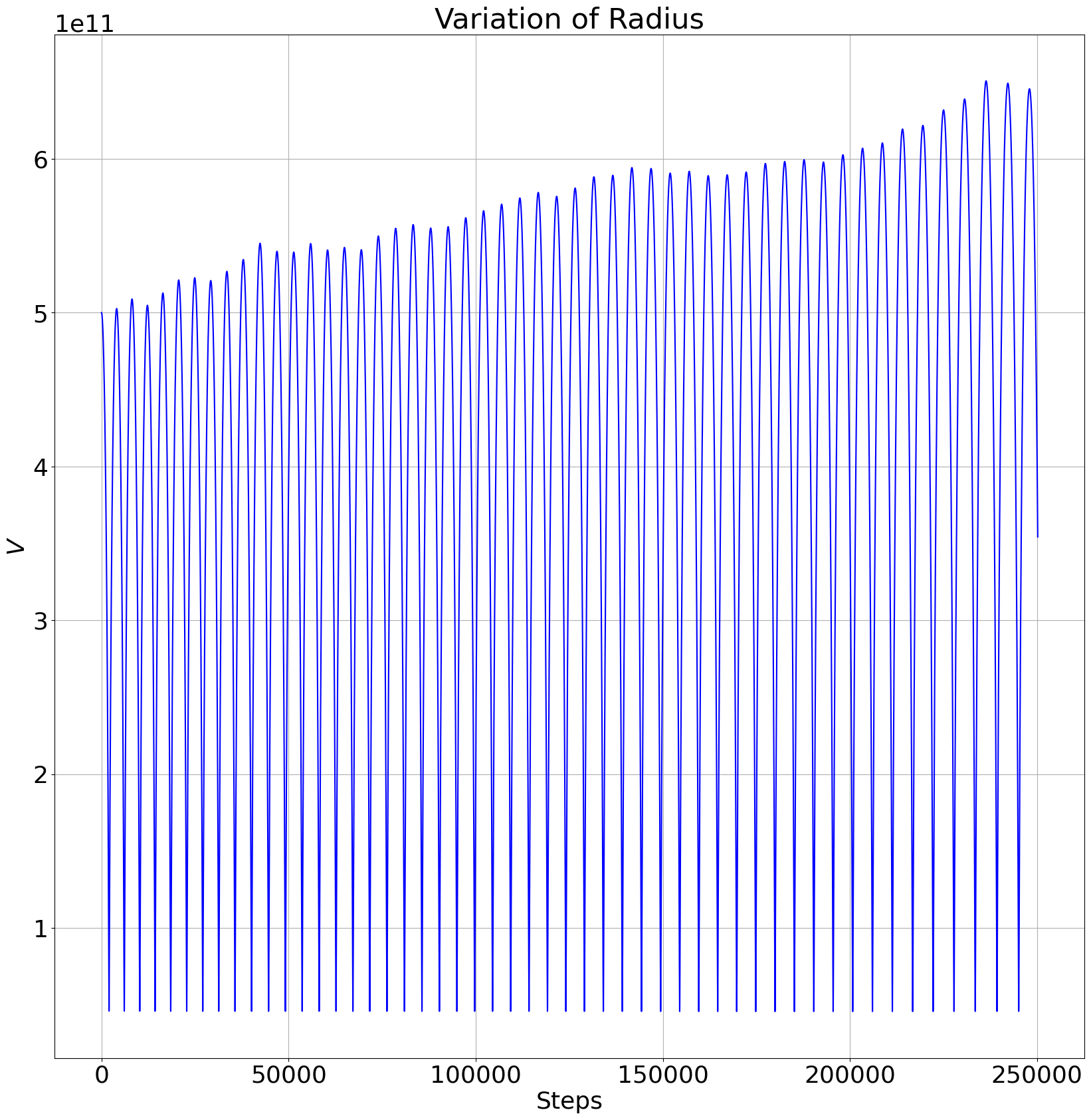}}
\caption{Velocity  9,500 $ms^{-1}$  (a) Position Plot  (b) Distance to Origin}
\label{cm-9500}
\end{figure}

\begin{figure} 
    \subfigure[]{\includegraphics[width=0.50\textwidth]{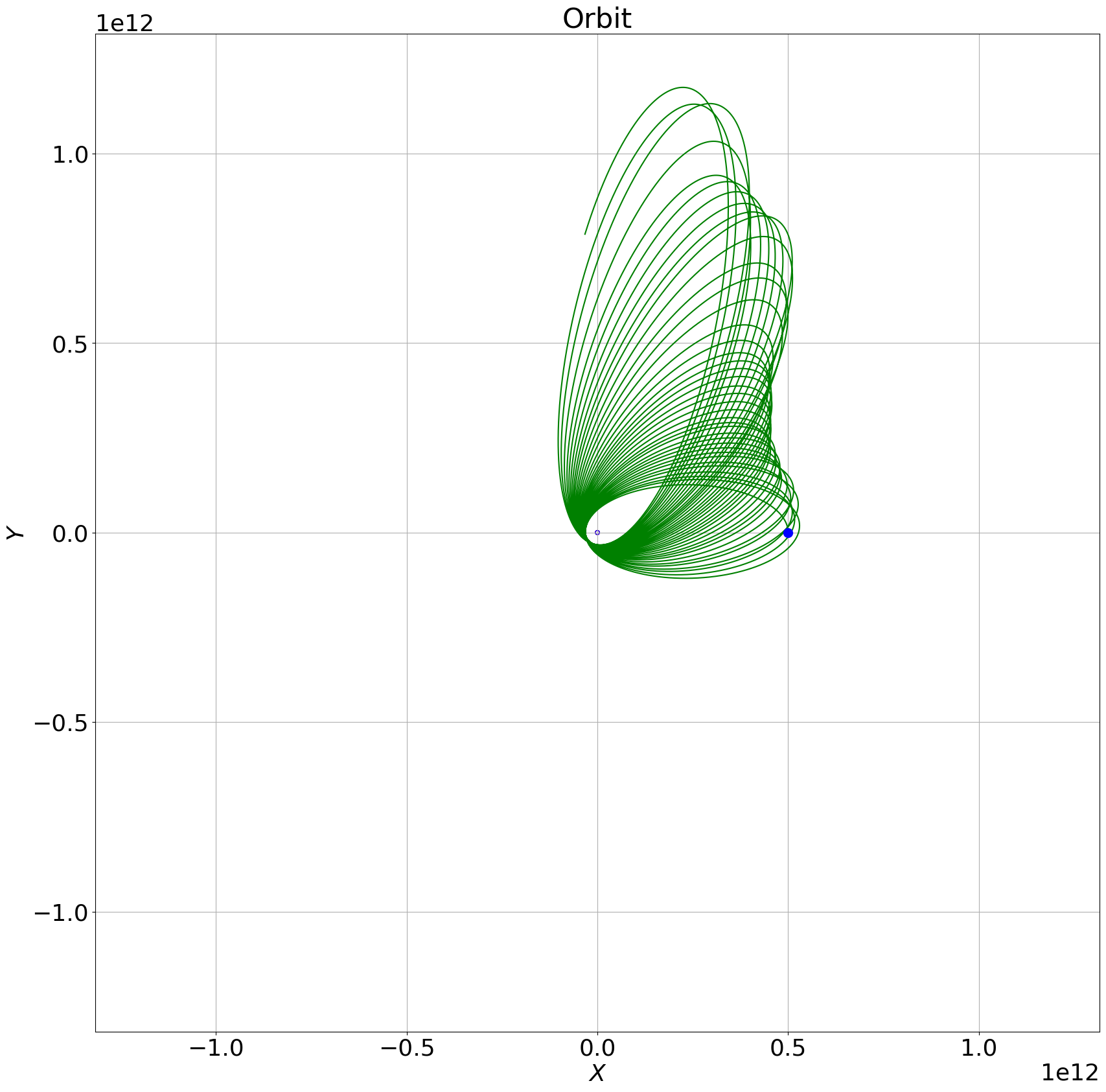}}
    \subfigure[]{\includegraphics[width=0.49\textwidth]{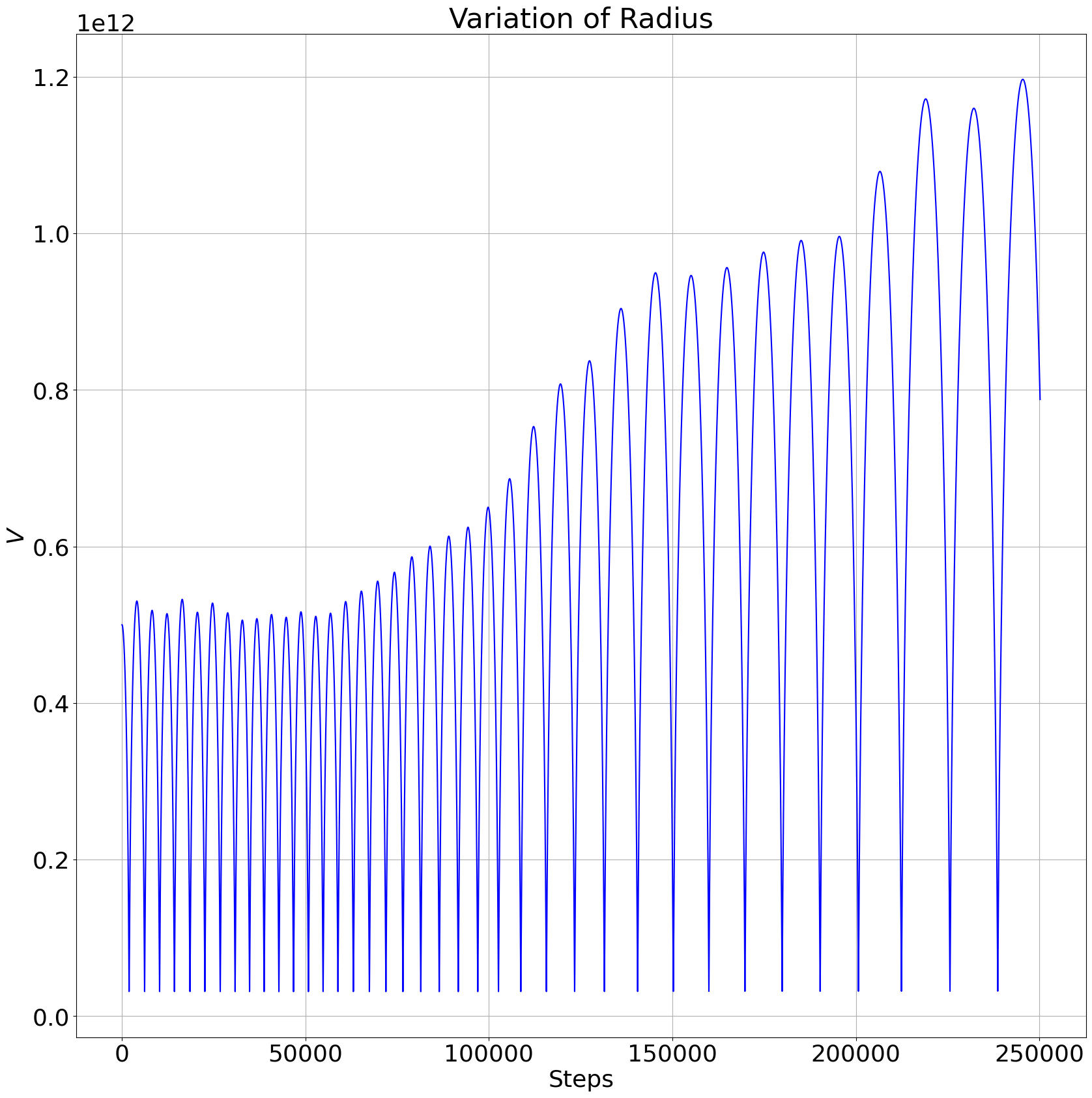}}
\caption{Velocity 8,000 $ms^{-1}$ (a) Position Plot  (b) Distance to Origin} 
\label{cm-8000}
\subfigure[]{\includegraphics[width=0.50\textwidth]{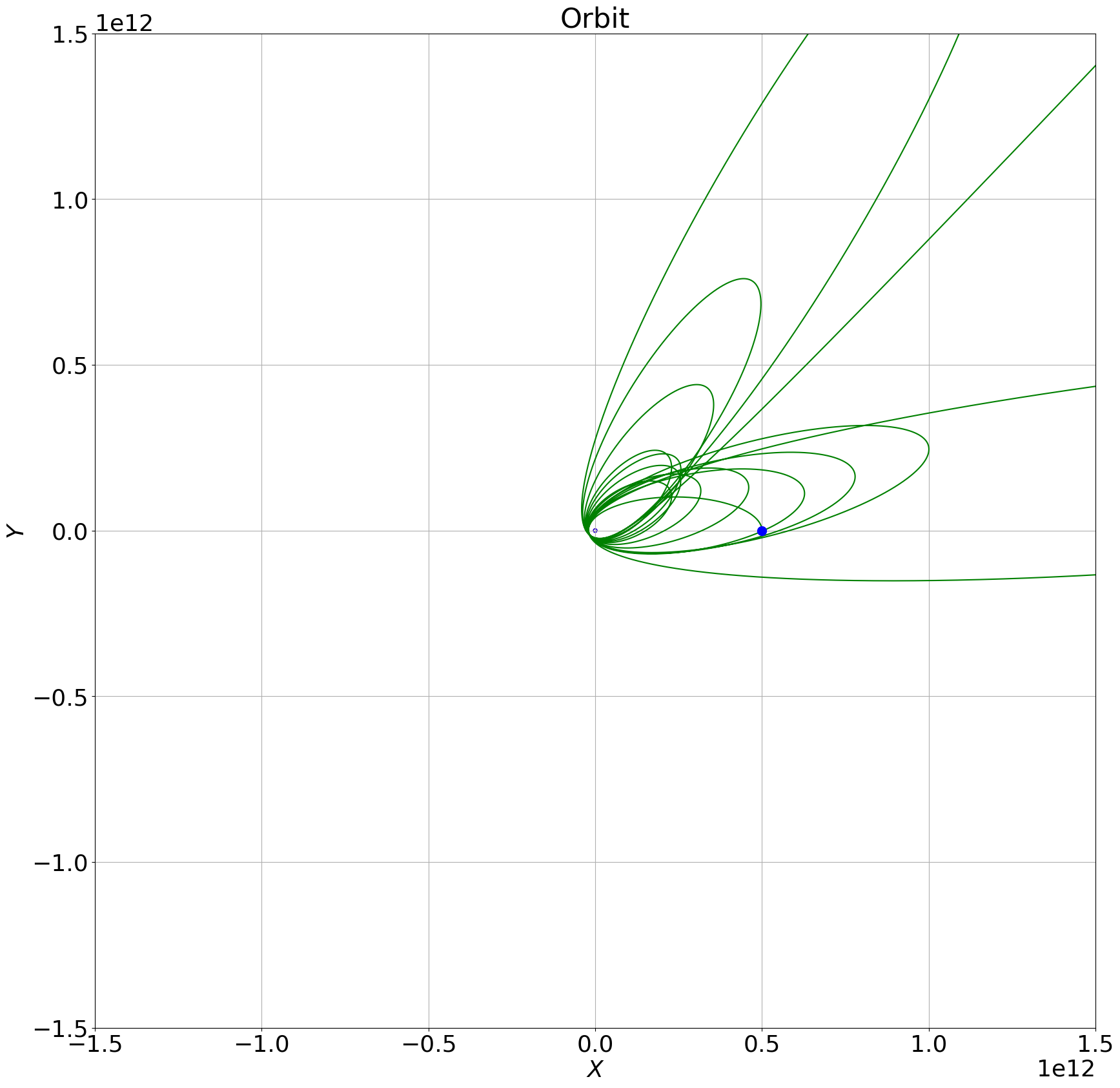}}
\subfigure[]{\includegraphics[width=0.49\textwidth]{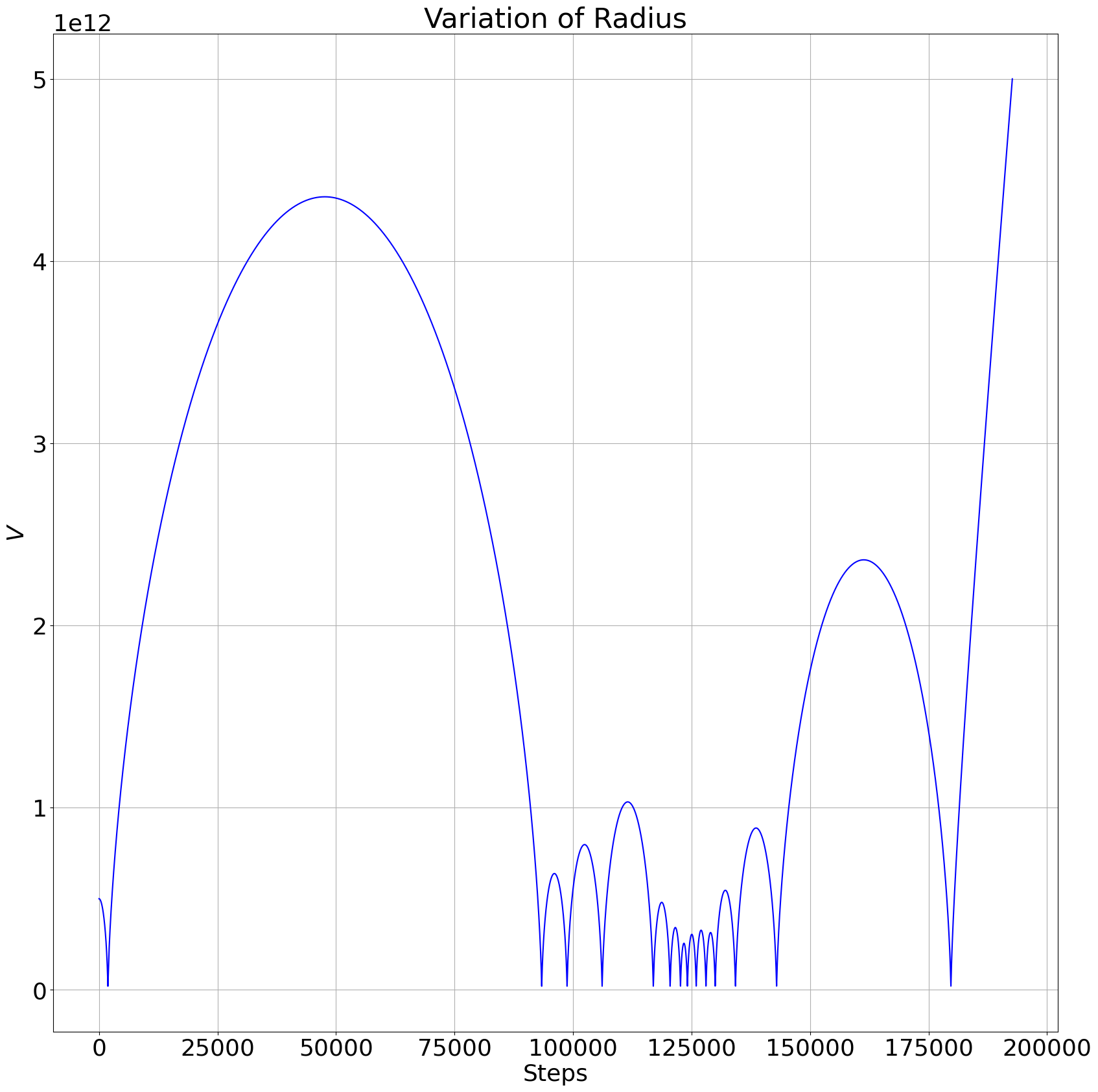}}
\caption{Velocity 6,500 $ms^{-1}$ (a) Position Plot  (b) Distance to Origin} 
\label{cm-6500}
\end{figure}

\begin{figure}[!ht]
\includegraphics[width=145mm]{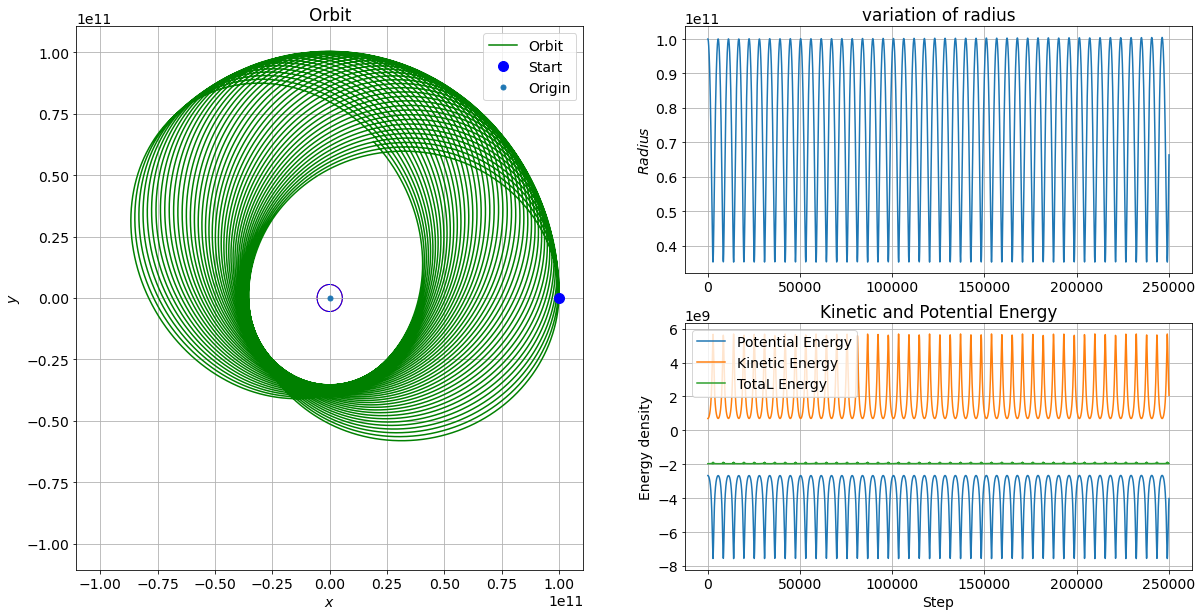}
\caption{Trajectory, Variation of Radius and Variation of Energy}
\label{CM with en}
\end{figure}

\begin{figure}[!ht]
\subfigure[]{\includegraphics[width=0.485\textwidth]{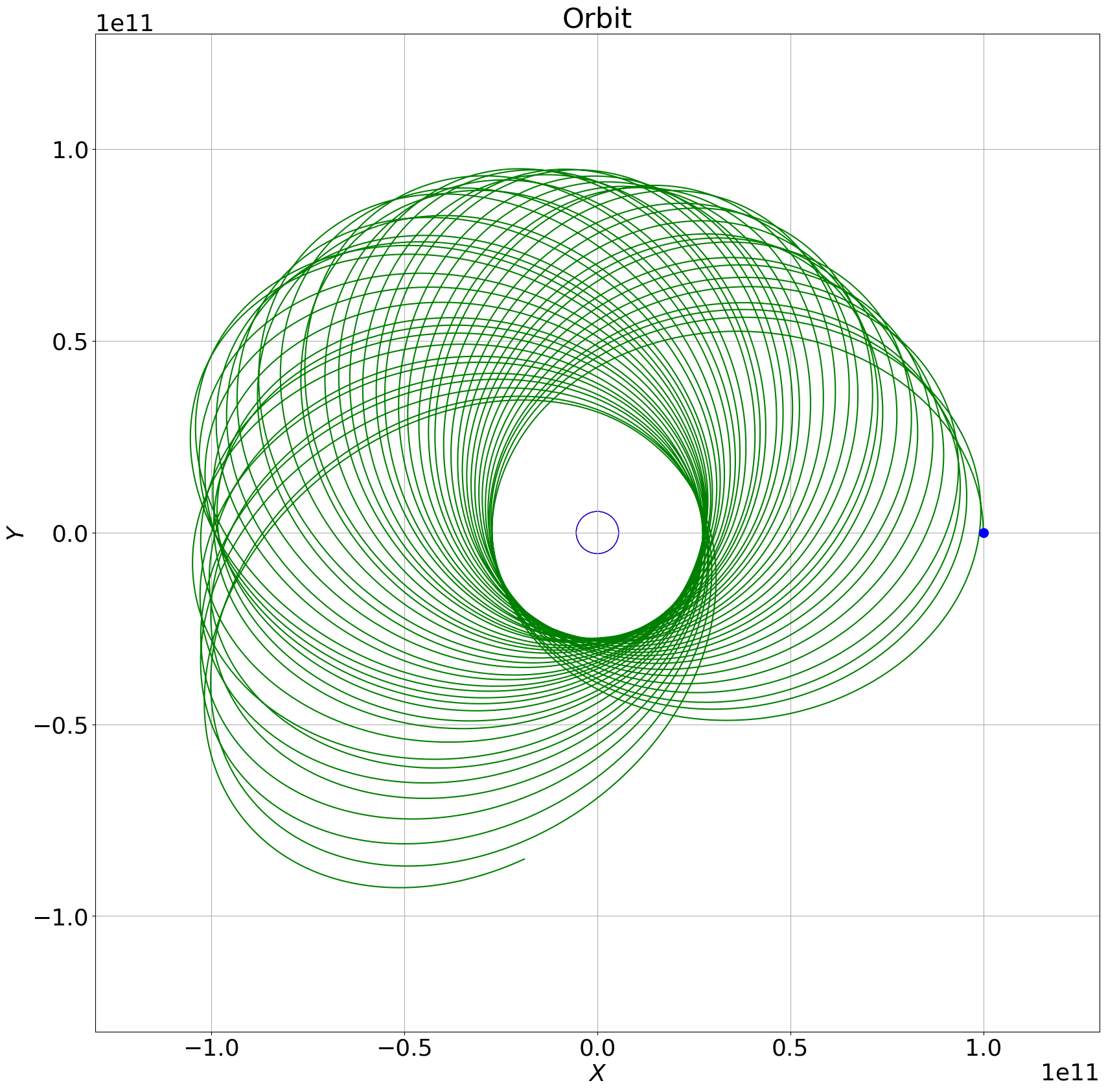}}
\subfigure[]{\includegraphics[width=0.50\textwidth]{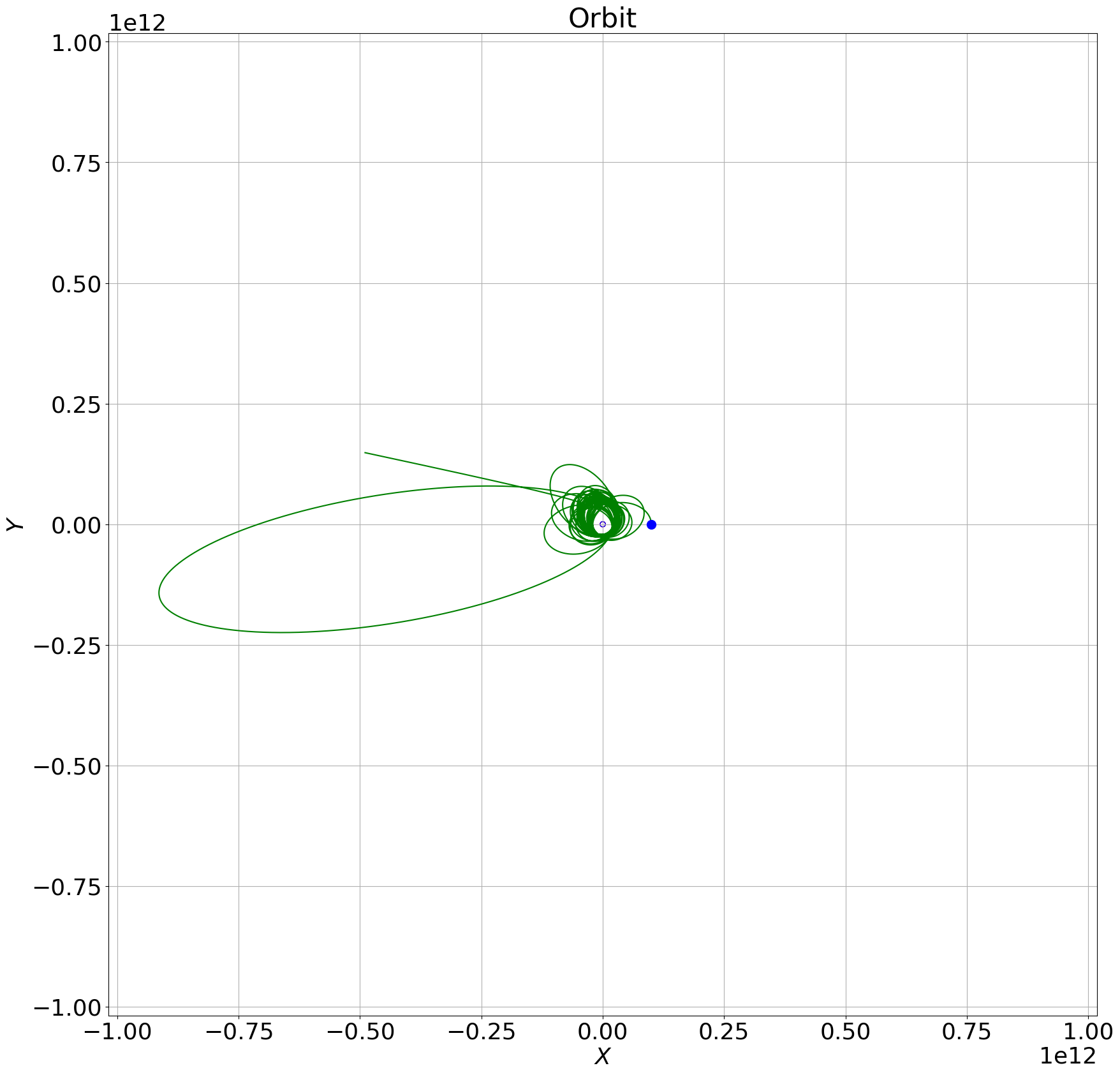}}
\caption{Test Mass Trajectory with Initial Y Velocity  (a) 34,000 $ms^{-1}$  (b) 30,000 $ms^{-1}$} 
\label{cm-34000}
\end{figure}

\begin{figure}[!ht]
\subfigure[]{\includegraphics[width=0.49\textwidth]{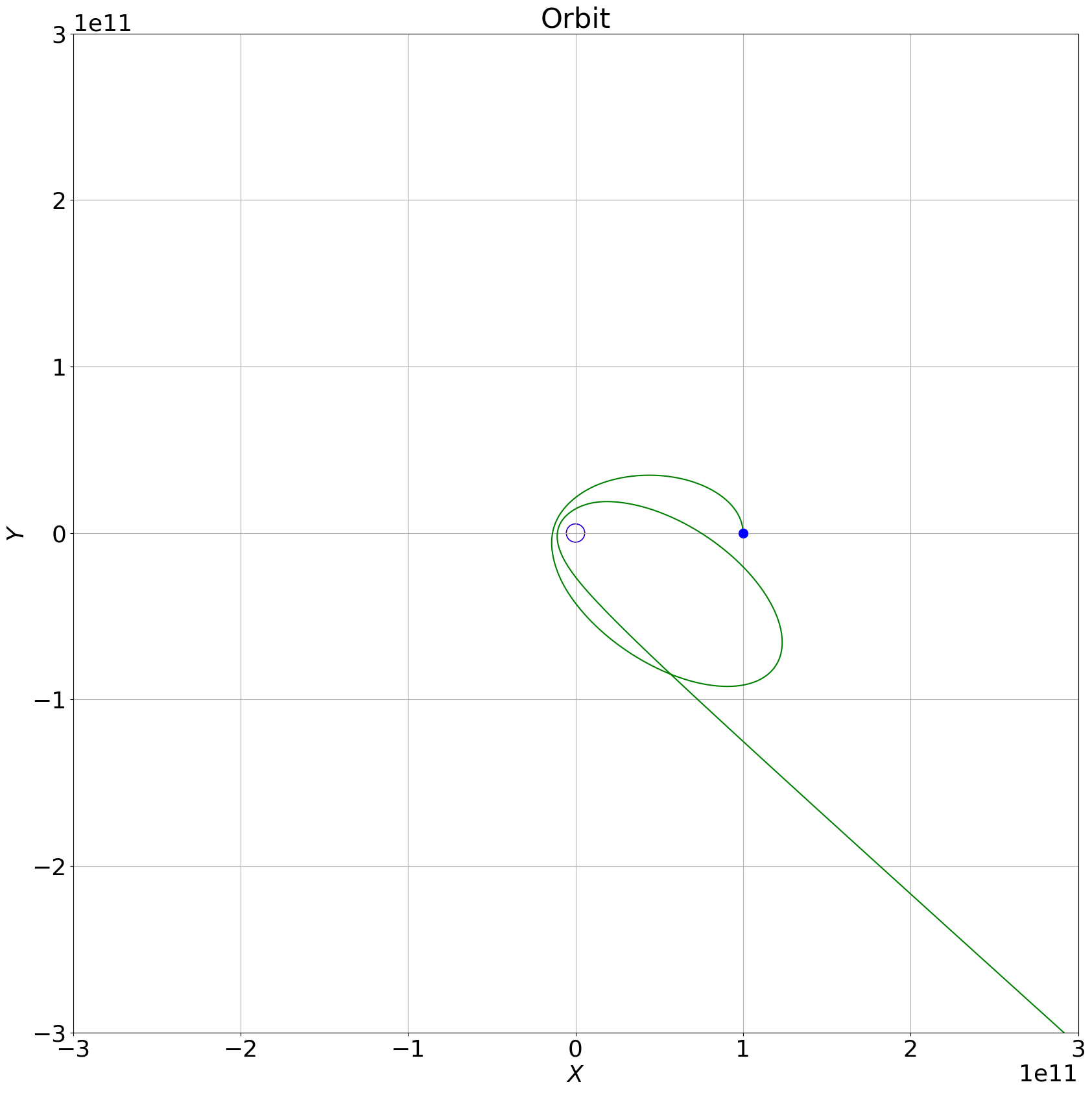}}
\subfigure[]{\includegraphics[width=0.49\textwidth]{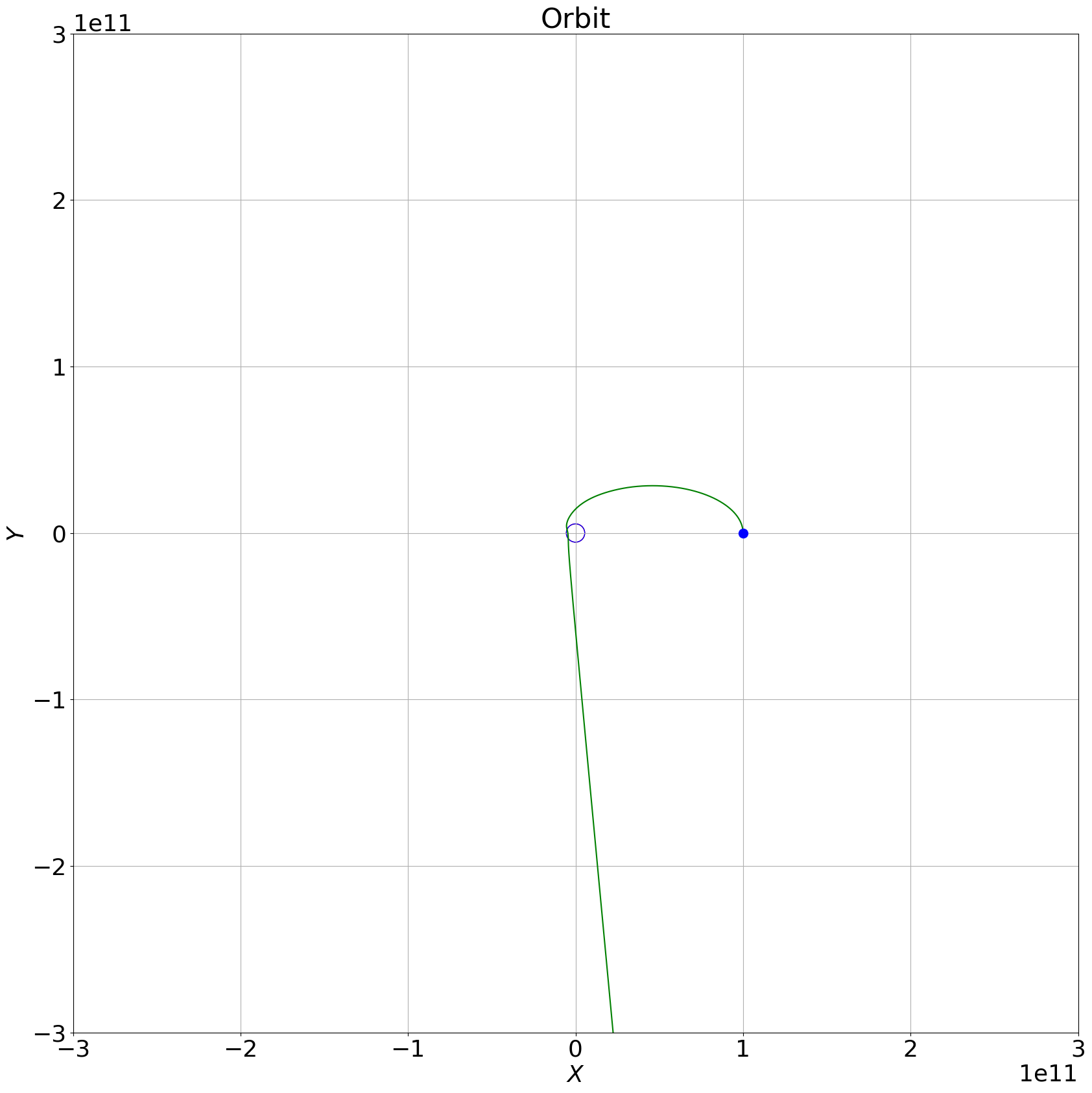}}
\caption{Initial Y Velocity  (a) 24,000 $ms^{-1}$  (b) 20,000 $ms^{-1}$} 
\label{cm-24000}
\end{figure}

\begin{figure}
\subfigure[]{\includegraphics[width=0.49\textwidth]{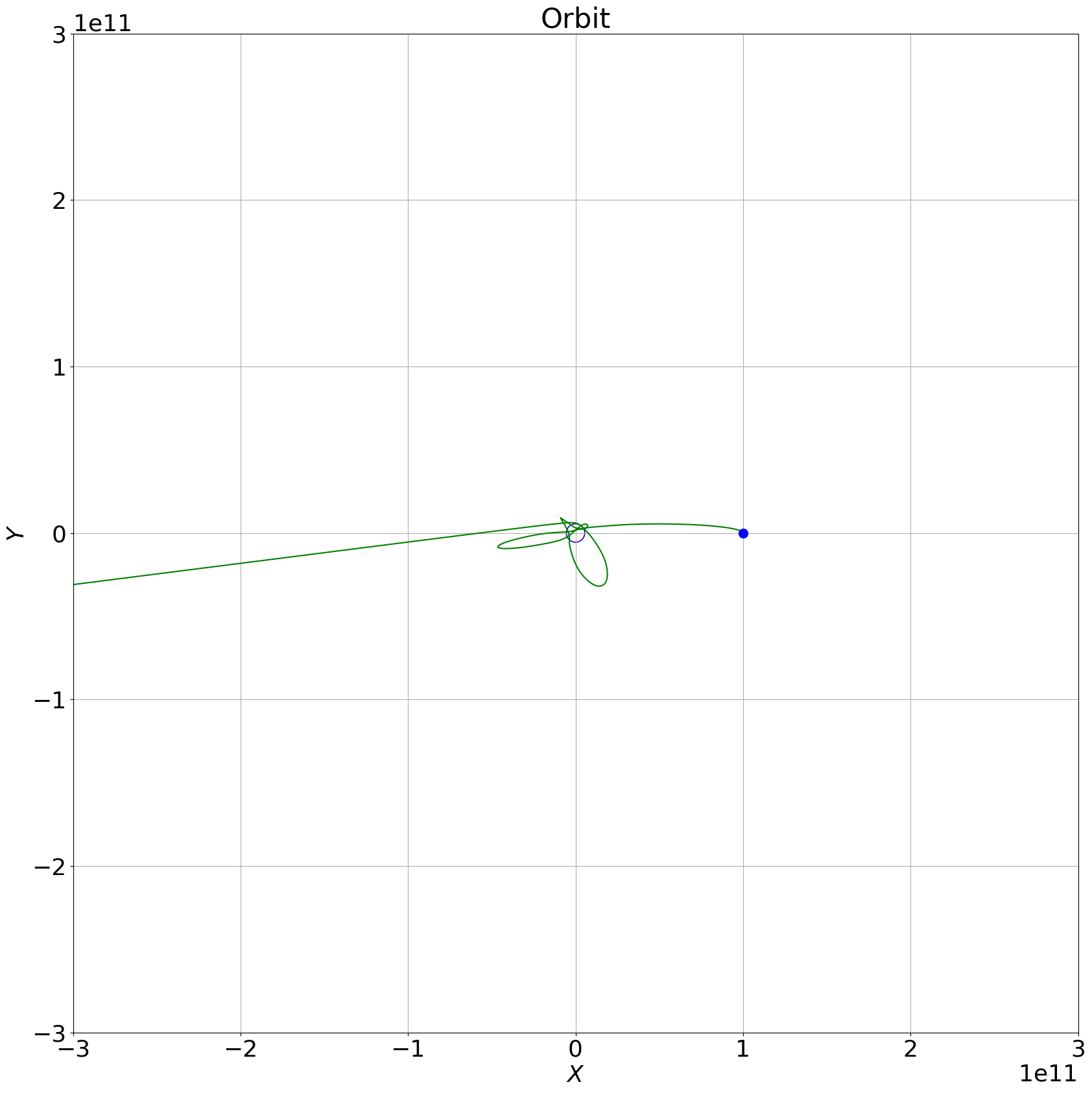}}
\subfigure[]{\includegraphics[width=0.49\textwidth]{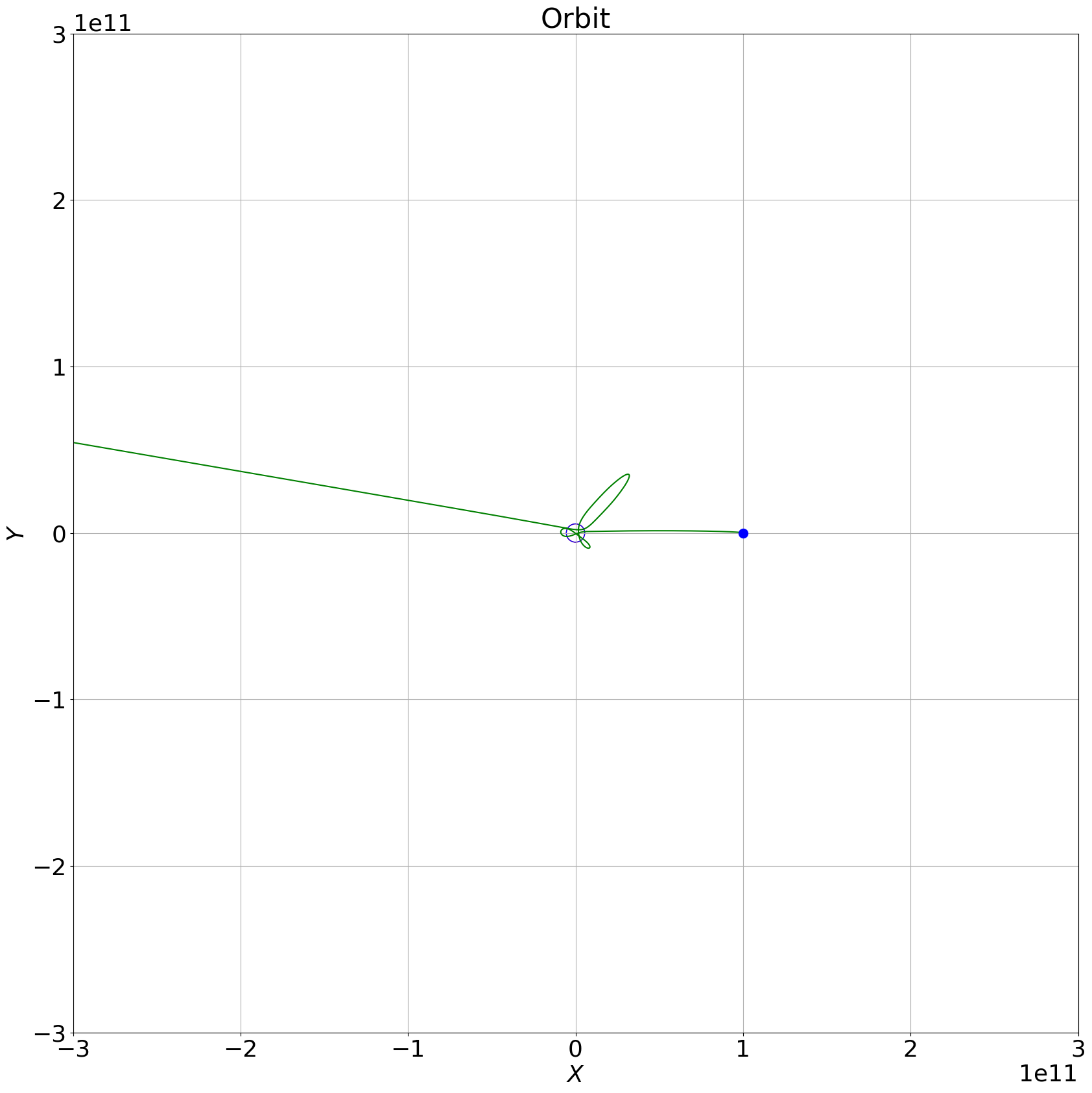}}
\caption{Initial Y Velocity  (a) 4,000 $ms^{-1}$  (b) 1,000 $ms^{-1}$} 
\label{cm-4000}
\end{figure}

\begin{figure}[!ht]
\subfigure[]{\includegraphics[width=0.50\textwidth]{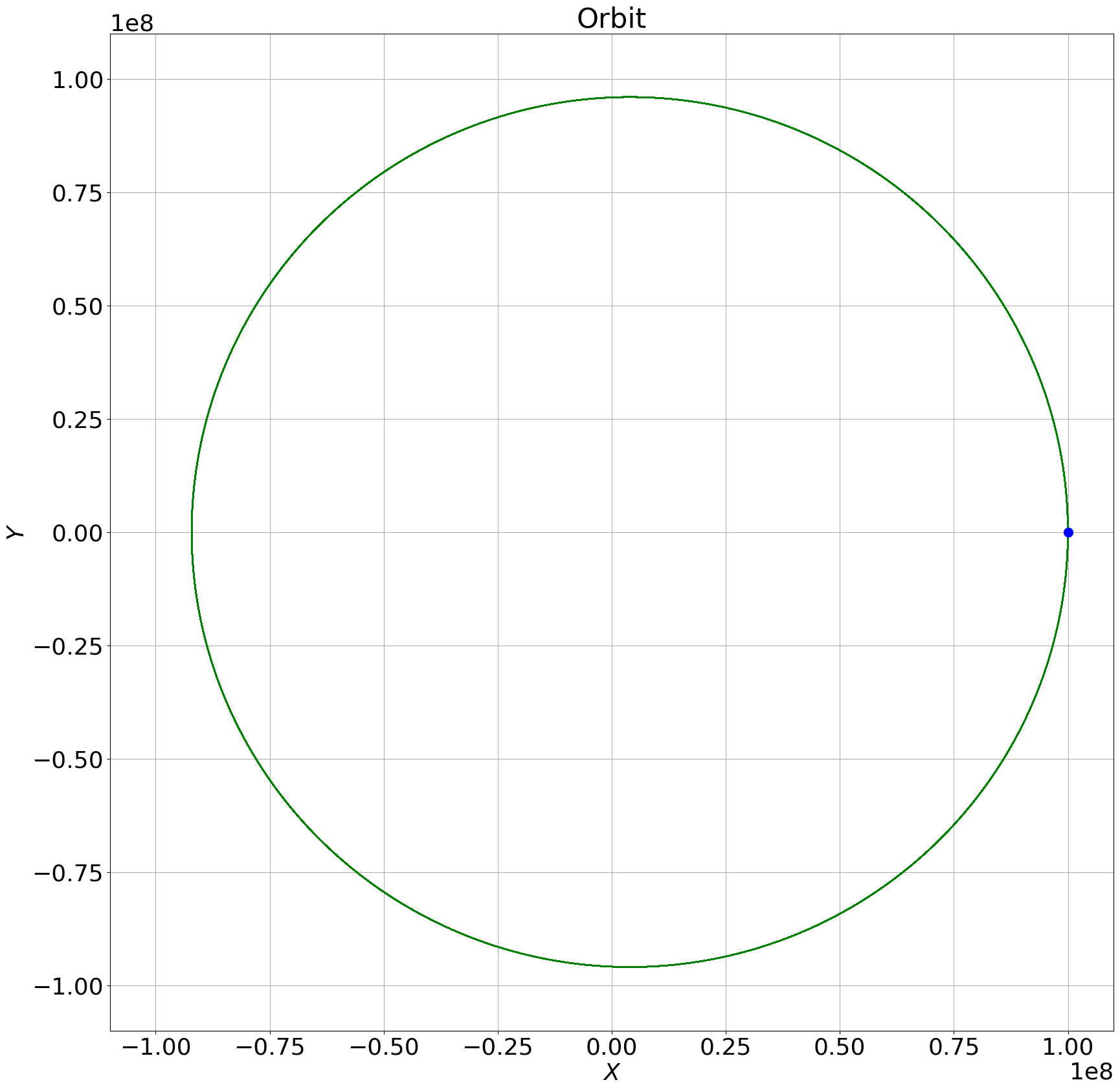}}
\subfigure[]{\includegraphics[width=0.49\textwidth]{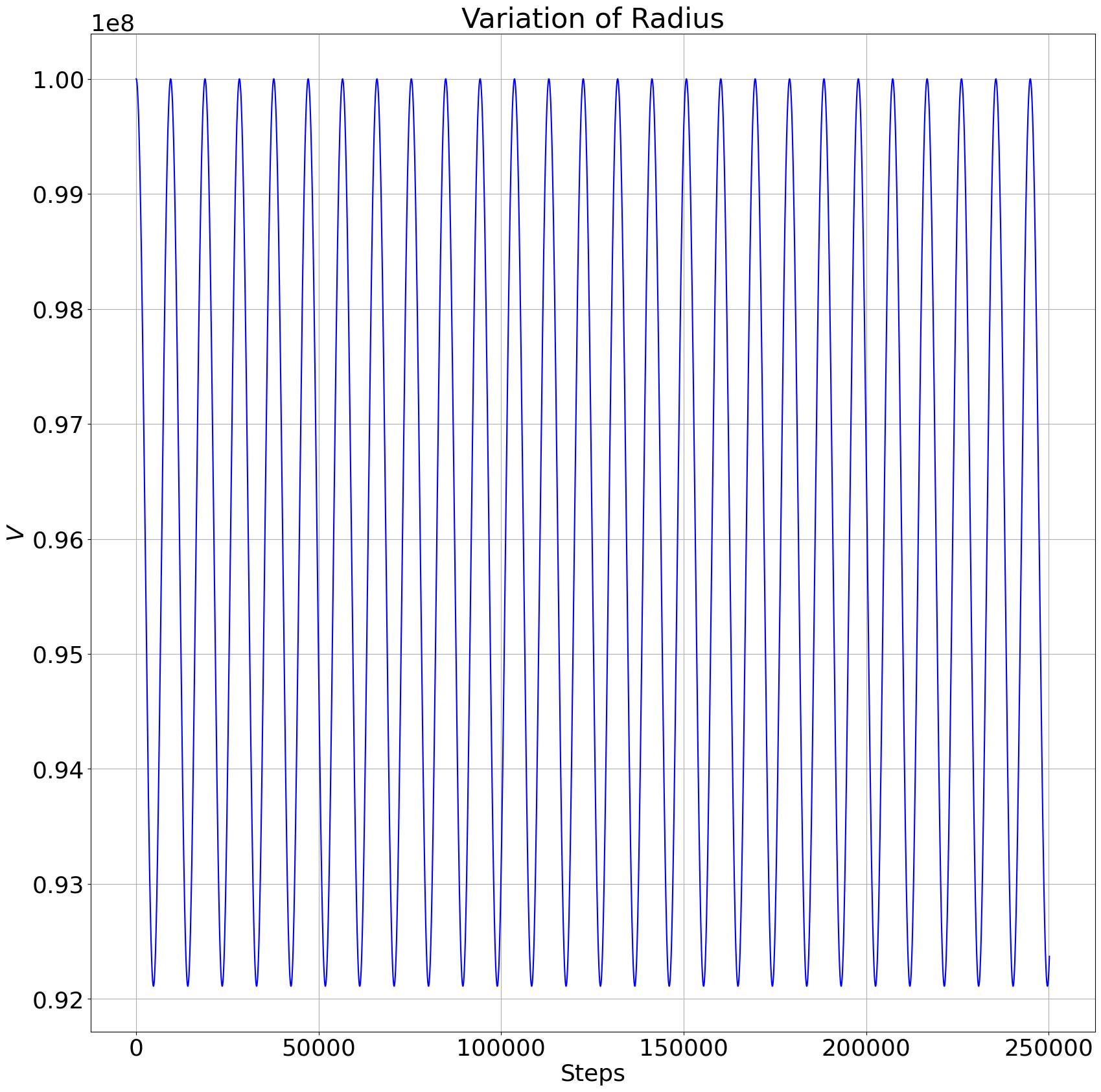}}
\caption{ (a) Initial Y Velocity 1,600,000 $ms^{-1}$ (b) Variation of Radius } 
\label{cm-1.6m}
\end{figure}

\begin{figure}[!ht]
\subfigure[]{\includegraphics[width=0.50\textwidth]{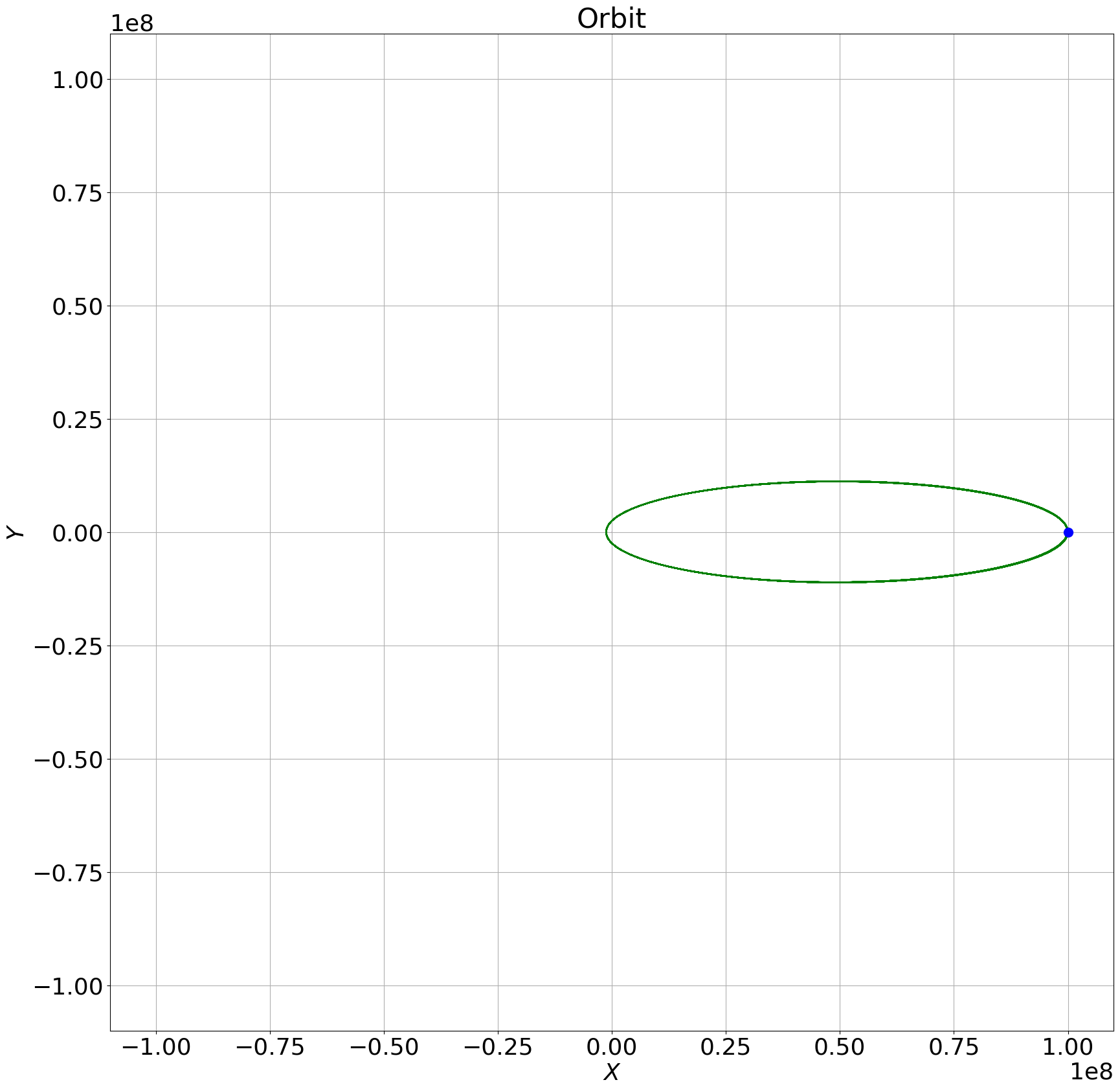}}
\subfigure[]{\includegraphics[width=0.485\textwidth]{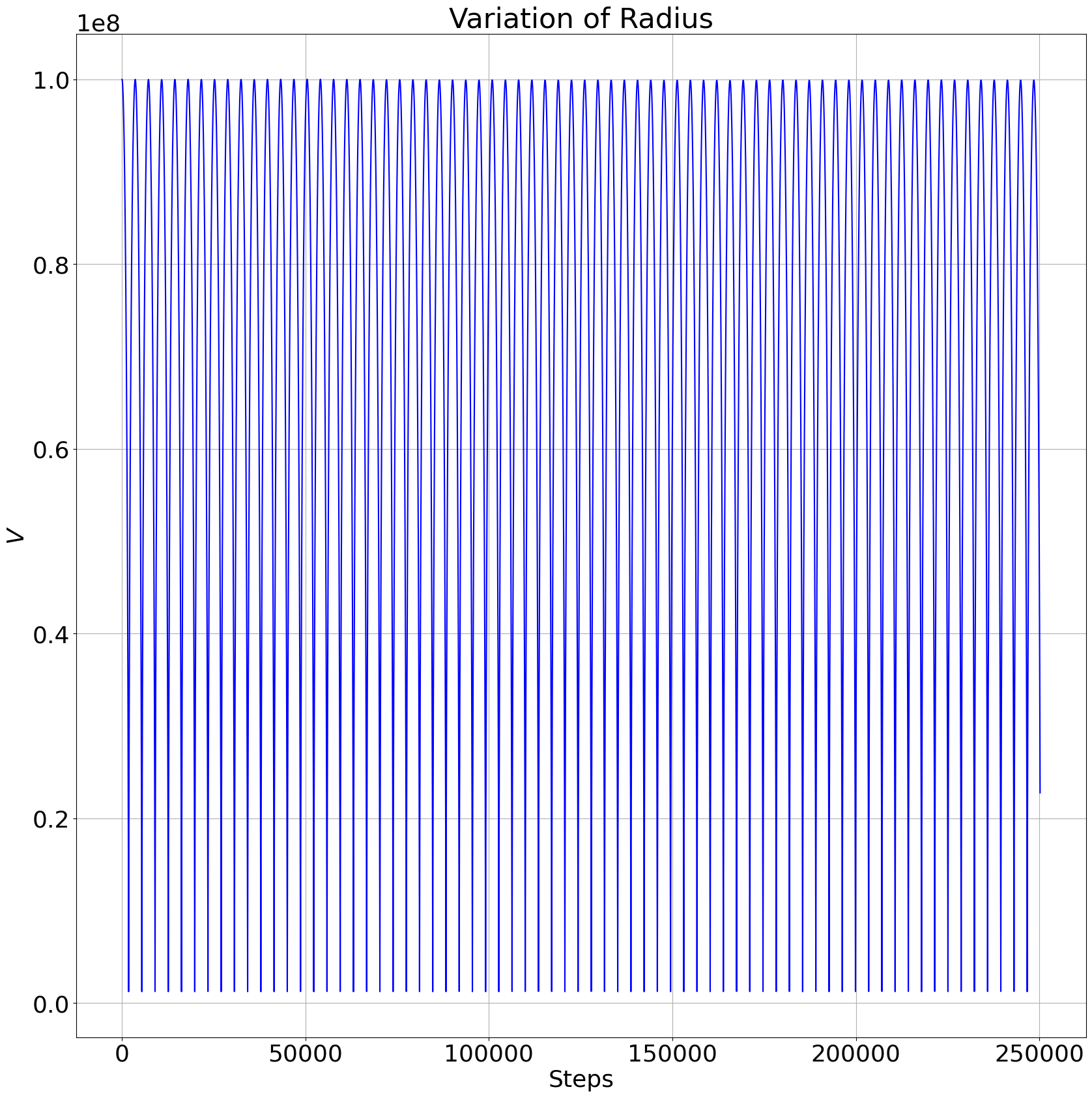}}
\caption{ (a ) Initial Y Velocity 256,000 $ms^{-1}$ (b) Variation of Radius } 
\label{cm-256k}
\end{figure}

\begin{figure}[!ht]
\subfigure[]{\includegraphics[width=0.50\textwidth]{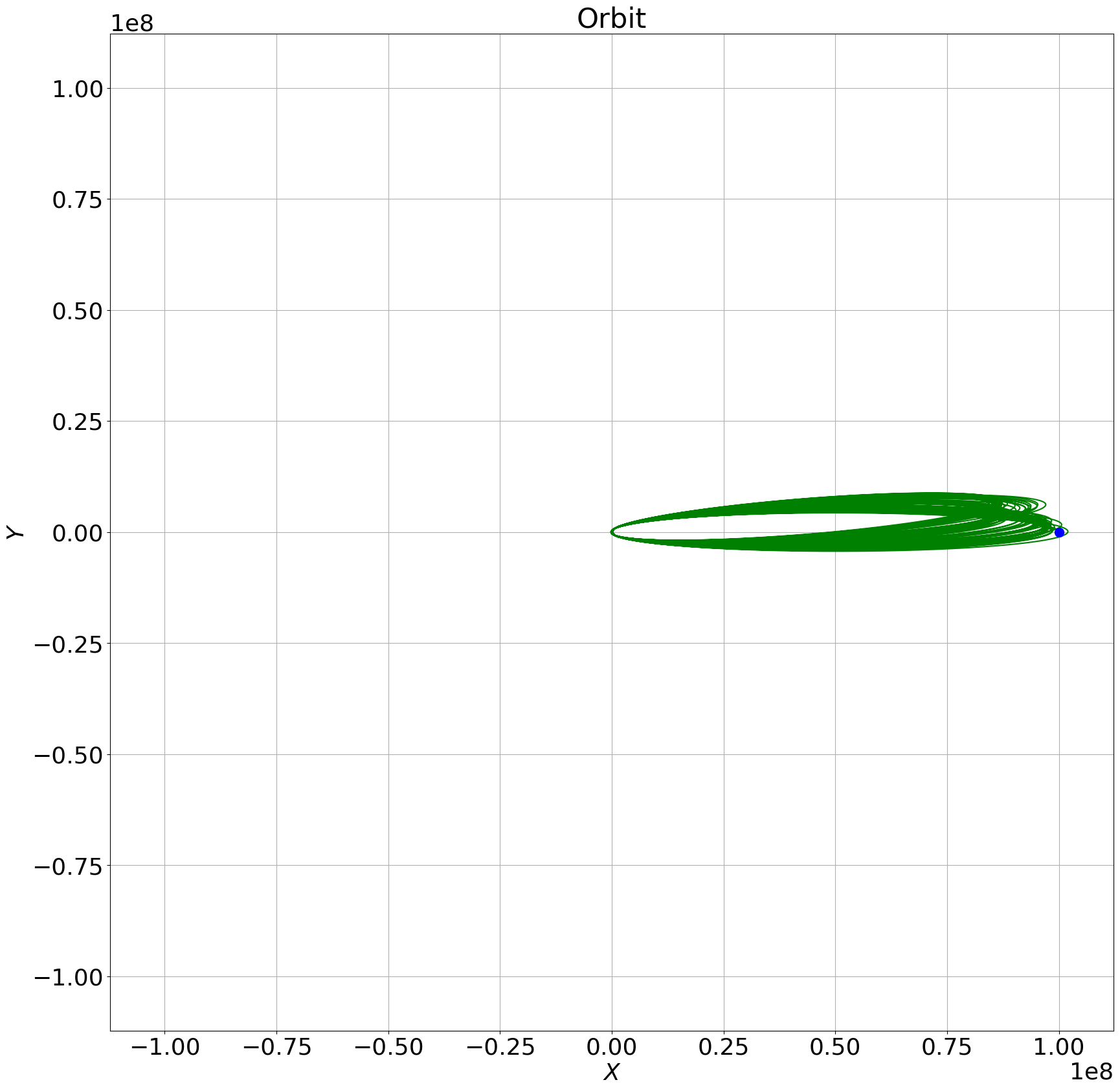}}
\subfigure[]{\includegraphics[width=0.485\textwidth]{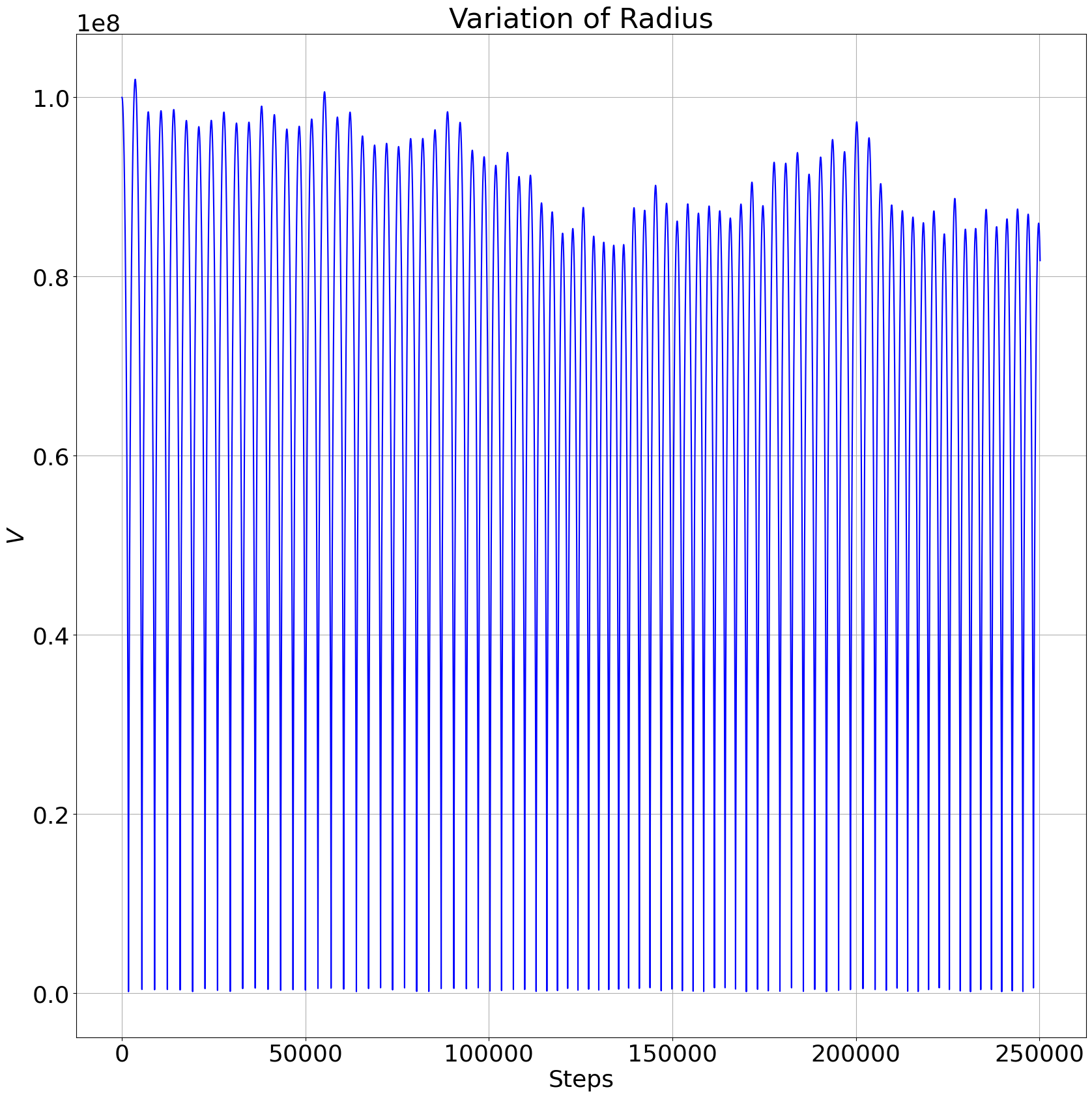}}
\caption{ (a ) Initial Y Velocity 100,000 $ms^{-1}$ (b) Variation of Radius } 
\label{cm-100k}
\end{figure}

\begin{figure}[!ht]
\subfigure[]{\includegraphics[width=0.50\textwidth]{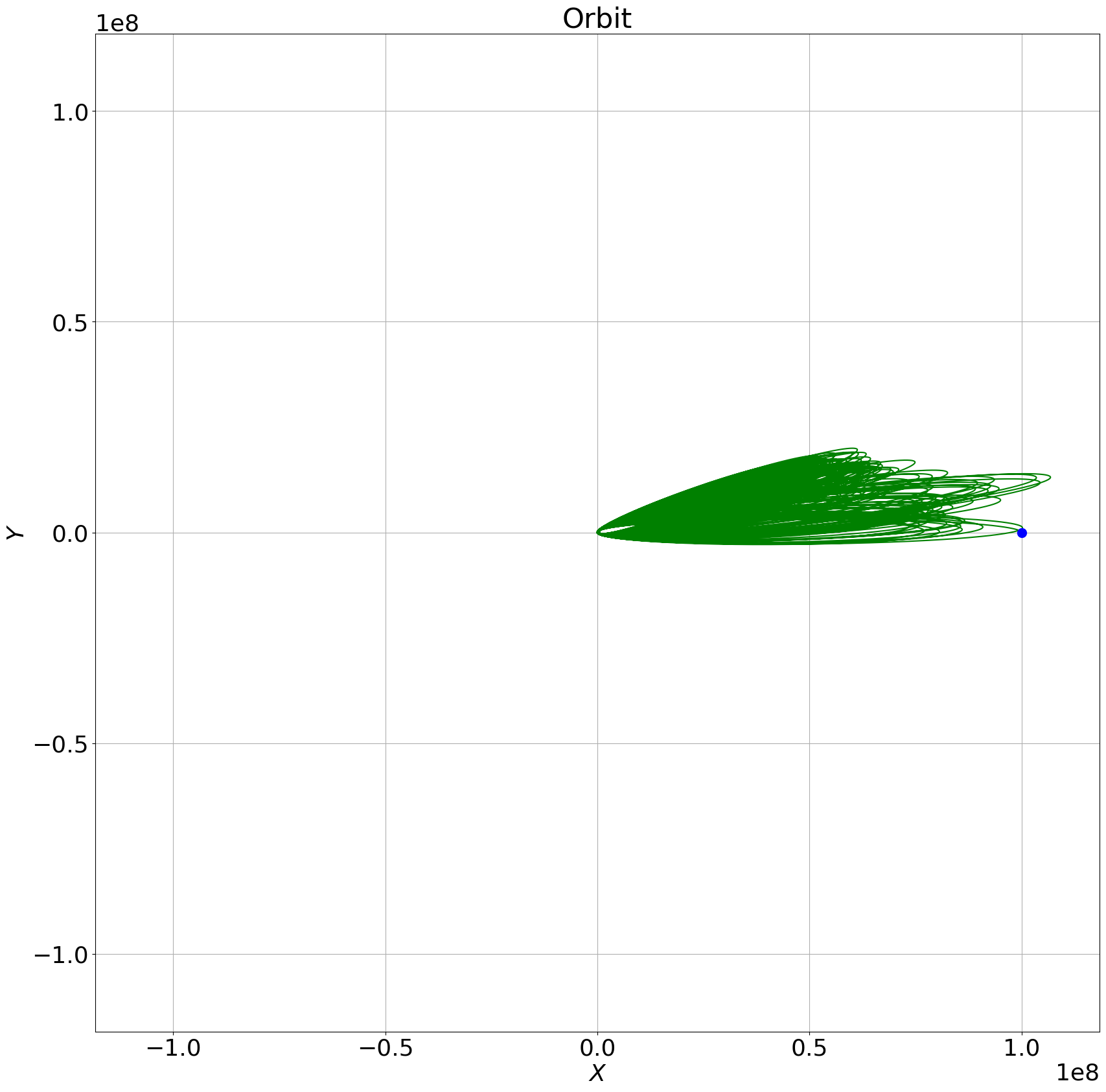}}
\subfigure[]{\includegraphics[width=0.49\textwidth]{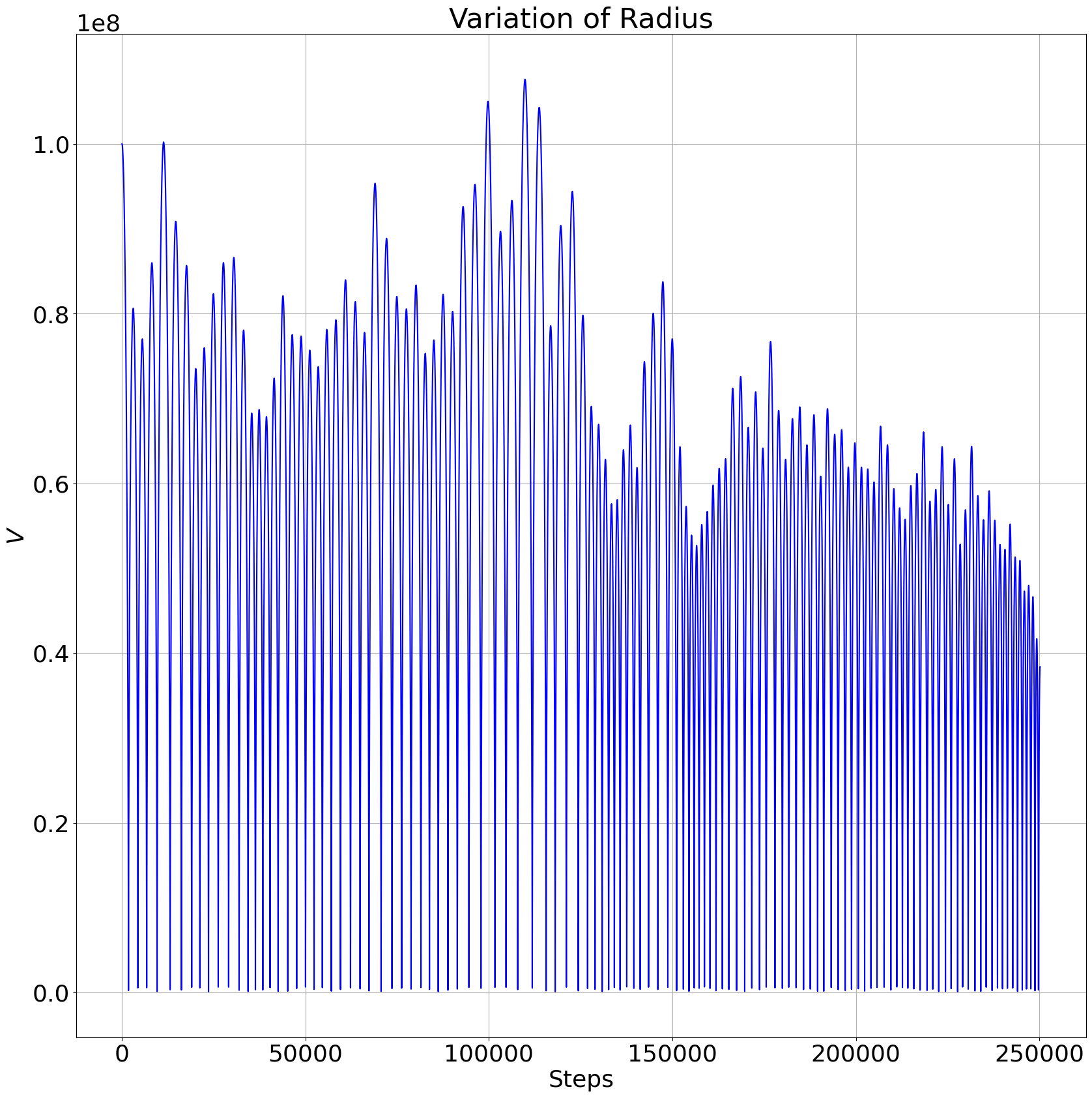}}
\caption{ (a ) Initial Y Velocity 76,000 $ms^{-1}$ (b) Variation of Radius } 
\label{cm-76k}
\end{figure}

\begin{figure}[!ht]
\subfigure[]{\includegraphics[width=0.49\textwidth]{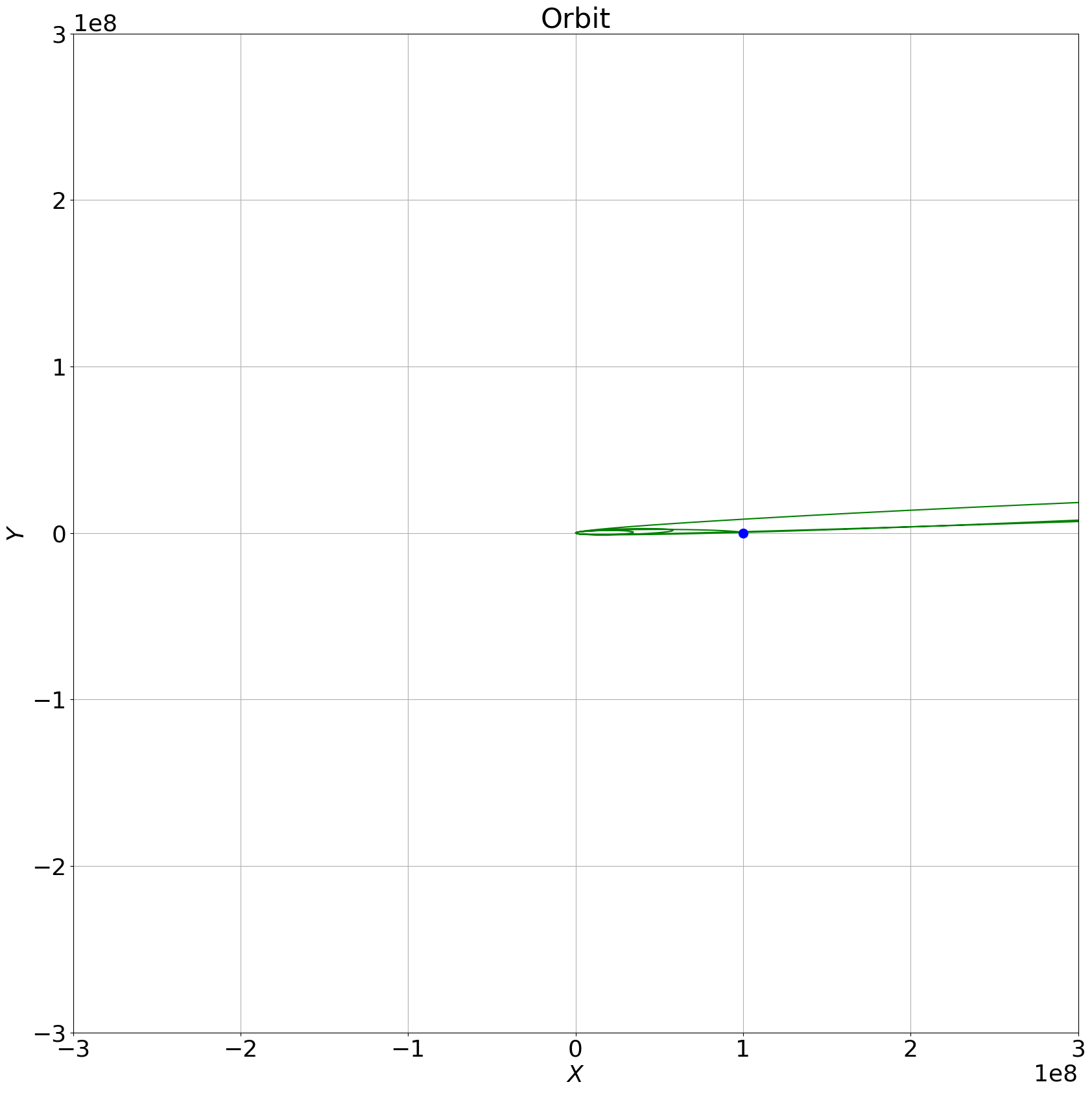}}
\subfigure[]{\includegraphics[width=0.485\textwidth]{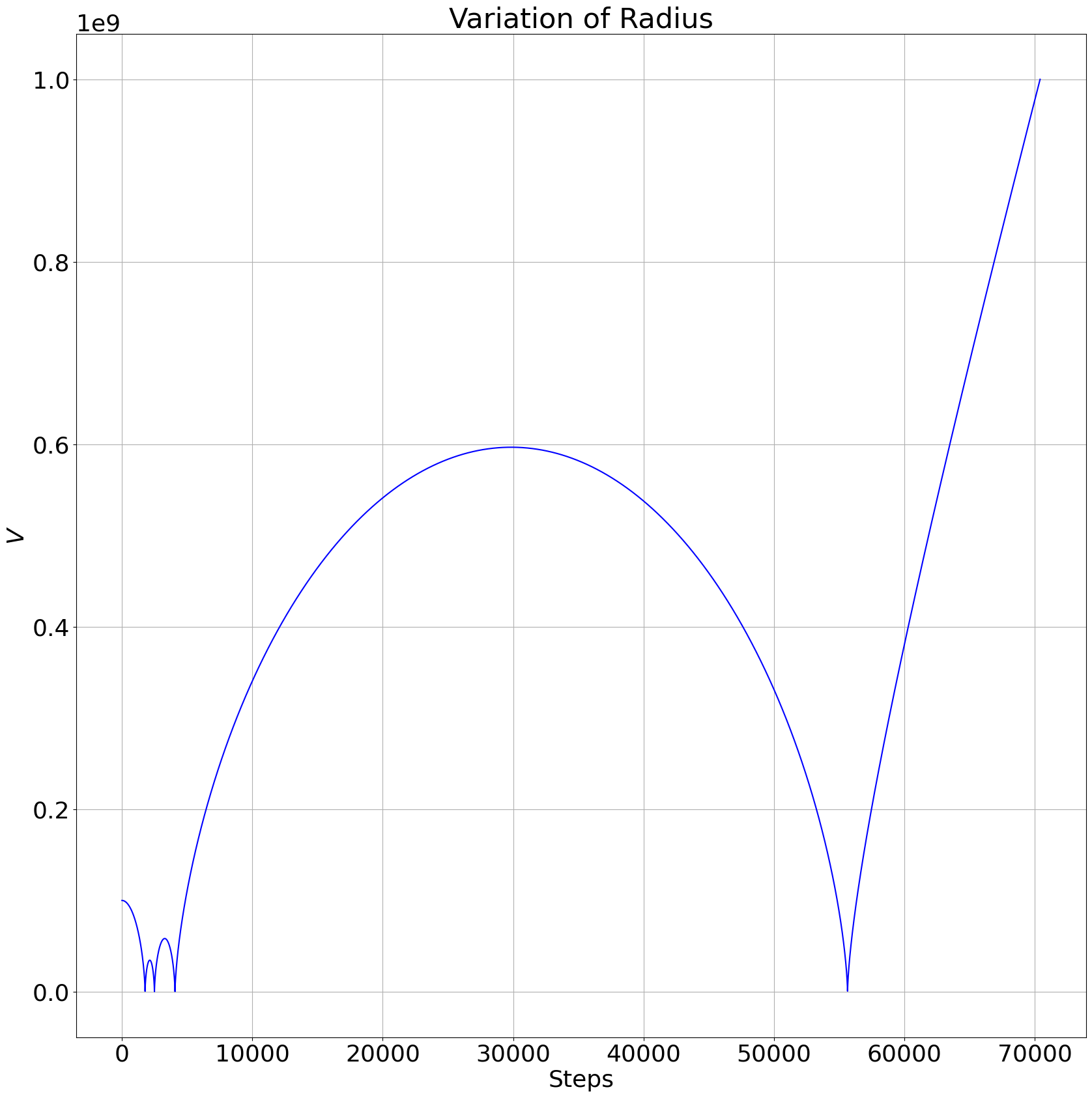}}
\caption{ (a ) Initial Y Velocity 50,000 $ms^{-1}$ (b) Variation of Radius } 
\label{cm-50.6k}
\end{figure}

In this paper, I will assume that none of the bodies carry electric charge (saving 3 degrees of freedom), that the two Primaries are in a stable orbit about their joint CoM,  that the sum of their masses is fixed and that they are not spinning (this is physically unlikely, but simplifies the analysis) leaving only 1 degree of freedom for the ratio of their masses and 1 degree of freedom for the distance between their centres of mass (saving a further 18 degrees of freedom). Finally, I will assume that the test mass moves only in the orbital plane of the binary system, that it is of unit mass and has no spin (saving a further 6 degrees of freedom). This still leaves us with six degrees of freedom: the ratio of the primary masses,  the distance between their centres of mass, two position coordinates and two velocity coordinates for the test mass.\\ \\
I will refer to the two large bodies as Primaries or $Mass1$ and $Mass2$ and will refer to the two together as the `binary' (short for binary system); I will also refer to the smaller body as the `test mass' or $Mass0$.\\ \\
Investigating the behaviour of test masses in these spacetimes is therefore reduced to postulating a model and a set of initial conditions within the constraints set and watching the modelled system evolve.  (In fact, in a spacetime governed by General Relativity it is necessary to know the positions of the primaries for some period of time prior to the initial time.)\\ \\
In this paper, I will compare the behaviour of the test mass when its motion is governed by:
\begin{itemize}
\item Newtonian Gravity
\item Newtonian Gravity plus a correction from the Schwarzschild metric
\item Newtonian Gravity plus the 1 PN, 2 PN and 2.5 PN corrections from Boekholt, Moerman and Portegies Zwart \cite{Boekholt_2021}
\item all of the above with the correction obtained from the propagation time of the field
\end{itemize}
making ten models in all.\\ \\
In addressing the `field propagation time'/'retarded potential', I have taken an engineering approach and separated the gravity models from the propagation time of the field and am therefore able to apply the same field propagation dynamics to all the gravitational models (0 PN to 2.5 PN). 
I have chosen to present all results in units of: metres, kilograms, seconds rather than natural units as I feel the extra effort in doing so is rewarded with greater physical intuition and understanding of scale.\\ \\
I will specifically be interested in the regions of phase space of the test mass where the precession of stable orbits, ejections and collisions occur and how these differ across the various models used. I discuss the metrics that I will use to compare the various models in Chapter \ref{chapresults}. \\ \\
I will also briefly comment on the differing impact of prograde and retrograde orbits but unfortunately, whilst I have written the code to allow for any ratio of masses between the primaries, I will not comment on the impact of varying that ratio in this paper.  \\ \\
In Newtonian mechanics, for test masses moving in the spacetime of a single large body, the trajectories are very simple (the Two Body problem). If the test mass has sufficient angular momentum not to collide with the large body, then the orbits described by the test mass are elliptical, stationary and in particular, there is no precession, there are no ejections of masses from the system (if the initial velocity is below escape velocity) and as angular momentum is conserved, collisions are rare as they require finely tuned initial conditions.\\ \\
Once I introduce a second large body, the trajectories exhibit considerably more complexity due mainly to the transfer of angular momentum between the bodies. As mentioned in the previous section I will consider a system of two heavy bodies orbiting each other with a third body of negligible mass traversing the spacetime of the two large bodies. \\ \\
In this paper, I have set the mass of each of the two large masses to be one Solar mass i.e. $2 \cdot 10^{30} kg$ (although I have written the code to allow for the ratio of masses of the primaries to be varied whilst the total mass remains the same using a parameter $\mu$). I set the initial conditions such that all three bodies start out lying on the $X$ axis and I allow the distance between the primaries ($R$), the $X$ coordinate of the test mass and the $Y$ velocity of the test mass to vary and unless stated otherwise, start the system with all three bodies rotating around the origin in an anti-clockwise direction. (See Appendix \ref{resonances} for explanation of why I use 1.1 times a power of 10 metres for values of $R$ rather than a simple power of 10)\\ \\
In this section, I set out some base case behaviours for the system in the purely Newtonian case. These types of behaviours then allow me to select a set of scenarios which will act as benchmarks against which I can compare the behaviour of the system when using different models of gravity with the same initial conditions. The scenarios I will examine are for 4 values of the separation between the binaries: $R$, up to 4 initial distances between the test mass and the origin, and up to 7 initial velocities (in the $Y$ direction) for a total of 72 scenarios which I will run using the ten model combinations. The various scenarios and in particular the initial velocities have been selected to illustrate the different types of behaviour: stable orbit, precession, chaotic, ejection/capture and the transitions between these regimes. However, there is nothing precise in the choice of these scenarios / initial conditions and other scenarios could easily be justified, but by keeping the scenarios common across all models, I will be able to measure the impact of the choice of model on the behaviour of the test mass. \\ 
\begin{figure}[!ht]
    \centering
    \subfigure[]{\includegraphics[width=0.49\textwidth]{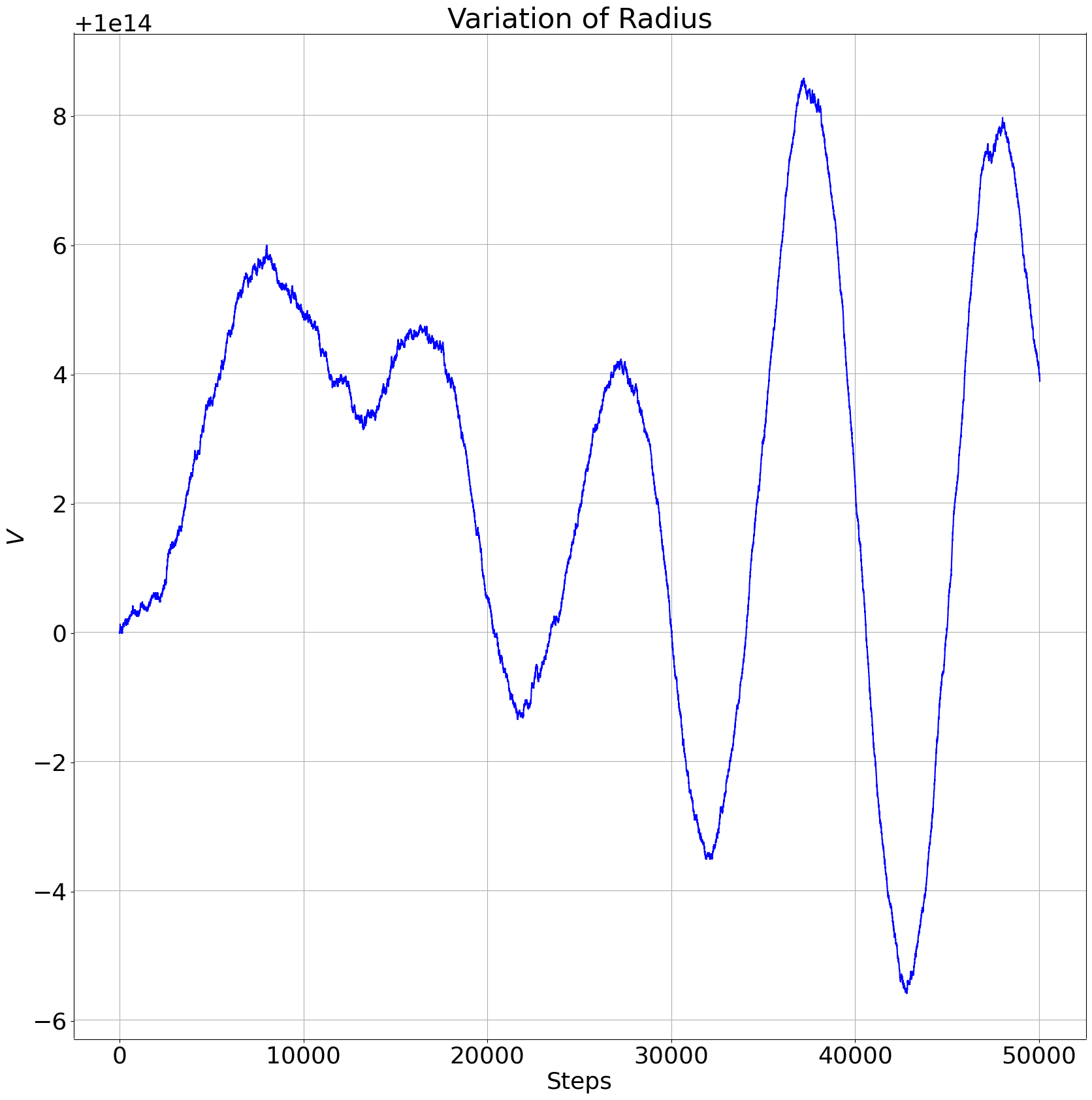}}
    \subfigure[]{\includegraphics[width=0.49\textwidth]{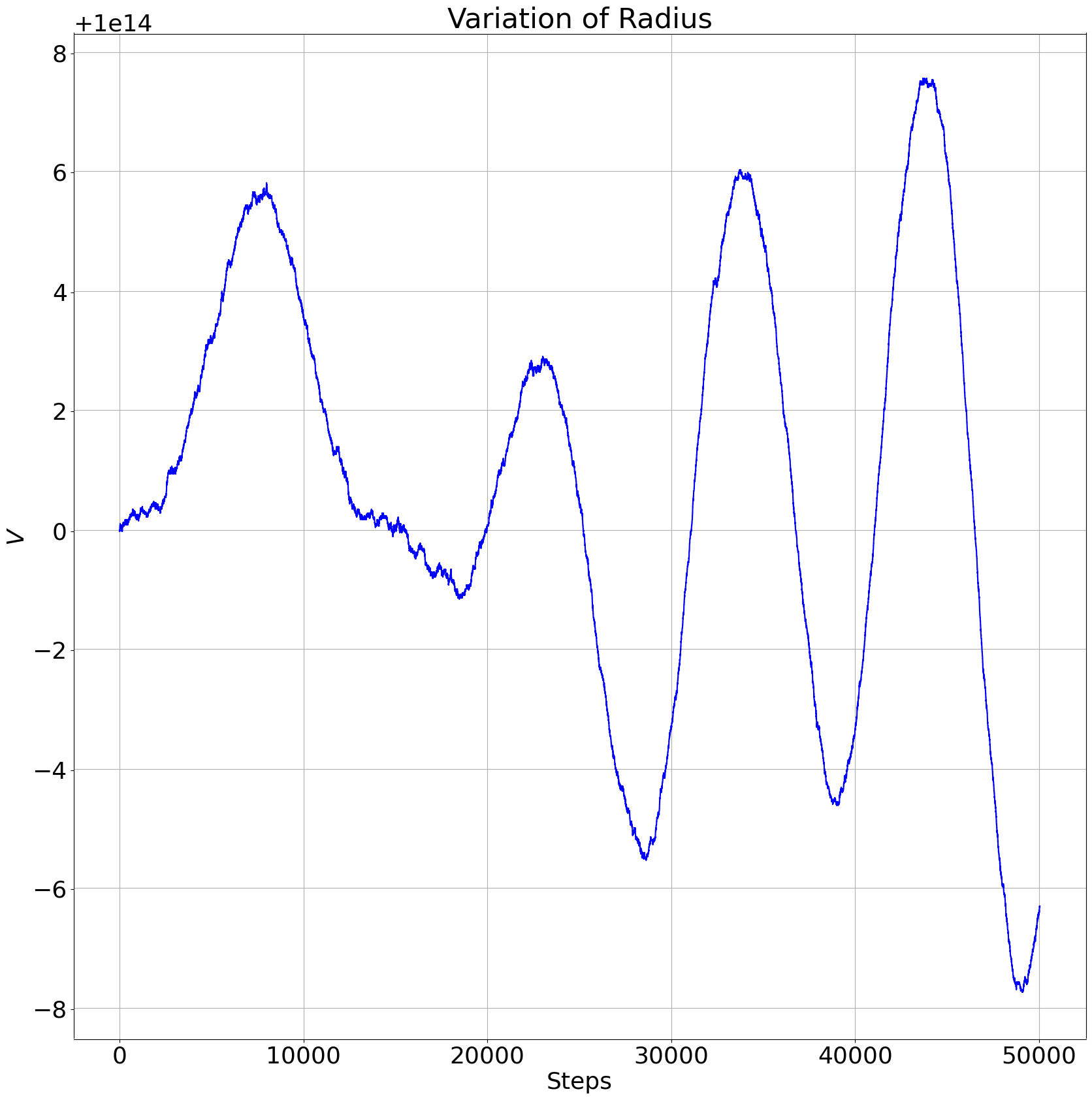}} \\ 
    \subfigure[]{\includegraphics[width=0.50\textwidth]{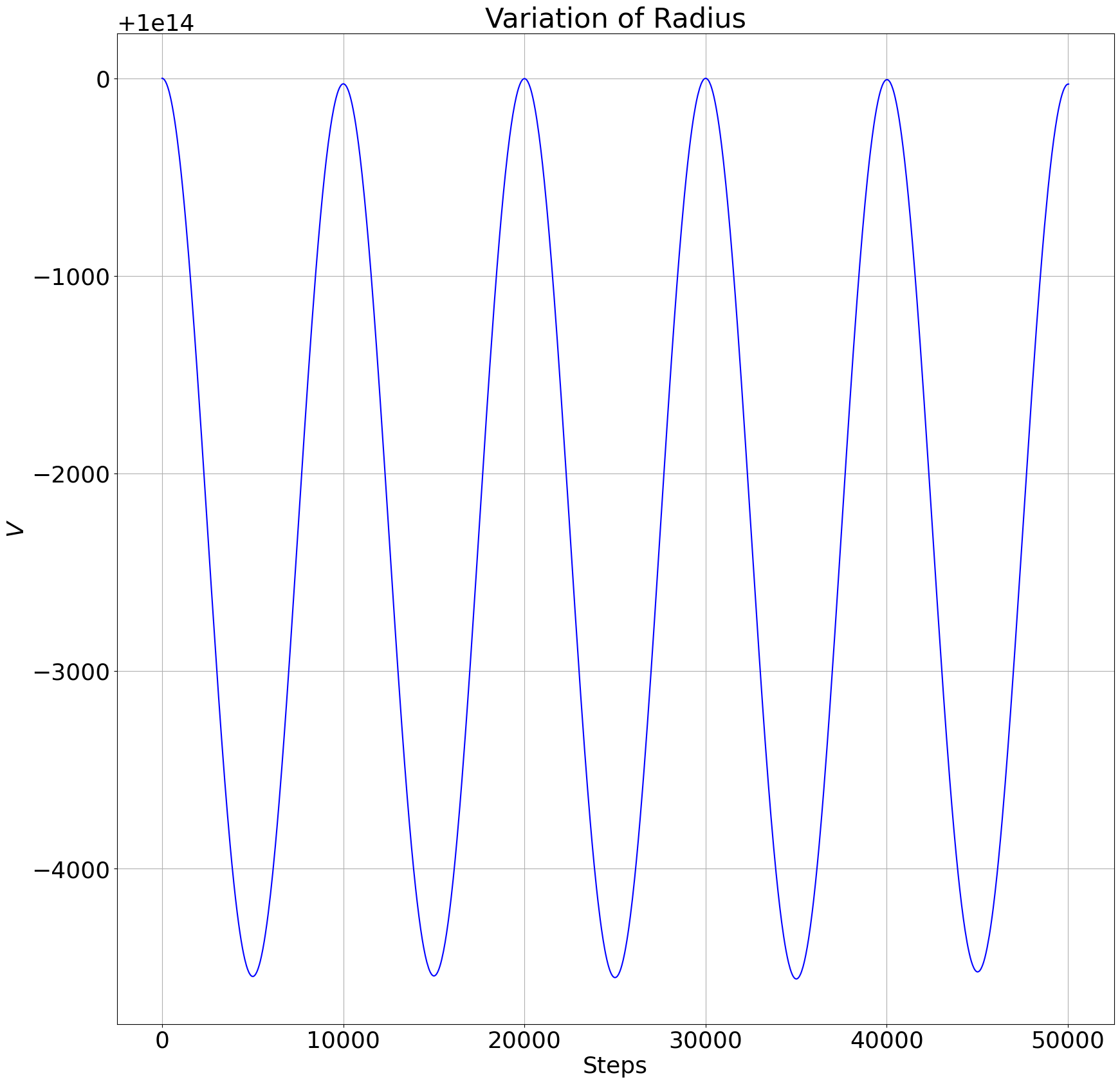}}
    \subfigure[]{\includegraphics[width=0.49\textwidth]{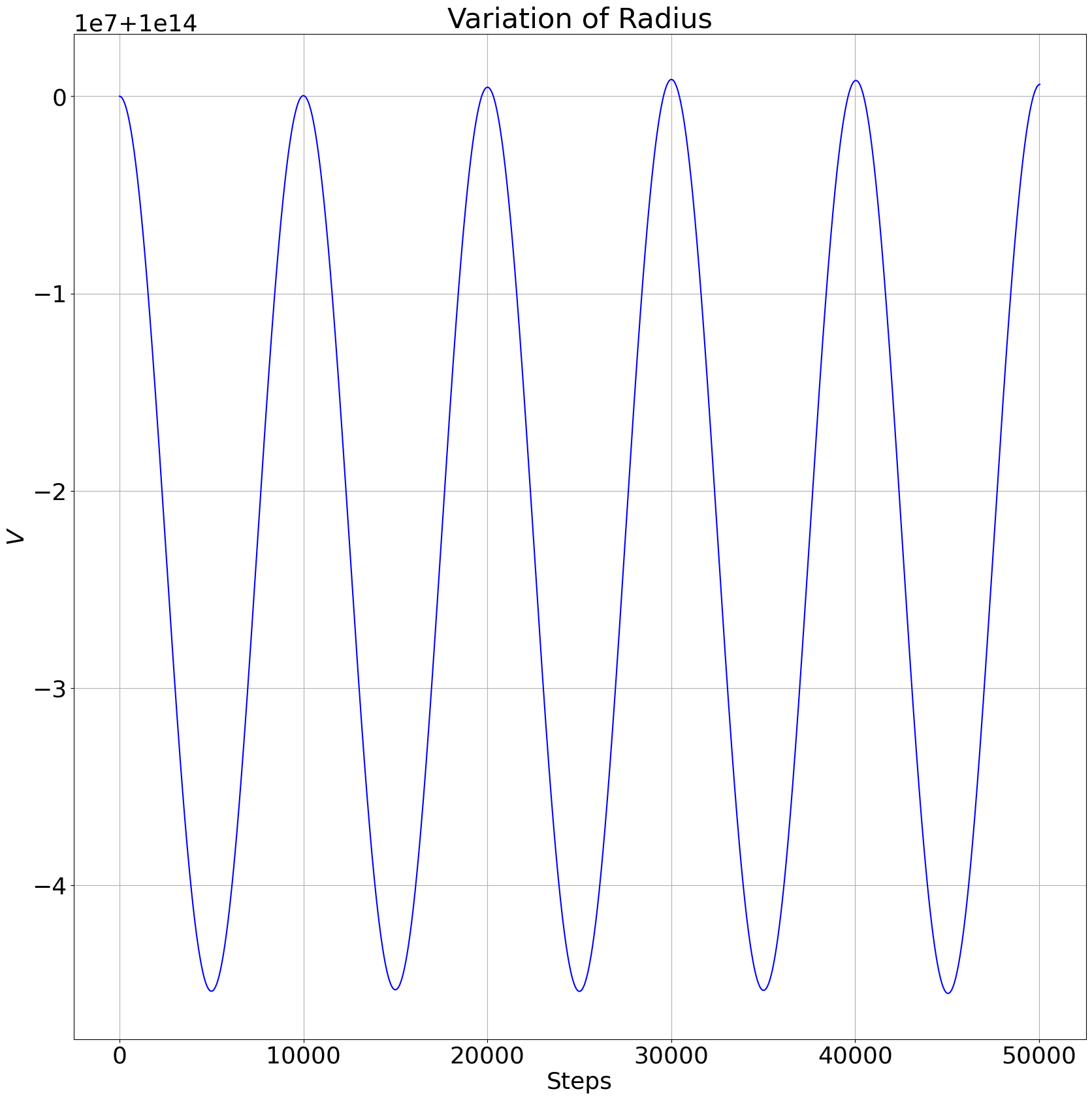}} 
    \caption{Separation = (a) $1.1 \cdot 10^5m$  (b) $1.1 \cdot 10^7 m$  (c) $1.1 \cdot 10^9 m$ (d) $1.1 \cdot 10^{11} m$ \\ Test Mass initial position: $x = 10^{14} m, \;  y = 0m$}
    \label{varyR}
\end{figure}
\noindent \\
To begin with, in Figure \ref{varyR}, I observe the oscillations of the orbit of the test mass caused by the spinning binary. There are three potential sources of variation of orbit:
\begin{itemize}
\item errors due to the numerical implementation (these are not programming errors, but structural errors due to the digital nature of numerical simulations)
\item oscillations of the gravitational field strength due to the spinning binary (with twice the frequency of the spinning binary), which in turn causes the orbital radius to oscillate
\item changes in the `energy' or velocity required for a circular orbit, meaning that the orbits become elliptical. (This change in required energy is caused by the different potential of the spinning binary relative to the potential created by a single static body) 
\end{itemize}
In Figure \ref{varyR}, we see the test mass start at $10^{14} m$ from the origin, with an initial velocity in the Y direction, with a magnitude corresponding to that required for a circular orbit in the two body case (computed to fourteen significant figures of accuracy: 1,633.9008537852$ms^{-1}$). The Figures show the change in the distance of the test mass from the origin at each time step of the simulation (where we have 10,000 steps per orbit of the test mass). The variation `v' is in metres for Figures (a), (b) and (c) and in $10^7$ metres for Figure (d). We see that when the primaries are close together (with separation of $1.1 \cdot 10^5 m$ in Figure (a) and $1.1 \cdot 10^7 m$ in Figure(b)), the test mass' orbital radius is extremely stable, with variations of the order of 1 to 10 metres; these variations are very largely due to numerical effects of the simulation. However as the separation between the primaries increases, we see that at $1.1 \cdot 10^9 m$ in Figure (c), the orbits start to oscillate with amplitude of the order of $10^3  m$ and with a separation of  $1.1 \cdot 10^{11}  m$ in Figure (d), the oscillations have an amplitude of the order of $10^7  m$ and are due to the fact that these orbits have become elliptical. As orbital periods scale as radius to the power of $3/2$, even in Figure \ref{varyR} (d), the gravitational field will oscillate with a frequency of approximately $2 \cdot \big(\frac{1000}{1.1} \big)^{1.5}$ per orbit of the test mass and will therefore not be visible in these Figures. \\  \\
In Figure \ref{elip-500}, keeping the separation of the primaries at $1.1 \cdot 10^{11} m$, as I alter the initial velocity of the test mass (still pointing in the Y direction), the orbits become elliptical and the distance of closest approach of the test mass to the binary CoM reduces. When this distance of closest approach falls to about 50 times the distance separating the binaries, we can just begin to see precession in the orbital plots as in Figure \ref{elip-500}, where the left panel shows approximately 200 elliptical orbits of the test mass around the binary (which is situated at the origin and the picture is seen from a position on the Z axis) and the right panel shows the distance of the test mass to the origin at each time step (this left panel = trajectory,  right panel = distance from origin, Figure combination will be used repeatedly below). \\ 
\noindent \\
As I continue to lower the initial velocity, we see in Table \ref{prec-rate-tab} and Figure \ref{prec-rate-fig} that the angular velocity of precession increases and that there is a clear relationship between the initial velocity and the rate of precession (where I have also shown a best fit regression line showing that the precession rate is proportional to the initial velocity to the power of $-4.036$ with an $R^2$ of  0.9993, but of course there are only 9 data points here, so this result should be considered tentative).  This relationship should be no surprise since lowering the initial velocity brings the test mass closer to the primaries and thus increases the amount of angular momentum that is temporarily transferred from the primaries to the test mass. Note that this interaction is more of a `kick', i.e. a one off change of angle of the major axis of the orbit rather than a permanent transfer of angular momentum from the binary to the test mass, similarly at this stage the data does not appear to suggest an increase in orbital radius, which is consistent with a change of direction rather than angular momentum. \\ 
\begin{table}[!ht]
\centering
\includegraphics[width=150mm]{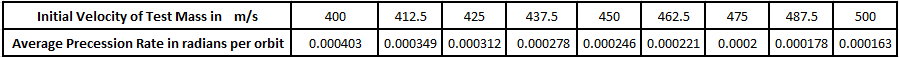}
\caption{Rate of Precession of Test Mass Orbit with Change of Initial Velocity}
\label{prec-rate-tab}
\end{table}
\noindent \\
Using smaller values for the separation between the primaries and suitable initial conditions for the test mass we begin to clearly see in Figures: \ref{cm-15000}, \ref{cm-9500}, \ref{cm-8000} and \ref{cm-6500} precession, chaos and ejection respectively. The latter is seen in Figure \ref{cm-6500} (d) with the final bounce of the radius becoming linear. To be clear, at this stage my description of Figure \ref{cm-8000} as chaotic is a purely visual observation based on a lack or regularity and is not a quantified statement. In these figures, the separation between the primaries is $1.1 \cdot 10^{10} m$ and the test mass starts at $5 \cdot 10^{12} m$ from the binary CoM. \\ 
\noindent \\
In order to explore scenarios where the interactions between the test mass and the binary are likely to be stronger, I again reduce the initial distance of the test mass to the origin so that it is now just 10 times (clearly 9 times, but I will continue to refer orders of magnitude for simplicity) the distance between the primaries ($ R = 1.1 \cdot 10^{10} m$ and the test mass starts at $X = 10^{11} m$). In Figure \ref{CM with en}, where the initial velocity is $37,500ms^{-1}$, where we see pronounced precession, we also see the orbit of the primaries for the first time (the small circle around the origin), previously this orbit(s) had been too small to see. We observe that the variation of the test mass' orbit is highly stable as are the changes in potential and kinetic energy and we also observe small variations in total energy during the interactions with the binary and observe that these changes are temporary and do not accumulate over multiple orbits, confirming our previous observation that no permanent transfer of energy is visible at this stage or with this set of initial conditions. In the lower right hand pane of Figure \ref{CM with en}, Kinetic Energy of the test mass is: $ \frac{1}{2} m v^2$ and the Potential Energy is the potential energy of a particle in Schwarzschild spacetime given by the expression:  $V = - \frac{G \cdot M}{\rho}  - \frac{G \cdot M \cdot l^2 \cdot c^2}{\rho^3} \ \ $, where $\rho$ is the distance from the particle the CoM of the system and $l$ is the angular momentum of the particle relative to the CoM of the system. I discuss the Schwarzschild corrections in Chapter \ref{chapschwar}. \\
\noindent \\
Finally, in Figures \ref{cm-34000}, \ref{cm-24000} and \ref{cm-4000}, (and keeping $ R = 1.1 \cdot 10^{10} m$ and $X = 10^{11} m$) I explore the lower energy (lower initial velocity cases) and find increasingly erratic orbits where virtually all runs/simulations lead to ejections, even those that appear aimed directly at the binary, this is because the orbital radius of the binary is very large compared to the Schwarzschild radii of the two primaries (3000 metres) therefore the odds are heavily in favour of an ejection rather than a capture / collision. In fact, in the Newtonian paradigm, collisions are more likely to occur for a binary star system than a binary black hole system of the same mass as the cross-sectional area of the stars is greater, thus enhancing the probability of collision. Capture is also a function of the ratio of the speed of the test mass to the speed of the binaries since the time that the test mass spends in the path of the binaries is given by:
\begin{align}
T = \frac{ R_S}{ v_{tm}}  \label{prob1}
\end{align}
where $T$ is the time taken to cross the path, $R_S$ is the Schwarzschild radius of the black holes and $v_{tm}$ is the velocity of the test mass. The probability of capture is therefore:
\begin{align}
\mathcal{P}_{cap} = \frac{4 \ T}{ P_{bin} } \label{prob2}
\end{align}
where $P_{bin}$ is the orbital period of the binary and where the factor of 4 comes from the fact that there are two black holes and the test mass crosses the orbit twice.\\ \\ 
Figures \ref{cm-34000}, \ref{cm-24000} and \ref{cm-4000}, show the trajectory of the test mass over 250,000 time steps, with the test mass starting at the blue dot on the X axis to the right of the origin and the purple circle around the origin shows the orbit of the binary. \\  \\
Figure \ref{cm-34000} shows two trajectories that perform a number of increasingly eccentric orbits before being ejected, Figure \ref{cm-24000} shows two trajectories that are very rapidly ejected by the binary whereas Figure \ref{cm-4000} shows two trajectories that perform several passes through the binary before being ejected.
Figure \ref{cm-24000}(b) is particularly notable as the test mass' trajectory passes inside the orbit of the binary and is still ejected (it does not however reach either of the primaries' event horizons, as these are now very small compared to the separation of the primaries. In all of the runs presented in this paper, the simulation exits if the test masses pass within the capture distance / Schwarzschild Radius of either primary). It is in fact the common behaviour for initial velocities that take the trajectory through the binary's orbit, as the velocities of the test mass required for capture are unlikely unless the test mass has had much of its energy dissipated (this is partly because equations \eqref{prob1} and \eqref{prob2} require low speeds in order to increase the probability of capture and partly because at higher velocities, angular momentum is likely to be higher, keeping the test mass away from the primaries). \\ 
\noindent \\
To study the strong field limit, I set the distance between the primaries to be $11,000m$ (which is just over distance to the innermost stable circular orbit of each primary). At this distance, their Newtonian angular velocity is 14162.394 radians per second which corresponds to a linear velocity of  $77,893,166.93 ms^{-1}$  or $ 0.2598c$, using the Schwarzschild correction (discussed later in Chapter \ref{chapschwar} equation \eqref{schwar}), the mass corrections now give us an angular velocity for the binary of 15859.1 radians per second or $87,224,967.05ms^{-1}$ or $0.291c$. I start the test mass at distances of $10^8m$, $10^7m$, $10^6m$, $10^5m$, and starting with an initial velocity equal to that required for a circular orbit, and I then lower the velocities progressively to observe the various regimes.  \\ \\
For the $10^8m$ cases, we see in Figure \ref{cm-1.6m} that we have a stable orbit when the initial velocity is 1,600,000 $ms^{-1}$ and that orbits remain stable (but elliptical) until the velocity is reduced to 256,000 $ms^{-1}$ Figure \ref{cm-256k}, I call these orbits stable as the right hand pane of these Figures shows stationary variations in radius and in the left hand pane, the orbits are drawn one on top of another. As the initial velocity is lowered further, we begin to observe variations in the orbit in that the trajectories no longer sit on top of one another and the change in radius becomes erratic (as seen in Figure \ref{cm-100k}). As we decrease the velocity further we observe highly erratic orbits as in Figure \ref{cm-76k} with initial velocity 76,000 $ms^{-1}$ and then when the initial velocity falls below approximately 50,000 $ms^{-1}$ we see ejections. For Figures \ref{cm-1.6m}, \ref{cm-256k}, \ref{cm-100k}, \ref{cm-76k} and \ref{cm-50.6k}, I again show the trajectory of the test mass in the left pane and the variation of the distance to the origin for each time step in the right pane and do not show the binary orbit in the trajectory panes.  \\ 
\noindent \\
Note that these transitions from one regime to another are not sharp and I could have picked other thresholds. In particular the transition from erratic orbits to ejections is very blurred as ejections increase in likelihood with the number of orbits simulated. It should also be noted that at this initial position ($10^8m$), no captures / collisions were observed (although these can be manufactured by careful/deliberate selection of the initial velocity).\\ \\
In this section, I have shown that a Newtonian 3 body system can produce precession, chaotic behaviour and ejections and found that collisions are very rare and I can now construct a set of benchmark behaviours (which in practice means a set of initial conditions) against which to measure the behaviour generated by other gravitational models.\\ \\

\section{Chaos} \label{chapchaos}
\begin{figure}[!ht]
\includegraphics[width=145mm]{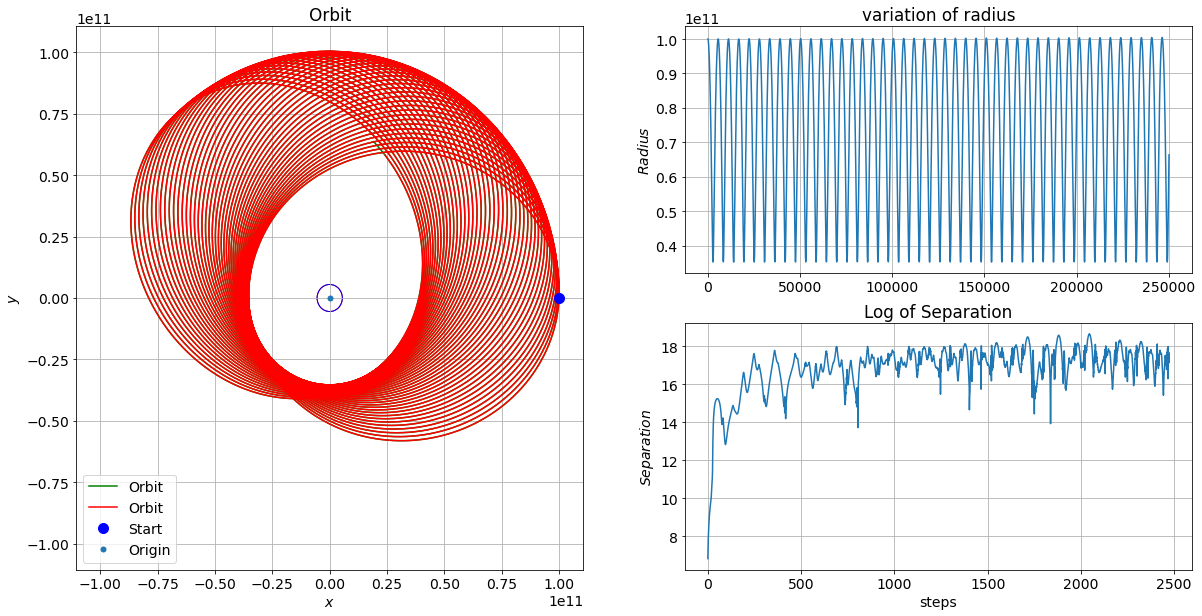}
\caption{Trajectory, Variation of Radius and Log of Separation, Initial velocity: $37,500ms^{-1}$ }
\label{cm chaos}
\end{figure}
\begin{figure}[!ht]
\includegraphics[width=145mm]{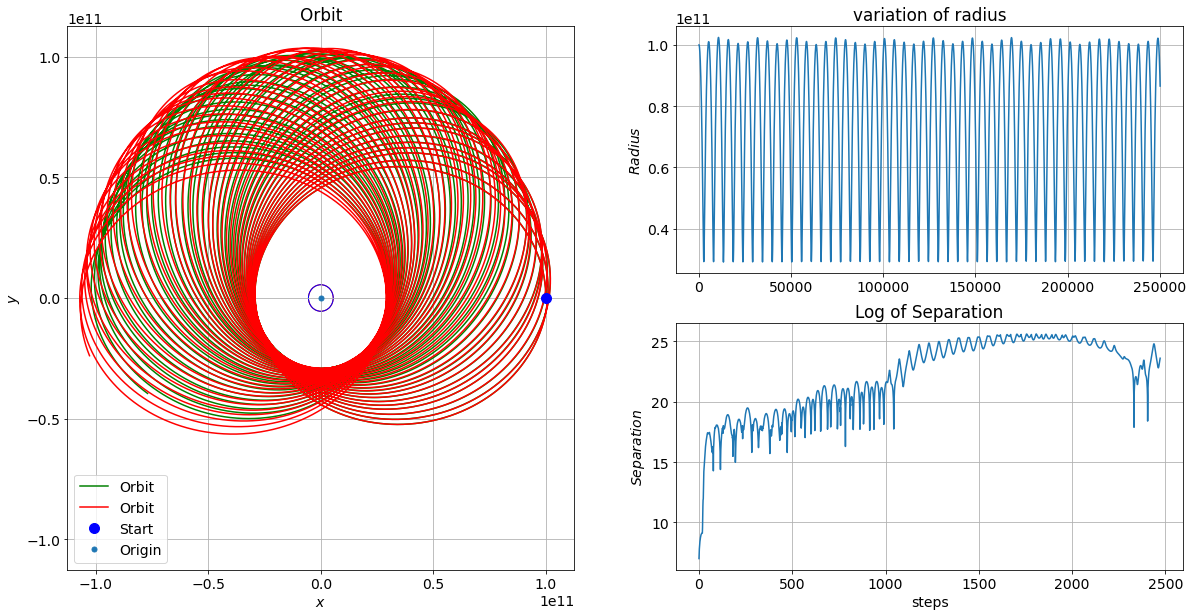}
\caption{Trajectory, Variation of Radius and Log of Separation, Initial velocity: $35,000ms^{-1}$.  The left panel shows both particle trajectories, one in red and one in green }
\label{cm chaos3.5}
\end{figure}
\begin{figure}[!ht]
\includegraphics[width=145mm]{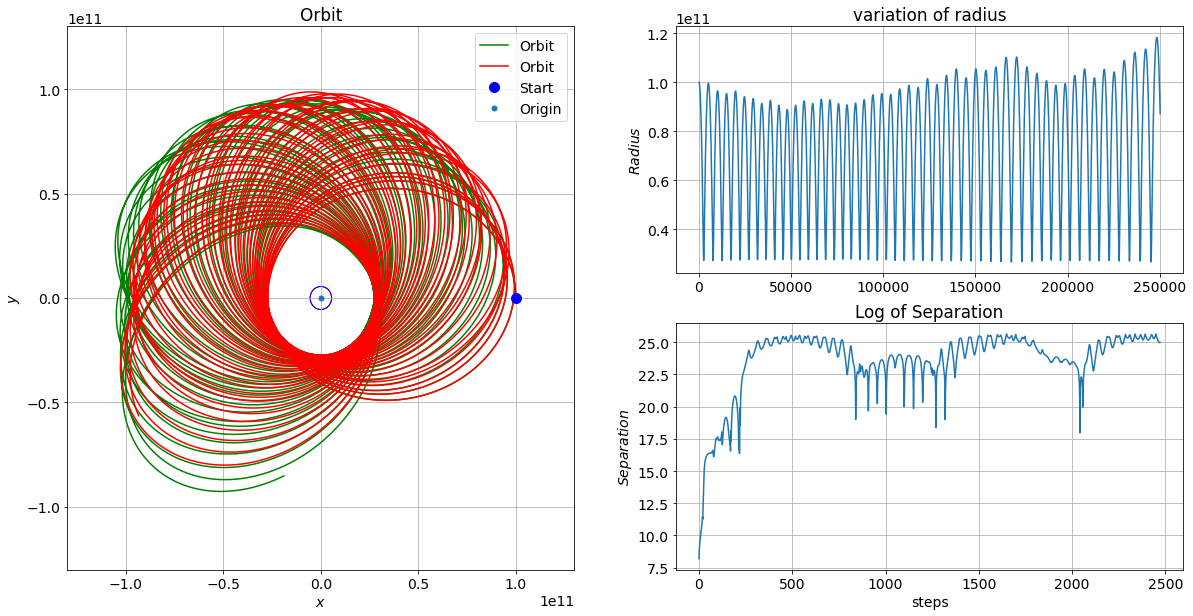}
\caption{Initial Velocity $34,000 ms^{-1}$}
\label{cm chaos 4}
\end{figure}
\begin{figure}[!ht]
\includegraphics[width=145mm]{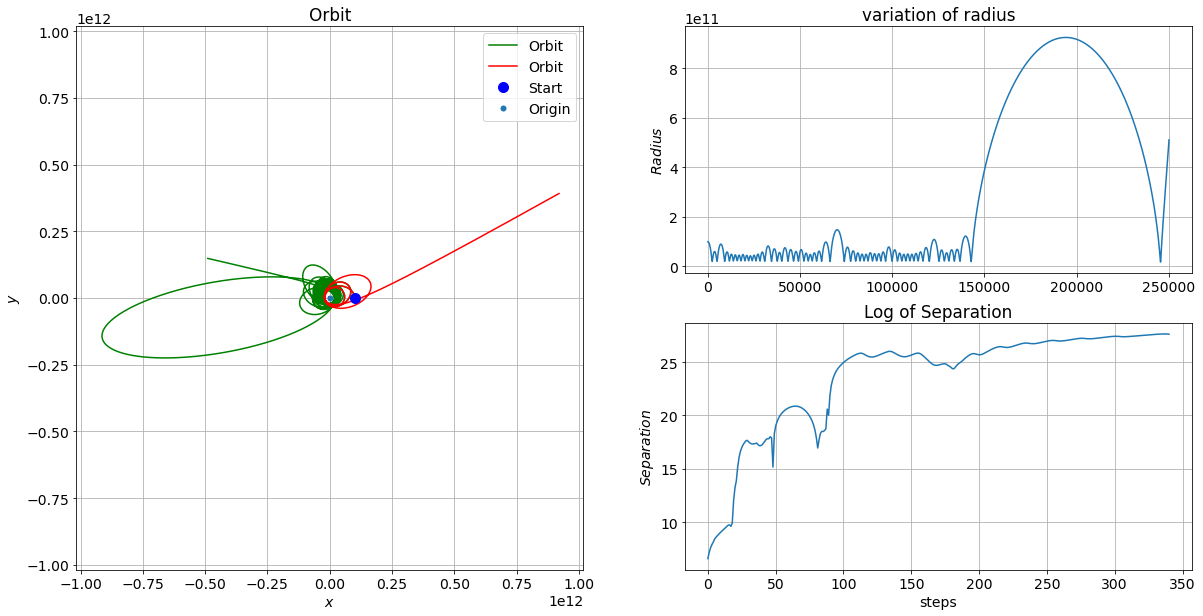}
\caption{Initial Velocity $30,000 ms^{-1}$}
\label{cm chaos 5}
\end{figure}

\noindent
In Figure \ref{cm chaos}, using the same initial conditions as in Figure \ref{CM with en}, I explicitly measure Chaos for the first time. In order to create a more intuitive feel for chaotic behaviour, I have introduced a slight variation of the standard Lyapunov Characteristic Exponent method commonly used in the literature as described in \cite{Tancredi_2001}. As per the standard method, I have simultaneously set off two test masses, the second having a distance from the origin a factor of ($1 + 10^{-10} $) greater than the first (in this scenario 10 metres apart) but the same total energy and direction as the first particle, and I then measure the evolution of the separation of these trajectories. I sample the spatial position of the first particle at regular intervals (in this case every 100 steps) and at each sample point I then search for the point of closest approach of the trajectory of the second mass and store this distance. Whilst clearly we expect geodesics to spread out in GR since the deviation is proportional to the Riemann tensor, what I am interested in is whether the rate of deviation varies for different models of gravity and hence comparing the Lyapunov exponent for different models under one common scenario.  (I will use this method throughout this paper)\\ \\ 
In this way, I generate a series of closest distances between two curves at regular intervals along the first curve, despite the fact that for a given point of closest approach, the time parameters (step numbers in the numerical process) may be very different along the path of the first mass and the path of the second mass. The actual exponent $L_e$ is then computed as usual and is given by: 
\begin{align}
L_e = \frac{1}{n} \sum_n abs \bigg( \frac{S[n] - S [n - 1 ]}{S[n - 1]} \bigg)
\end{align}
where $S[n] $ is the distance of separation between the two curves at time step $n$ along the first curve. By computing absolute separations between the curves, one avoids having multiple exponents (one for each dimension) and therefore avoids having to go through periodic Gram-Schmidt re-orthonormalisation procedures which would otherwise be necessary. (Gram-Schmidt re-orthonormalisation procedure constructs an orthonormal basis from any set of basis vectors; this is necessary here since the curvature of space means that the direction of greatest increase in deviation changes along the geodesic) \\ \\
In Figures \ref{cm chaos}, \ref{cm chaos3.5}, \ref{cm chaos 4} and \ref{cm chaos 5}, the left panel shows the trajectory of the test mass in the X / Y plane (in green) and its companion (in red), the top right panel shows the distance of the test mass to the origin at each time step and the bottom right panel shows the log of the separation between the two test masses every 100 time steps and all three panels represent 250,000 time steps or approximately 50 orbits.\\ \\ 
I then take these separations and plot their time series.  I will also measure the step number (effectively a time index) at which the separation reaches $10^3$, $10^4$, $10^5$ and $10^6$ times the initial separation and I will measure the standard deviation of the separation series. In fact I will take six readings of standard deviation of the log of the separations:
\begin{itemize}
\item the standard deviation of the whole separation series (log)
\item the standard deviation of the separation series from the start to the index point at which the separation reaches $10^3$ times the initial separation (log)
\item the standard deviation of the separation series from the index point at which the separation reaches $10^3$ times the initial separation to the index point at which the separation reaches $10^4$ times the initial separation (log)
\item the standard deviation of the separation series from the index point at which the separation reaches $10^4$ times the initial separation to the index point at which the separation reaches $10^5$ times the initial separation (log)
\item the standard deviation of the separation series from the index point at which the separation reaches $10^5$ times the initial separation to the index point at which the separation reaches $10^6$ times the initial separation (log)
\item the standard deviation of the separation series from the index point at which the separation reaches $10^6$ times the initial separation to the end of the separation series (log)
\end{itemize}
The combination of the modified Lyapunov exponent, the time indices at which the 4 thresholds of separation are reached and the six standard deviations of the logs of the deviation measures are most of the metrics that I will use to compare the behaviour of the behavioural differences between various gravity models.\\ \\ 
For chaotic trajectories I would expect that after some time, the separation between the particles would grow exponentially. In Figure \ref{cm chaos}, I have plotted both trajectories, the first in green and the second in red, but as the separation between the two trajectories remains bounded we only see the red orbit which completely covers the green line at the resolution of the Figure. I also show the log (base $e$) of the separation, which after rising initially stabilises at a constant value around 17 (which corresponds to approximately 40 million metres  or 0.04$\%$ of the semi major axis of the orbit). I then make a small change in initial conditions, again reducing the initial velocity, this time  from $37,500ms^{-1}$ to $35,000ms^{-1}$ and now we see a second trajectory in Figure \ref{cm chaos3.5} in red showing how sensitive the system is to initial conditions. We also see that the separation grows meaningfully and exhibits some structure.\\
\noindent \\
In Figures \ref{cm chaos 4} and  \ref{cm chaos 5}, I have replicated the initial conditions from Figure \ref{cm-34000}, and we see that with the same initial starting positions as used in Figure \ref{cm chaos} and relatively small differences of initial velocity (37,500$ms^{1}$ vs 34,000$ms^{1}$ and 30,000$ms^{1}$), we observe very different paths and much greater and more volatile separations which diverge in Figure \ref{cm chaos 5} and would diverge in Figure \ref{cm chaos 4} if the simulation were run for a few more iterations. \\

\chapter{Results} \label{chapresults}
For each of the ten models I consider, I have run 72 scenarios, and for each of these I compute 16 different metrics and produce two graphical figures. The first of the two figures  (for each of the 72 cases) is the trajectory(ies) of the test mass in the X/Y plane as seen from a position on the positive Z axis. The trajectories show the behaviours of two test masses each started $(10^{-10} \cdot X_0)$ apart, where $ X_0$ is the initial distance from the test mass to the CoM of the binary (which is chosen as the origin); the two masses have equal total energy, are set off parallel to each other and their paths are respectively coloured red and green. Where only a red path is visible in the graphic, this is because the paths are indistinguishable. The second figure shows the distance of the first test mass to the origin at each time step (which is determined by one iteration of the numerical process).\\ \\
The 16 computed metrics are:
\begin{itemize}
\item the greatest distance from the origin reached by the test mass during the simulation
\item the point of closest approach of the particles to the origin at any time during the simulation
\item the status of the particle at the end of the simulation, which can be one of three possible states: in orbit, ejected or captured
\item the step number of the ejection or capture event, or the total number of steps in the simulation if no event occurred
\item the modified Lyapunov exponent
\item the precession rate of the first test mass
\item the remaining 10 metrics all relate to the standard deviation of the log of the separation of the two test masses, sampled every 100 steps, as discussed in Chapter \ref{chapchaos}
	\begin{itemize}
	\item the standard deviation over the whole path
	\item the step number at which the separation has grown by a factor of $10^3$: T3 
	\item the standard deviation from step 0 to T3
	\item the step number at which the separation has grown by a factor of $10^4$: T4
	\item the standard deviation from T3 to T4
	\item the step number at which the separation has grown by a factor of $10^5$: T5
	\item the standard deviation from T4 to T5
	\item the step number at which the separation has grown by a factor of $10^6$: T6
	\item the standard deviation from T5 to T6
	\item the standard deviation from T6 to the end of the simulation
	\end{itemize}
\end{itemize}
As discussed earlier, all simulations involve releasing the test mass from a point on the X axis with velocity in the positive Y direction for prograde orbits and in the negative Y direction for retrograde orbits. The angular velocity of the binary is determined by its mass, separation and by which model is being used and is always in an anticlockwise direction. However, 9 of the 72 scenarios have the test mass released very close to $Mass1$ (2 percent of the separation of the binaries) and hence the test mass orbits $Mass1$ which itself is in orbit with/around $Mass2$. Whilst I do run the 0.5 PN (Schwarzschild correction) for these last 9 scenarios, it is pretty clear that this spacetime does not satisfy the Schwarzschild conditions, so any results obtained from these runs are more of a curiosity than data from which to draw insight.\\ \\
Appendix \ref{results-tables} contains the tabulated results of this project. The first forty tables each present data for one of the separations considered for each of the models considered. For instance, the results in Table \ref{Newt-no-prop-4} are for the Newtonian model with the field propagation time effect turned off in the $10^{4}m$ separation case. \\ \\
In Table \ref{Newt-no-prop-4} (and all the tables in Appendix \ref{results-tables} up to Table \ref{2.5-WF-13}) each box represents a range of release velocities at a given release distance and these velocities are shown in the second column beside each row of results; the release distances are also shown on the second column above the velocities. In the case of Table \ref{Newt-no-prop-4}, the top box is for $x = 10^{7}$ i.e. the test mass was released at  $10^{7}m$ from the origin at velocities ranging from 5,160,000$ms^{-1}$ to 200,000$ms^{-1}$. The minimum orbit is given in metres, the precession rate is given in radians per orbit and the other metrics are dimensionless.\\ \\
Due to the large amount of data generated, most of the Tables and Figures are provided in the Appendices rather than in this chapter.\\
\section{The Weak(er) Field Case}
In Figures \ref{results1} and \ref{results2}, I show the trajectories and oscillations of the distance of the test mass to the origin over 250,175 time steps which equates to approximately 25 orbits (although for other velocities, the relationship between time steps and number of orbits will vary) for all 10 models in the weakest field case studied. The weakest field case studied in this paper is for a separation of the binaries of $R = 1.1 \cdot 10^{13}m$ and a distance from the test mass to the CoM of the binary (the origin) being approximately 1,000 times greater that the separation of the primaries: $X = 10^{16}m$. In Figure \ref{results1} the test mass is set off with a velocity close to that required for a circular orbit whilst in \ref{results2} the test mass is set off with a highly elliptical orbit (lower initial velocity). As we would expect in the weak field case, these two pages of graphics are very dull and it is hard to make out any differences in the pictures between any of the gravity models. Note that, as mentioned previously, I refer to the Newtonian model as 0 PN and the Schwarzschild Model as 0.5 PN.\\ \\
This visual comparison of trajectories and distance oscillation whilst not terribly enlightening in this case, does I feel provide a good grasp of the physics and indeed was one of the ways of getting more confidence in the output of the code.\\  \\ 
A more quantitative comparison between the models (for the case shown in Figure \ref{results2}) is provided in Table \ref{tabresults2}, where we observe that when field propagation time is not taken into account, the 1 PN, 2 PN and 2.5 PN metrics are identical but when field propagation is taken into account there are slight differences. The 1 PN model, even in this weak field case does show differences from the Newtonian model and the 0.5 PN also shows differences. \\ \\
The fact that the 1 PN, 2 PN and 2.5 PN models give identical results in the weak field case is what we would have expected, and it is perhaps a little surprising that we can see any difference between the Newtonian and 1 PN case. However we would expect this difference to fall as the field is weakened further. This was indeed the case when the models were run with the test mass 1,000 times further out and with correspondingly lower initial velocity: in that case the difference in minimum orbit was only visible at the 11\textsuperscript{th} significant figure as against 4\textsuperscript{th} significant figure in Table \ref{tabresults2} and the Lyapunov exponent was different in the  3\textsuperscript{rd} significant figure rather than the 2\textsuperscript{nd}. 
\begin{table}[!ht]
\centering
\includegraphics[width =1 \textwidth]{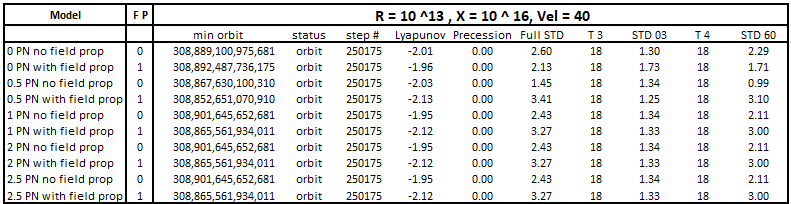}
\caption{Results table for all models where R = 1.1 $\cdot 10^{10}m$, X = $10^{10}m$, Velocity = $40ms^{-1}$\\ Tables \ref{tabresults2}, \ref{tabresults3}, \ref{tabresults4}, \ref{tabresults5} and \ref{tabresults6} show from left to right: The minimum orbit reached by the test mass, its status at the end of the run, the step number when the run ended, the Lyapunov exponent, the Precession Rate, the standard deviation of the log of the separation of the two test masses over the full run, the step number at which the test mass separation increased by a factor of $10^3$ (T3), the STD of the log of the separation up to T3, T4 and the STD from T6 to the end of the run}
\label{tabresults2}
\end{table}
\section{The Strong(er) Field Case}
In contrast, at the strong(er) field end of the spectrum, when the binary separation is $1.1 \cdot 10^{4}m$ and the test mass is released from $10^5m$, we see in Figure \ref{results3}, highly differentiated behaviour between the models. Both Newtonian models (with and without field propagation time effects) result in the test mass being captured by one of the black holes whereas the 1 PN models and 2 PN models are clearly exhibiting the repulsive force which I described in section \ref{PNapprox}, and we see that the field propagation time effect in all cases is imparting angular momentum (and energy) to the test mass resulting in greater distances of closest approach or a delayed capture. We can also begin to see here that the 0.5 PN model does not exhibit the repulsive behaviour seen in the 1, 2 and 2.5 PN models.  The corresponding numerical results are shown in Figure \ref{tabresults3}, and again the 2 PN and 2.5 PN cases are indistinguishable.\\ 
\begin{table}[!ht]
\includegraphics[width =1 \textwidth]{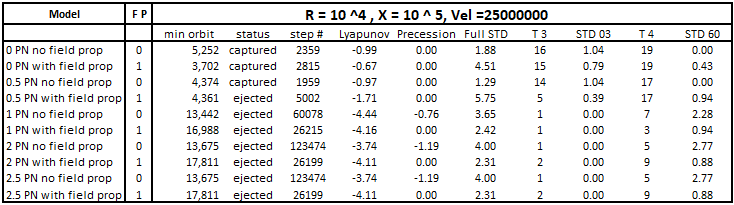}
\caption{Results table for all models where R = 1.1 $\cdot 10^{4}m$, X = $10^{5}m$, Velocity = 25,000,000$ms^{-1}$}
\label{tabresults3}
\end{table}
\noindent \\ 
In the interests of brevity, I will not show or discuss the full comparison of 10 models for all 72 scenarios, but only discuss a subset where I hope there is something to be gleaned from the Figures or Tables.\\ \\ 
Staying with the strong field case (R = 1.1 $\cdot 10^{4}m$, X = $10^{5}m$), and allowing the initial velocity to increase, we observe in Figures \ref{results4}, \ref{results5} and \ref{results6} and Tables \ref{tabresults4}, \ref{tabresults5} and \ref{tabresults6} a range of phenomenon. In Figure \ref{results4} we see strong precession even in the purely Newtonian case as discussed in section \ref{initchoice}, and we see that the rate of precession increases for the 1, 2 and 2.5 PN models (in fact from table \ref{tabresults4} we see that the rate of precession has approximately tripled). \\ \\
We continue to observe that the 0.5 PN model (Schwarzschild correction) behaves completely differently, this is not surprising as we know that it only has an increased attractive force and no repulsive element. Unsurpringly therefore, we observe that the field propagation time effect is strongly repulsive in the 1, 2 and 2.5 PN cases and somewhat less strongly in the 0 and 0.5 PN cases. We observe similar phenomena in Figure \ref{results5} and Table \ref{tabresults5} but in Figure \ref{results6} and Table \ref{tabresults6} we really begin to observe the magnitude of the transfer of angular momentum from the binary to the test mass that is present when the model is run with field propagation time effects turned on, as we see the orbits of the test mass spiral outwards for all five gravity models. \\
\begin{table}[!ht]
\includegraphics[width =1 \textwidth]{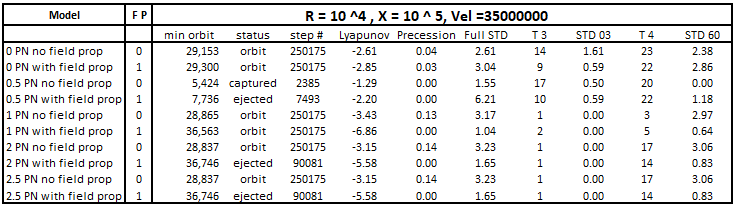}
\caption{Results table for all models where R = 1.1 $\cdot 10^{4}m$, X = $10^{5}m$, Velocity = 35,000,000$ms^{-1}$}
\label{tabresults4}
\end{table}
\begin{table}[!ht]
\includegraphics[width =1 \textwidth]{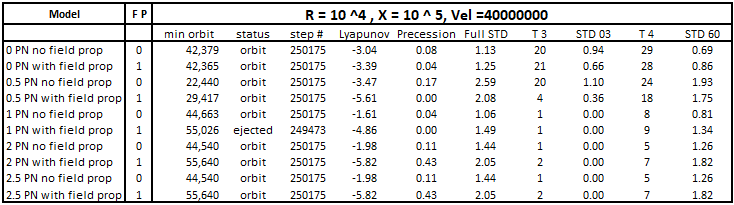}
\caption{Results table for all models where R = 1.1 $\cdot 10^{4}m$, X = $10^{5}m$, Velocity = 40,000,000$ms^{-1}$}
\label{tabresults5}
\end{table}
\begin{table}[!ht]
\includegraphics[width =1 \textwidth]{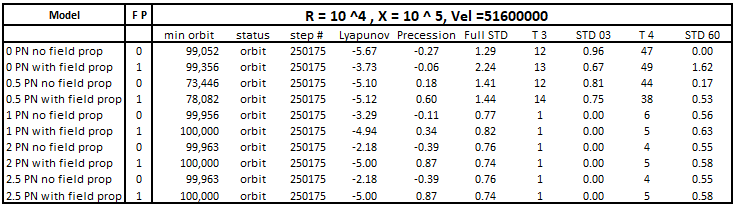}
\caption{Results table for all models where R = 1.1 $\cdot 10^{4}m$, X = $10^{5}m$, Velocity = 51,600,000$ms^{-1}$}
\label{tabresults6}
\end{table}  
\noindent \\ 
In Tables \ref{av-res-0PN}, \ref{av-res-0.5PN}, \ref{av-res-1PN}, \ref{av-res-2PN}, \ref{av-res-2.5PN}, I have computed average statistics for each of the five gravity models and presented these separately for the with and without field propagation time effects in order to allow an overall comparison between the models. I have separated the scenarios into two categories, those where the test mass is still in orbit at the end of the run and those where the test mass has either been captured or ejected, as these categories represent such different behaviour that it is not always enlightening to average across these categories. From these Tables we can make a number of observations:
\begin{itemize}
\item where ejection or capture occurs, the average time to the event for all the gravity models except the Schwarzschild based model (0.5 PN) is greater when field propagation time effects are turned on than when they are turned off. As ejections tend to occur just after very close approaches to one of  the primaries, and captures tautologically occur just after a very close approach to one of the primaries; the observed behaviour is consistent with our previous observation that field propagation time effects provide extra angular momentum / energy to the test mass and thus disfavour or delay the close approaches
\item where the test mass remains in orbit, the minimum orbital radius for all the gravity models except the Schwarzschild based model (0.5 PN) is greater when field propagation time effects are turned on than when they are turned off, but the effect is very small
\item where the test mass remains in orbit, the Lyapunov exponent is more negative for all the gravity models except the Newtonian model when field propagation time effects are turned on than when they are turned off, suggesting that these orbits are less chaotic, again consistent with being kept away from close contact with the primaries
\item where the test mass remains in orbit, there seems to be no obvious pattern in the rate of separation between the two test masses 
\item there is a consistent increase in the number of runs ending in ejection or capture as we move up the gravity models, i.e. the Newtonian Models has the fewest ejections or captures, followed by the Schwarzschild model followed by the 1 PN and followed by the 2 and 2.5 PN model. This feature is true both when the field propagation time effects are turned on or off
\end{itemize}
It is important to note that for many runs (strictly speaking all runs), the boundary between stable orbit and capture or ejection is not crisp and many runs that are still in orbit after a given number of orbits may be captured or ejected in a longer simulation. \\ \\
In the spectrum of scenarios between the strong and weak field cases, no new phenomena emerge and the evolution between the behaviours at the two extremes of field strength seems progressive. We do see some scenarios were the difference between the 1 PN and 2 PN models look very pronounced such as for the case of $R = 10^ 4m$,  $X = 10^7m$ and velocity $=$ 700,000$ms^{-1}$ but in actuality, this is one situation where, due to the chaotic nature of the system, a very small difference in position or velocity (due to a very slightly different acceleration) results in a fractionally closer approach to one of the primaries, which in turn leads to a greater transfer of angular momentum and hence a meaningfully different trajectory thereafter, as we see in Figure \ref{intermediate}. \\ \\
The most significant result remains the substantial transfer of angular momentum / energy in the strong(er) field cases due to the field propagation time effects, a phenomenon that we do not observe when we do not consider field propagation time.\\

\section{Orbits Around One of the Primaries}
\begin{figure}[!ht]
    \centering
    \subfigure[]{\includegraphics[width=0.50\textwidth]{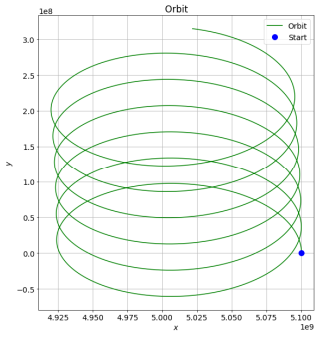}}
    \subfigure[]{\includegraphics[width=0.324\textwidth]{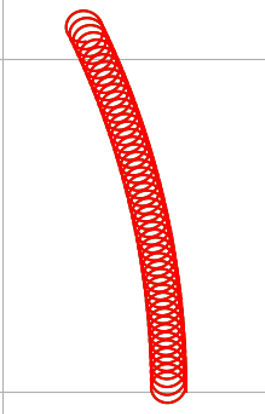}} 
    \caption{Test Mass in Stable Orbit around one of the Primaries (a) 6 Orbits,   (b) circa 100 Orbits. \\ (b) Shows both test masses but these do not deviate enough for the green mass to be seen}
    \label{mass1orb}
\end{figure}

In the special cases of orbits around one of the black holes: for nearly circular orbits as shown in Figure \ref{mass1orb}, the behaviour is stable and no ejections or captures were observed although the 1, 2 and 2.5 PN models exhibited some low frequency oscillation not observed in the Newtonian model; the 0.5 PN model has more pronounced oscillations but as previously discussed, this model is not expected to be appropriate to such a configuration. The field propagation time effects are very muted across all gravitational models which is what we would expect since in this case any deviation in field strength and direction would be constant (the deviation would only be constant in terms of angle away from the CoM and not along a constant direction in space). In the stronger field case, when the primaries are separated by $10^7m$ and the test mass orbits at a distance of $10^5m$ from one of the black holes, we do see captures and ejections for all models in the case of highly elliptical models when using the field propagation time effect. However, in these cases the distances of closest approach to the black hole were of the order of a few Schwarzschild radii and it is not clear without further work whether there was sufficient resolution in the model precision used to handle such extreme scenarios. \\
\section{Statistical and Averaged Results}  
A review of all the tables in Appendix \ref{results-tables} and Table \ref{luap-tab} in particular, as well as visual inspection of the (large number of) Figures suggests that whilst there is a small increase in the Lyapunov exponent from the Newtonian model to the 1 PN and then 2 $\&$ 2.5 PN models and a little larger increase to the 0.5 PN model, none of these increases are substantial (of the order of 1$\%$ or 2$\%$) and there is a somewhat greater decrease in chaotic behaviour for all models when the field propagation time effects is included, which is of order 0.5$\%$ to circa 8$\%$.  Table \ref{luap-tab} shows the sums of Lyapunov exponents over the 63 scenarios for each model (there are only 63 scenarios as I have excluded the 9 scenarios that orbit one of the primaries rather than orbiting the binary system as a whole). Note that the Lyapunov exponent is usually a negative number so a larger number may be just `less' negative. 
\begin{table}[!ht]
\centering
\includegraphics[width =0.4 \textwidth]{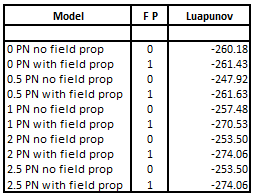}
\caption{Sum of Lyapunov Exponents over the 63 Scenarios}
\label{luap-tab}
\end{table}
\section{Outward Drift and Long Duration Simulation}
The proposal made in this paper that \emph{the momentum transfer due to field propagation time is a strong effect} needs to be quantified and then tested for its observational impact. Early analysis suggests that the rate of increase of orbital radius of the test mass due to field propagation time falls inversely with the distance from the binary but is positively correlated with the separation of the primaries; this feels intuitively reasonable, but more work needs to be done here. Tables \ref{drift-out} show extremely rough measures of the amounts of outward drift from the simulations (more precise estimates would require a little additional coding and some specific runs, which would be an important follow up to this paper).  The Tables are only shown for two gravity models: the Newtonian case and the 1 PN case as the 2 PN and 2.5 PN models give very similar results to the 1 PN results.\\
\begin{table}[!ht]
  \subtable[0 PN: $\%$ Change in Radius per Orbit]{\includegraphics[width=0.48\textwidth]{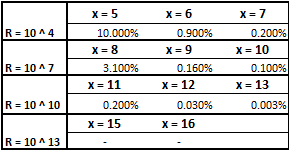}}
  \subtable[0 PN: $metres$ Change in Radius per Orbit]{\includegraphics[width=0.49\textwidth]{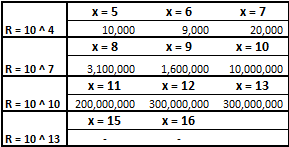}} \\
  \subtable[1 PN: $\%$ Change in Radius per Orbit]{\includegraphics[width=0.48\textwidth]{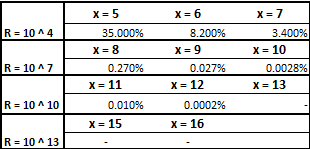}}
  \subtable[1 PN: $metres$ Change in Radius per Orbit]{\includegraphics[width=0.49\textwidth]{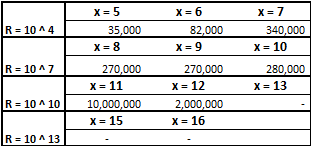}} 
  \caption{Outward Drift per Orbit}
    \label{drift-out}
\end{table}
\noindent \\
To give a concrete example, I set the model to run for two solar mass black holes separated by 110 million kilometres orbited by a test mass 10 billion kilometres from the origin and allowed the system to run for 1,000 orbits (at this distance, an orbit takes approximately 385 Earth years). I have run each of the 5 gravity models both with and without the field propagation time effect and we see from Figure \ref{long-run} and Table \ref{tab-long-run} that:
\begin{itemize}
\item in the no field propagation mode, the orbits are all relatively stable, in that their variation is of order $10^{-4} \cdot X_0$, where $X_0$ is the initial radius
\item  in the no field propagation mode, we observe a very slow drift outwards in radius for the 1 PN model, however the data set is too short (only 400,000 Earth years) to be conclusive; this drift is not present in the 2 and 2.5 PN models
\item   in the no field propagation mode, in the Newtonian model there is slow growth in amplitude of variation but again the data set is too short to be conclusive
\item all five of the models with field propagation time effects included show a very steady drift outwards in radius such that at the end of 1,000 orbits, the orbits have all grown by 0.3$ \%$ and with strikingly similar rate of growth
\item the growth rate in orbits for the field propagation mode is very steady as, to the naked eye, it is hard to see any deviation from a straight line in the upwardly sloping radius
\end{itemize}
\noindent \\
Note that the above runs took 30 hours of continuous processing on the AWS server.\\ \\
From Table \ref{drift-out} and the results of the `Long Run', a tentative hypothesis is that for a given separation of the binaries the orbits increase by a constant (very approximately) distance at each orbit (this proposition is clearly only supported by two of the rows in Table \ref{drift-out}, and for three of the other rows, the drifts are within the same order of magnitude, so more work is required here); and that within a certain range of binary separations, the outward drift is positively correlated with the binary separation. This seems plausible in the sense that, as the orbits widen the effect of the binary gets smaller, but the wider orbits last longer and hence the test mass `feels' the effect of the binary for longer (during each orbit), and therefore the distance/spatial  and time derivatives of the effect somewhat offset each other.\\ \\
If the growth of an orbit is a constant distance per orbit, then the rate of growth of orbits is inversely proportional to the orbital period and remembering that the period $T$ of a body in orbit with an orbital radius $r$ is given by: $ T \propto r^{3/2}$ then 
\begin{align}
&\text{}& \dv{r}{t}&\propto \frac{1}{T} &\text{}\\
&\Rightarrow&  \dv{r}{t}&\propto \frac{1}{r^{3/2}} &\text{} \\
&\Rightarrow&  \int_{R_0}^{R_{Max}} r^{3/2} \dd r&\propto \int^{T_{End}}_{T_{0}} \dd t  &\text{} \\
&\Rightarrow&  R_{Max}&\propto t^{2/5}  &\text{}
\end{align}
would see orbits grow approximately as $t^{2/5}$ where $t$ is time. I conclude from these observations that this phenomenon is not incompatible with cosmological observations, as extrapolation would suggest that starting from the initial conditions in the `Long Run' above, that over one billion years the orbit would grow of the order of 10$\%$ and only two or three times more over the lifetime of the Universe, and therefore we would still observe many bodies in orbit around binary systems but with a lower density of bodies in tighter orbits than other models might predict. However, much tighter orbits would be swept out very quickly. It should be possible to test the \emph{reduced density of bodies in tighter orbits} proposition with sufficient observational data.  These calculations are deliberately very imprecise as the assumption of a constant growth per orbit is, most probably at best, good to within an order of magnitude.\\
\begin{table}[!ht]
\includegraphics[width =1 \textwidth]{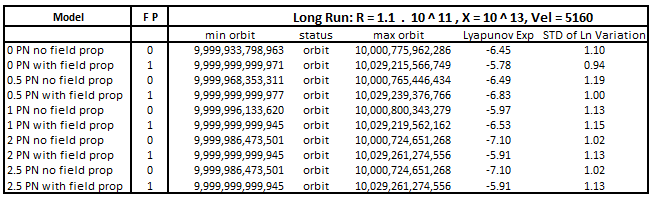}
\caption{Results table for all models where R = 1.1 $\cdot 10^{11}m$, X = $10^{13}m$, Velocity = 5,160$ms^{-1}$}
\label{tab-long-run}
\end{table}

\chapter{Conclusions and Next Steps}
\section{Conclusions}
In this paper, I set out to explore three related ideas:
\begin{itemize}
\item Whether the various GR models were major causes of chaotic behaviour in the Restricted Relativistic Three Body Problem over and above that exhibited in the Newtonian case
\item Whether the corrections to the Newtonian potential obtained from the Schwarzschild metric could be used as a low (processing) cost substitute for the traditional Post Newtonian models used in the literature
\item Whether including the field propagation time effects has much impact on the dynamics of the Restricted Relativistic Three Body system
\end{itemize}
None of the data obtained in the exercise suggested that any of the gravitational models were major sources of chaotic behaviour over and above the already chaotic behaviour of the Newtonian Three Body system. \\ \\
With few exceptions, the Schwarzschild based model was not a good proxy for the behaviour generated by the Post Newtonian models. This is perhaps not surprising, since there is little (no?) evidence of the community using this approximation in the way proposed. \\ \\
I did however find that the impact of field propagation time / retarded potentials in the strong field case was repulsive in the prograde case and attractive in the retrograde case, and that in the strong field case, the magnitude of the effect was very high, with orbits increasing by tens of percentage points per orbit. I built an initial model and tried to show in Figure \ref{doppler} a graphical interpretation for the momentum transfer as a kind of `Doppler effect' and then in Figure \ref{projected} a more sophisticated model allowing for the gravitational equivalent of the Liénard Wiechert potential and whilst these simple models did provide some insights, I ultimately concluded that these simple models could not in fact fully replicate the observed effects in the full simulation. The magnitude of this effect is inversely related to the distance of the body / test mass to the centre of mass of the binary system and also positively related to the separation of the primaries. Even in the weaker field case, the effect is measurable over large timescales, and I therefore also propose that the effect should, in theory, be observable in the astronomical data. \\ \\
If these results can be reproduced, then they may be of interest to researchers studying accretion discs of binary systems and possibly other phenomenon with rapidly moving gravitational sources.\\

\section{Next Steps}
There are three next steps which can be somewhat run in parallel or even by different groups of people:
\begin{itemize}
\item Reproduce the results presented here as regards the effect of the time of propagation of the gravitational field
\item Run more simulations so that an empirical relationship (from the simulations) between the various parameters (masses and separation of the binary, location and velocity of test mass) and the amount of angular momentum / energy transferred can be established when time of propagation of the gravitational field is accounted for
\item Try to develop an analytical model to predict the amount of momentum / energy transfer due to the field propagation time effect.  This is probably best done initially for a test mass in a stable orbit around the binary, rather than in the general case
\end{itemize} 
In addition and if there is interest, I could tidy up the code and make it available through Open Source to future students as a learning / teaching aide.

\appendix
\chapter{The Model and its Implementation}
I have written the code used in all of the above simulations and as presented at the end of this paper as a multifunctional piece of code - a sort of Swiss Army knife.\\ \\
The code handles:
\begin{itemize}
\item Two different integration methods: Euler and Dormand Prince
\item Various gravitational models
	\begin{itemize}
	\item Newtonian
	\item Schwarzschild Corrections
	\item 1 PN
	\item 2 PN
	\item 2.5 PN
	\item Allows for Field Propagation time effects to be applied to all 5 of the above
	\end{itemize}
\item Is fully 3 dimensional as far as the test mass is concerned (although I have only shown results from the planar restricted system in this paper)
\item Has measures for speed of precession of the test mass
\item Has modified measure of chaos and Lyapunov Exponent
\item Can run as single-threaded routine which publishes results to the console, or as a multi-threaded batch mode, with a bulk scenario generator and which publishes output files to the file directory      
\item Has many graphing options      
\end{itemize}
\noindent \\
The Model was written in Python 3.8 using the Spyder 4.2.5 IDE. \\ \\ 
All runs were initially done on a Microsoft laptop running Windows 10 operating system with an 11\textsuperscript{th} Gen Intel(R) Core(TM) i7-1185G7 @ 3.00GHz, 2995 MHz, 4 Core(s), 8 Logical Processor(s) and 16GB of RAM, however this proved to be insufficiently powerful to run multiple scenarios of the 2 PN and 2.5 PN.  As a result, the final `production' runs were done on an AWS server running an Intel XEON Platinum 8124 CPU with 18 cores and 36 logical processors and 72GB of RAM.

\chapter{Accuracy of Integrator versus CPU run time}\label{accuracy}
\begin{figure}
    \centering
    \subfigure[]{\includegraphics[width=0.49\textwidth]{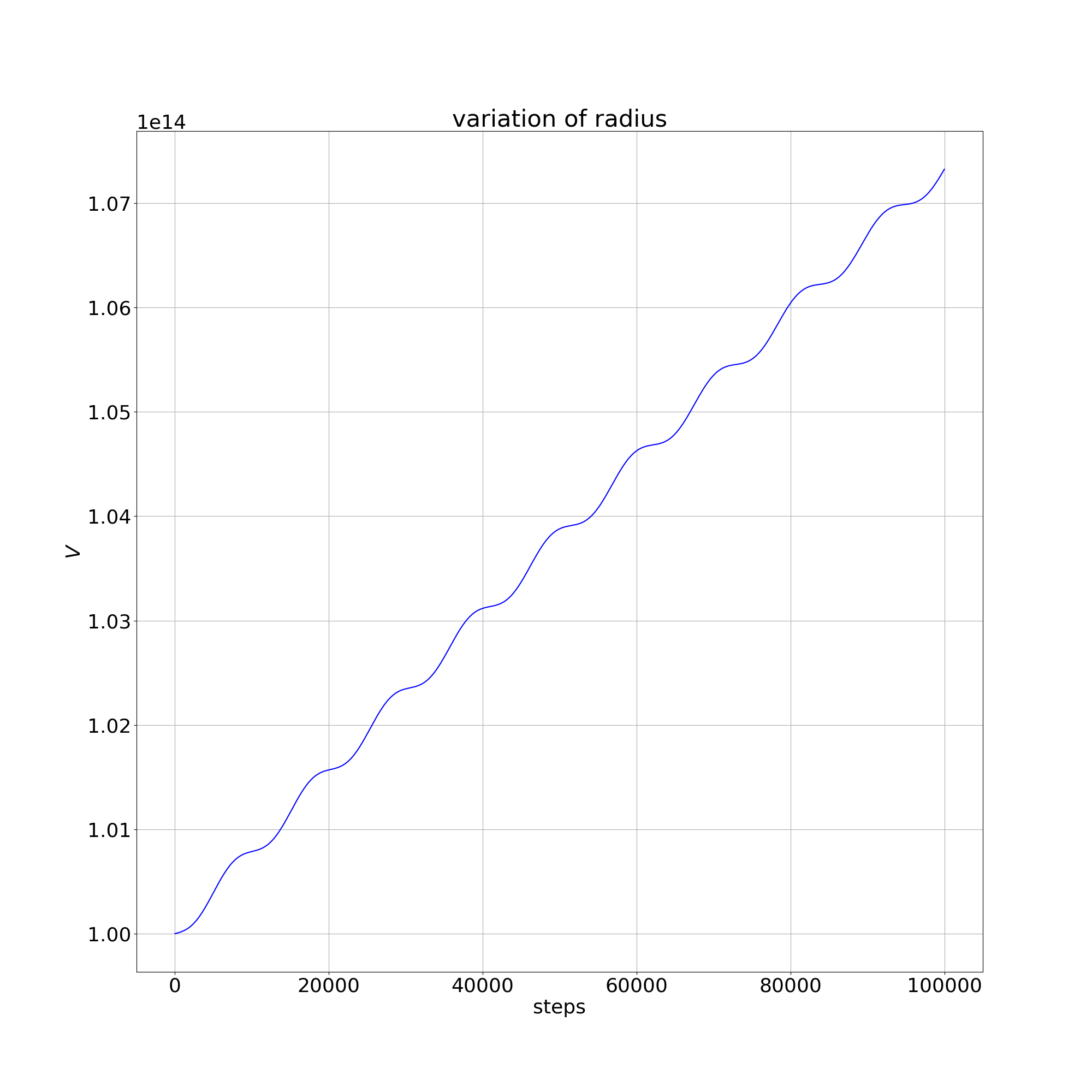}}
    \subfigure[]{\includegraphics[width=0.49\textwidth]{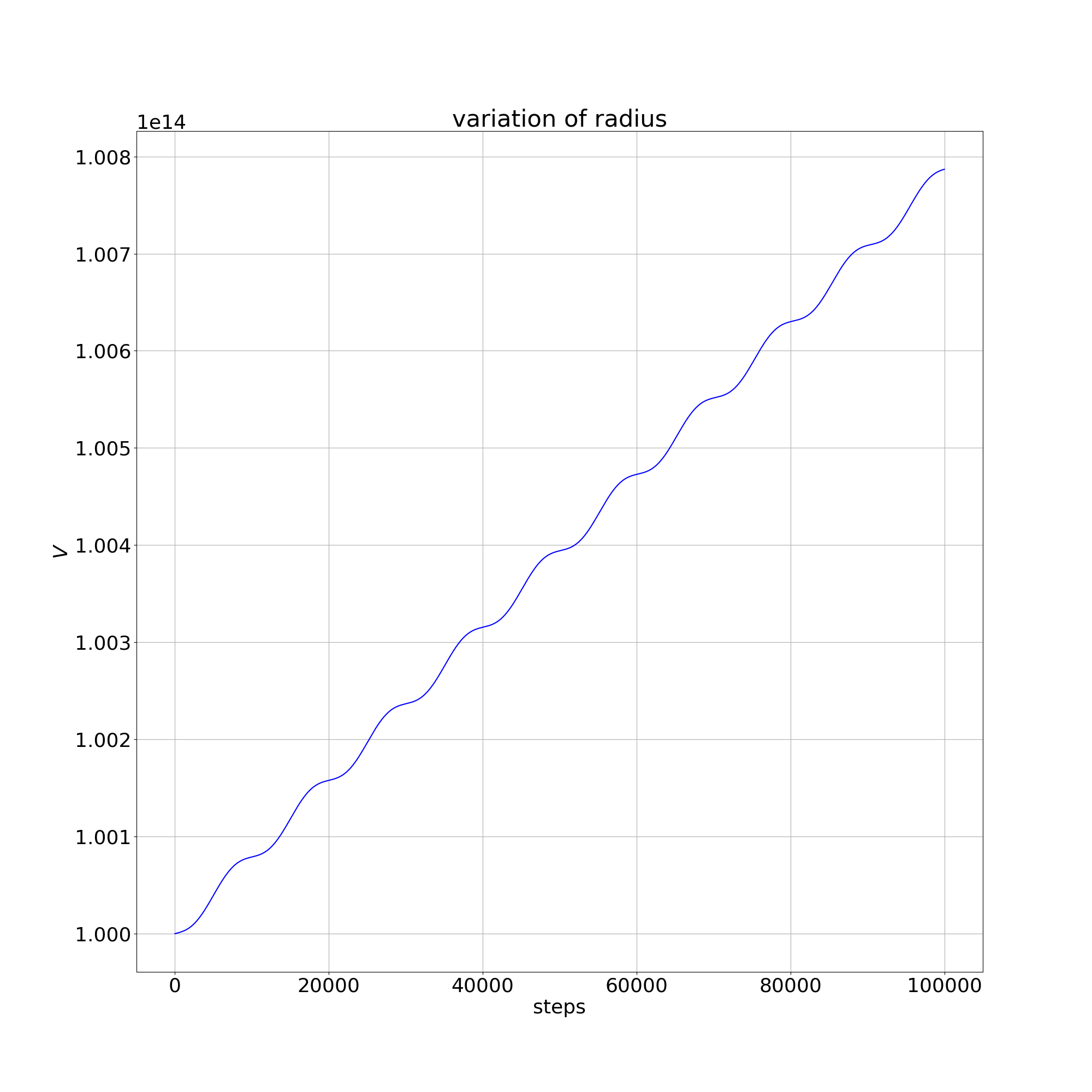}} \\ 
    \subfigure[]{\includegraphics[width=0.49\textwidth]{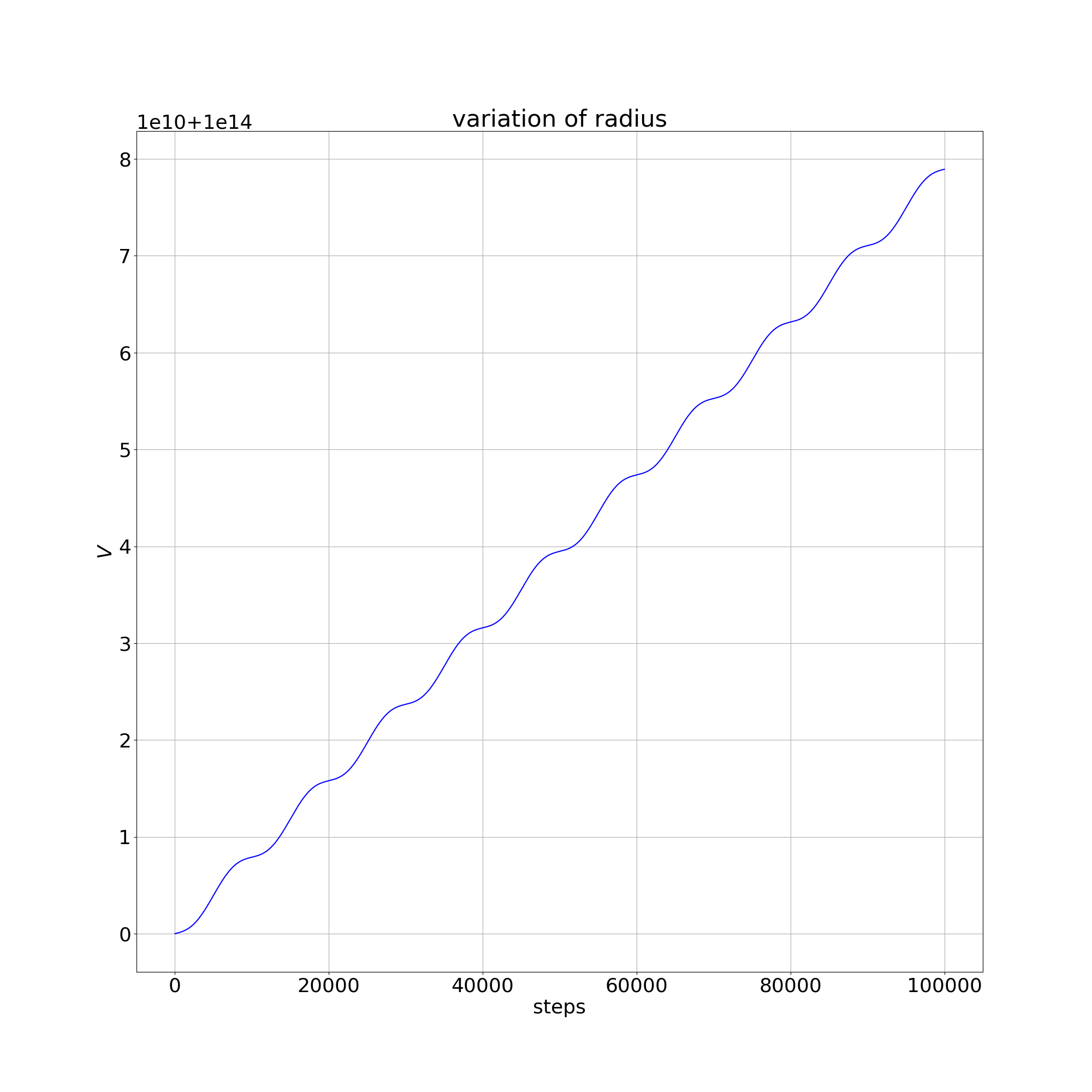}}
    \subfigure[]{\includegraphics[width=0.49\textwidth]{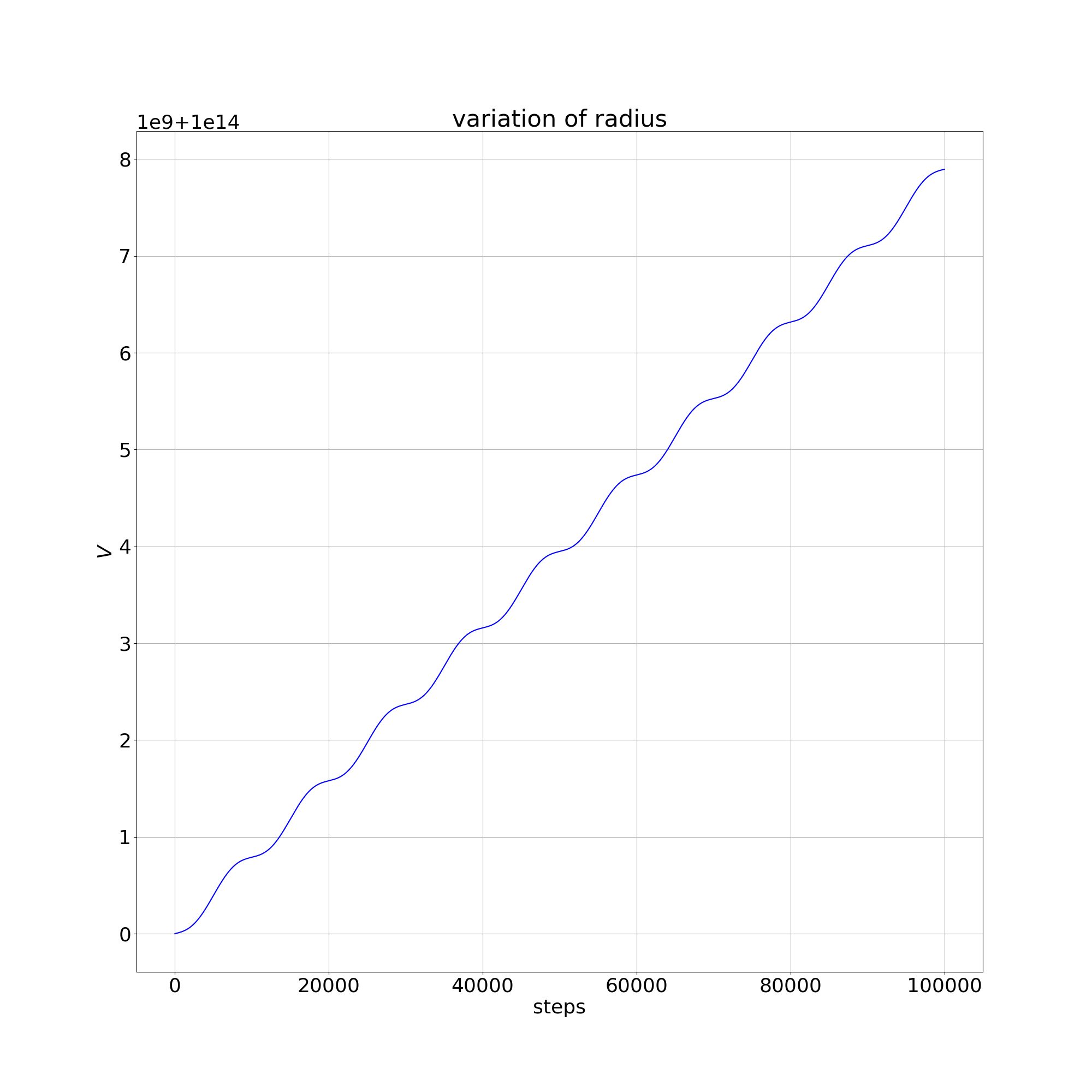}} 
    \caption{Steps per orbit = (a) 10,000  (b) 100,000  (c) 1,000,000 (d) 10,000,000}
    \label{fluct}
\end{figure}

\begin{figure}[!ht]
\includegraphics[width=0.99\textwidth]{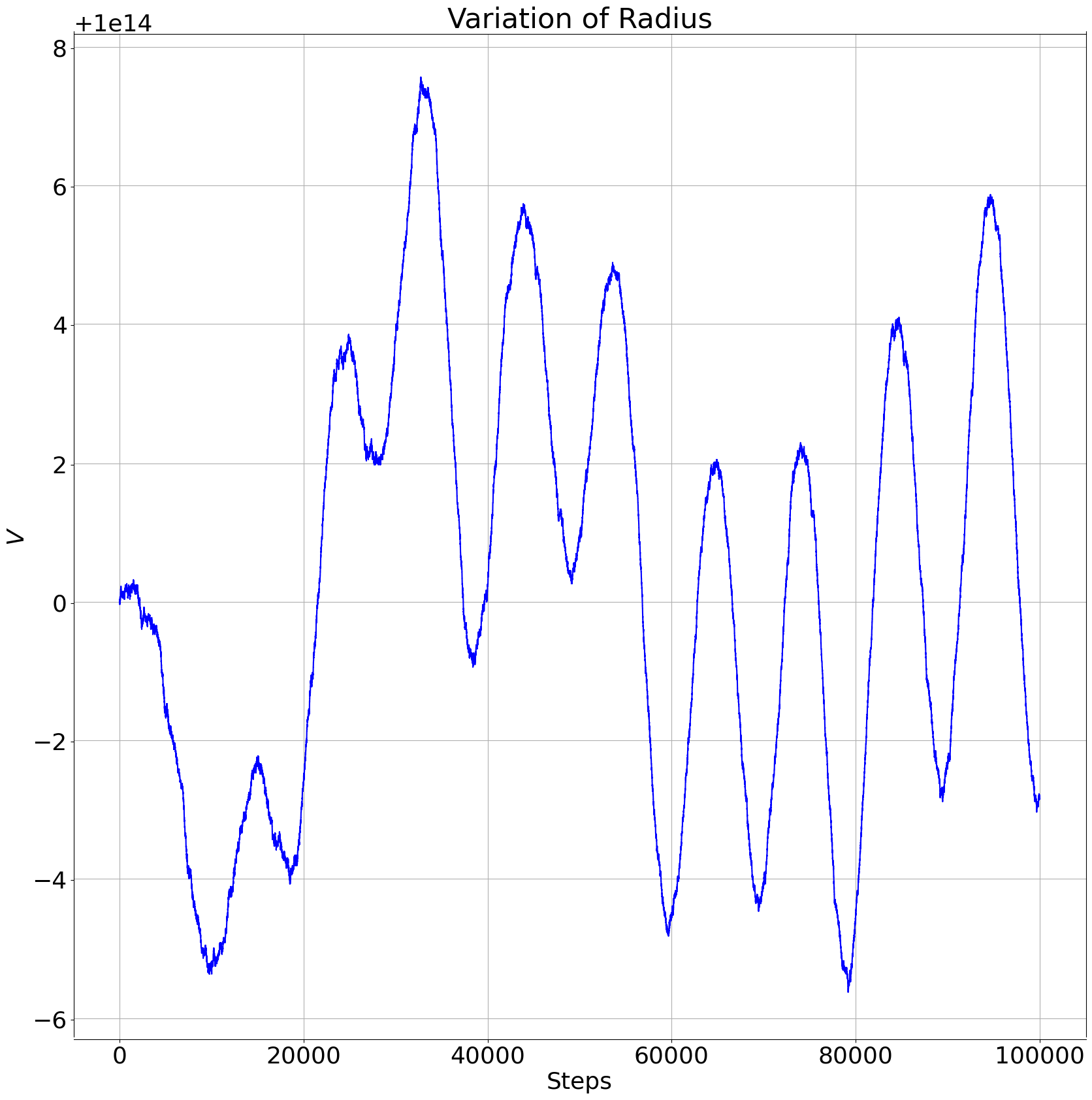}
\caption{Dormand Prince in Newtonian Framework for 10 orbits }
\label{dp}
\end{figure}

When attempting to solve differential equations numerically, it is necessary to use an `integration' technique, i.e. to start with a set of initial conditions, apply the derivatives for a small increment of time (or whatever the appropriate parameter is) in order to compute new values of the variables a short time later and then keep repeating the process over the desired parameter range. The simplest approach to this procedure is to use Euler's method, however, there are also a range of `Integrators' such as Heun's method and in particular tools based on the Runge-Kutta method.  The object of these `Integrators' is to provide accurate estimates of the path of the solution to the differential equation, which in the current case describes the path of the test mass.\\ \\
I have implemented two integrators: Euler's method and Dormand Prince implementation of Runge Kutta with adaptive step sizes.  A simple test of this code is to compute the velocity required to maintain a stable circular orbit for a given radius in a Newtonian framework. By setting the distance between the primaries to be small compared with the distance between the system CoM and the test mass, the binary can be approximated to a single static body i.e. the Newtonian two body problem.\\ \\ 
Starting with Euler's method and setting the mass of the two primaries to be equal at $2 \cdot 10^{30} kg$ each (one solar mass each), the distance between the primaries to be $10^6$ metres (approximately 160 times their combined Schwarzschild radii) and starting the test mass at position $X = 10^{14}m$. Using the Newtonian expression for velocity $ V =  (G \cdot (M1 + M2) / R )^{0.5} $ gives us an initial value for velocity equal to $1,633.90085378520$ metres per second, one can see from Figure \ref{fluct} an oscillation of the radius of the order of $10^9$ metres, when using a step size of 1,000,000 steps per orbit.  In fact not only is there an oscillation but there is a marked drift outwards of the orbit. This drift is unsurprising since circular orbits have constant acceleration and therefore any first order approximation will necessarily fail to account for the curvature and produce a constant error related in size to the step size. \\ \\
By searching the velocity space it is possible to find an initial velocity which does not cause the particle to drift out, but this change of initial velocity, whilst able to correct for the systematic drift, does so at the expense of increasing the amplitude of oscillations and is a generally unsatisfactory procedure in the sense of trying to compensate for an error / bias by introducing another offsetting bias.  \\ \\
By reducing the step size, i.e. increasing the number of steps per orbit, one can reduce the amplitude of oscillation of the orbit, with each power of 10 increase in the number of steps producing a factor of 10 reduction in the oscillations and drift. We see in Figure \ref{fluct} (a) the step size is 10,000 steps per orbit and the drift in radius is of the order of $1\%$ per orbit and the oscillations are an order of magnitude smaller than the drift; in (b) the step size is 100,000 steps per orbit and the drift in radius is of the order of $0.1\%$ per orbit and again the oscillations are an order of magnitude smaller than the drift; the same pattern repeats in (c) and (d) where each of the drift and amplitude of oscillations reduce inversely proportionally to the number of steps per orbit. The CPU run time for the 10,000,000 step run was approximately 930 seconds (for 3 orbits only) and the amplitude of oscillations was of the order of $10^{-6}$ radii or $10^8$ metres. One has to conclude that the Euler method is too inefficient to conduct precise simulations and therefore we have to look to other tools.
\\ \\
I compare these results with the Dormand Prince Runge Kutta method based integrator: keeping all of same parameters as those used in the Euler simulations and running 10,000 steps per orbit (this is not a like for like comparison as the Dormand Prince method has an extra parameter which allows an adaptive step size so 10,000 steps for Dormand Prince is more akin to 100,000 for Euler) and again considering the Newtonian framework, we observe almost no drift and the oscillations are of the order of $10^{-13}$ radii or $10$ metres, i.e. 7 orders of magnitude better precision than using Euler's method and the run time for 3 orbits was 22 seconds, so this second integrator is about 0.5 billion times more efficient in terms of accuracy per CPU cycle. Running the same parameters for 1,000 orbits does increase the level of noise of the orbital radius. The variation of radius graph is shown in Figure \ref{dp} for 10 orbits and using 10,000 steps per orbit. \\
\\ 
Note that although in the above test I am measuring the radii to as much as 14 significant figures, the values for the natural constants $c$ and $G$ are only specified to 9 and 6 significant figures respectively. However, as I have used the same specification of natural constants in constructing the Newtonian benchmarks, the comparisons between the numerical procedures and the analytical procedure can legitimately be done to arbitrary precision although any forecast made can only be compared to observational data to the level of precision of the natural constants.\\ \\
Two other features of the Dormand Prince method worth noting are:
\begin{itemize}
\item  the method requires the user or implementer to specify an accuracy threshold for the algorithm to use when deciding whether to adapt the step size, the parameter does not translate directly into a measure of accuracy and has no direct physical significance. In all of the simulations runs in this paper, this parameter: $eps$ was set to $10^{-10}$
\item the run times per orbit mentioned above were for smooth orbits where the adaptive part of the algorithm did not need to add extra steps. For highly accelerated trajectories, the algorithm may run many more steps, and for short sections of the trajectory (in particular when the test mass is very close to one of the primaries) the numbers of steps per unit of physical distance can increase by several orders of magnitude and therefore run much slower
\end{itemize}
Another important point to note is that Python `Float' objects are double precision by default and store numbers to 53 binary places of precision which corresponds to 16 significant (decimal)  figures. \cite{pythonfloat} \\ \\
If one assumes 10,000 steps per orbit multiplied by the average adaptive step number for Dormand Prince (which I initially set at 10) and assumes that there are of the order of 1,000 calculations per Dormand Prince iteration, then we have $10^8$ calculations per orbit.  So over a 100 orbit run we would have of order $10^{10}$ computations, suggesting that a worst case could have errors of $10^{-6}$ from machine precision after 100 orbits which is at the limit of what is acceptable for the purposes of this paper and longer runs would need to use the Python $Decimal$ function but this would increase memory usage and CPU run times.
\begin{align}
\big( 1 + 10^{-16}  \big) ^{10} = 1 + 10^{-6} + \mathcal{O}(10^{-7})
\end{align}

\chapter{Unwanted Resonances}\label{resonances}
\begin{figure}[!ht]
    \centering
    \subfigure[]{\includegraphics[width=0.49\textwidth]{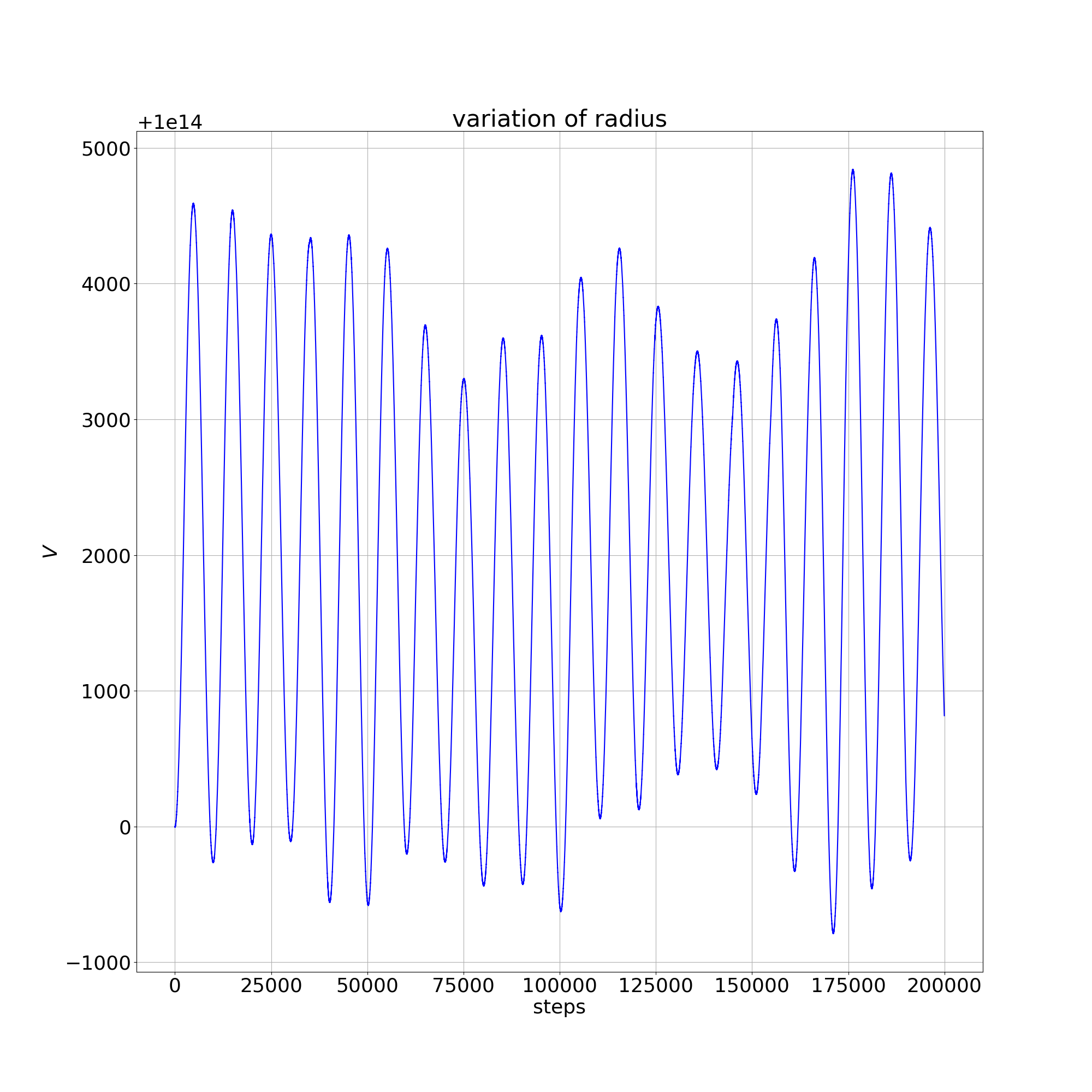}}
    \subfigure[]{\includegraphics[width=0.49\textwidth]{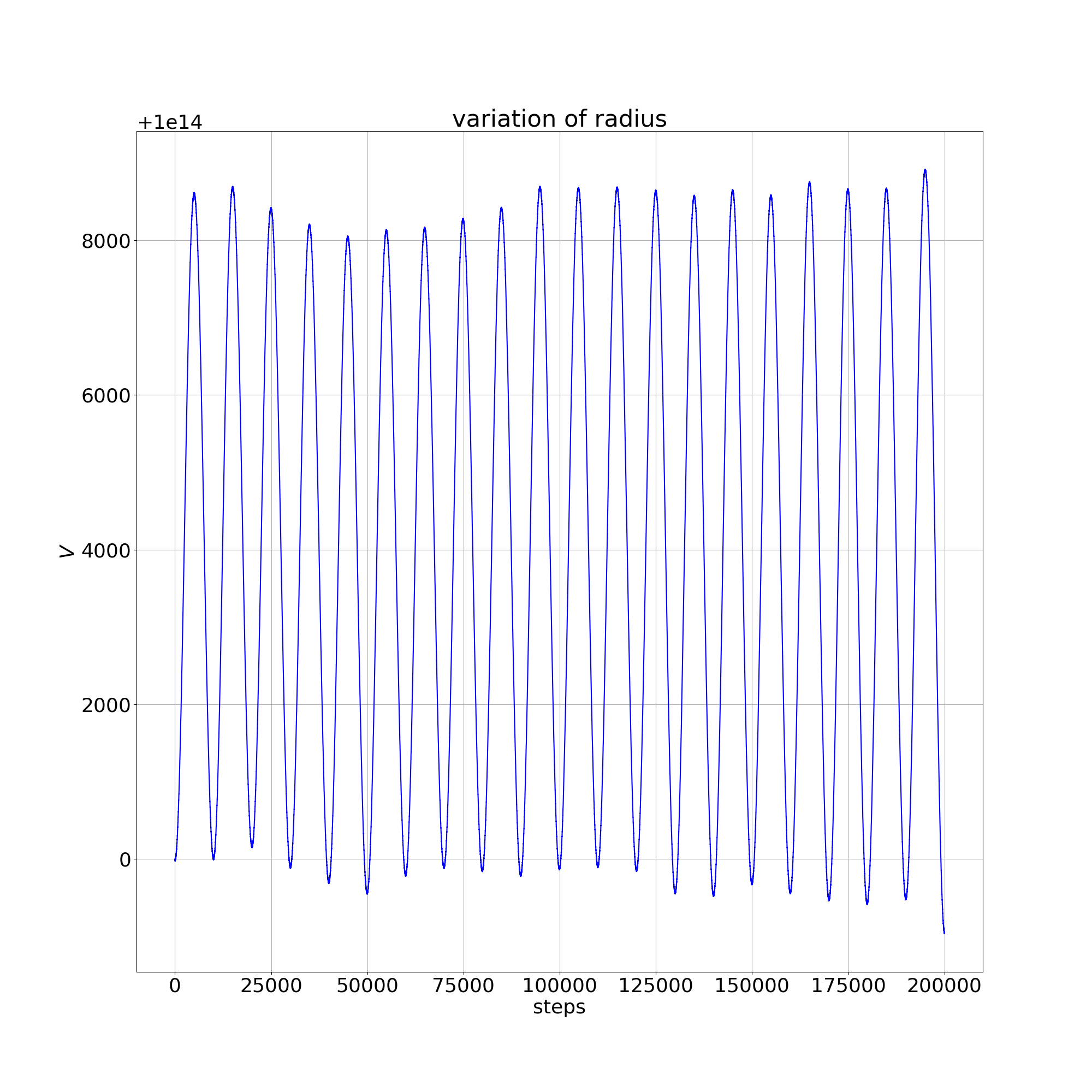}} \\ 
    \subfigure[]{\includegraphics[width=0.49\textwidth]{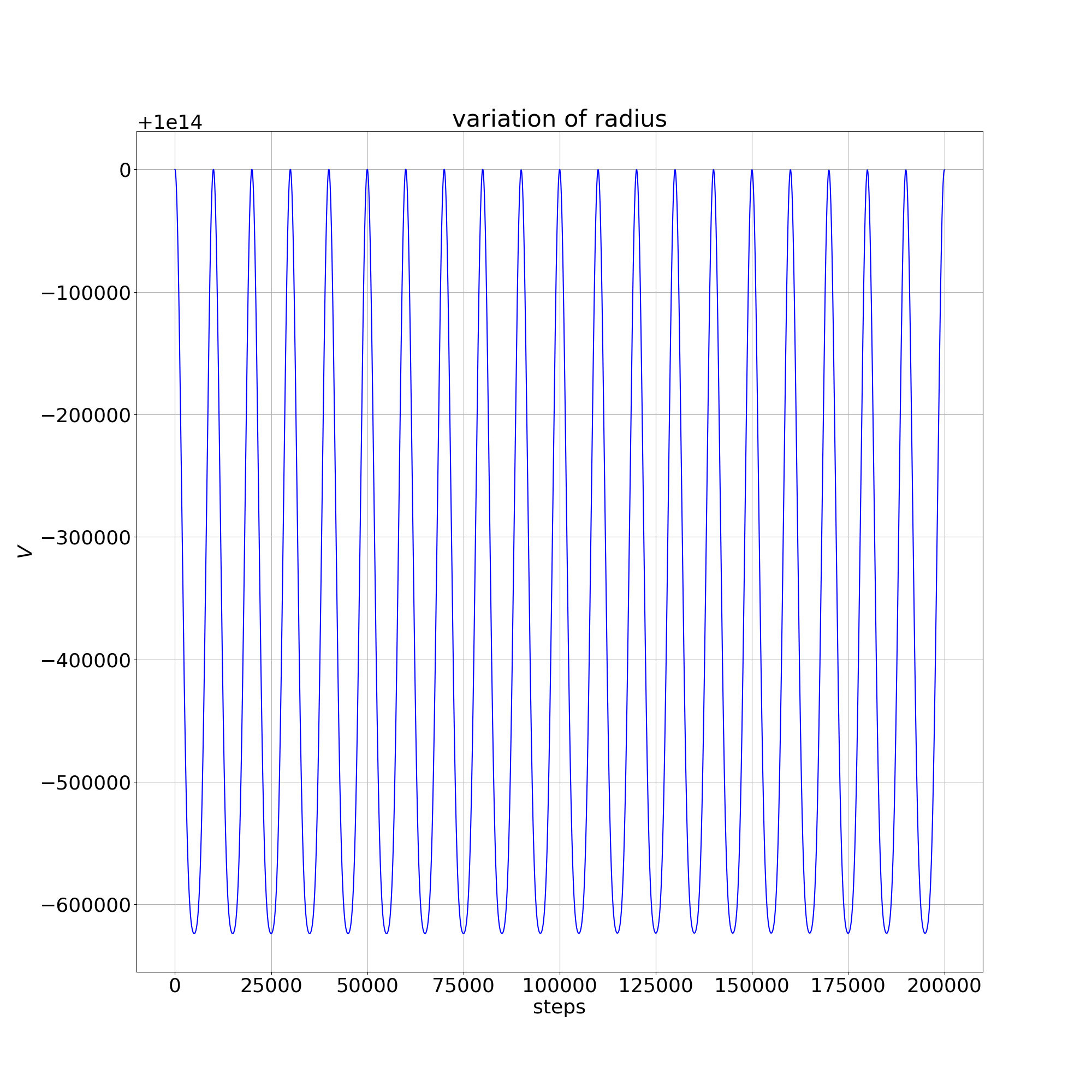}}
    \subfigure[]{\includegraphics[width=0.49\textwidth]{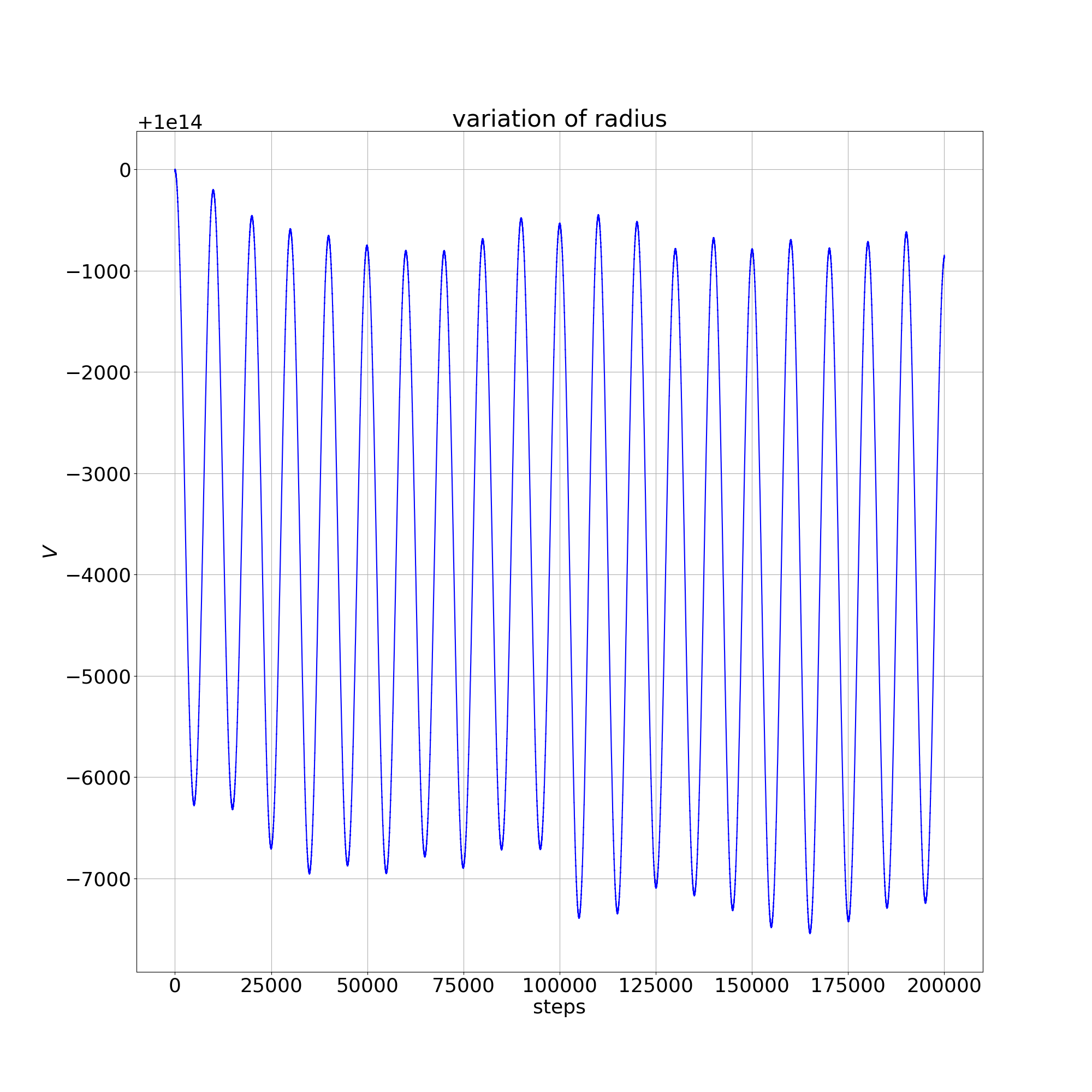}} \\
    \subfigure[]{\includegraphics[width=0.49\textwidth]{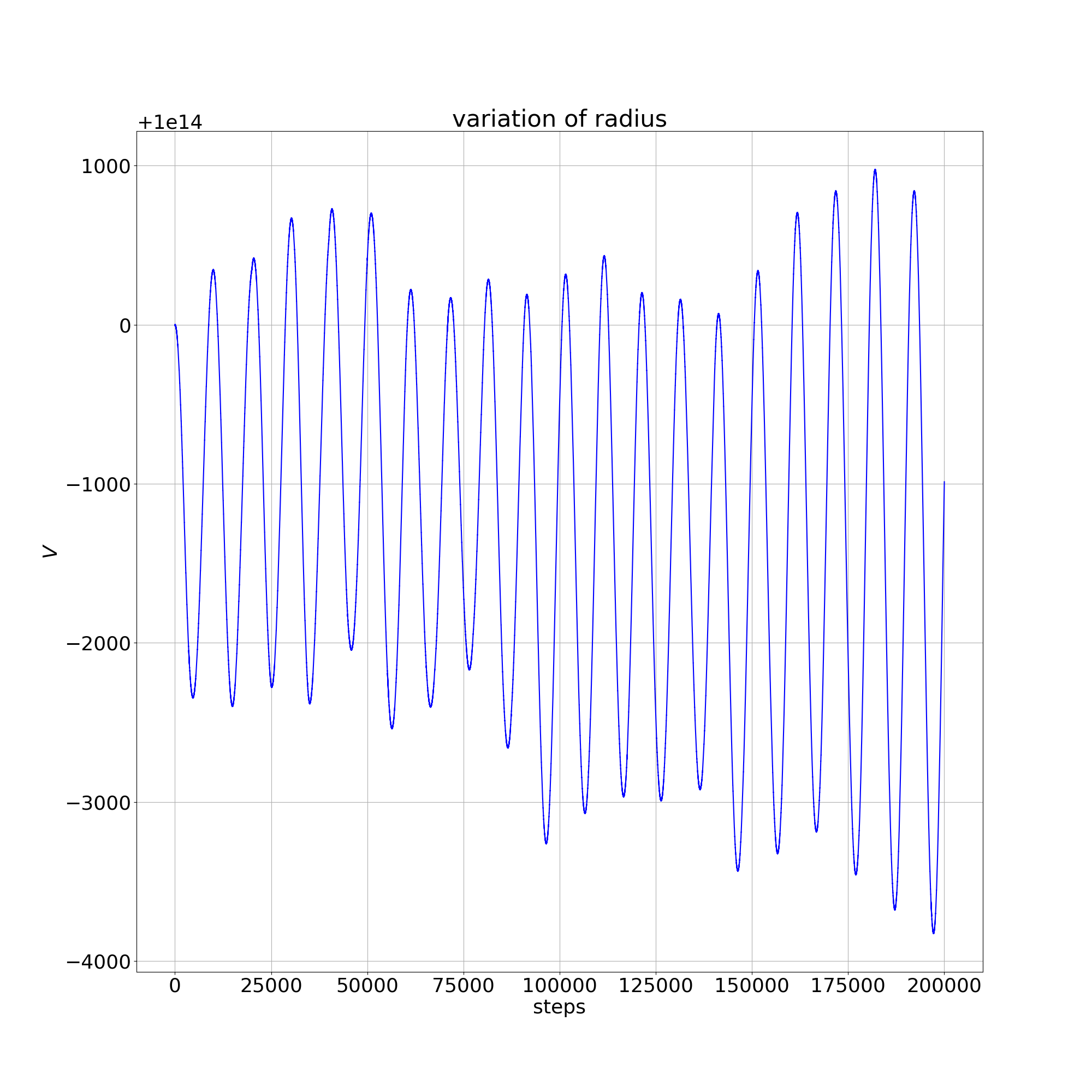}} 
    \subfigure[]{\includegraphics[width=0.49\textwidth]{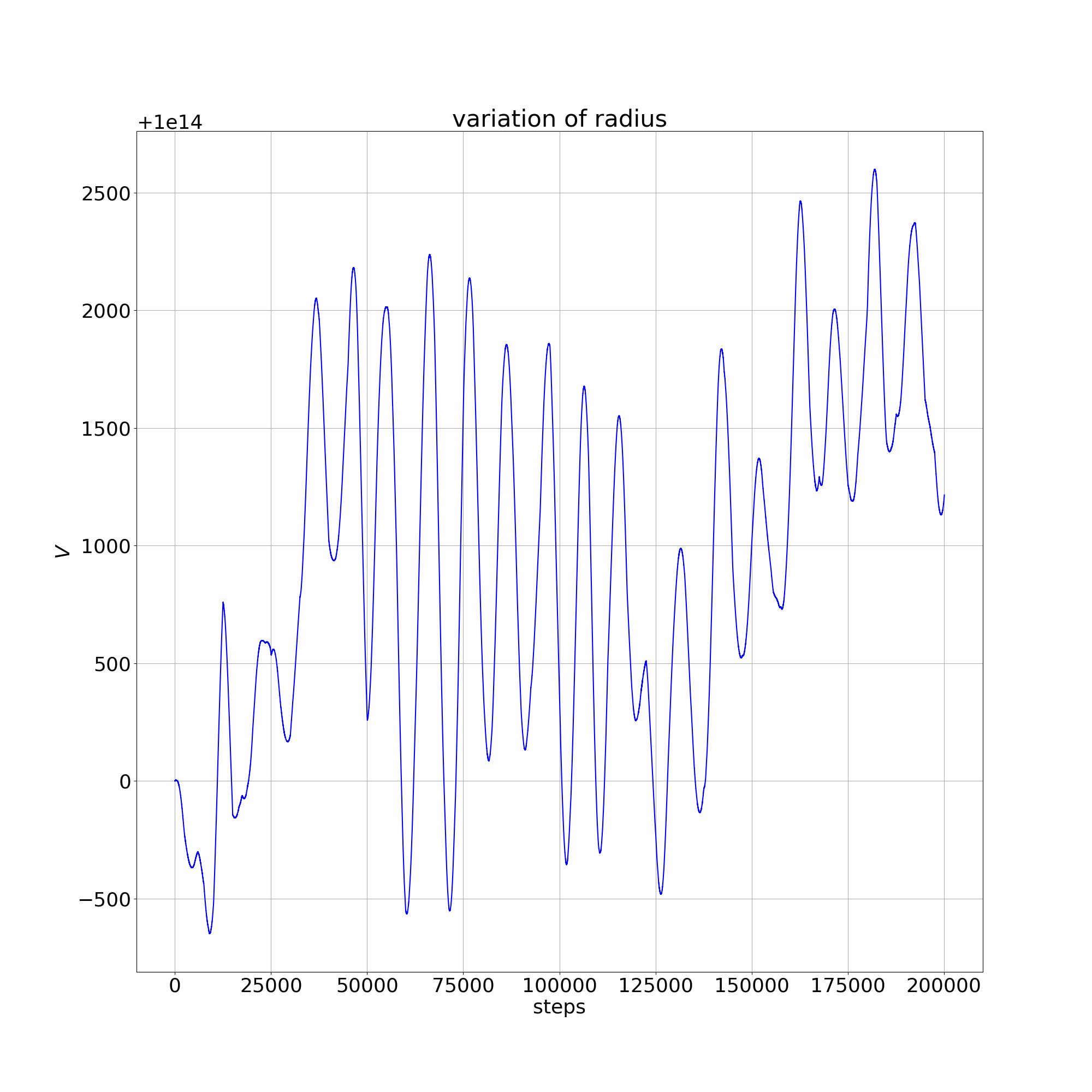}} 
    \caption{Steps = (a):9998  (b)9999  (c)10000 (d)10001  (d)10002  (e)10007}
    \label{rad drift}
\end{figure}

\begin{figure}[!ht]
\centering
\includegraphics[width=120mm]{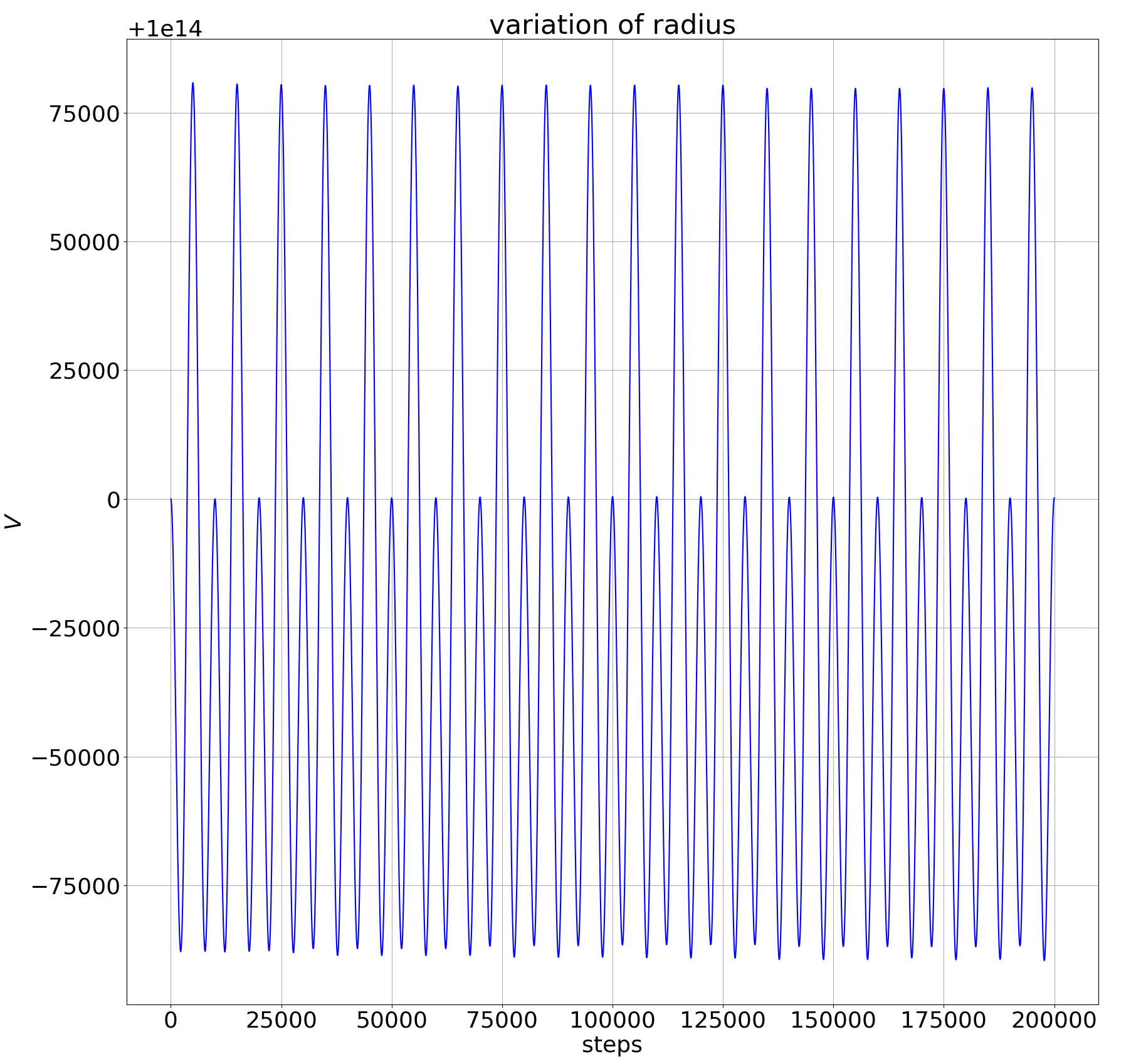}
\caption{Steps =10000 with retuned velocity to dampen oscillation}
\label{resonan}
\end{figure}

When choosing initial conditions, great care must be taken to avoid unexpected artefacts. In particular the relationship between `the distance between the primaries' and `the distance from the CoM of the primaries to the test mass' and `the size of steps selected for the test mass step in the numerical integrator'.\\ \\
As orbital periods are inversely proportional to the orbital radius to the power of 1.5, choosing initial conditions such as $10^{10}$ m for the distance between primaries and $10^{14}$ m for the distance from the test to the CoM of the primaries means that the primaries will orbit each other $10^6$ times for each orbit of the test mass around the binary system. If one then chooses $10,000$ steps per orbit for the numerical integrator and $0.1$ for the Dormand Prince adaptive step, then the primaries will have performed exactly 10 orbits for each time their position is `sampled' by the integrator: in other words the primaries will appear to be static and the computation will look like the computation done in a frame rotating with the line through the centres of mass of the two primaries. In the above example, for ordinary Newtonian gravity and not allowing for the time of propagation of the field (i.e. the simplest possible configuration of the Dormand Prince set-up) choosing the initial velocity for a stable circular orbit and allowing the number of iterations per orbit (1 / step size) to vary over the range $9998, 9999, 10000, 10001, 10002$,  I obtain the oscillation of distance of the test mass from the CoM of the system shown in Figure \ref{rad drift}. \\ \\
Note that this resonance effect did not show up when the distance between the primaries was $10^6$ metres since, in that configuration, the Binary system acts like a point mass. At the wider distance between the primaries, as the system no longer looks like a point mass, and therefore the orbital velocity of the test mass is no longer expected to be exactly equal to the simple Newtonian calculation and a process of iterative search is used to establish the velocity that produces a stable orbit.\\ \\
Figure \ref{rad drift} shows the variation in the distance from the system CoM to the test mass during 20 orbits for a system whose initial conditions were tuned to give a circular orbit when the $step$ parameter was set at 10,007 steps per orbit (10,007 being the first prime number larger than 10,000). If one looks closely at the Y axis one sees that the orbital radius variation for 5 of the 6 figures is of the order of $10^3$ m whereas in Figure \ref{rad drift} (c), the case of $step = 10,000$ has variation of the order of $10^6$ m.\\
\\
By slightly adjusting the initial velocity for the case  $step = 10,000$ one can reduce the orbital variation by approximately an order of magnitude. In so doing one observes a new harmonic at twice the orbital frequency, Figure \ref{resonan} . This is to be expected, since in using $step = 10,000$, the Integrator samples the positions of the primaries at exactly the same point of their orbit each time which creates the effect of a static binary system. The test mass orbiting this `Bar Bell' shape will feel two regions of higher gravity (along the axis of the two primaries) and two regions of lower gravity (when the test mass is $\pi$ / 2 away from the axis of the primaries and therefore equidistant from the two bodies).\\ \\
This effect is entirely spurious and unwanted in the non-rotating frame simulation, and it is therefore a good idea to use prime numbers for the number of steps per orbit and indeed to avoid having the various distance be multiples of one another. I have therefore throughout the paper set the separation between the binaries as a multiple of 1.1 .

\chapter{Results: Tables} \label{results-tables}
\begin{sidewaystable}[!ht]
\centering
\includegraphics[width =1.1 \textwidth]{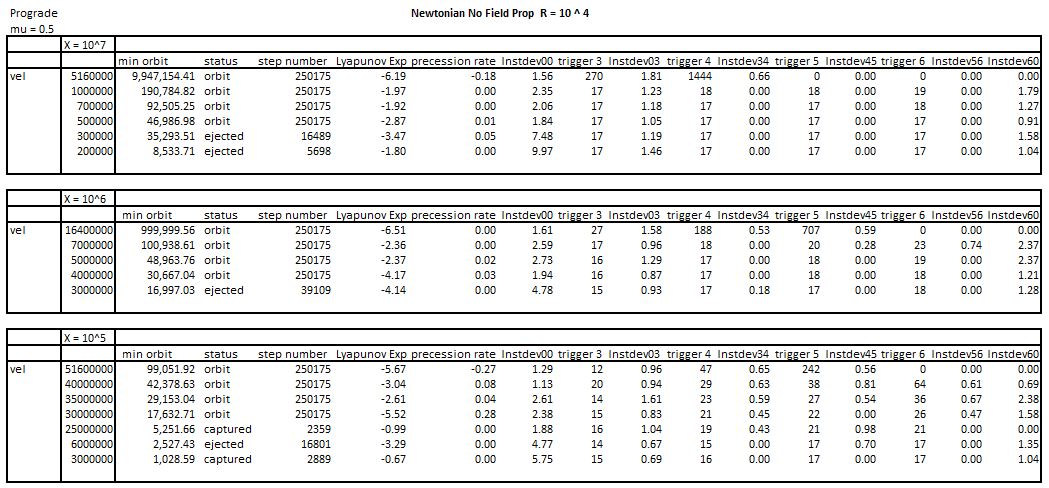}
\caption{Results table for the Newtonian model with no field propagation\\Separation of primaries = $10^{4}m$}
\label{Newt-no-prop-4}
\end{sidewaystable}
\begin{sidewaystable}[!ht]
\centering
\includegraphics[width =1.1 \textwidth]{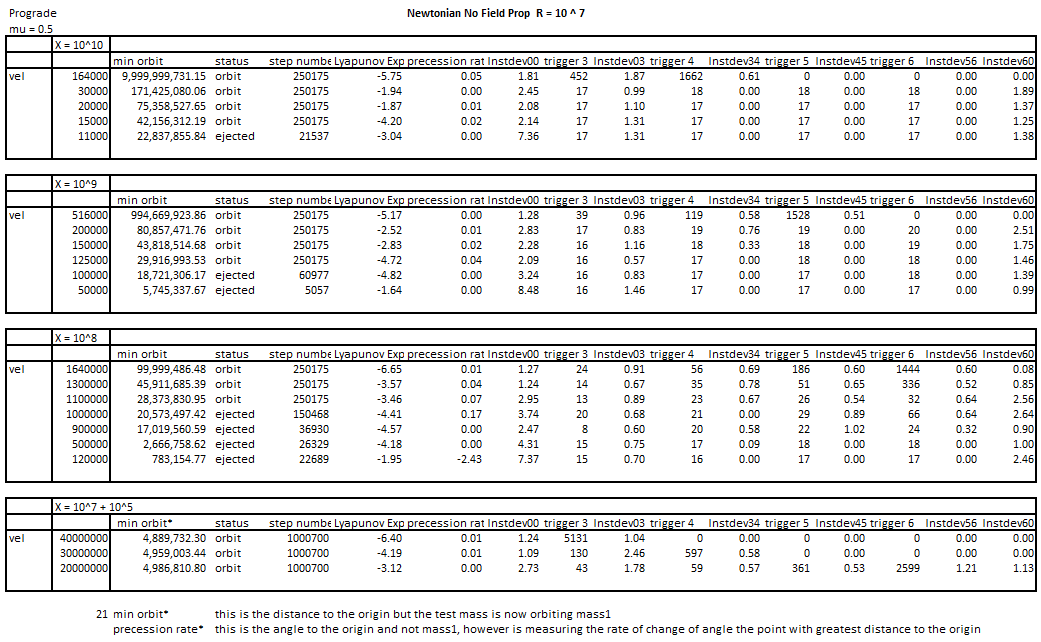}
\caption{Results table for the Newtonian model with no field propagation\\Separation of primaries = $10^{7}m$}
\label{Newt-no-prop-7}
\end{sidewaystable}
\begin{sidewaystable}[!ht]
\centering
\includegraphics[width =1.1 \textwidth]{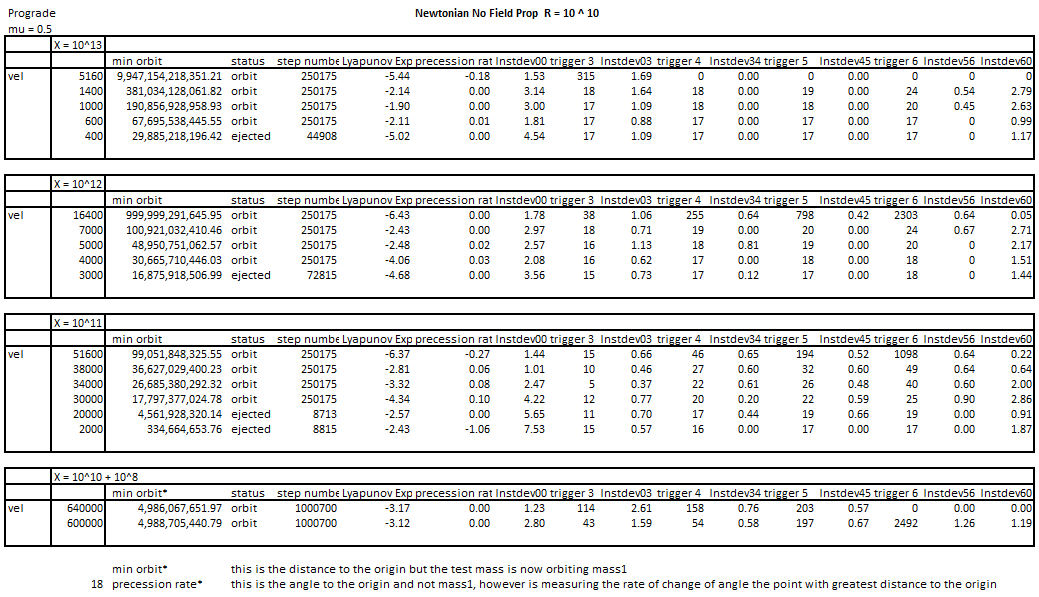}
\caption{Results table for the Newtonian model with no field propagation\\Separation of primaries = $10^{10}m$}
\label{Newt-no-prop-10}
\end{sidewaystable}
\begin{sidewaystable}[!ht]
\centering
\includegraphics[width =1.1 \textwidth]{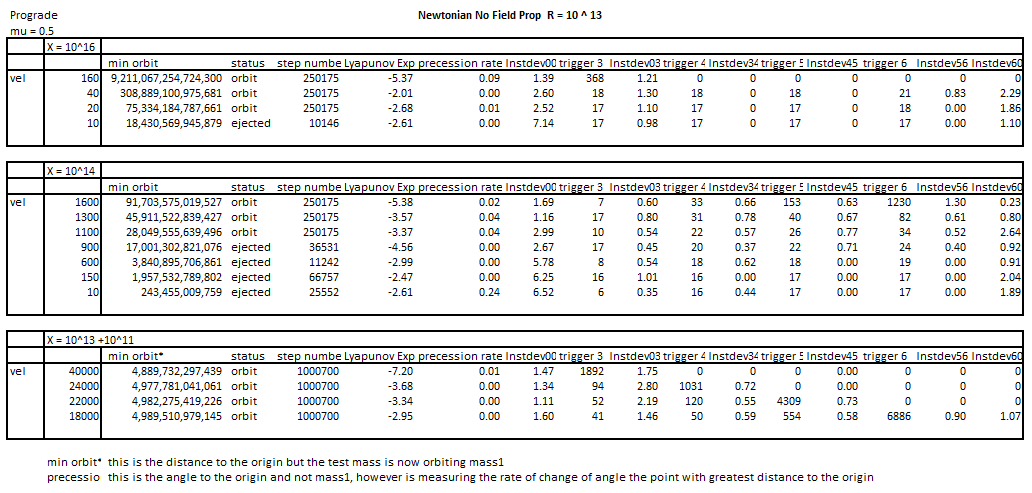}
\caption{Results table for the Newtonian model with no field propagation\\Separation of primaries = $10^{13}m$}
\label{Newt-no-prop-13}
\end{sidewaystable}

\begin{sidewaystable}[!ht]
\centering
\includegraphics[width =1.1 \textwidth]{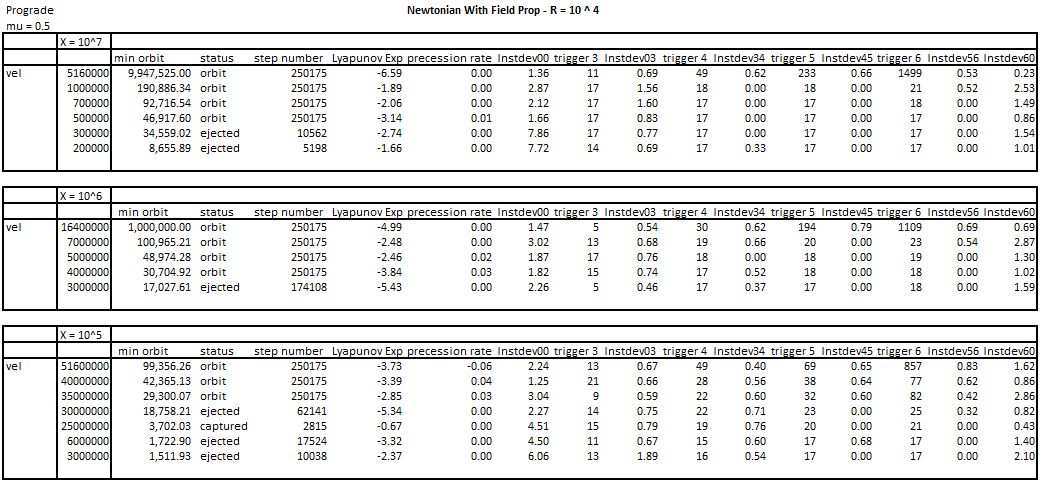}
\caption{Results table for the Newtonian model WITH field propagation\\Separation of primaries = $10^{4}m$}
\label{Newt-WF-4}
\end{sidewaystable}
\begin{sidewaystable}[!ht]
\centering
\includegraphics[width =1.1 \textwidth]{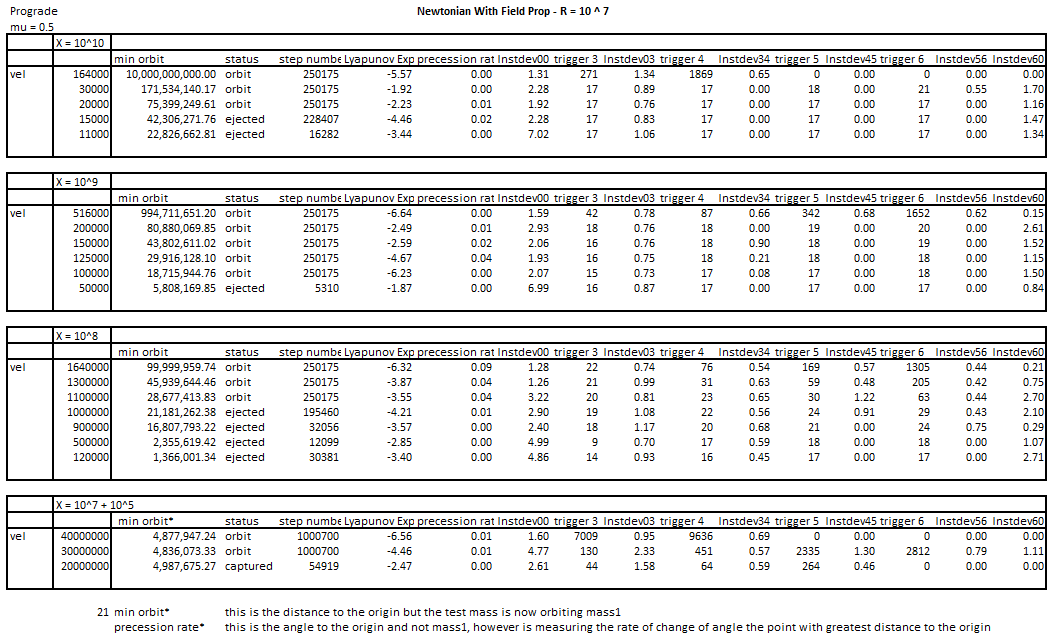}
\caption{Results table for the Newtonian model WITH field propagation\\Separation of primaries = $10^{7}m$}
\label{Newt-WF-7}
\end{sidewaystable}
\begin{sidewaystable}[!ht]
\centering
\includegraphics[width =1.1 \textwidth]{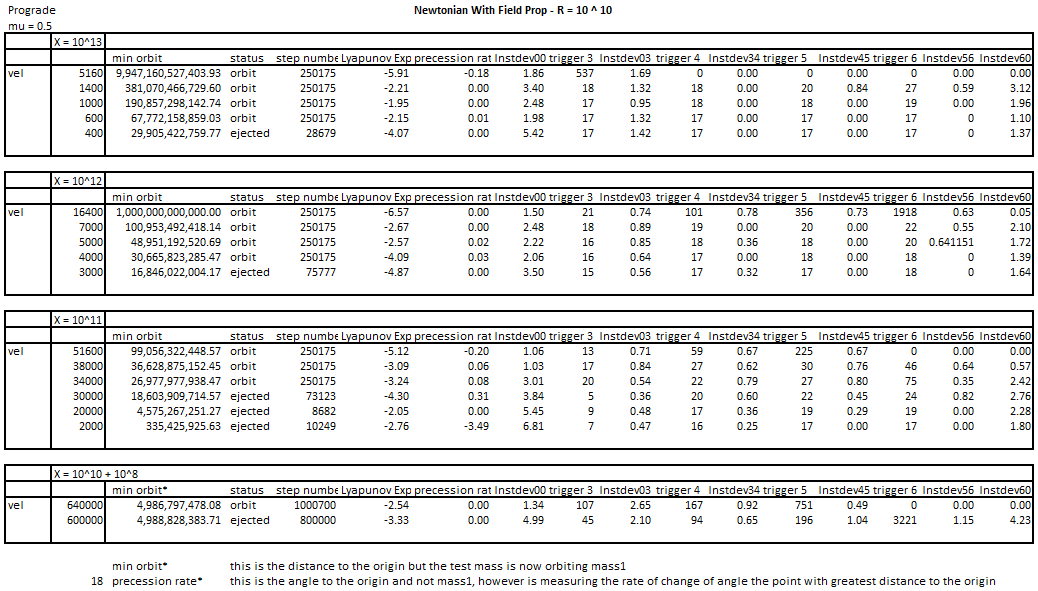}
\caption{Results table for the Newtonian model WITH field propagation\\Separation of primaries = $10^{10}m$}
\label{Newt-WF-10}
\end{sidewaystable}
\begin{sidewaystable}[!ht]
\centering
\includegraphics[width =1.1 \textwidth]{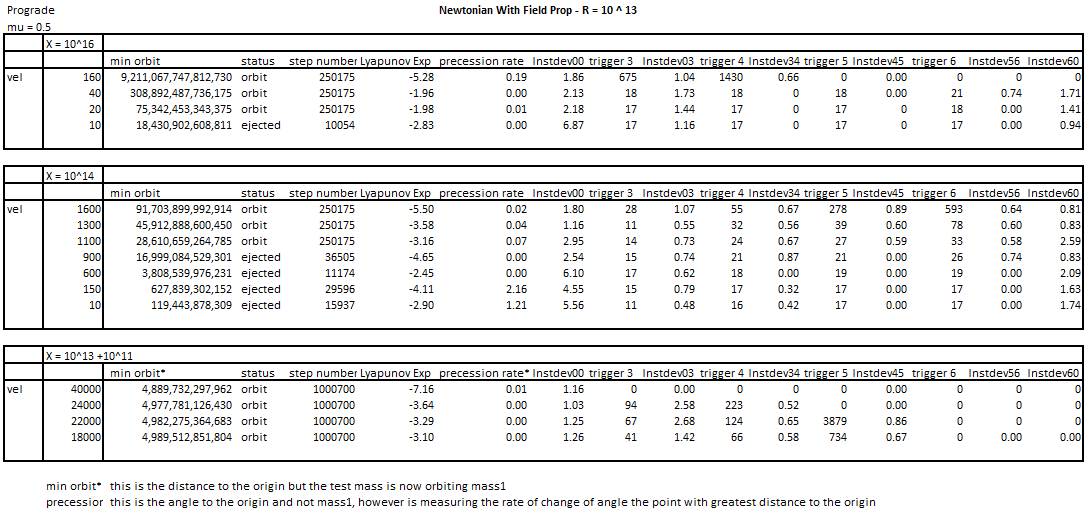}
\caption{Results table for the Newtonian model WITH field propagation\\Separation of primaries = $10^{13}m$}
\label{Newt-WF-13}
\end{sidewaystable}

\begin{sidewaystable}[!ht]
\centering
\includegraphics[width =1.1 \textwidth]{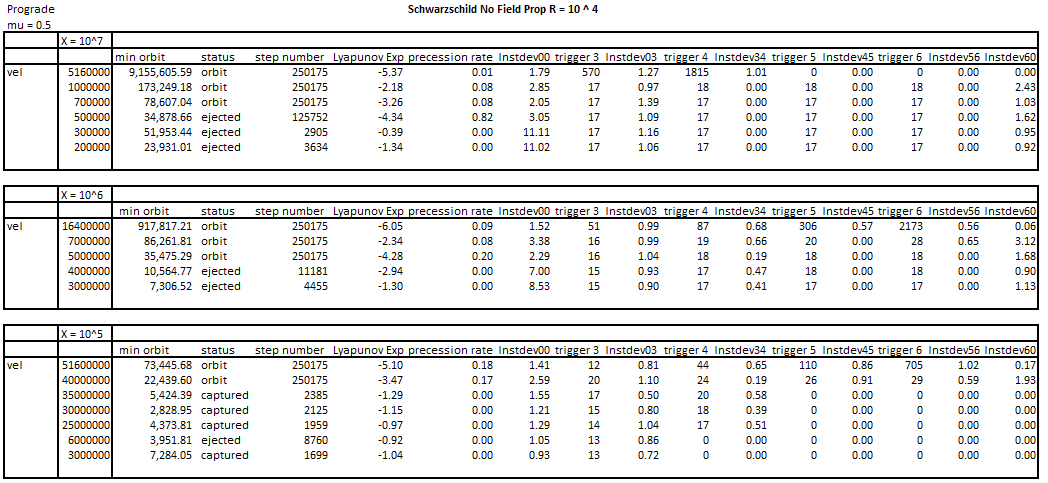}
\caption{Results table for the Schwarzschild Correction with no field propagation\\Separation of primaries = $10^{4}m$}
\label{0.5-NF-4}
\end{sidewaystable}
\begin{sidewaystable}[!ht]
\centering
\includegraphics[width =1.1 \textwidth]{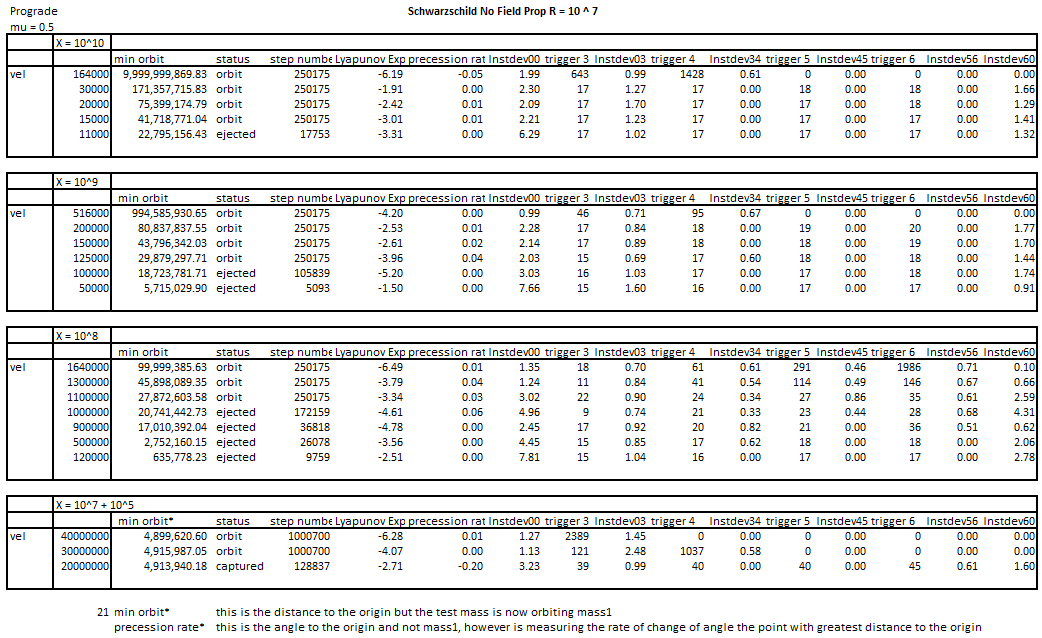}
\caption{Results table for the Schwarzschild Correction with no field propagation\\Separation of primaries = $10^{7}m$}
\label{0.5-NF-7}
\end{sidewaystable}
\begin{sidewaystable}[!ht]
\centering
\includegraphics[width =1.1 \textwidth]{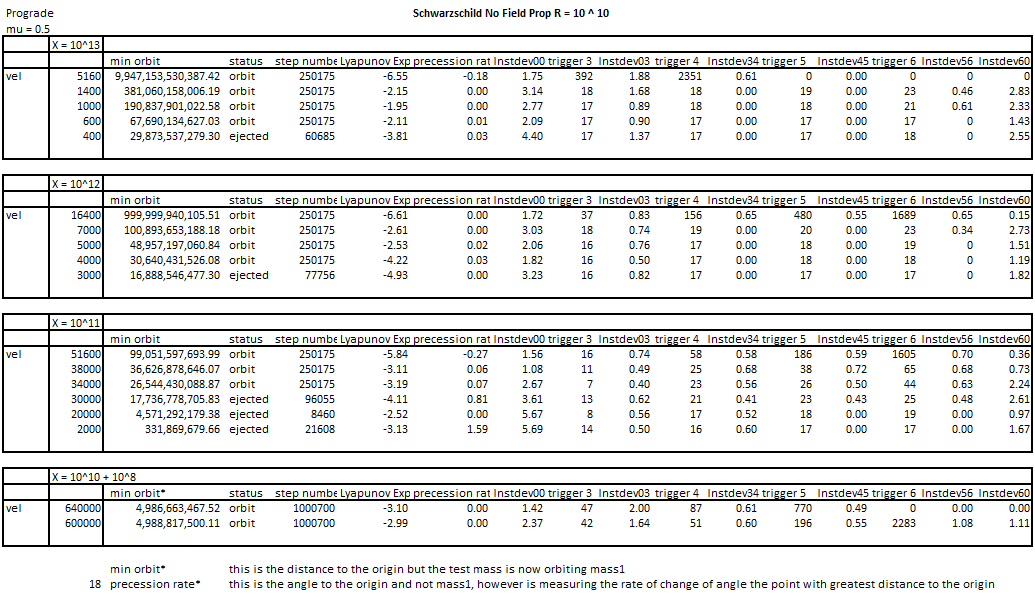}
\caption{Results table for the Schwarzschild Correction with no field propagation\\Separation of primaries = $10^{10}m$}
\label{0.5-NF-10}
\end{sidewaystable}
\begin{sidewaystable}[!ht]
\centering
\includegraphics[width =1.1 \textwidth]{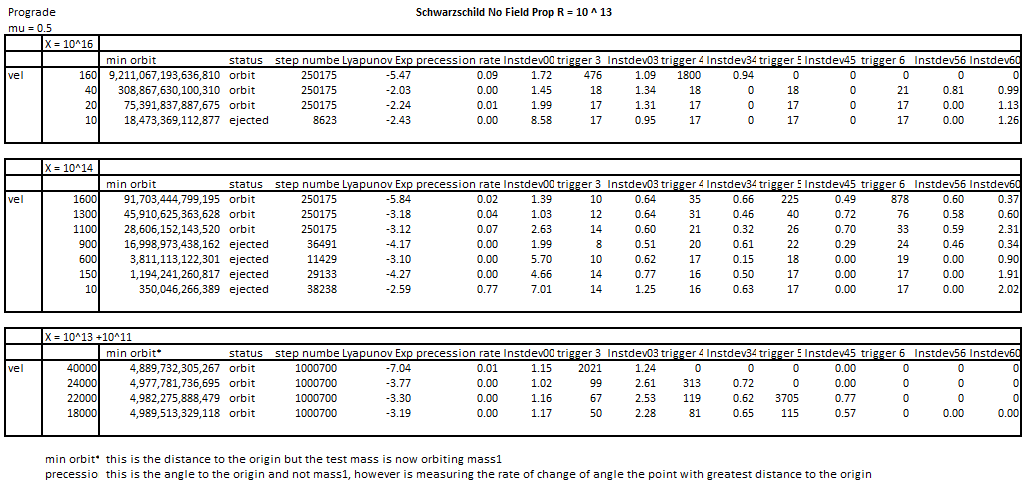}
\caption{Results table for the Schwarzschild Correction with no field propagation\\Separation of primaries = $10^{13}m$}
\label{0.5-NF-13}
\end{sidewaystable}

\begin{sidewaystable}[!ht]
\centering
\includegraphics[width =1.1 \textwidth]{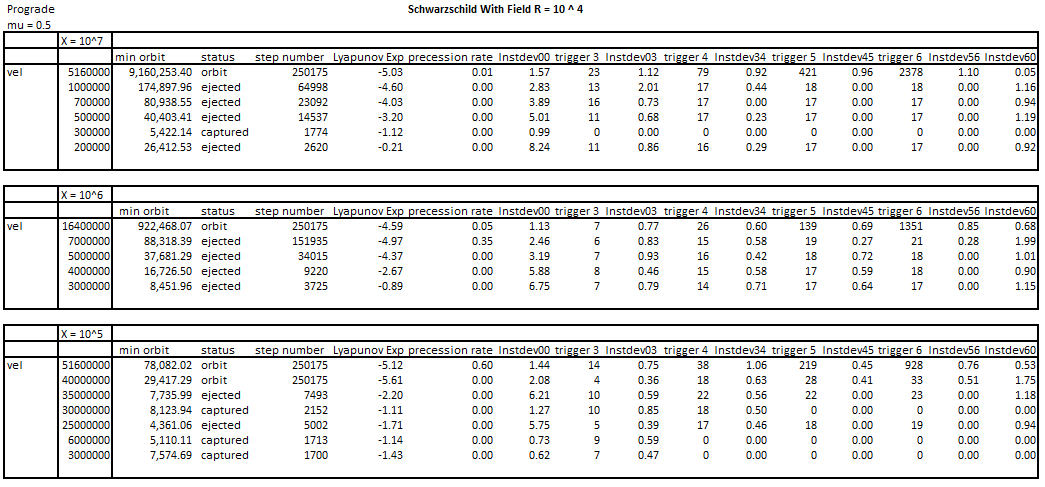}
\caption{Results table for the Schwarzschild Correction WITH field propagation\\Separation of primaries = $10^{4}m$}
\label{0.5-WF-4}
\end{sidewaystable}
\begin{sidewaystable}[!ht]
\centering
\includegraphics[width =1.1 \textwidth]{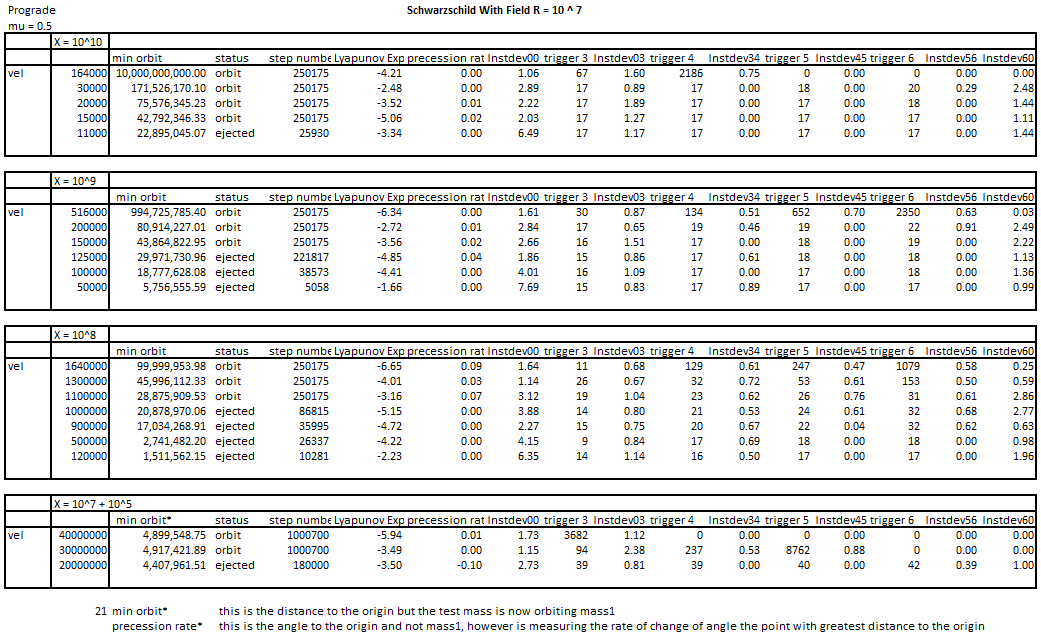}
\caption{Results table for the Schwarzschild Correction WITH field propagation\\Separation of primaries = $10^{7}m$}
\label{0.5-WF-7}
\end{sidewaystable}
\begin{sidewaystable}[!ht]
\centering
\includegraphics[width =1.1 \textwidth]{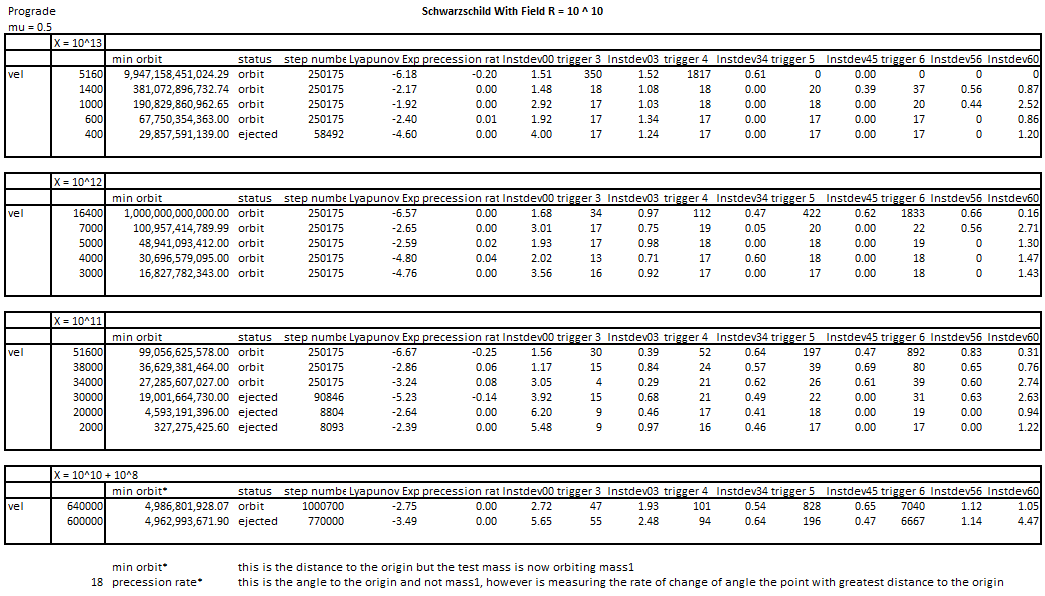}
\caption{Results table for the Schwarzschild Correction WITH field propagation\\Separation of primaries = $10^{10}m$}
\label{0.5-WF-10}
\end{sidewaystable}
\begin{sidewaystable}[!ht]
\centering
\includegraphics[width =1.1 \textwidth]{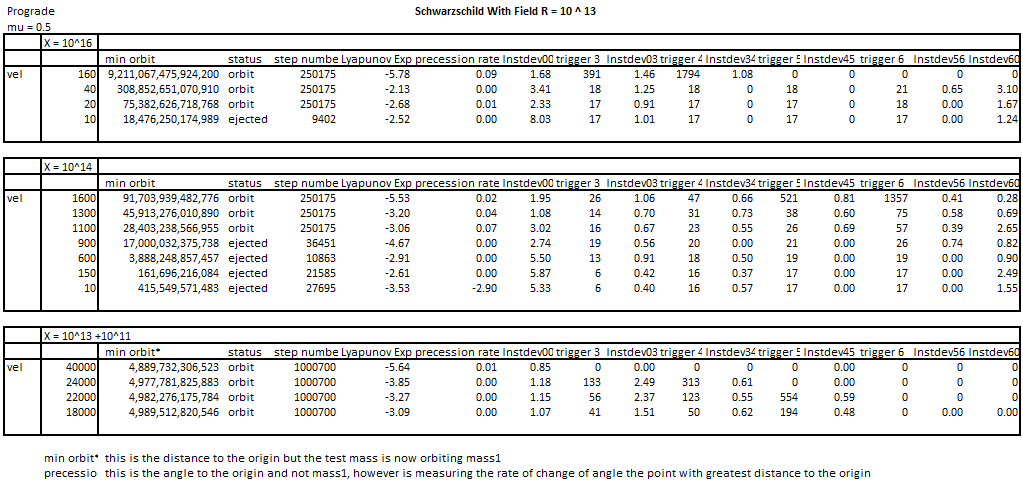}
\caption{Results table for the Schwarzschild Correction WITH field propagation\\Separation of primaries = $10^{13}m$}
\label{0.5-WF-13}
\end{sidewaystable}

\begin{sidewaystable}[!ht]
\centering
\includegraphics[width =1.1 \textwidth]{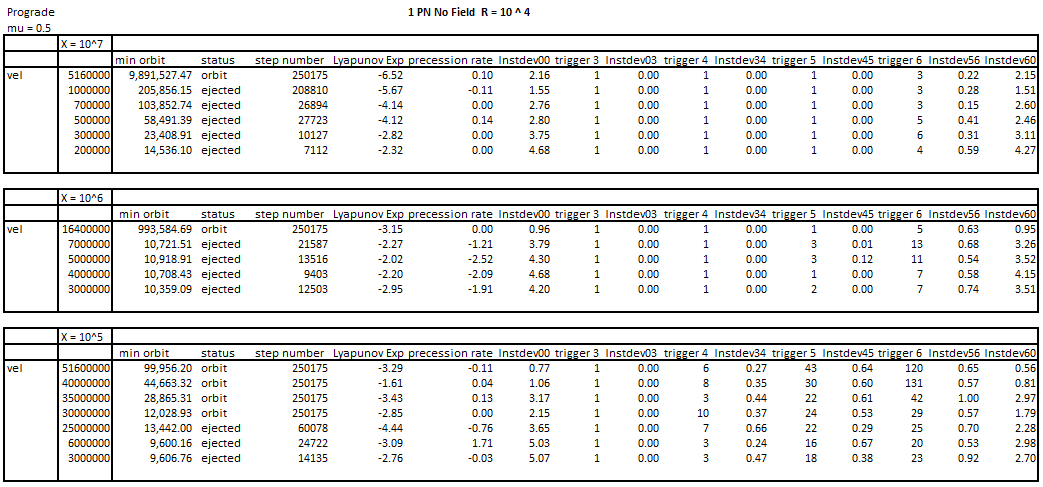}
\caption{Results table for the 1 PN model with no field propagation\\Separation of primaries = $10^{4}m$}
\label{1-NF-4}
\end{sidewaystable}
\begin{sidewaystable}[!ht]
\centering
\includegraphics[width =1.1 \textwidth]{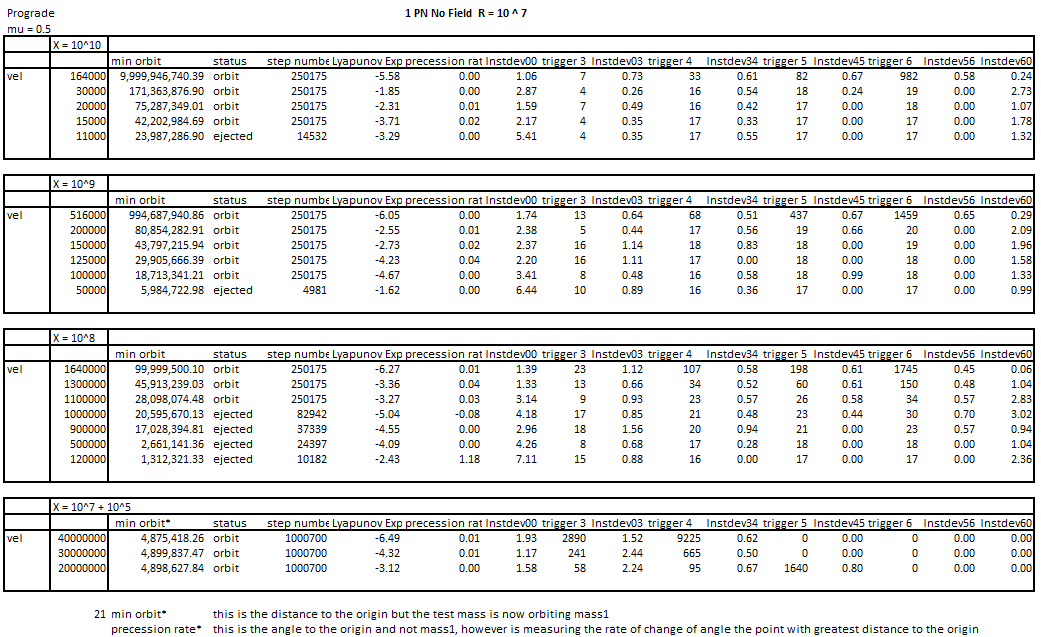}
\caption{Results table for the 1 PN model with no field propagation\\Separation of primaries = $10^{7}m$}
\label{1-NF-7}
\end{sidewaystable}
\begin{sidewaystable}[!ht]
\centering
\includegraphics[width =1.1 \textwidth]{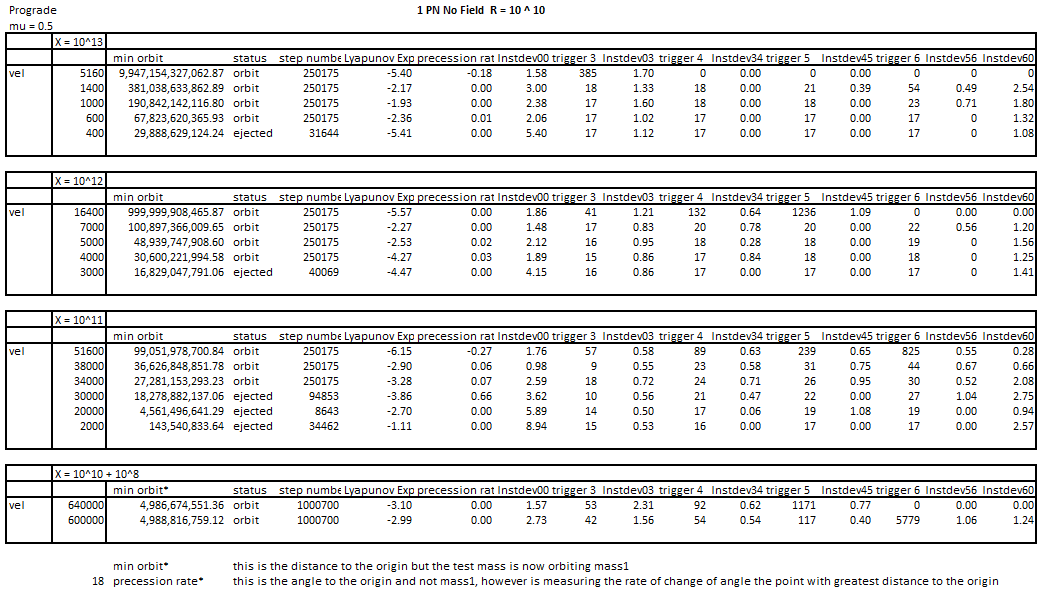}
\caption{Results table for the 1 PN model with no field propagation\\Separation of primaries = $10^{10}m$}
\label{1-NF-10}
\end{sidewaystable}
\begin{sidewaystable}[!ht]
\centering
\includegraphics[width =1.1 \textwidth]{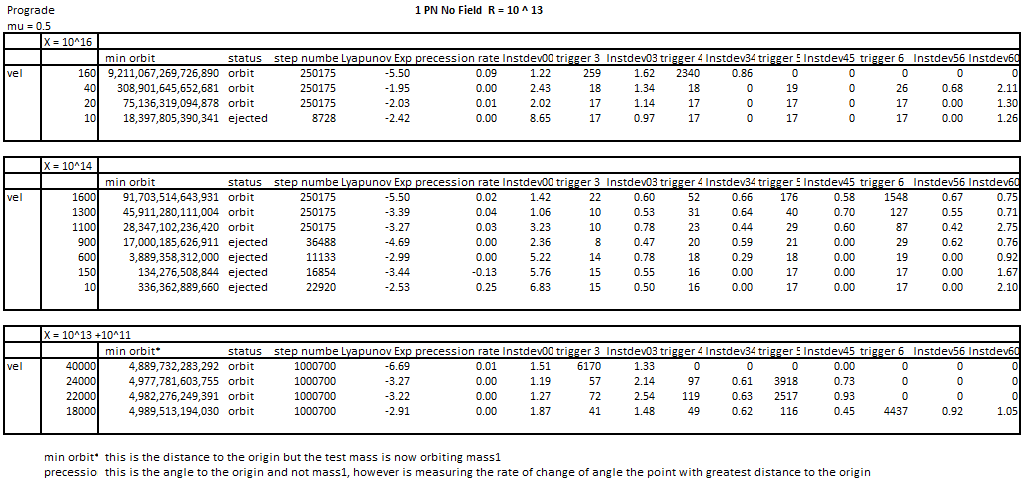}
\caption{Results table for the 1 PN model with no field propagation\\Separation of primaries = $10^{13}m$}
\label{1-NF-13}
\end{sidewaystable}

\begin{sidewaystable}[!ht]
\centering
\includegraphics[width =1.1 \textwidth]{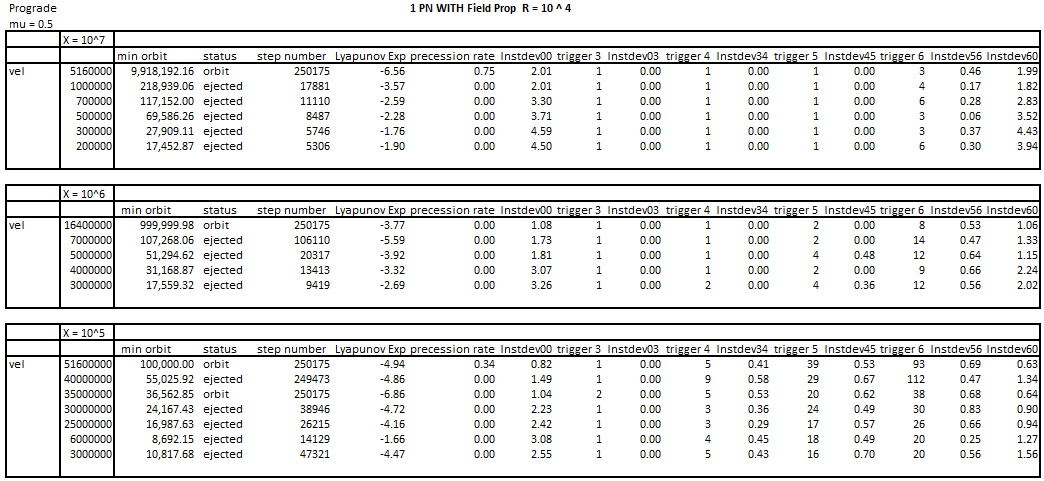}
\caption{Results table for the 1 PN model WITH field propagation\\Separation of primaries = $10^{4}m$}
\label{1-WF-4}
\end{sidewaystable}
\begin{sidewaystable}[!ht]
\centering
\includegraphics[width =1.1 \textwidth]{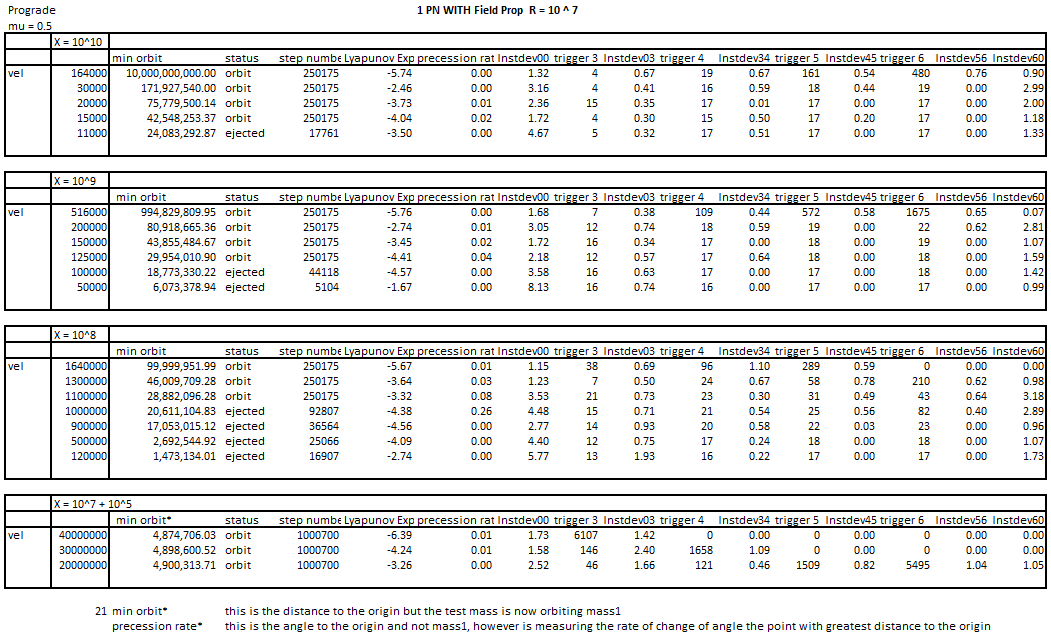}
\caption{Results table for the 1 PN model WITH field propagation\\Separation of primaries = $10^{7}m$}
\label{1-WF-7}
\end{sidewaystable}
\begin{sidewaystable}[!ht]
\centering
\includegraphics[width =1.1 \textwidth]{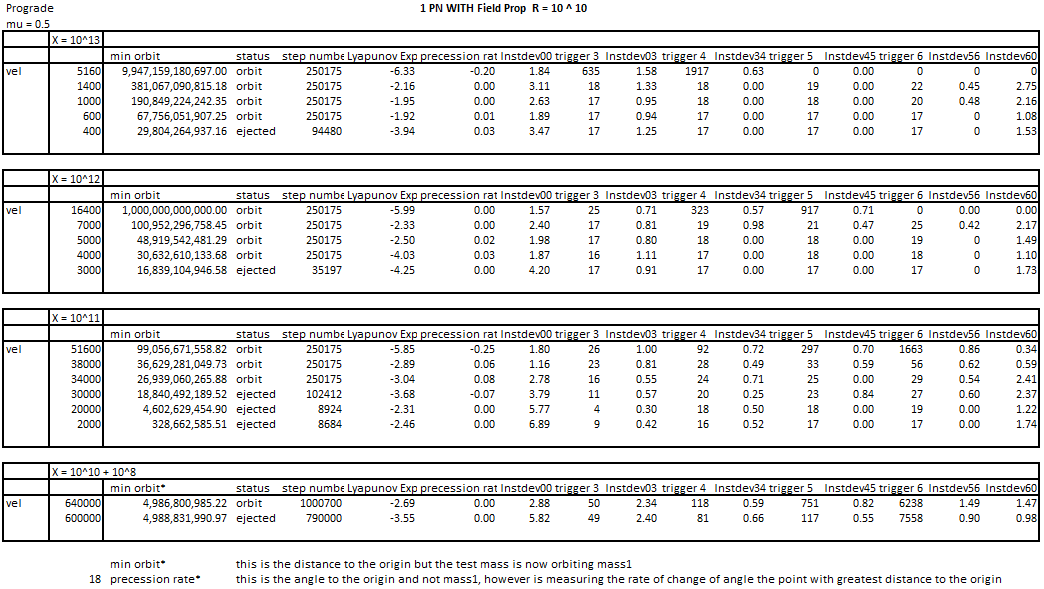}
\caption{Results table for the 1 PN model WITH field propagation\\Separation of primaries = $10^{10}m$}
\label{1-WF-10}
\end{sidewaystable}
\begin{sidewaystable}[!ht]
\centering
\includegraphics[width =1.1 \textwidth]{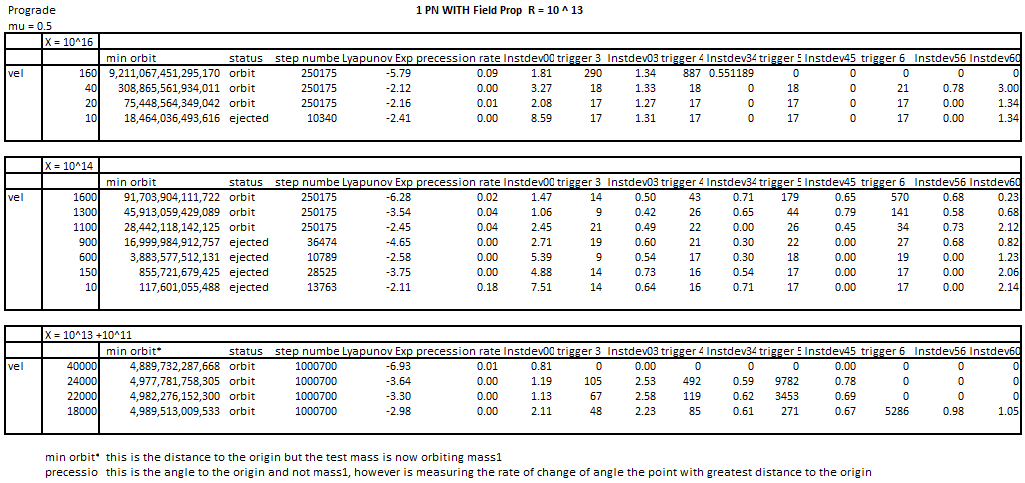}
\caption{Results table for the 1 PN model WITH field propagation\\Separation of primaries = $10^{13}m$}
\label{1-WF-13}
\end{sidewaystable}

\begin{sidewaystable}[!ht]
\centering
\includegraphics[width =1.1 \textwidth]{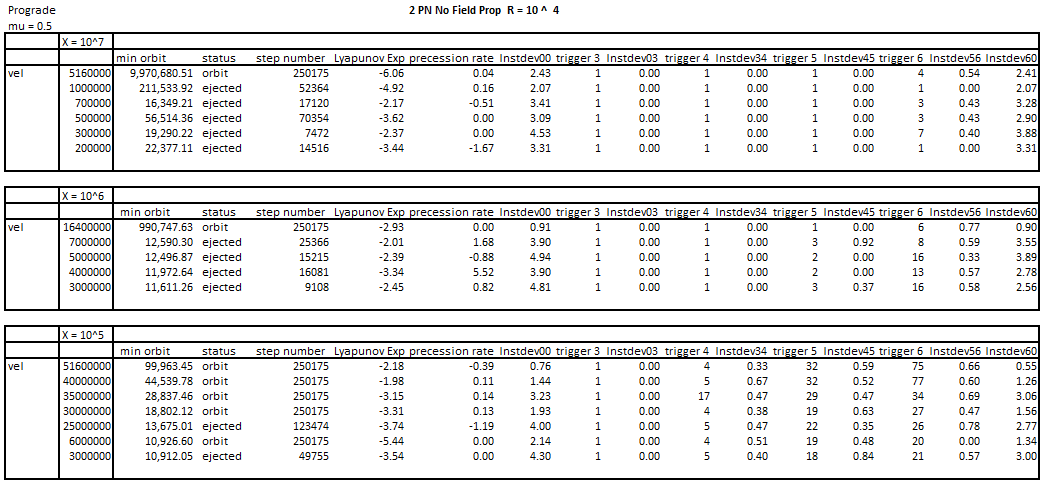}
\caption{Results table for the 2 PN model with no field propagation\\Separation of primaries = $10^{4}m$}
\label{2-NF-4}
\end{sidewaystable}
\begin{sidewaystable}[!ht]
\centering
\includegraphics[width =1.1 \textwidth]{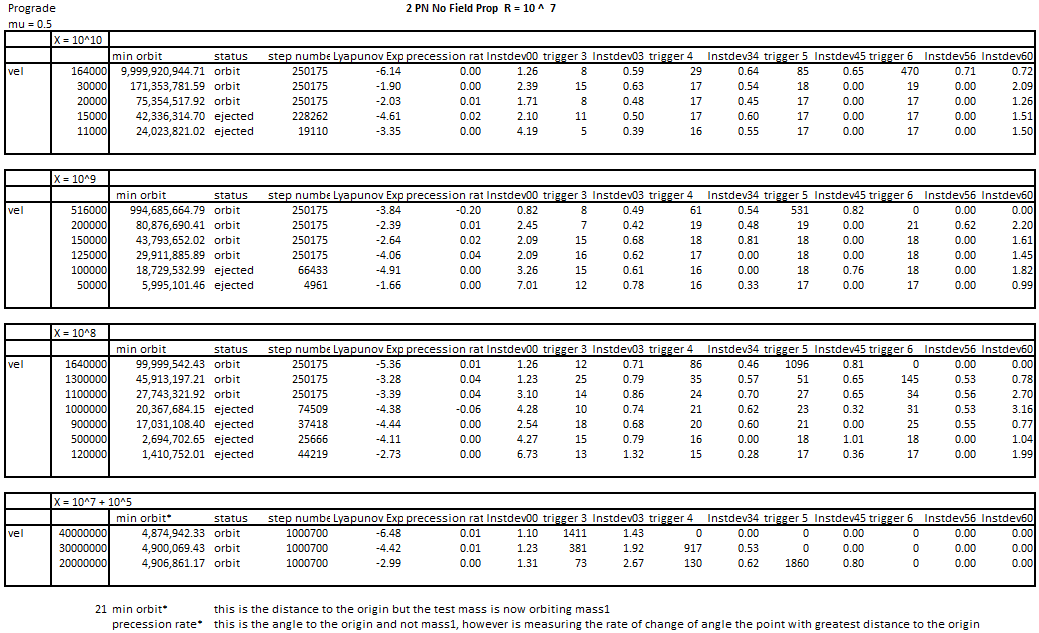}
\caption{Results table for the 2 PN model with no field propagation\\Separation of primaries = $10^{7}m$}
\label{2-NF-7}
\end{sidewaystable}
\begin{sidewaystable}[!ht]
\centering
\includegraphics[width =1.1 \textwidth]{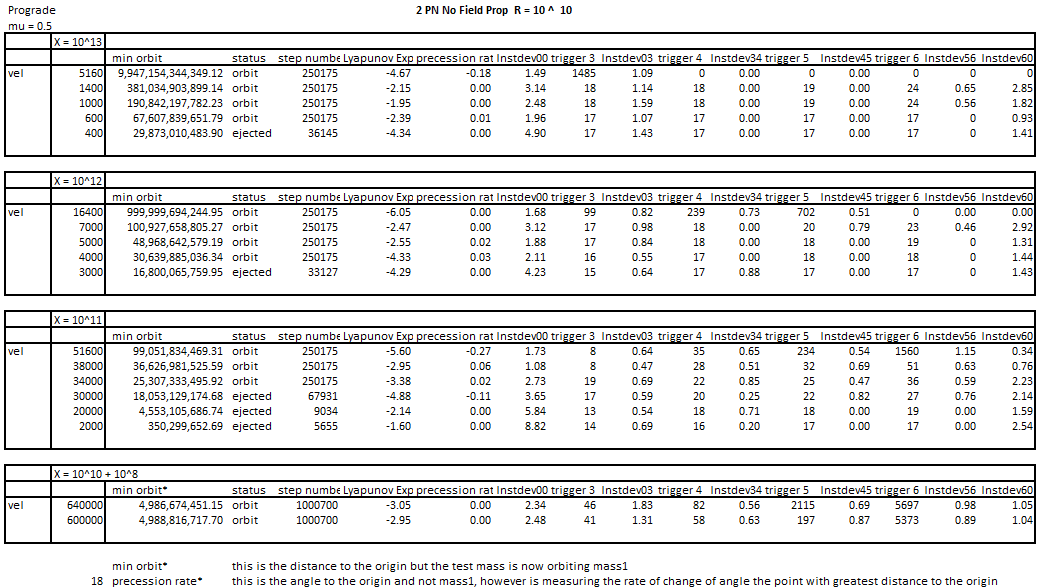}
\caption{Results table for the 2 PN model with no field propagation\\Separation of primaries = $10^{10}m$}
\label{2-NF-10}
\end{sidewaystable}
\begin{sidewaystable}[!ht]
\centering
\includegraphics[width =1.1 \textwidth]{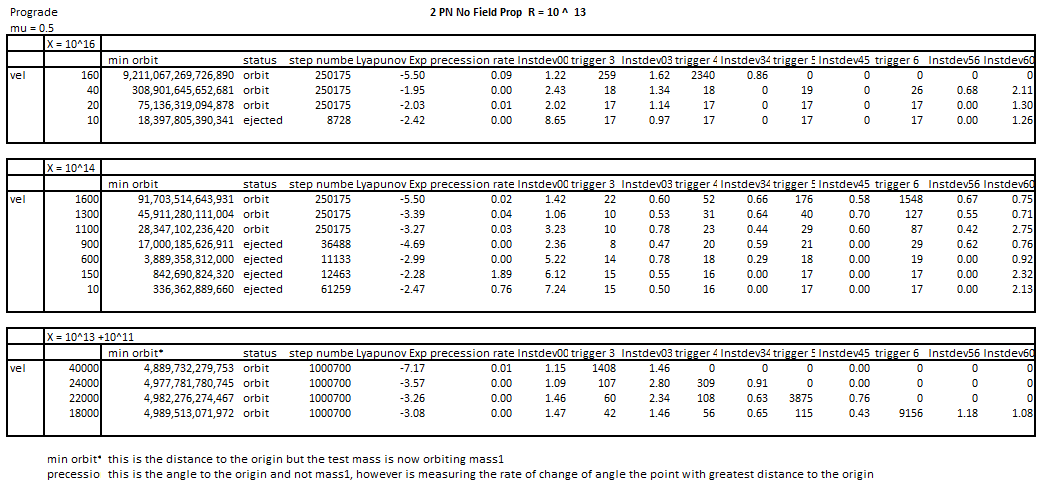}
\caption{Results table for the 2` PN model with no field propagation\\Separation of primaries = $10^{13}m$}
\label{2-NF-13}
\end{sidewaystable}

\begin{sidewaystable}[!ht]
\centering
\includegraphics[width =1.1 \textwidth]{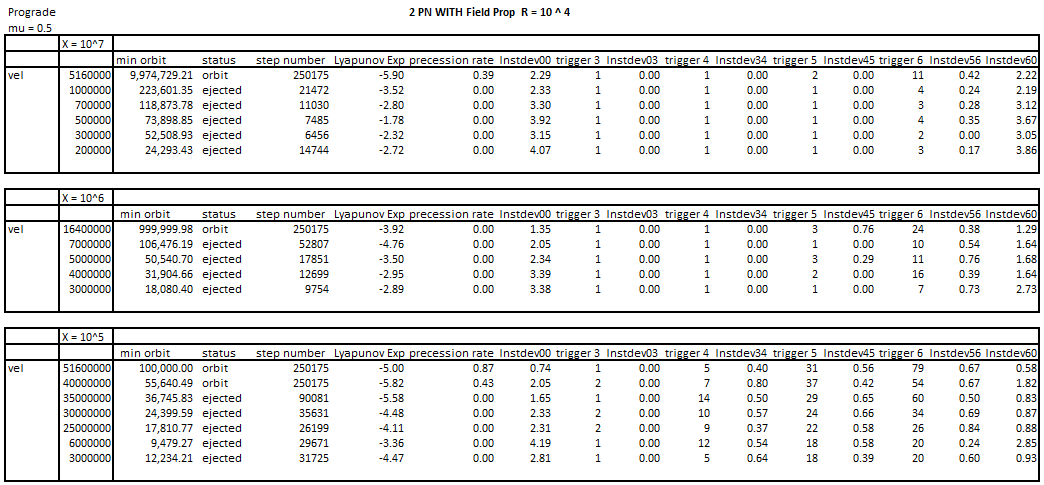}
\caption{Results table for the 2 PN model WITH field propagation\\Separation of primaries = $10^{4}m$}
\label{2-WF-4}
\end{sidewaystable}
\begin{sidewaystable}[!ht]
\centering
\includegraphics[width =1.1 \textwidth]{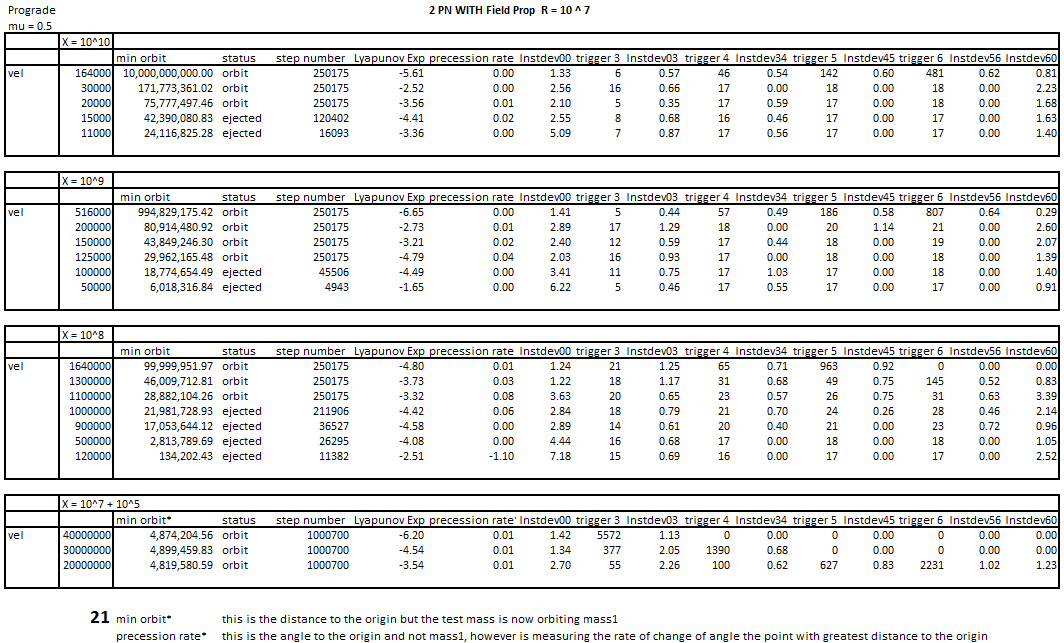}
\caption{Results table for the 2 PN model WITH field propagation\\Separation of primaries = $10^{7}m$}
\label{2-WF-7}
\end{sidewaystable}
\begin{sidewaystable}[!ht]
\centering
\includegraphics[width =1.1 \textwidth]{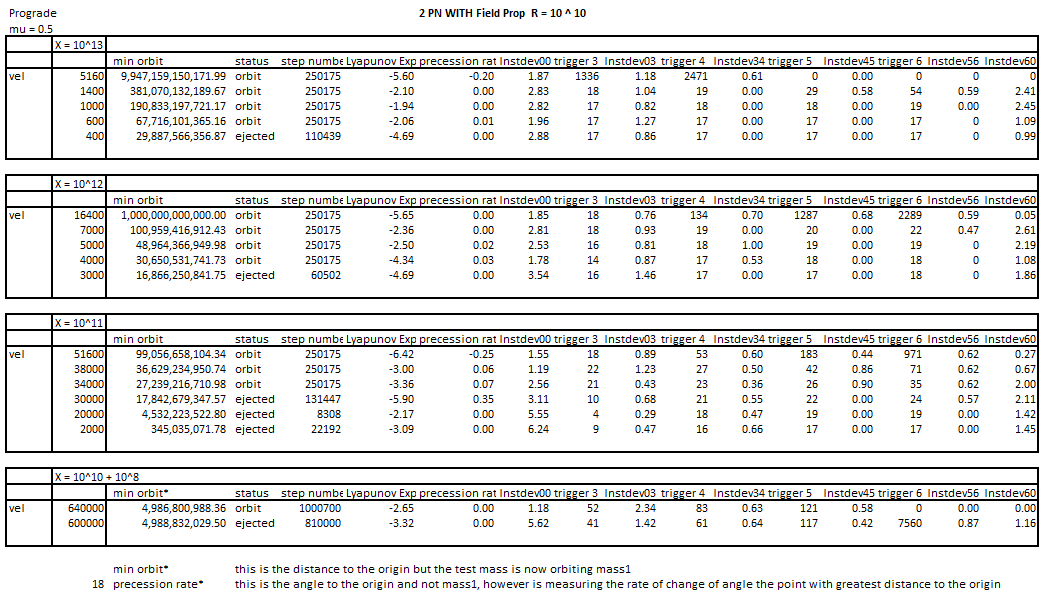}
\caption{Results table for the 2 PN model WITH field propagation\\Separation of primaries = $10^{10}m$}
\label{2-WF-10}
\end{sidewaystable}
\begin{sidewaystable}[!ht] 
\centering
\includegraphics[width =1.1 \textwidth]{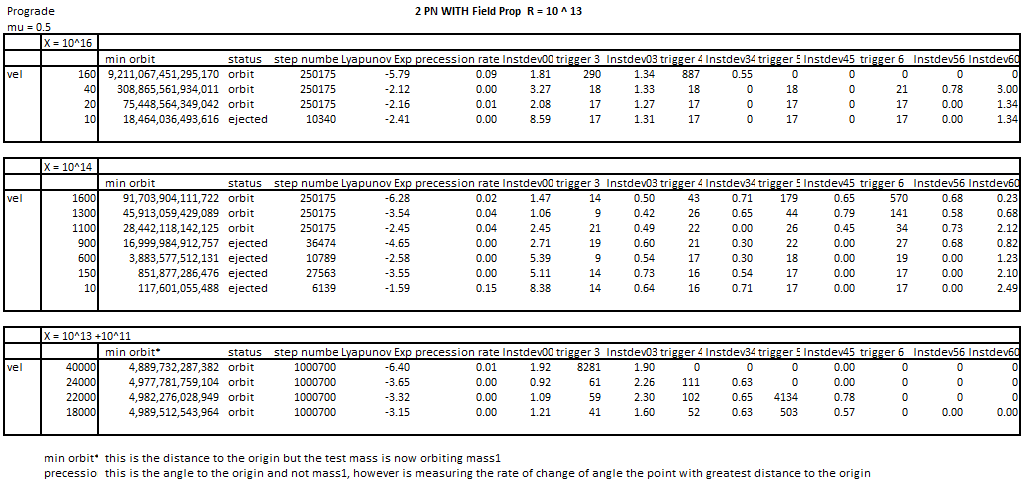}
\caption{Results table for the 2 PN model WITH field propagation\\Separation of primaries = $10^{13}m$}
\label{2-WF-13}
\end{sidewaystable}

\begin{sidewaystable}[!ht]
\centering
\includegraphics[width =1.1 \textwidth]{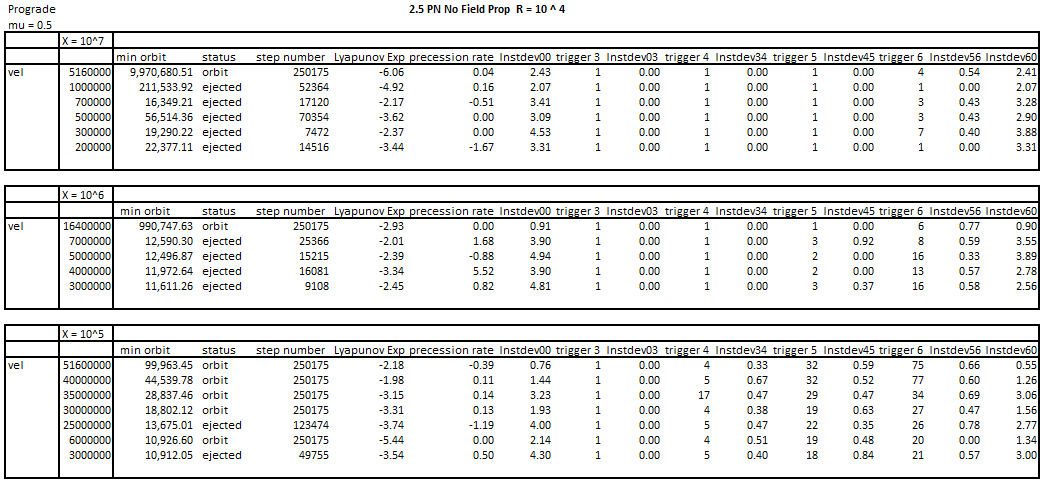}
\caption{Results table for the 2.5 PN model with no field propagation\\Separation of primaries = $10^{4}m$}
\label{2.5-NF-4}
\end{sidewaystable}
\begin{sidewaystable}[!ht]
\centering
\includegraphics[width =1.1 \textwidth]{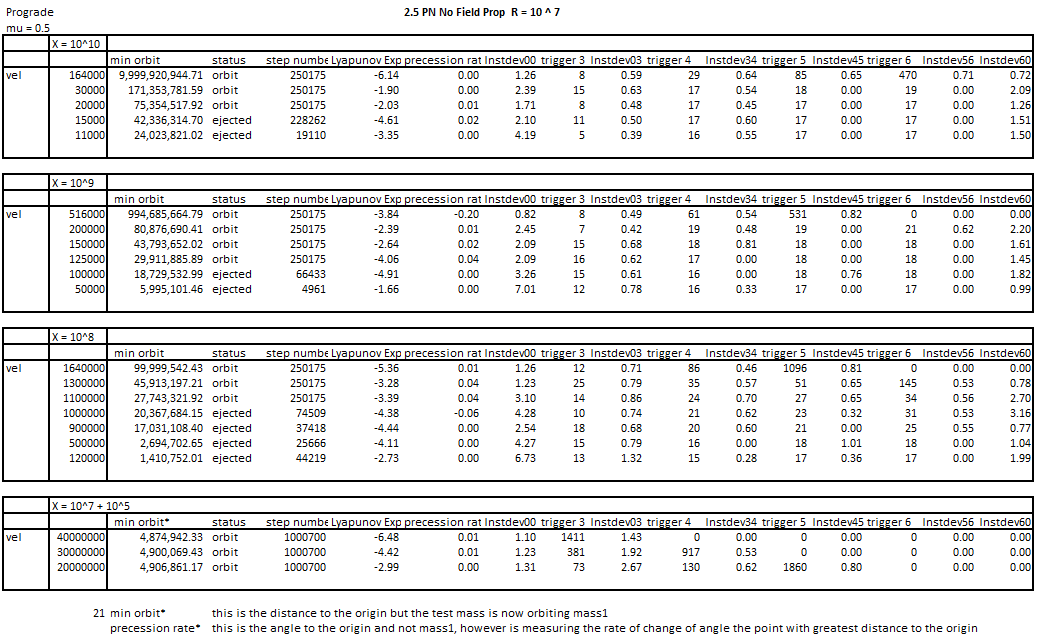}
\caption{Results table for the 2.5 PN model with no field propagation\\Separation of primaries = $10^{7}m$}
\label{2.5-NF-7}
\end{sidewaystable}
\begin{sidewaystable}[!ht]
\centering
\includegraphics[width =1.1 \textwidth]{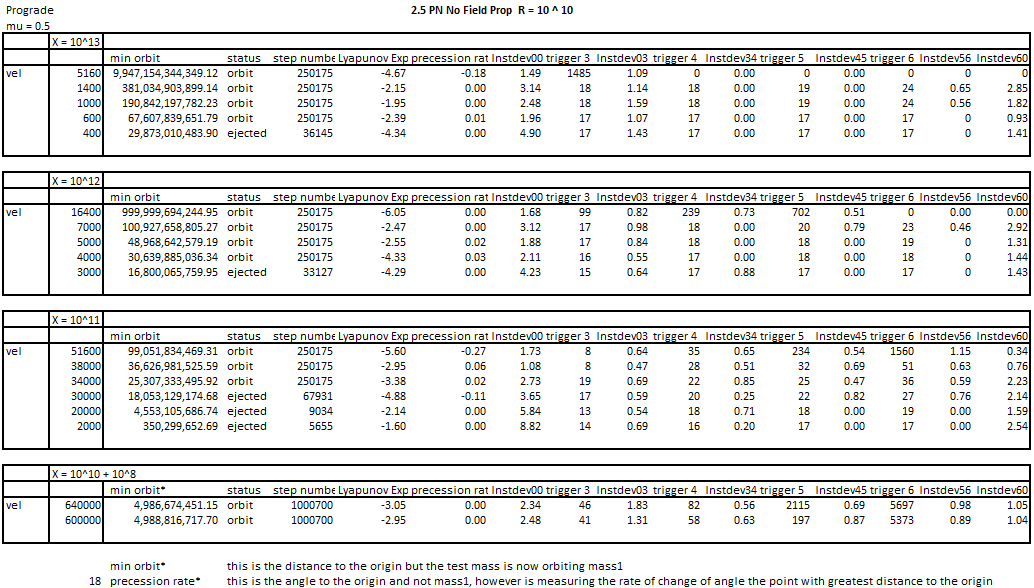}
\caption{Results table for the 2.5 PN model with no field propagation\\Separation of primaries = $10^{10}m$}
\label{2.5-NF-10}
\end{sidewaystable}
\begin{sidewaystable}[!ht]
\centering
\includegraphics[width =1.1 \textwidth]{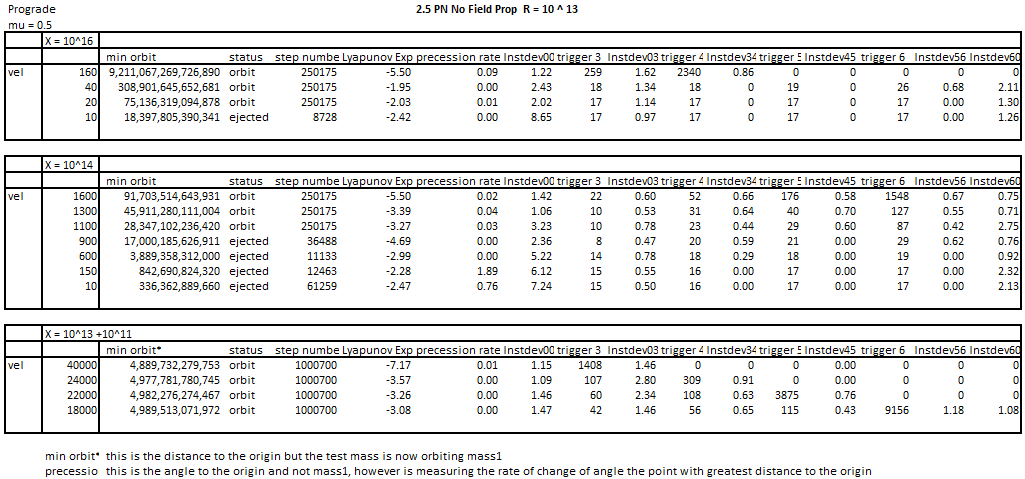}
\caption{Results table for the 2.5 PN model with no field propagation\\Separation of primaries = $10^{13}m$}
\label{2.5-NF-13}
\end{sidewaystable}

\begin{sidewaystable}[!ht]
\centering
\includegraphics[width =1.1 \textwidth]{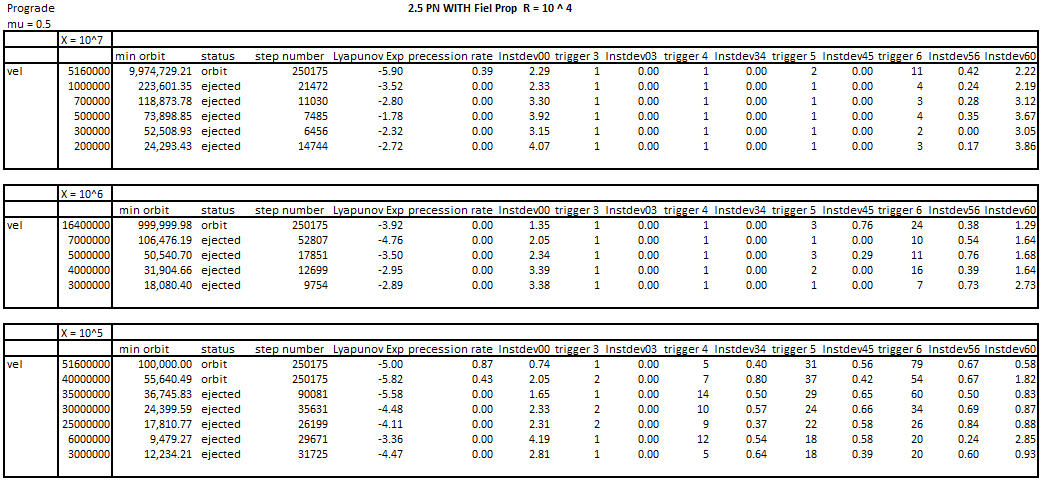}
\caption{Results table for the 2.5 PN model WITH field propagation\\Separation of primaries = $10^{4}m$}
\label{2.5-WF-4}
\end{sidewaystable}
\begin{sidewaystable}[!ht]
\centering
\includegraphics[width =1.1 \textwidth]{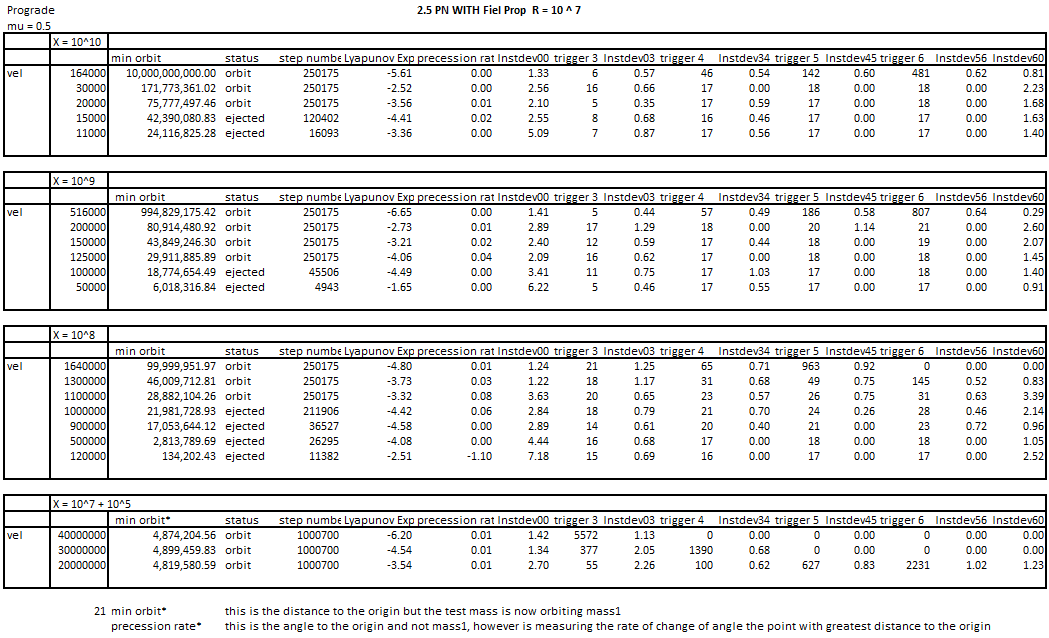}
\caption{Results table for the 2.5 PN model WITH field propagation\\Separation of primaries = $10^{7}m$}
\label{2.5-WF-7}
\end{sidewaystable}
\begin{sidewaystable}[!ht]
\centering
\includegraphics[width =1.1 \textwidth]{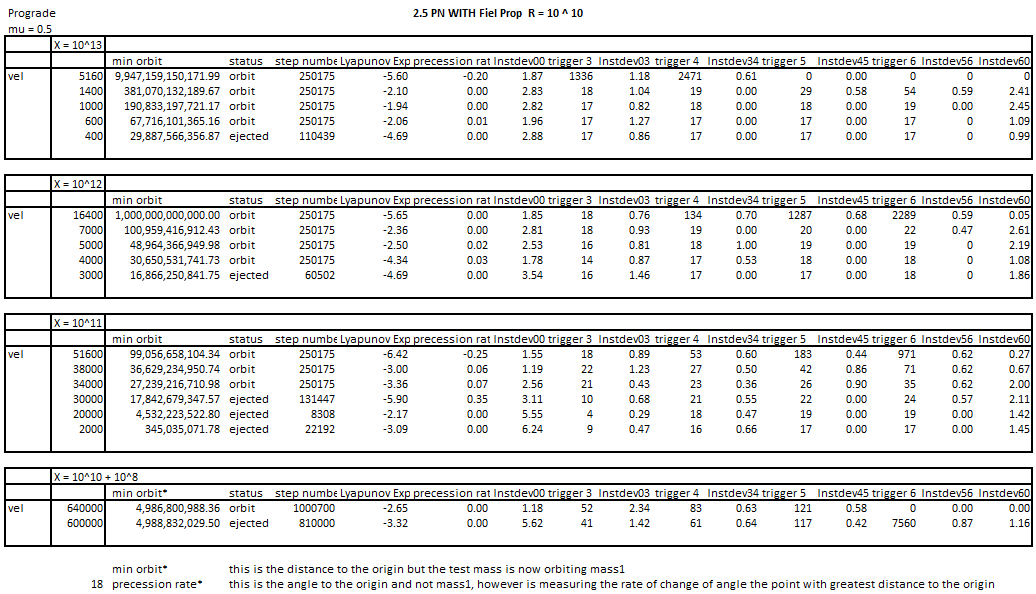}
\caption{Results table for the 2.5 PN model WITH field propagation\\Separation of primaries = $10^{10}m$}
\label{2.5-WF-10}
\end{sidewaystable}
\begin{sidewaystable}[!ht]
\centering
\includegraphics[width =1.1 \textwidth]{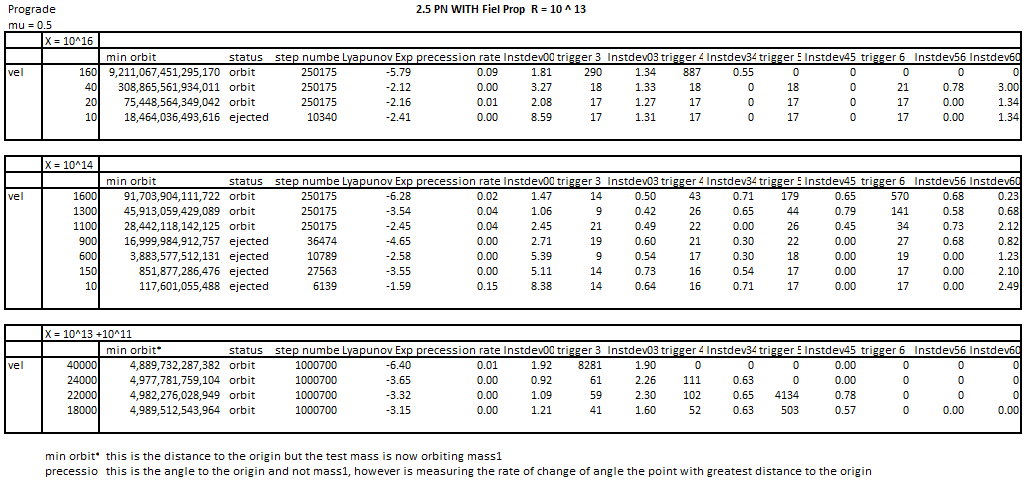}
\caption{Results table for the 2.5 PN model WITH field propagation\\Separation of primaries = $10^{13}m$}
\label{2.5-WF-13}
\end{sidewaystable}

\begin{table}[!ht]
\centering
\includegraphics[width =1.1 \textwidth]{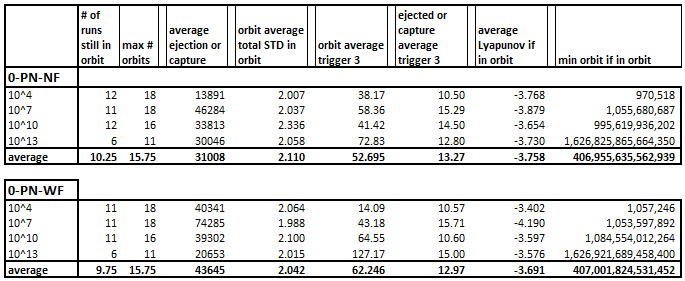}
\caption{Statistics from Results table for the 0 PN model}
\label{av-res-0PN}
\end{table}
\begin{table}[!ht]
\centering
\includegraphics[width =1.1 \textwidth]{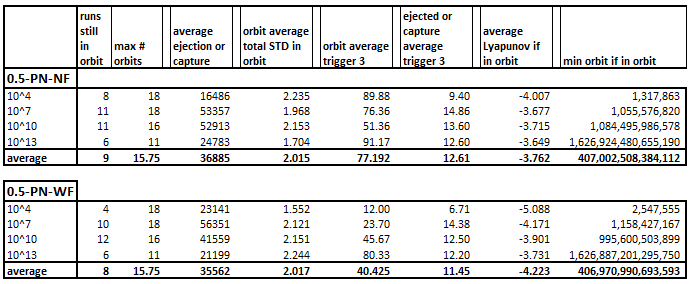}
\caption{Statistics from Results table for the 0.5 PN model}
\label{av-res-0.5PN}
\end{table}
\begin{table}[!ht]
\centering
\includegraphics[width =1.1 \textwidth]{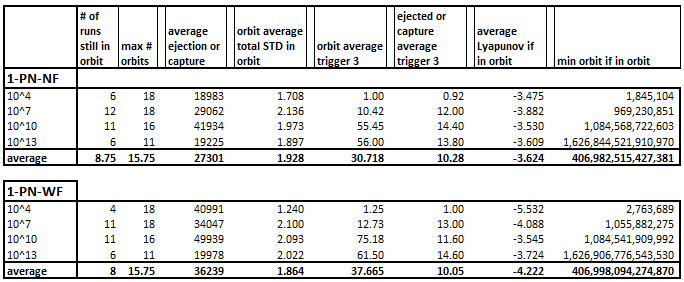}
\caption{Statistics from Results table for the 1 PN model}
\label{av-res-1PN}
\end{table}
\begin{table}[!ht]
\centering
\includegraphics[width =1.1 \textwidth]{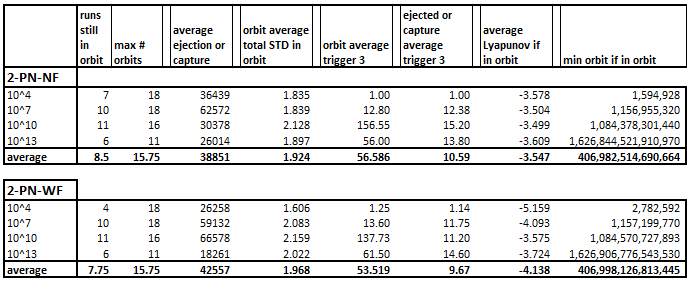}
\caption{Statistics from Results table for the 2 PN model}
\label{av-res-2PN}
\end{table}
\begin{table}[!ht]
\centering
\includegraphics[width =1.1 \textwidth]{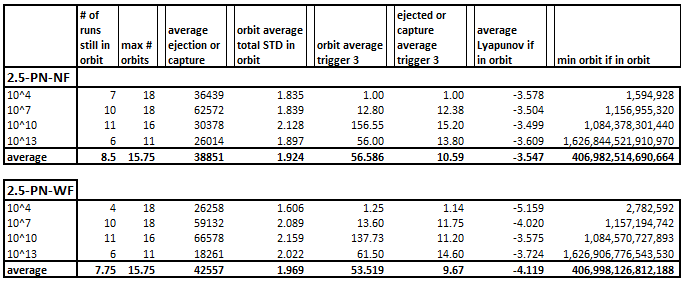}
\caption{Statistics from Results table for the 2.5 PN model}
\label{av-res-2.5PN}
\end{table}

\chapter{Results: Figures} \label{results-figures}
\begin{figure}[!ht]
    \centering
    \subfigure[0 PN-no-Prop]{\includegraphics[width=0.24\textwidth]{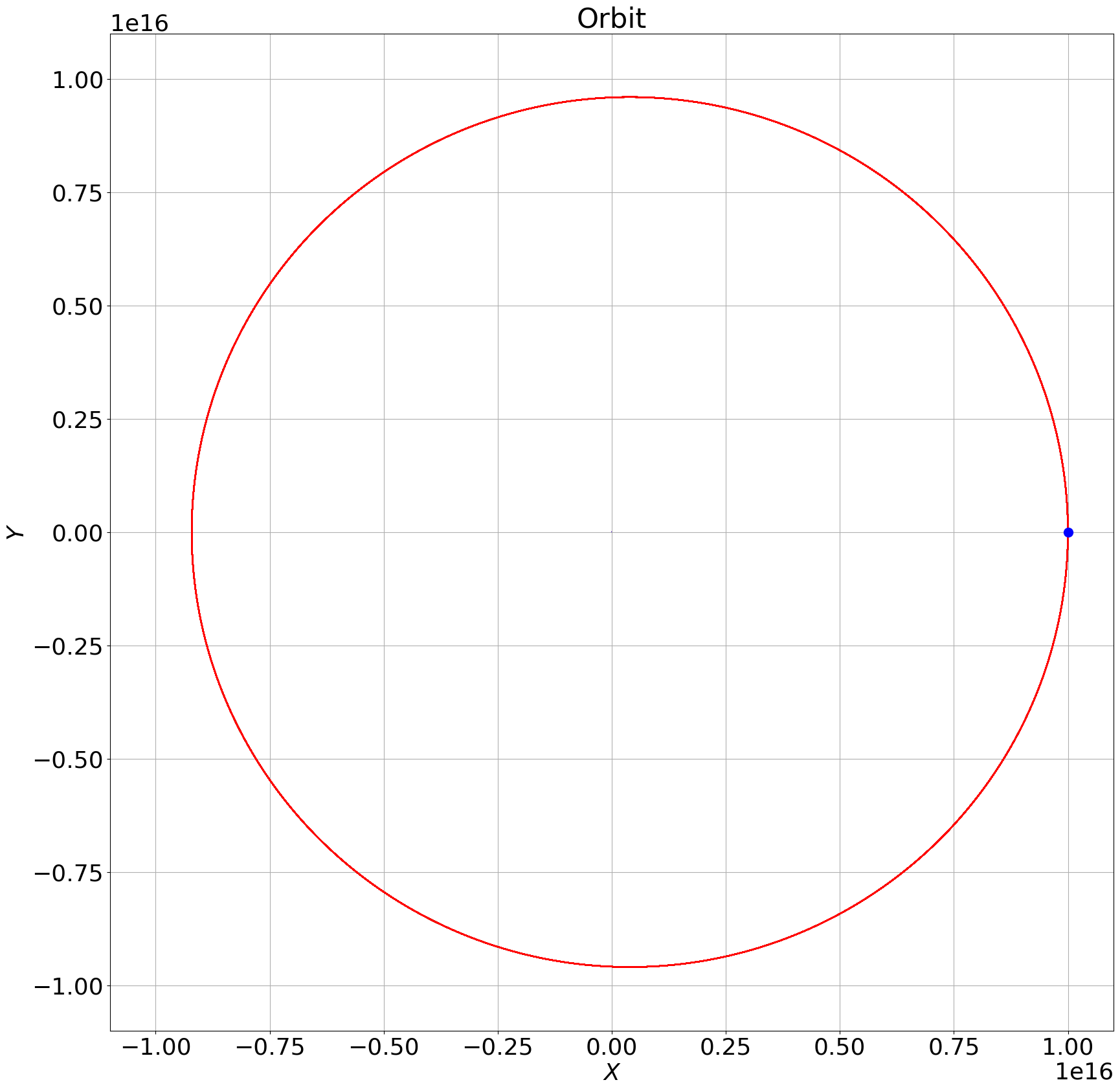}}
    \subfigure[0 PN-no-Prop]{\includegraphics[width=0.24\textwidth]{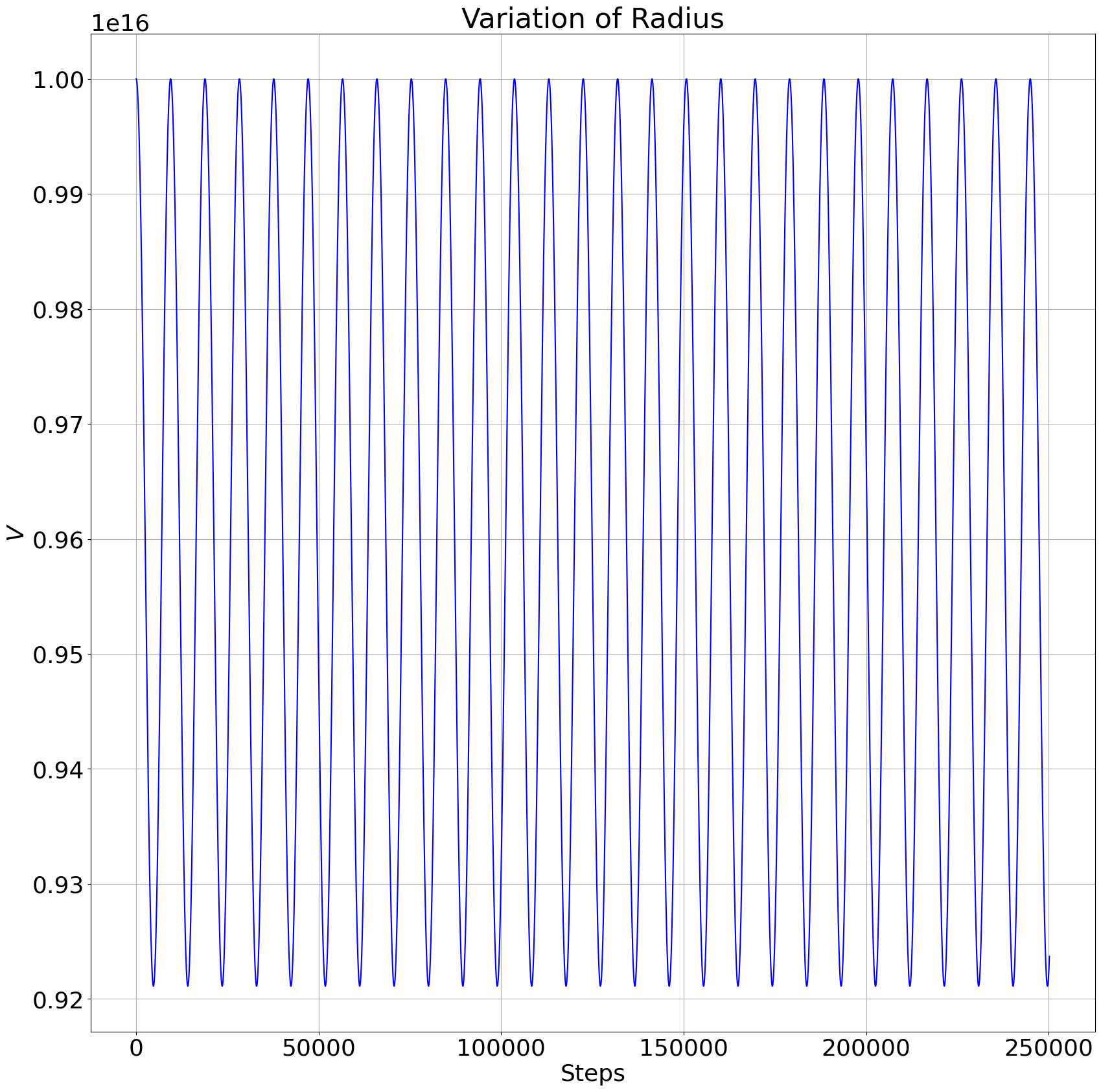}} 
    \subfigure[0 PN-with-Prop]{\includegraphics[width=0.24\textwidth]{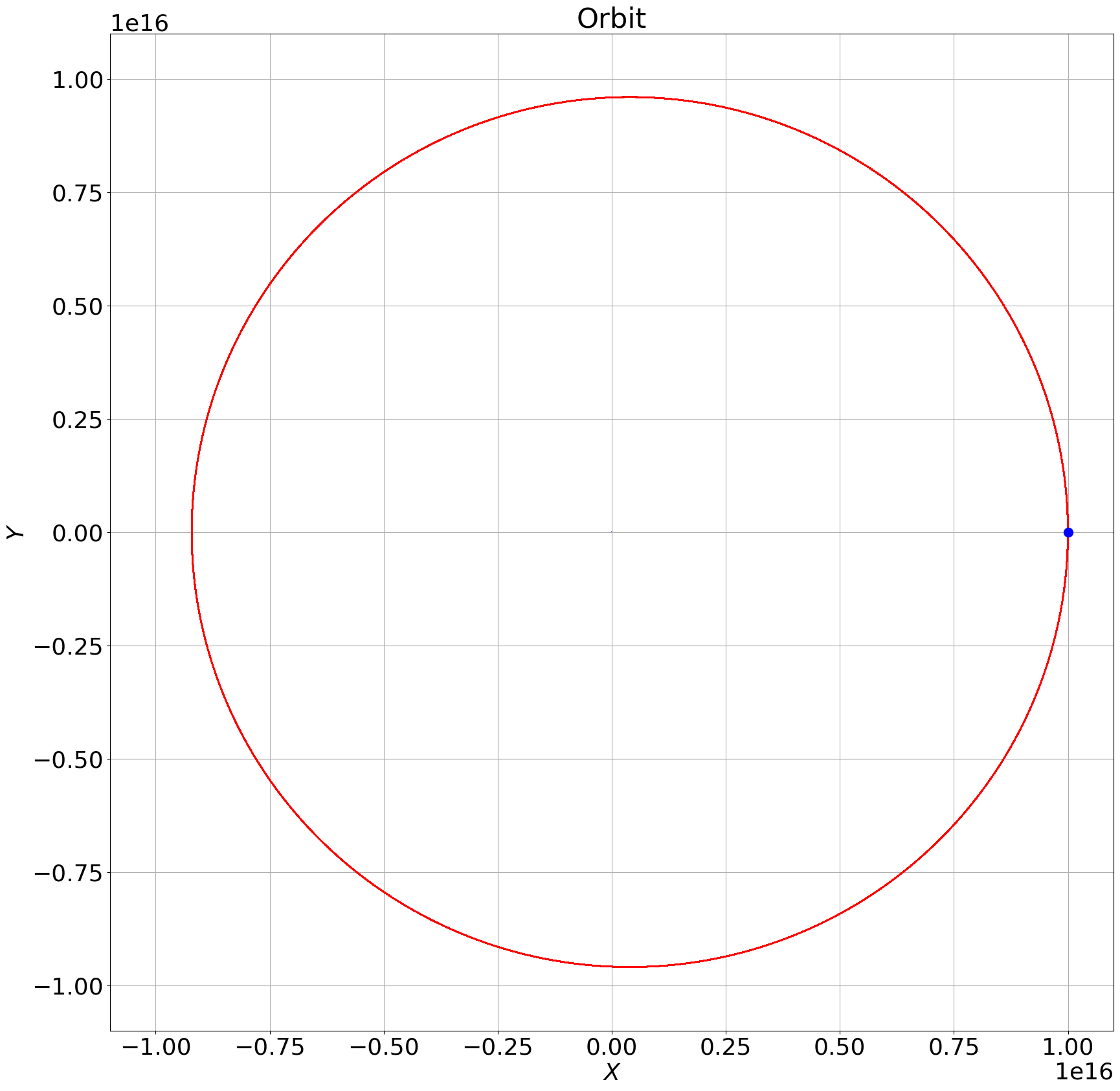}}
    \subfigure[0 PN-with-Prop]{\includegraphics[width=0.24\textwidth]{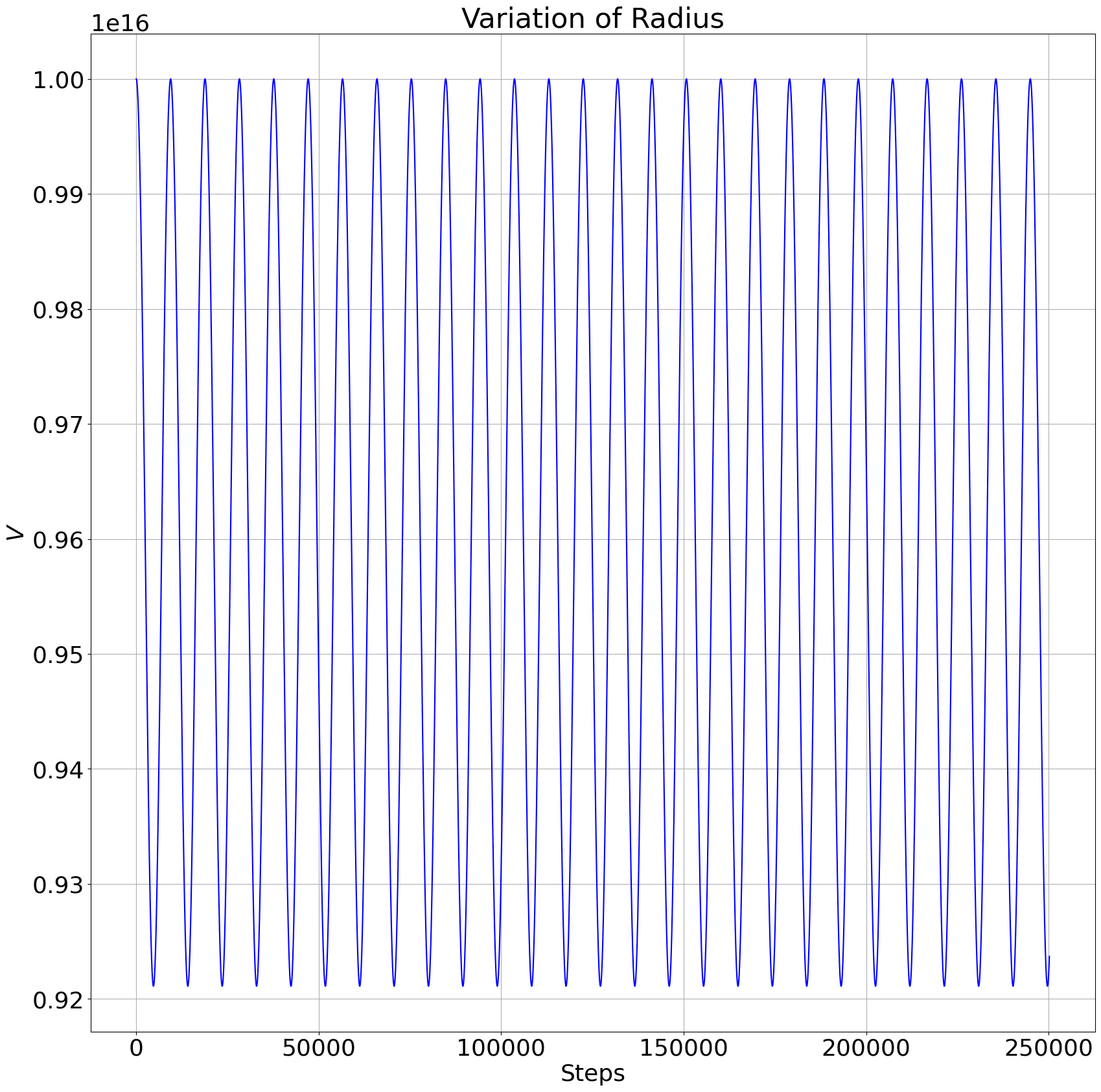}} \\
    \subfigure[0.5 PN-no-Prop]{\includegraphics[width=0.24\textwidth]{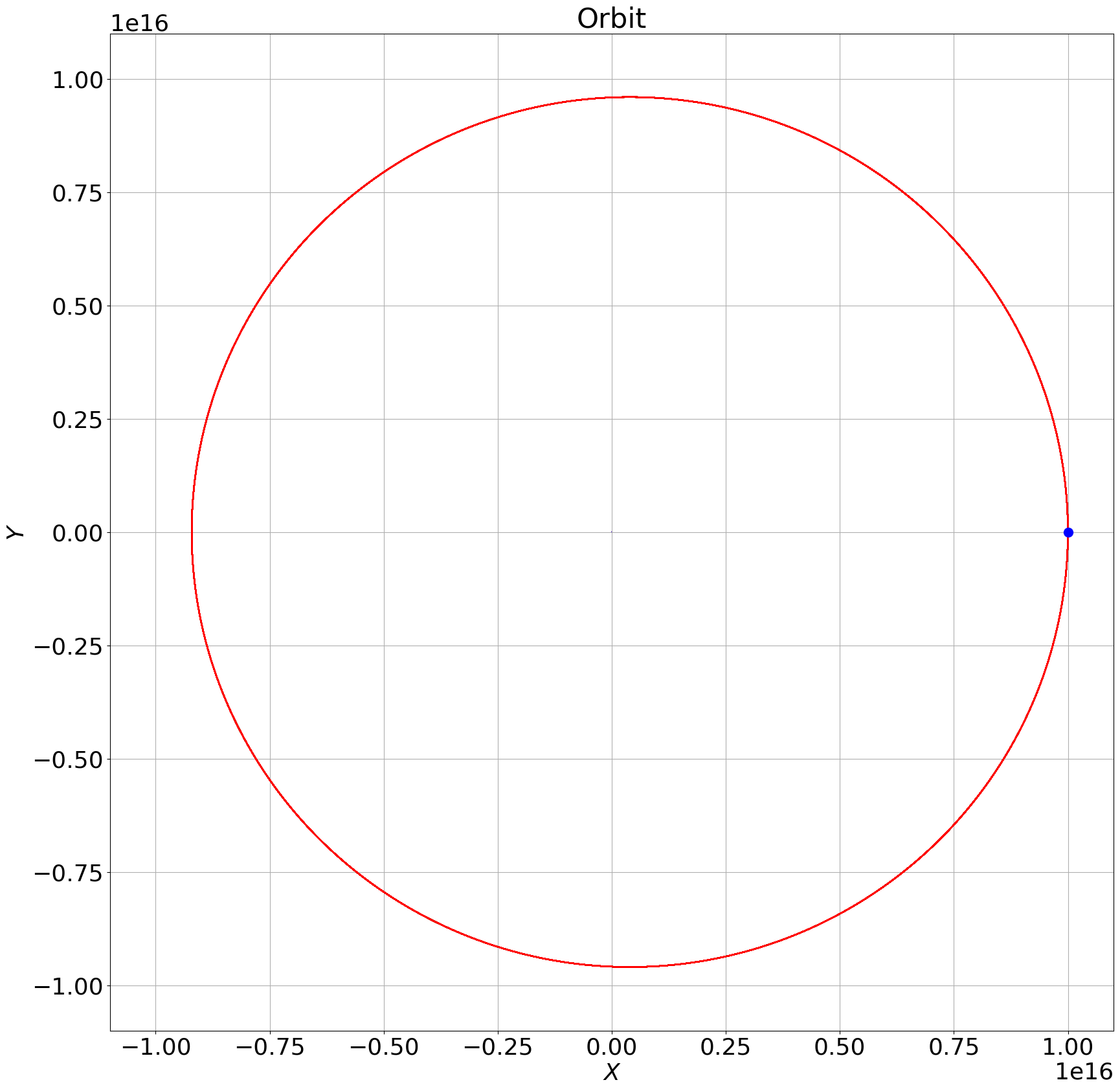}}
    \subfigure[0.5 PN-no-Prop]{\includegraphics[width=0.24\textwidth]{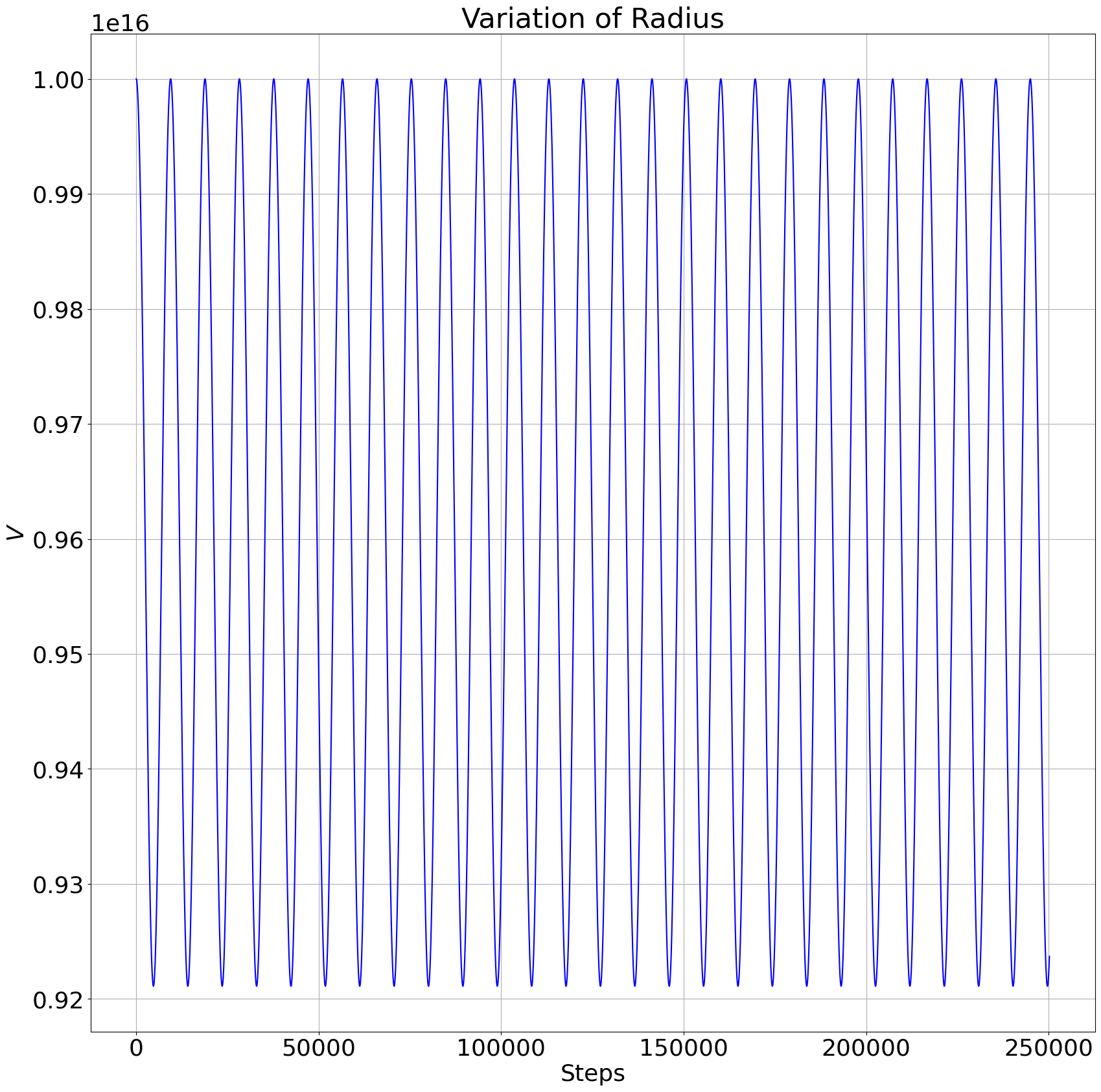}} 
    \subfigure[0.5 PN-with-Prop]{\includegraphics[width=0.24\textwidth]{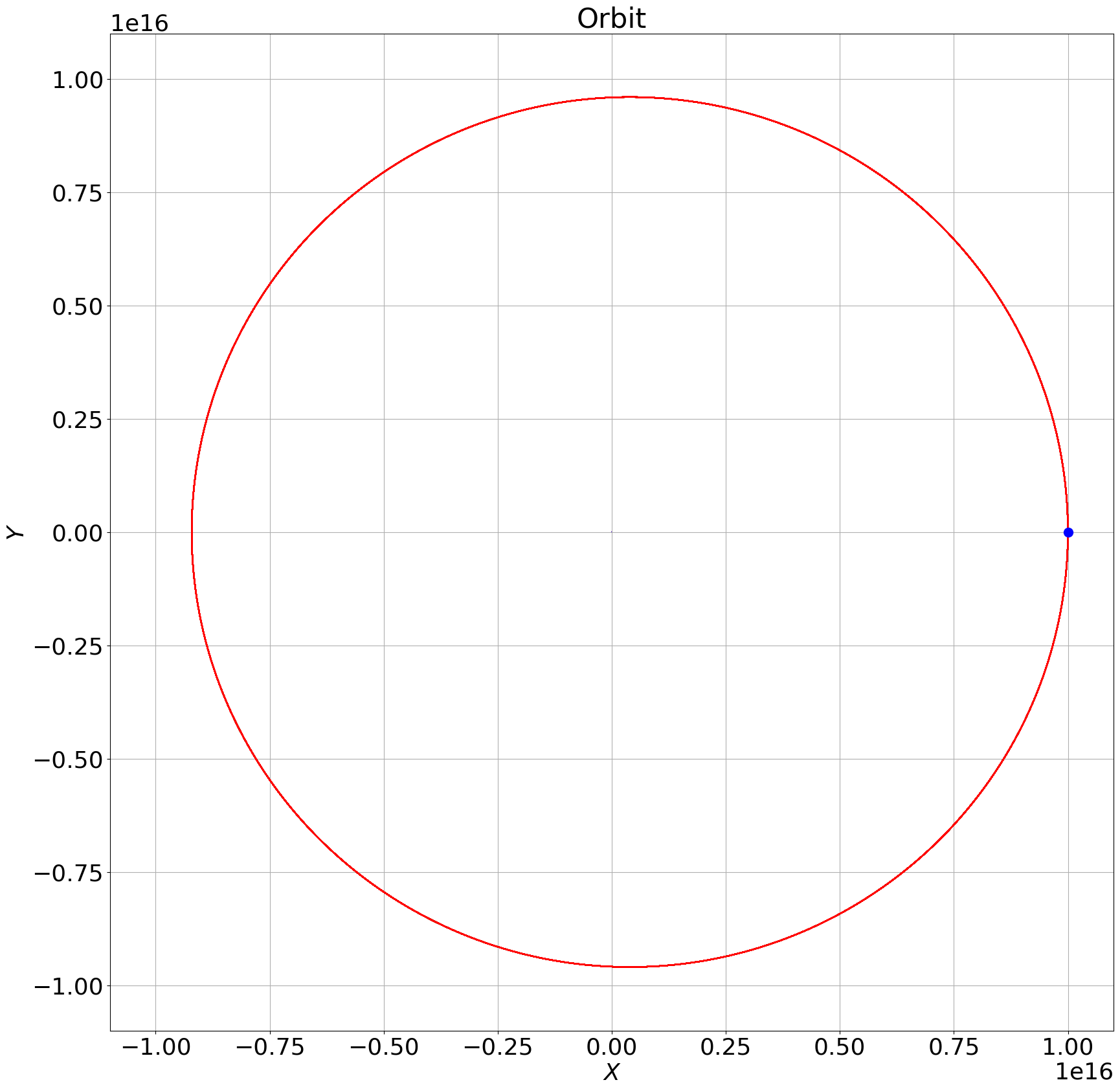}}
    \subfigure[0.5 PN-with-Prop]{\includegraphics[width=0.24\textwidth]{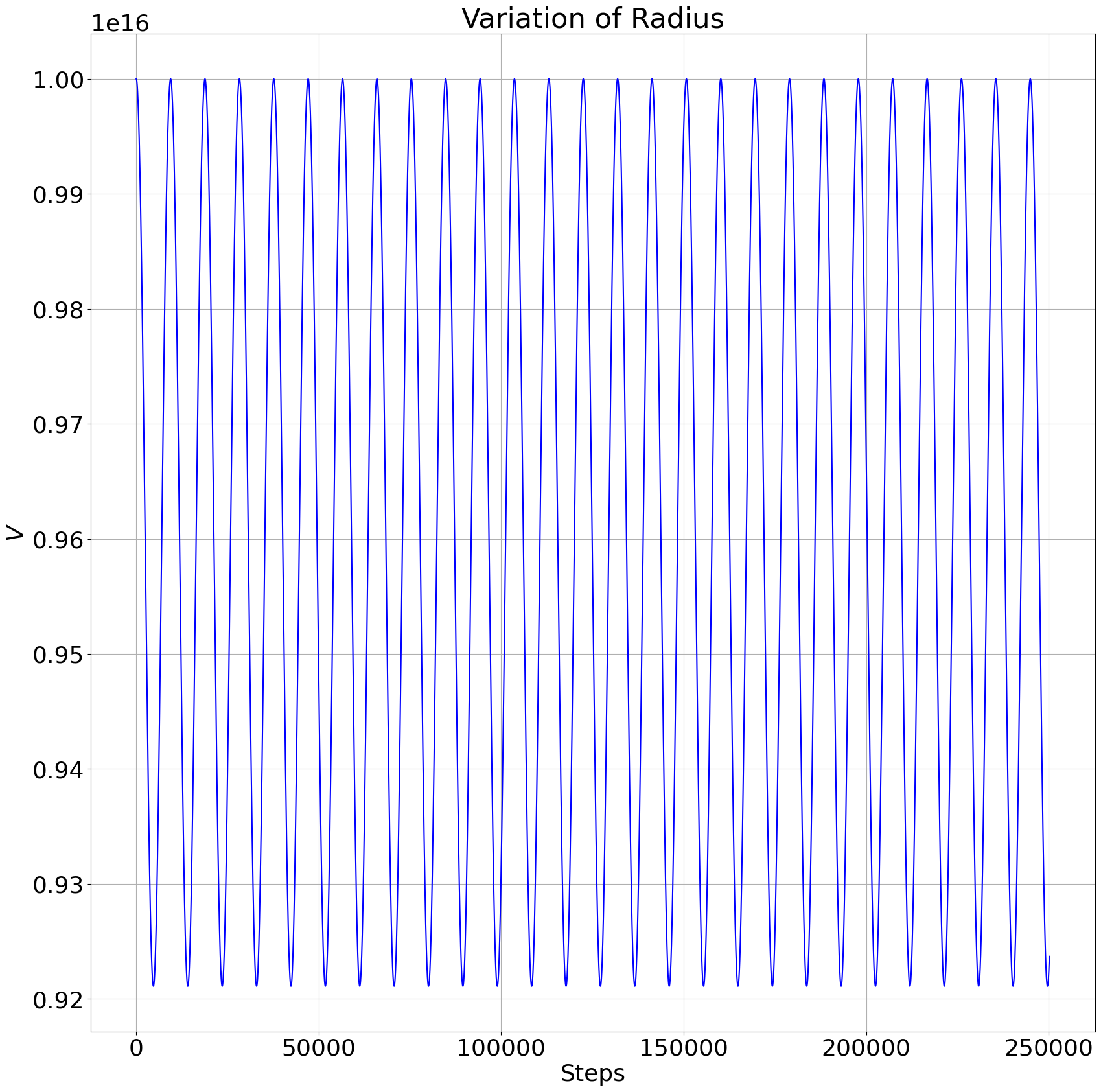}} \\ 
    \subfigure[1 PN-no-Prop]{\includegraphics[width=0.24\textwidth]{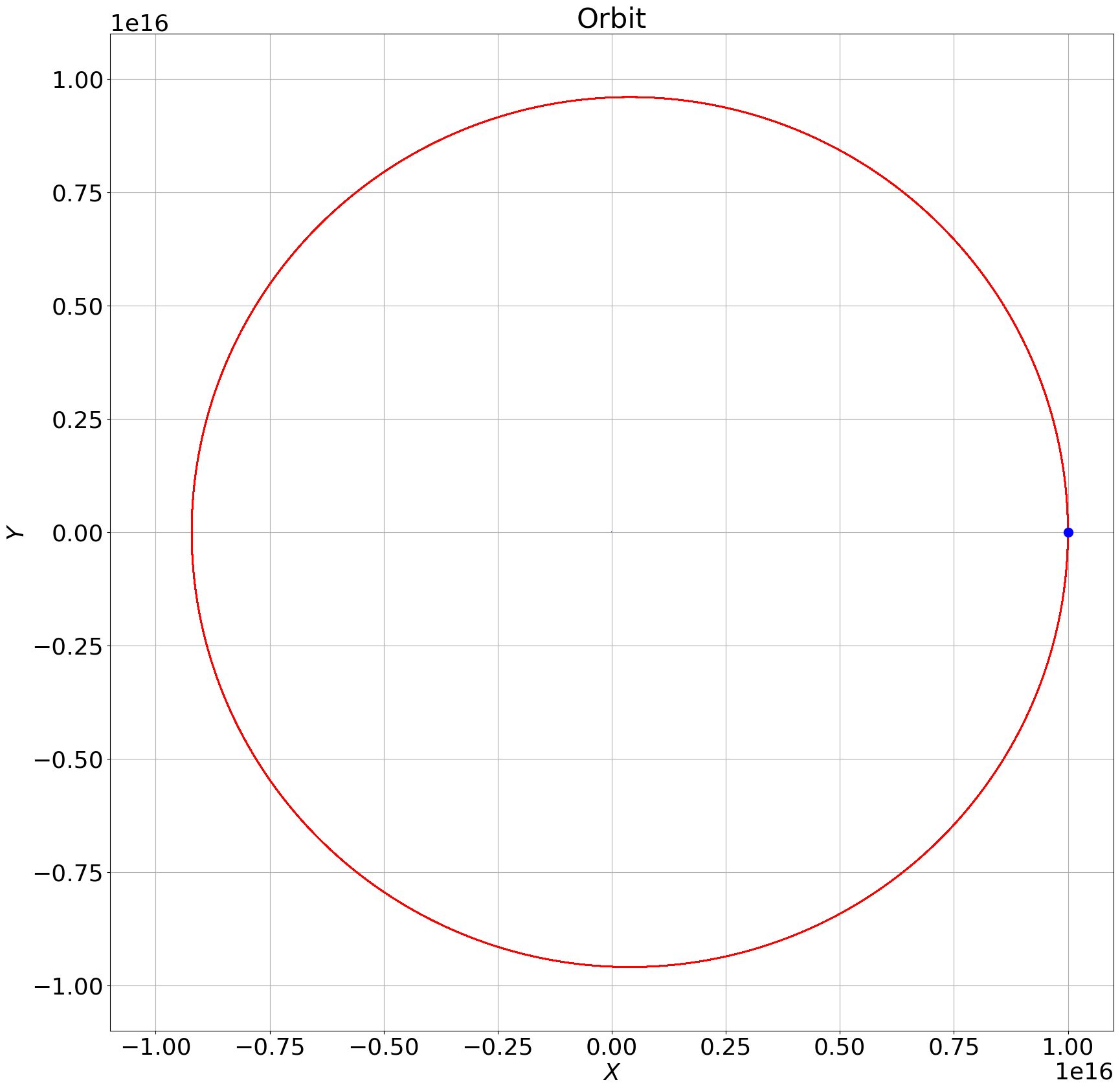}}
    \subfigure[1 PN-no-Prop]{\includegraphics[width=0.24\textwidth]{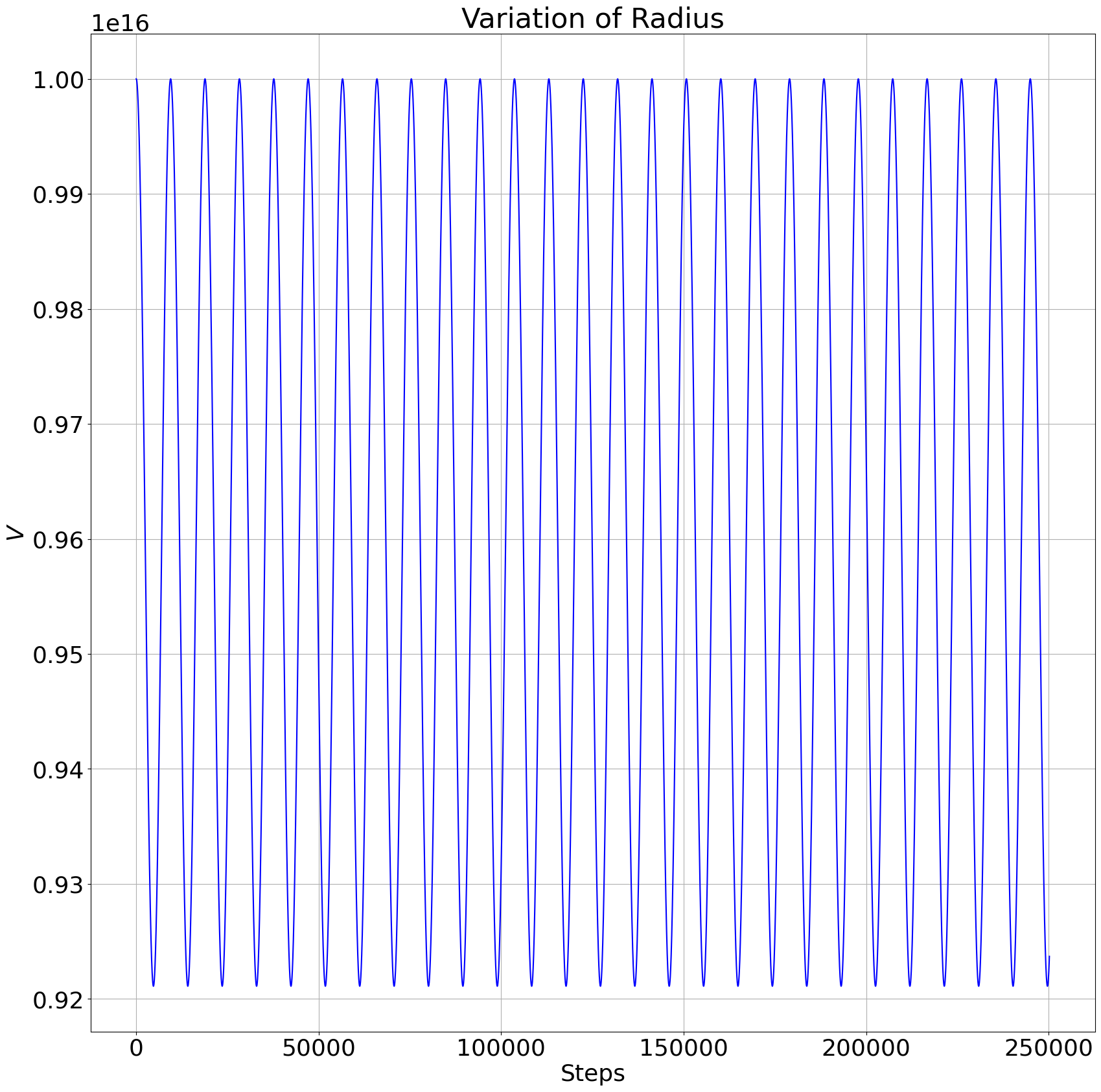}} 
    \subfigure[1 PN-with-Prop]{\includegraphics[width=0.24\textwidth]{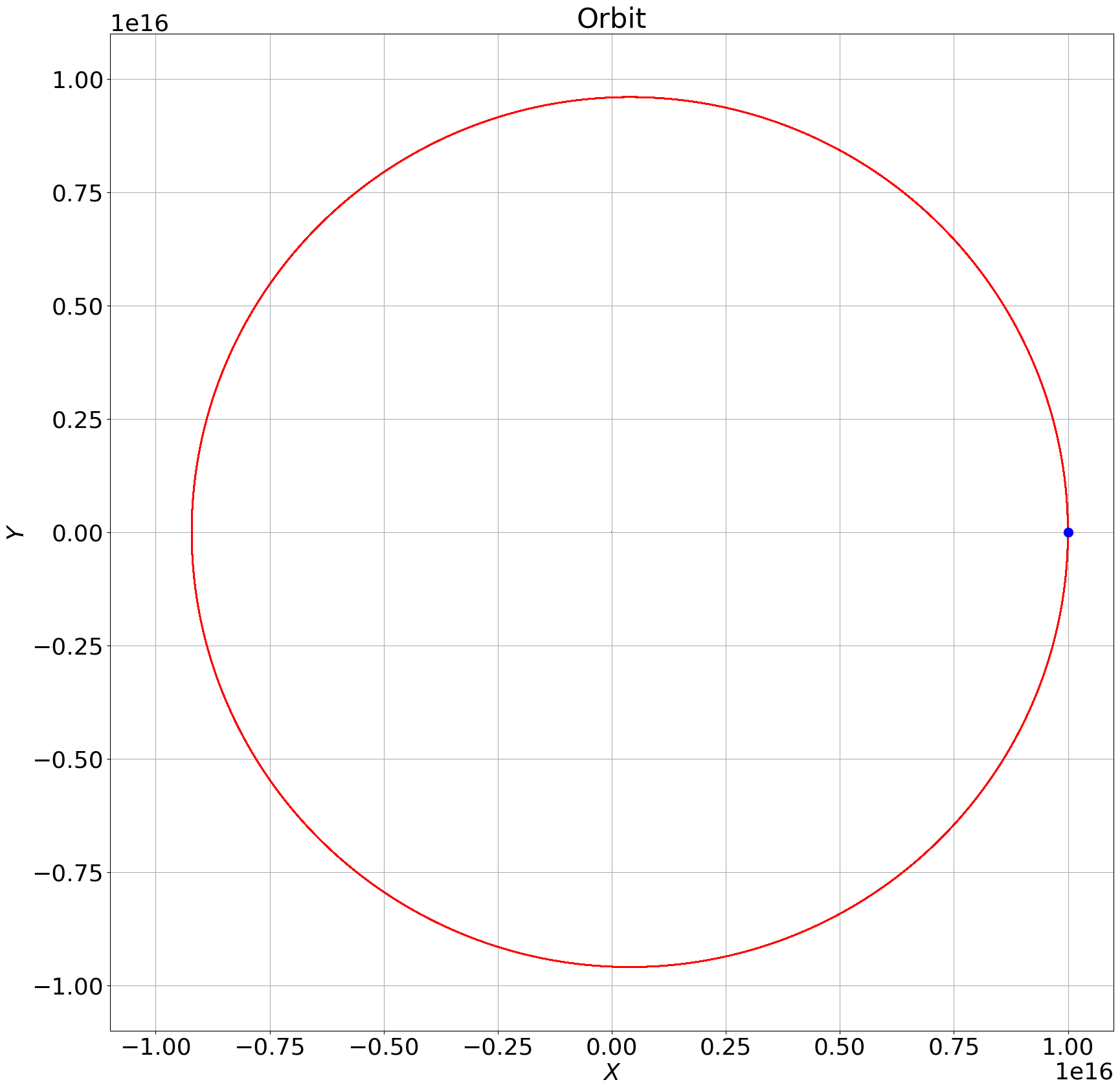}}
    \subfigure[1 PN-with-Prop]{\includegraphics[width=0.24\textwidth]{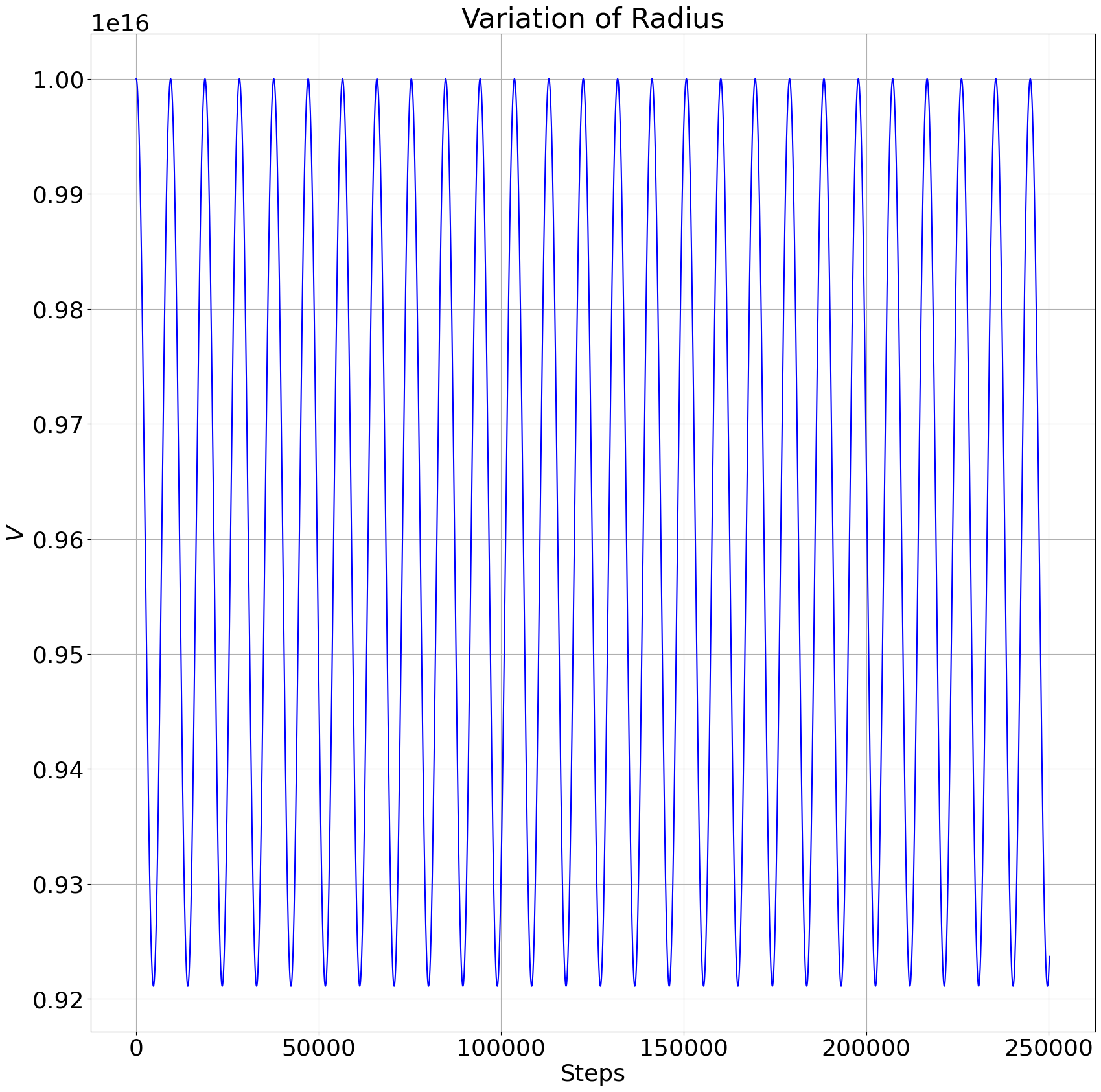}} \\
    \subfigure[2 PN-no-Prop]{\includegraphics[width=0.24\textwidth]{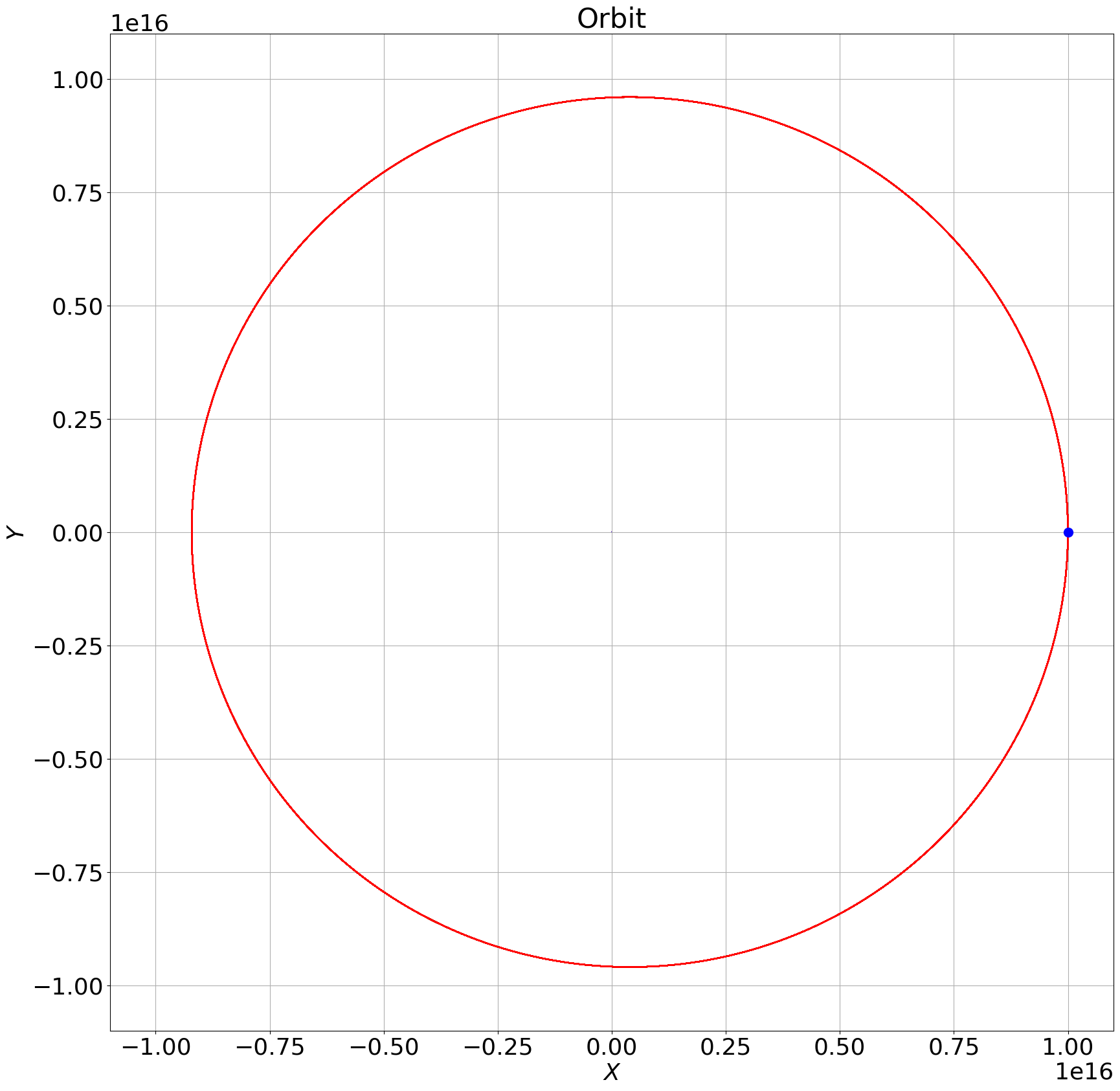}}
    \subfigure[2 PN-no-Prop]{\includegraphics[width=0.24\textwidth]{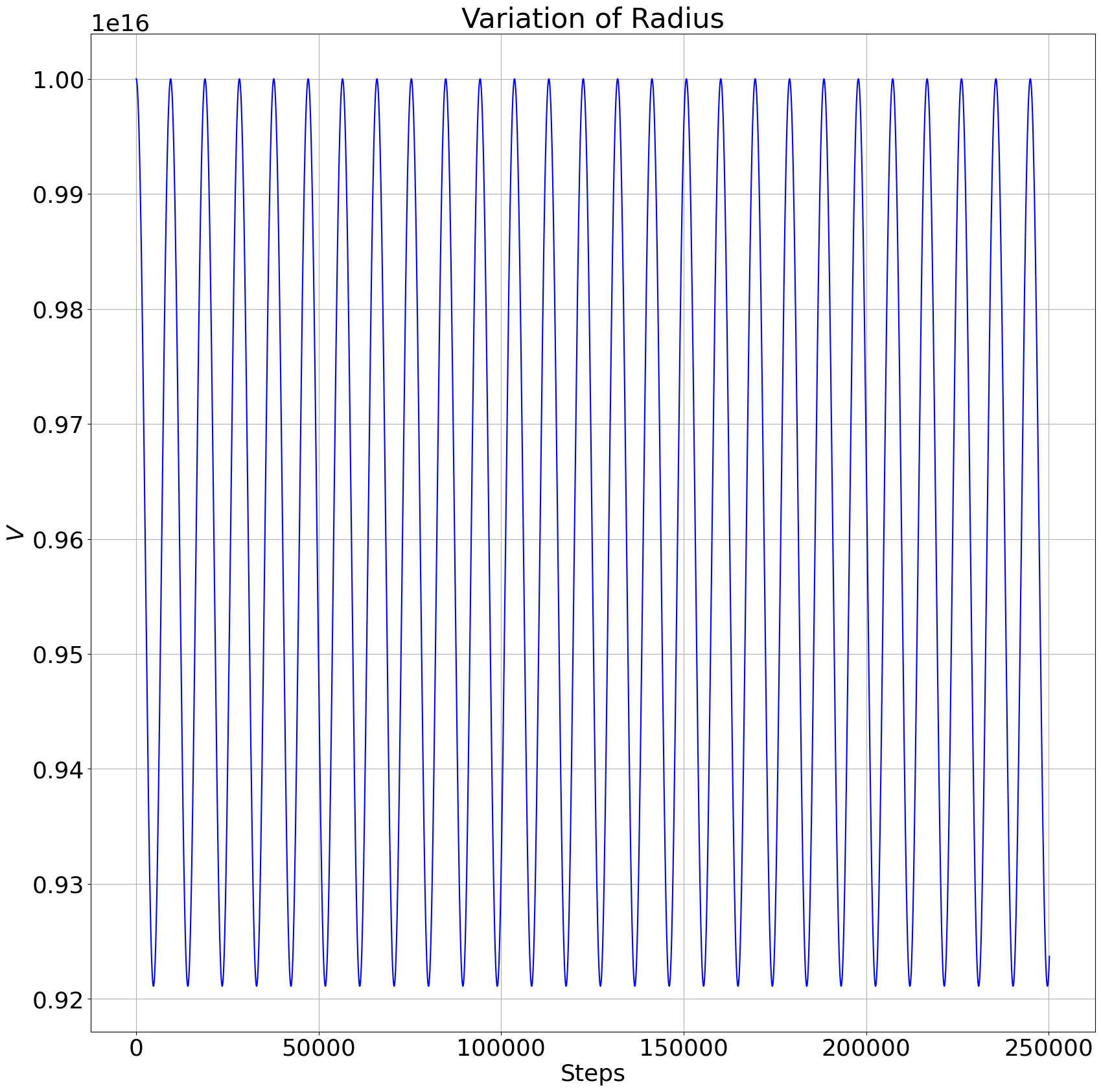}} 
    \subfigure[2 PN-with-Prop]{\includegraphics[width=0.24\textwidth]{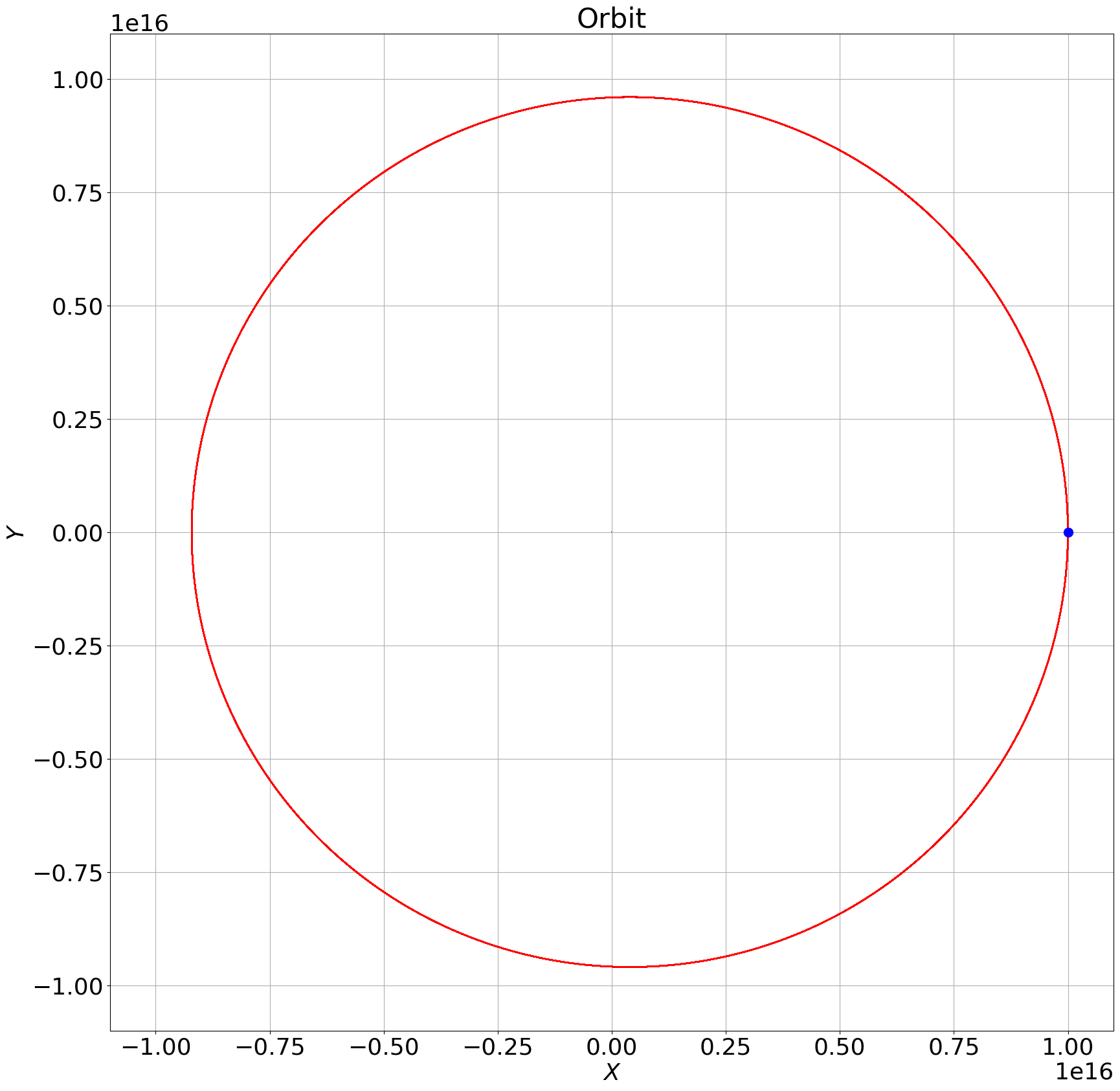}}
    \subfigure[2 PN-with-Prop]{\includegraphics[width=0.24\textwidth]{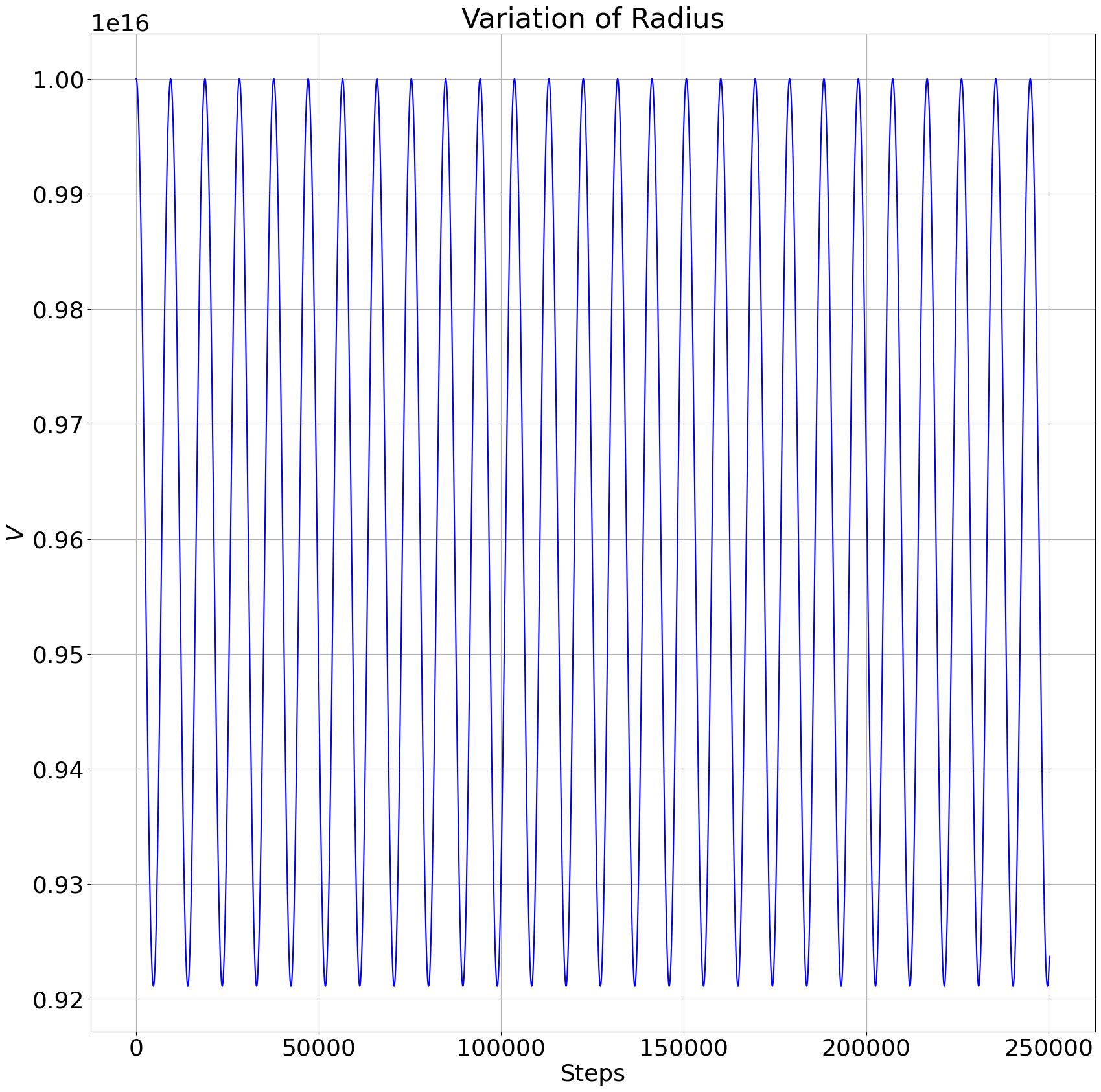}} \\ 
    \subfigure[2.5 PN-no-Prop]{\includegraphics[width=0.24\textwidth]{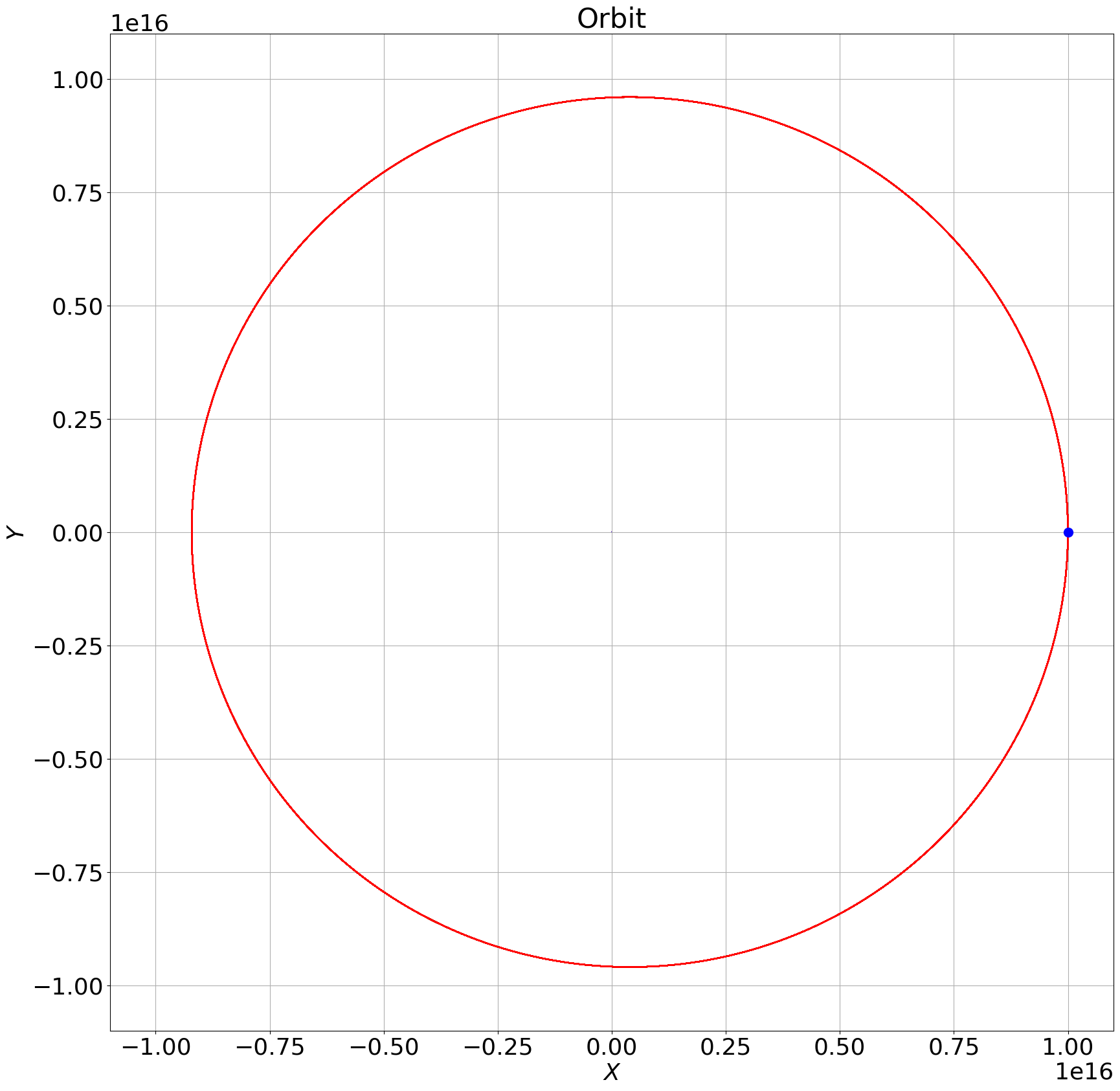}}
    \subfigure[2.5 PN-no-Prop]{\includegraphics[width=0.24\textwidth]{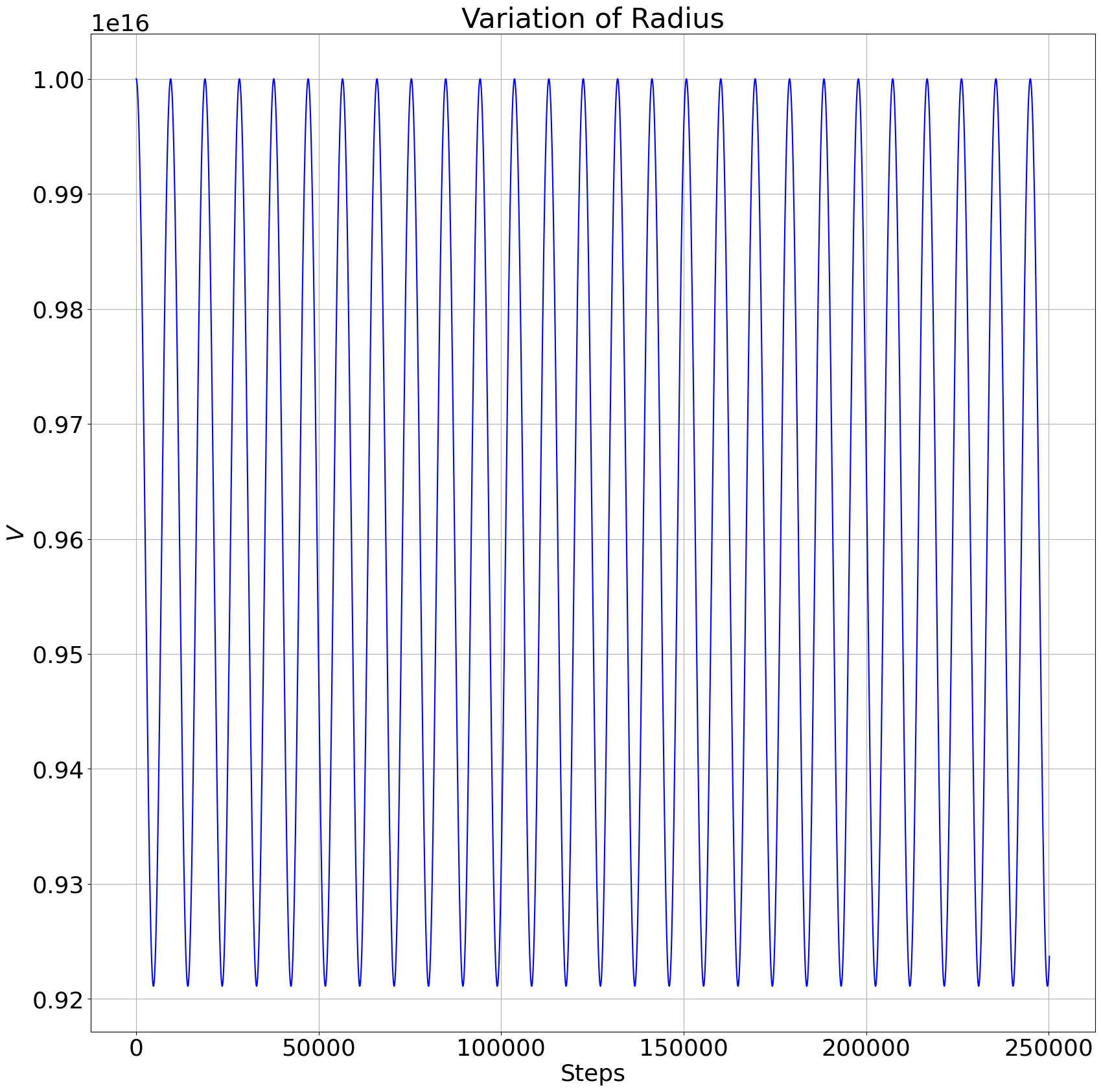}} 
    \subfigure[2.5 PN-with-Prop]{\includegraphics[width=0.24\textwidth]{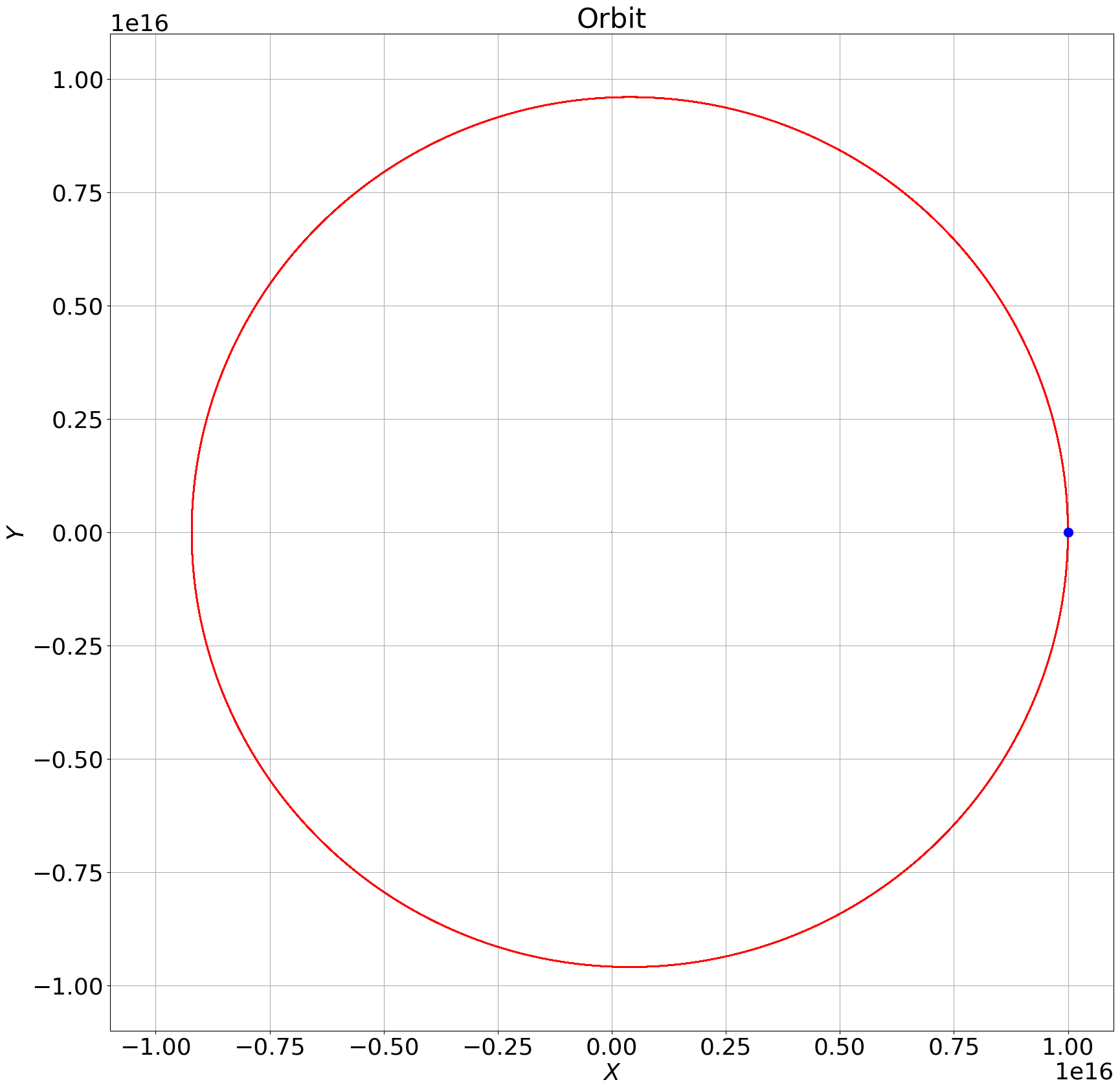}}
    \subfigure[2.5 PN-with-Prop]{\includegraphics[width=0.24\textwidth]{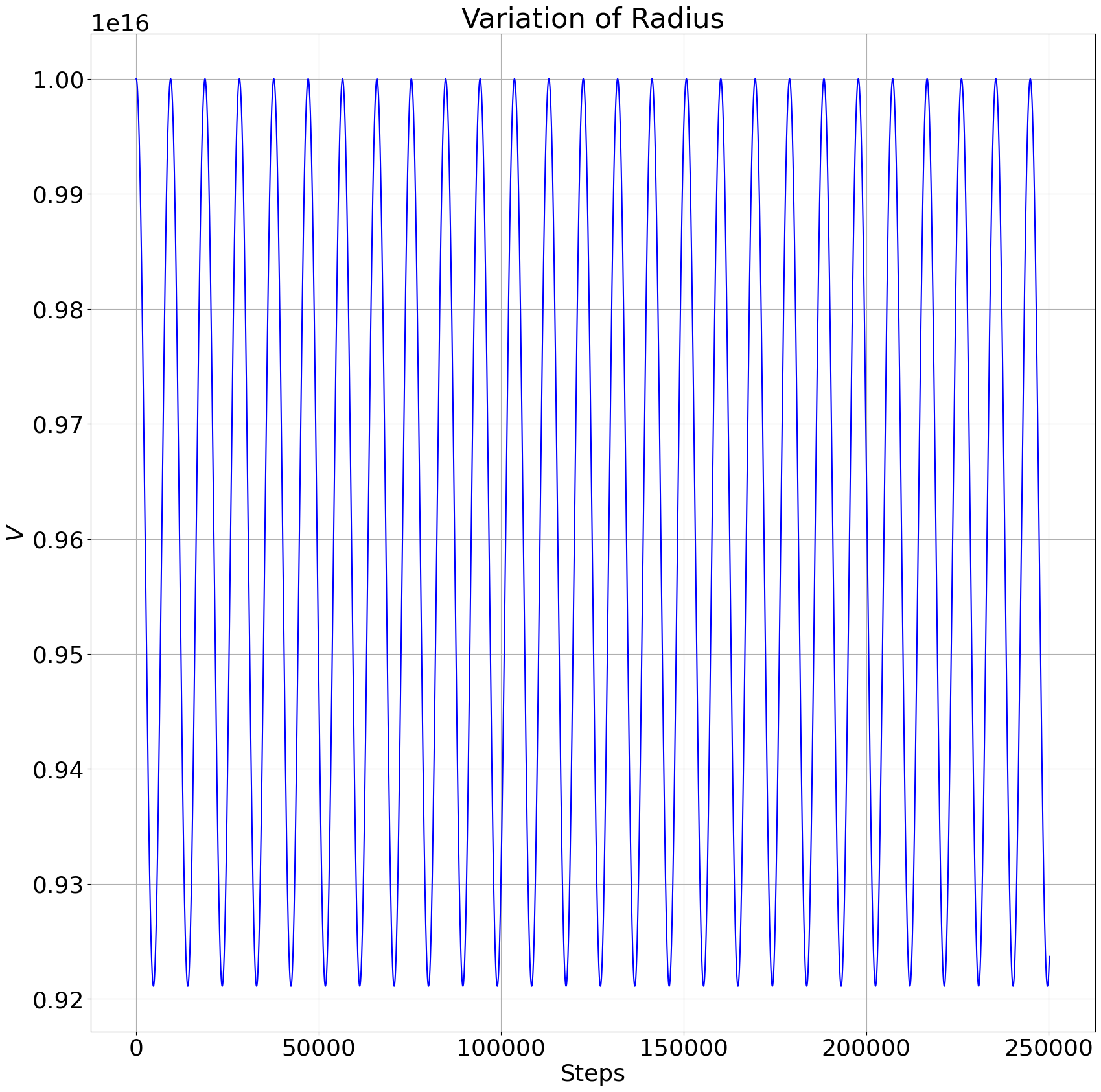}} \\ 
    \caption{Weak Field, Circular. R = $1.1 \cdot 10^{13}m$, X = $10^{16}5m$, Velocity = 160$ms^{-1}$}
    \label{results1}
\end{figure}
\begin{figure}[!ht]
    \centering
    \setcounter{subfigure}{0}
    \subfigure[0 PN-no-Prop]{\includegraphics[width=0.24\textwidth]{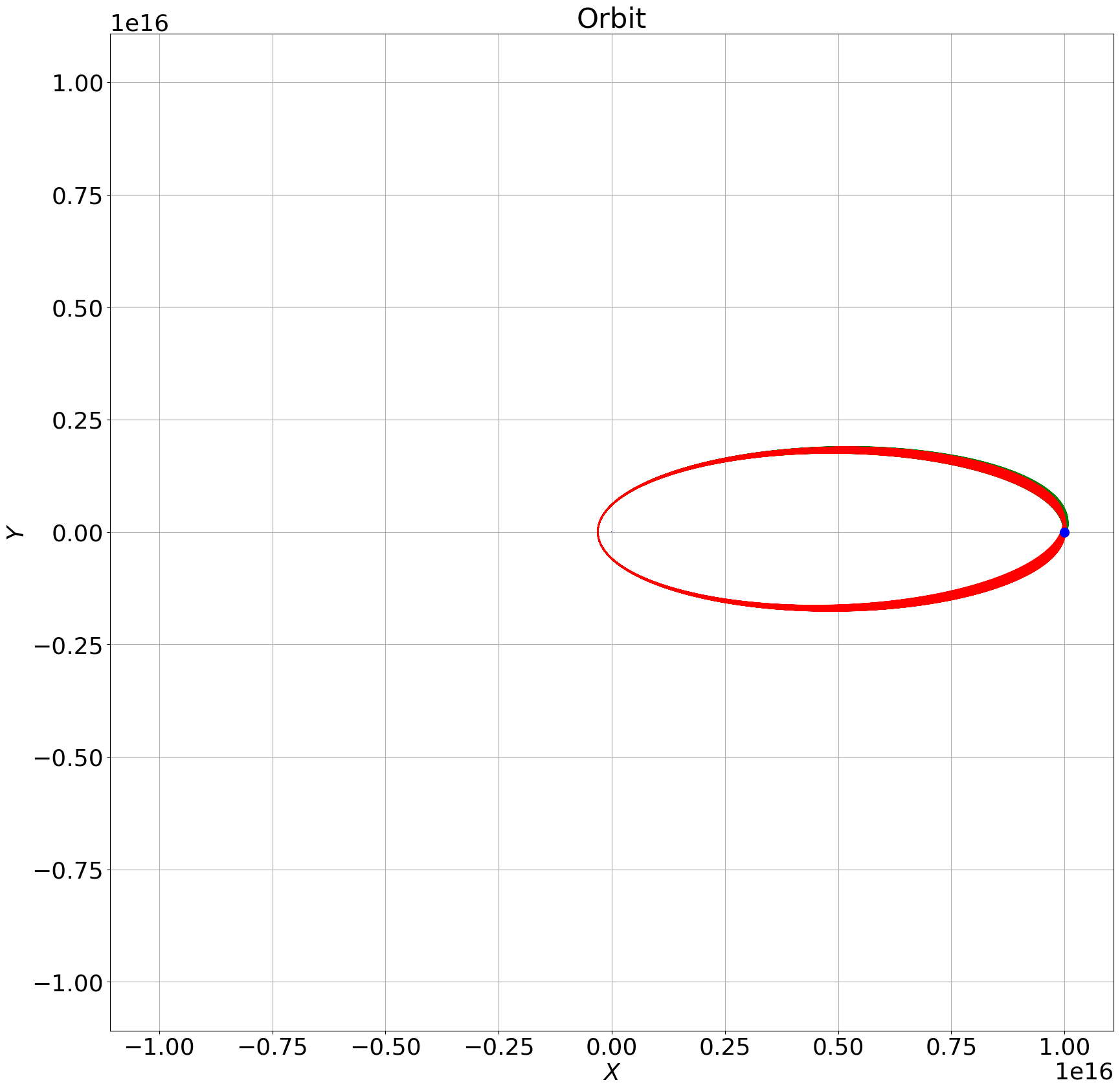}}
    \subfigure[0 PN-no-Prop]{\includegraphics[width=0.24\textwidth]{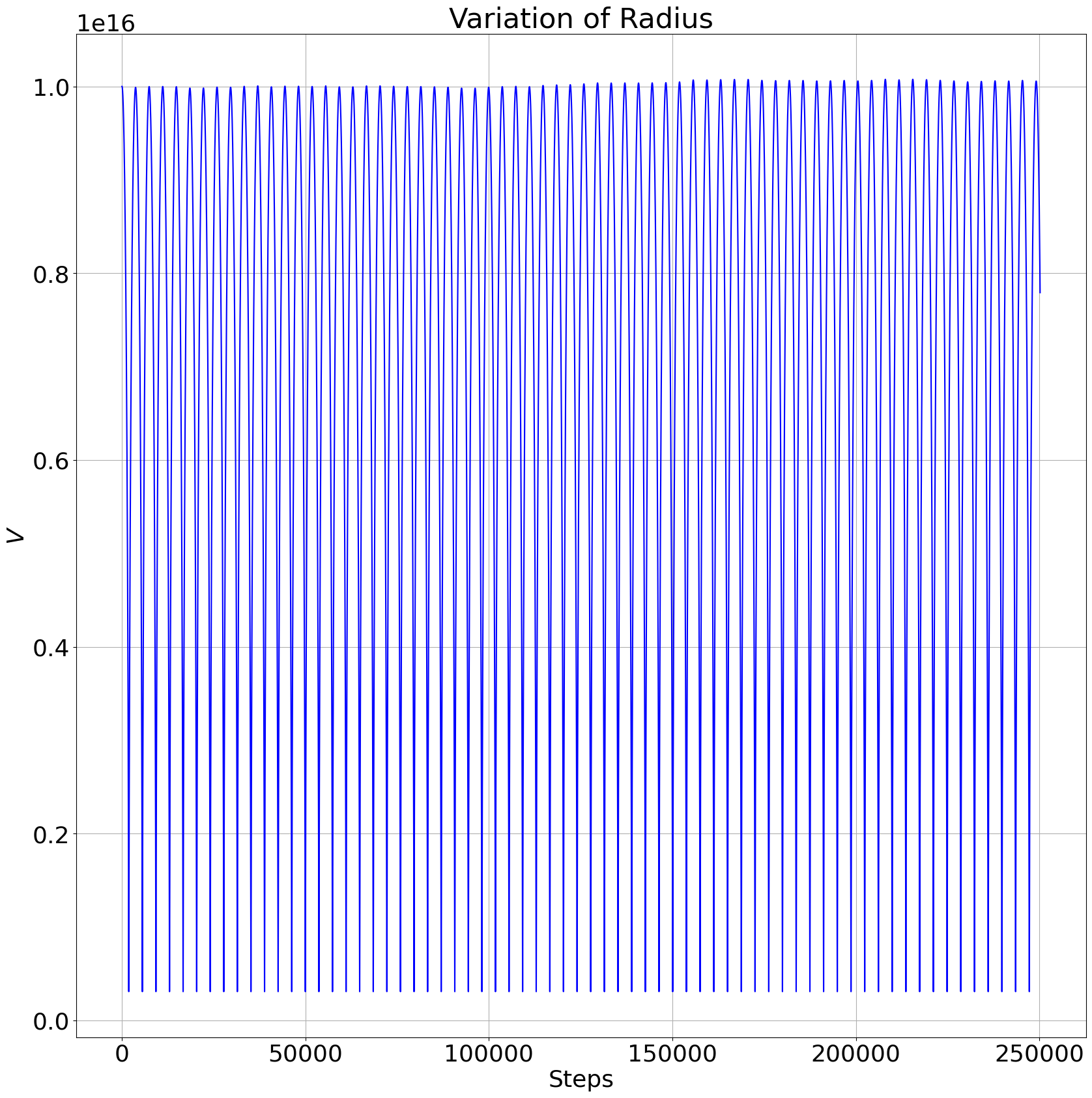}} 
    \subfigure[0 PN-with-Prop]{\includegraphics[width=0.24\textwidth]{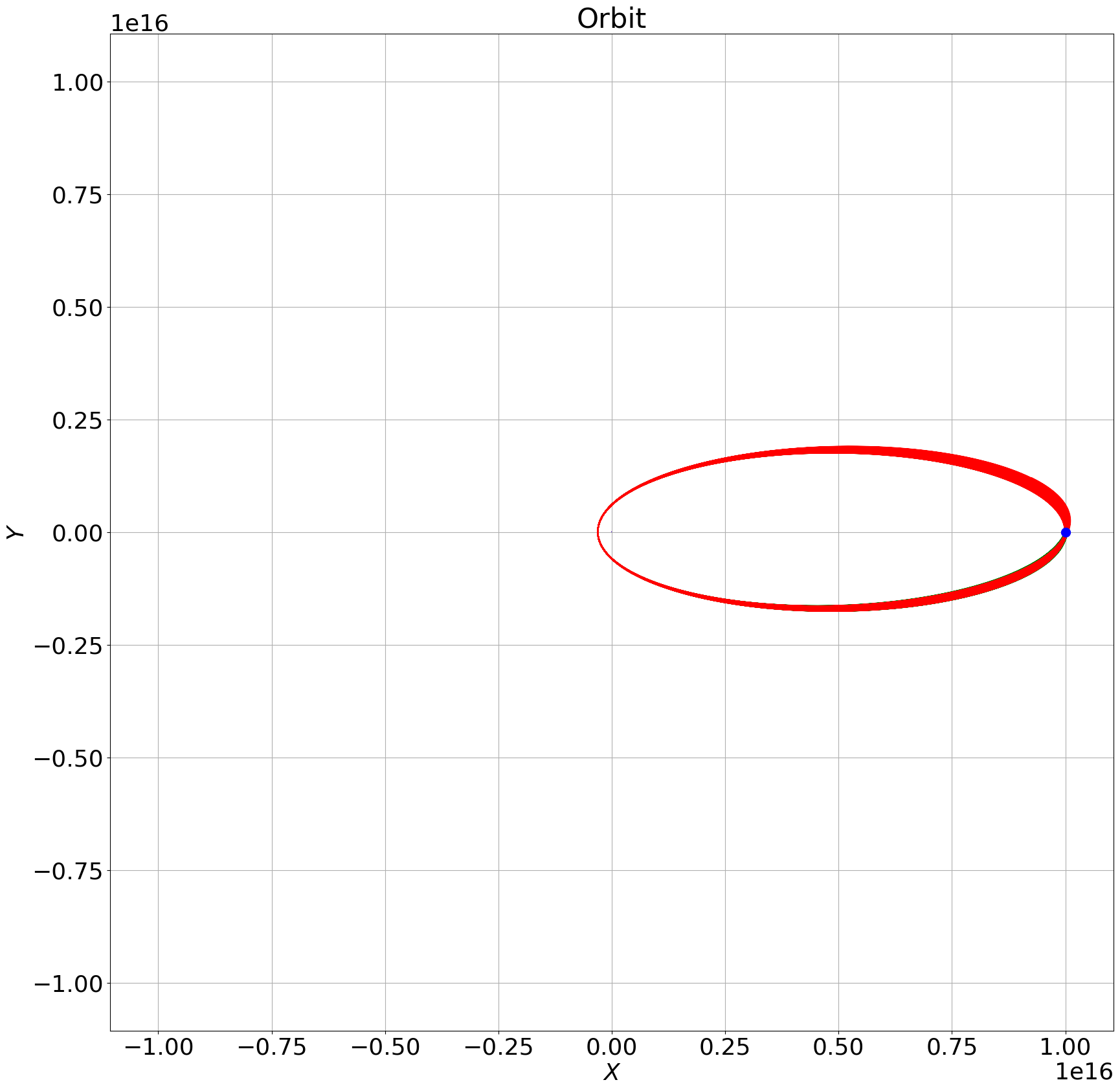}}
    \subfigure[0 PN-with-Prop]{\includegraphics[width=0.24\textwidth]{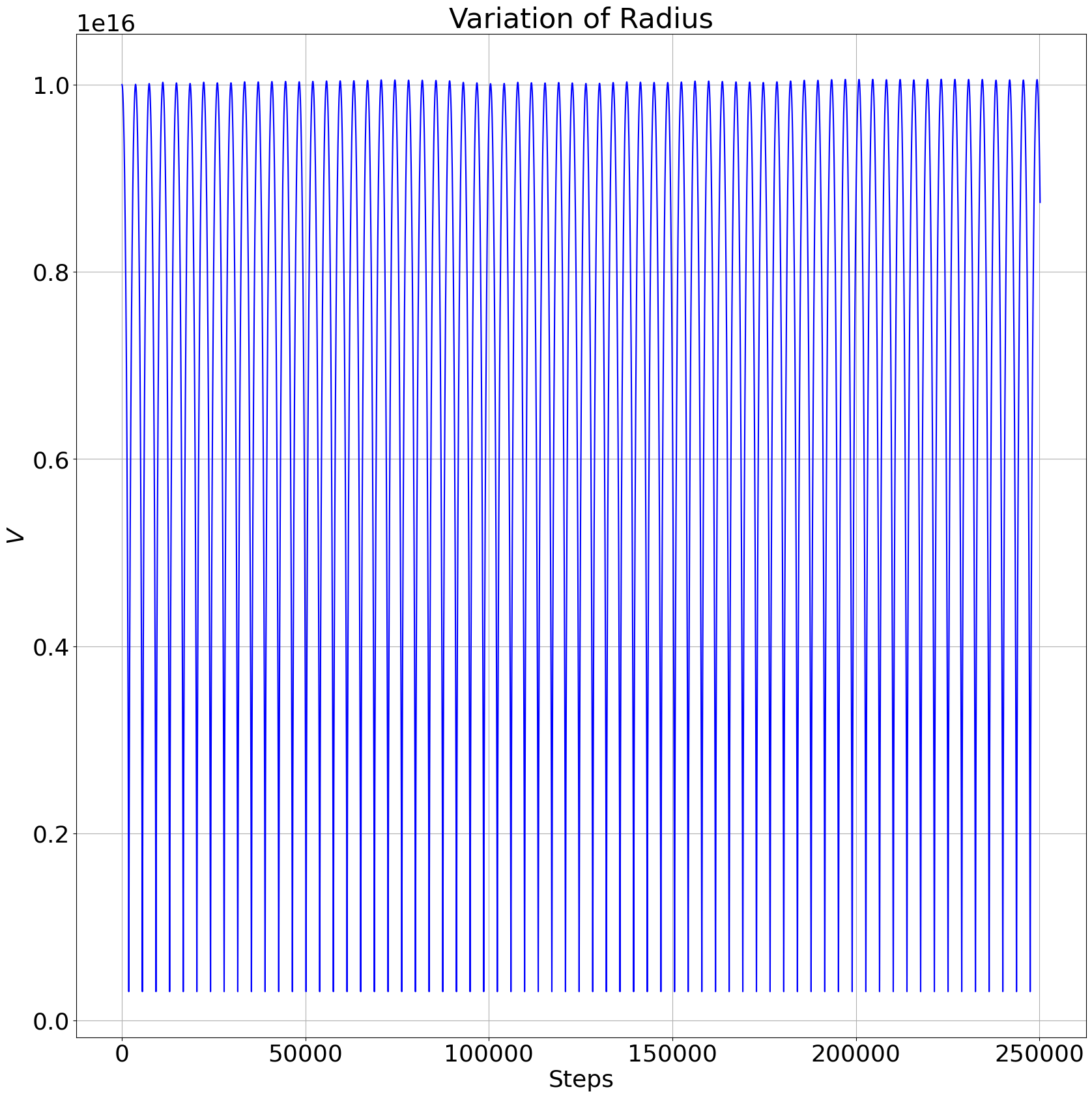}} \\
    \subfigure[0.5 PN-no-Prop]{\includegraphics[width=0.24\textwidth]{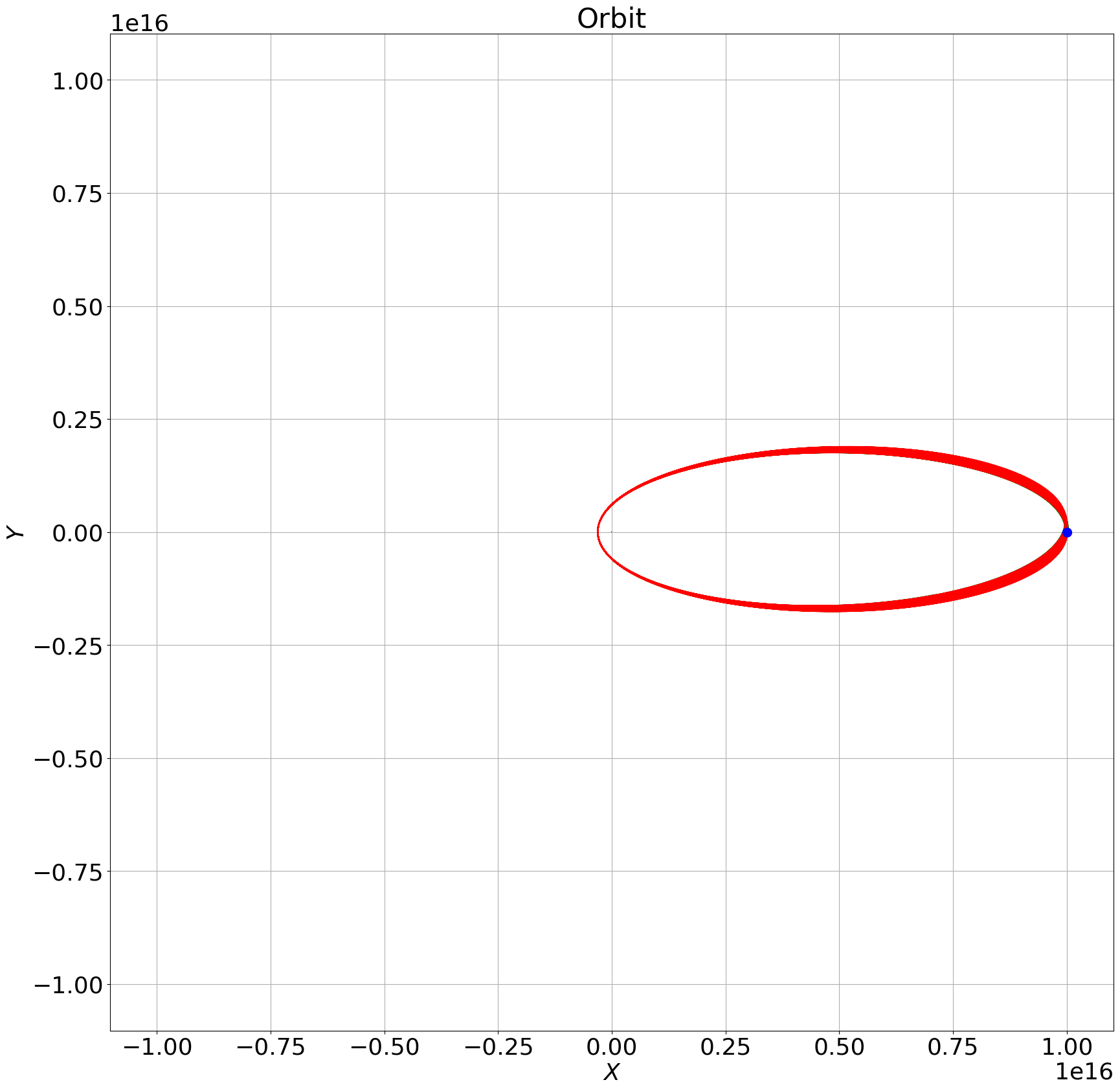}}
    \subfigure[0.5 PN-no-Prop]{\includegraphics[width=0.24\textwidth]{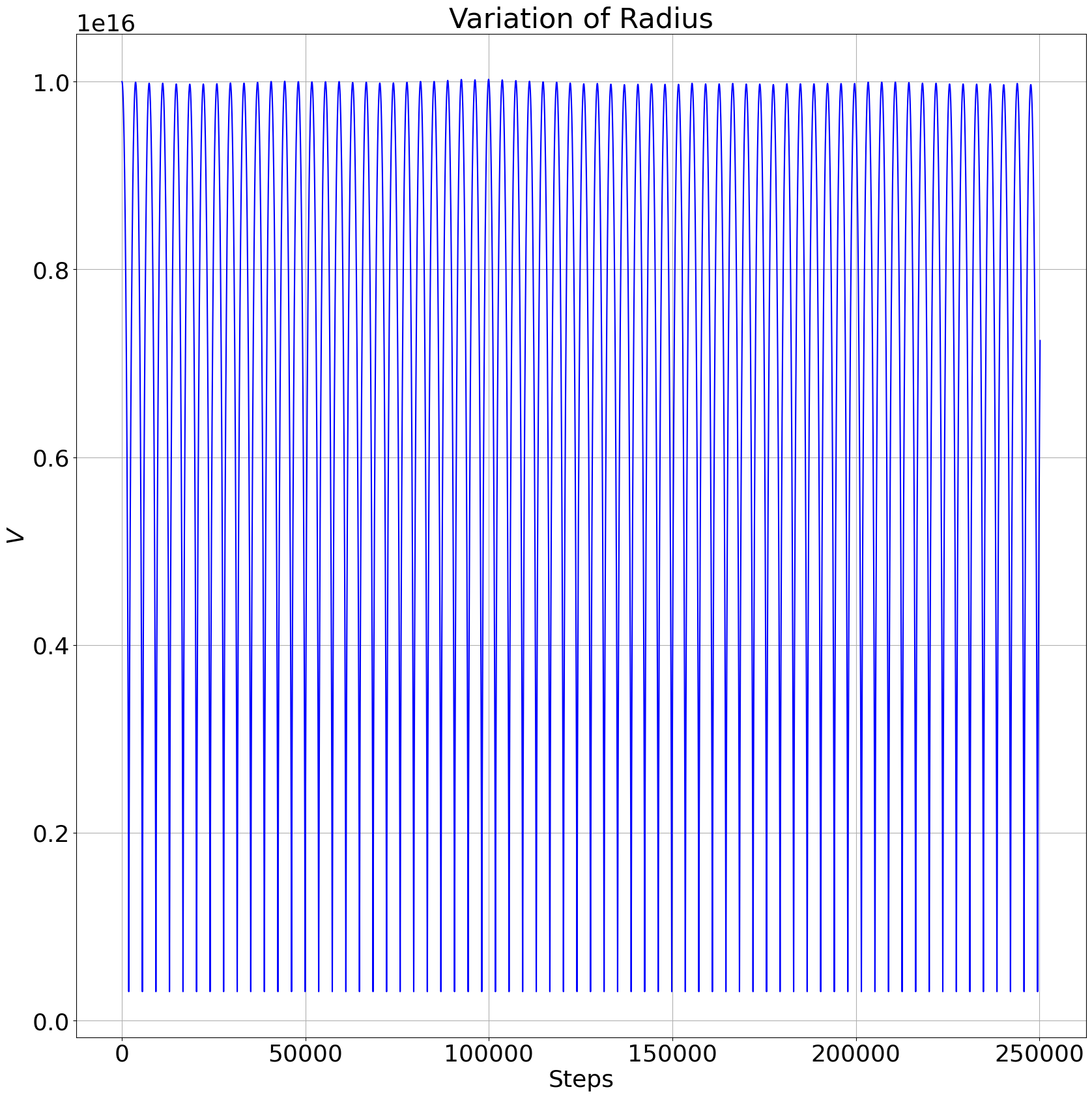}} 
    \subfigure[0.5 PN-with-Prop]{\includegraphics[width=0.24\textwidth]{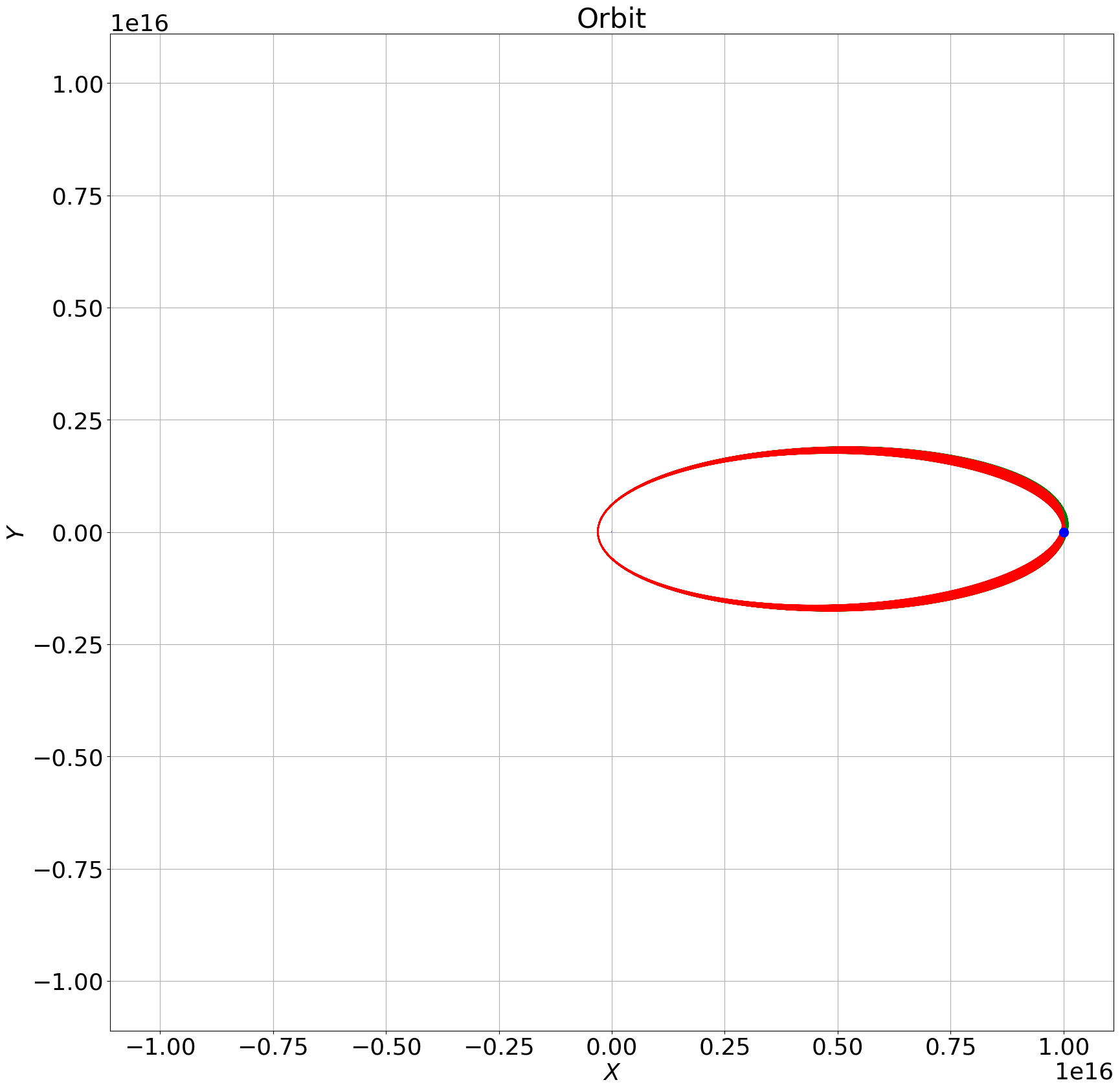}}
    \subfigure[0.5 PN-with-Prop]{\includegraphics[width=0.24\textwidth]{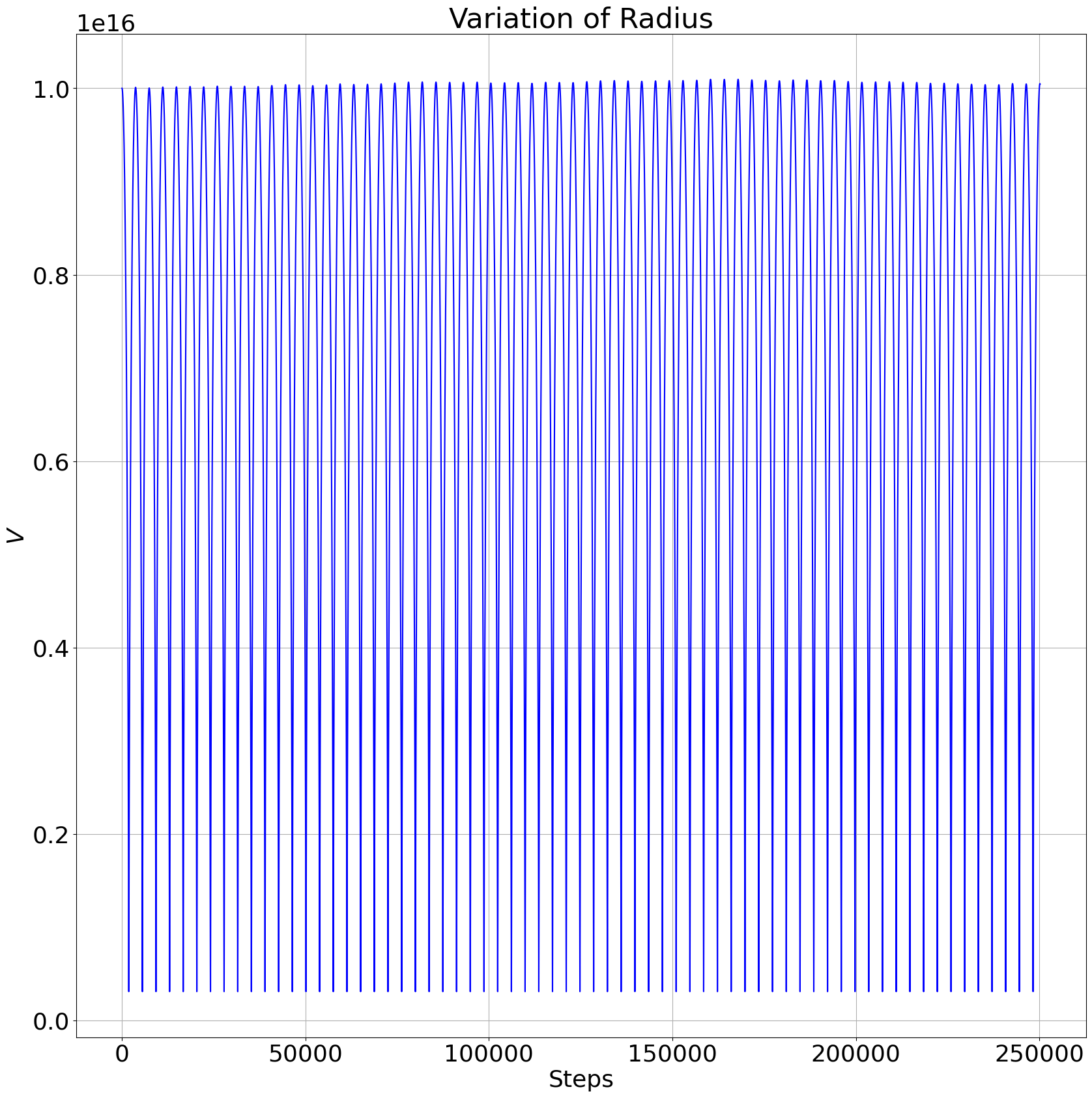}} \\ 
    \subfigure[1 PN-no-Prop]{\includegraphics[width=0.24\textwidth]{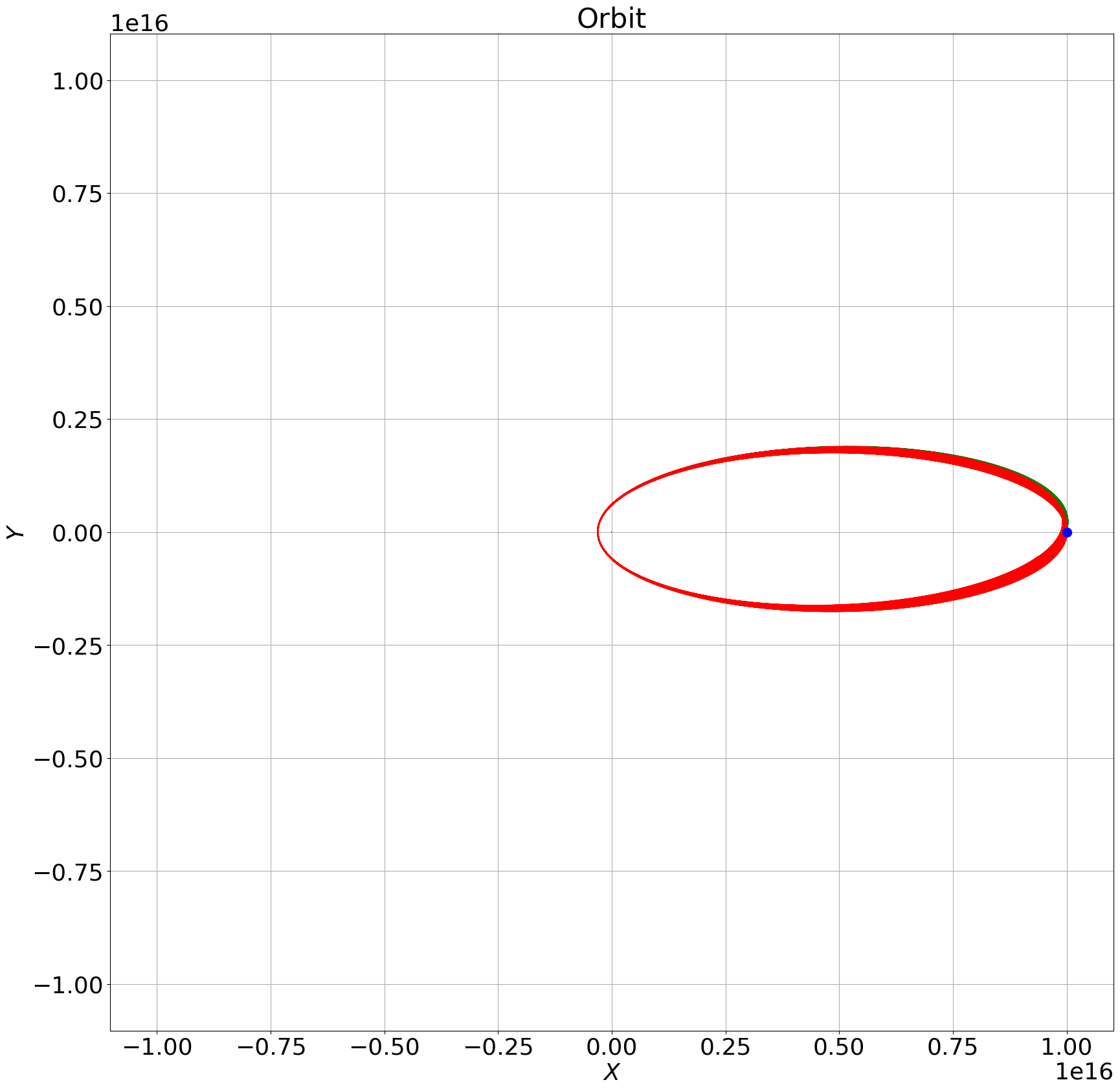}}
    \subfigure[1 PN-no-Prop]{\includegraphics[width=0.24\textwidth]{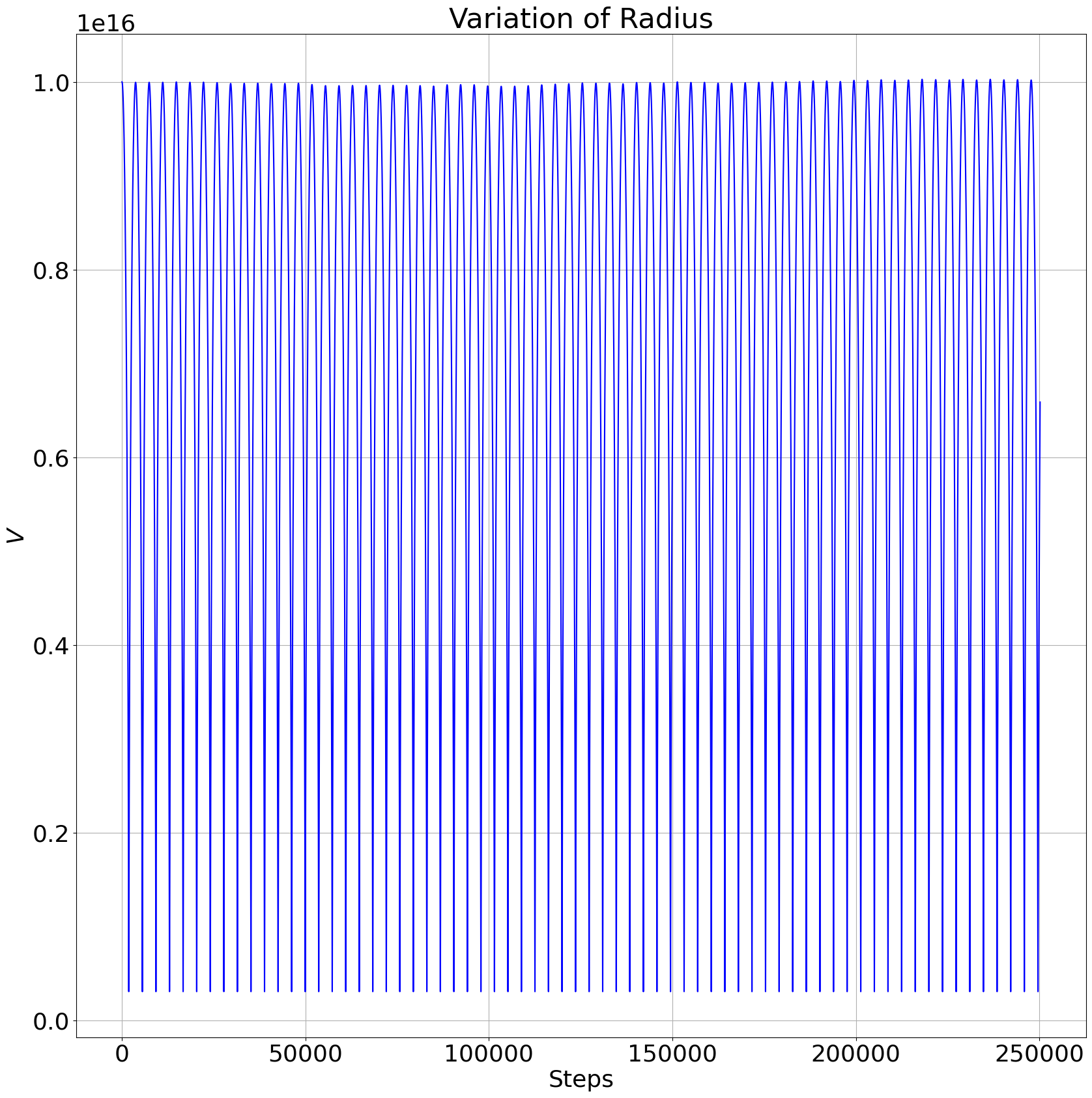}} 
    \subfigure[1 PN-with-Prop]{\includegraphics[width=0.24\textwidth]{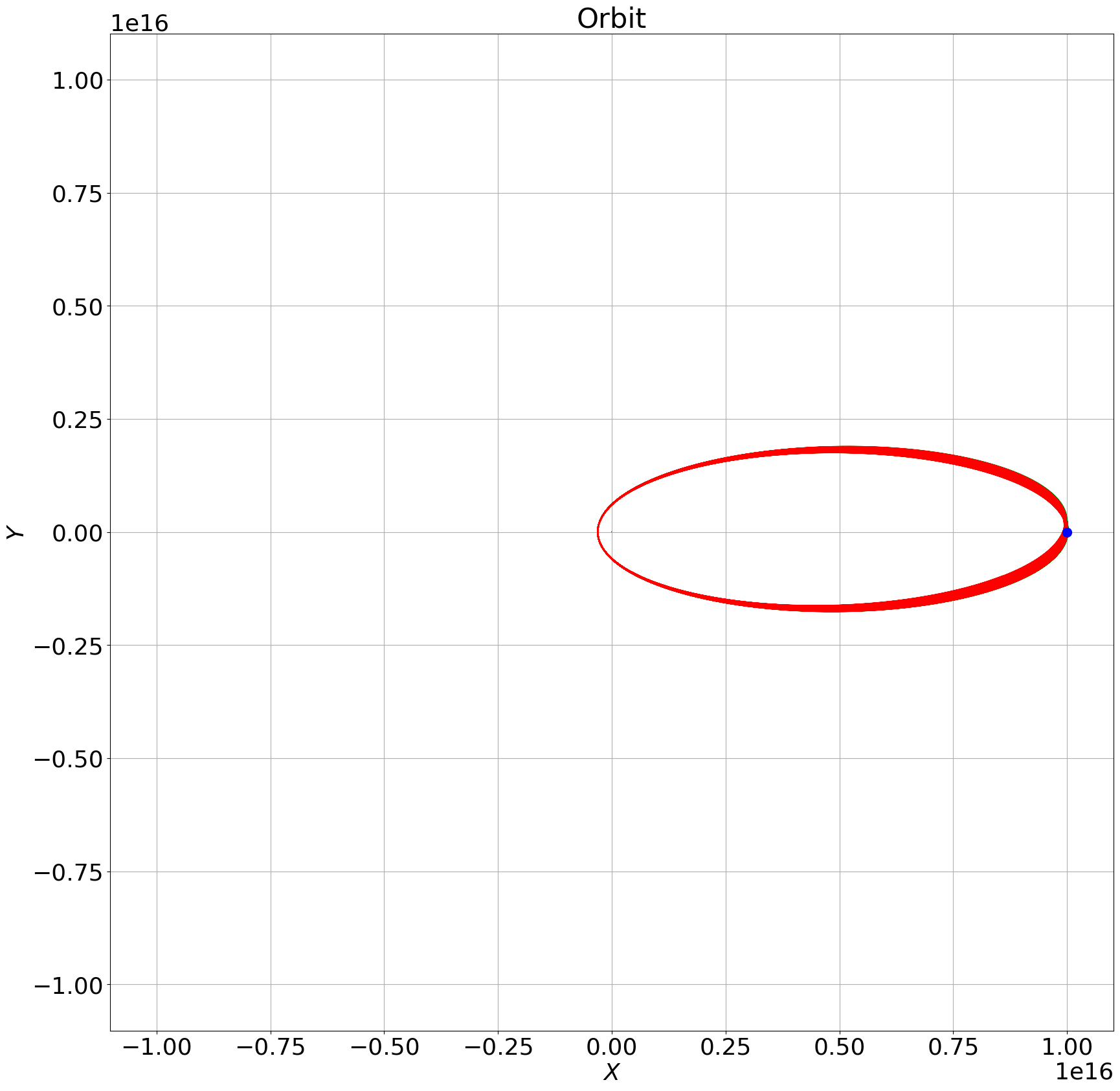}}
    \subfigure[1 PN-with-Prop]{\includegraphics[width=0.24\textwidth]{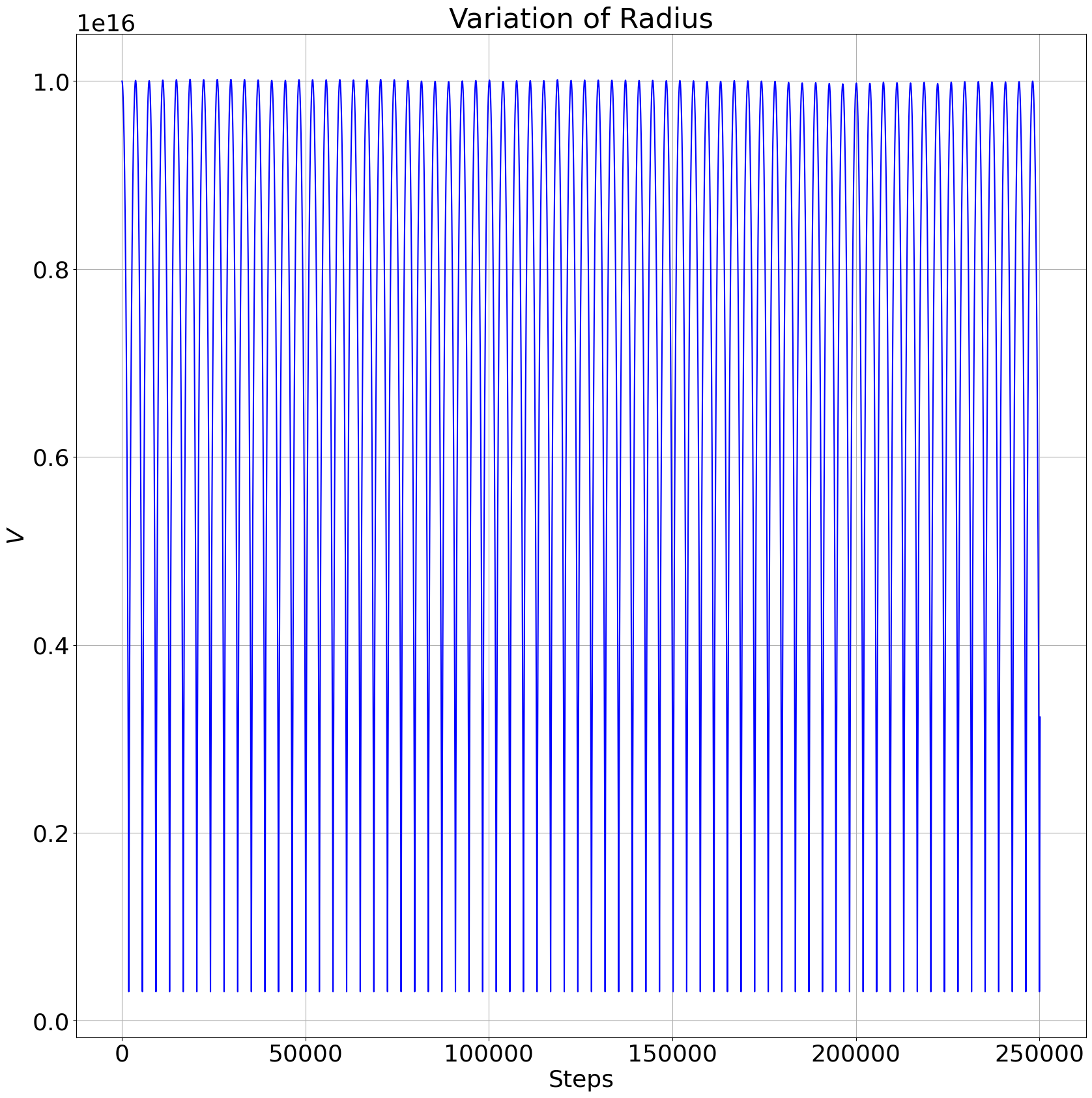}} \\
    \subfigure[2 PN-no-Prop]{\includegraphics[width=0.24\textwidth]{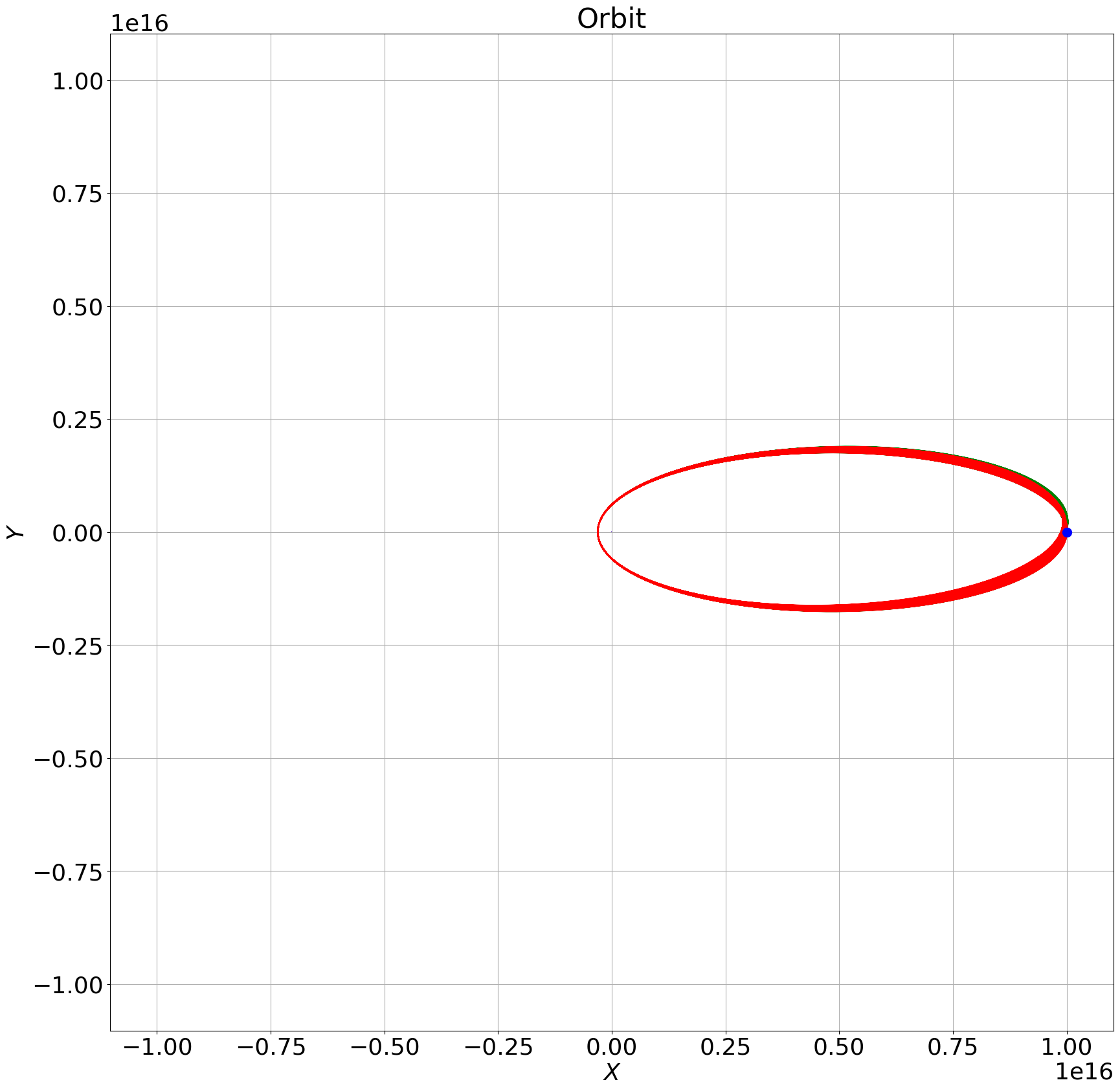}}
    \subfigure[2 PN-no-Prop]{\includegraphics[width=0.24\textwidth]{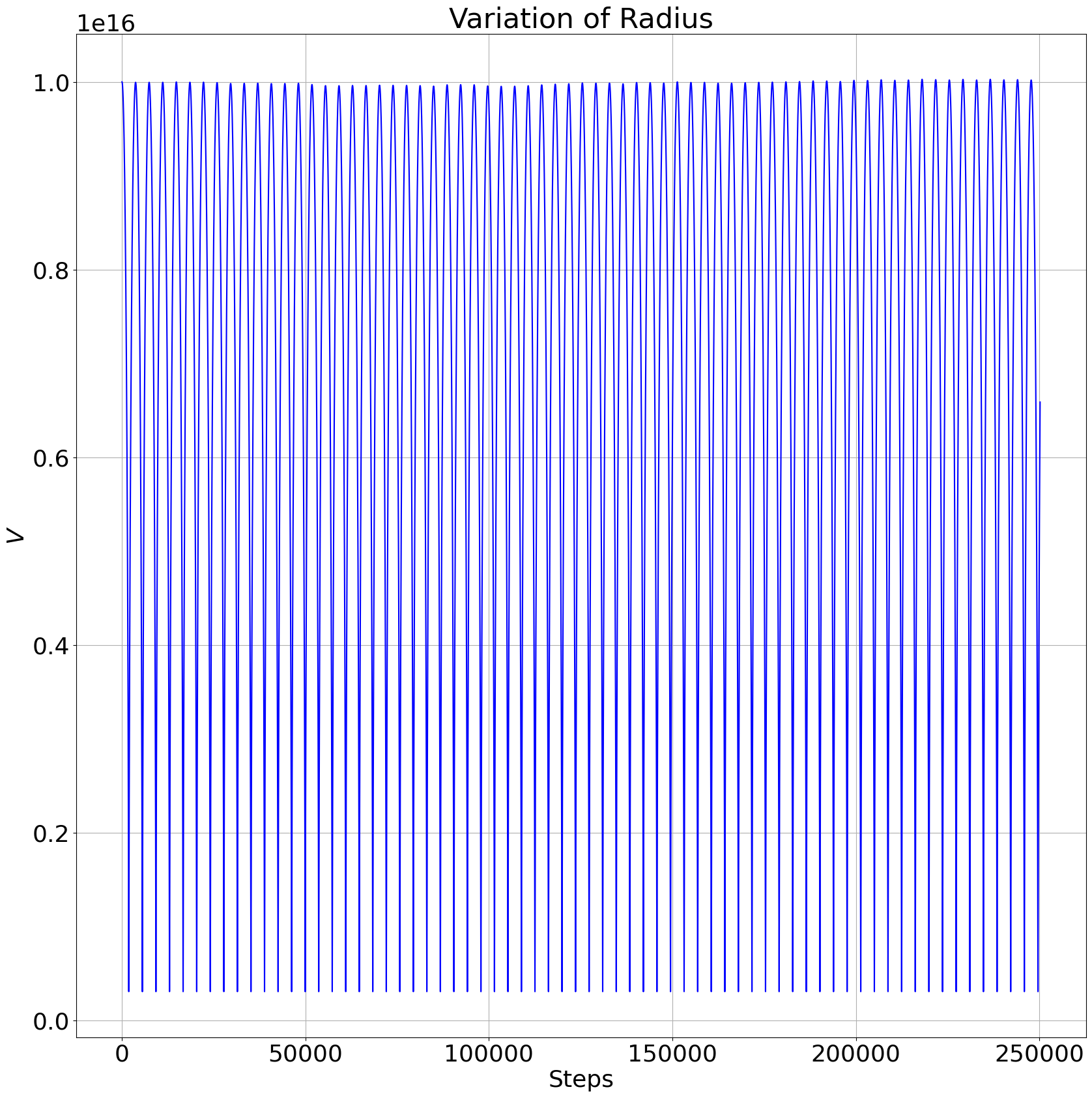}} 
    \subfigure[2 PN-with-Prop]{\includegraphics[width=0.24\textwidth]{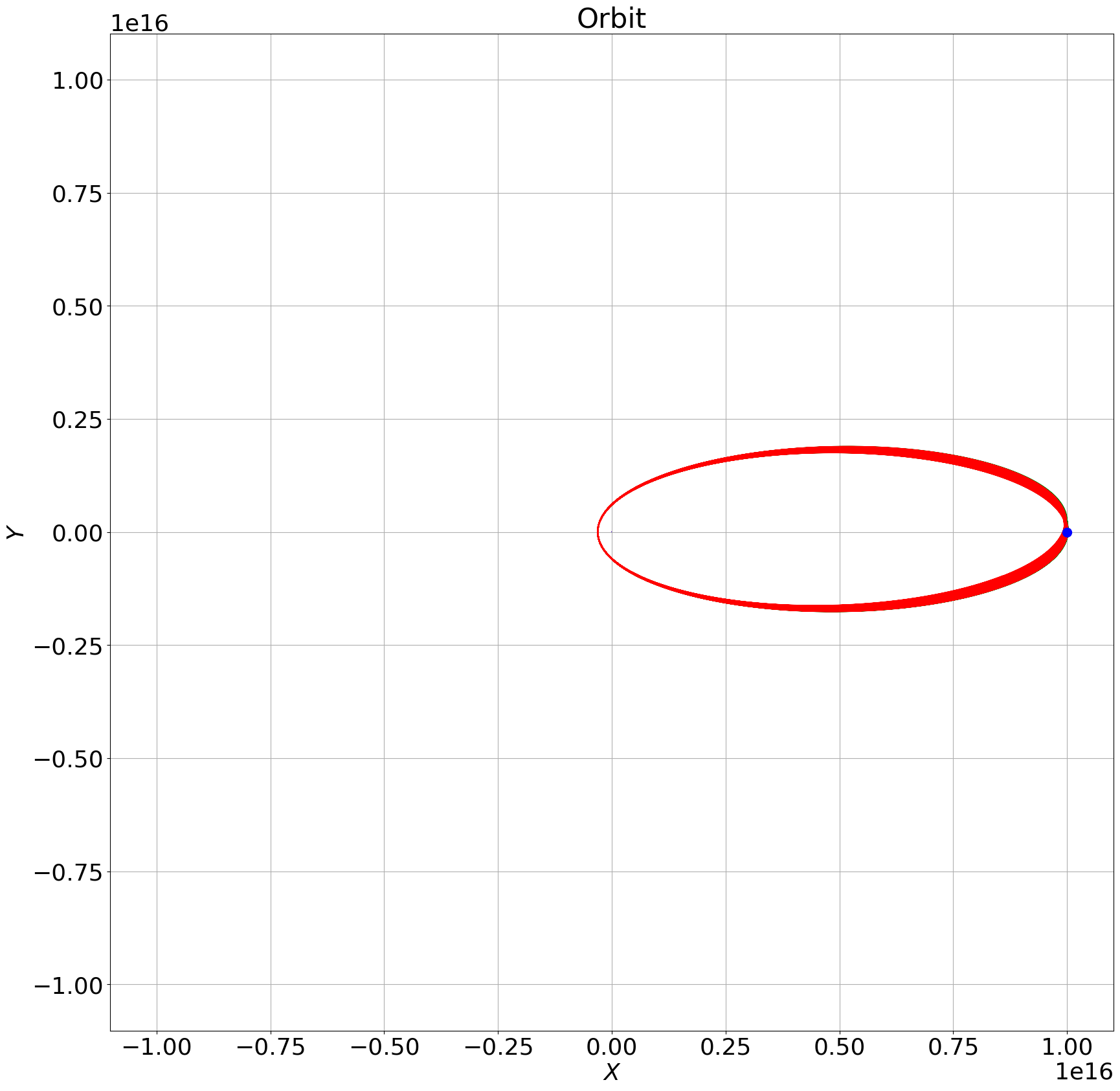}}
    \subfigure[2 PN-with-Prop]{\includegraphics[width=0.24\textwidth]{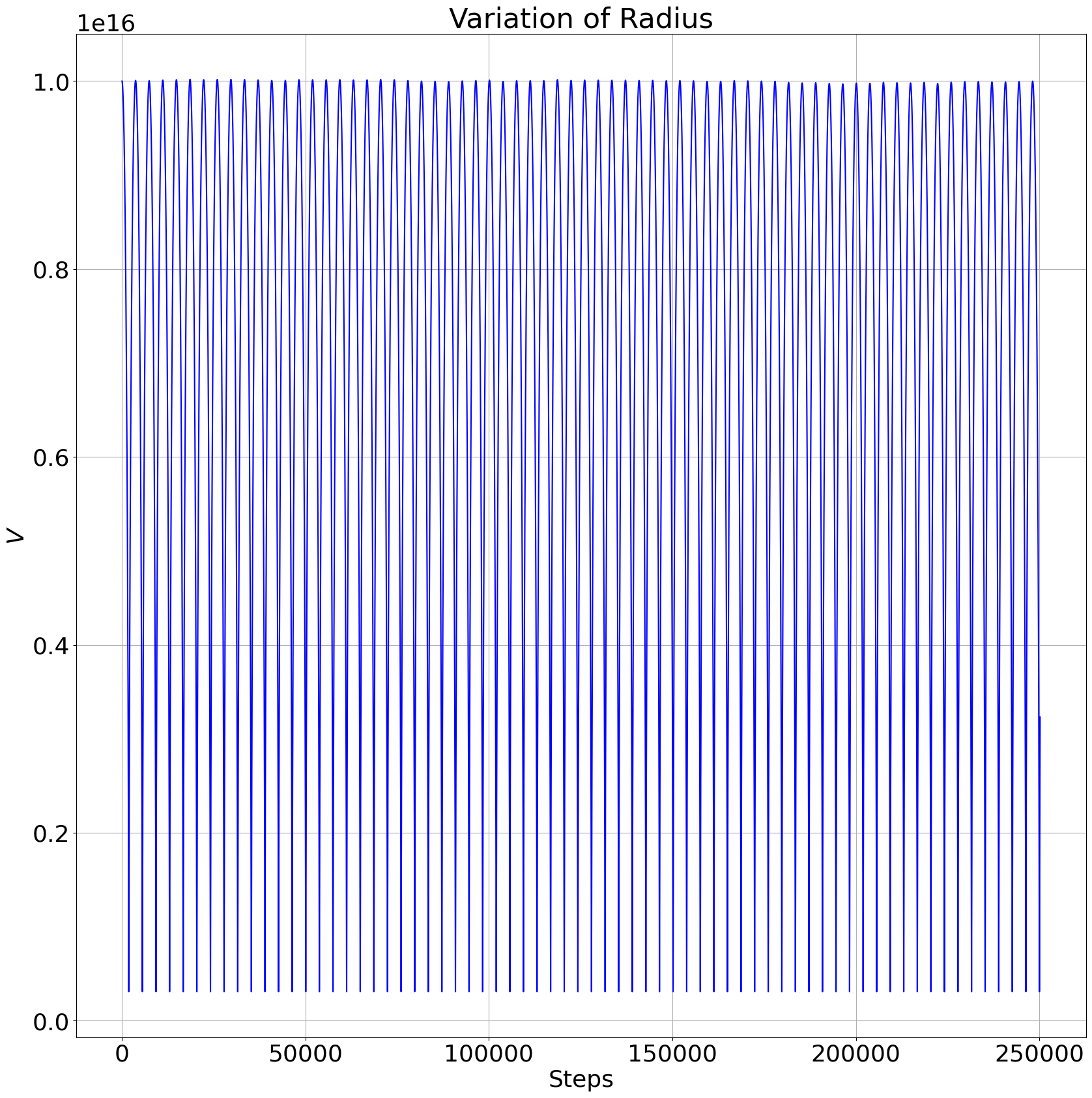}} \\ 
    \subfigure[2.5 PN-no-Prop]{\includegraphics[width=0.24\textwidth]{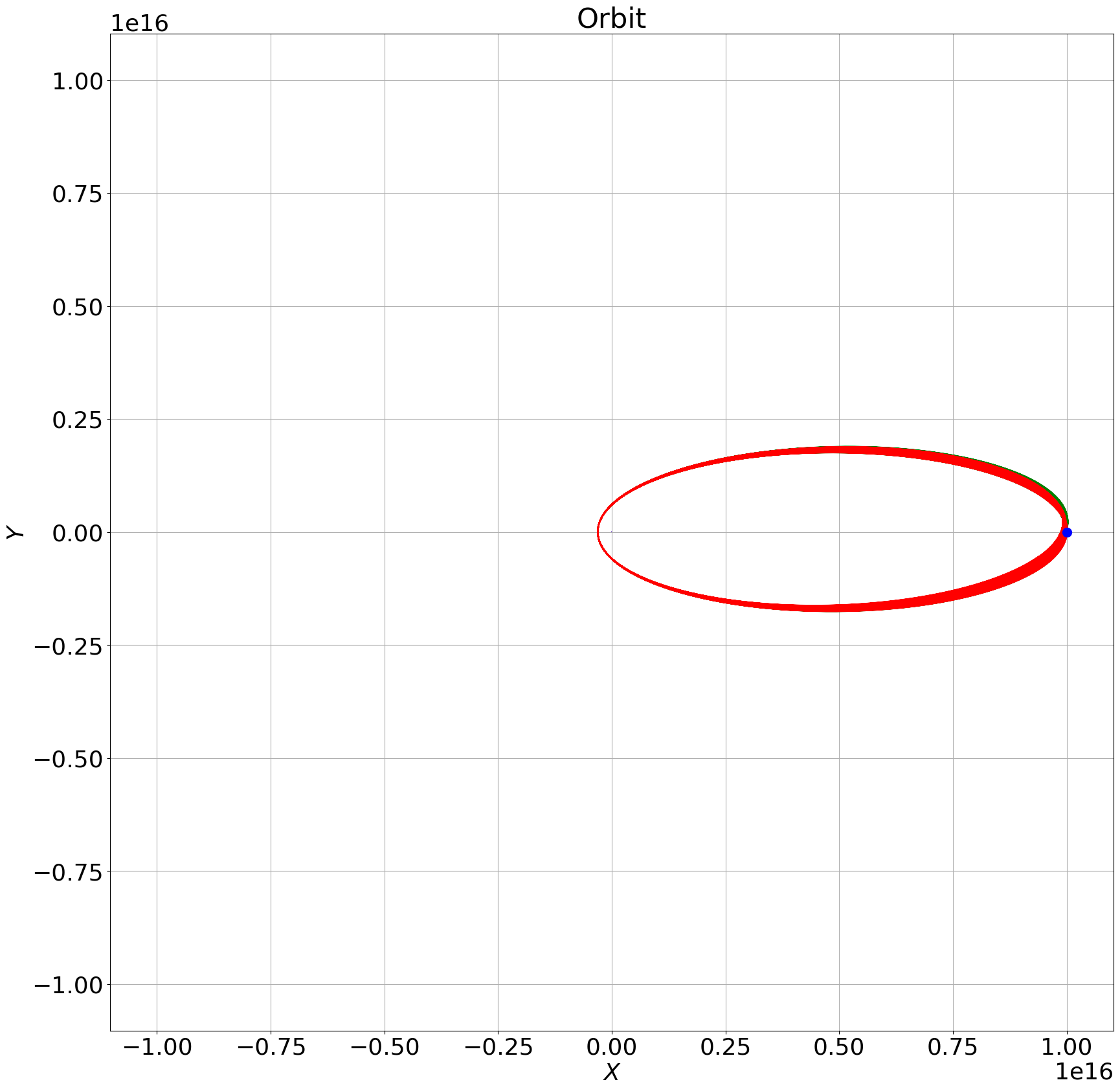}}
    \subfigure[2.5 PN-no-Prop]{\includegraphics[width=0.24\textwidth]{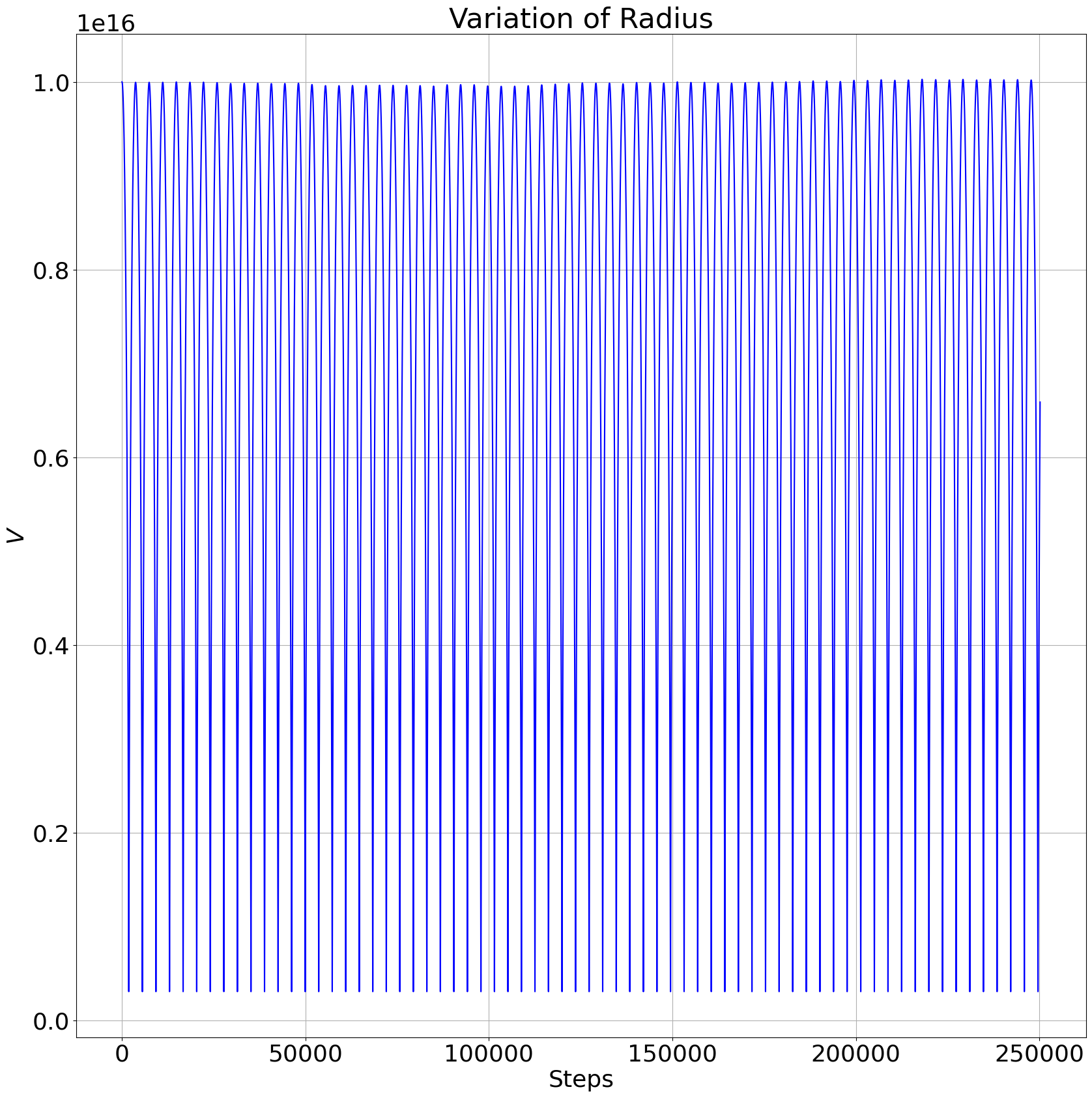}} 
    \subfigure[2.5 PN-with-Prop]{\includegraphics[width=0.24\textwidth]{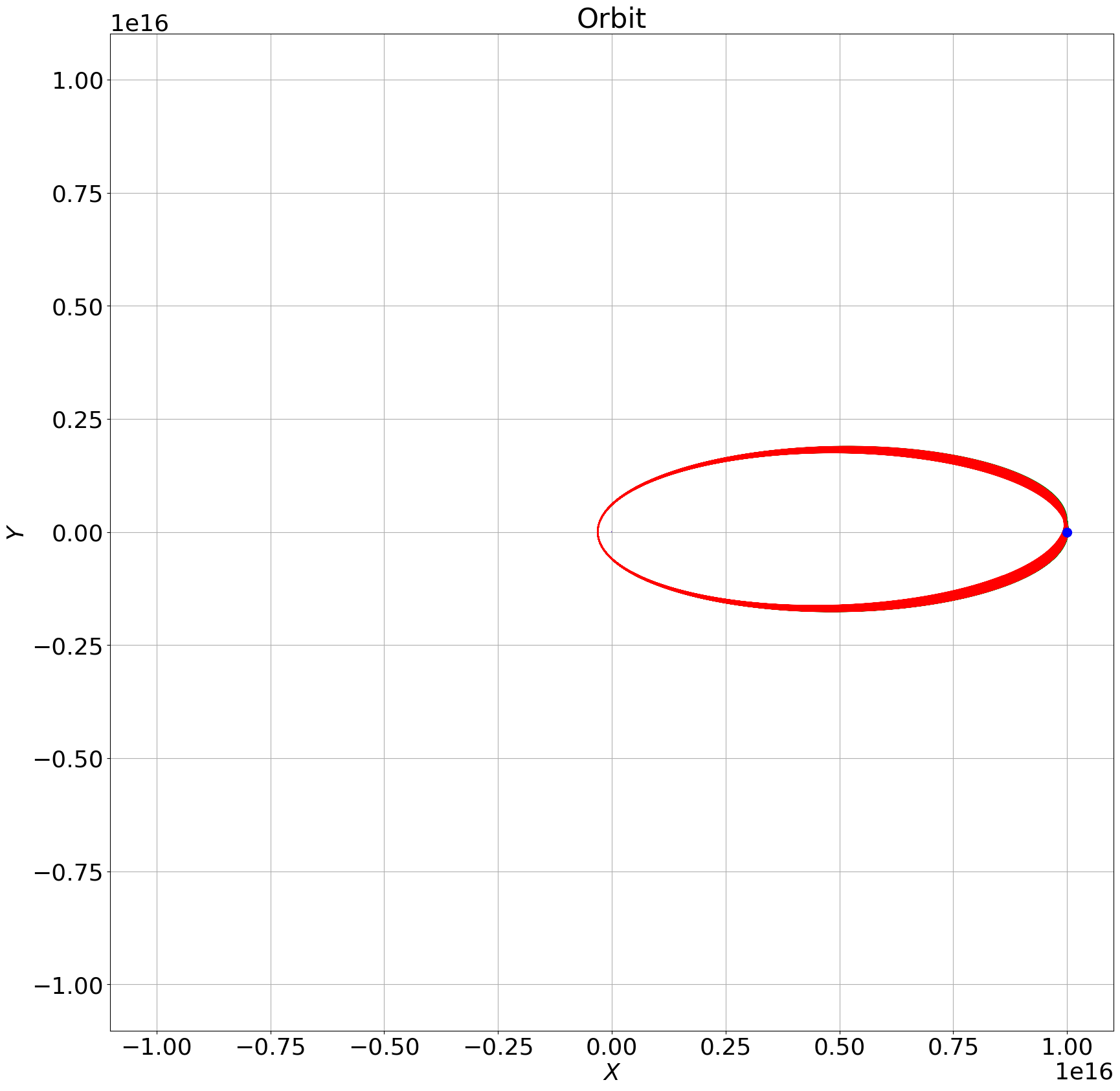}}
    \subfigure[2.5 PN-with-Prop]{\includegraphics[width=0.24\textwidth]{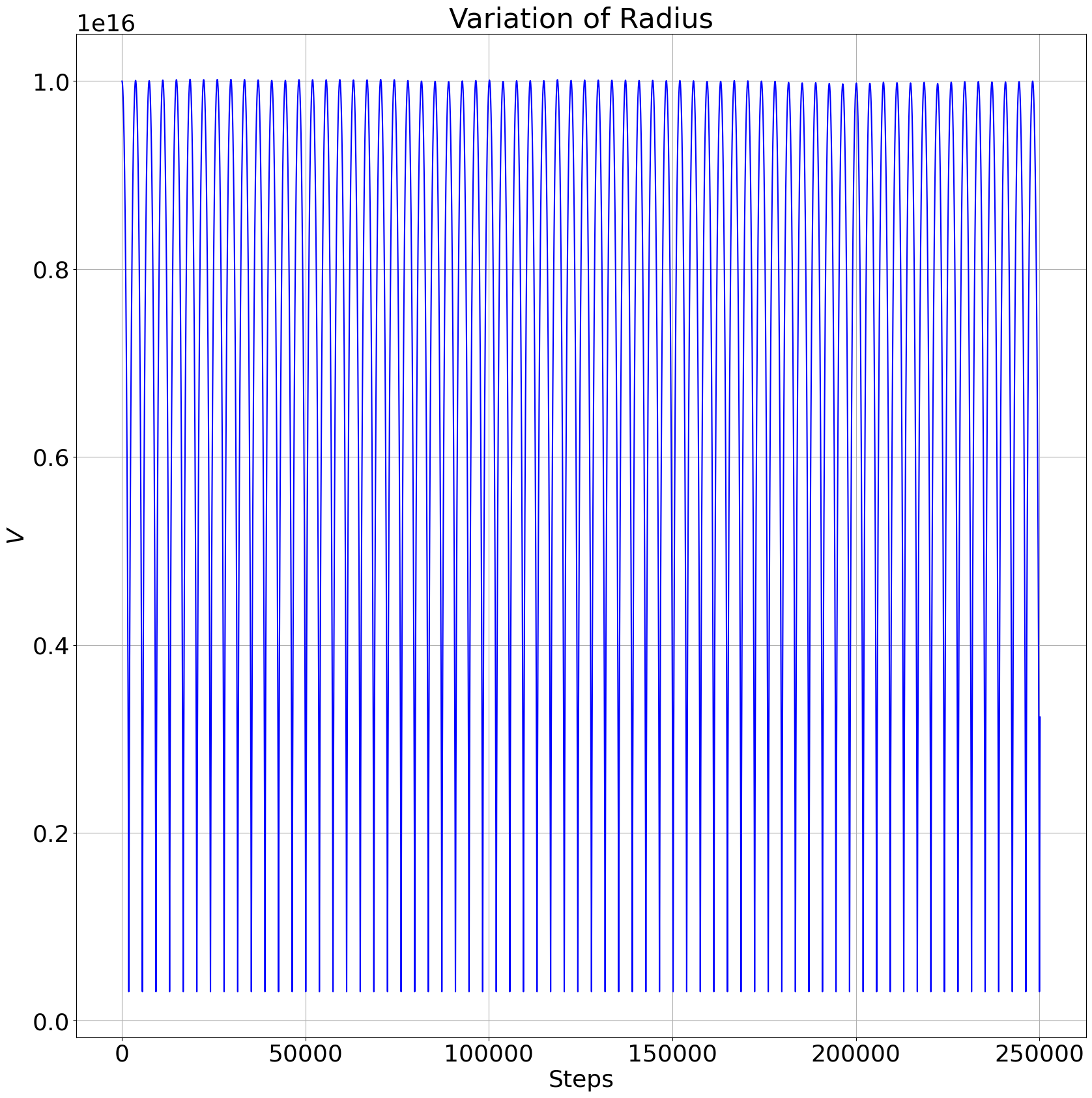}} \\ 
    \caption{Weak Field, Elliptical. R = $1.1 \cdot 10^{13}m$, X = $10^{16}m$, Velocity = 40$ms^{-1}$}
    \label{results2}
\end{figure}
\begin{figure}[!ht]
    \centering
    \setcounter{subfigure}{0}
    \subfigure[0 PN-no-Prop]{\includegraphics[width=0.24\textwidth]{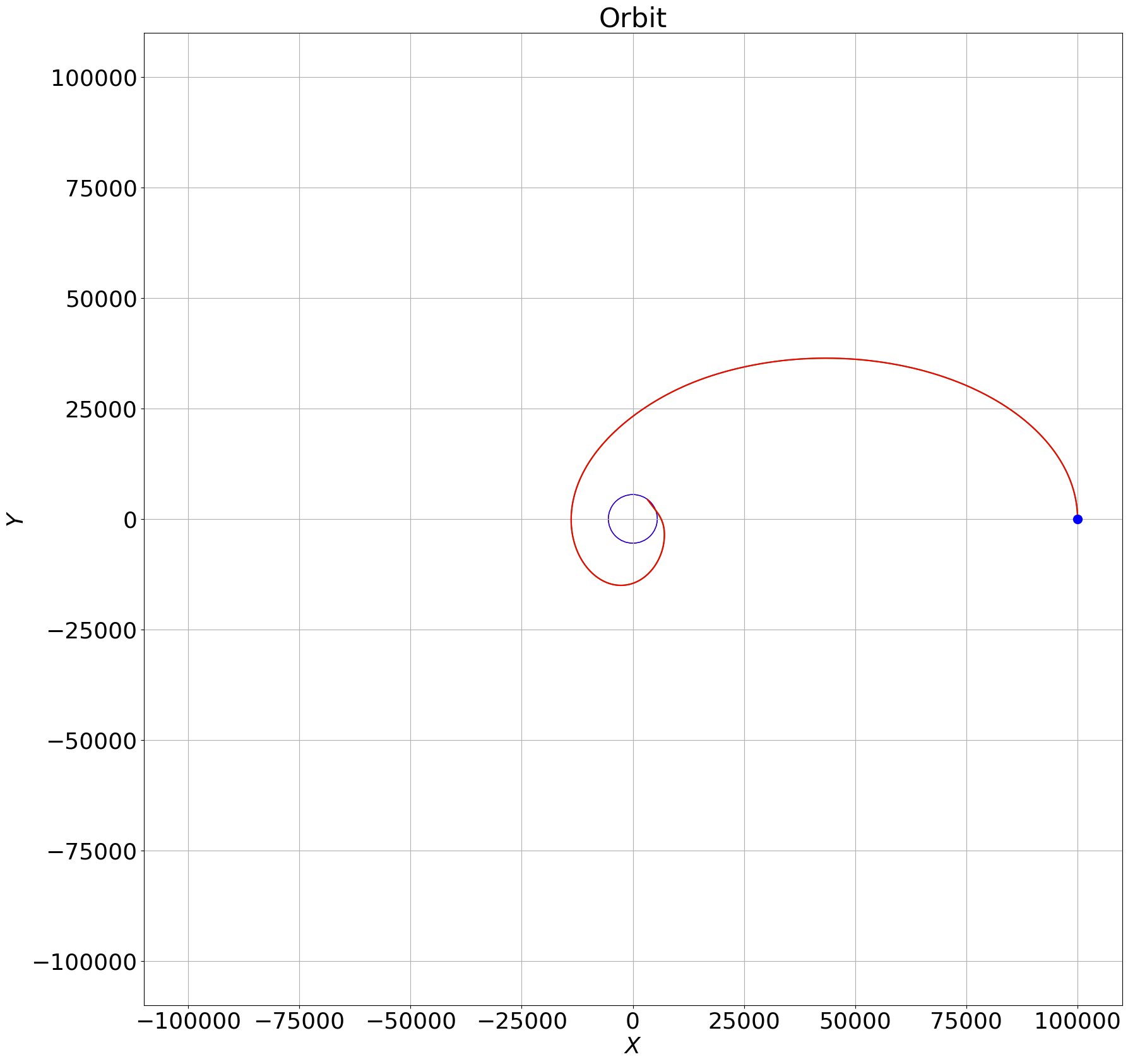}}
    \subfigure[0 PN-no-Prop]{\includegraphics[width=0.24\textwidth]{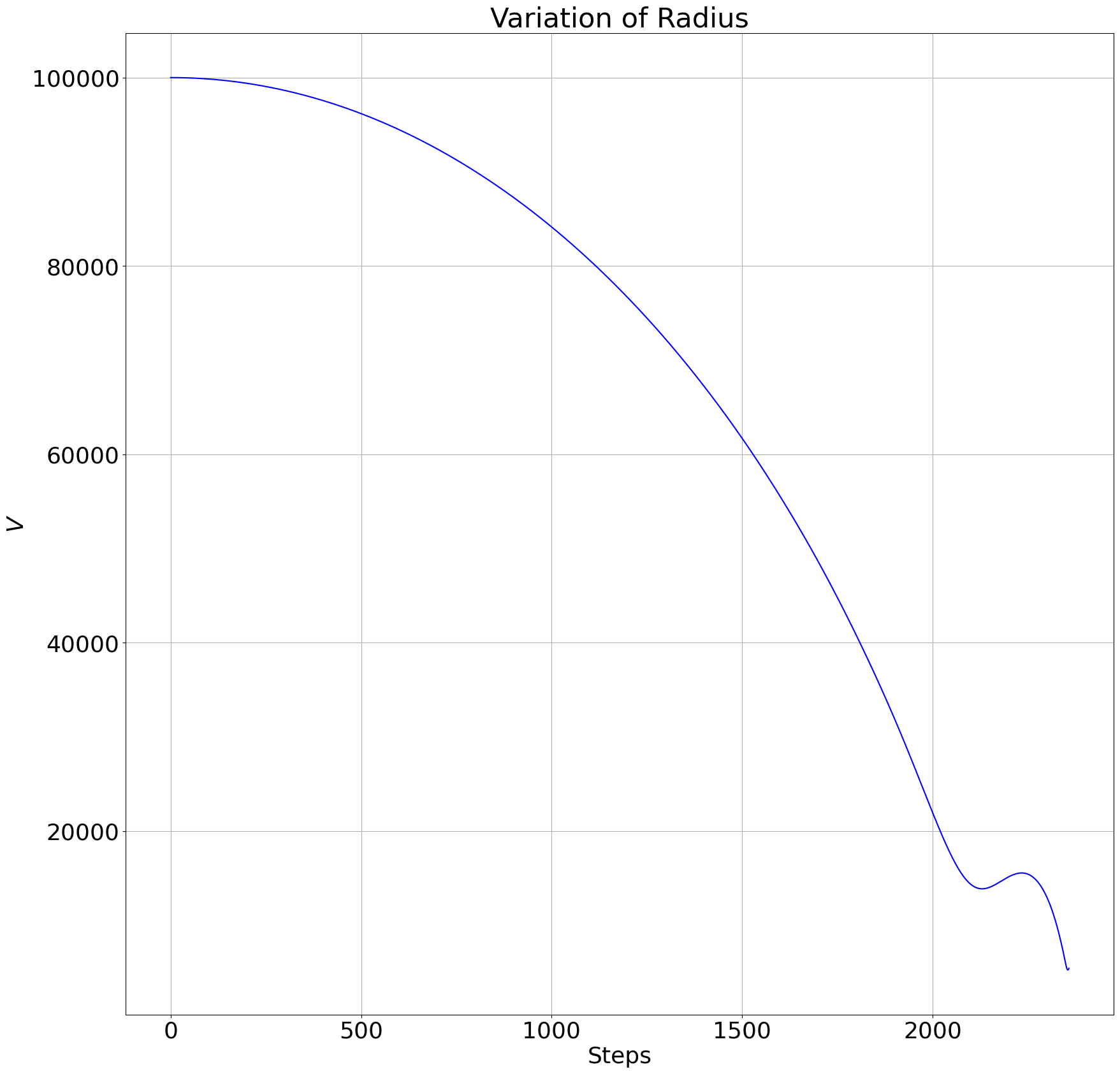}} 
    \subfigure[0 PN-with-Prop]{\includegraphics[width=0.24\textwidth]{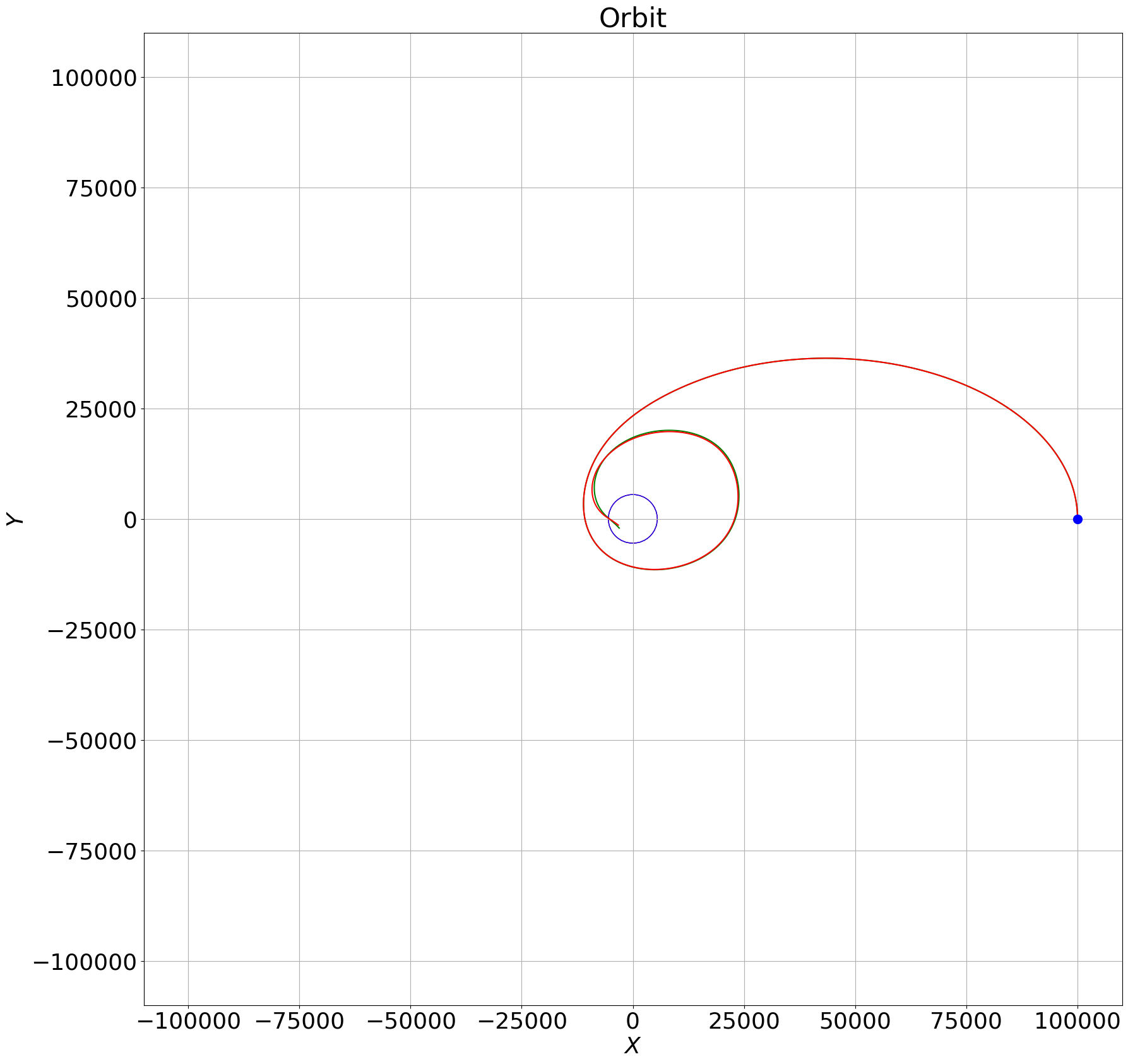}}
    \subfigure[0 PN-with-Prop]{\includegraphics[width=0.24\textwidth]{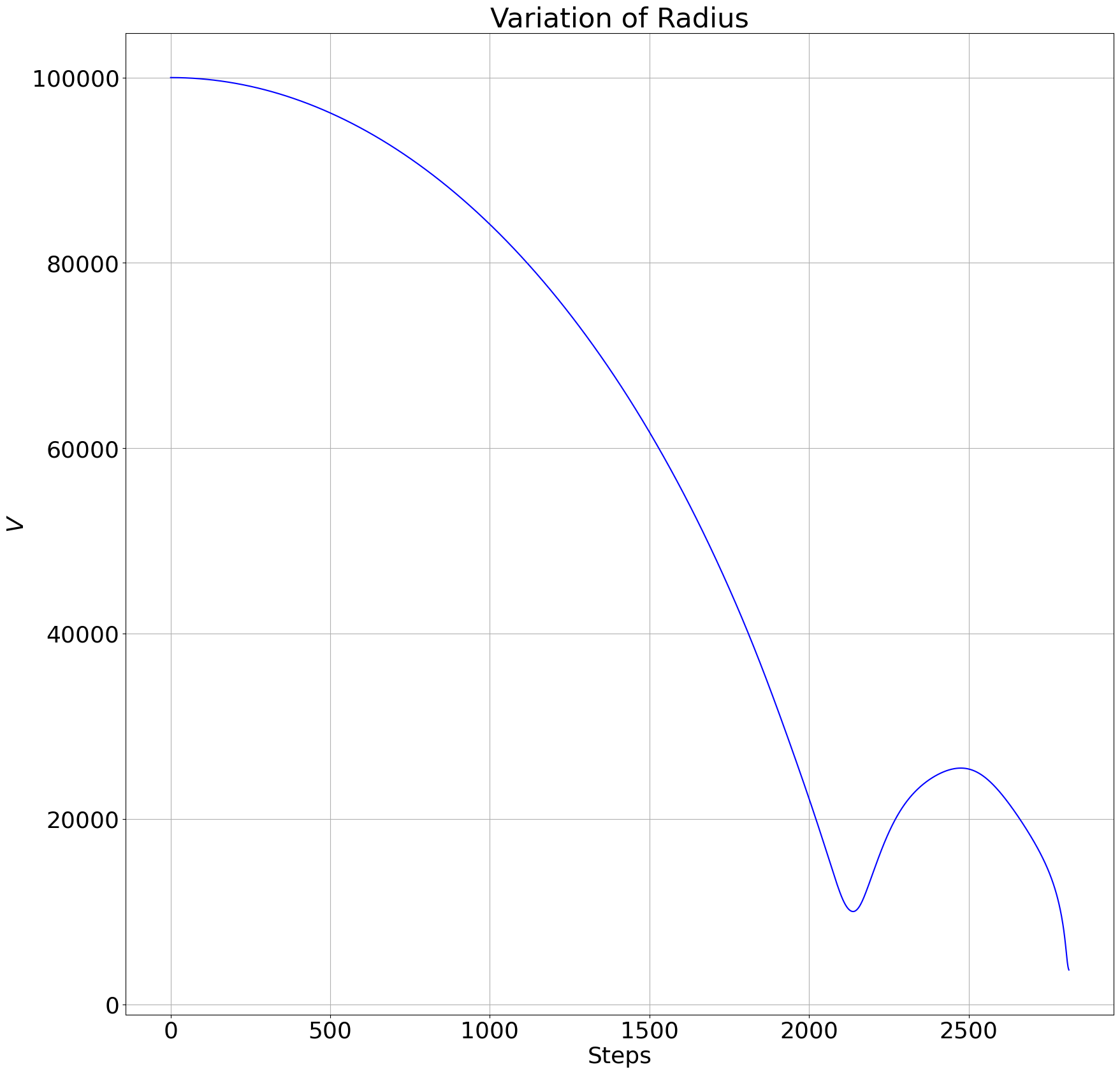}} \\
    
    \subfigure[0.5 PN-no-Prop]{\includegraphics[width=0.24\textwidth]{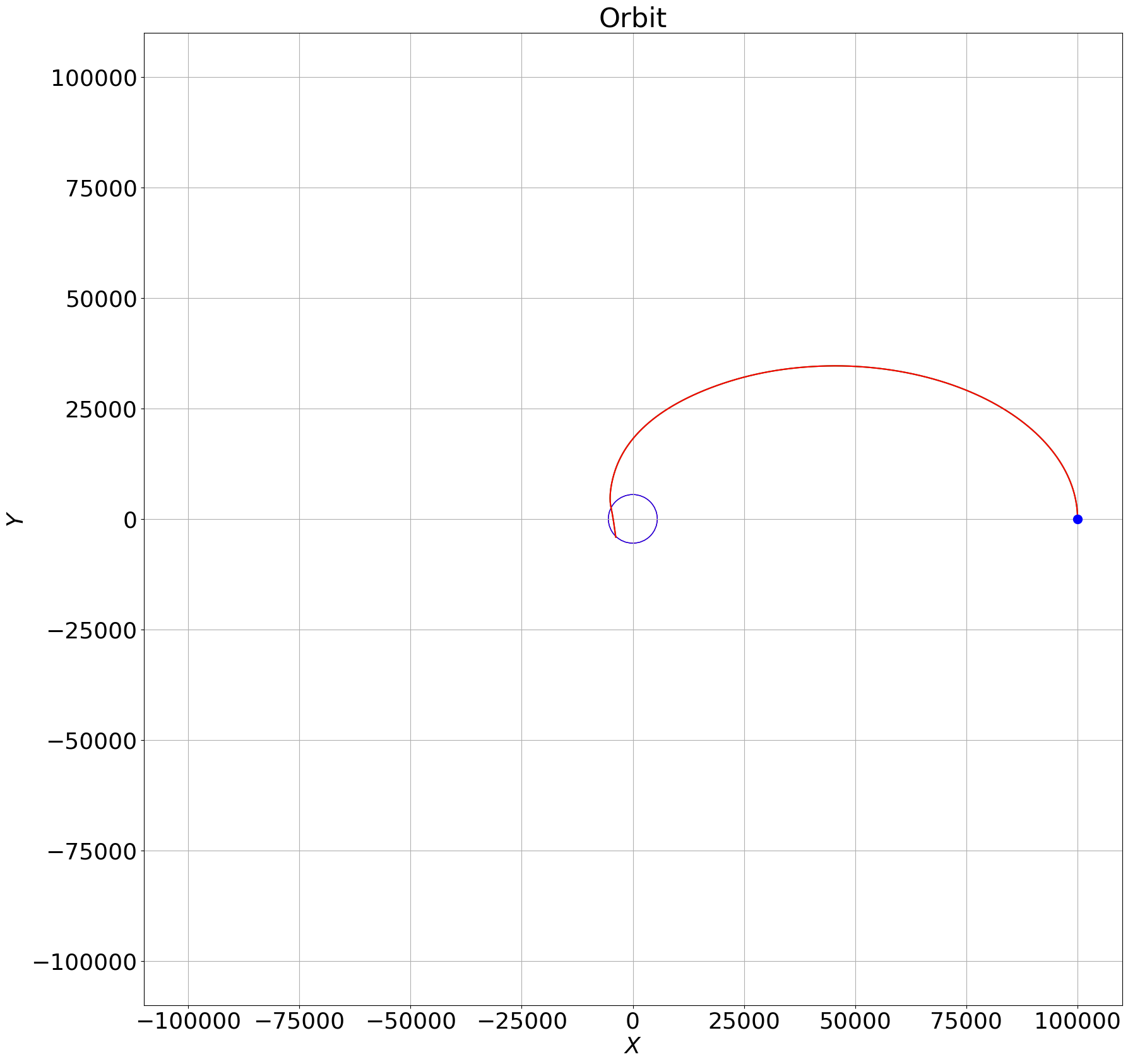}}
    \subfigure[0.5 PN-no-Prop]{\includegraphics[width=0.24\textwidth]{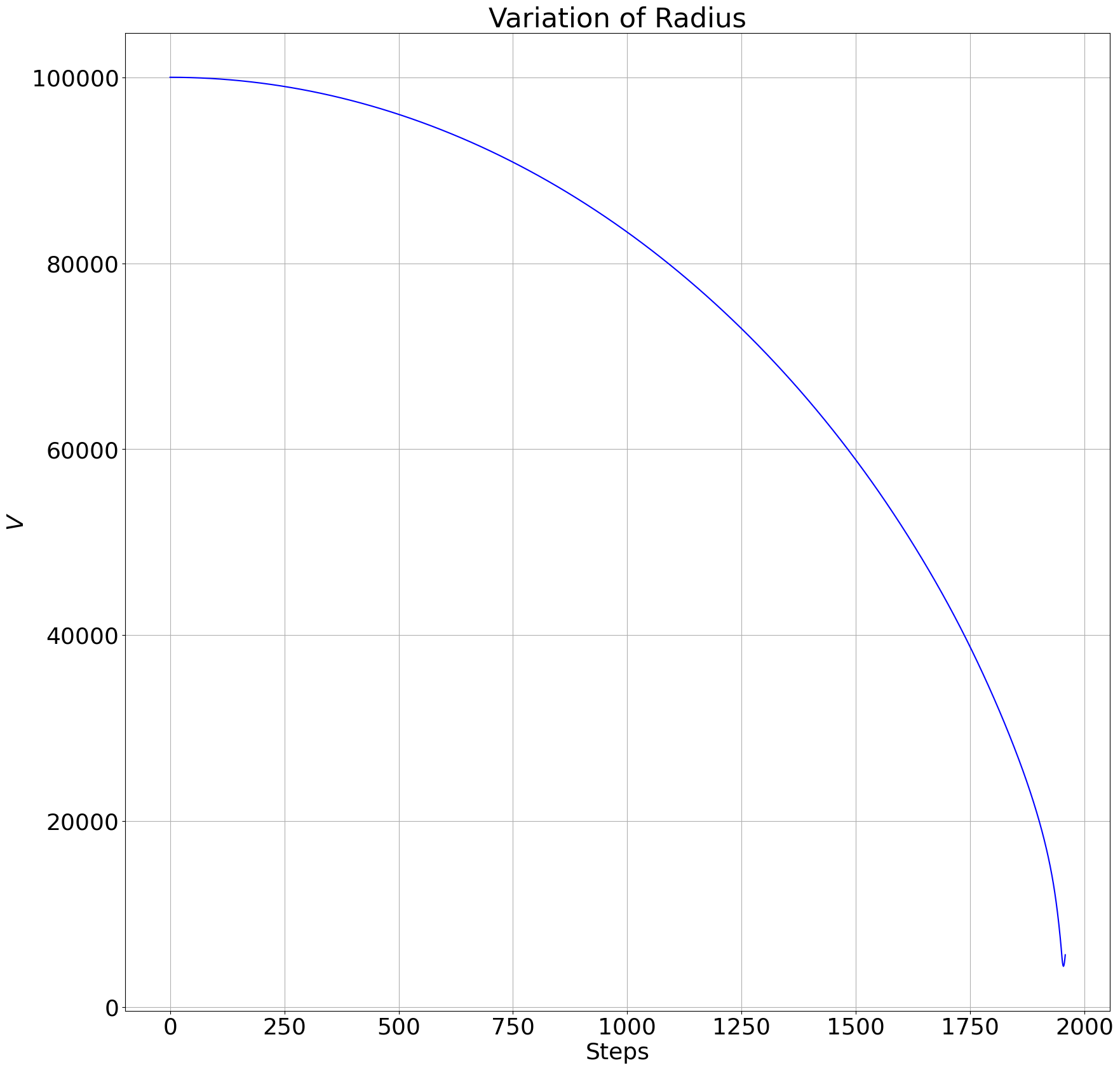}} 
    \subfigure[0.5 PN-with-Prop]{\includegraphics[width=0.24\textwidth]{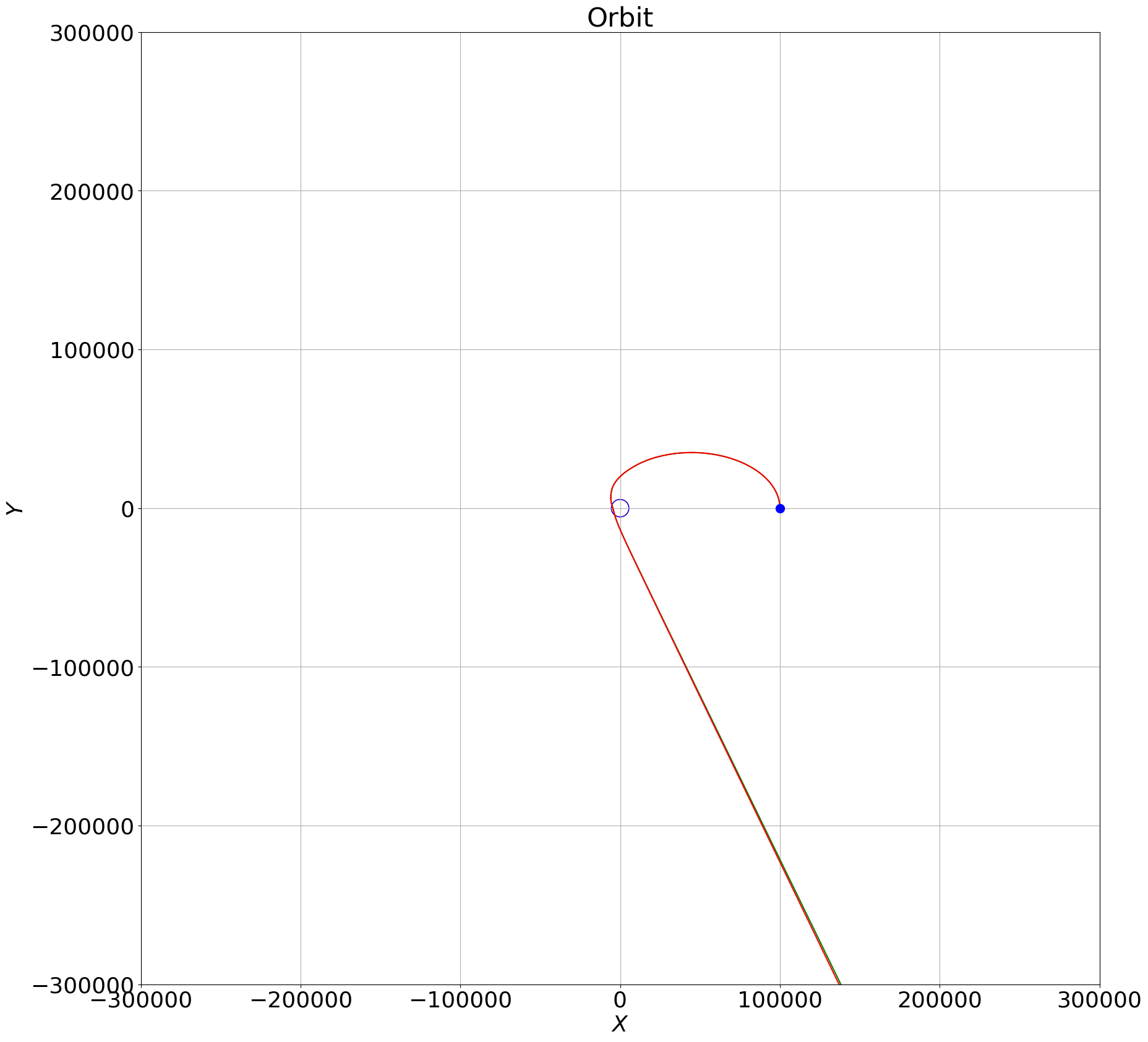}}
    \subfigure[0.5 PN-with-Prop]{\includegraphics[width=0.24\textwidth]{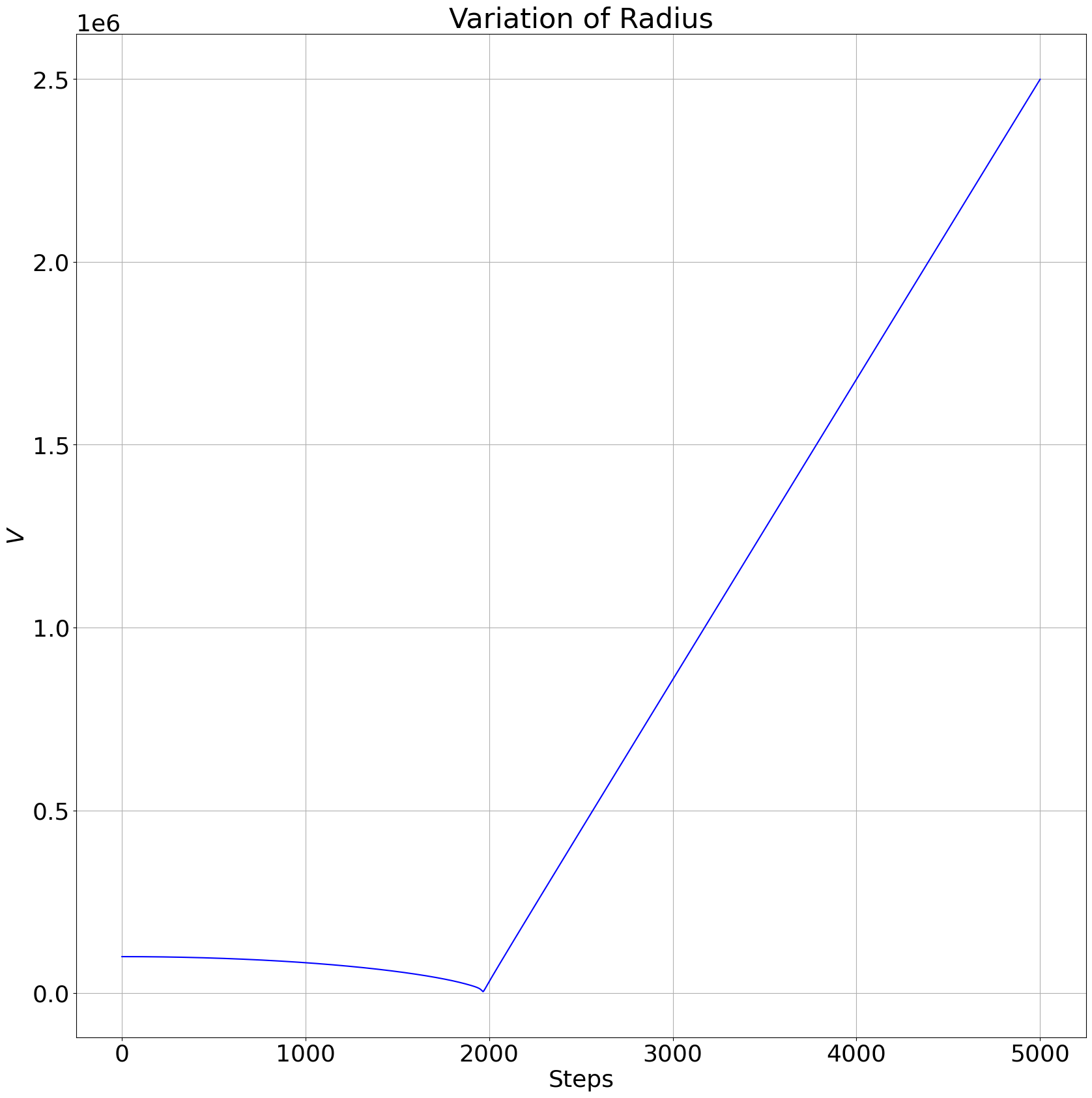}} \\
    
    \subfigure[1 PN-no-Prop]{\includegraphics[width=0.24\textwidth]{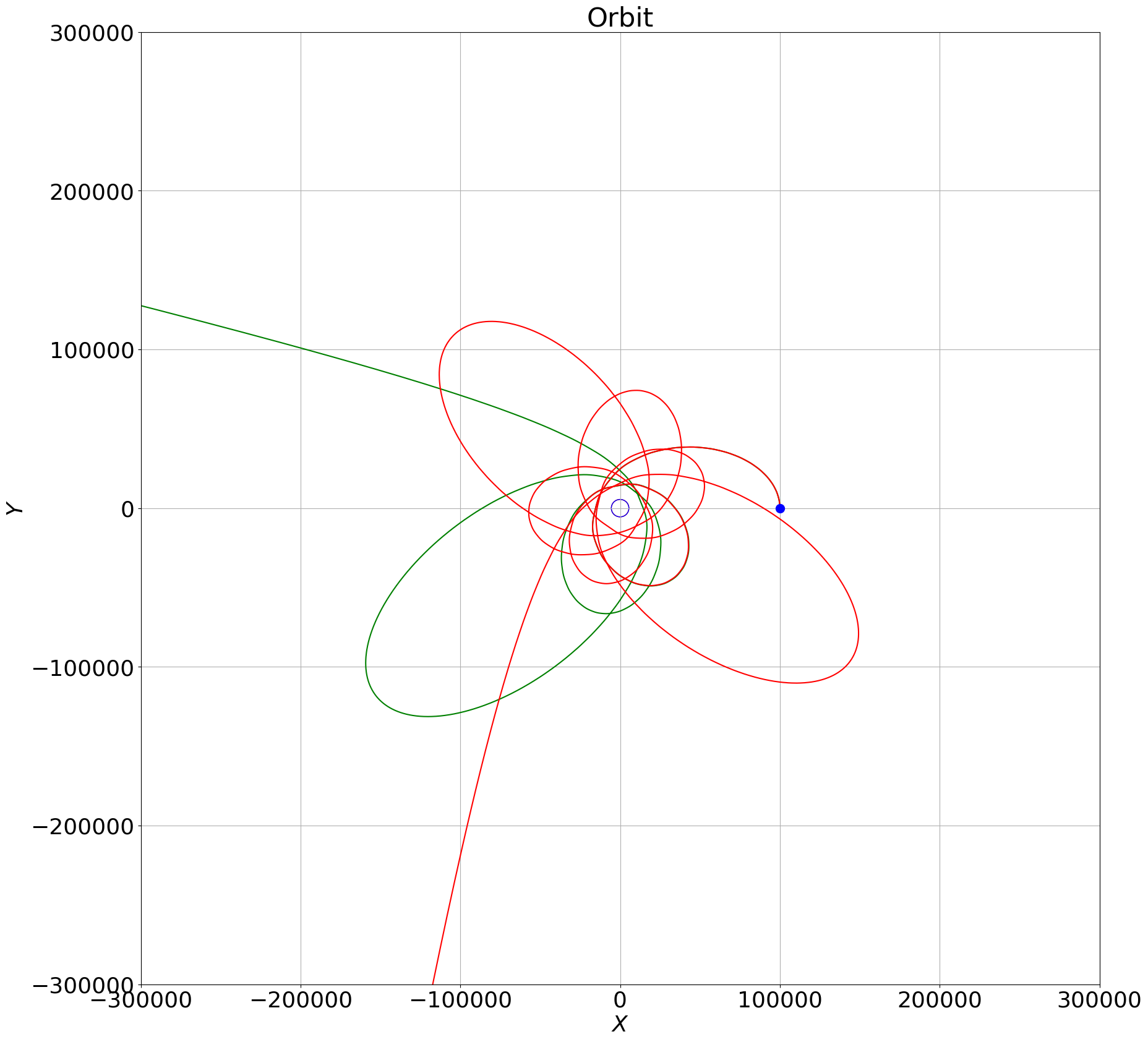}}
    \subfigure[1 PN-no-Prop]{\includegraphics[width=0.24\textwidth]{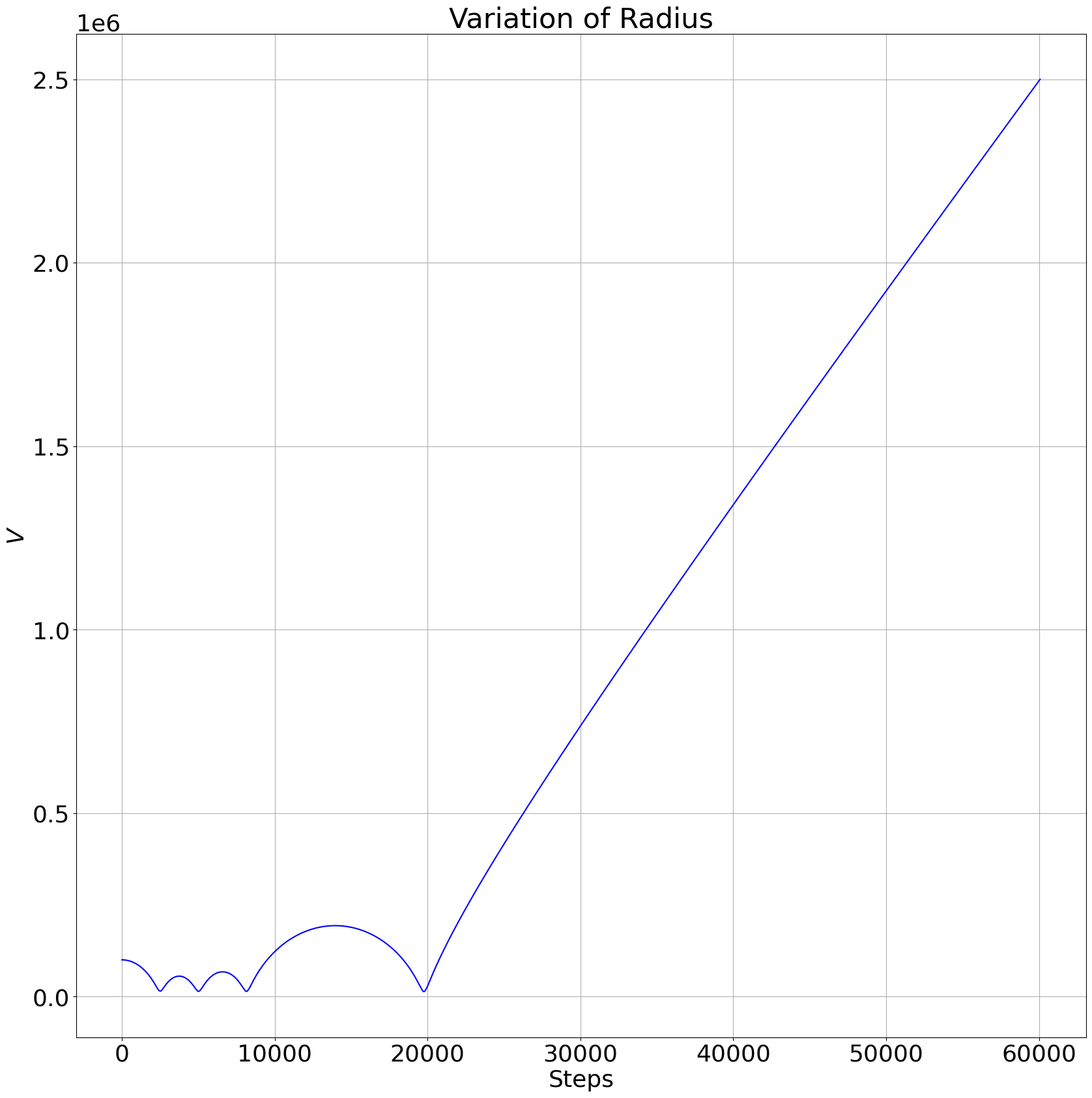}} 
    \subfigure[1 PN-with-Prop]{\includegraphics[width=0.24\textwidth]{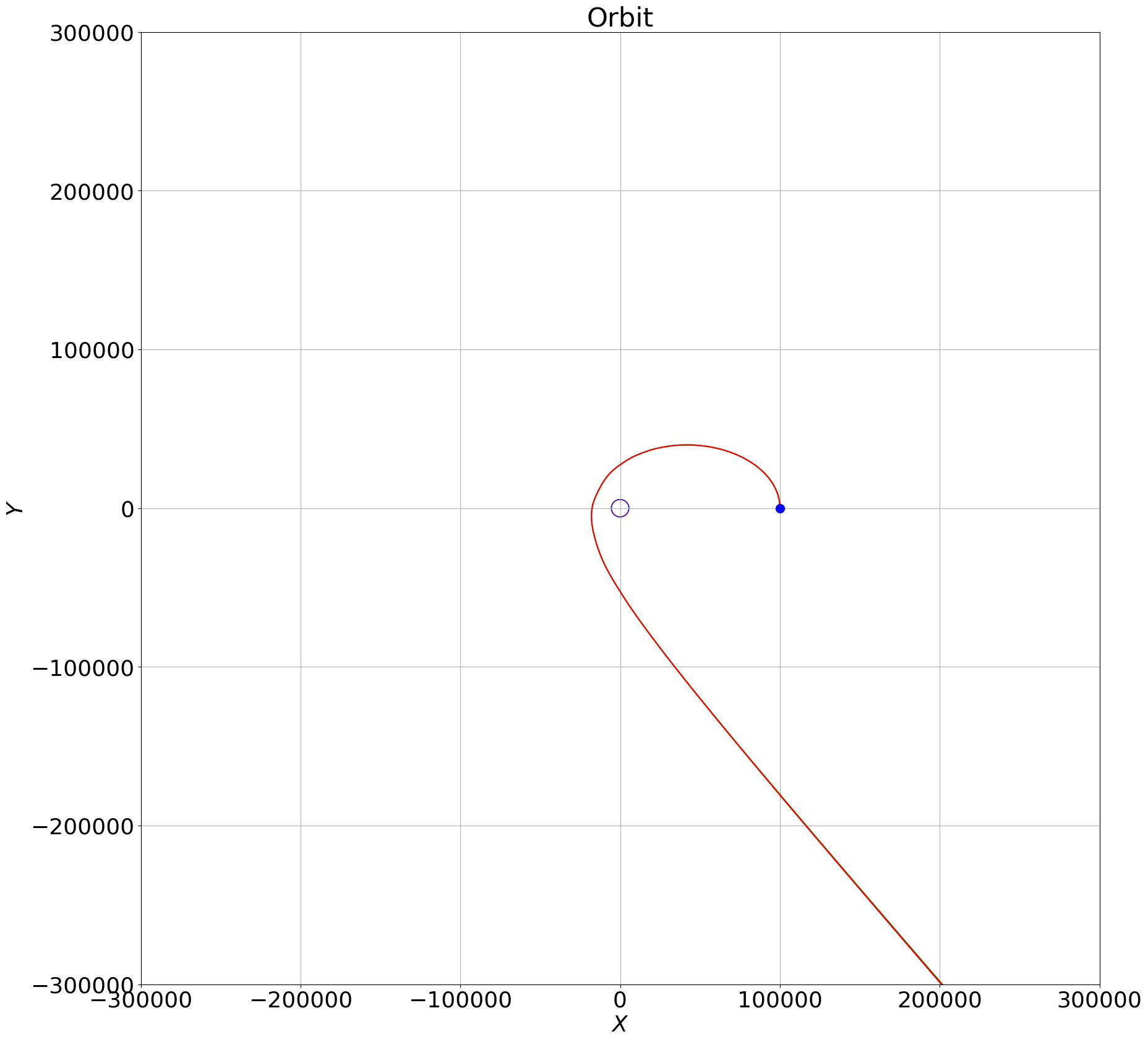}}
    \subfigure[1 PN-with-Prop]{\includegraphics[width=0.24\textwidth]{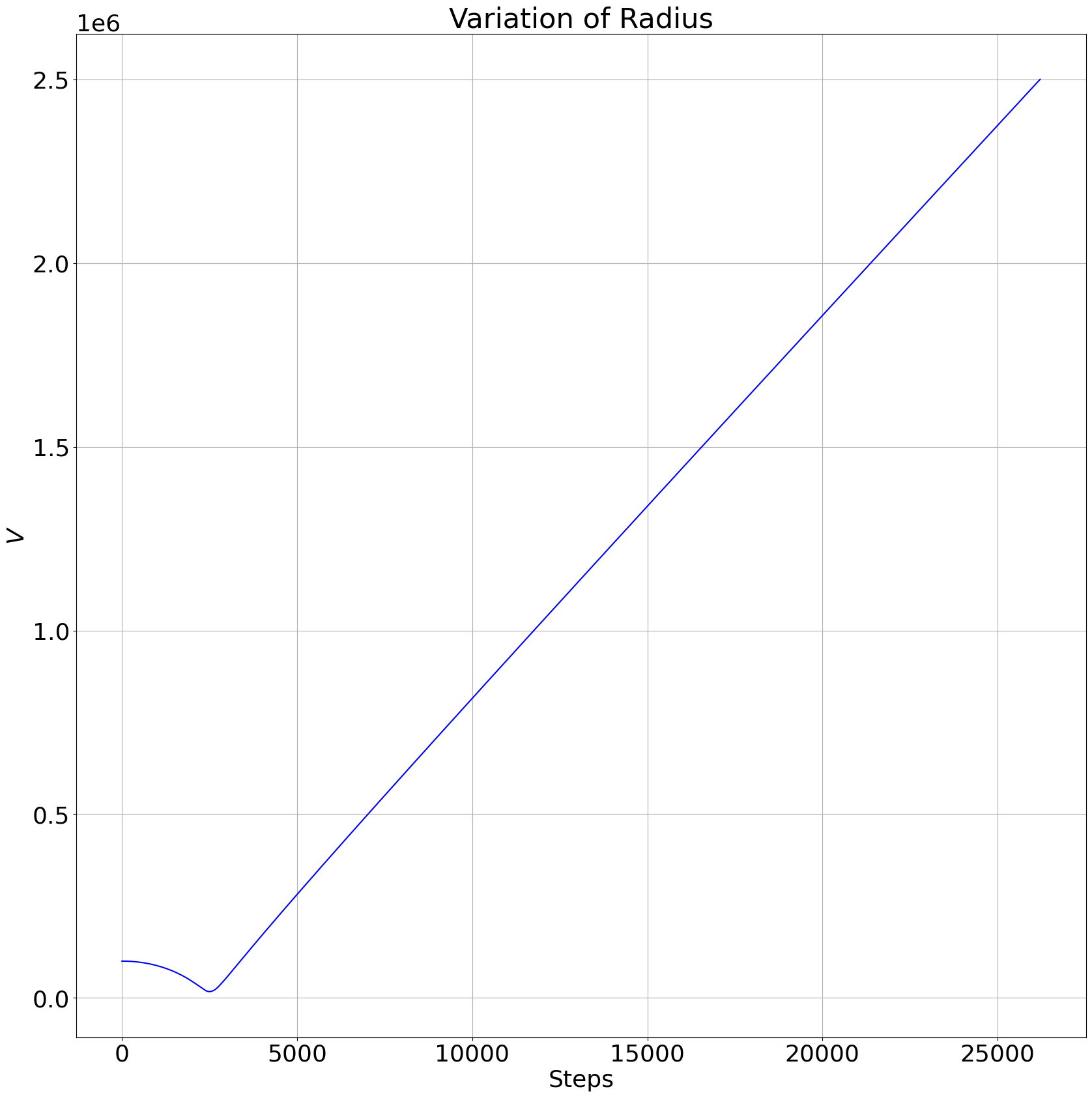}} \\
    
    \subfigure[2 PN-no-Prop]{\includegraphics[width=0.24\textwidth]{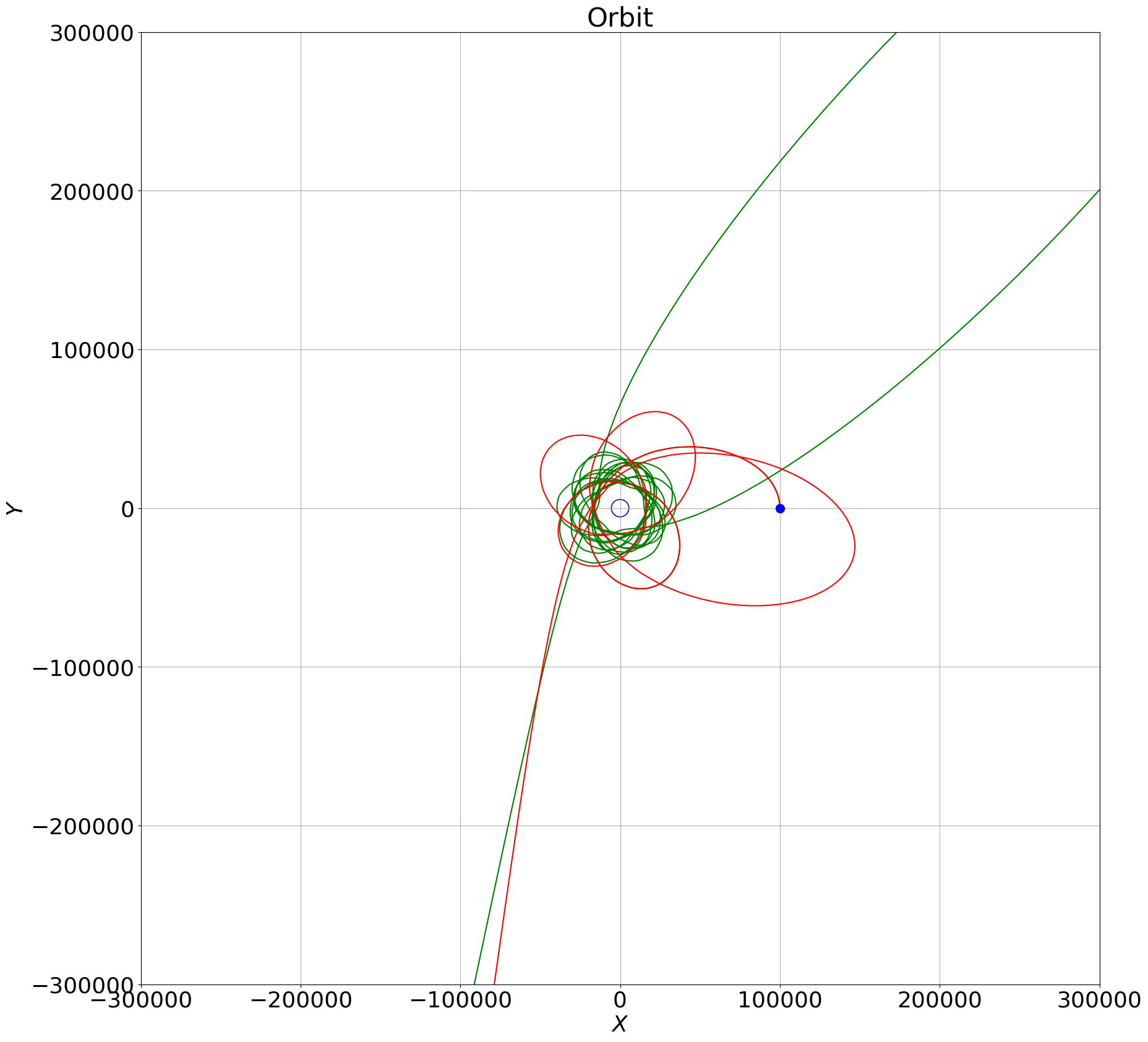}}
    \subfigure[2 PN-no-Prop]{\includegraphics[width=0.24\textwidth]{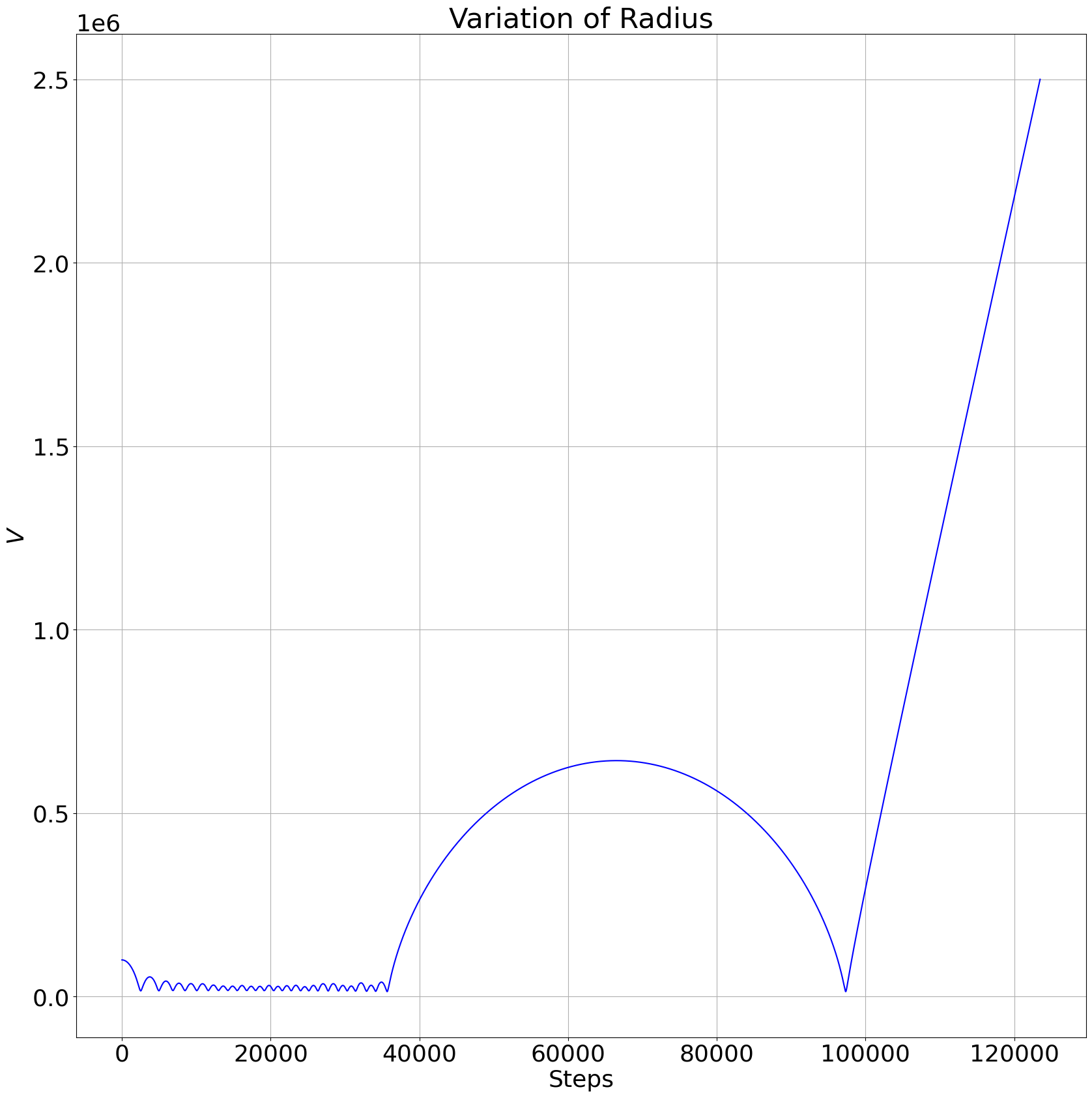}} 
    \subfigure[2 PN-with-Prop]{\includegraphics[width=0.24\textwidth]{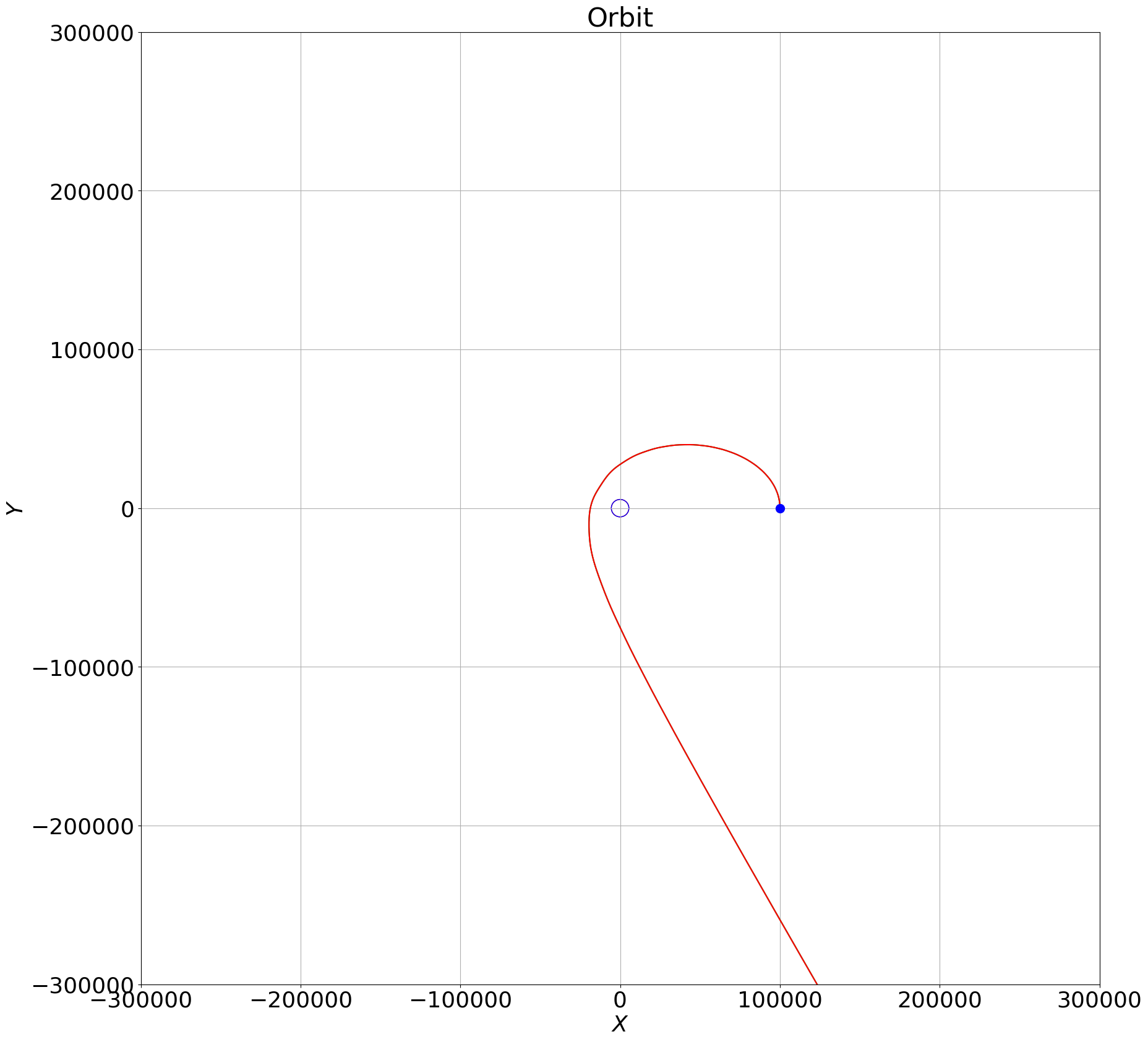}}
    \subfigure[2 PN-with-Prop]{\includegraphics[width=0.24\textwidth]{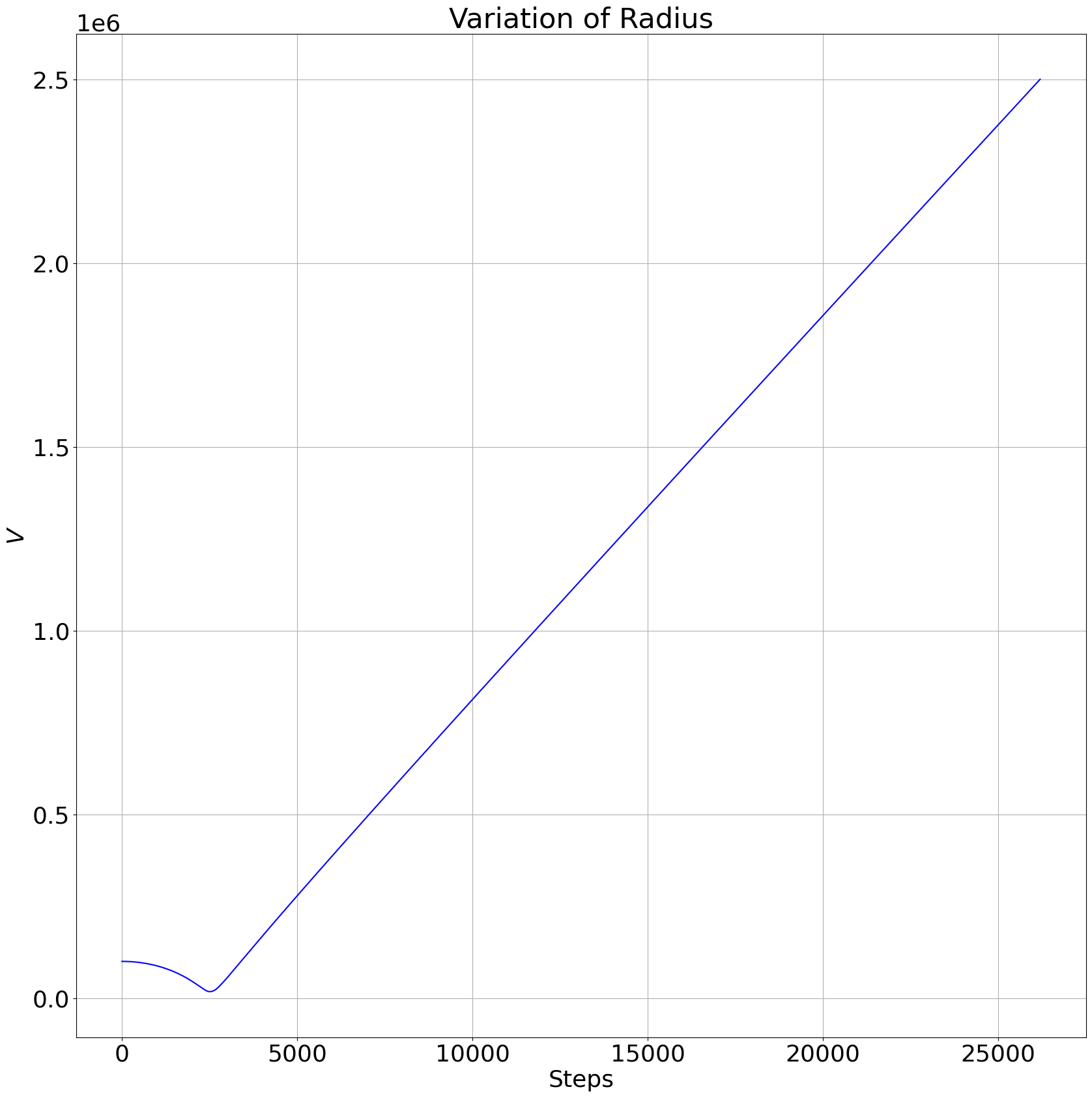}} \\
    
    \subfigure[2.5 PN-no-Prop]{\includegraphics[width=0.24\textwidth]{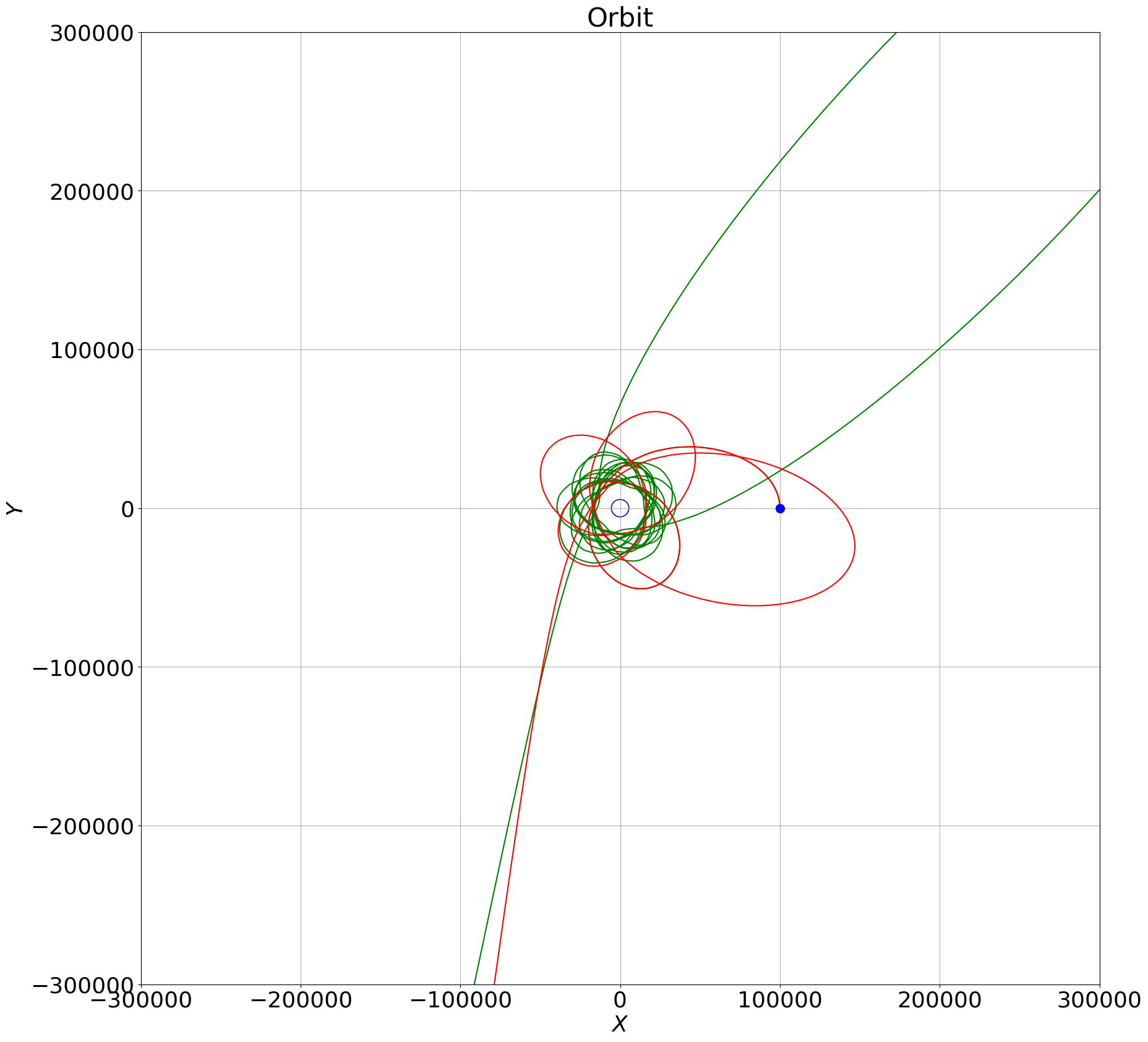}}
    \subfigure[2.5 PN-no-Prop]{\includegraphics[width=0.24\textwidth]{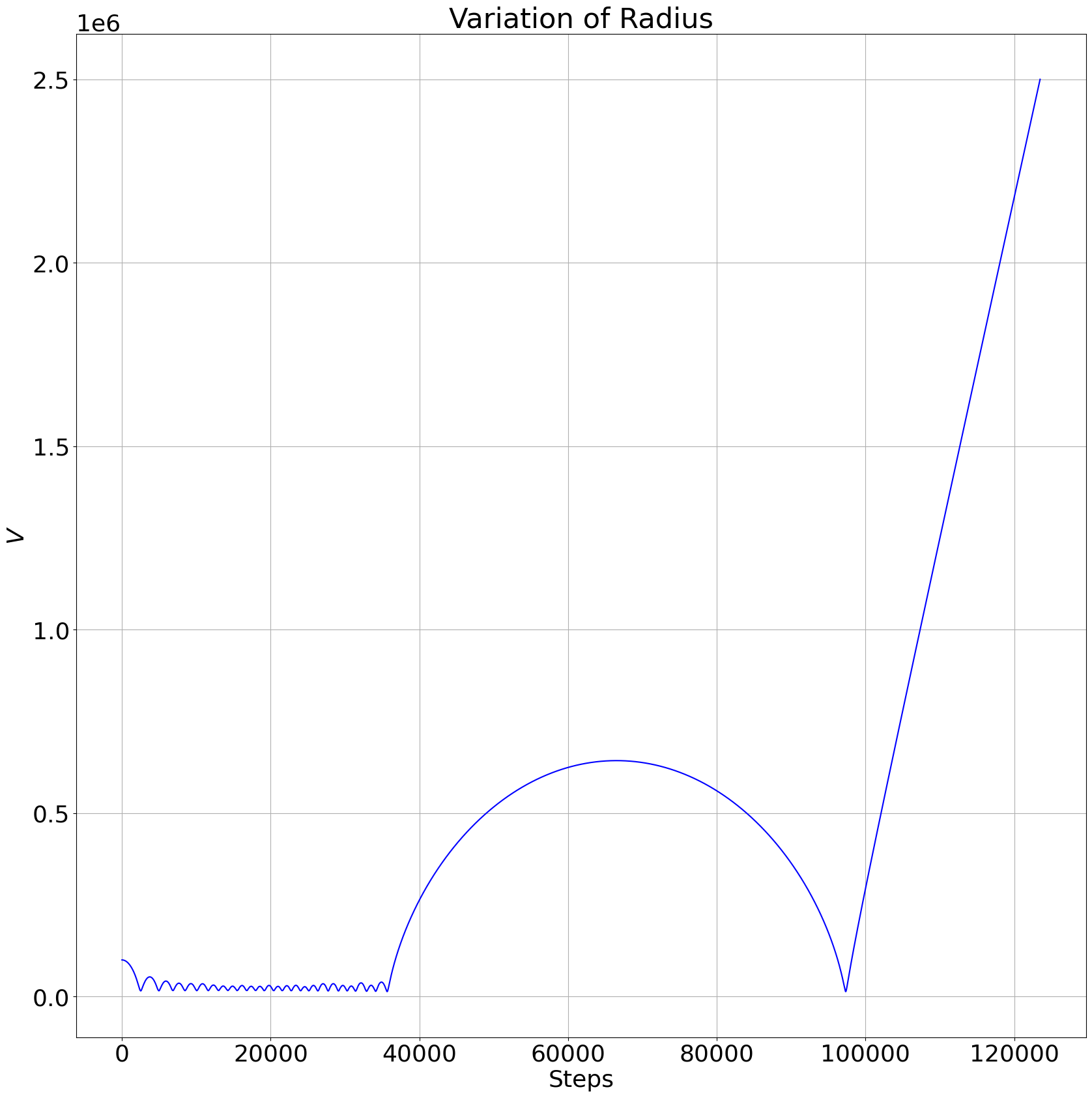}} 
    \subfigure[2.5 PN-with-Prop]{\includegraphics[width=0.24\textwidth]{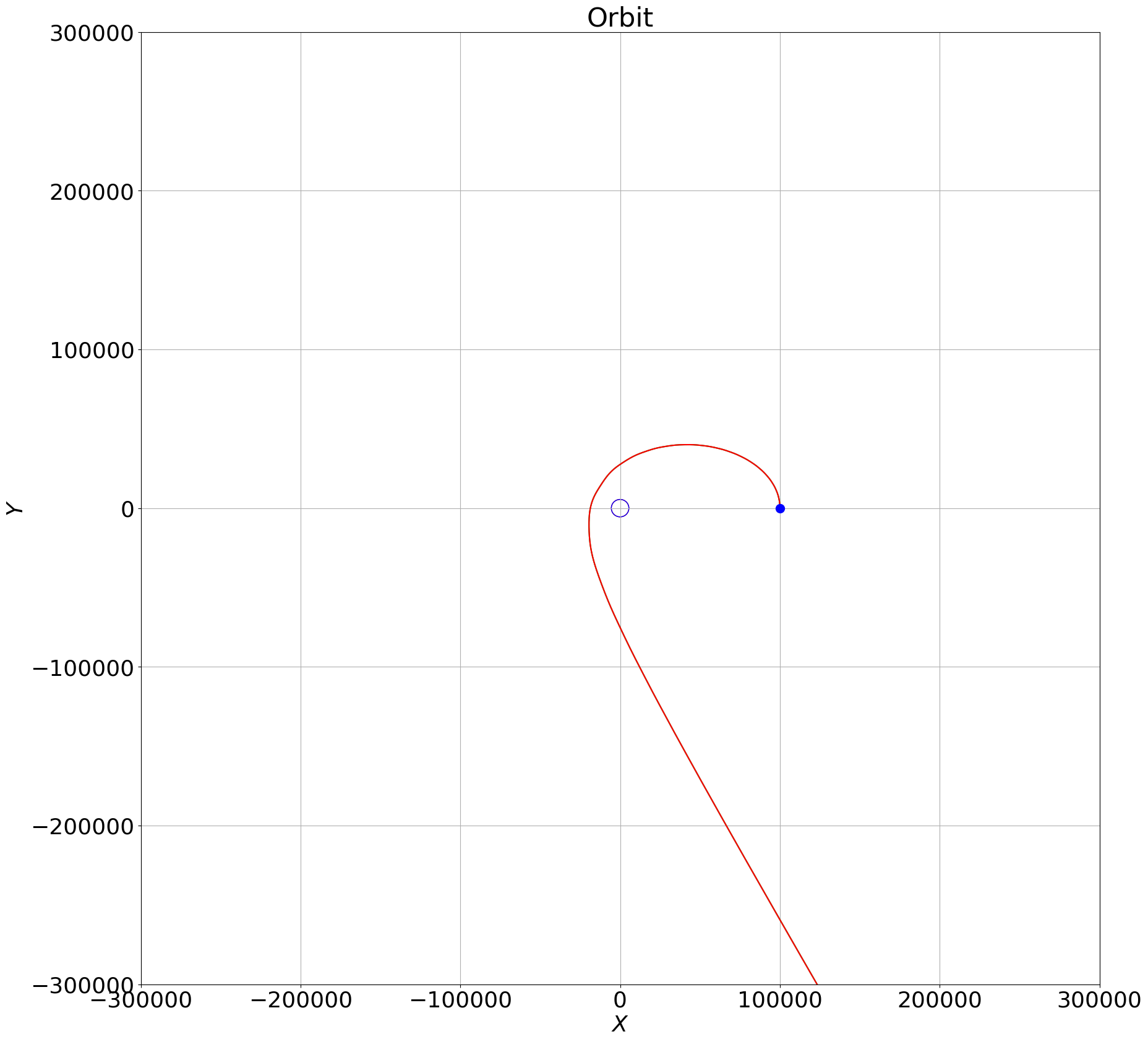}}
    \subfigure[2.5 PN-with-Prop]{\includegraphics[width=0.24\textwidth]{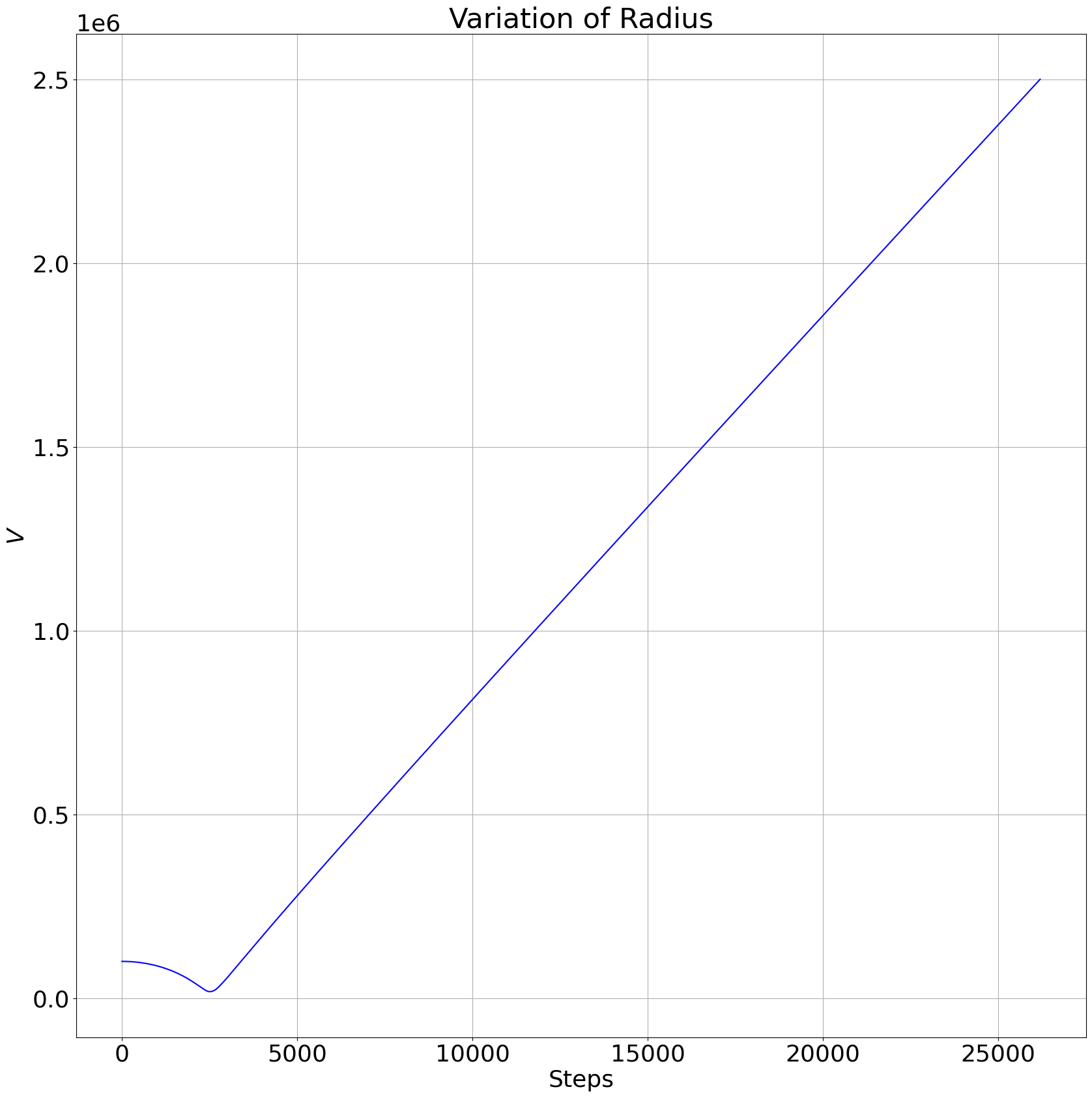}} \\
    \caption{Strong(er) Field, R = $1.1 \cdot 10^4m$, X = $10^5m$, Velocity = 25,000,000$ms^{-1}$}
    \label{results3}
\end{figure}

\begin{figure}[!ht]
    \centering
    \setcounter{subfigure}{0}
    \subfigure[0 PN-no-Prop]{\includegraphics[width=0.24\textwidth]{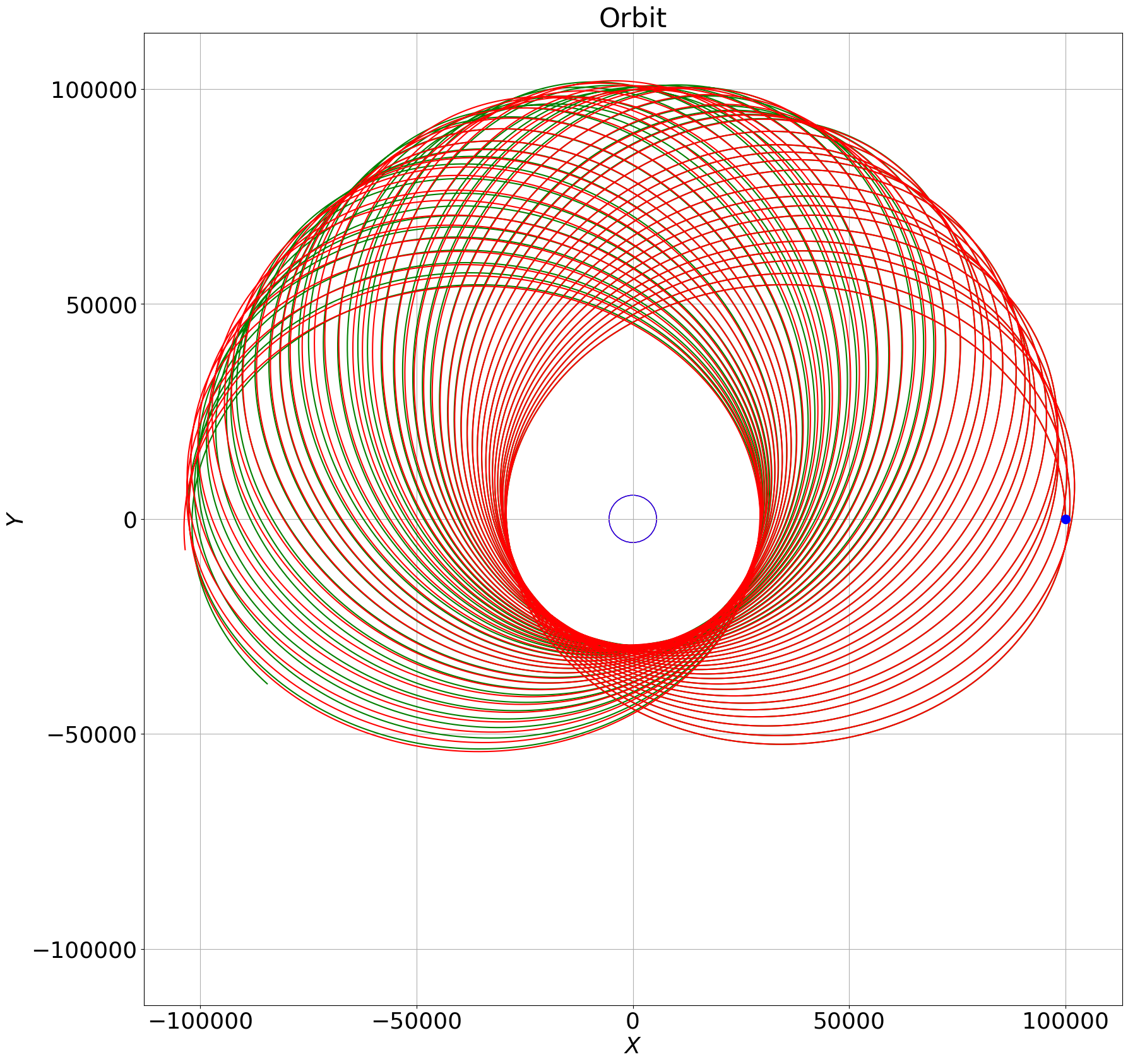}}
    \subfigure[0 PN-no-Prop]{\includegraphics[width=0.24\textwidth]{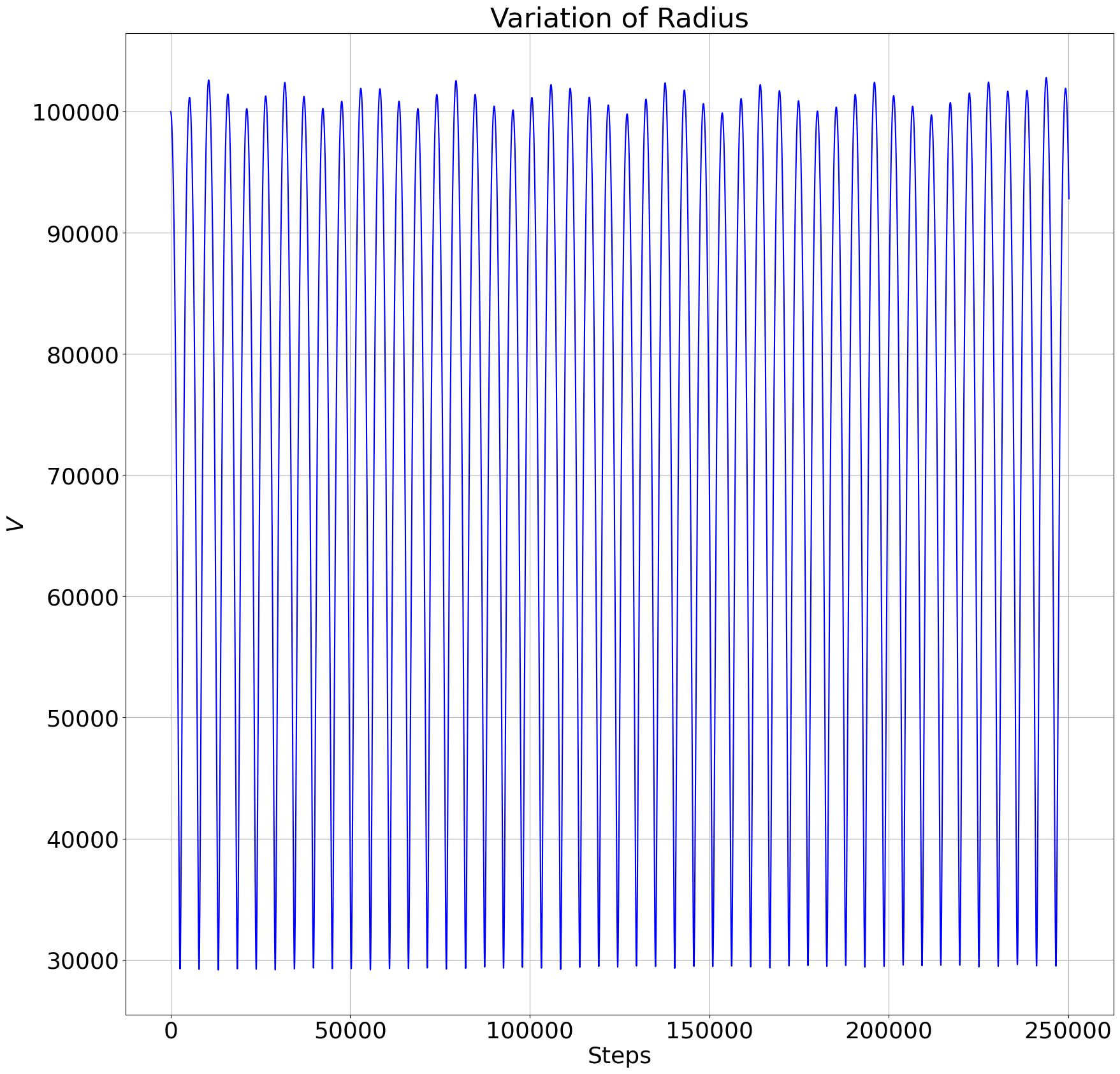}} 
    \subfigure[0 PN-with-Prop]{\includegraphics[width=0.24\textwidth]{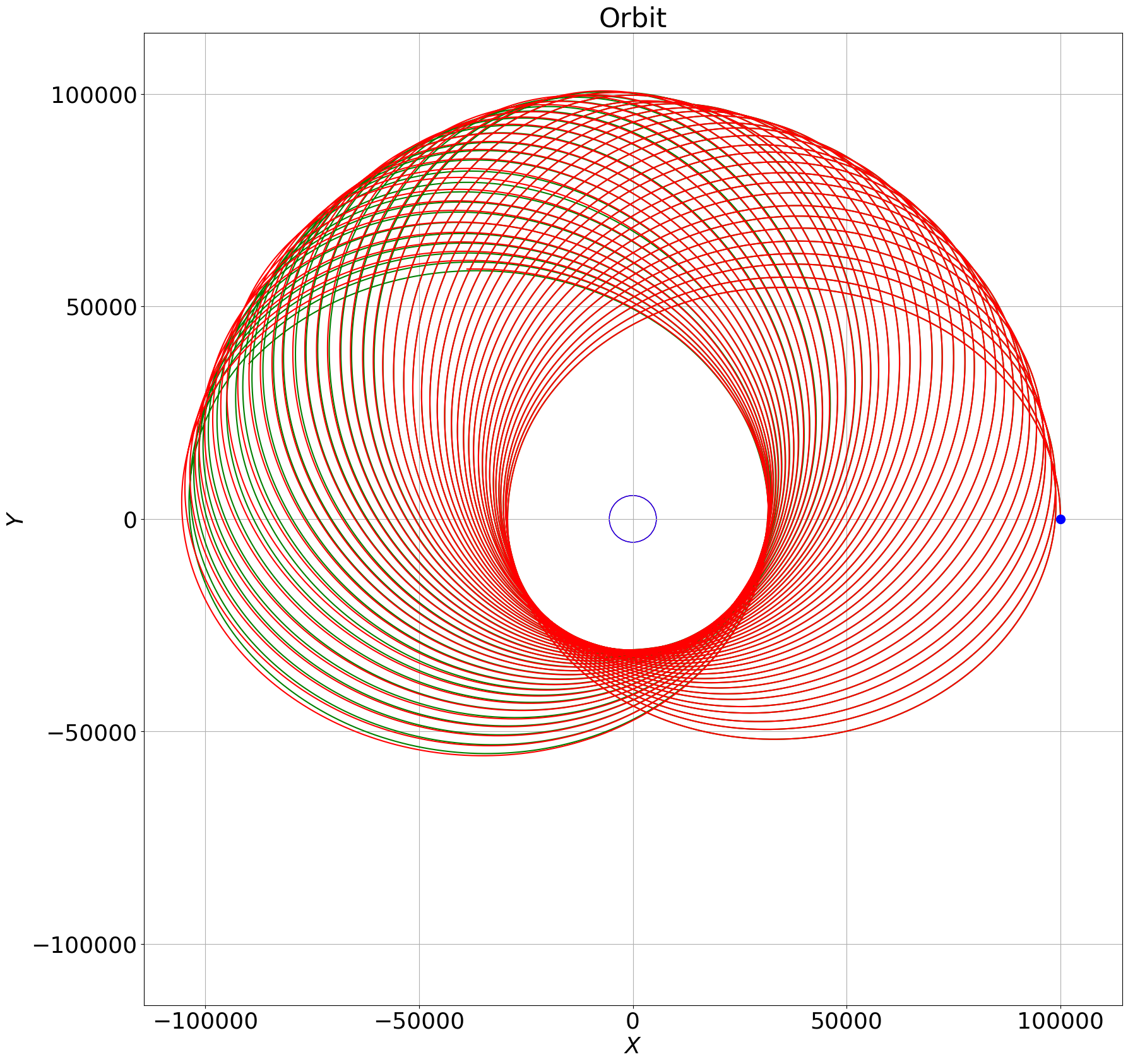}}
    \subfigure[0 PN-with-Prop]{\includegraphics[width=0.24\textwidth]{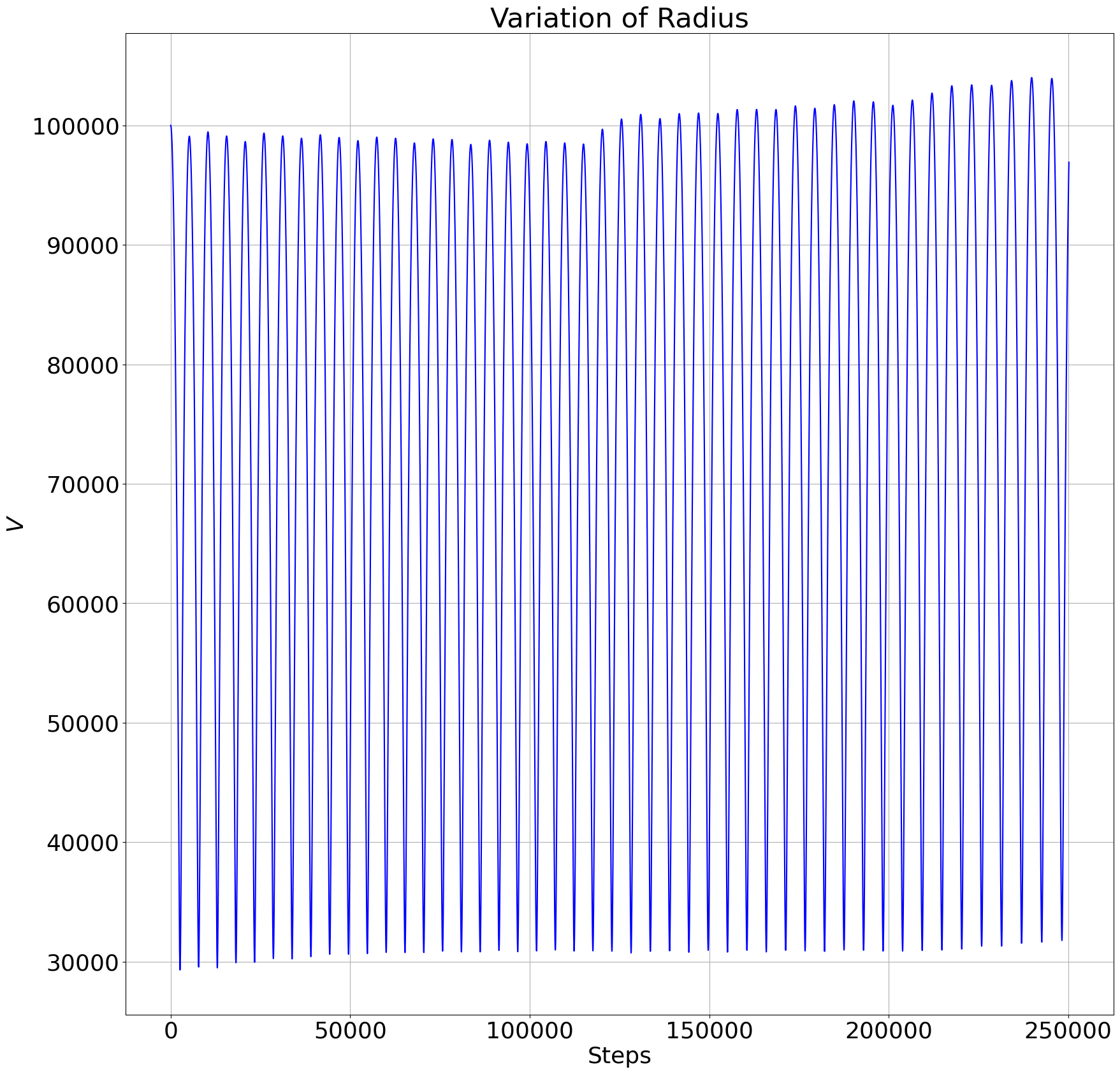}} \\
    
    \subfigure[0.5 PN-no-Prop]{\includegraphics[width=0.24\textwidth]{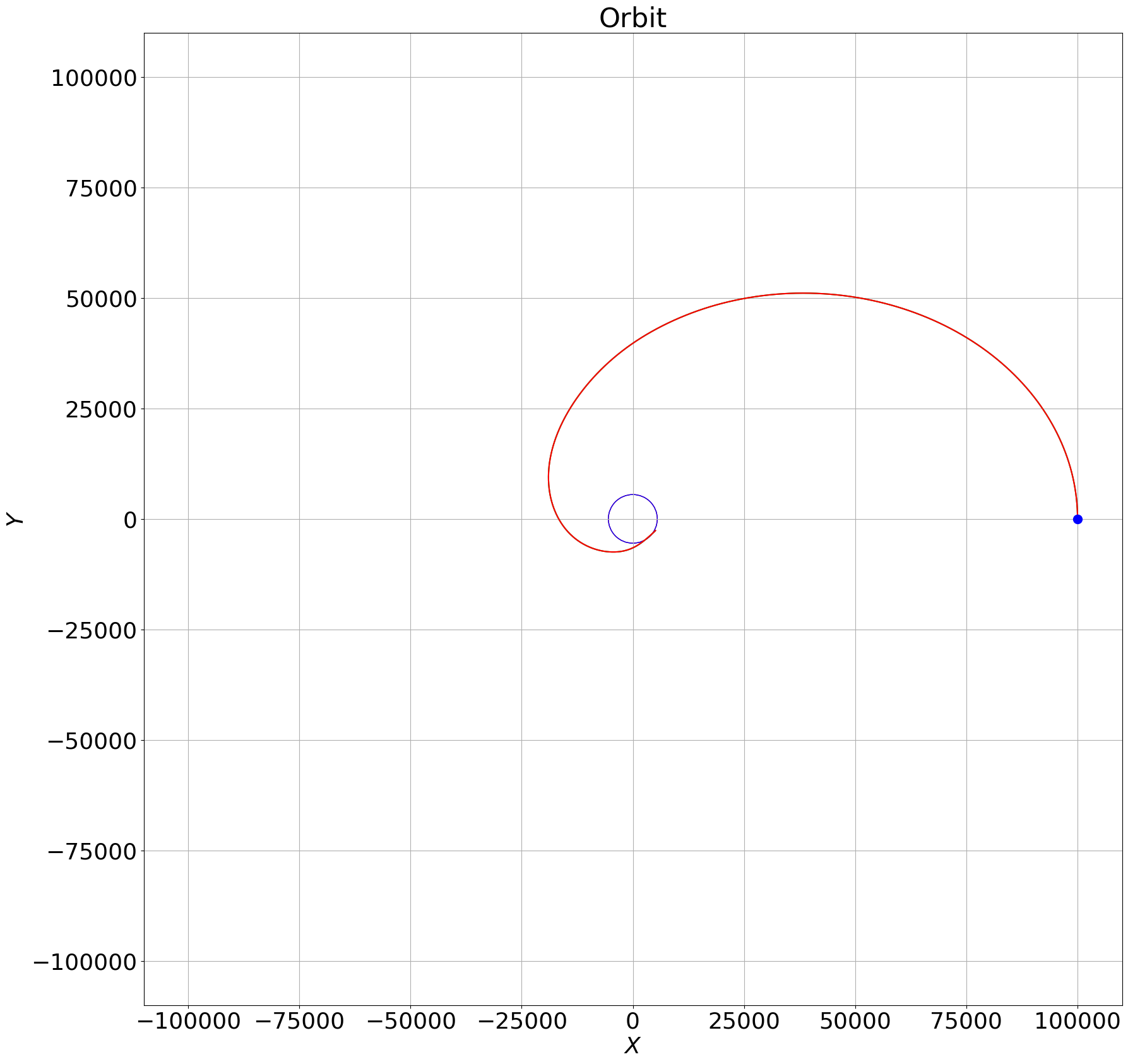}}
    \subfigure[0.5 PN-no-Prop]{\includegraphics[width=0.24\textwidth]{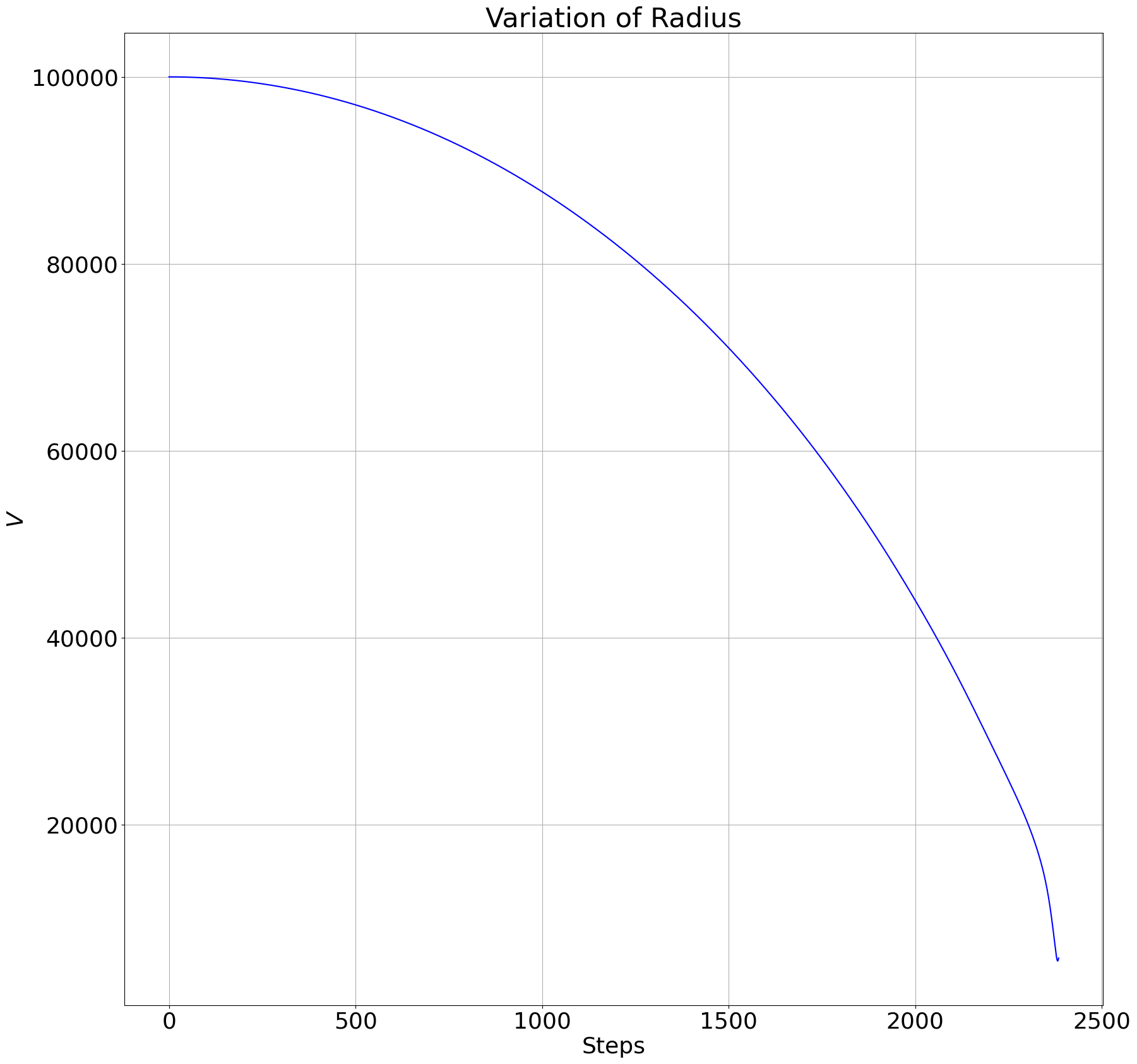}} 
    \subfigure[0.5 PN-with-Prop]{\includegraphics[width=0.24\textwidth]{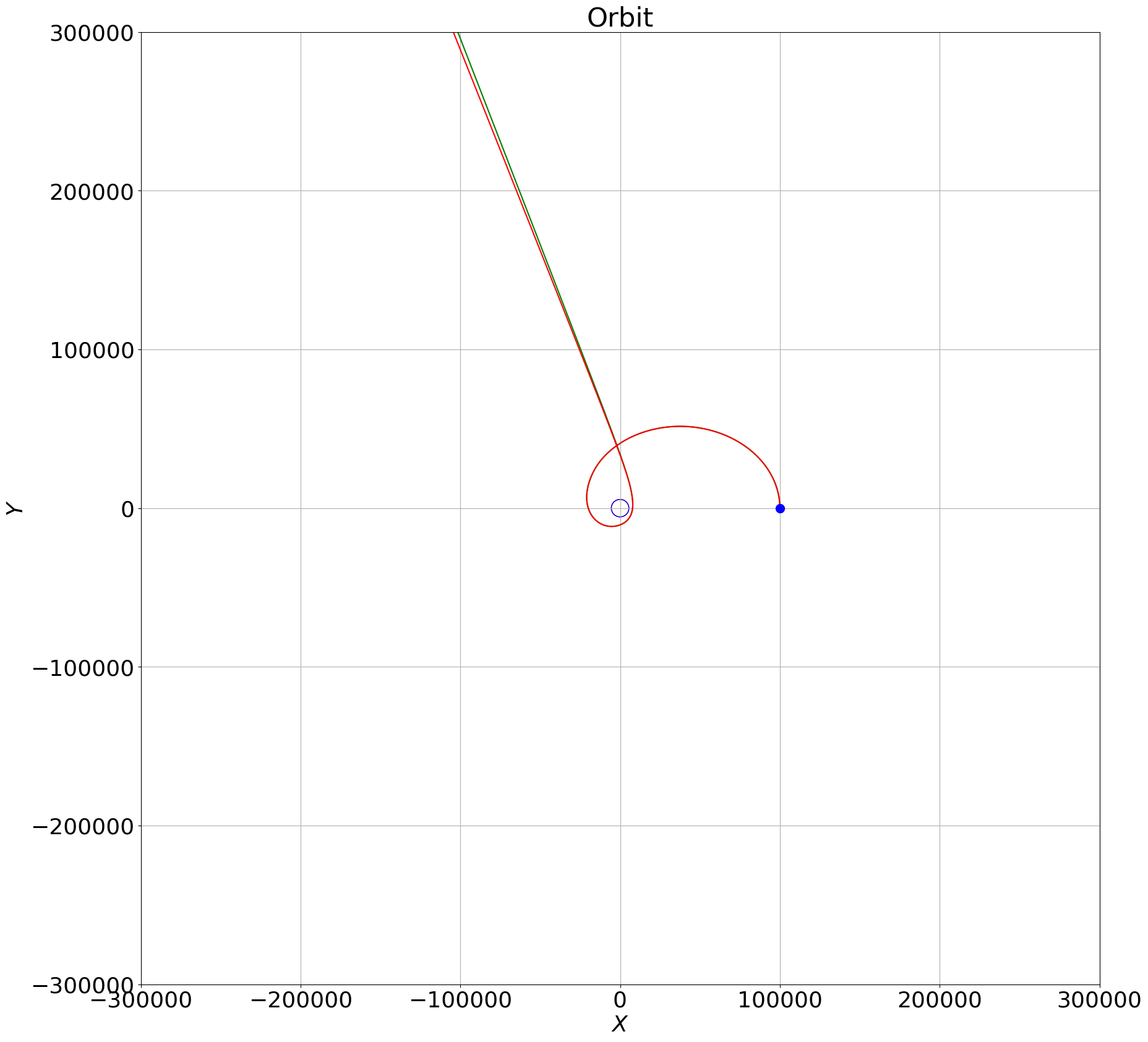}}
    \subfigure[0.5 PN-with-Prop]{\includegraphics[width=0.24\textwidth]{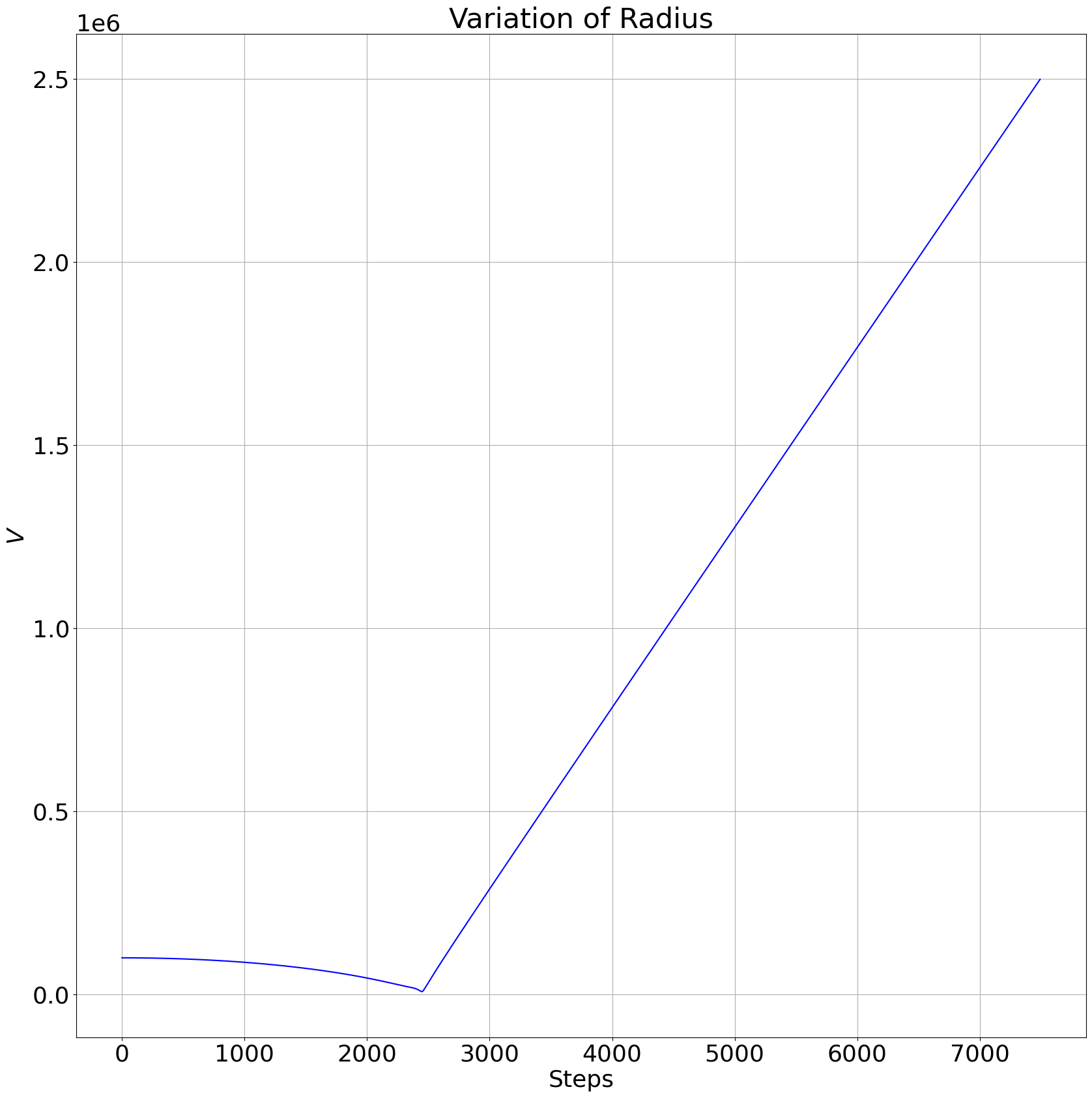}} \\
    
    \subfigure[1 PN-no-Prop]{\includegraphics[width=0.24\textwidth]{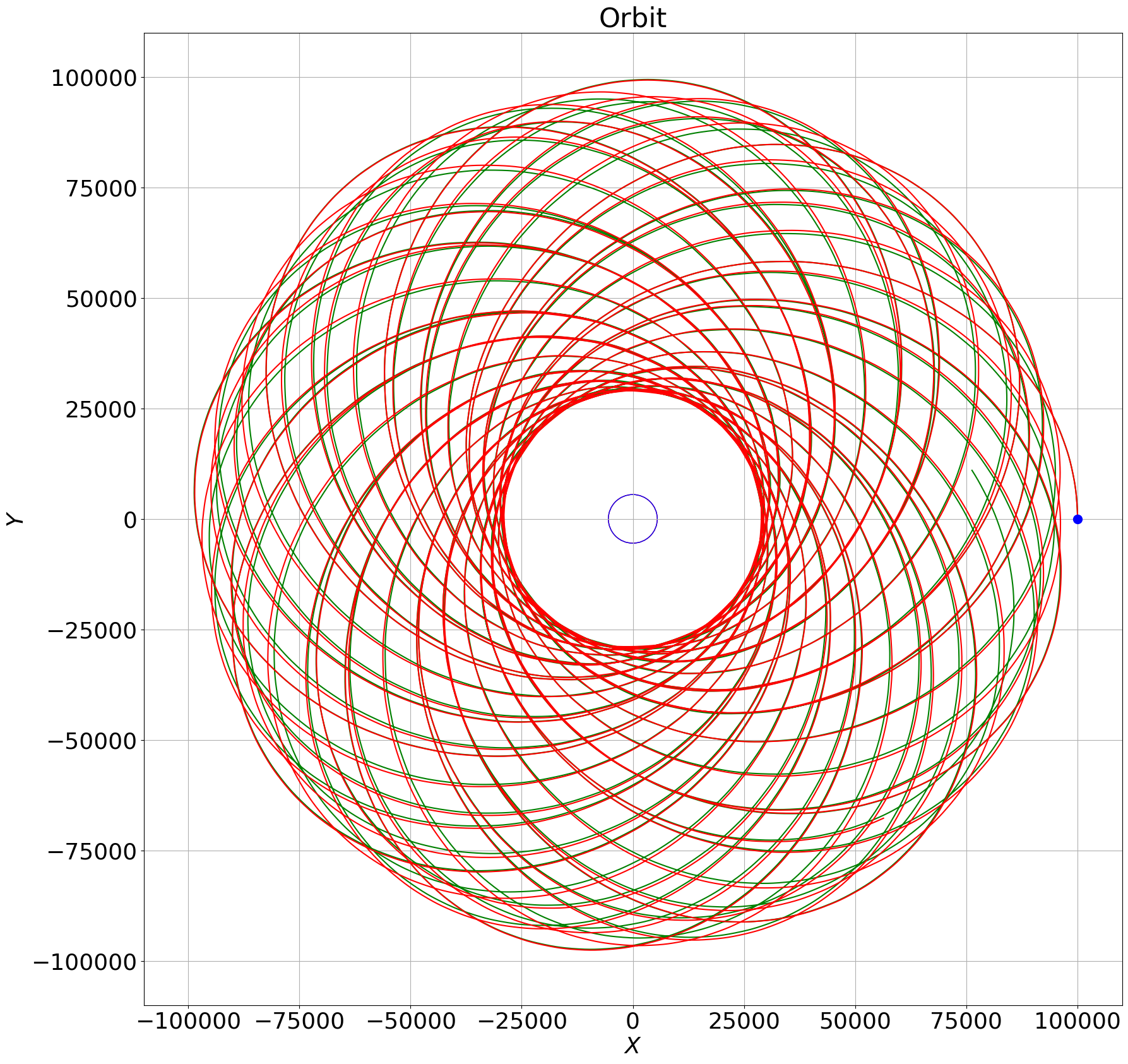}}
    \subfigure[1 PN-no-Prop]{\includegraphics[width=0.24\textwidth]{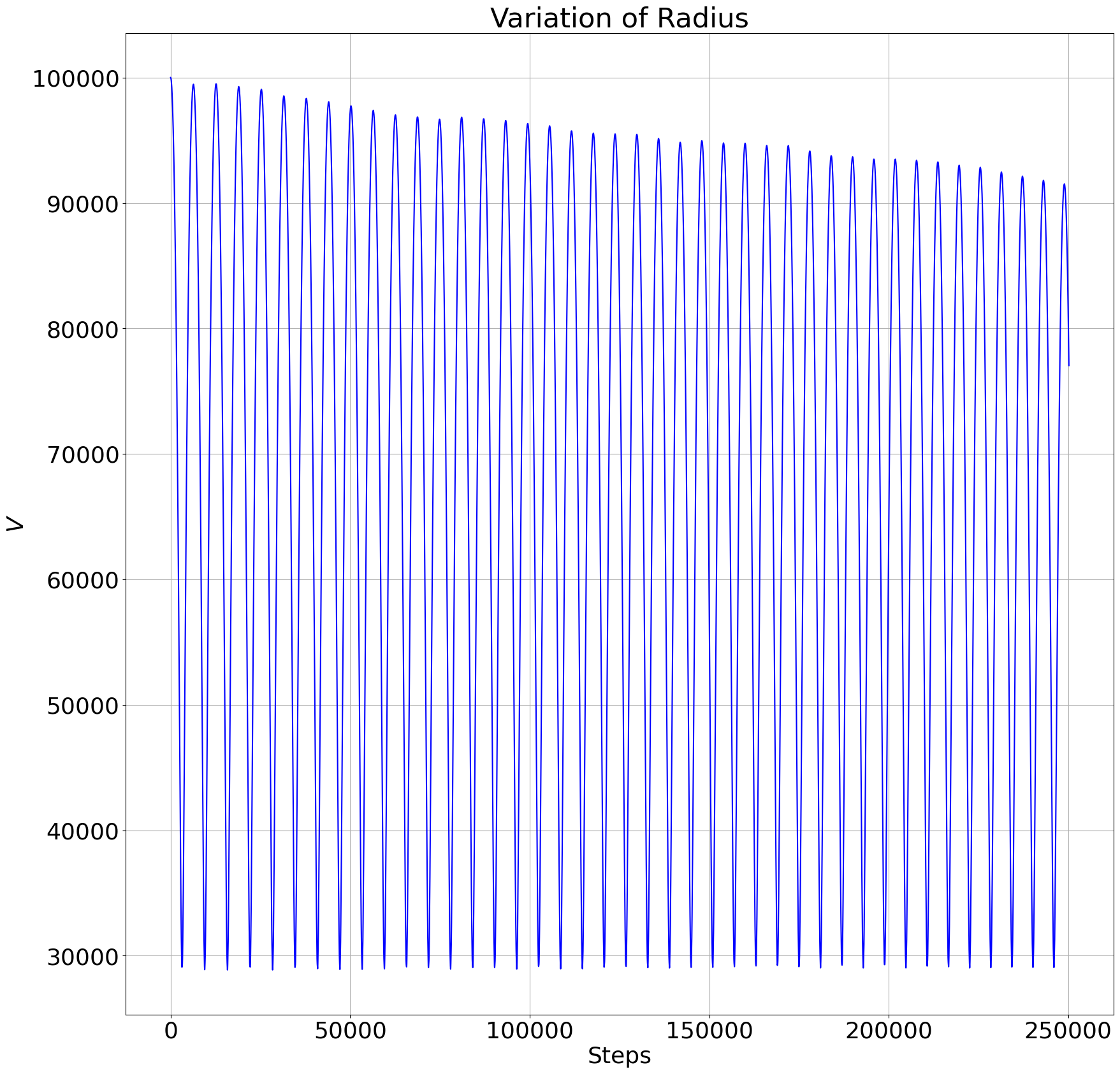}} 
    \subfigure[1 PN-with-Prop]{\includegraphics[width=0.24\textwidth]{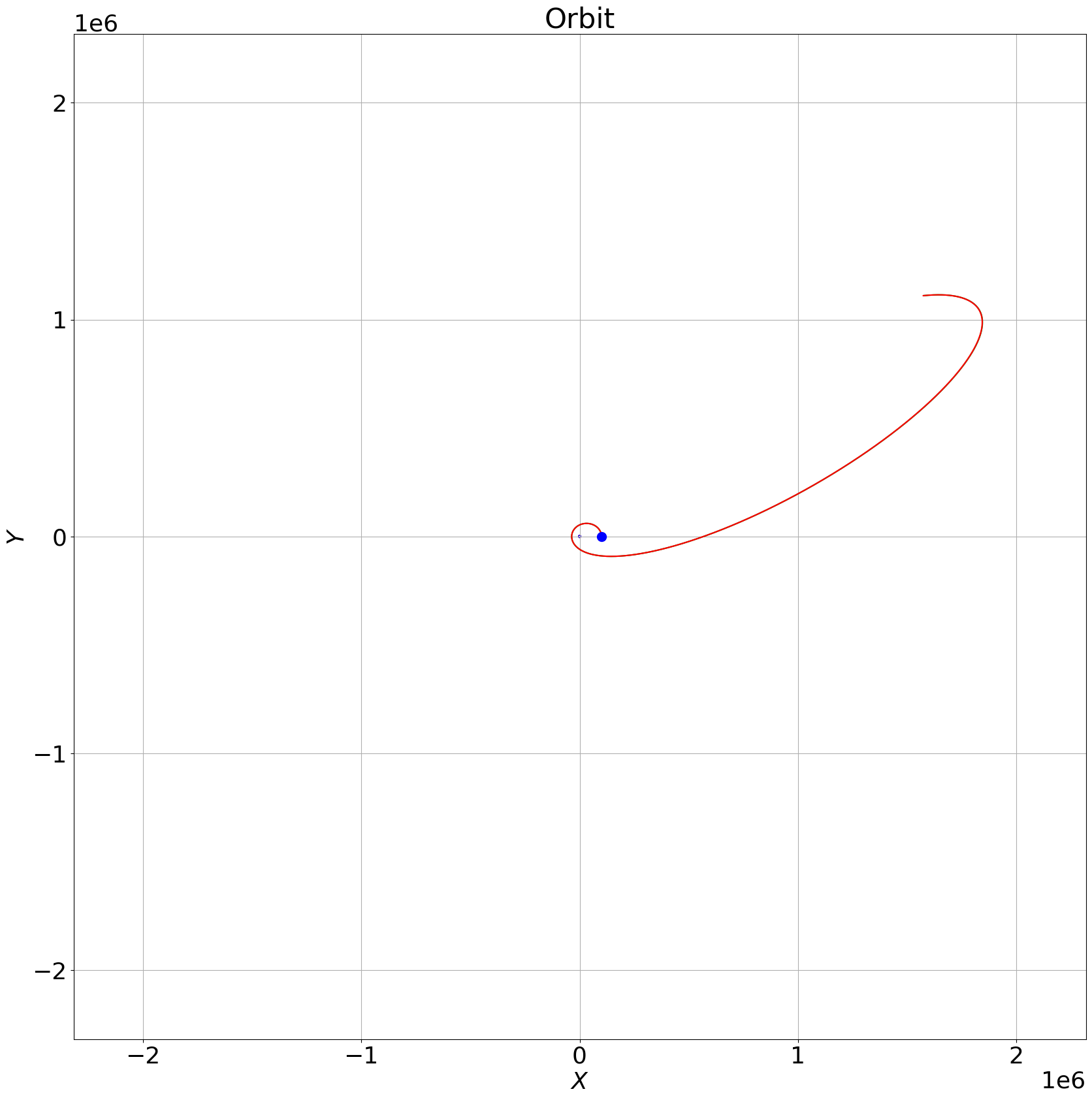}}
    \subfigure[1 PN-with-Prop]{\includegraphics[width=0.24\textwidth]{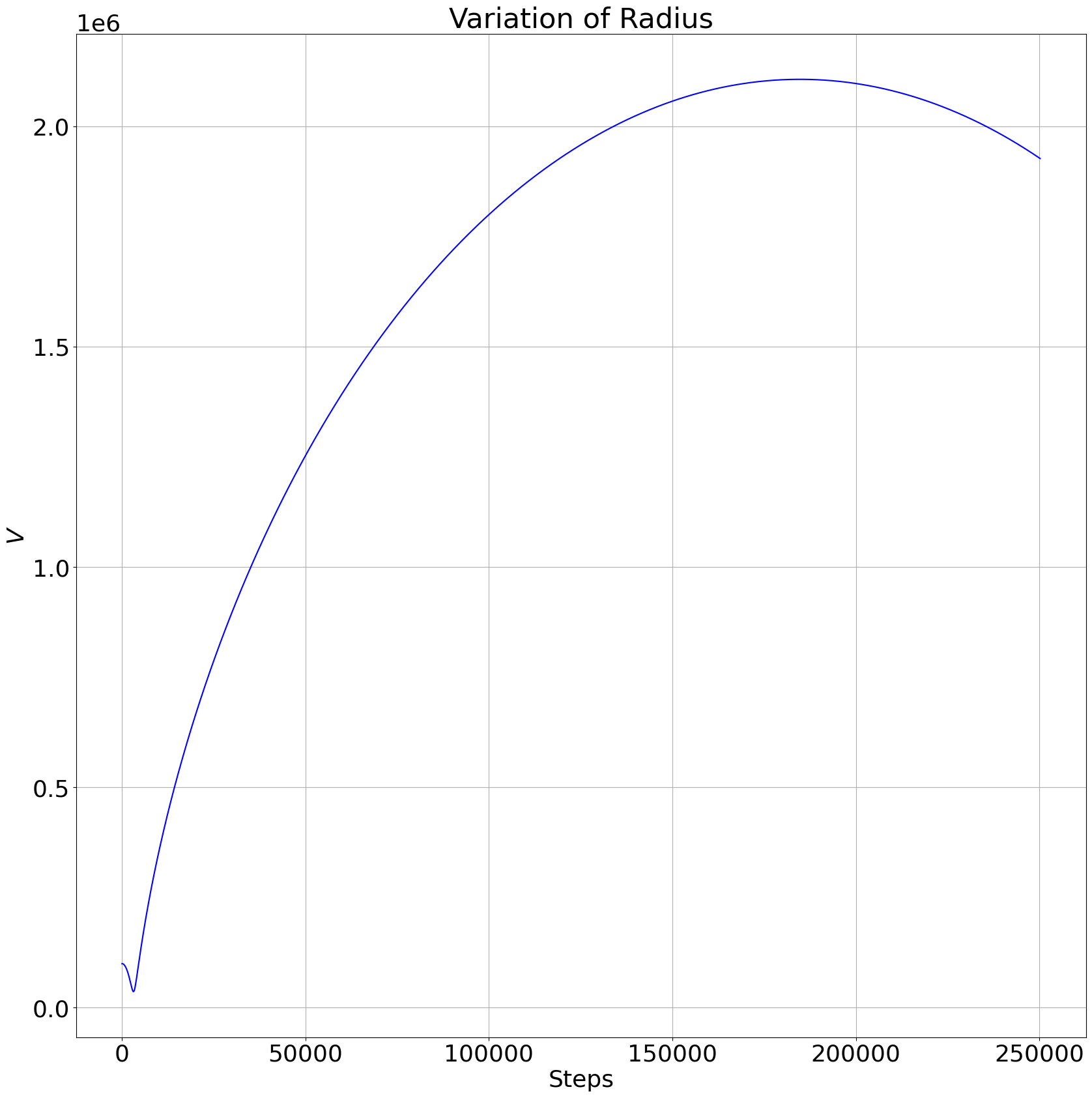}} \\
    
    \subfigure[2 PN-no-Prop]{\includegraphics[width=0.24\textwidth]{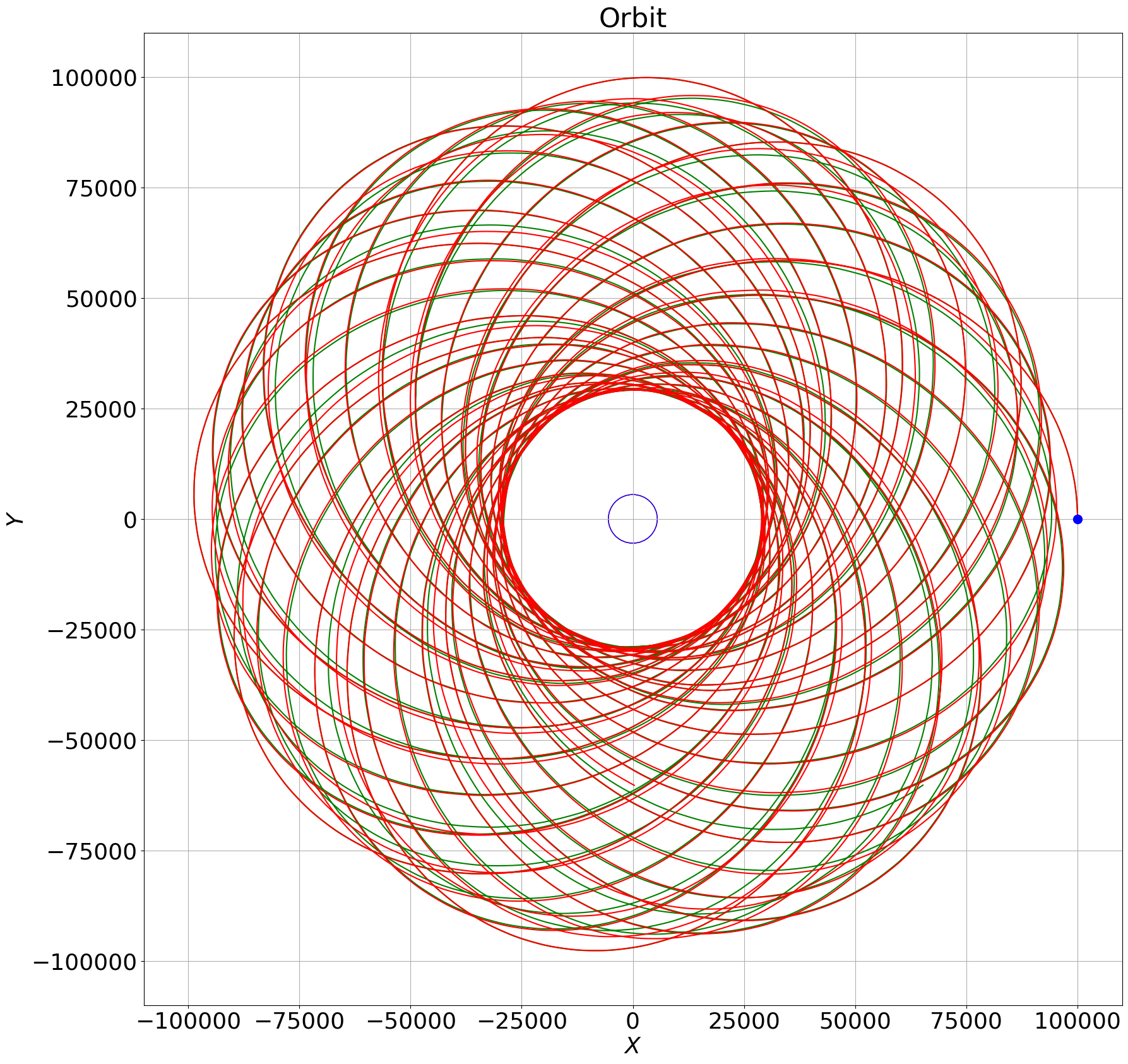}}
    \subfigure[2 PN-no-Prop]{\includegraphics[width=0.24\textwidth]{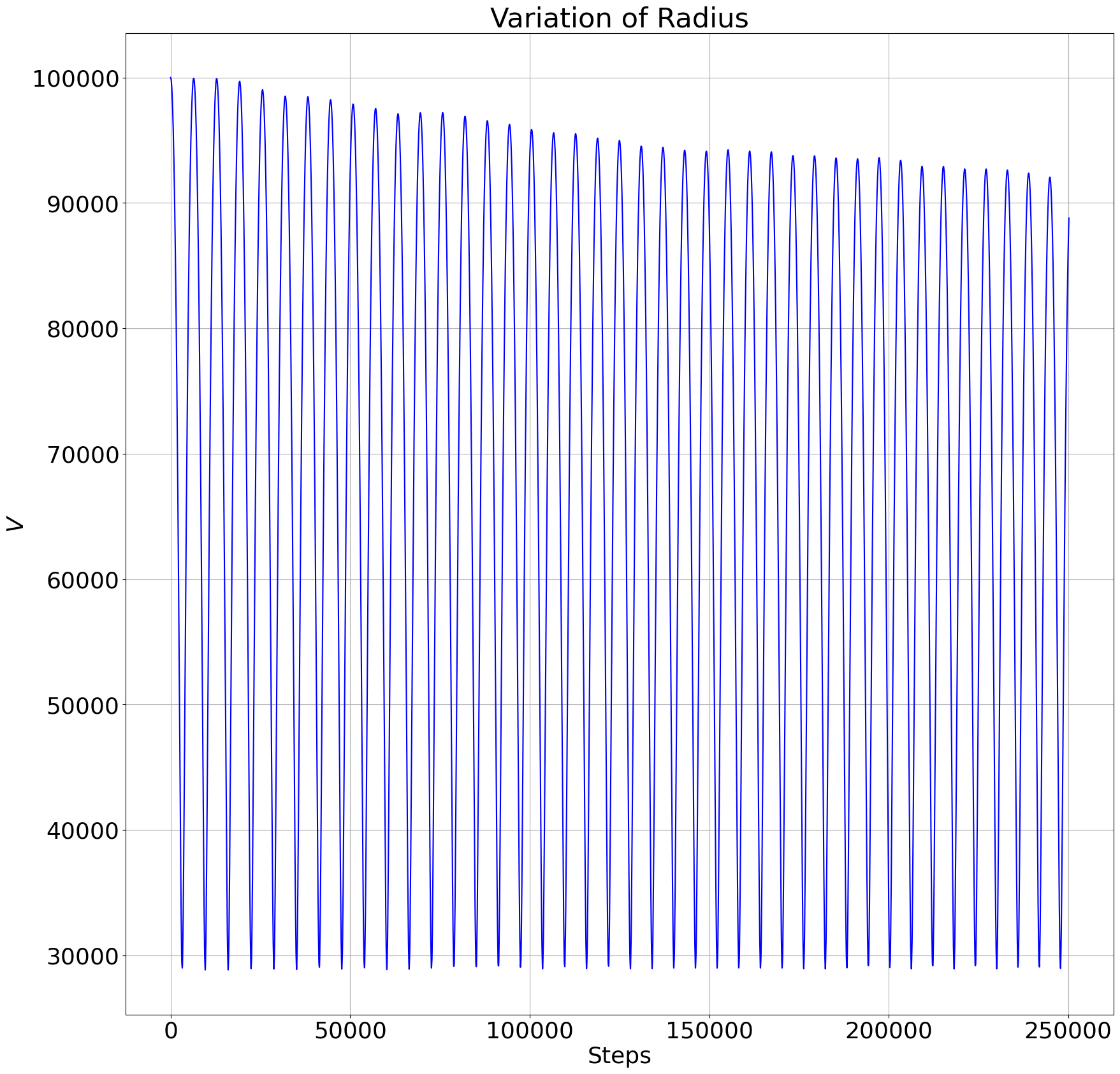}} 
    \subfigure[2 PN-with-Prop]{\includegraphics[width=0.24\textwidth]{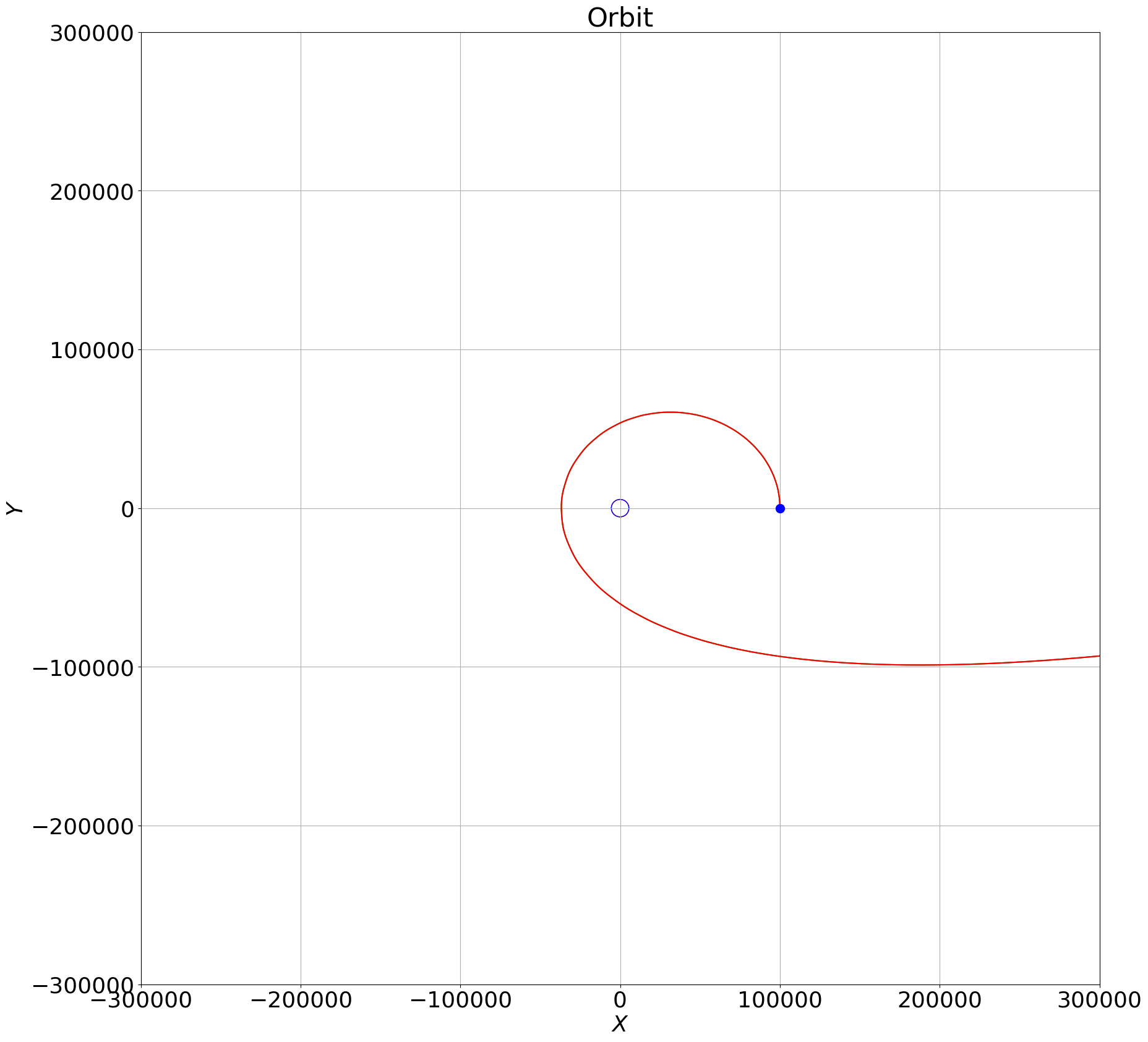}}
    \subfigure[2 PN-with-Prop]{\includegraphics[width=0.24\textwidth]{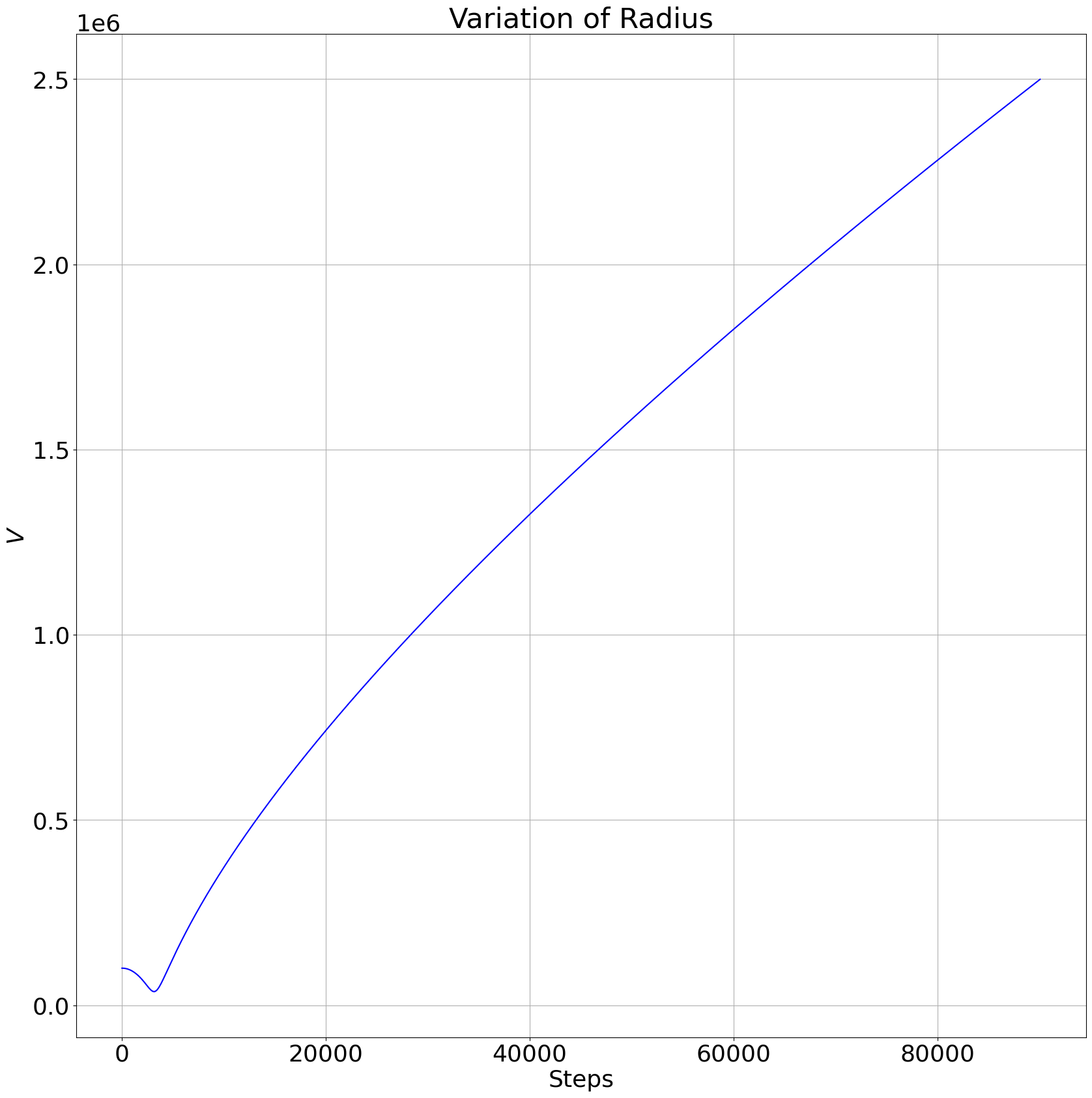}} \\
    
    \subfigure[2.5 PN-no-Prop]{\includegraphics[width=0.24\textwidth]{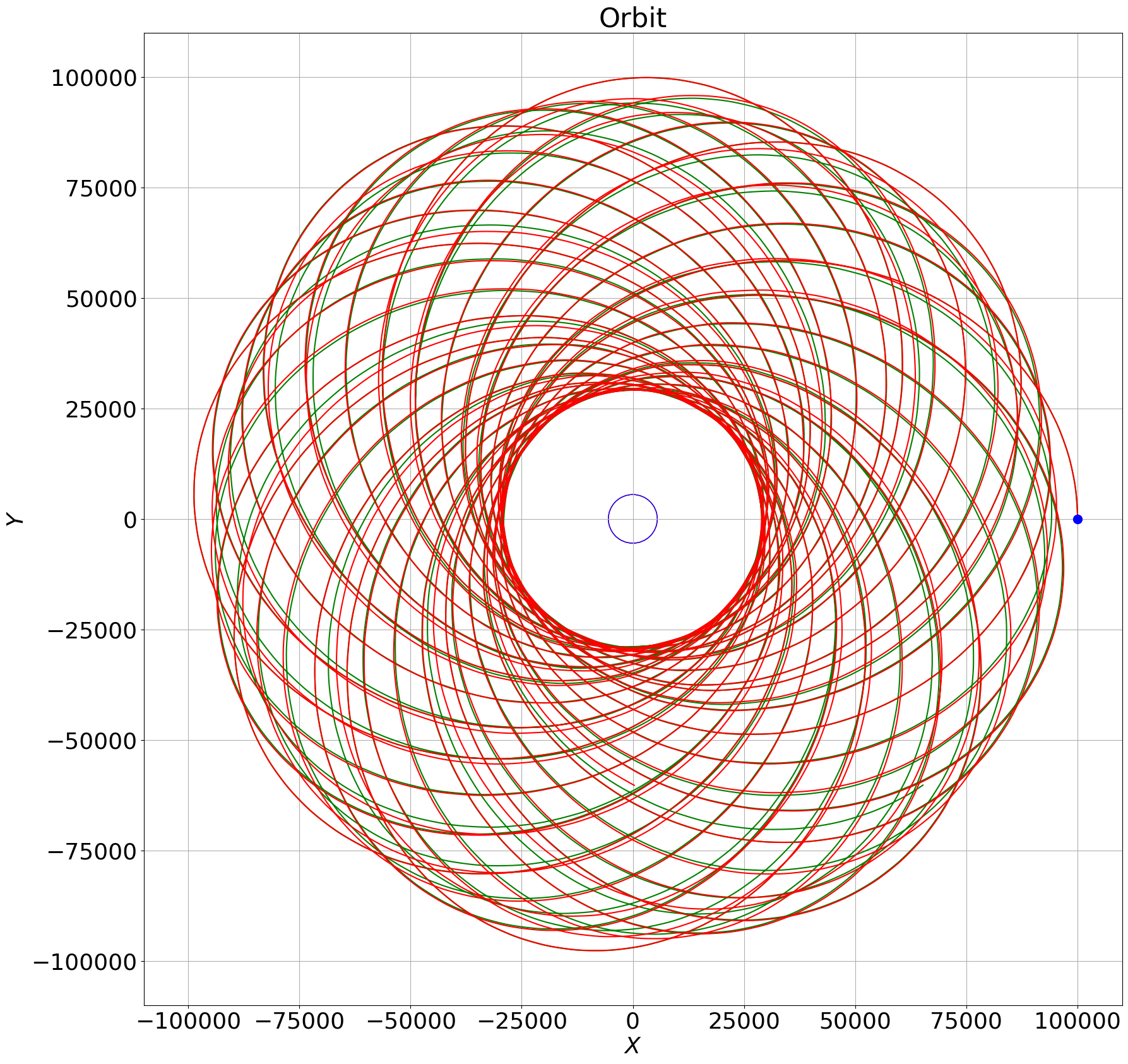}}
    \subfigure[2.5 PN-no-Prop]{\includegraphics[width=0.24\textwidth]{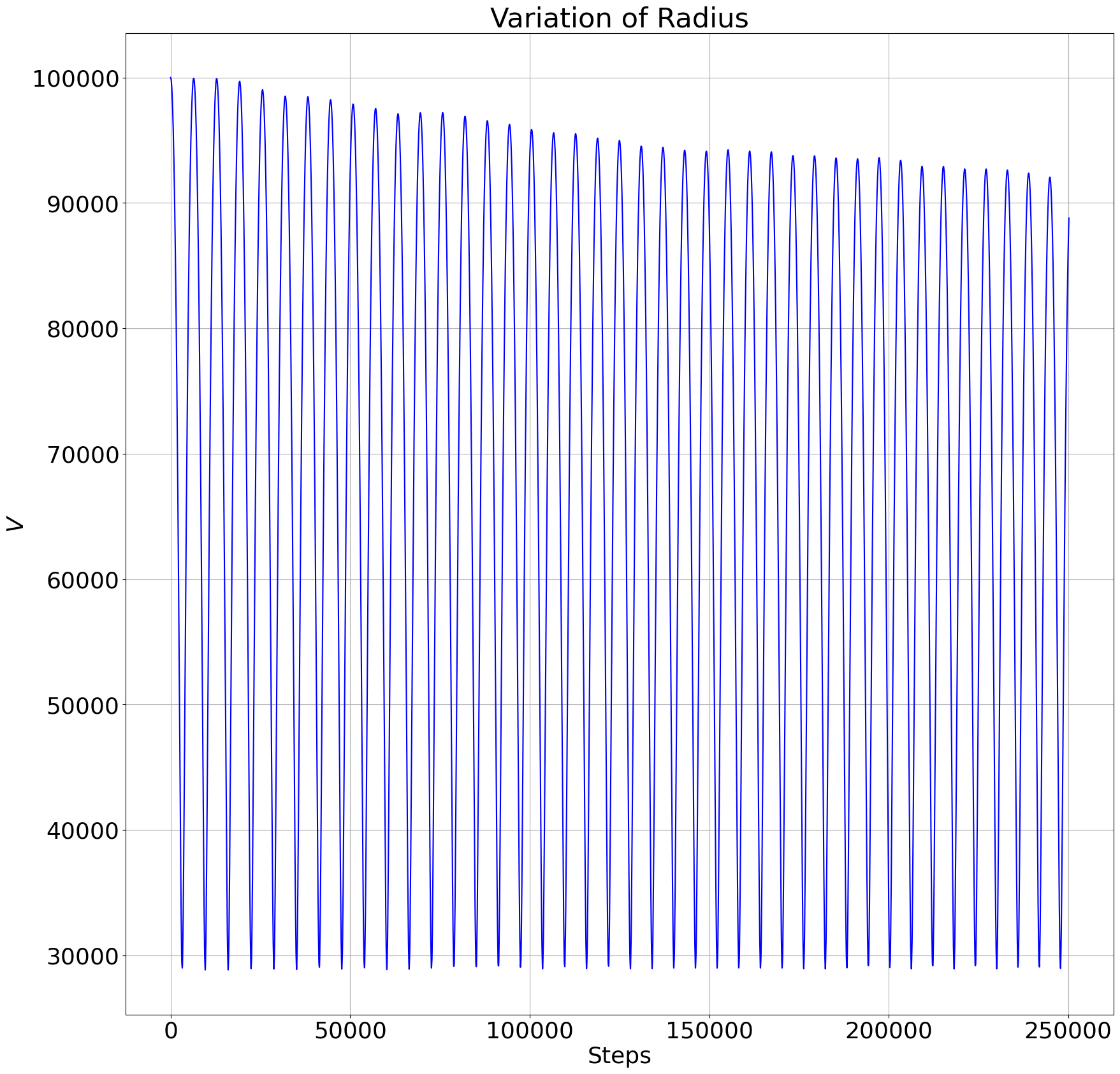}} 
    \subfigure[2.5 PN-with-Prop]{\includegraphics[width=0.24\textwidth]{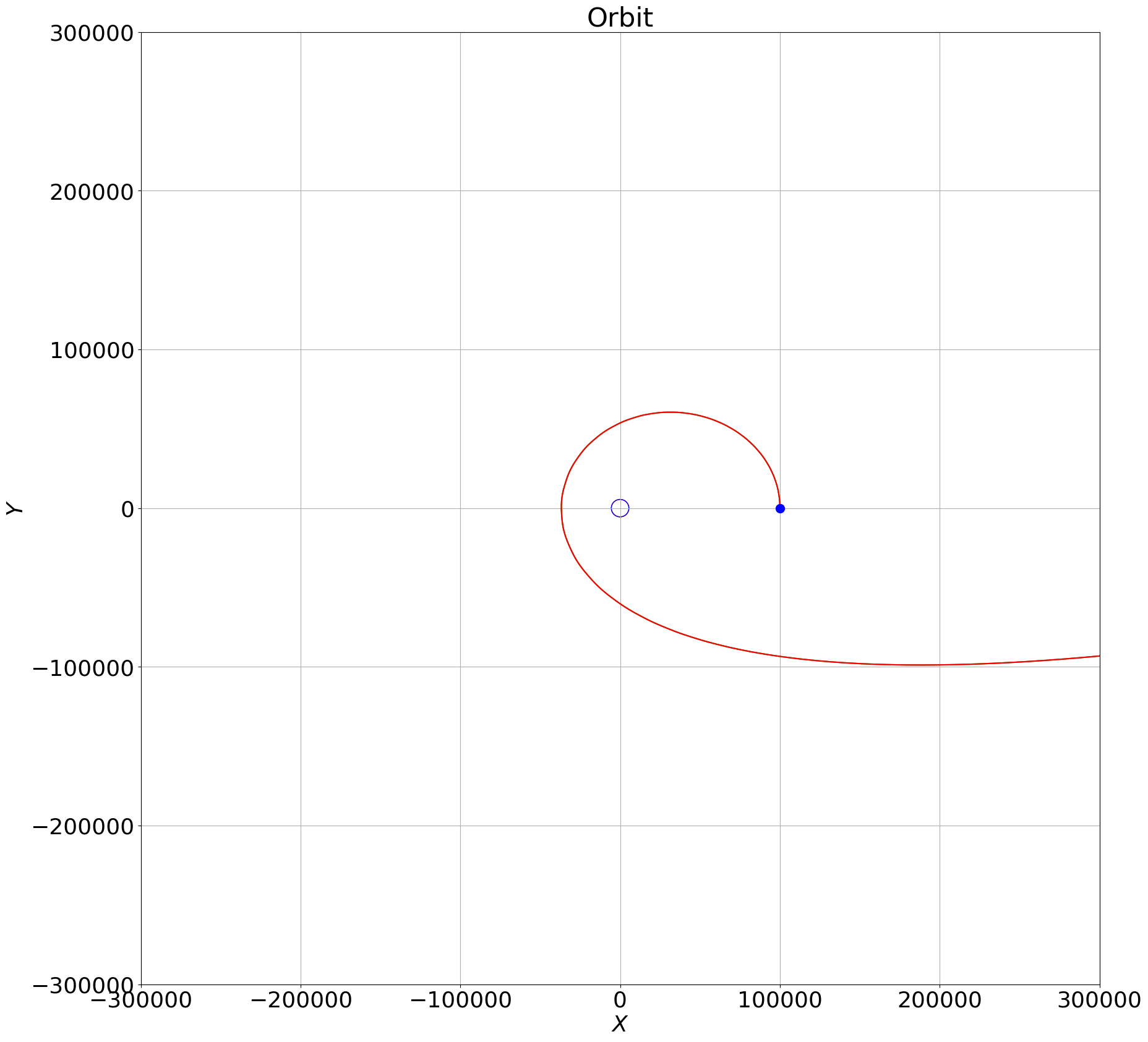}}
    \subfigure[2.5 PN-with-Prop]{\includegraphics[width=0.24\textwidth]{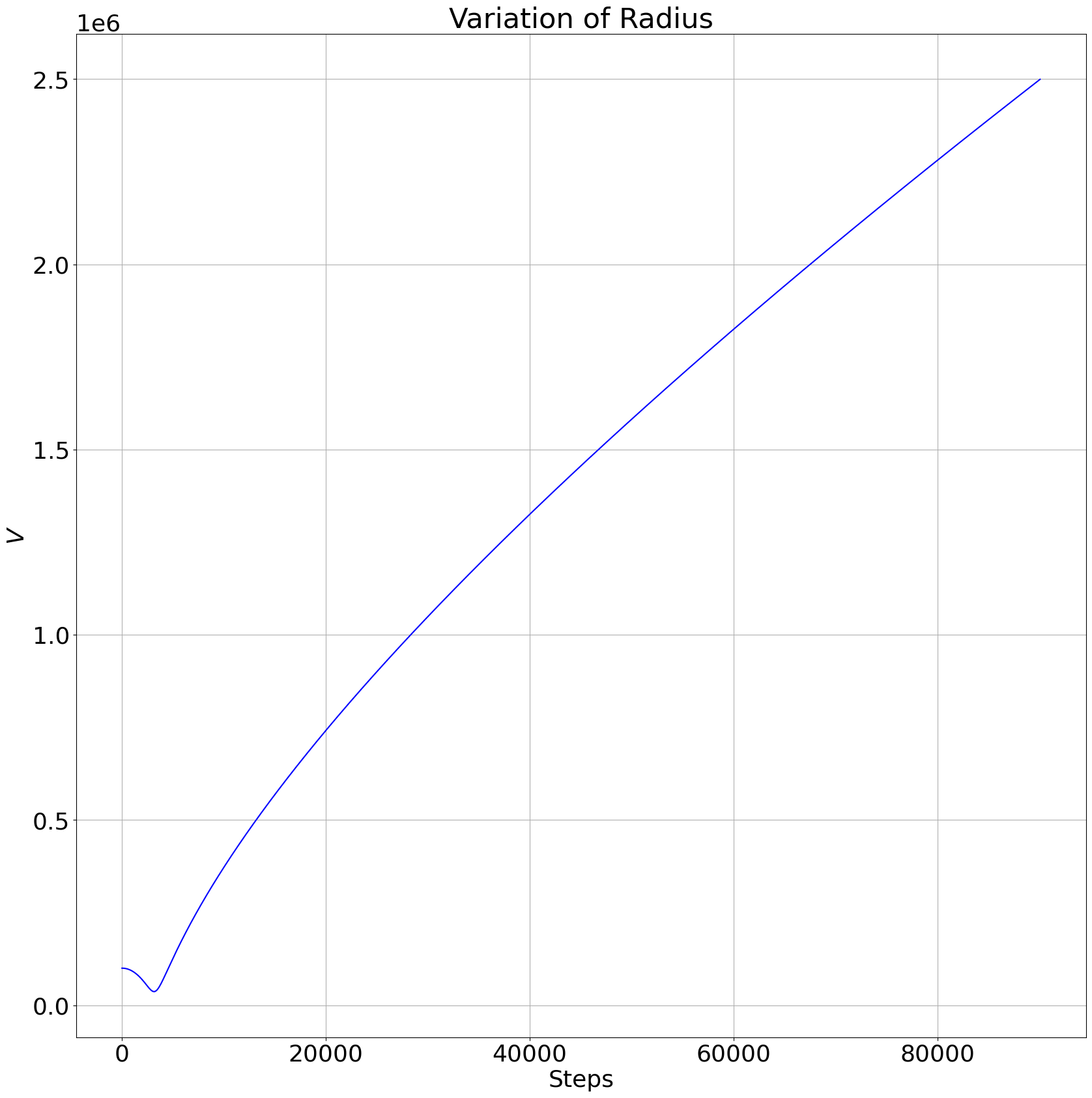}} \\
    \caption{Strong(er) Field, R = $1.1 \cdot 10^4m$, X = $10^5m$, Velocity = 35,000,000$ms^{-1}$}
    \label{results4}
\end{figure}

\begin{figure}[!ht]
    \centering
    \setcounter{subfigure}{0}
    \subfigure[0 PN-no-Prop]{\includegraphics[width=0.24\textwidth]{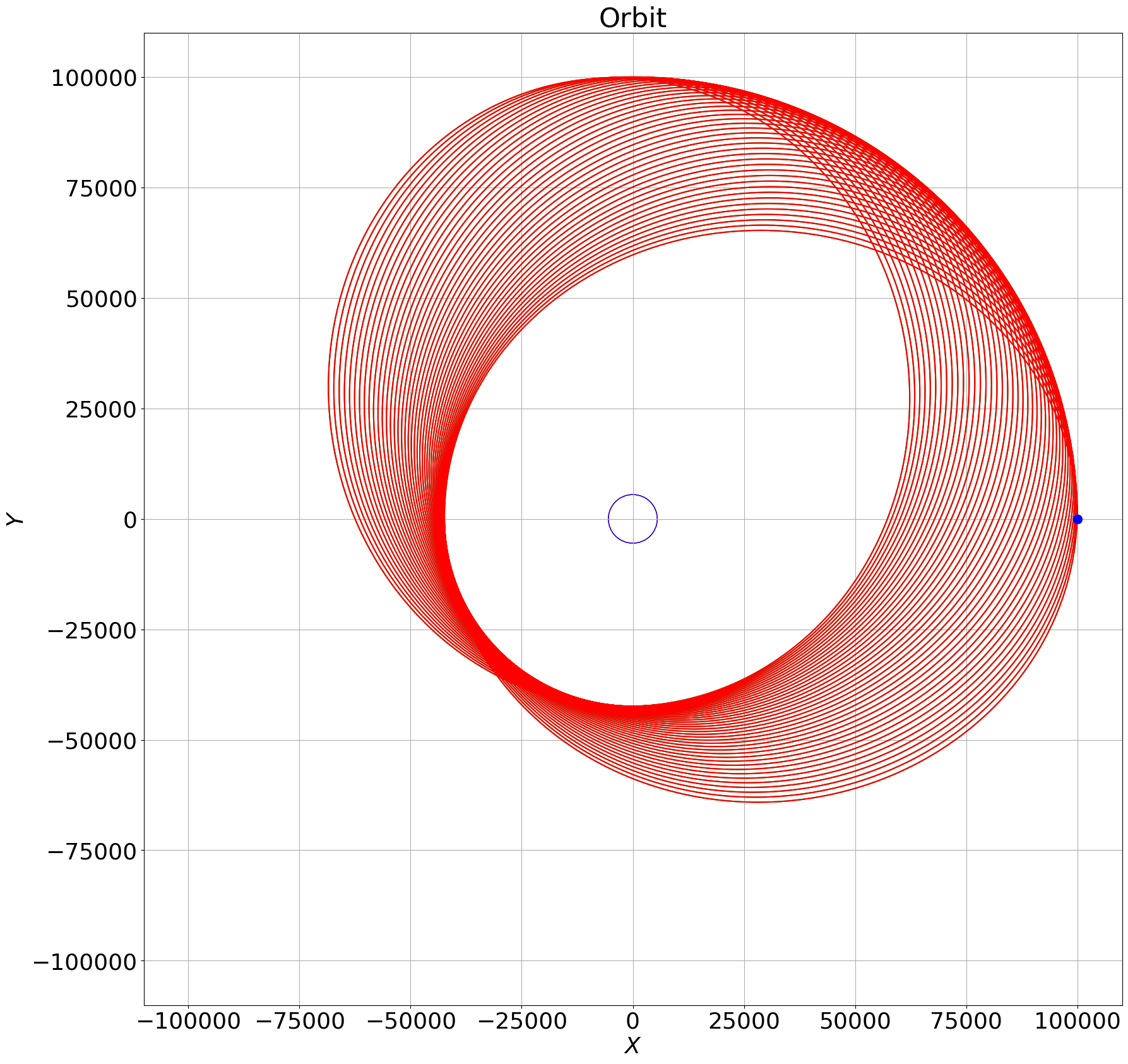}}
    \subfigure[0 PN-no-Prop]{\includegraphics[width=0.24\textwidth]{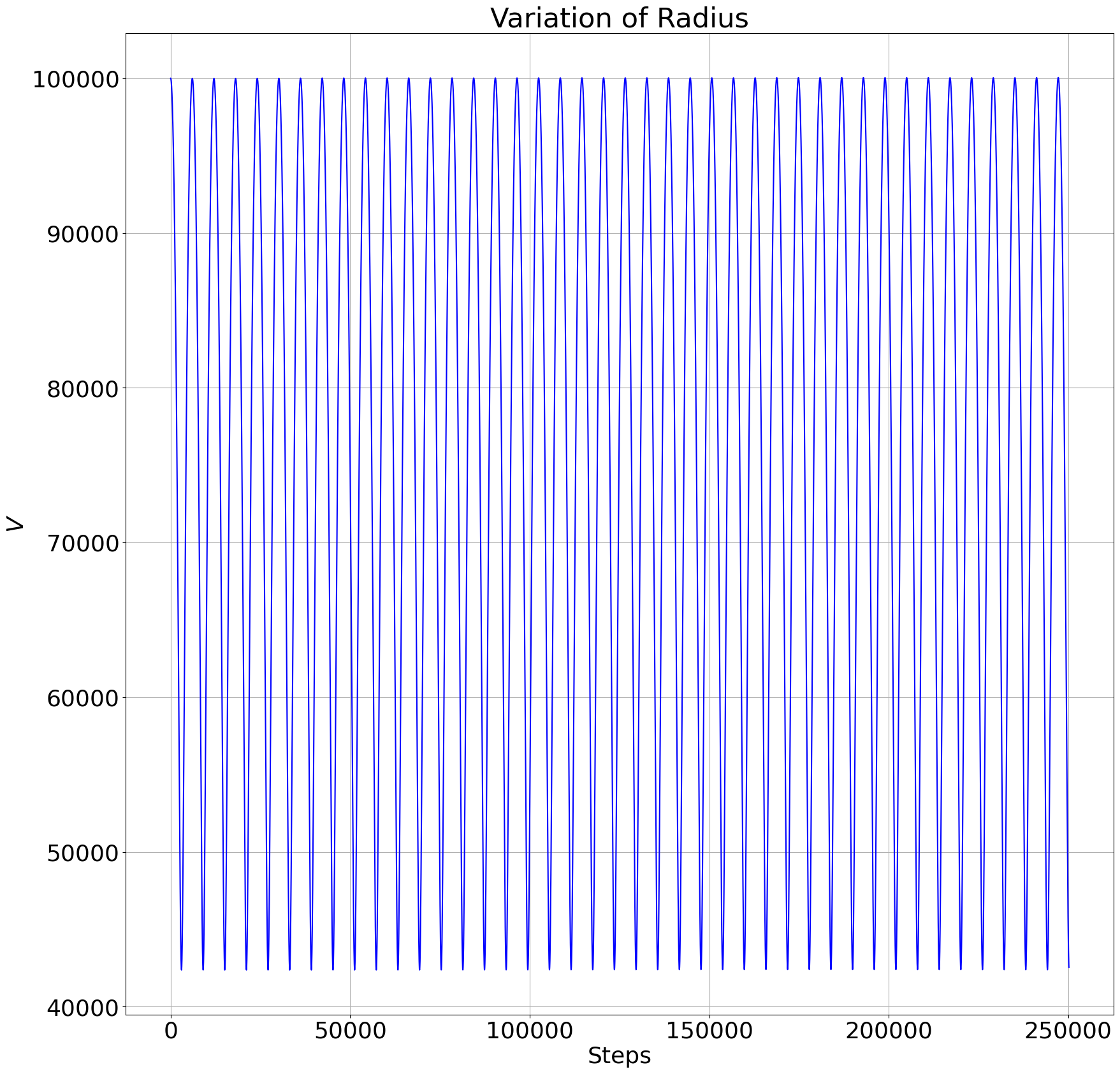}} 
    \subfigure[0 PN-with-Prop]{\includegraphics[width=0.24\textwidth]{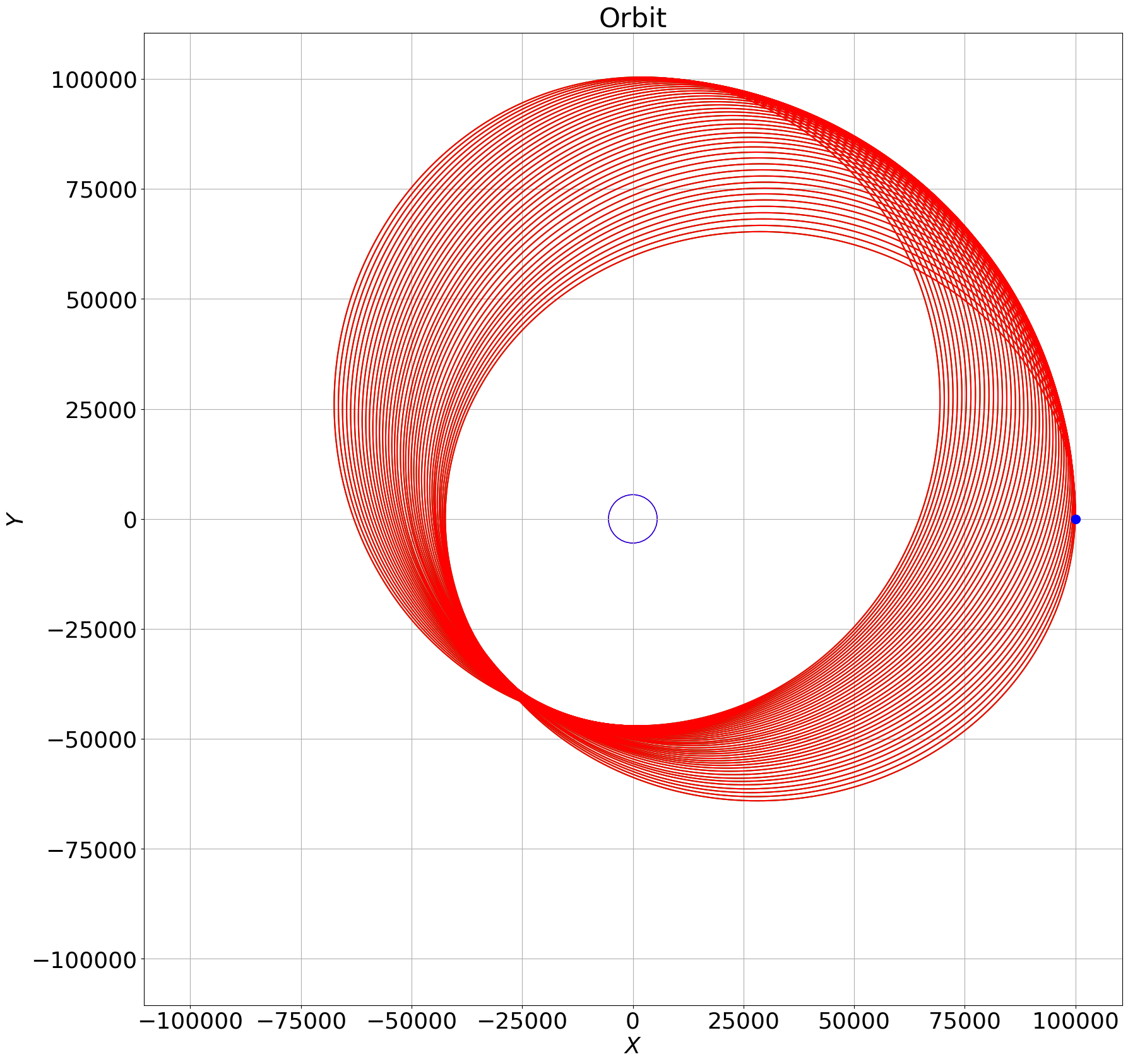}}
    \subfigure[0 PN-with-Prop]{\includegraphics[width=0.24\textwidth]{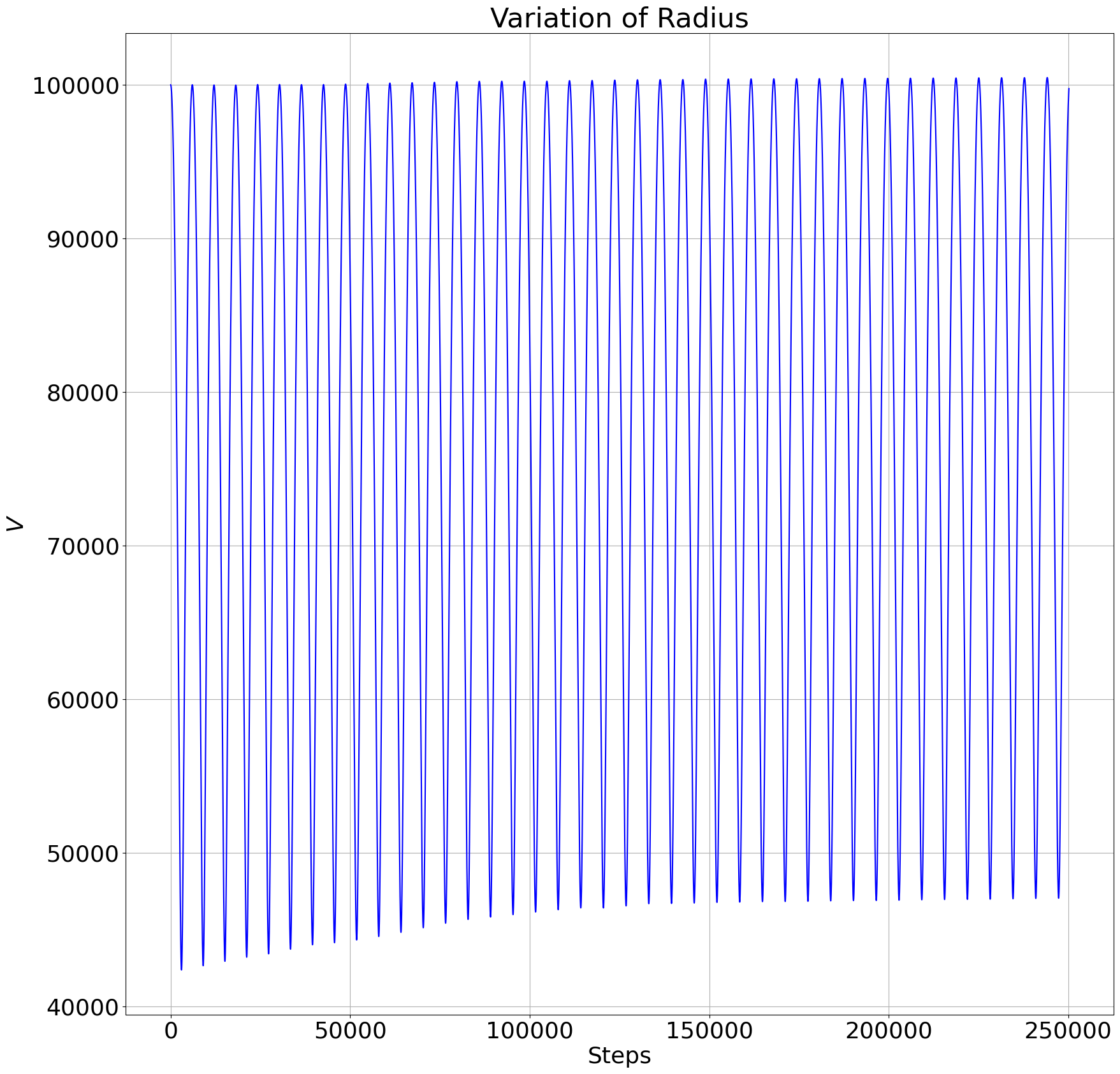}} \\
    
    \subfigure[0.5 PN-no-Prop]{\includegraphics[width=0.24\textwidth]{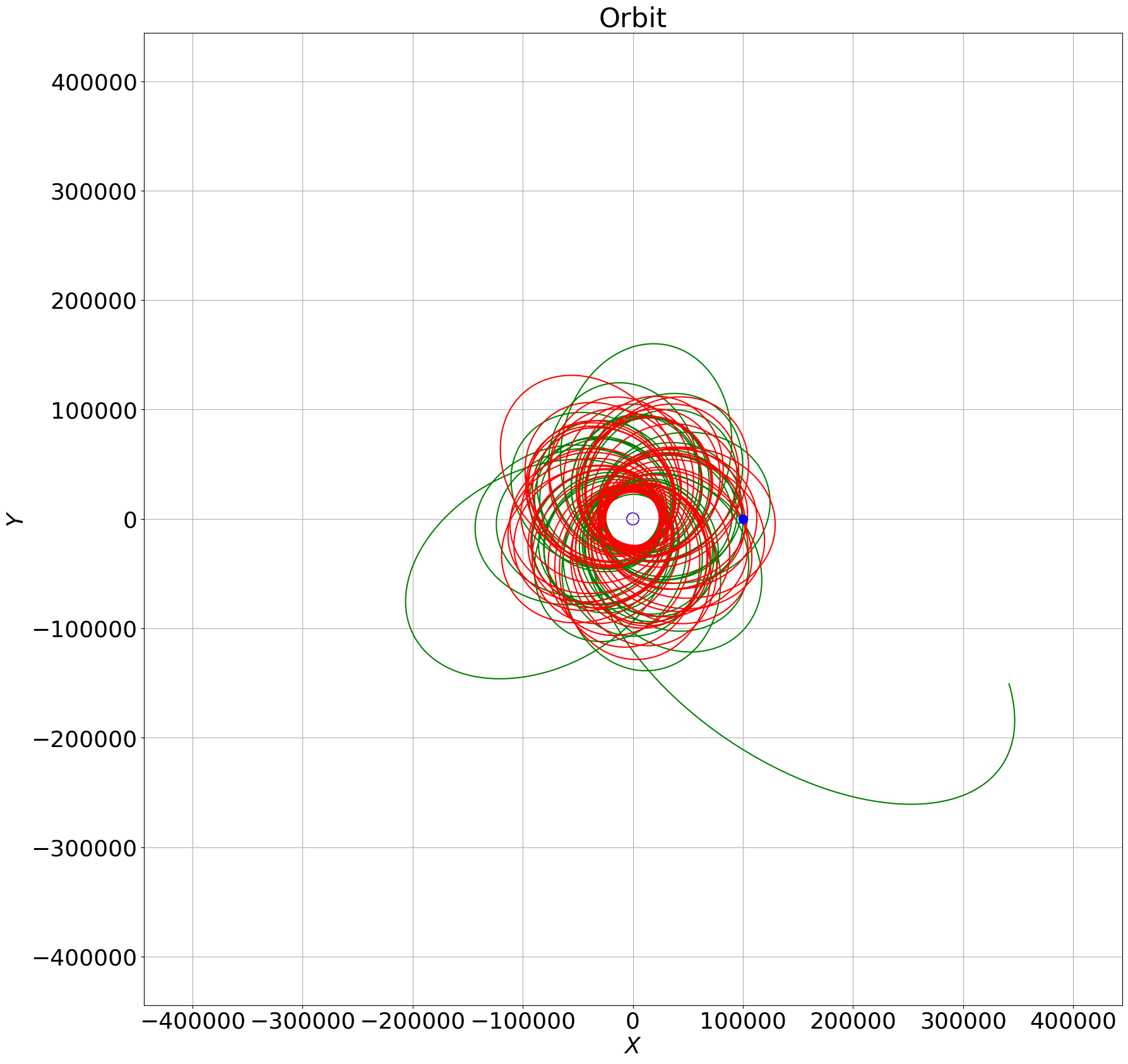}}
    \subfigure[0.5 PN-no-Prop]{\includegraphics[width=0.24\textwidth]{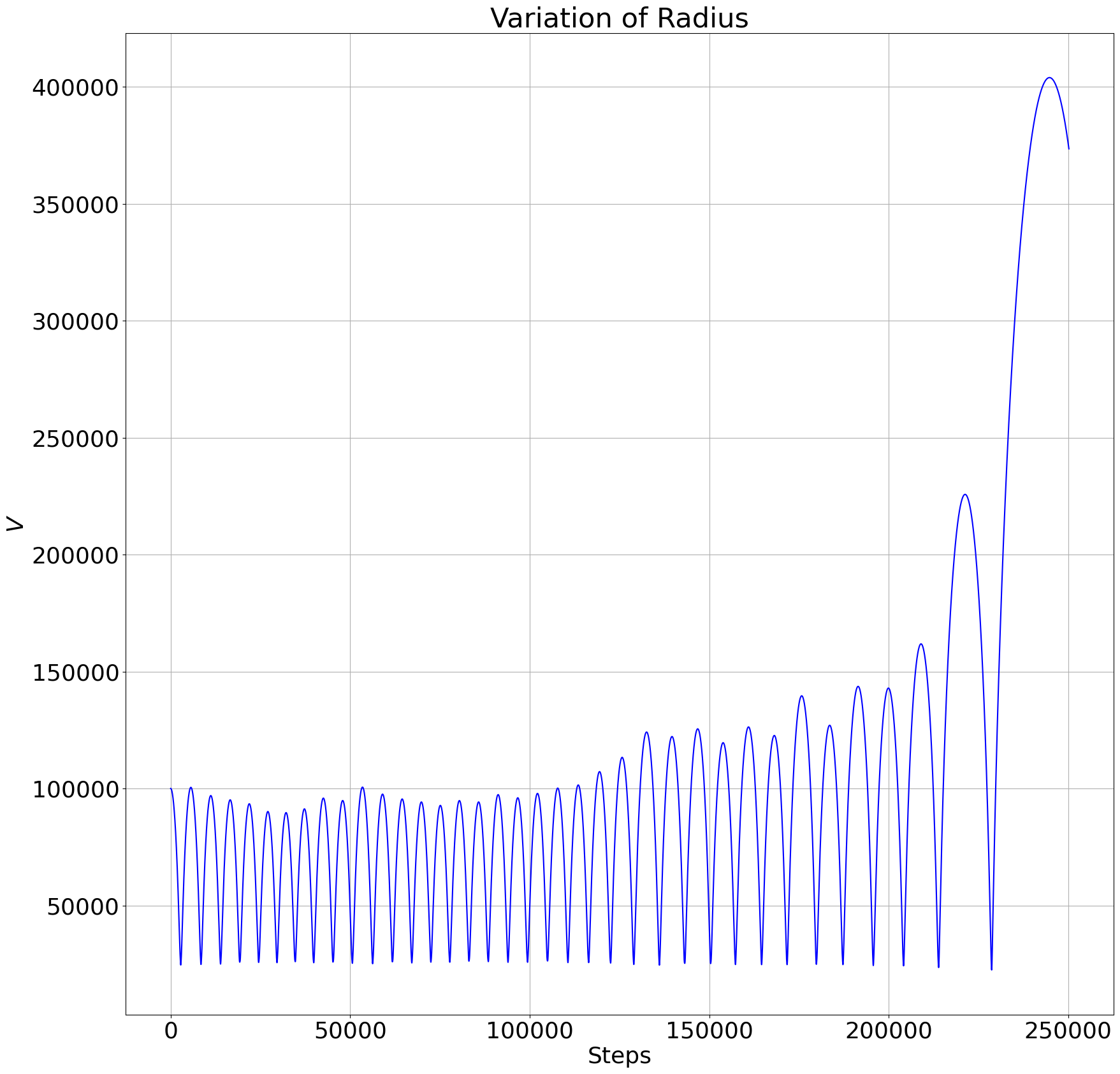}} 
    \subfigure[0.5 PN-with-Prop]{\includegraphics[width=0.24\textwidth]{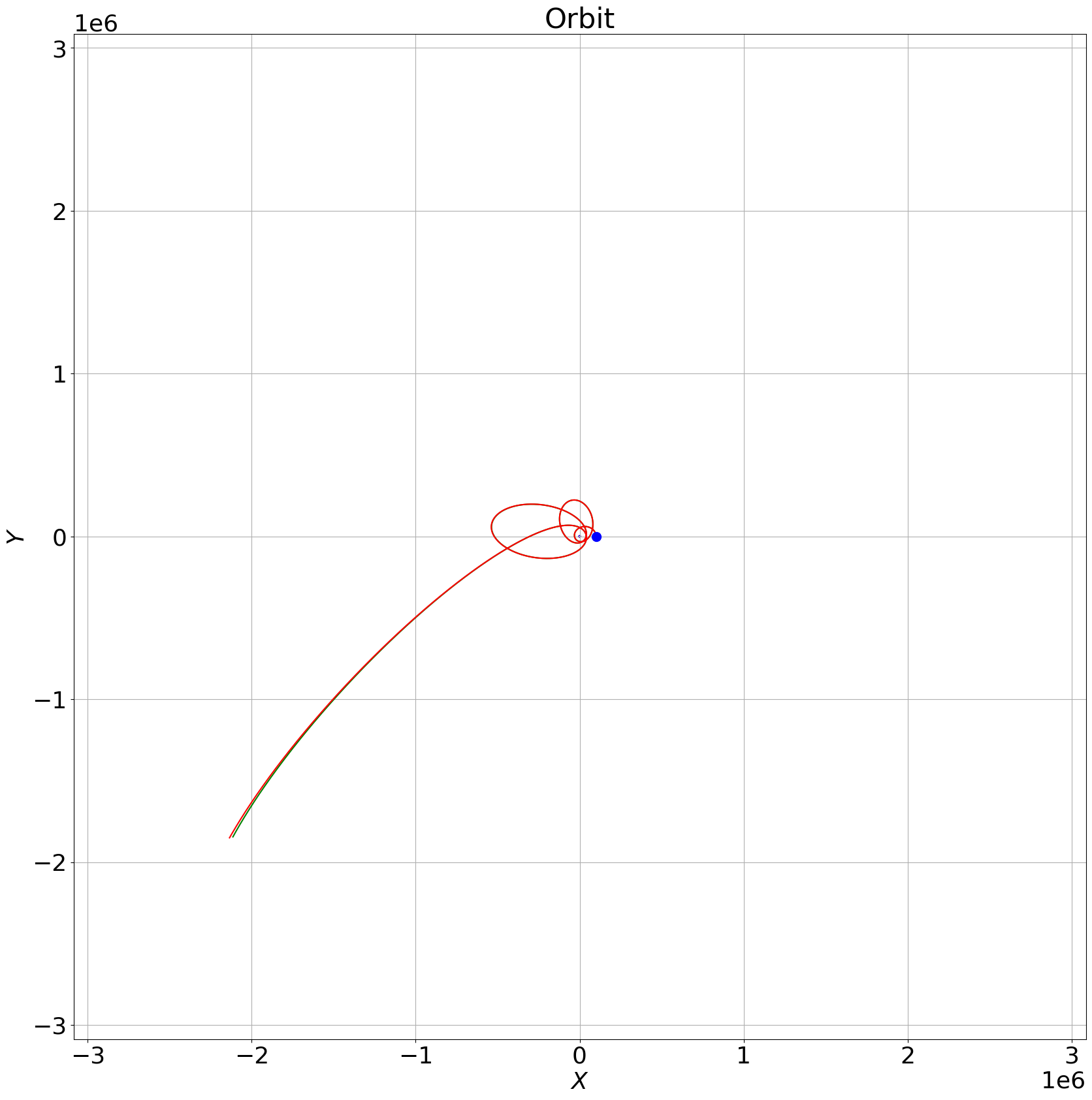}}
    \subfigure[0.5 PN-with-Prop]{\includegraphics[width=0.24\textwidth]{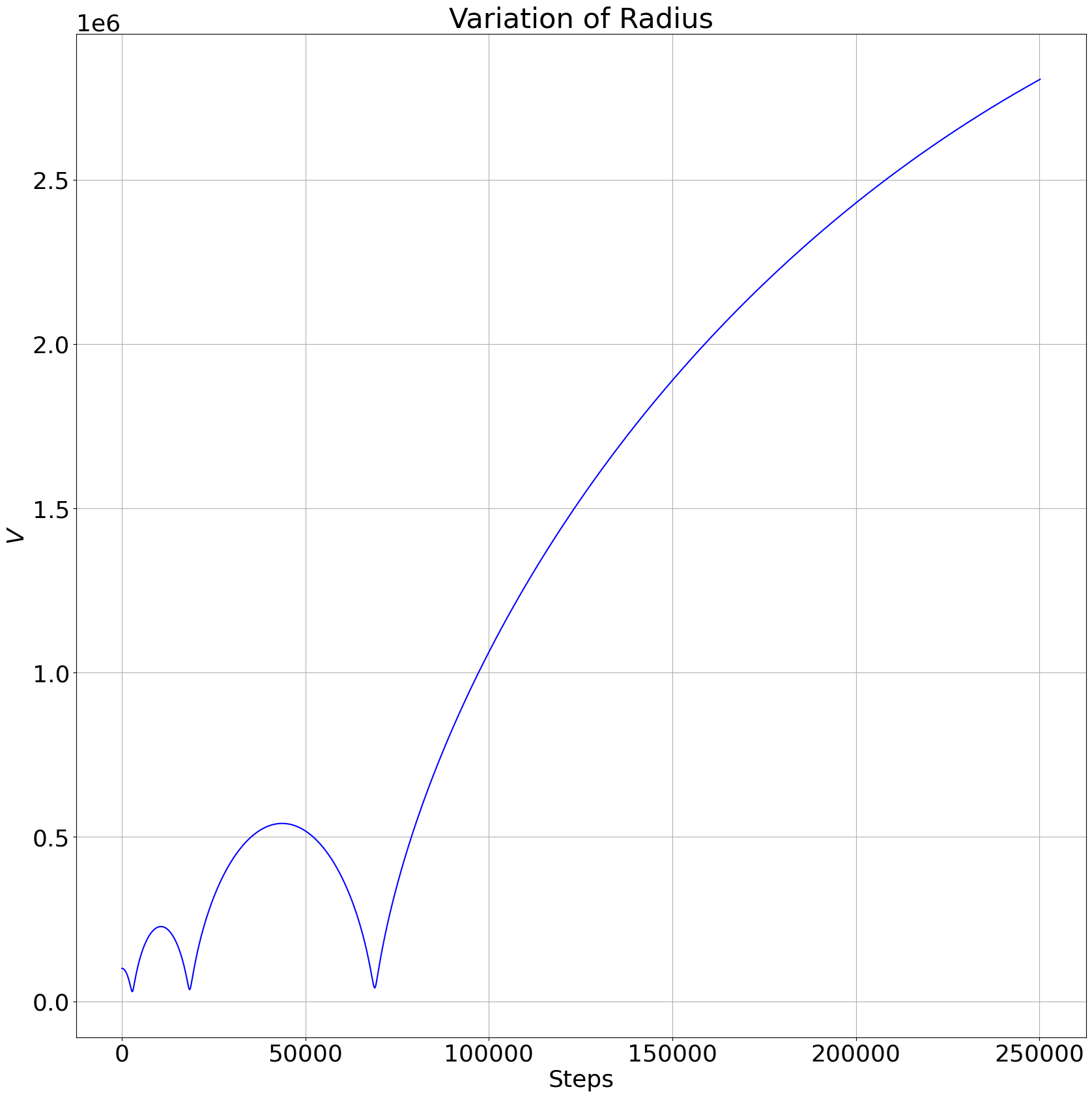}} \\
    
    \subfigure[1 PN-no-Prop]{\includegraphics[width=0.24\textwidth]{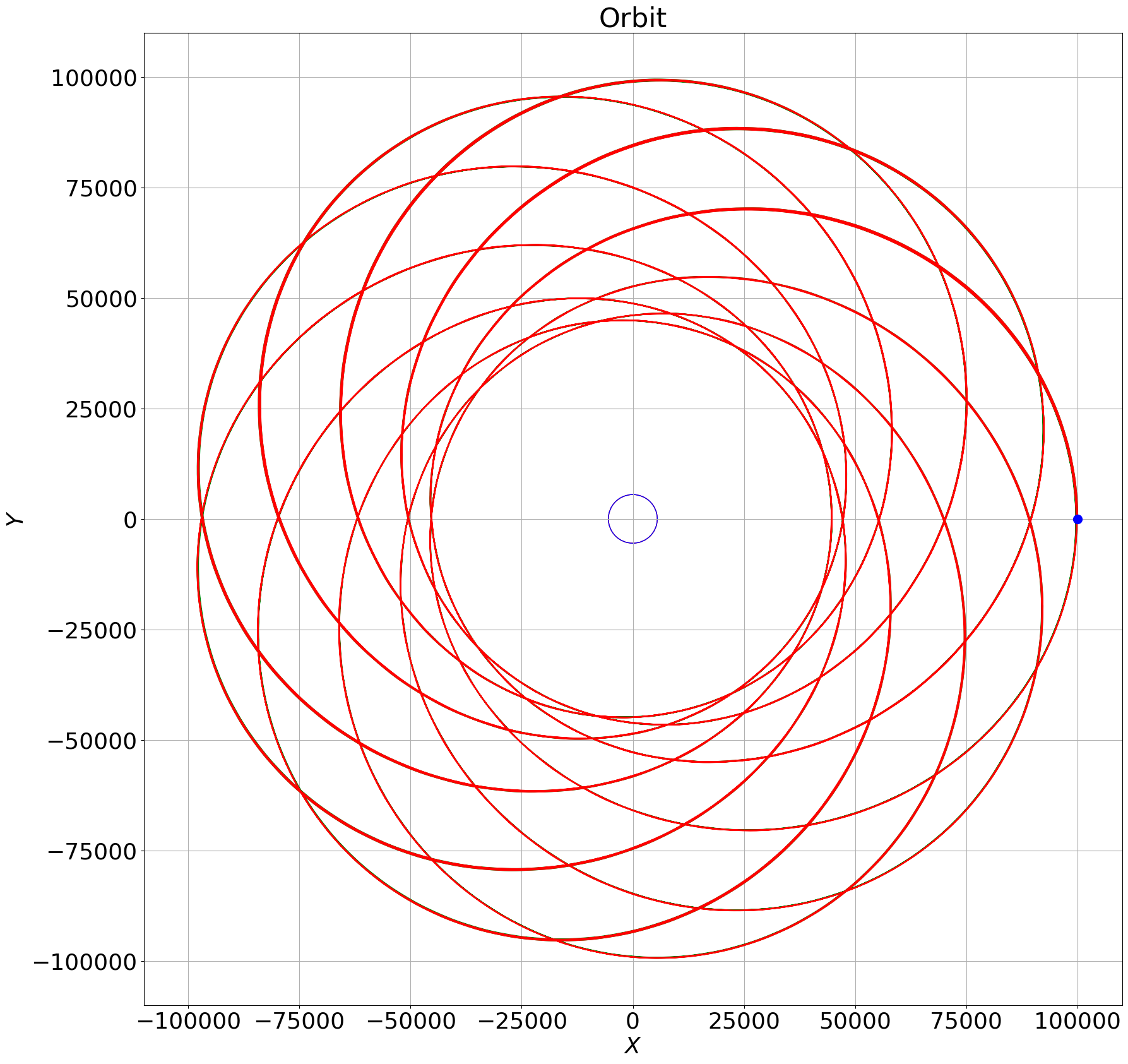}}
    \subfigure[1 PN-no-Prop]{\includegraphics[width=0.24\textwidth]{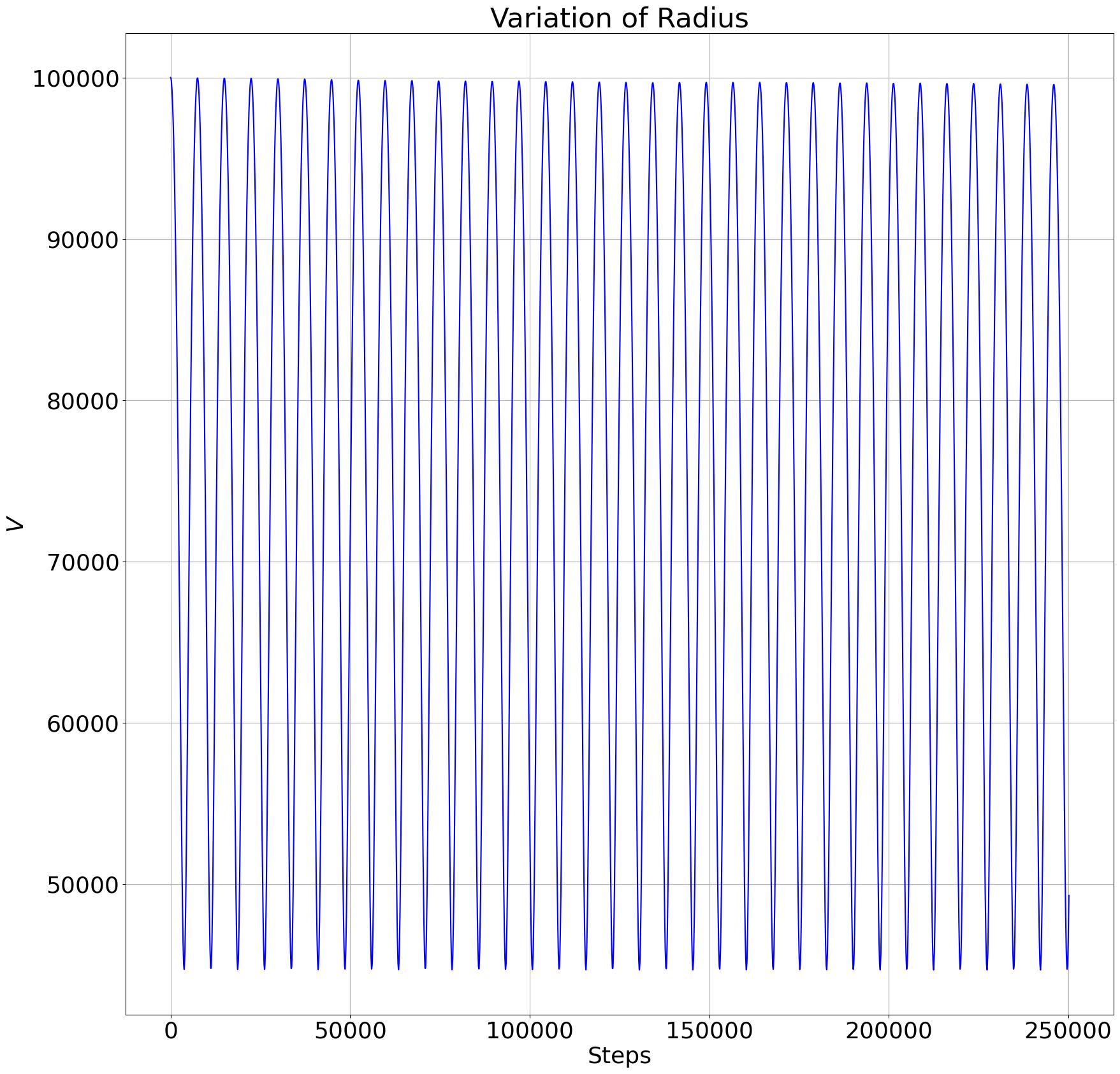}} 
    \subfigure[1 PN-with-Prop]{\includegraphics[width=0.24\textwidth]{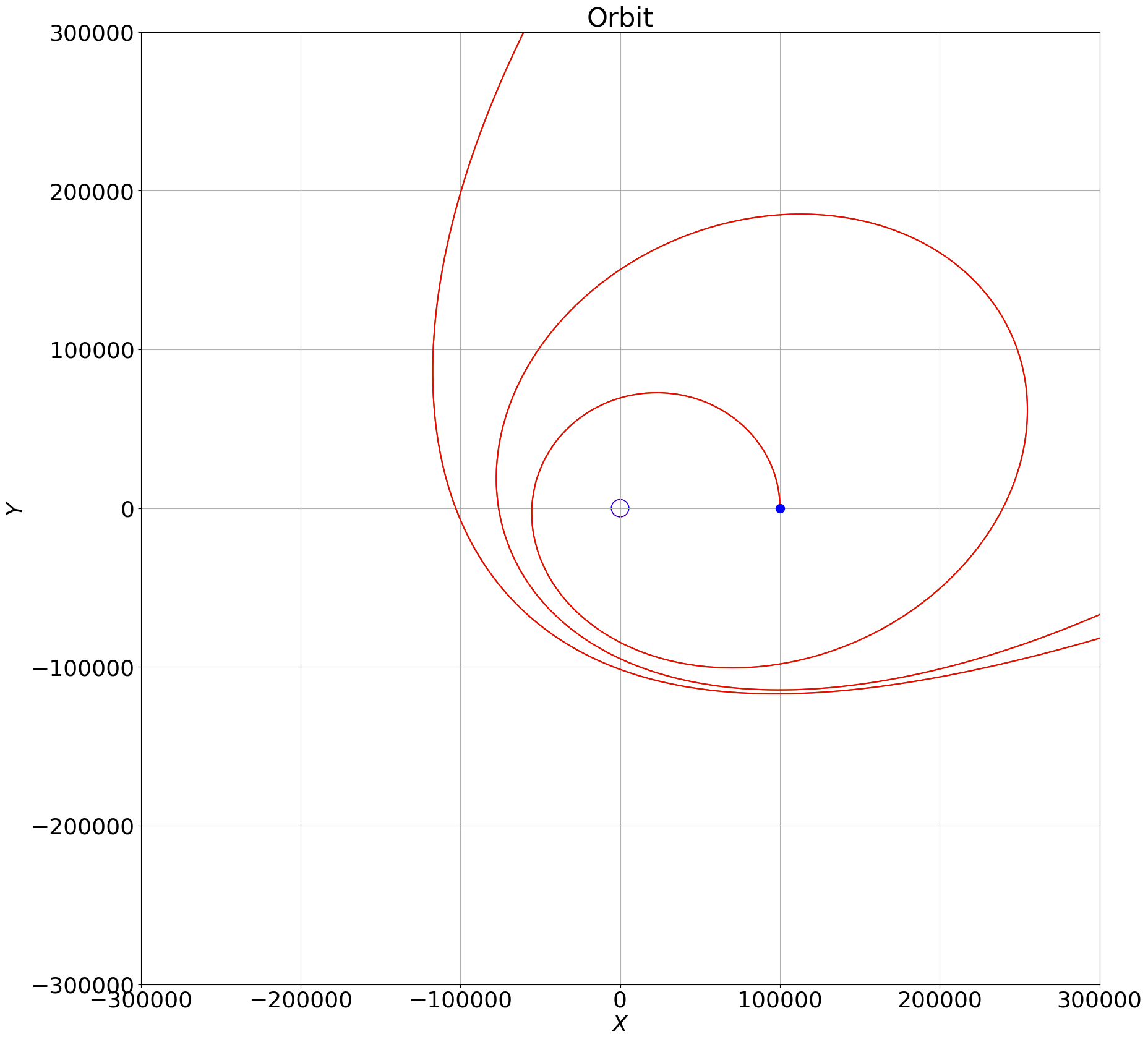}}
    \subfigure[1 PN-with-Prop]{\includegraphics[width=0.24\textwidth]{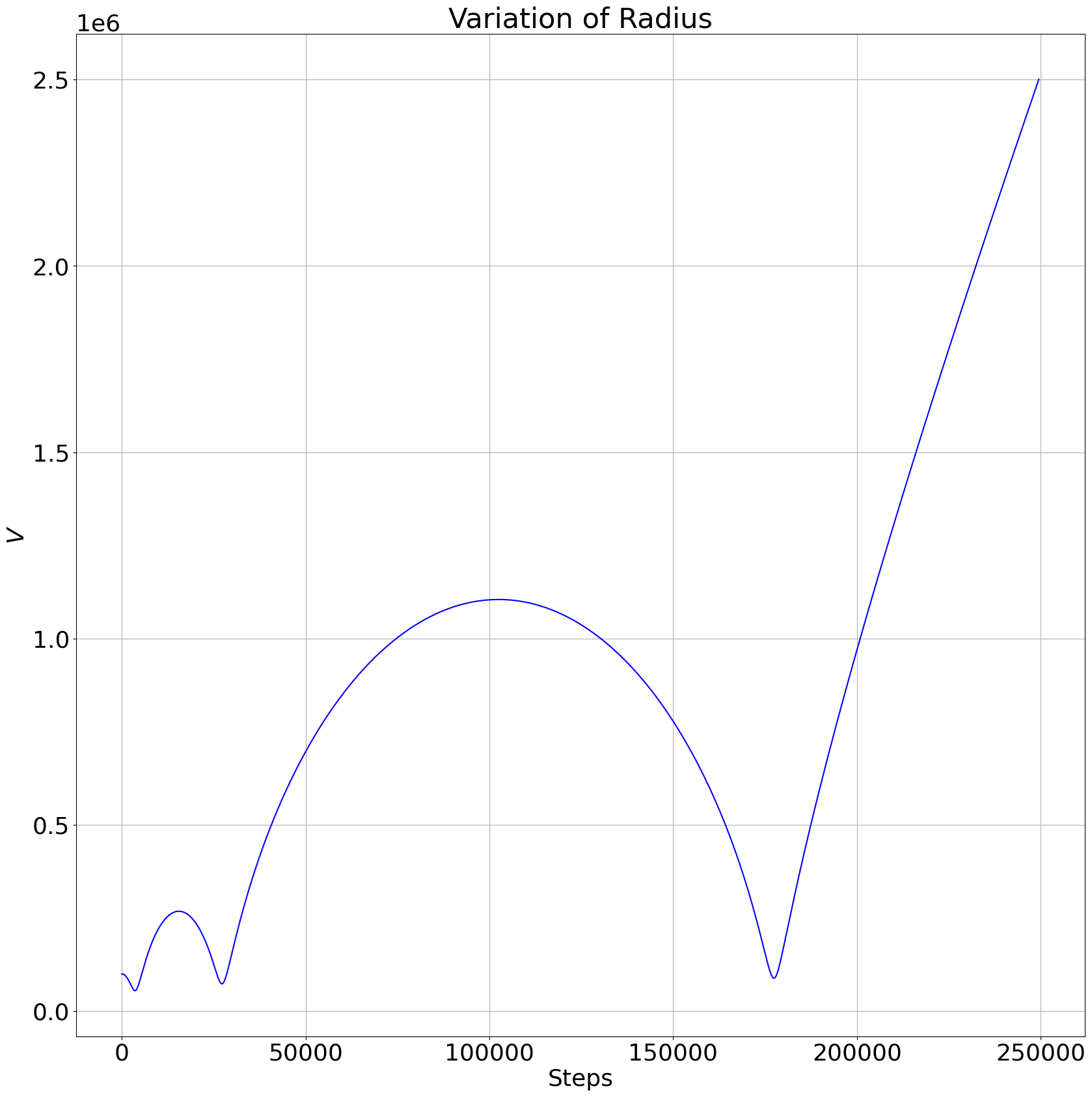}} \\
    
    \subfigure[2 PN-no-Prop]{\includegraphics[width=0.24\textwidth]{2PN-NF-R4-x5-Vel40000000.0-theta.png}}
    \subfigure[2 PN-no-Prop]{\includegraphics[width=0.24\textwidth]{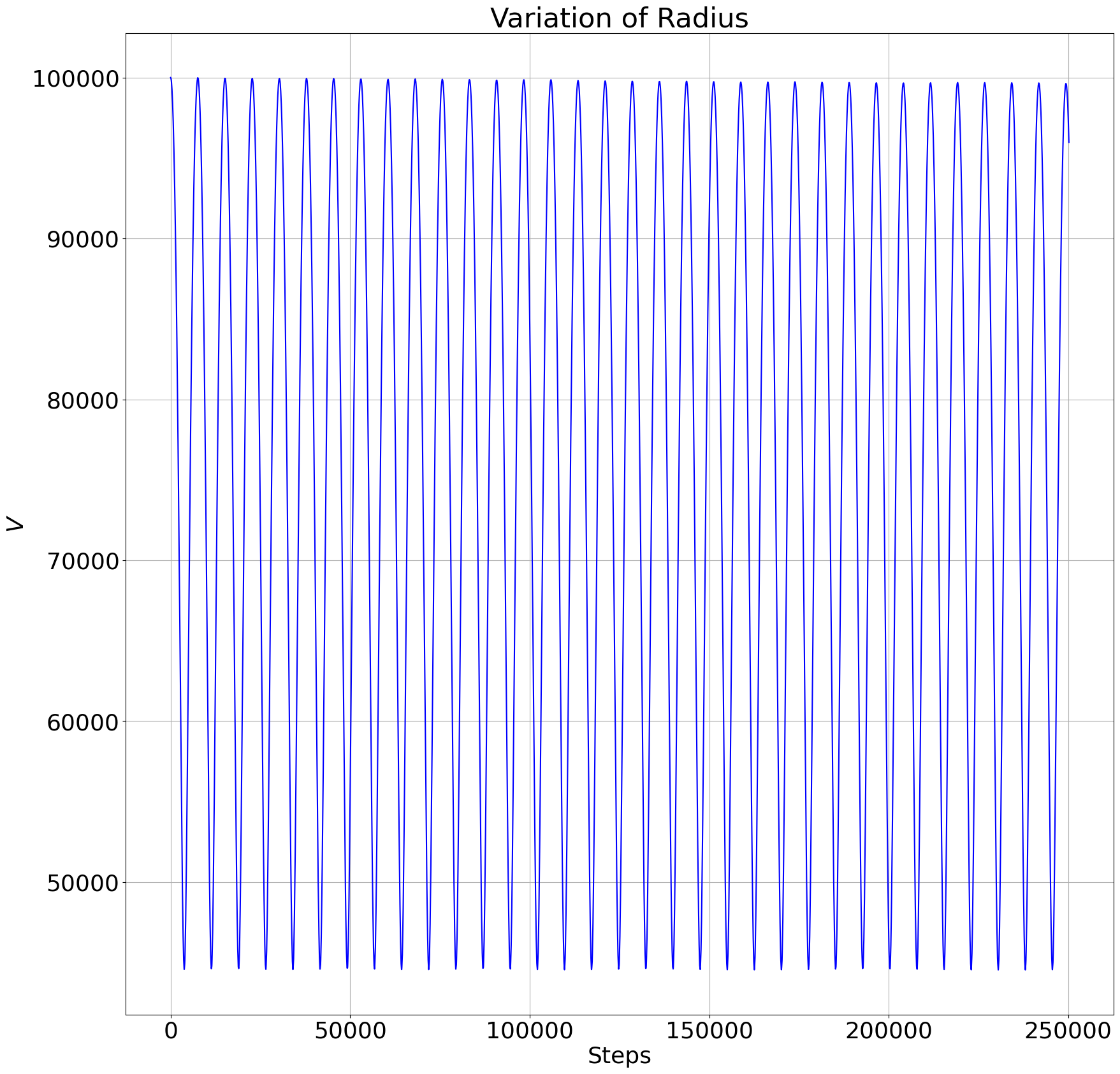}} 
    \subfigure[2 PN-with-Prop]{\includegraphics[width=0.24\textwidth]{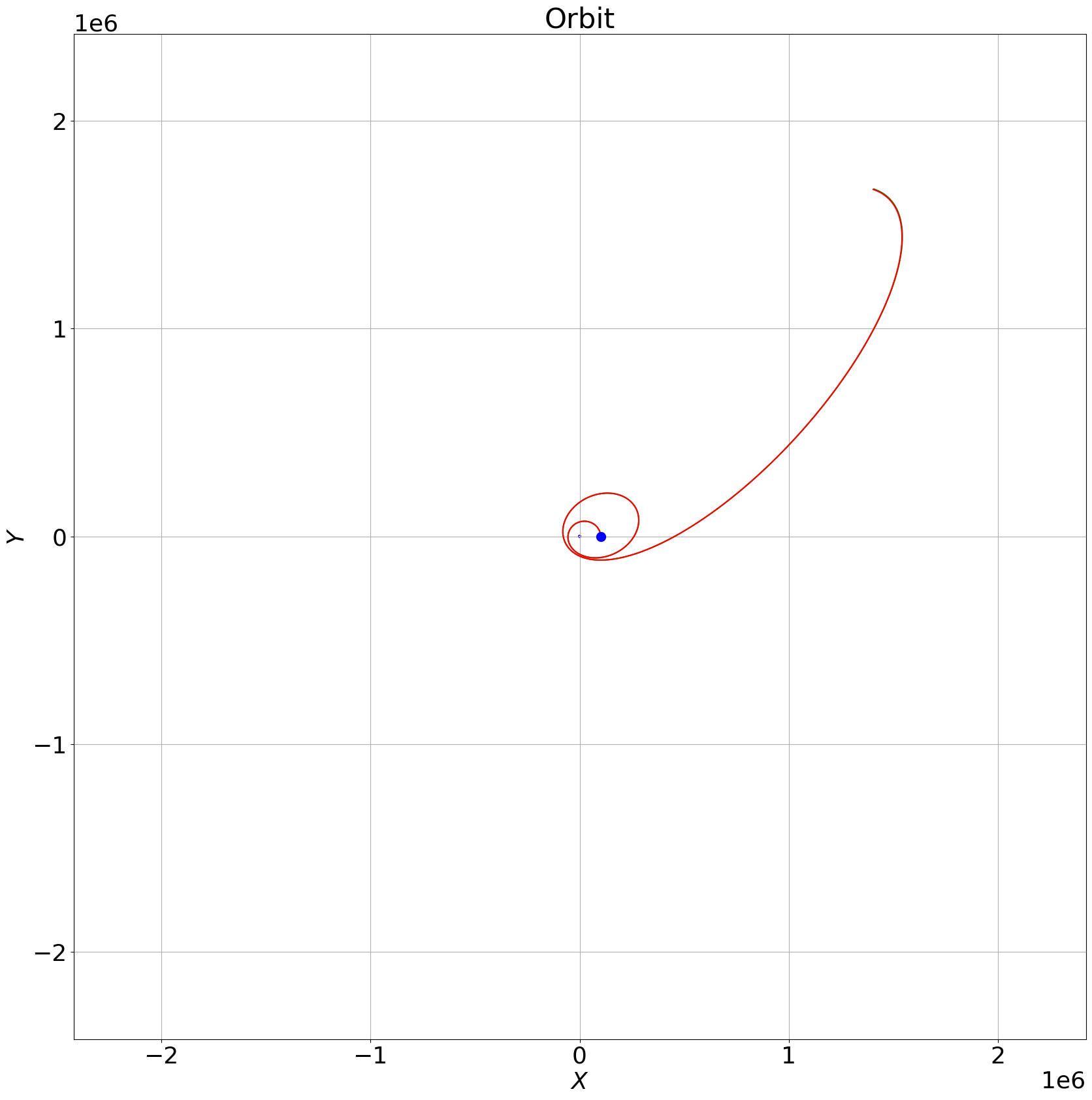}}
    \subfigure[2 PN-with-Prop]{\includegraphics[width=0.24\textwidth]{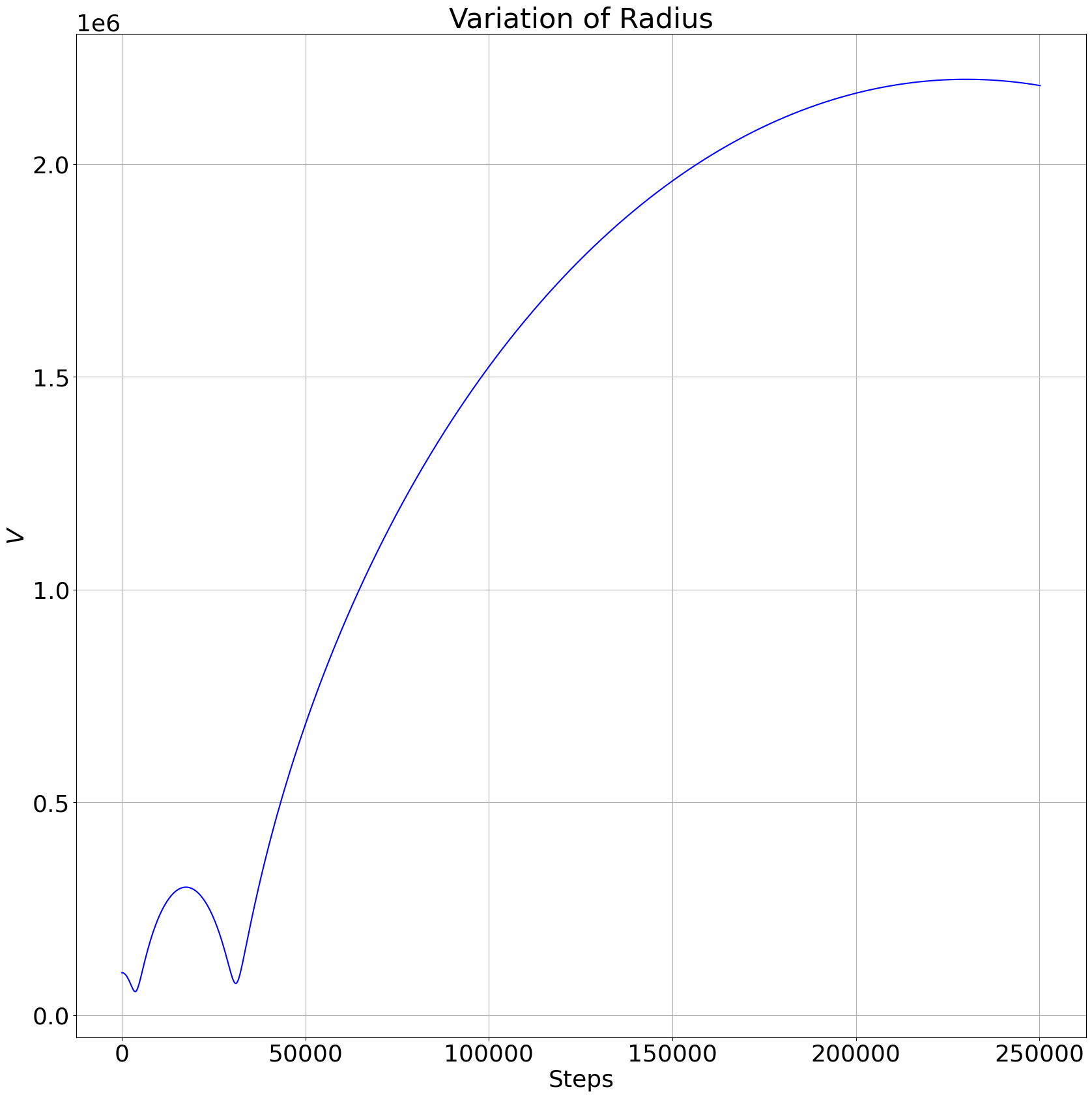}} \\
    
    \subfigure[2.5 PN-no-Prop]{\includegraphics[width=0.24\textwidth]{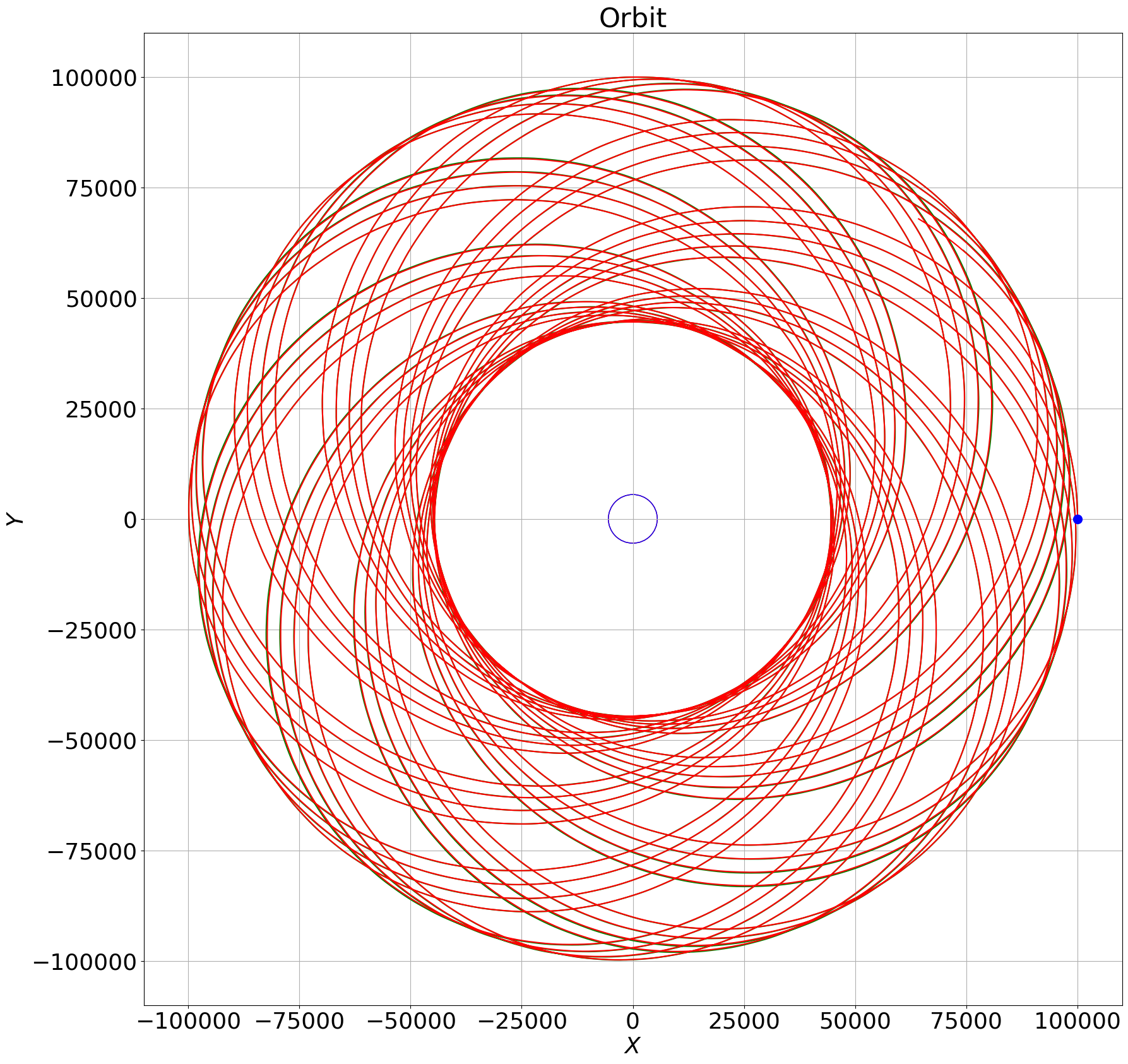}}
    \subfigure[2.5 PN-no-Prop]{\includegraphics[width=0.24\textwidth]{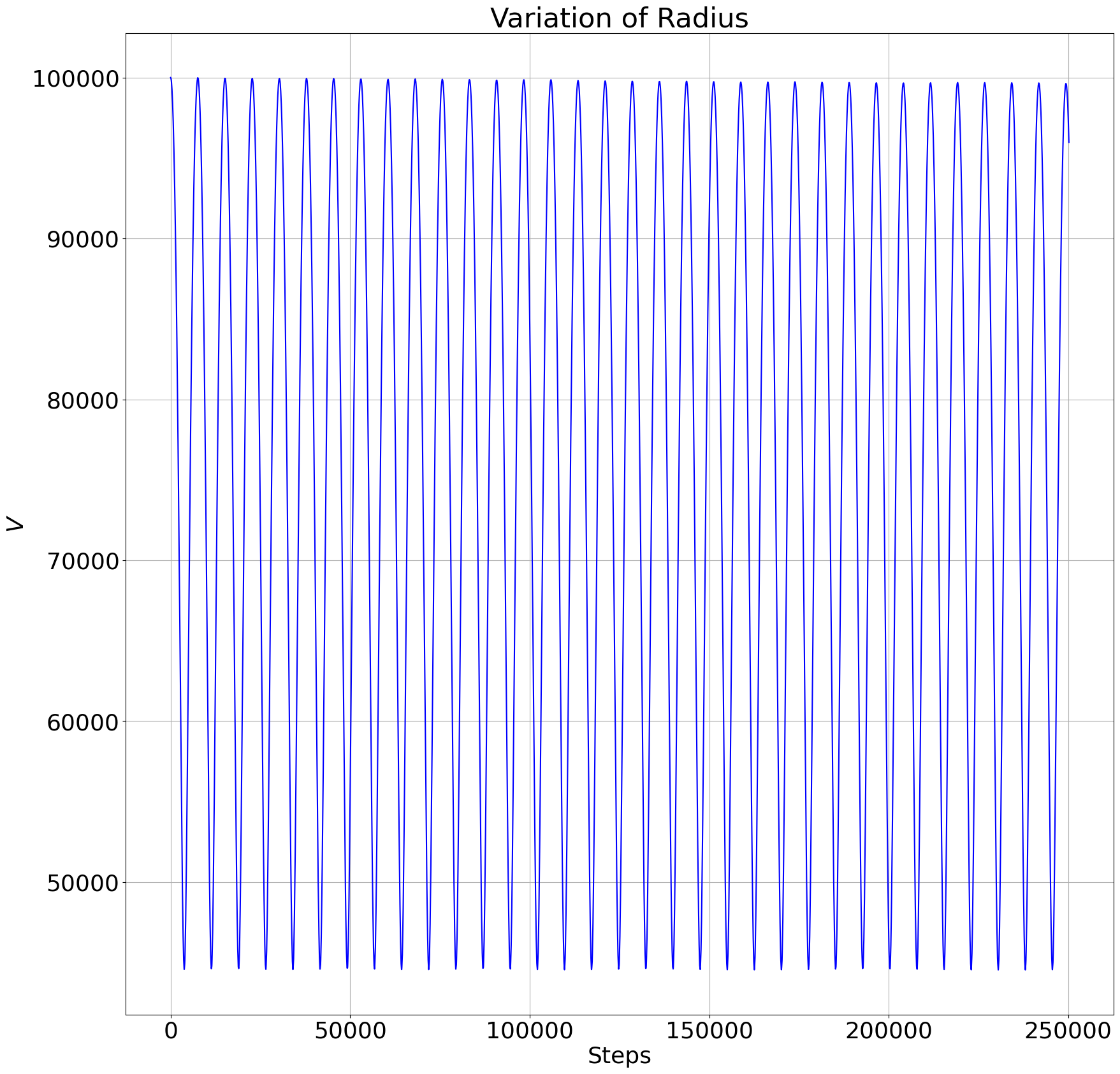}} 
    \subfigure[2.5 PN-with-Prop]{\includegraphics[width=0.24\textwidth]{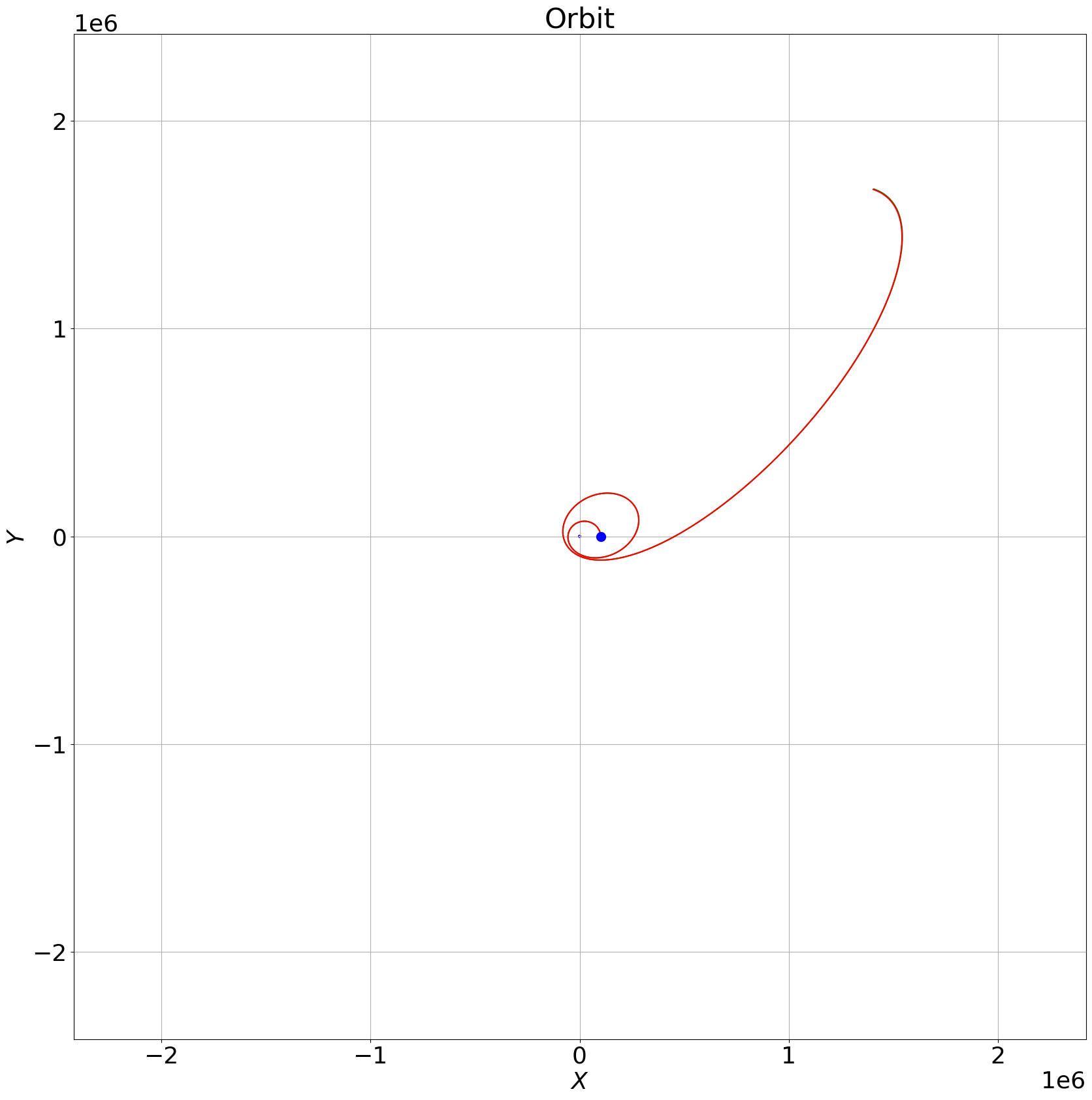}}
    \subfigure[2.5 PN-with-Prop]{\includegraphics[width=0.24\textwidth]{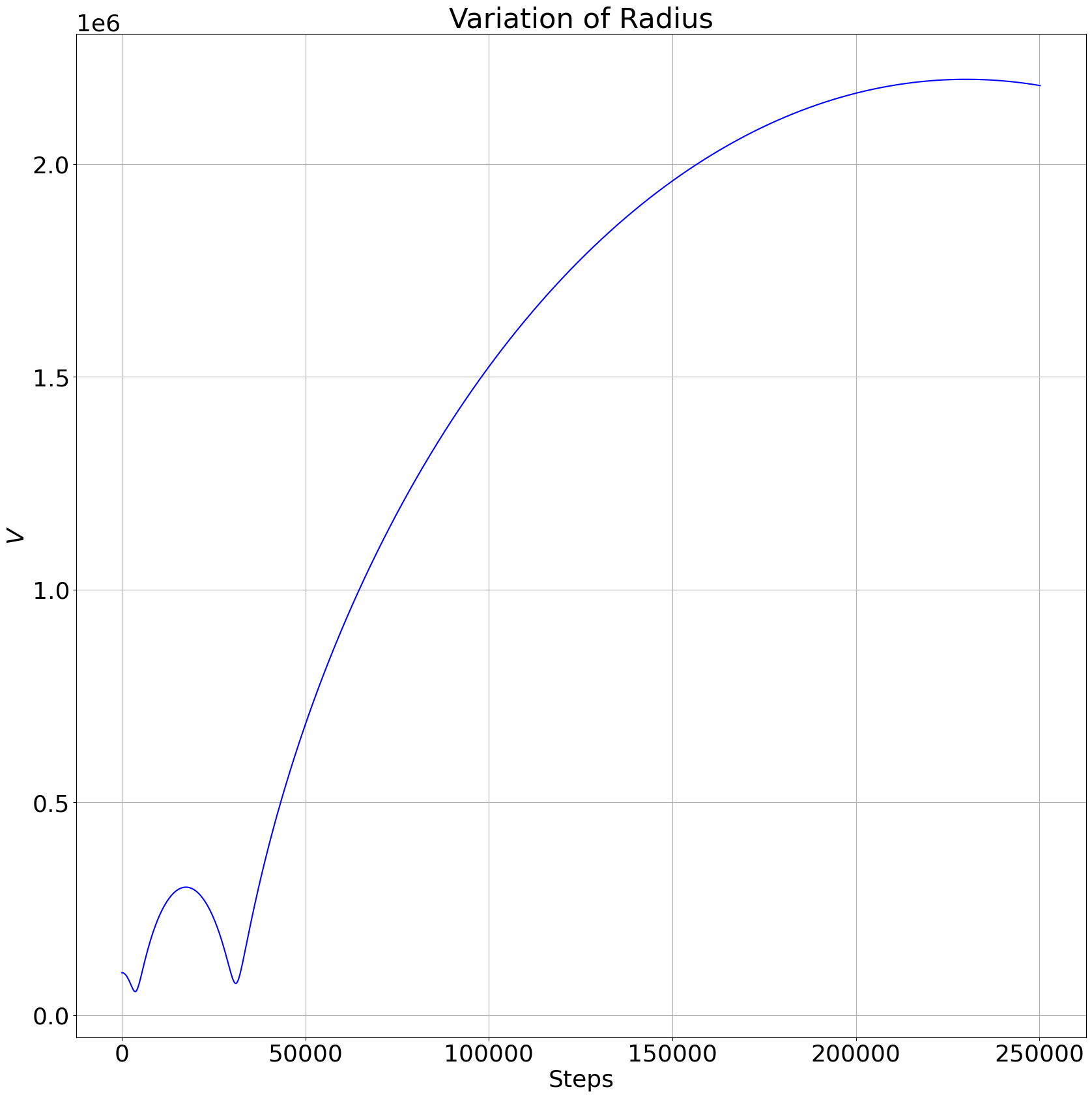}} \\
    \caption{Strong(er) Field, R = $1.1 \cdot 10^4m$, X = $10^5m$, Velocity = 40,000,000$ms^{-1}$}
    \label{results5}
\end{figure}

\begin{figure}[!ht]
    \centering
    \setcounter{subfigure}{0}
    \subfigure[0 PN-no-Prop]{\includegraphics[width=0.24\textwidth]{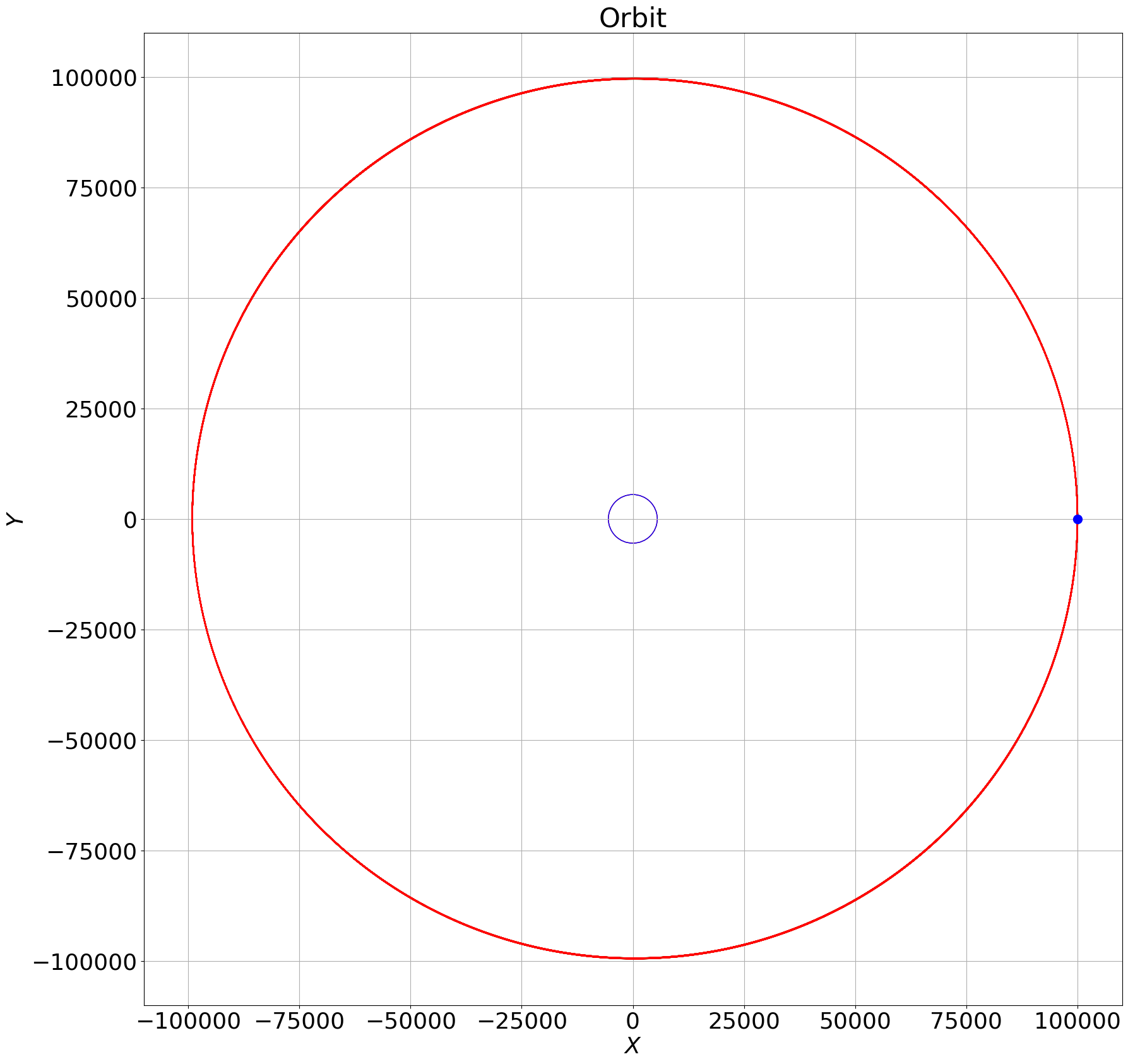}}
    \subfigure[0 PN-no-Prop]{\includegraphics[width=0.24\textwidth]{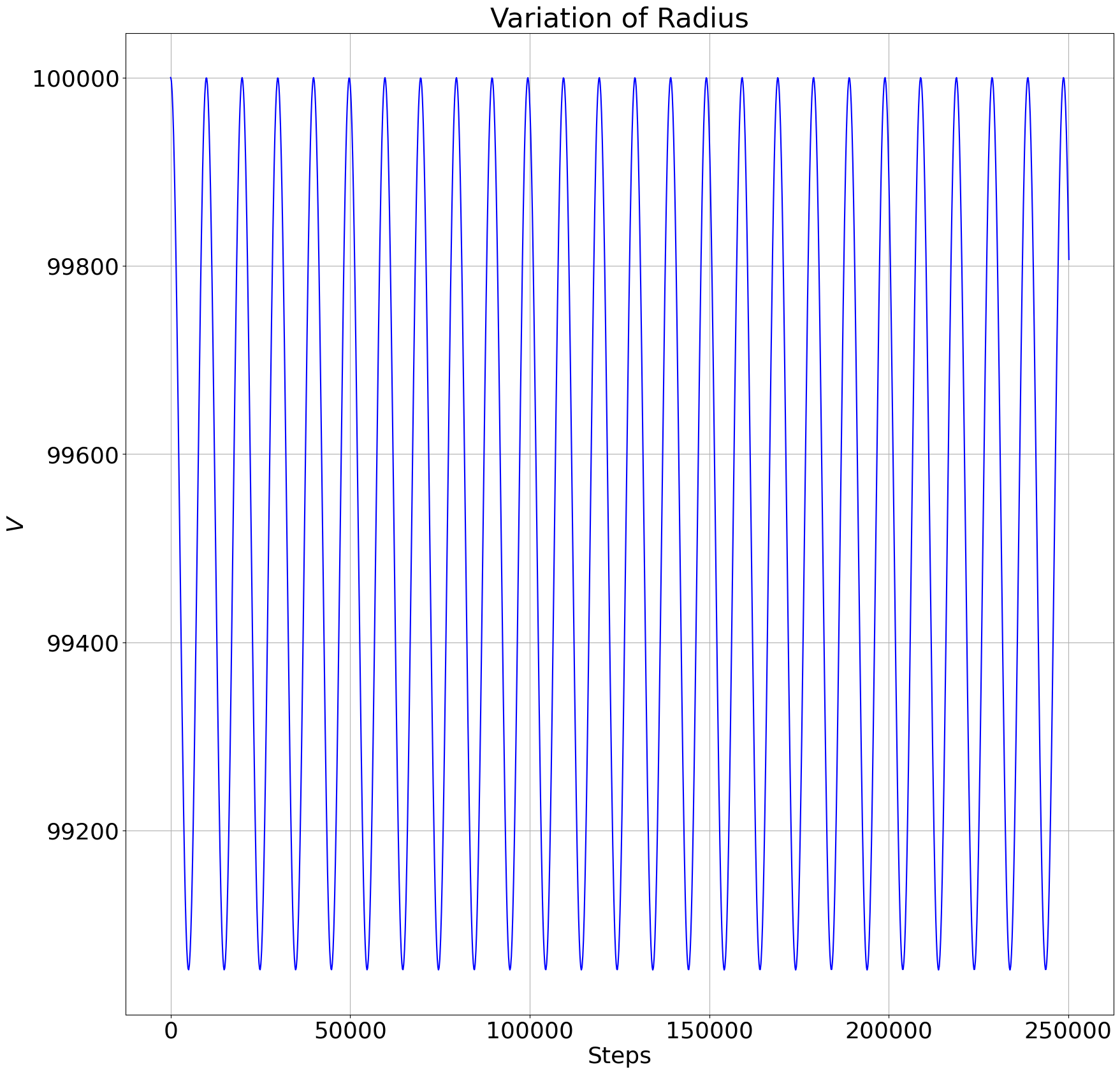}} 
    \subfigure[0 PN-with-Prop]{\includegraphics[width=0.24\textwidth]{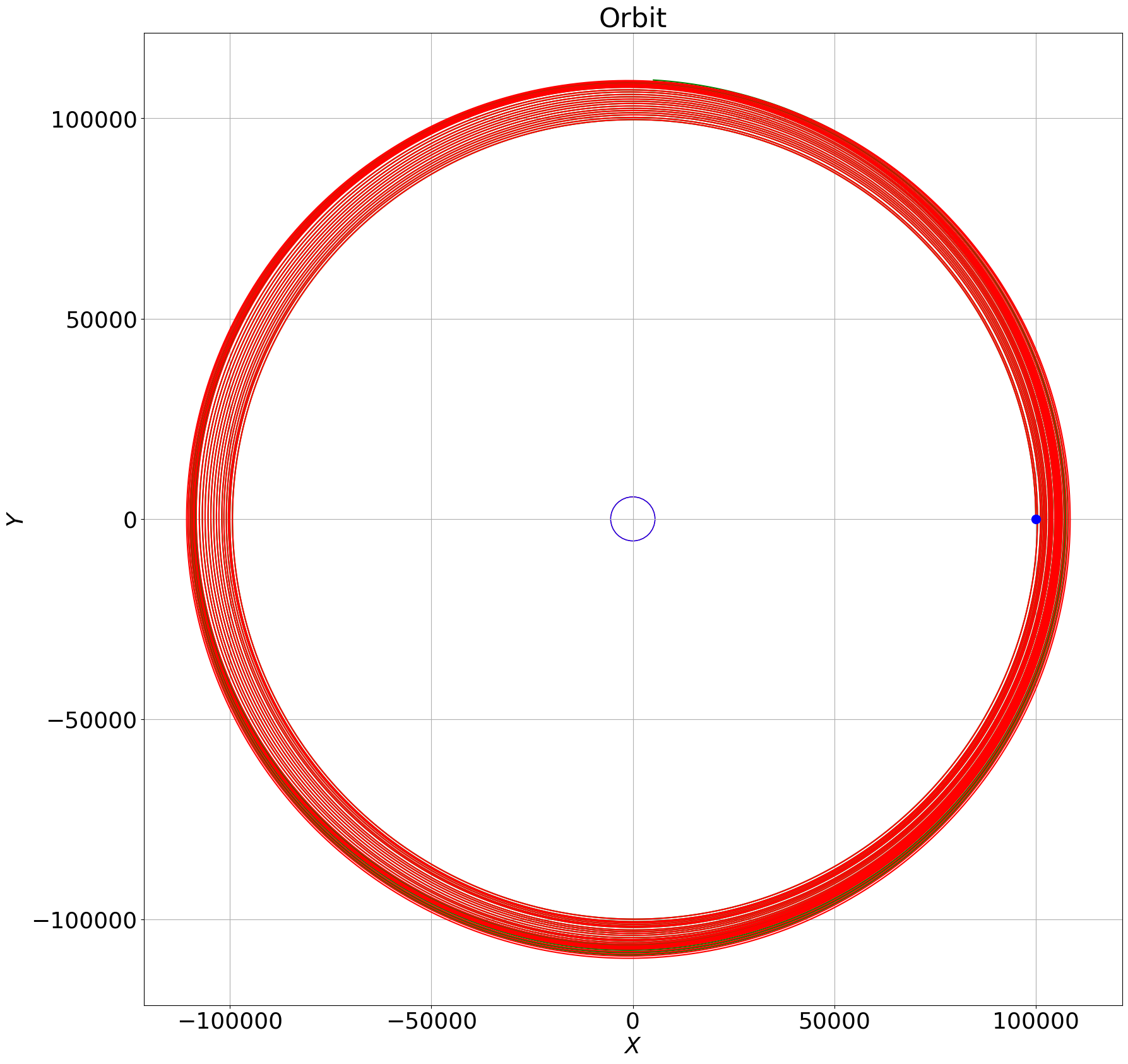}}
    \subfigure[0 PN-with-Prop]{\includegraphics[width=0.24\textwidth]{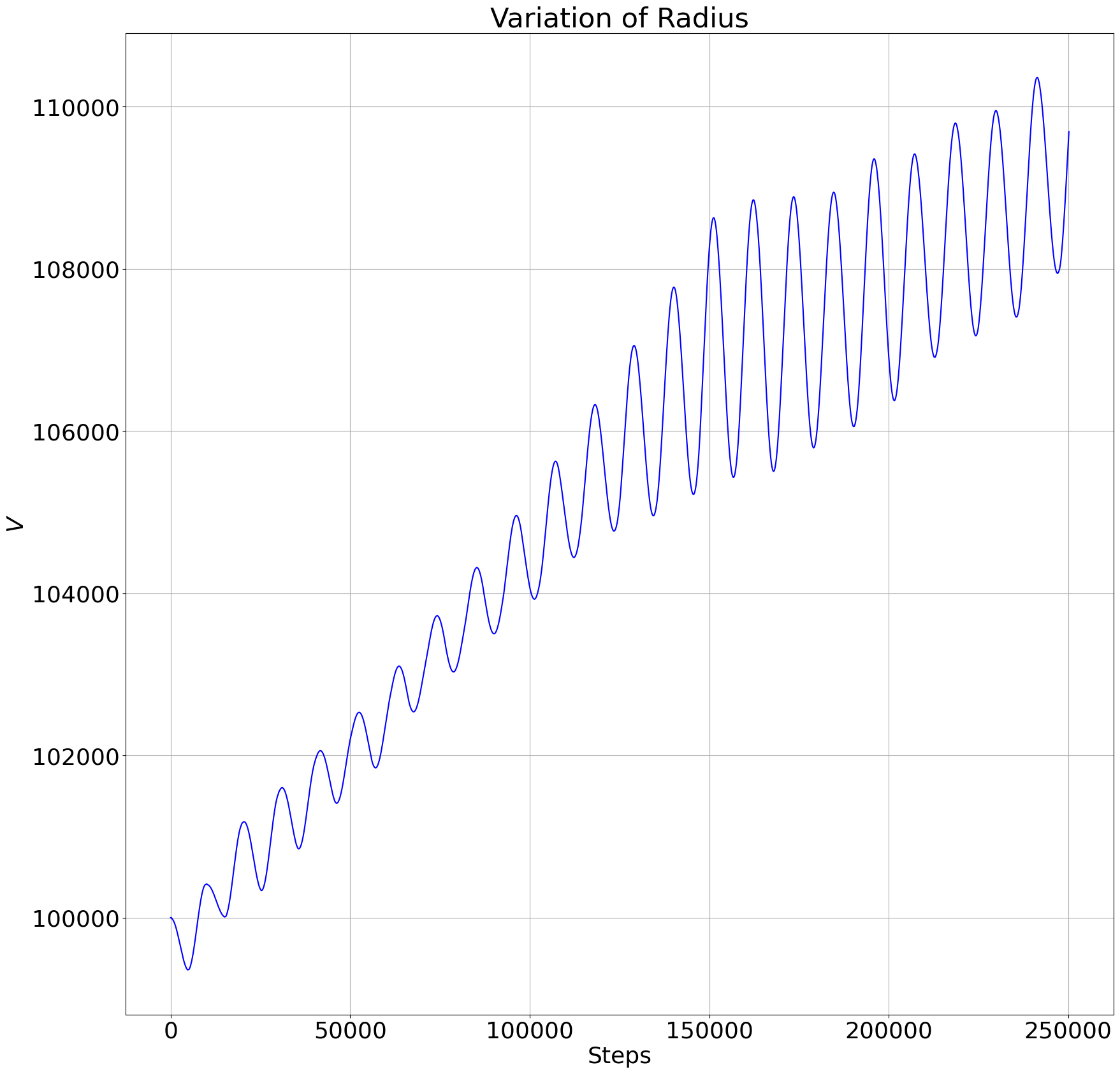}} \\
    
    \subfigure[0.5 PN-no-Prop]{\includegraphics[width=0.24\textwidth]{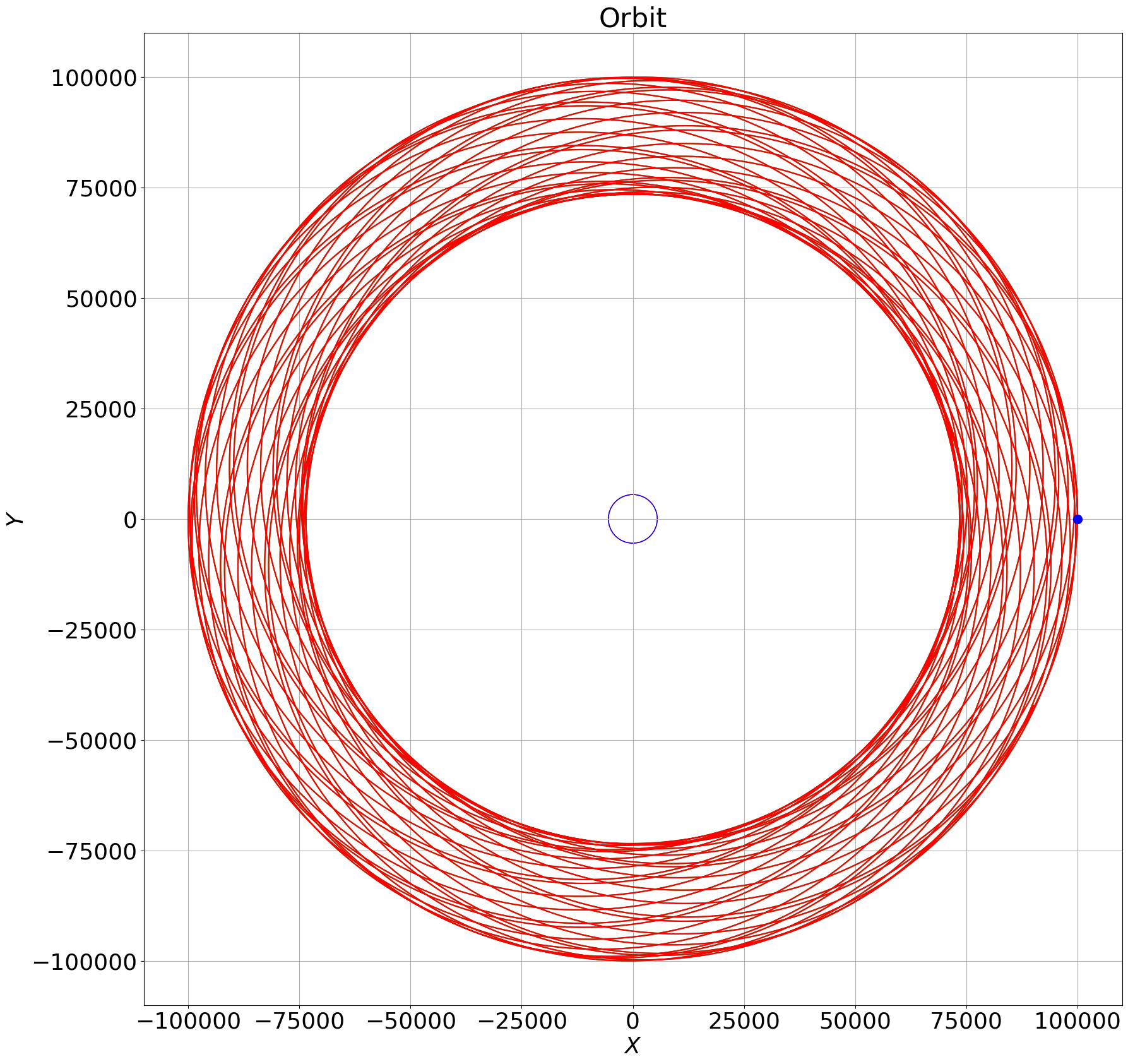}}
    \subfigure[0.5 PN-no-Prop]{\includegraphics[width=0.24\textwidth]{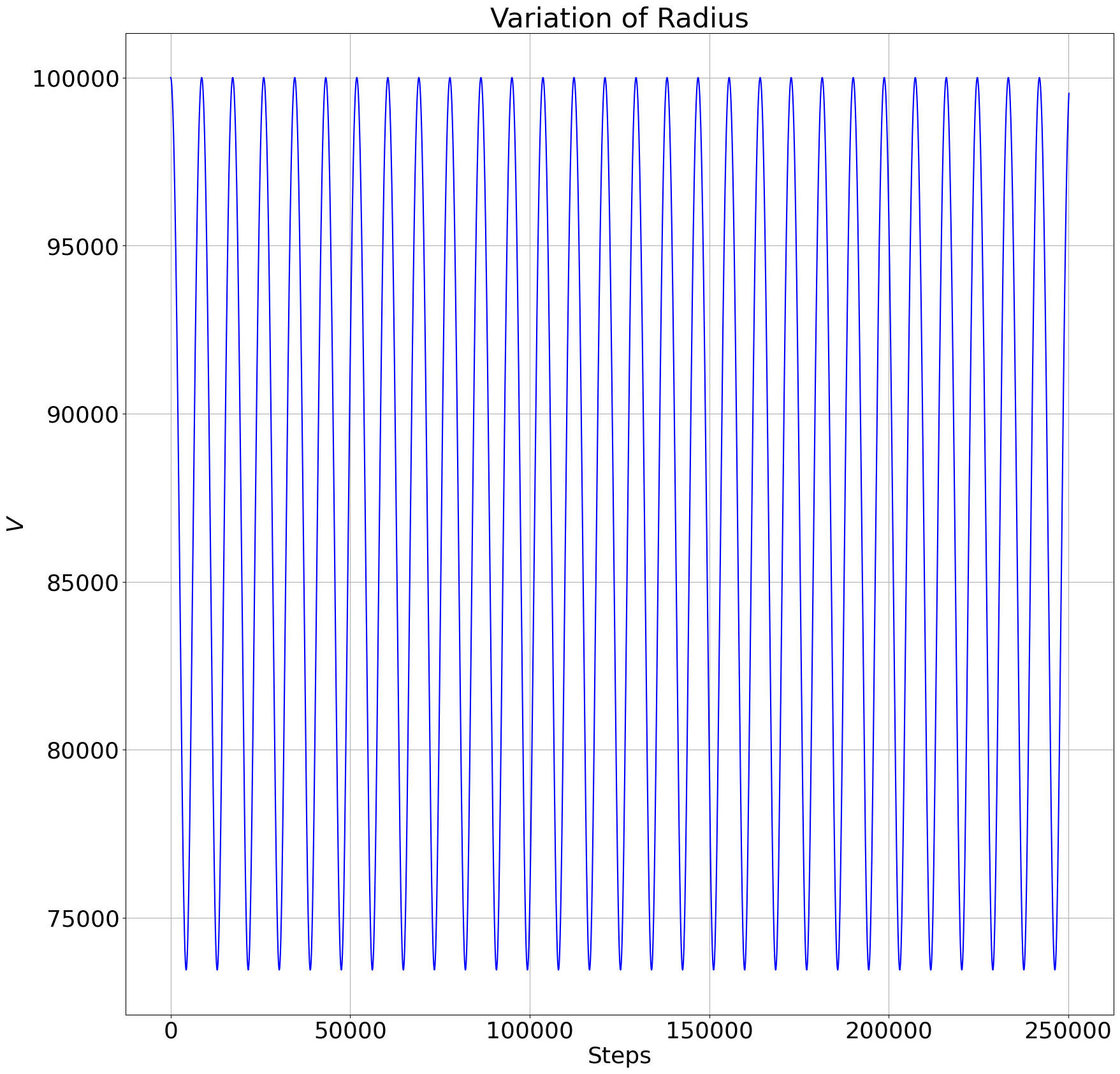}} 
    \subfigure[0.5 PN-with-Prop]{\includegraphics[width=0.24\textwidth]{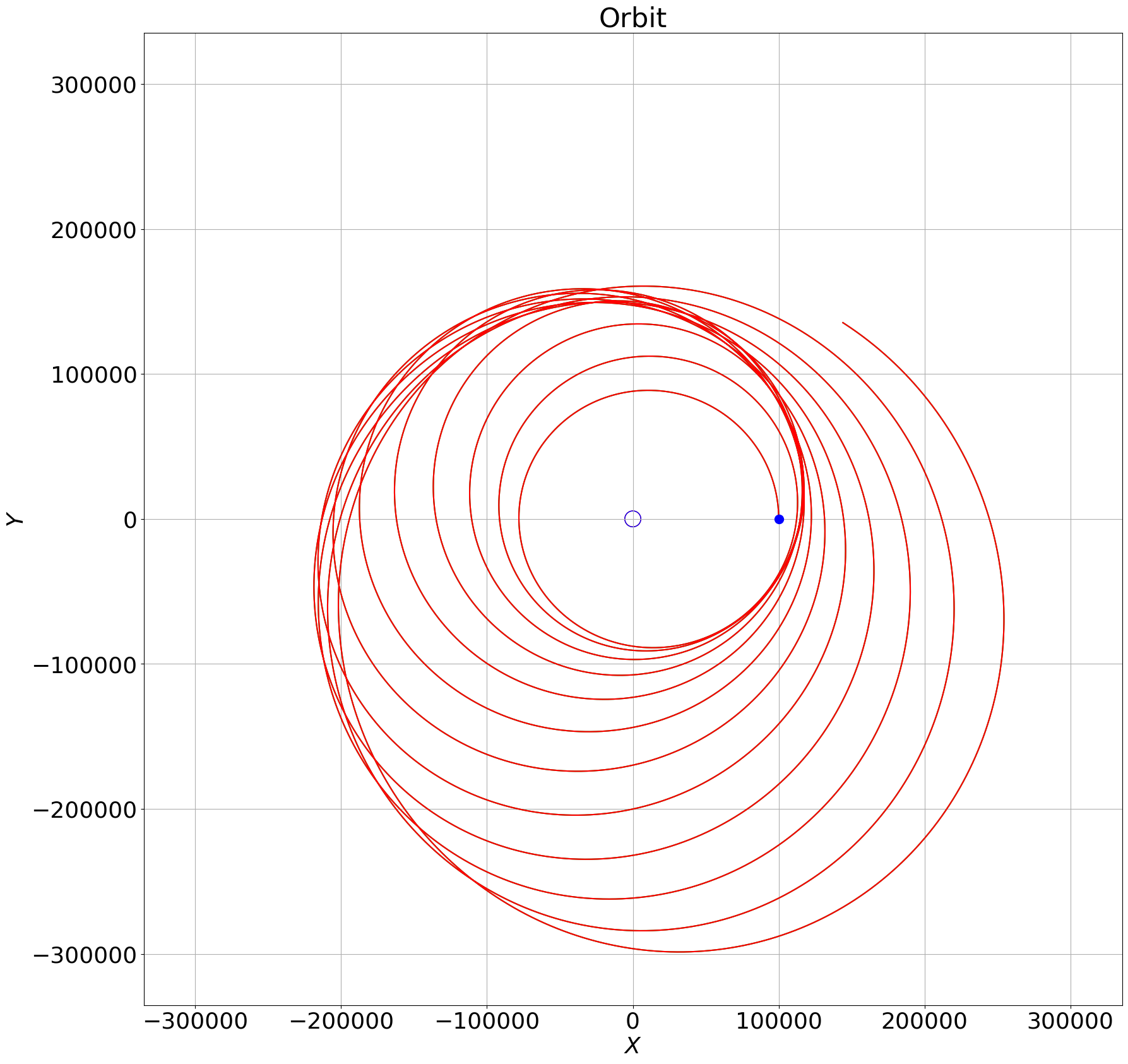}}
    \subfigure[0.5 PN-with-Prop]{\includegraphics[width=0.24\textwidth]{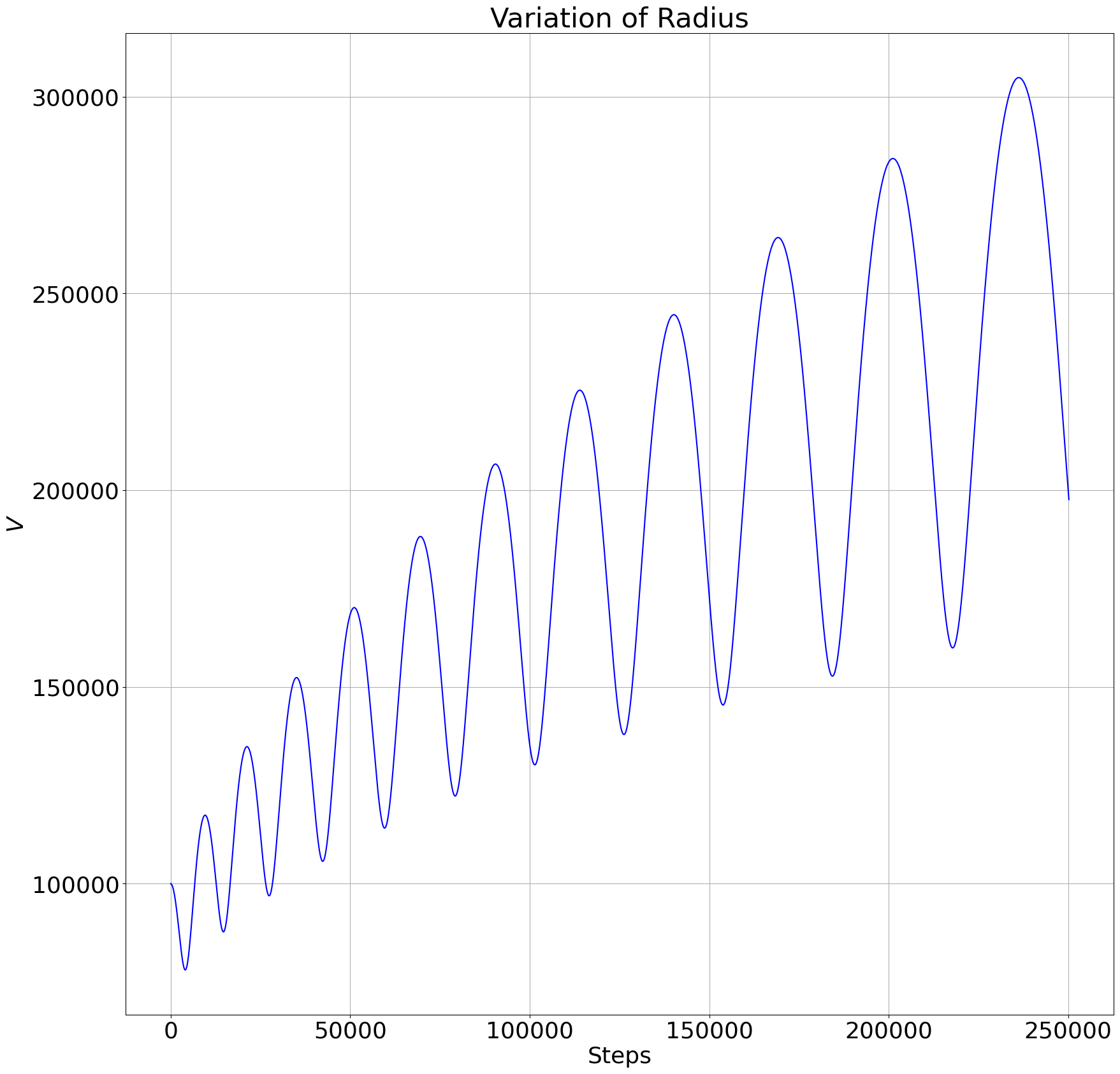}} \\
    
    \subfigure[1 PN-no-Prop]{\includegraphics[width=0.24\textwidth]{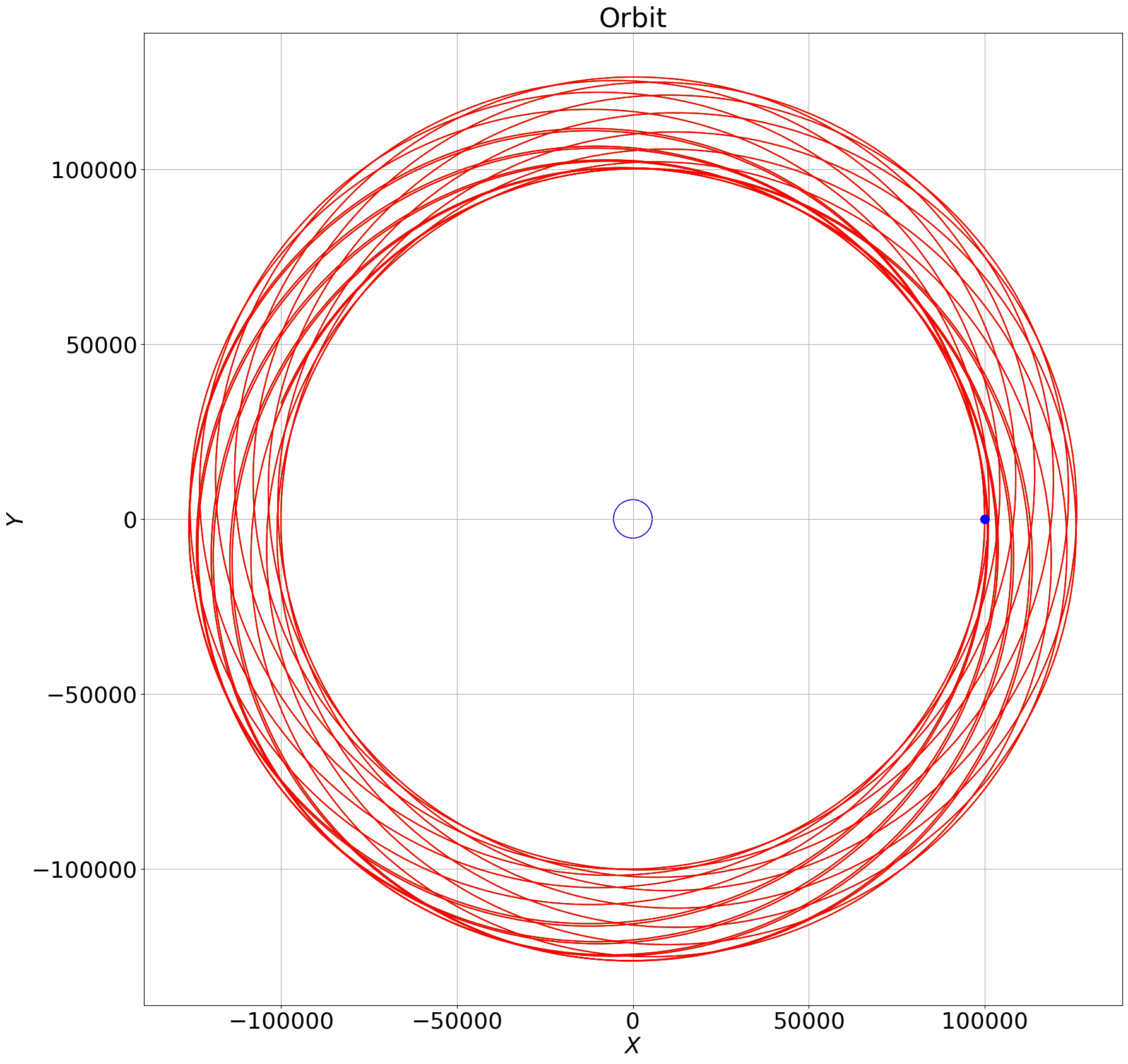}}
    \subfigure[1 PN-no-Prop]{\includegraphics[width=0.24\textwidth]{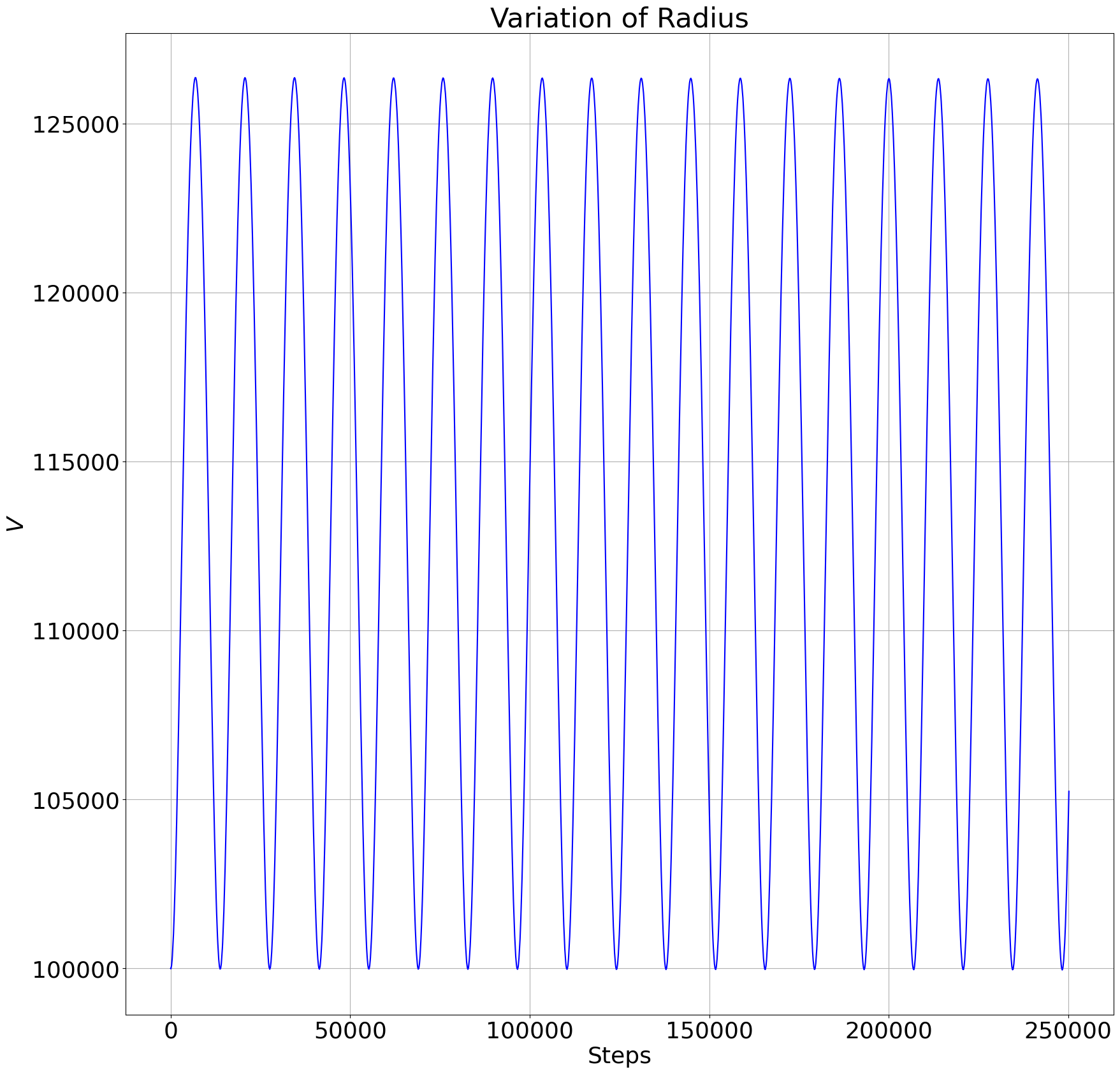}} 
    \subfigure[1 PN-with-Prop]{\includegraphics[width=0.24\textwidth]{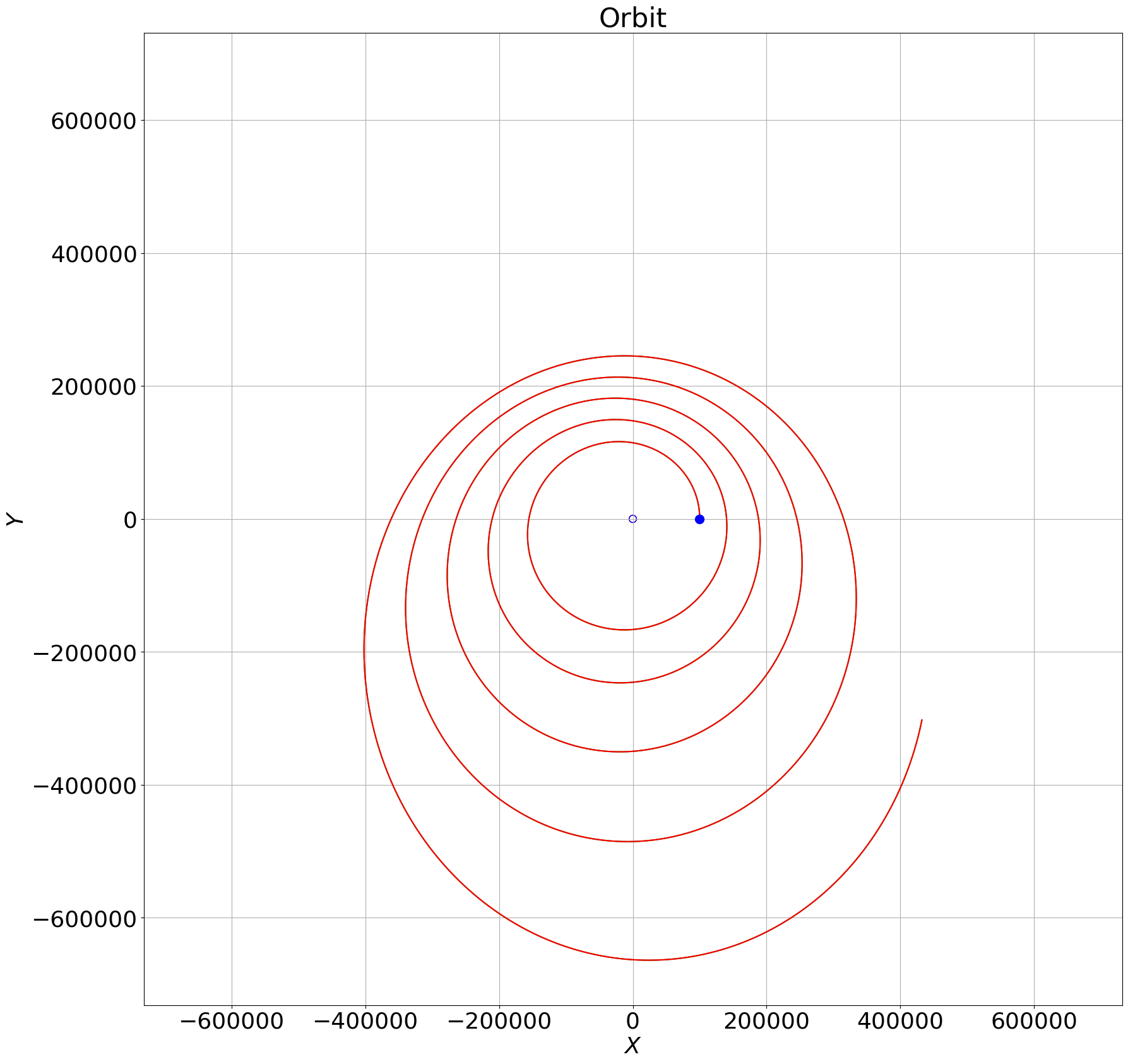}}
    \subfigure[1 PN-with-Prop]{\includegraphics[width=0.24\textwidth]{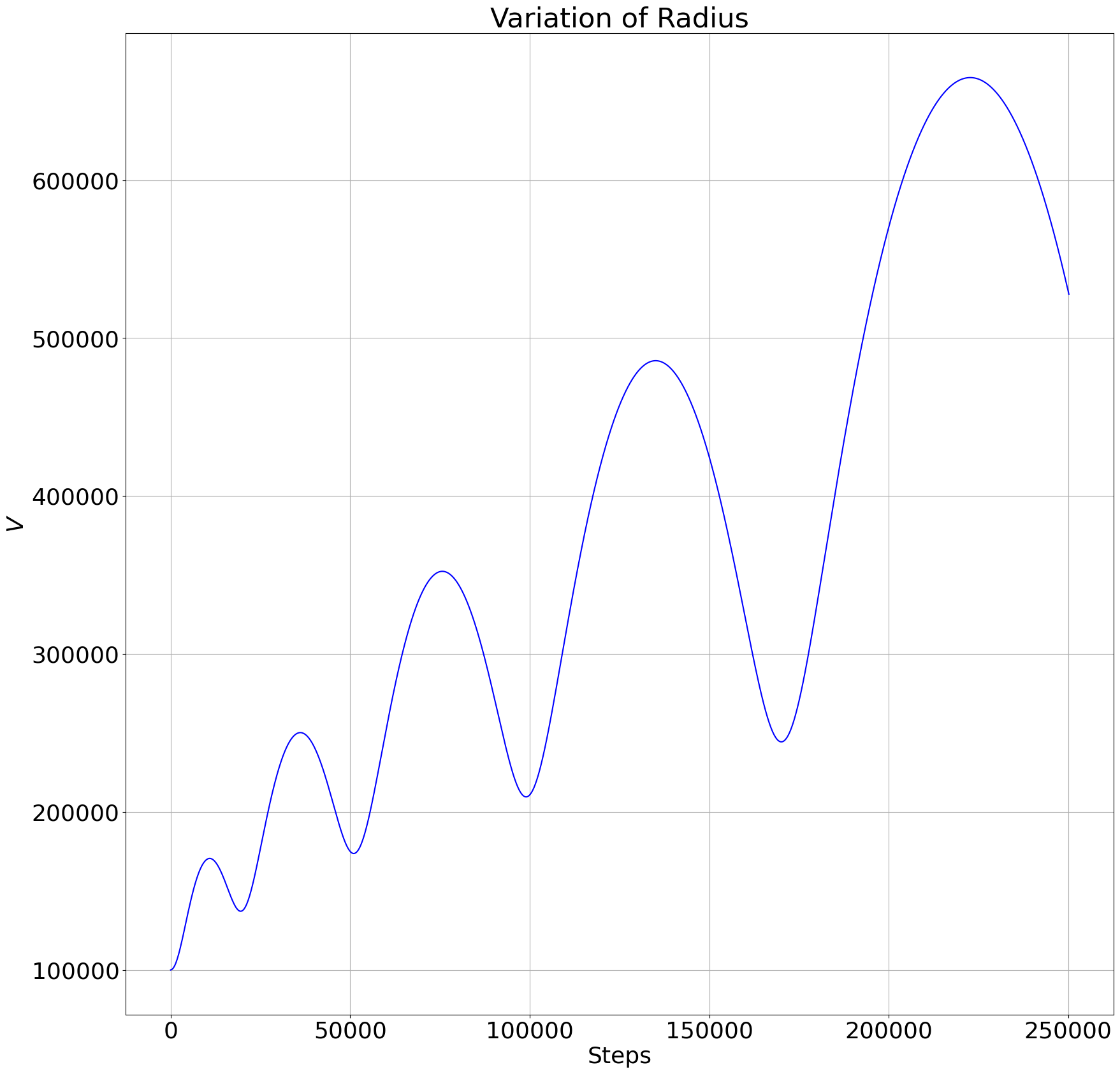}} \\
    
    \subfigure[2 PN-no-Prop]{\includegraphics[width=0.24\textwidth]{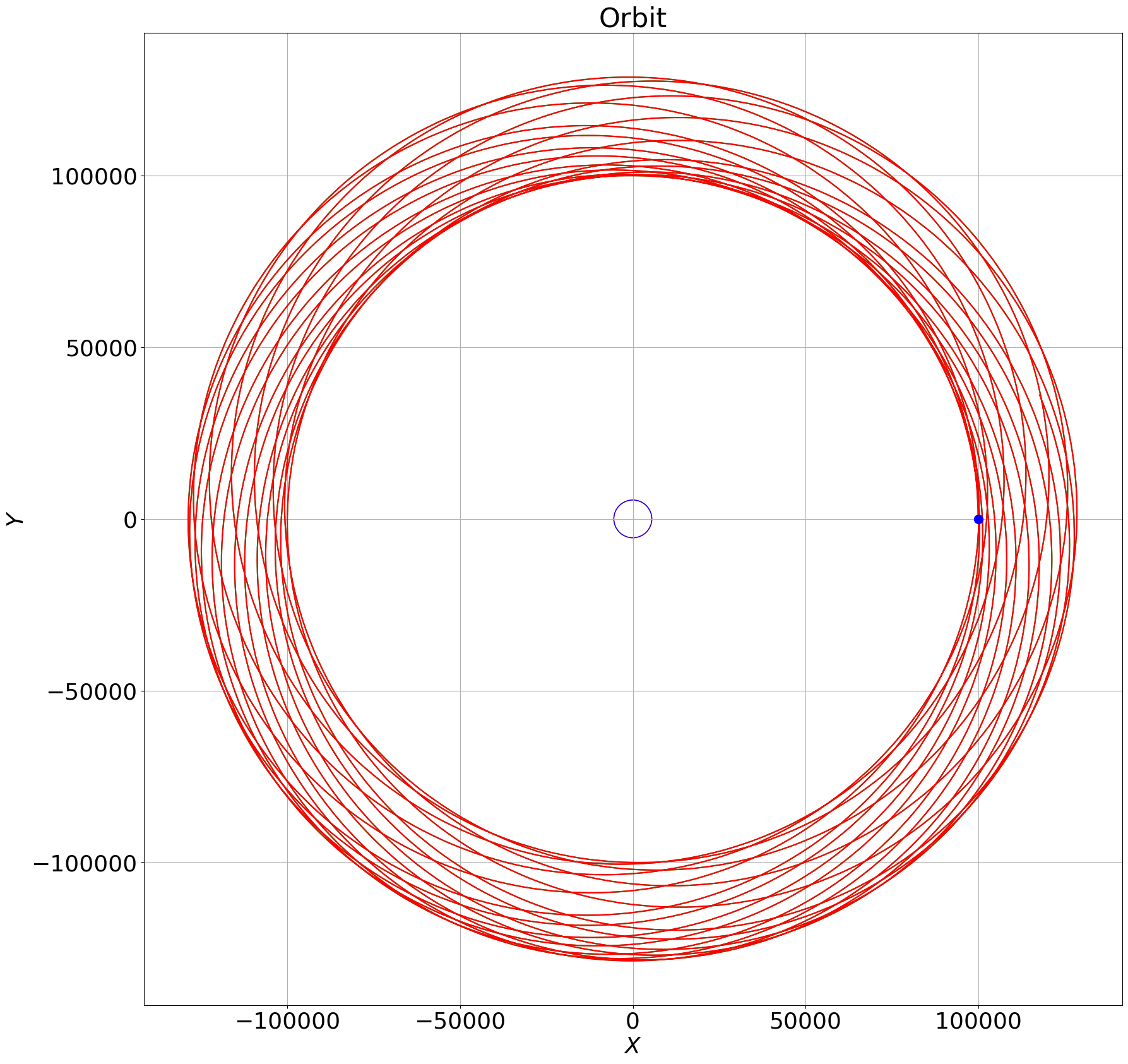}}
    \subfigure[2 PN-no-Prop]{\includegraphics[width=0.24\textwidth]{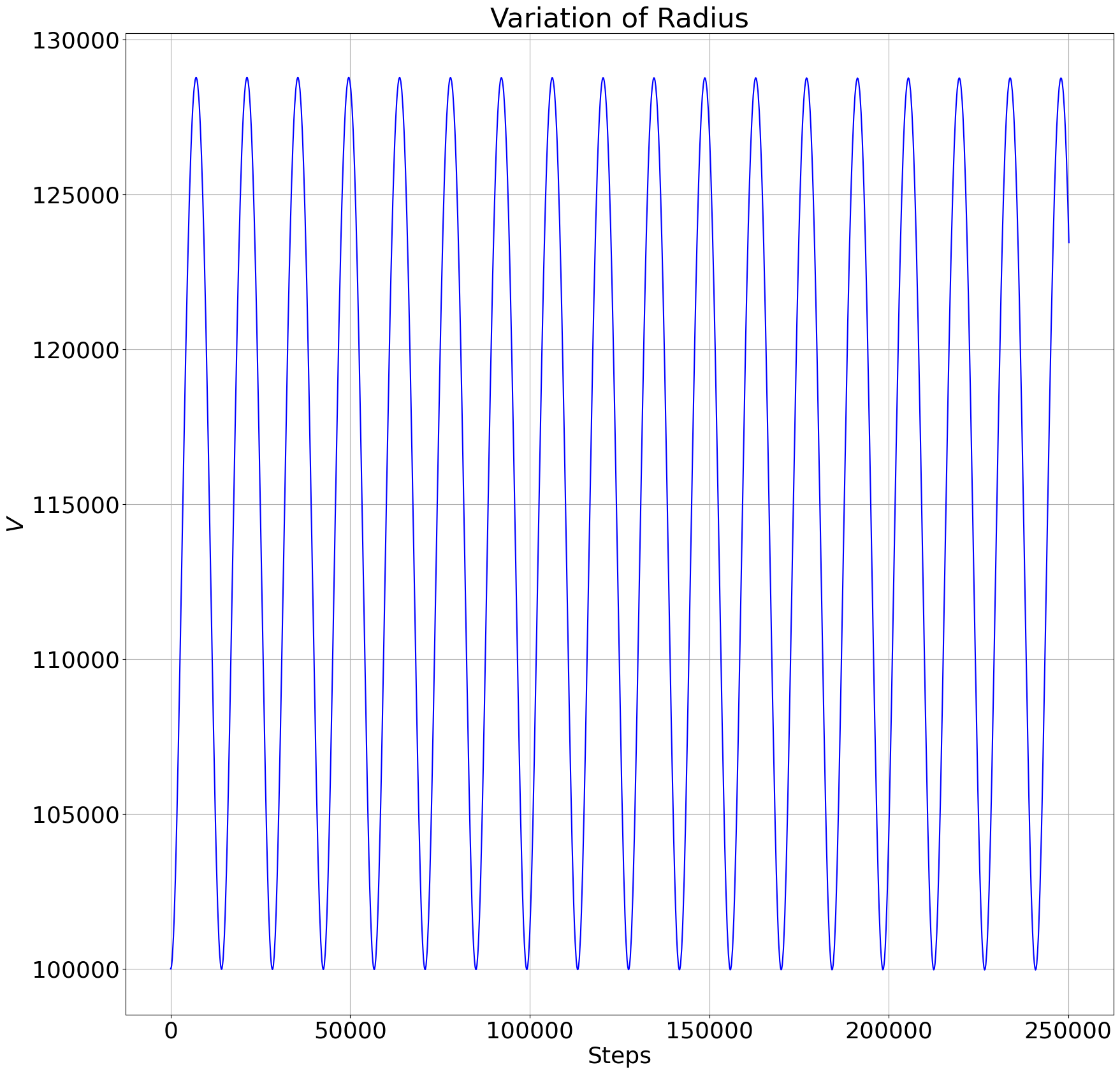}} 
    \subfigure[2 PN-with-Prop]{\includegraphics[width=0.24\textwidth]{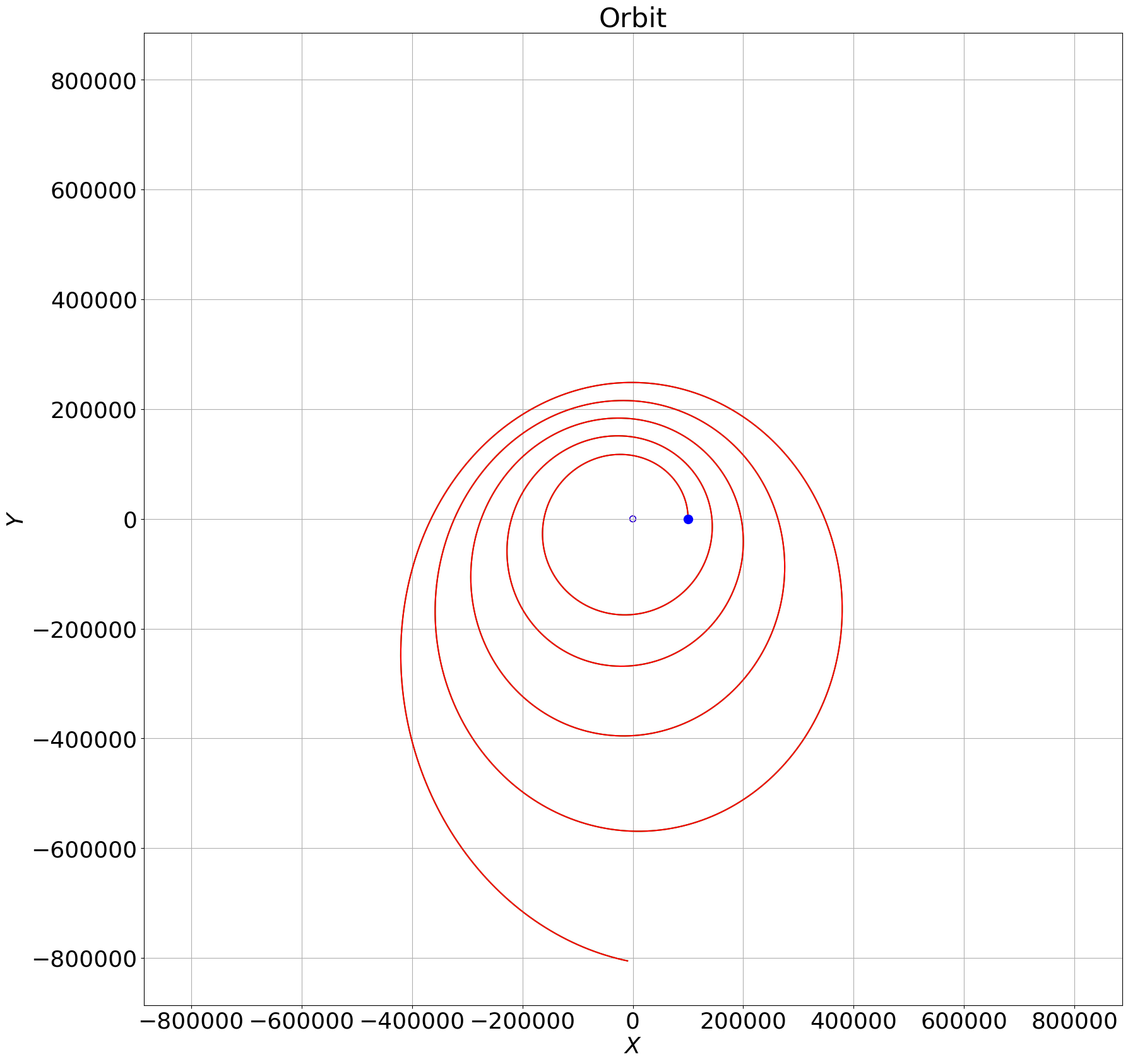}}
    \subfigure[2 PN-with-Prop]{\includegraphics[width=0.24\textwidth]{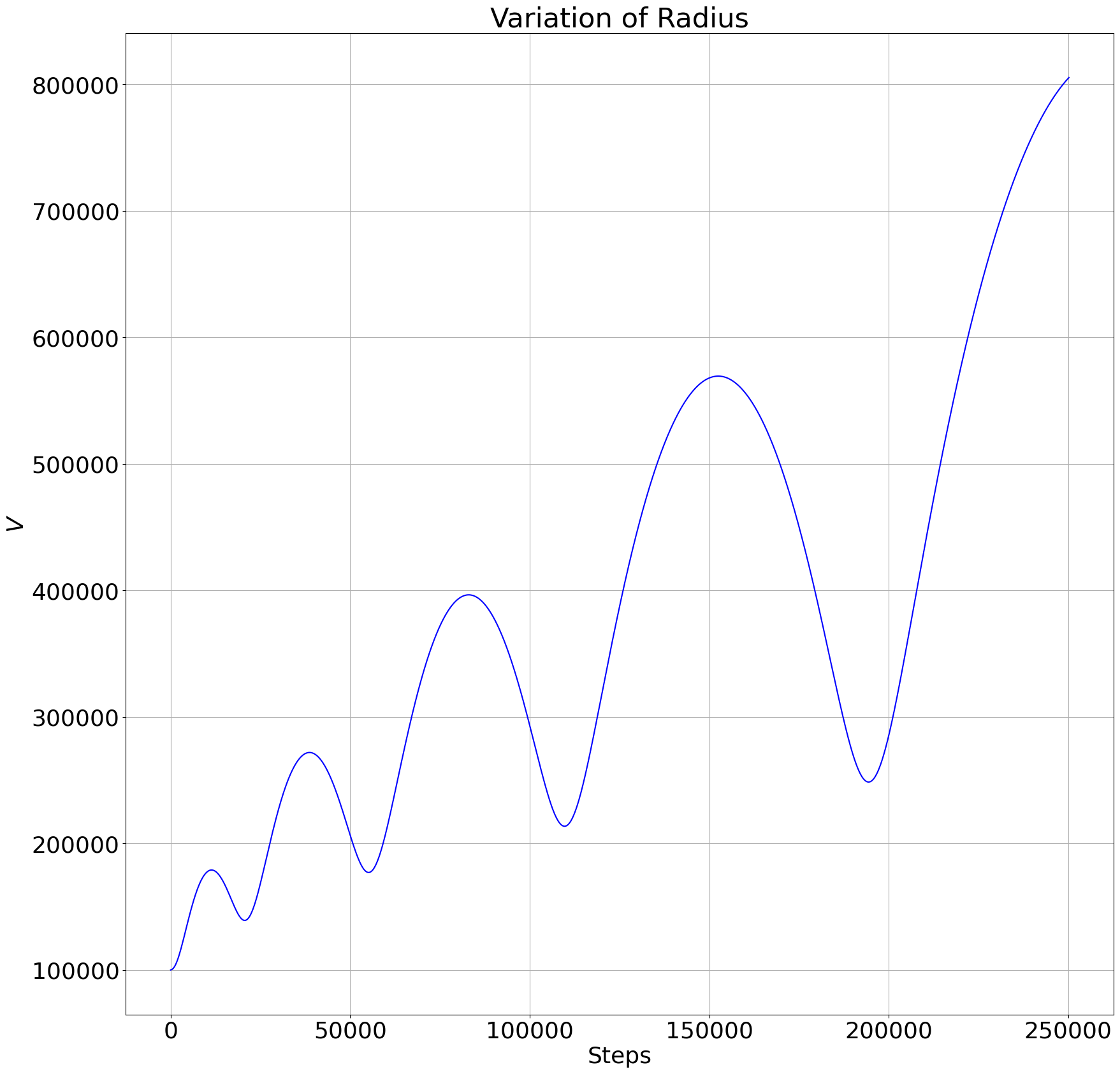}} \\
    
    \subfigure[2.5 PN-no-Prop]{\includegraphics[width=0.24\textwidth]{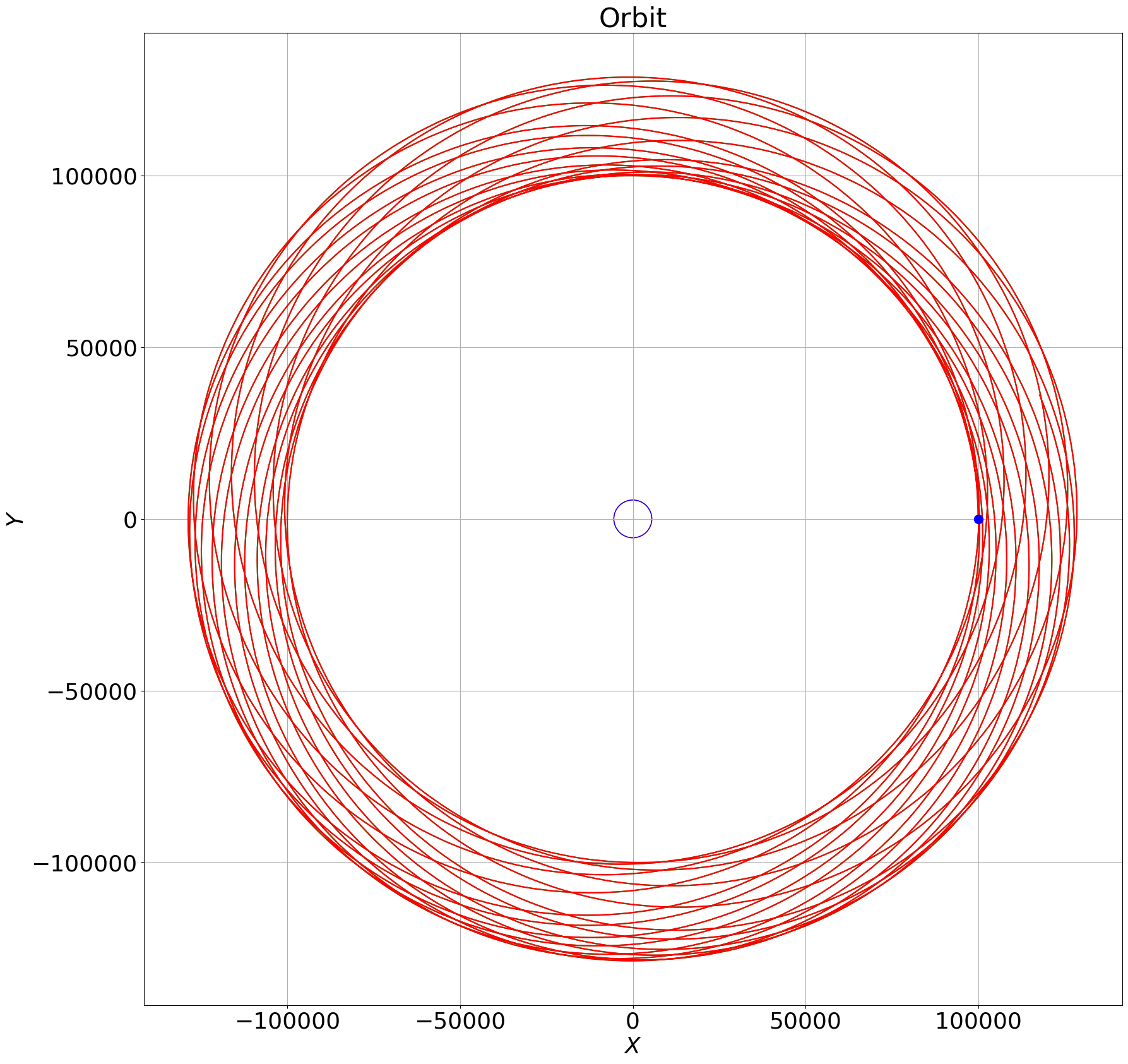}}
    \subfigure[2.5 PN-no-Prop]{\includegraphics[width=0.24\textwidth]{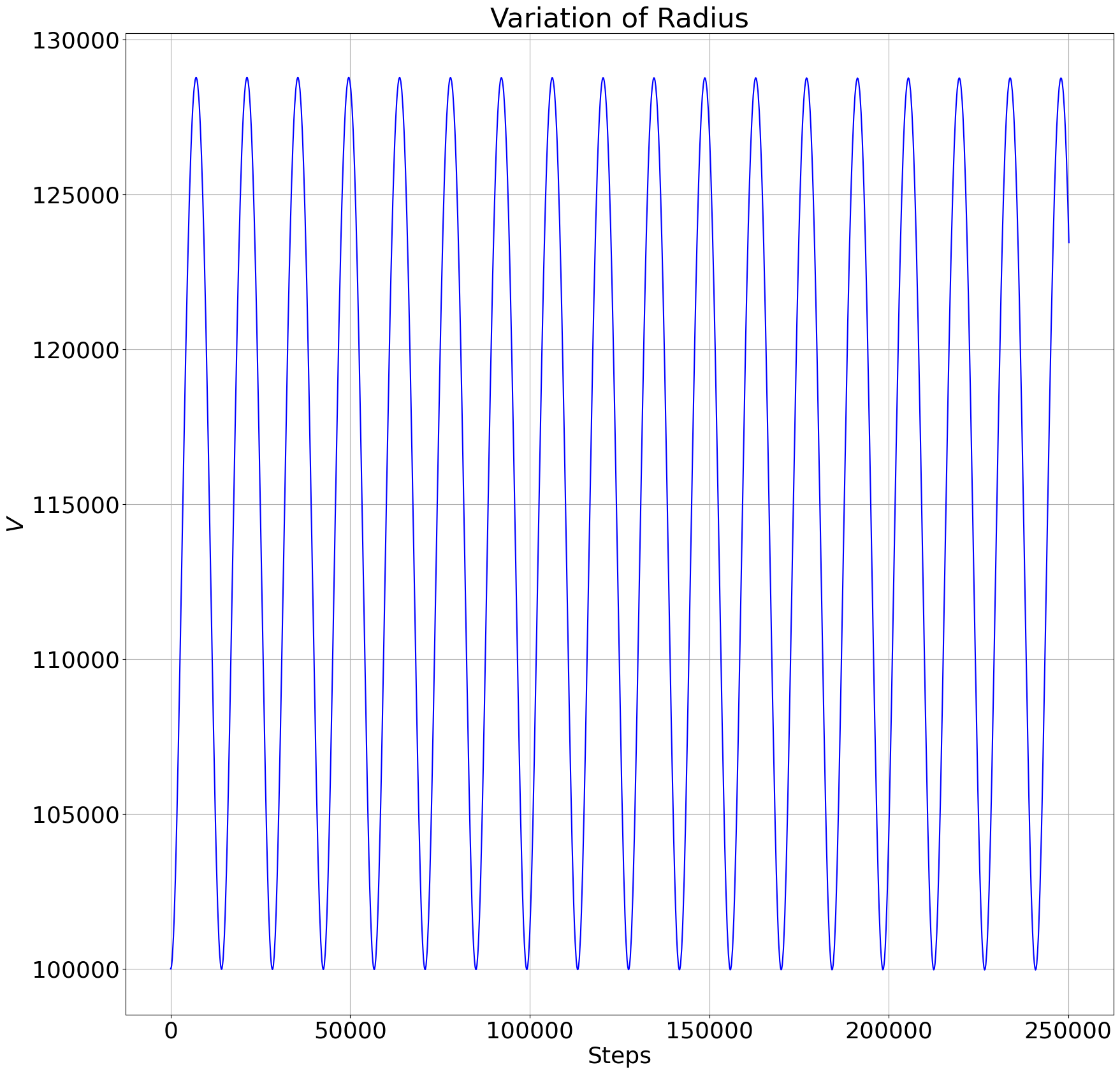}} 
    \subfigure[2.5 PN-with-Prop]{\includegraphics[width=0.24\textwidth]{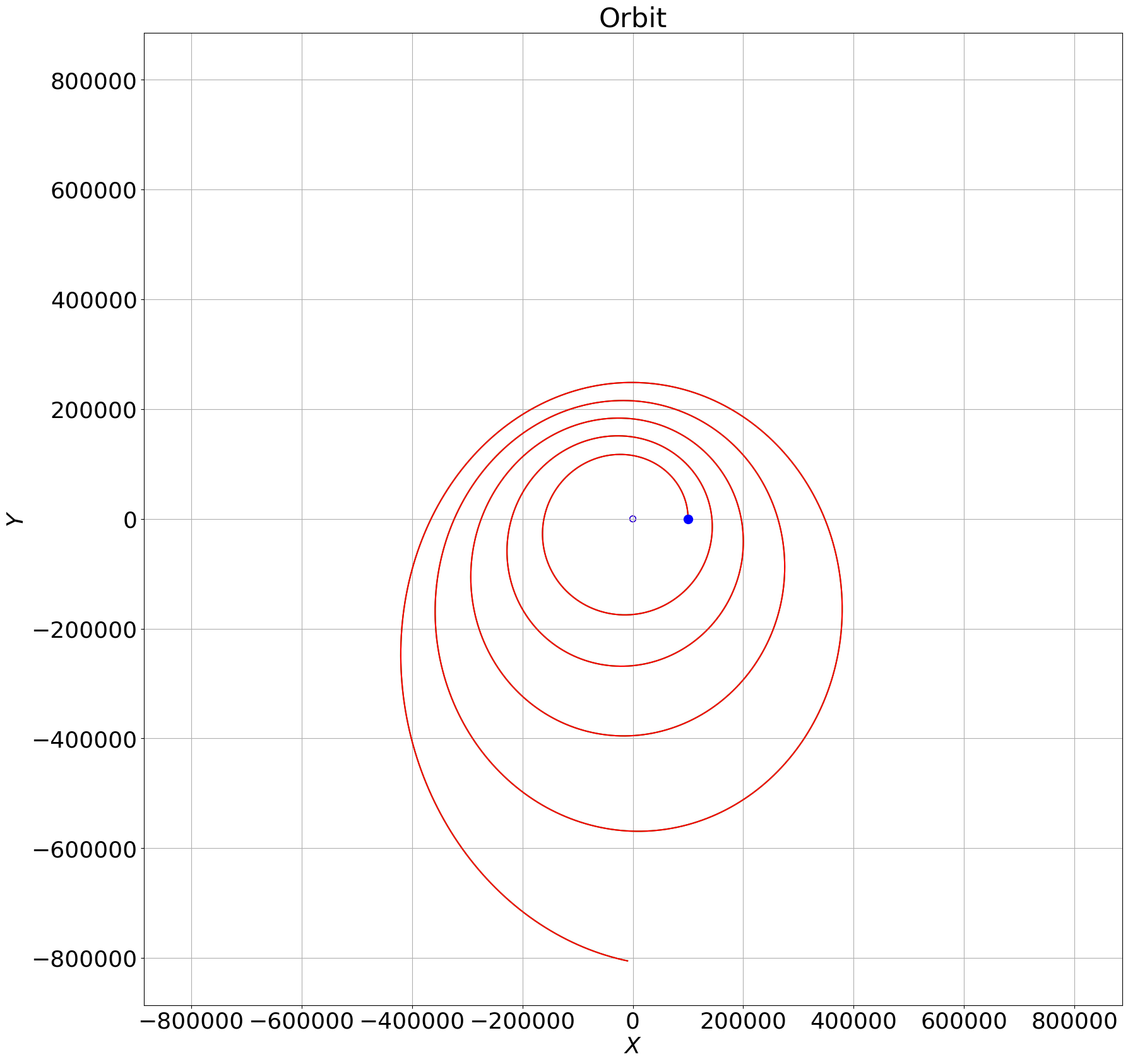}}
    \subfigure[2.5 PN-with-Prop]{\includegraphics[width=0.24\textwidth]{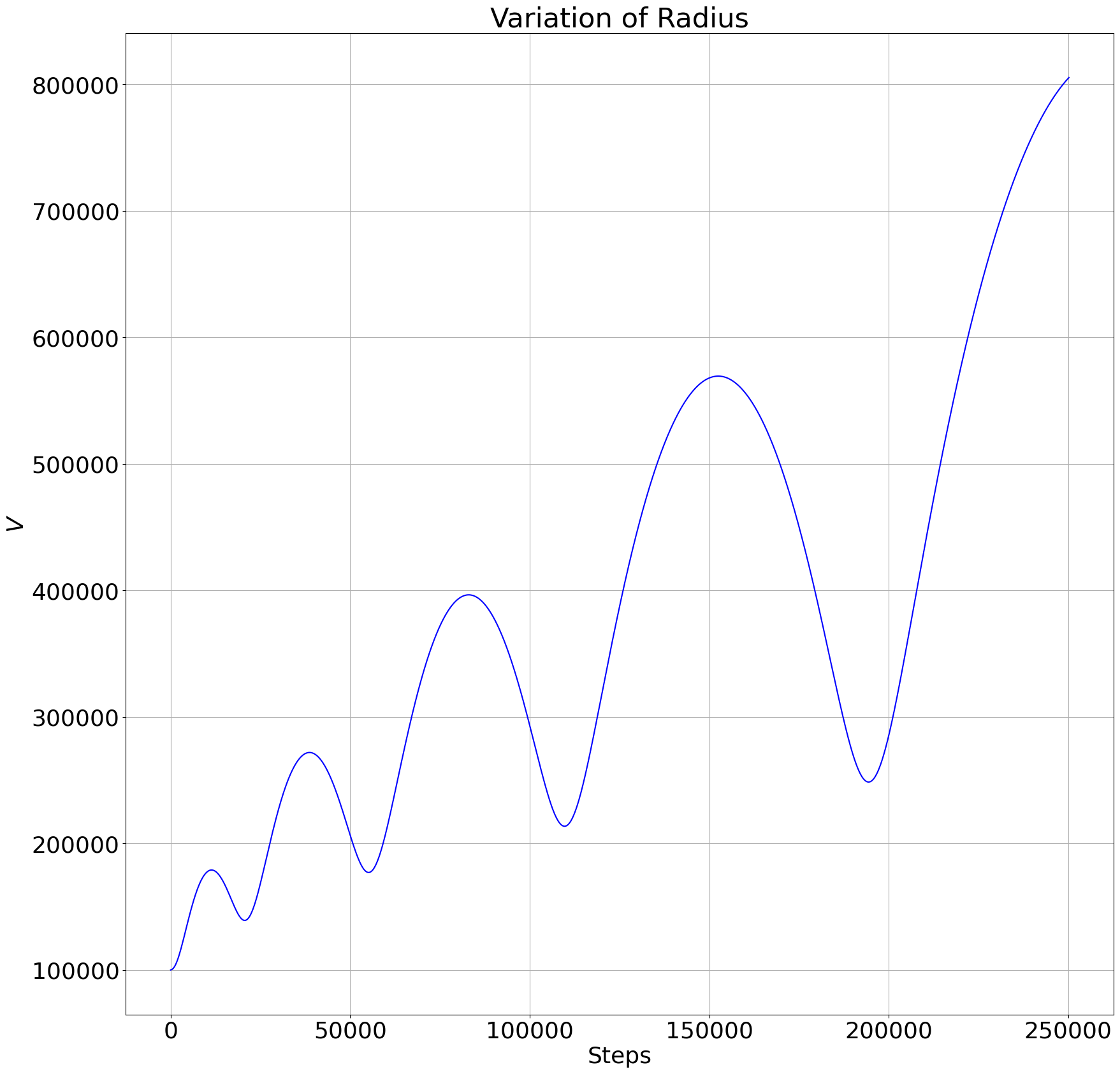}} \\
    \caption{Strong(er) Field, R = $1.1 \cdot 10^4m$, X = $10^5m$, Velocity = 51,600,000$ms^{-1}$}
    \label{results6}
\end{figure}

\begin{figure}[!ht]
    \centering
    \setcounter{subfigure}{0}
    \subfigure[0 PN-no-Prop]{\includegraphics[width=0.24\textwidth]{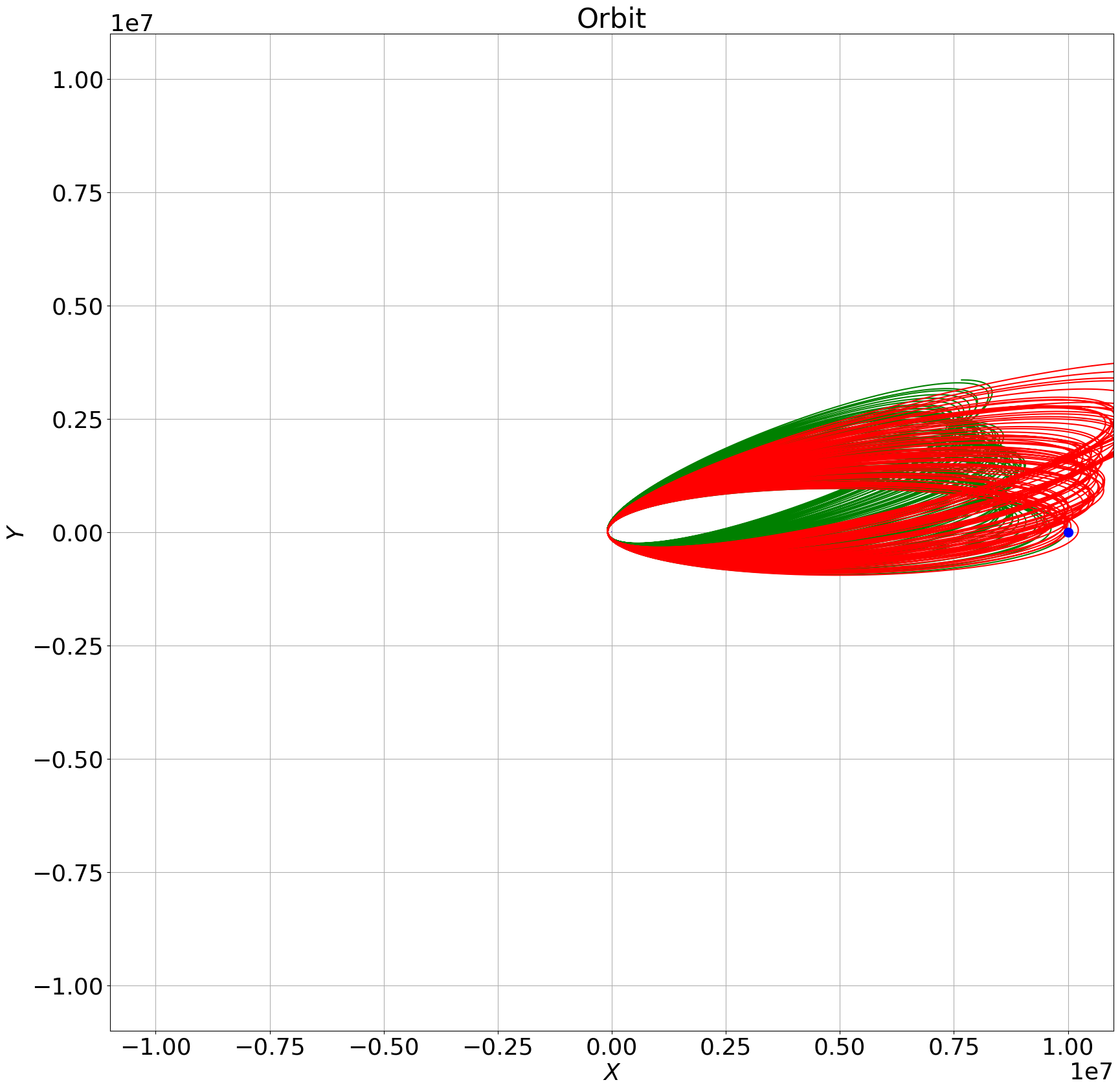}}
    \subfigure[0 PN-no-Prop]{\includegraphics[width=0.24\textwidth]{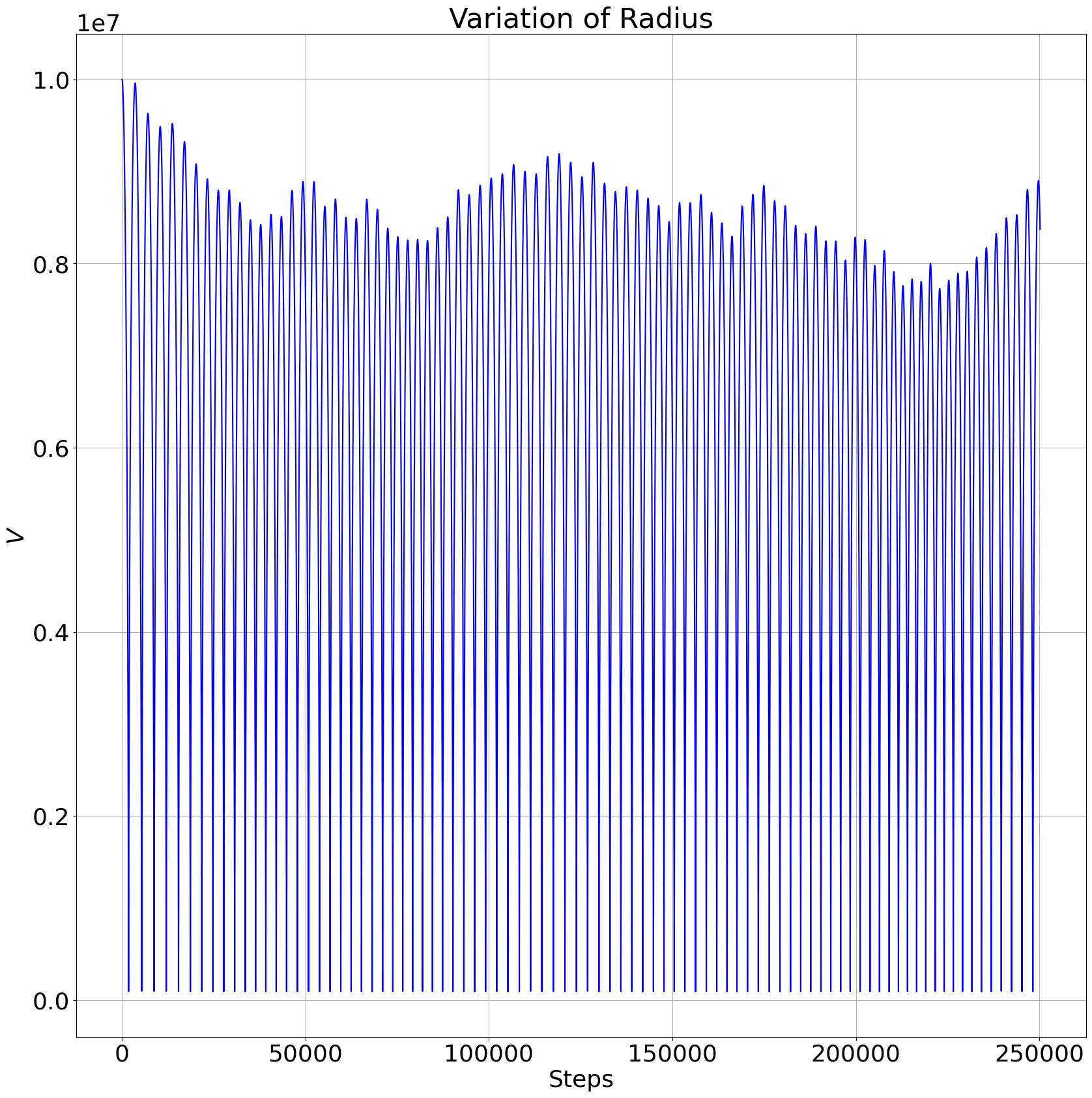}} 
    \subfigure[0 PN-with-Prop]{\includegraphics[width=0.24\textwidth]{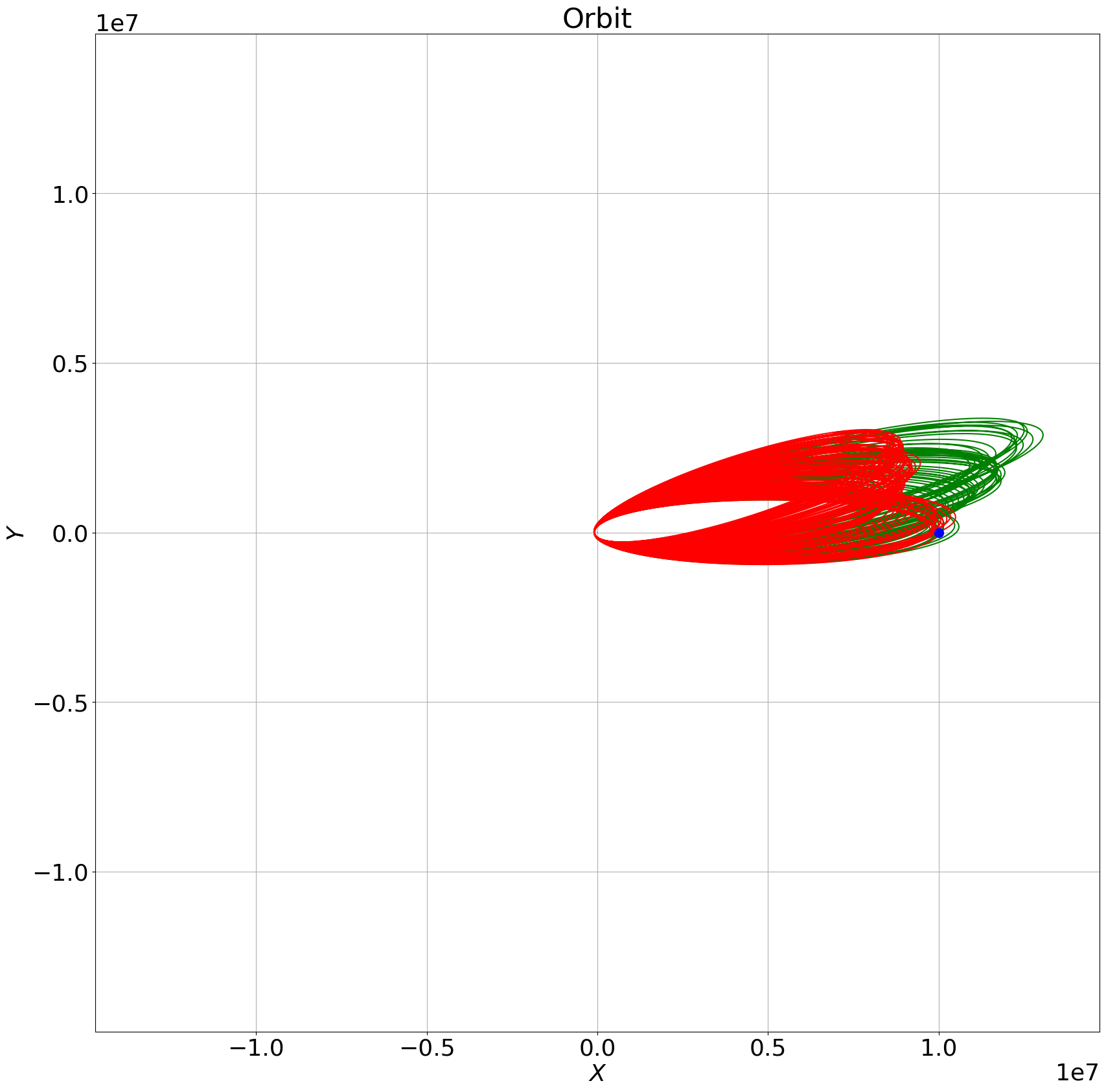}}
    \subfigure[0 PN-with-Prop]{\includegraphics[width=0.24\textwidth]{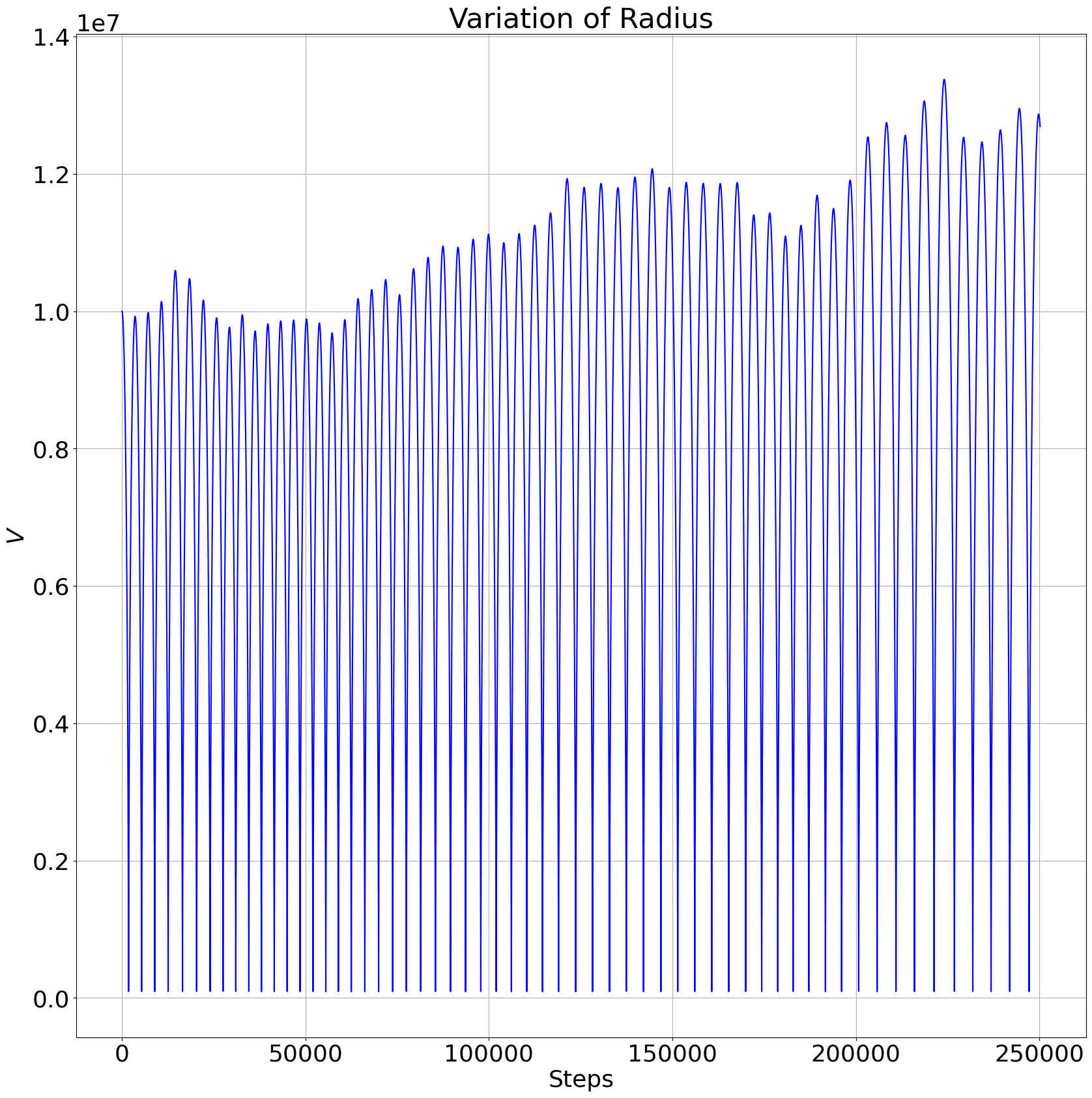}} \\
    
    \subfigure[0.5 PN-no-Prop]{\includegraphics[width=0.24\textwidth]{0.5PN-NF-R4-x7-Vel700-theta.png}}
    \subfigure[0.5 PN-no-Prop]{\includegraphics[width=0.24\textwidth]{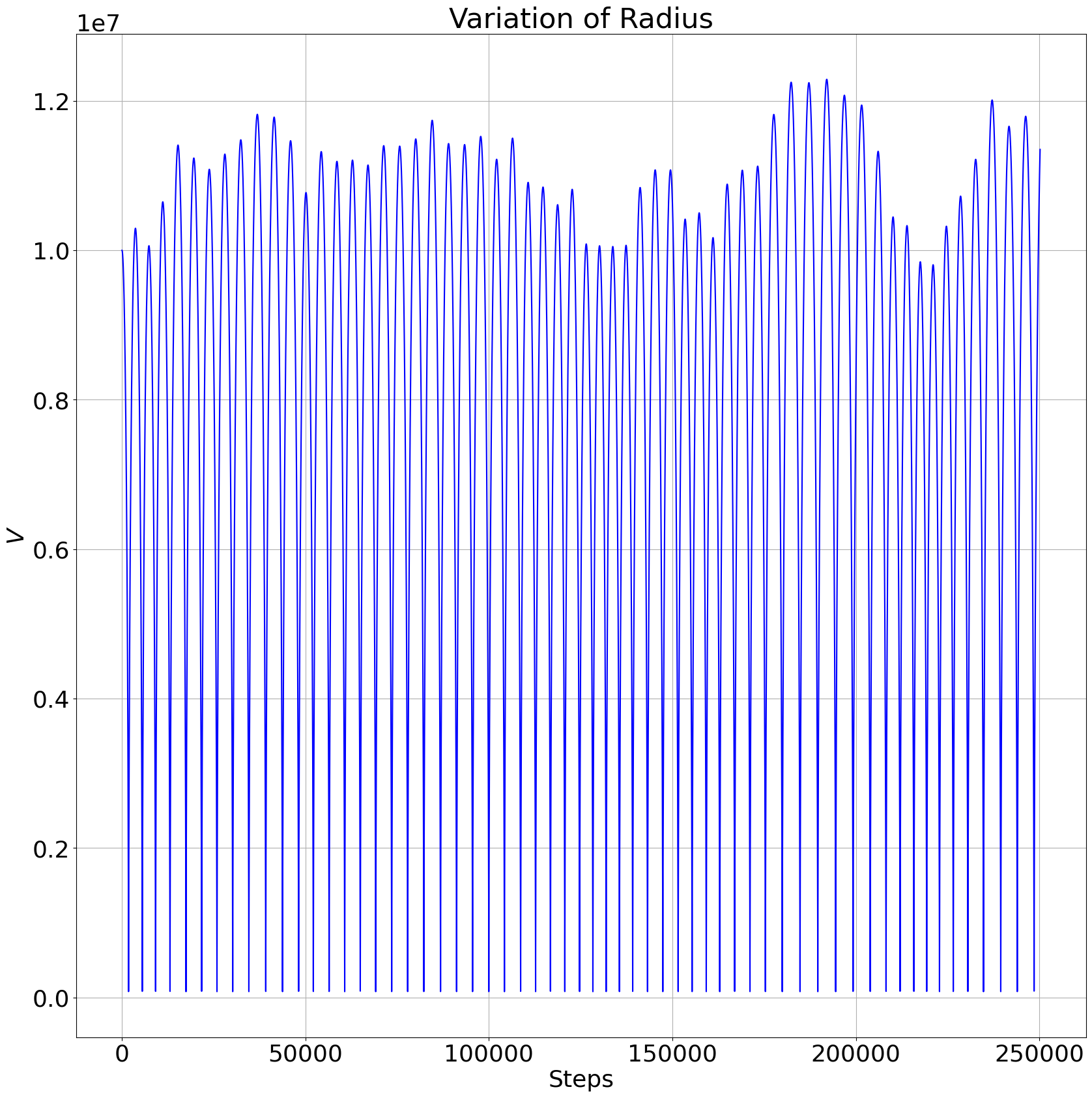}} 
    \subfigure[0.5 PN-with-Prop]{\includegraphics[width=0.24\textwidth]{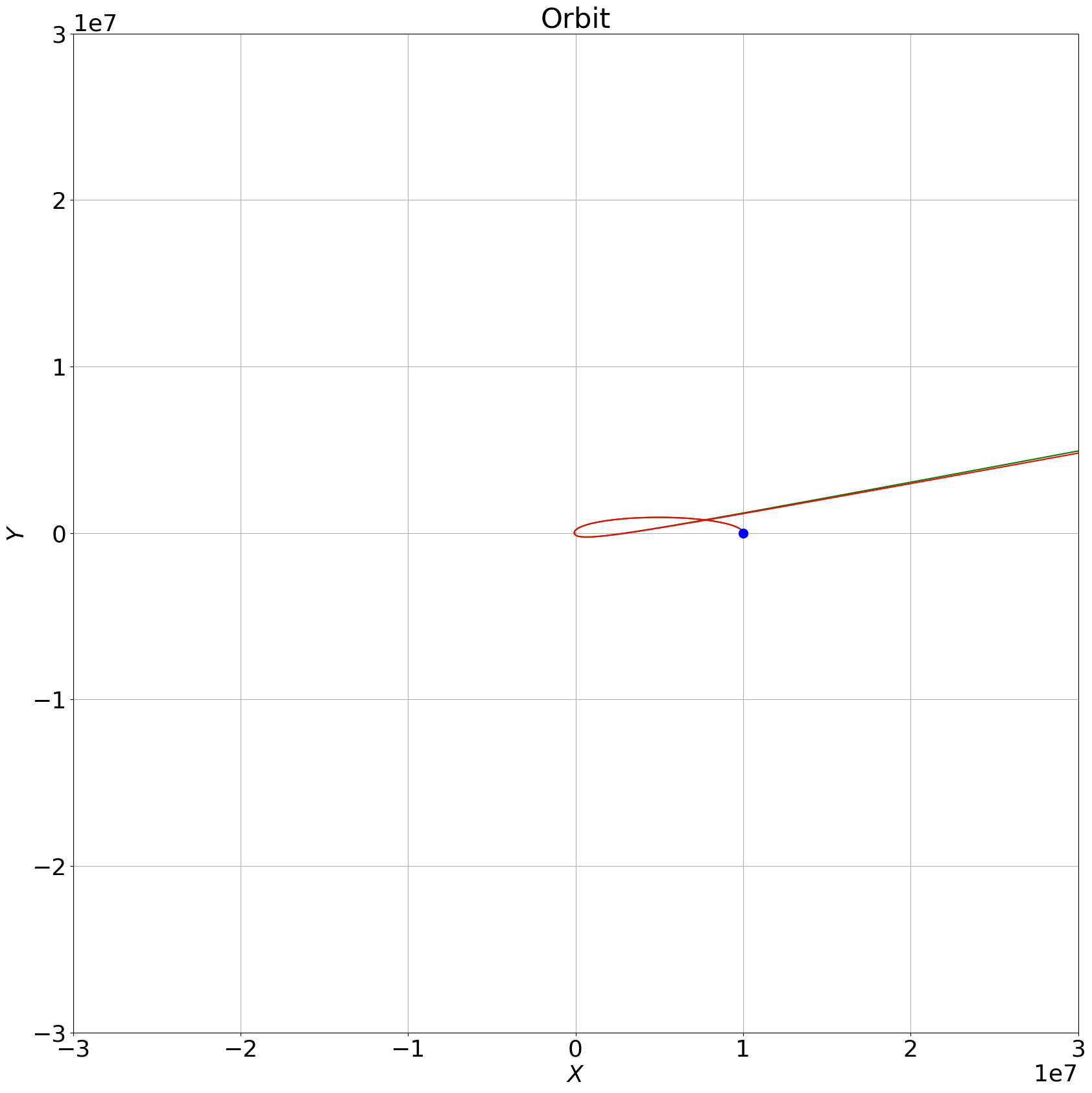}}
    \subfigure[0.5 PN-with-Prop]{\includegraphics[width=0.24\textwidth]{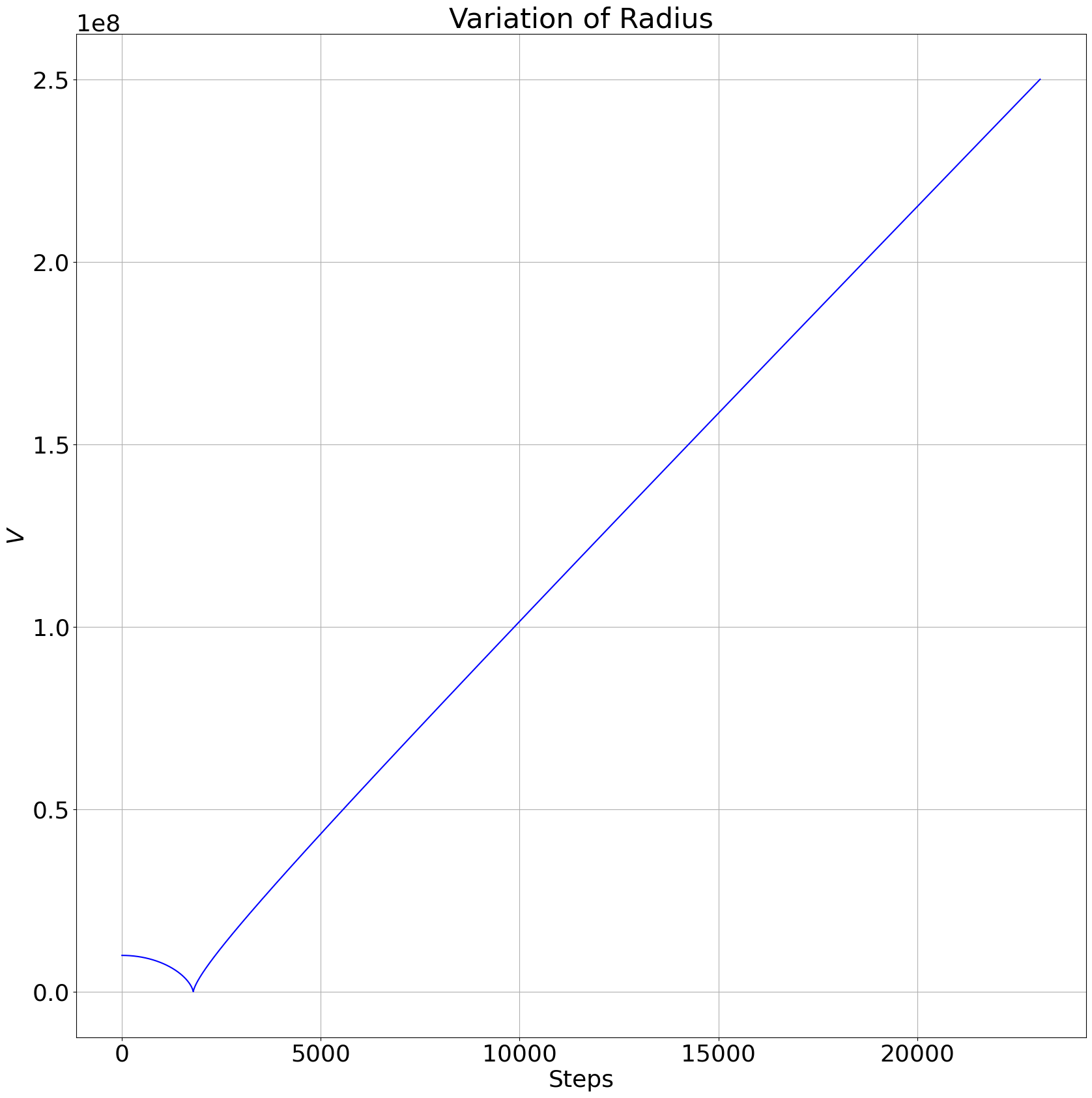}} \\
    
    \subfigure[1 PN-no-Prop]{\includegraphics[width=0.24\textwidth]{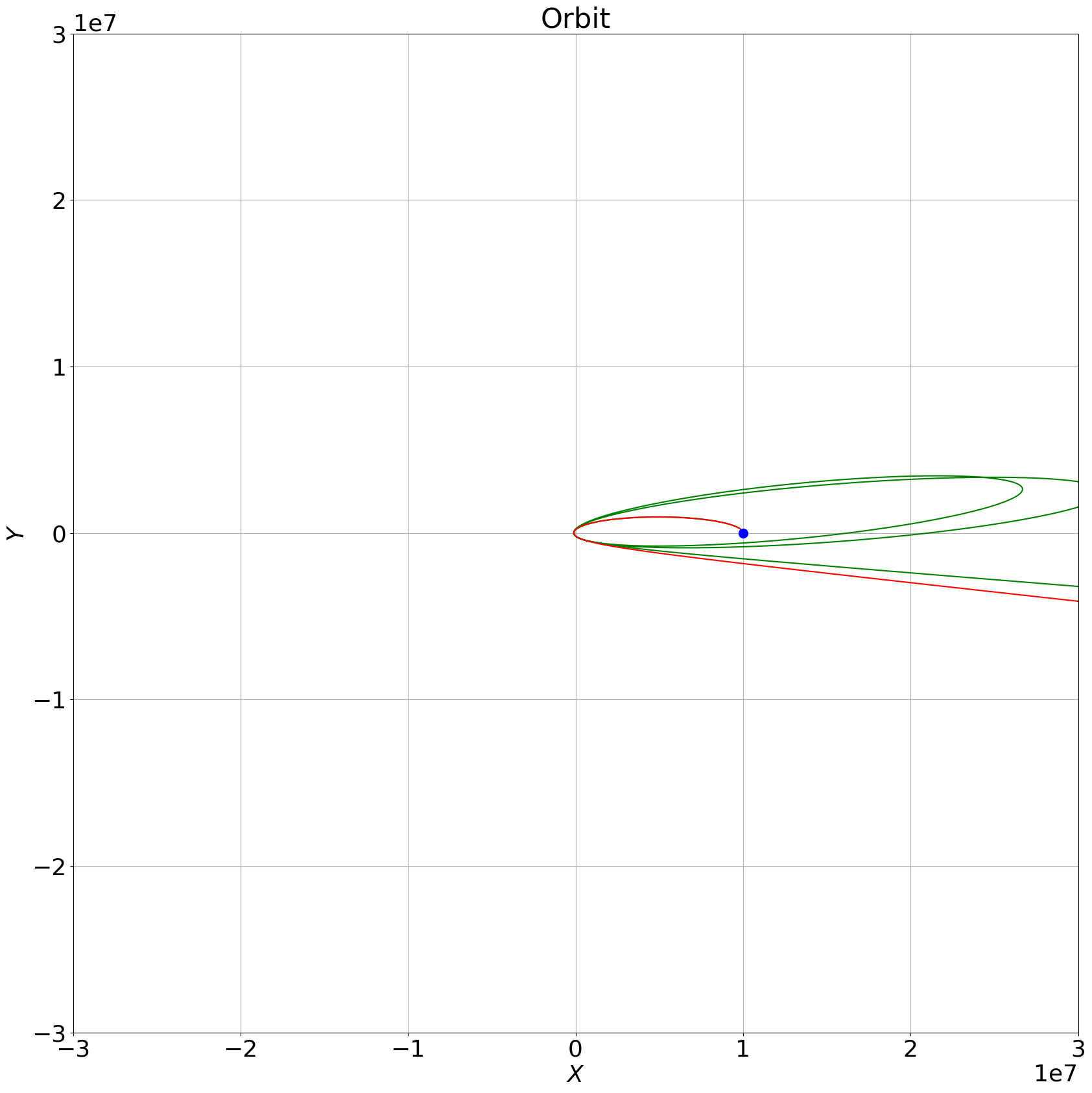}}
    \subfigure[1 PN-no-Prop]{\includegraphics[width=0.24\textwidth]{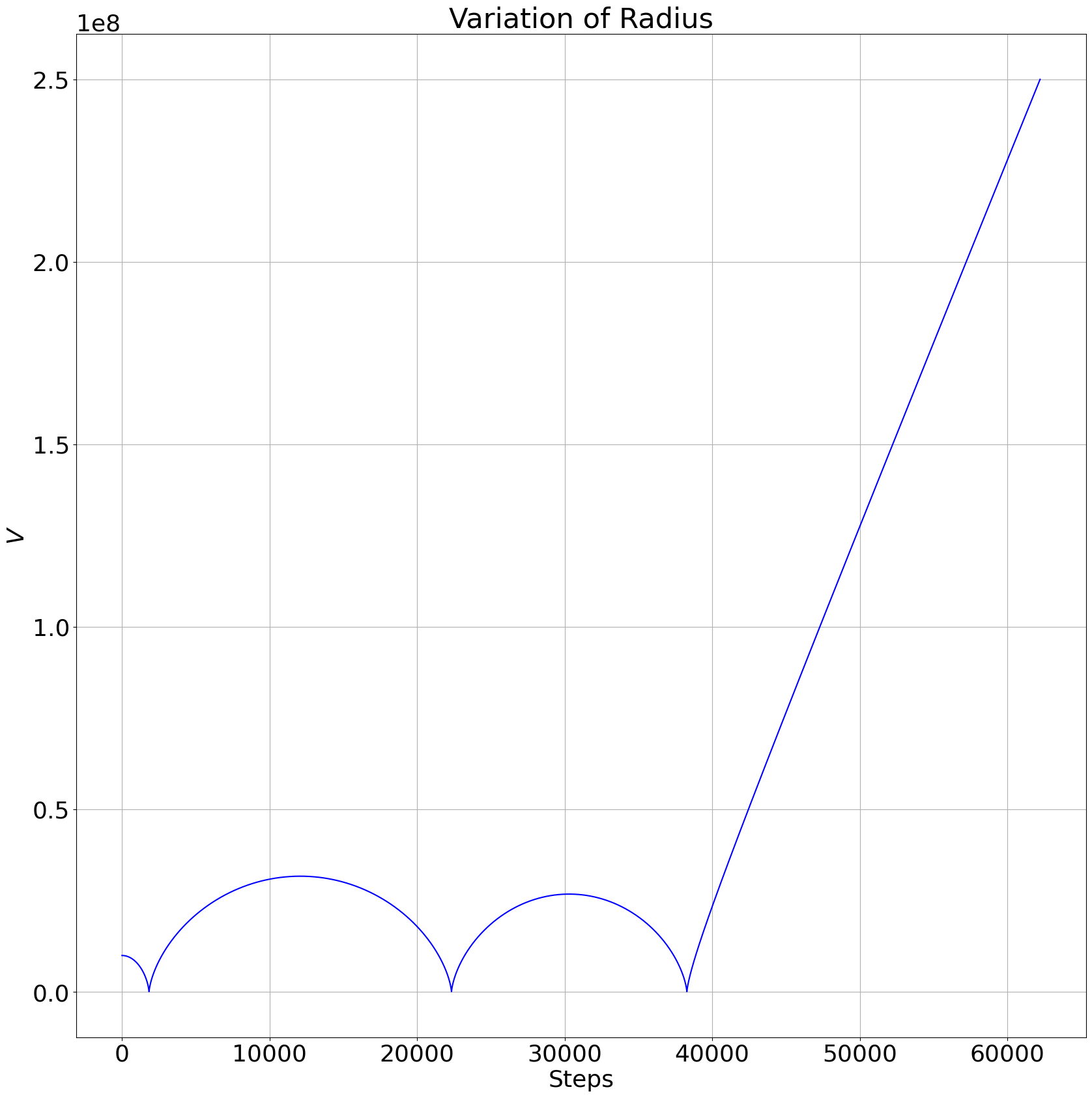}} 
    \subfigure[1 PN-with-Prop]{\includegraphics[width=0.24\textwidth]{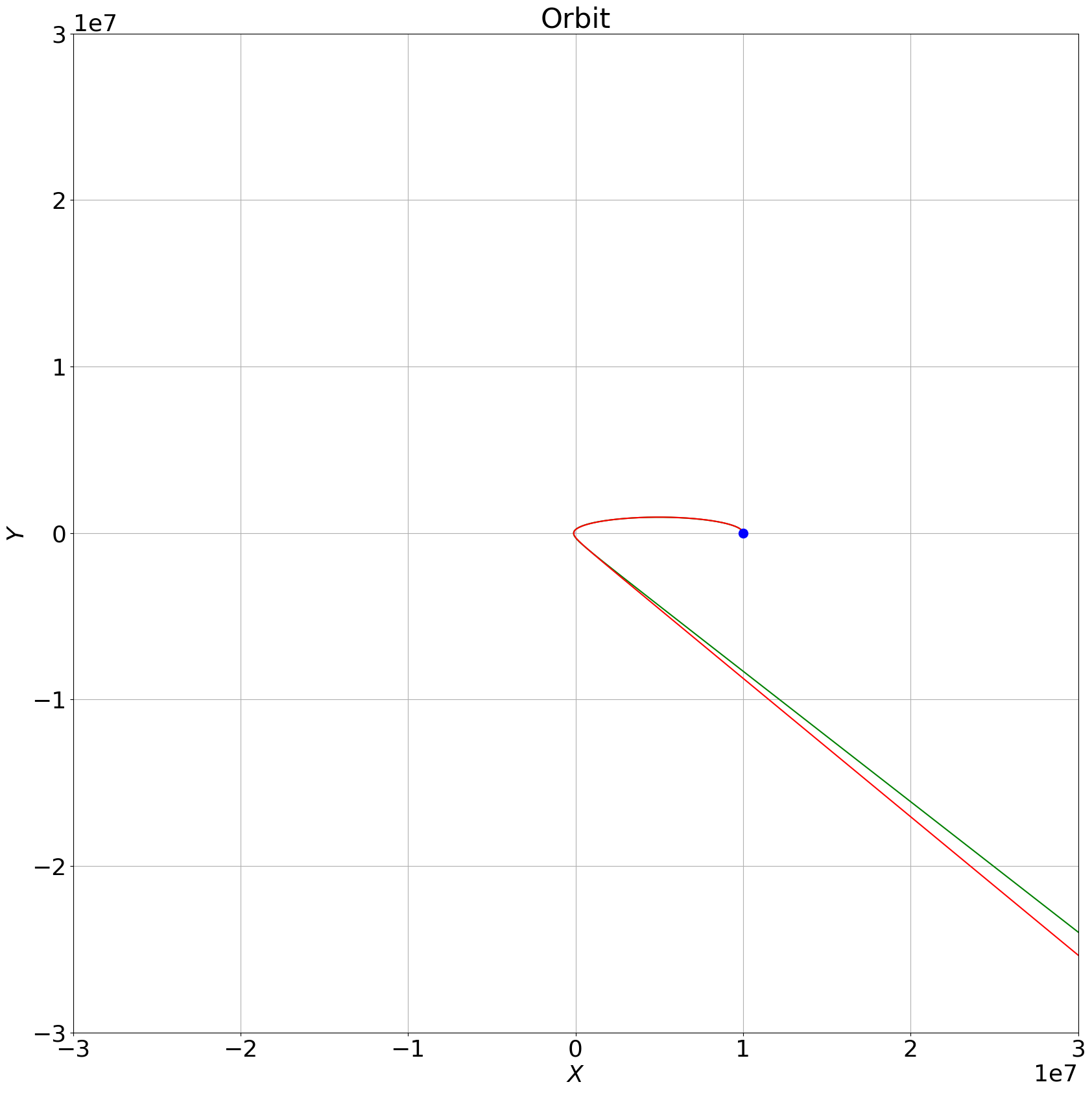}}
    \subfigure[1 PN-with-Prop]{\includegraphics[width=0.24\textwidth]{1PN-NF-R4-x7-Vel700-radius.png}} \\
    
    \subfigure[2 PN-no-Prop]{\includegraphics[width=0.24\textwidth]{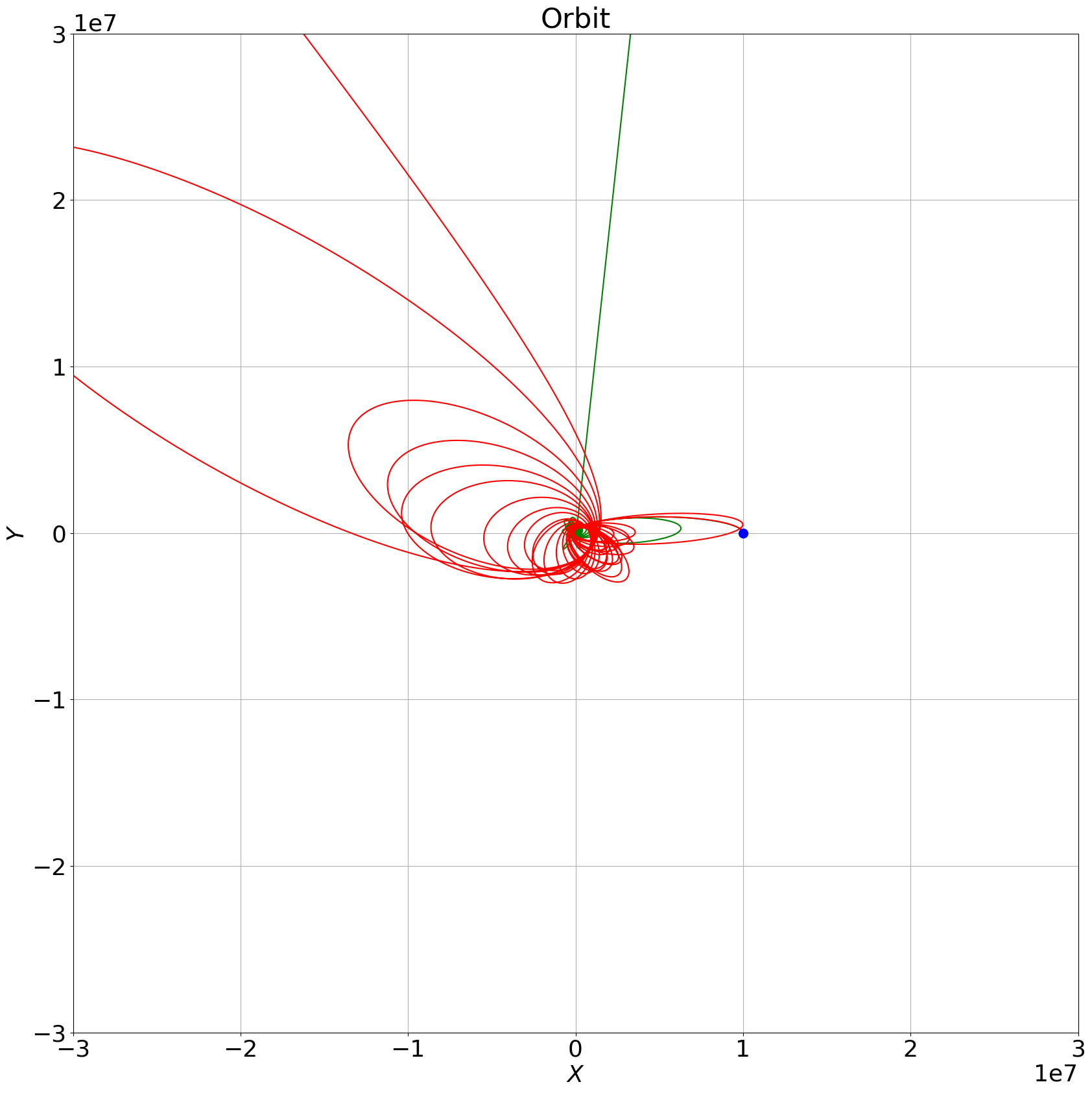}}
    \subfigure[2 PN-no-Prop]{\includegraphics[width=0.24\textwidth]{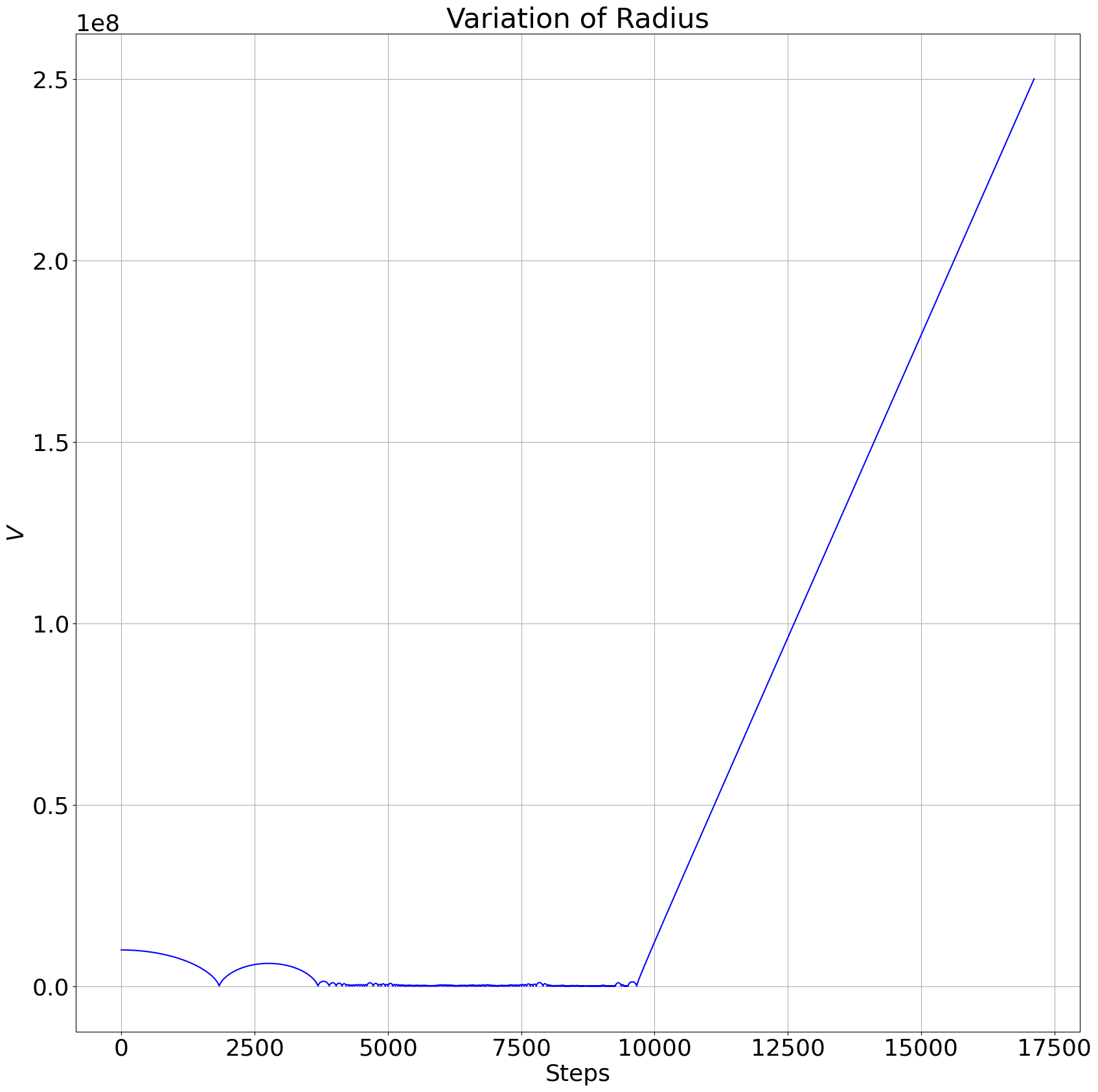}} 
    \subfigure[2 PN-with-Prop]{\includegraphics[width=0.24\textwidth]{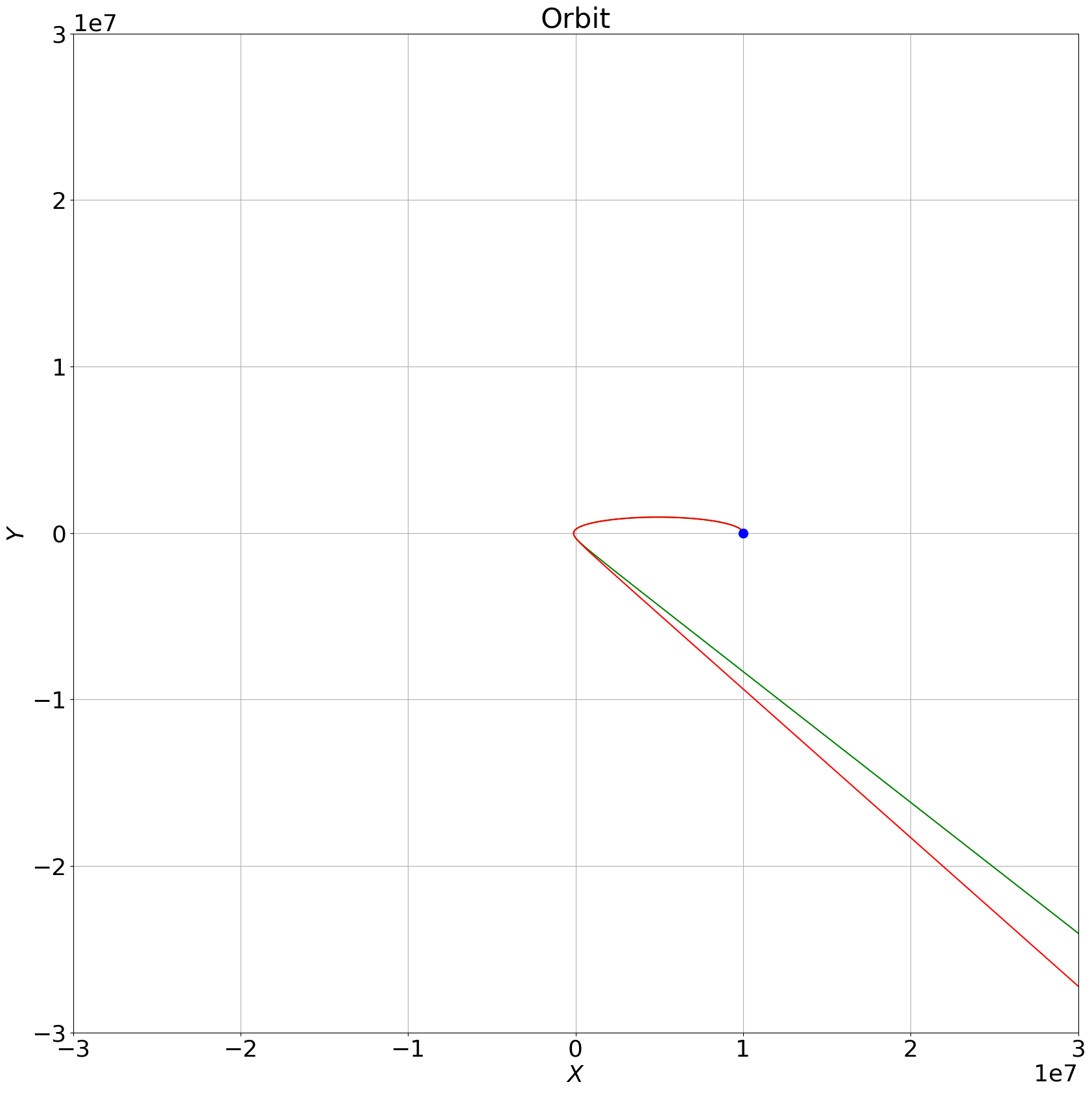}}
    \subfigure[2 PN-with-Prop]{\includegraphics[width=0.24\textwidth]{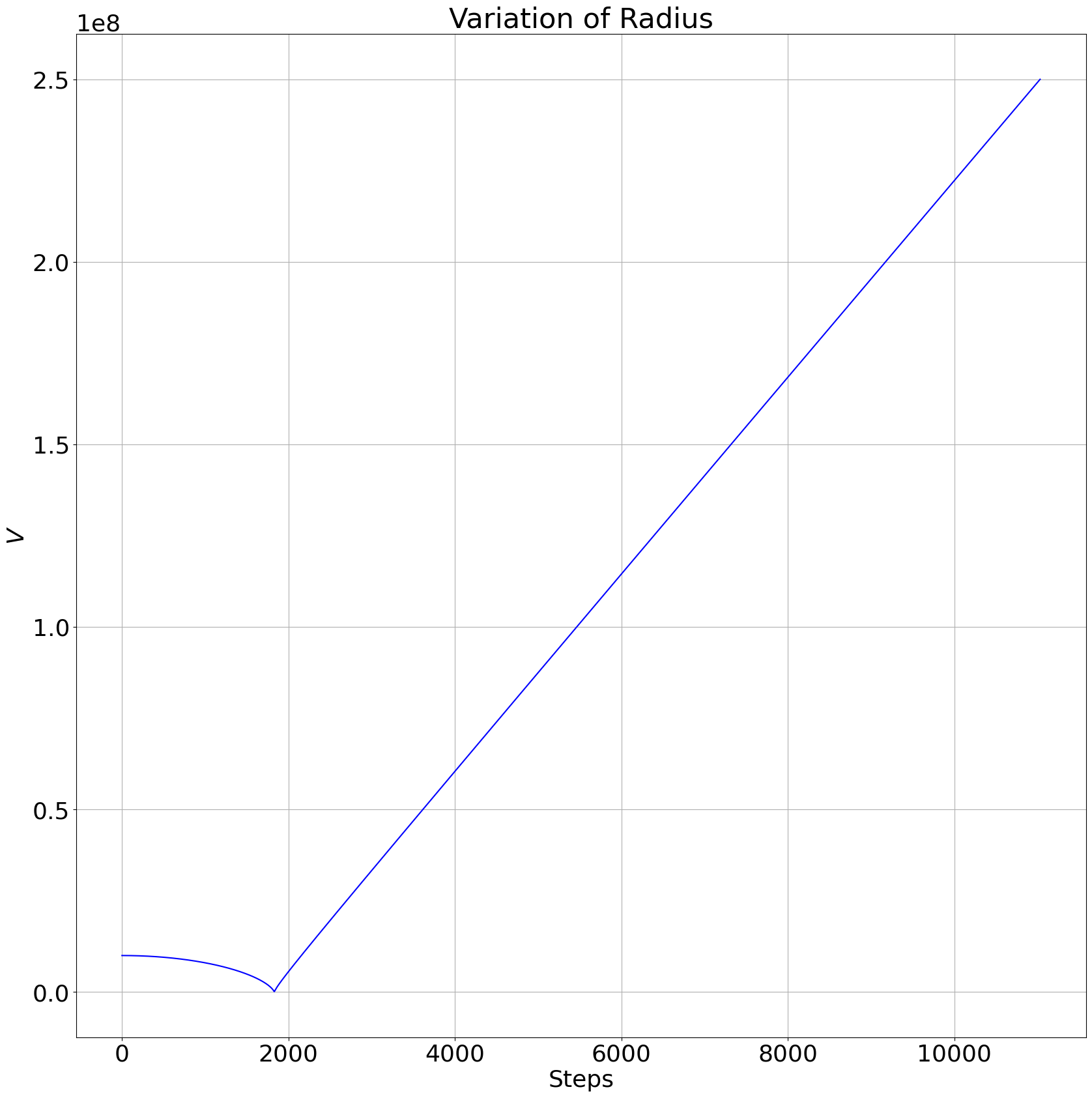}} \\
    
    \subfigure[2.5 PN-no-Prop]{\includegraphics[width=0.24\textwidth]{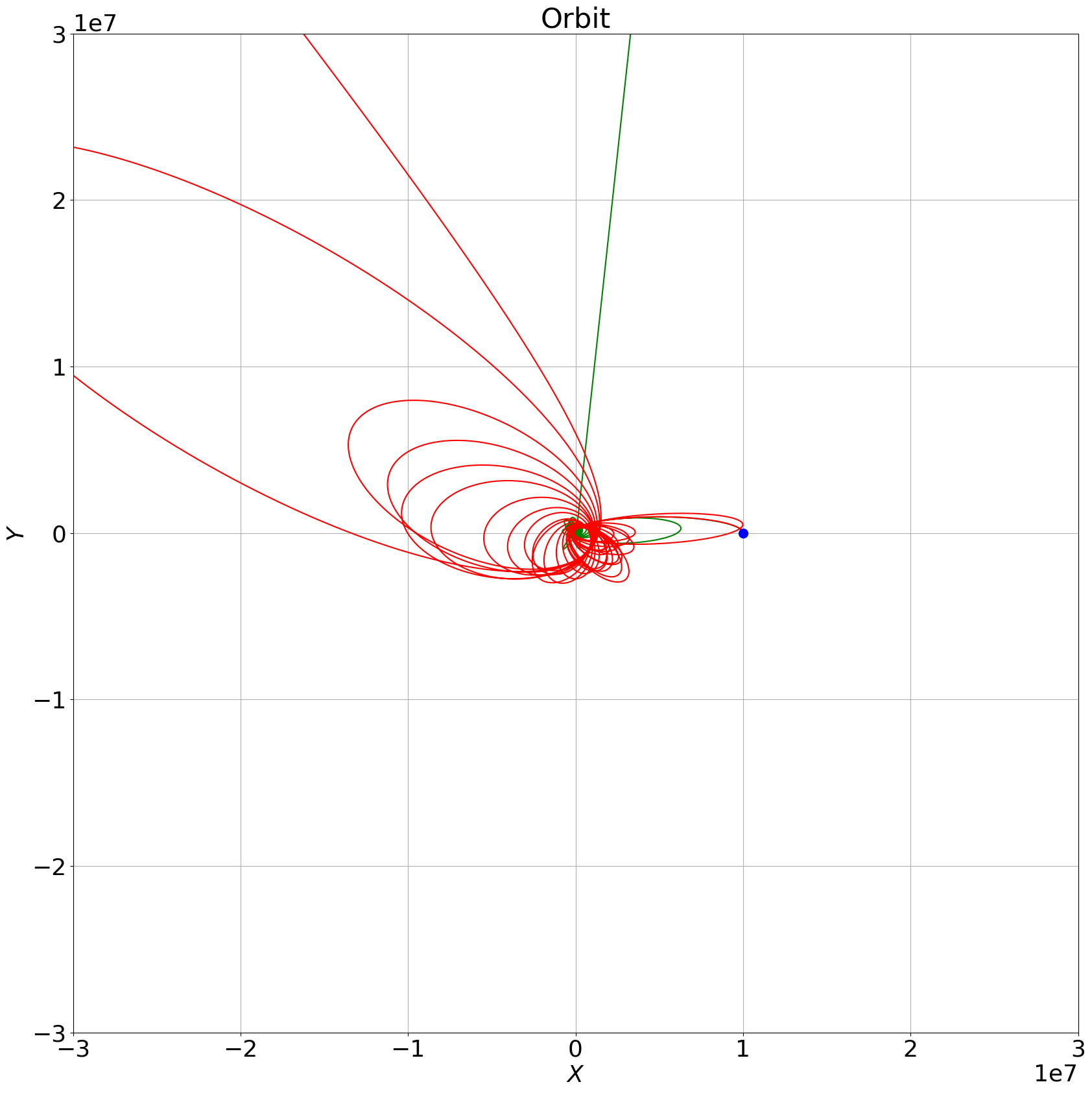}}
    \subfigure[2.5 PN-no-Prop]{\includegraphics[width=0.24\textwidth]{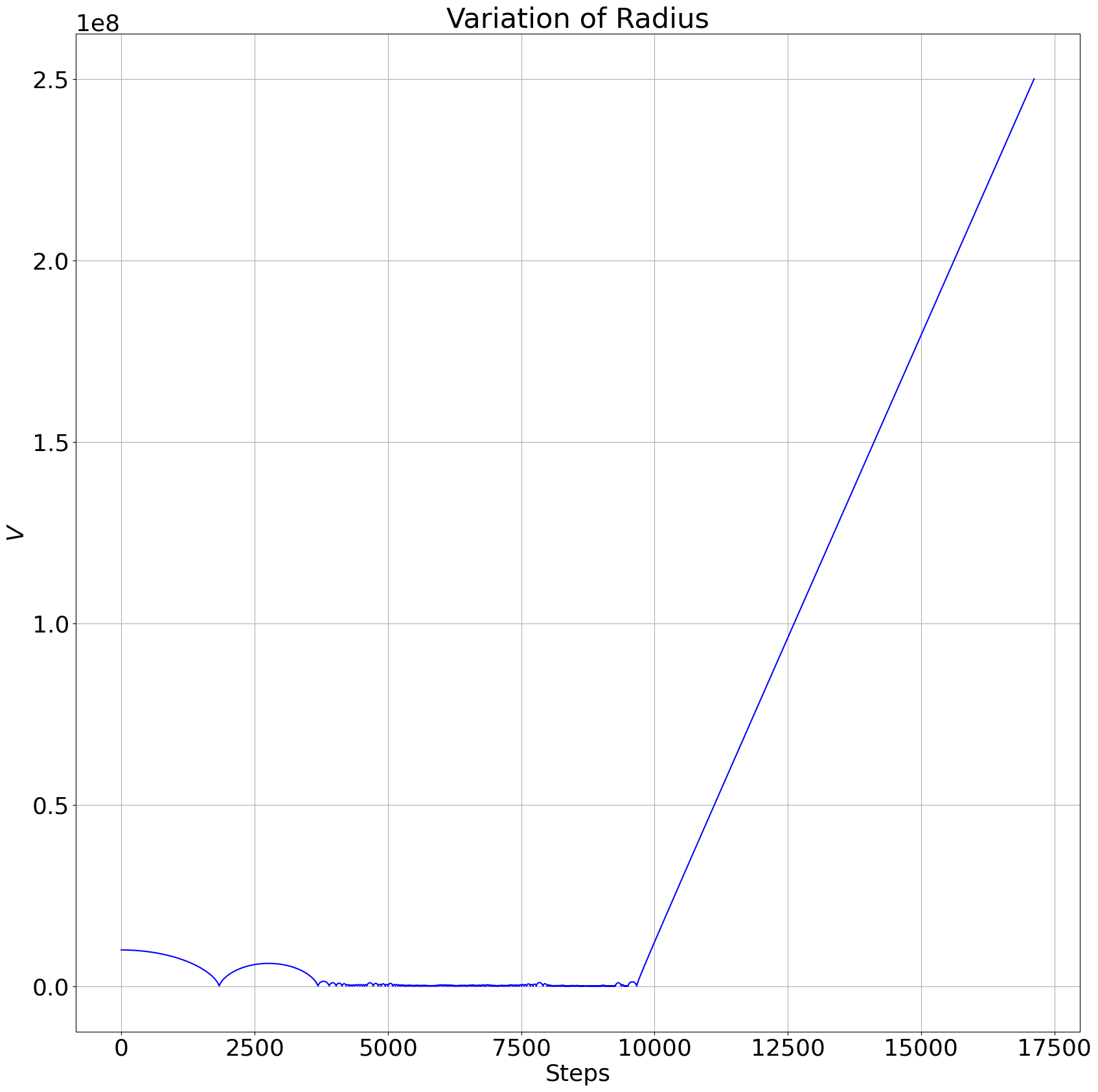}} 
    \subfigure[2.5 PN-with-Prop]{\includegraphics[width=0.24\textwidth]{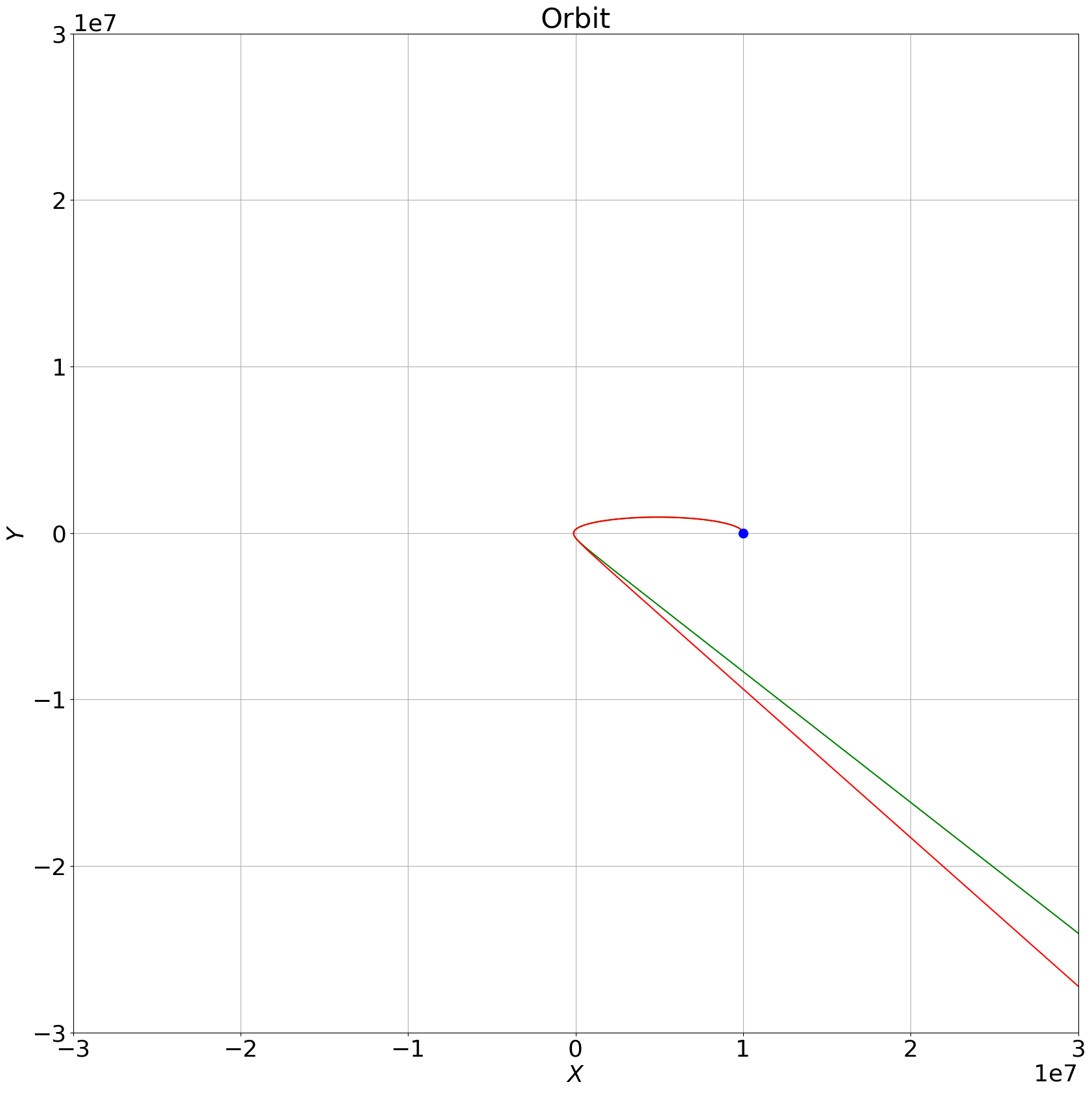}}
    \subfigure[2.5 PN-with-Prop]{\includegraphics[width=0.24\textwidth]{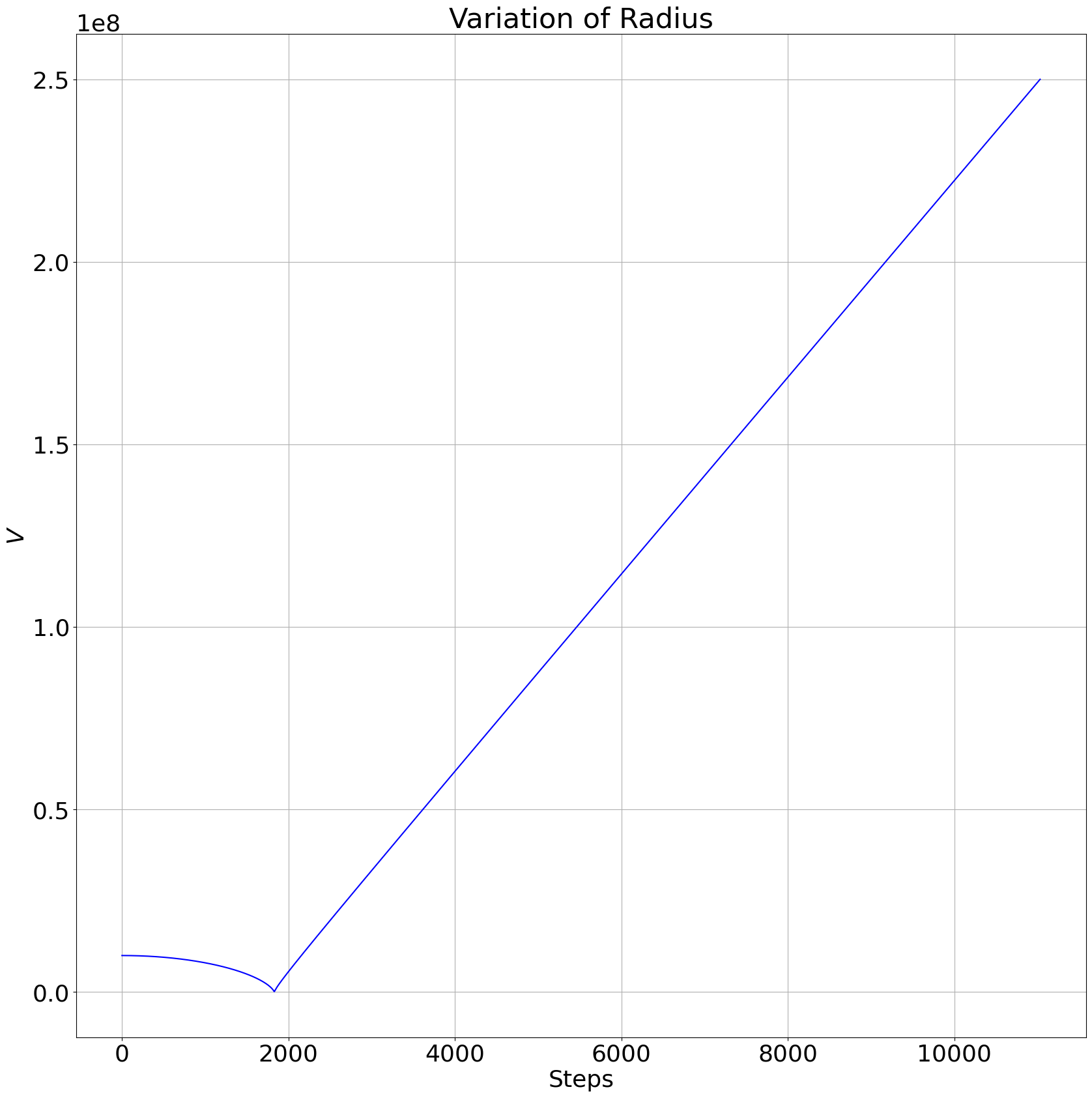}} \\
    \caption{Intermediate Field, R = $1.1 \cdot 10^4m$, X = $10^7m$, Velocity = 700,000$ms^{-1}$}
    \label{intermediate}
\end{figure}
\begin{figure}[!ht]
    \centering
    \setcounter{subfigure}{0}
    \subfigure[0 PN-no-Prop]{\includegraphics[width=0.24\textwidth]{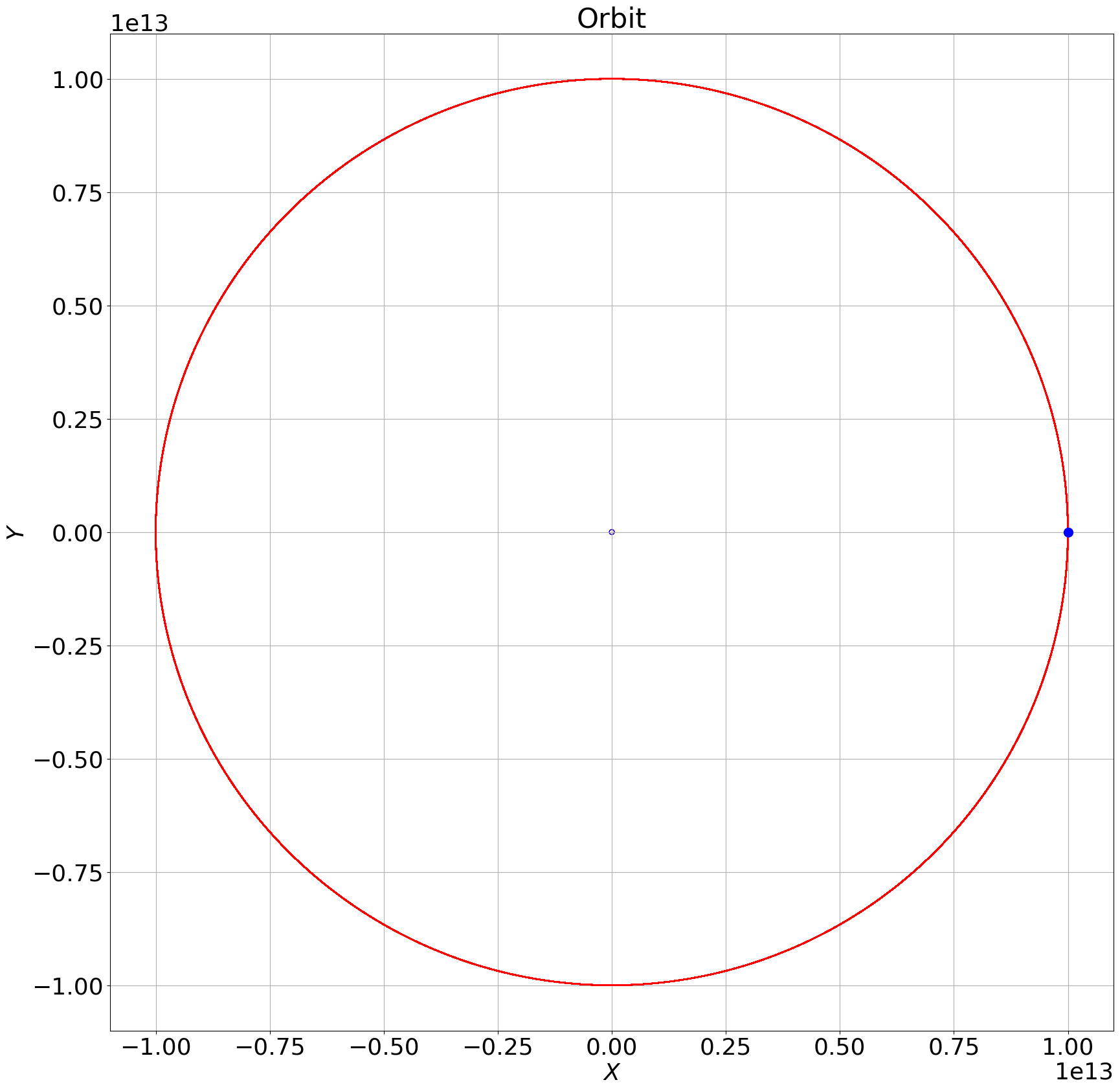}}
    \subfigure[0 PN-no-Prop]{\includegraphics[width=0.24\textwidth]{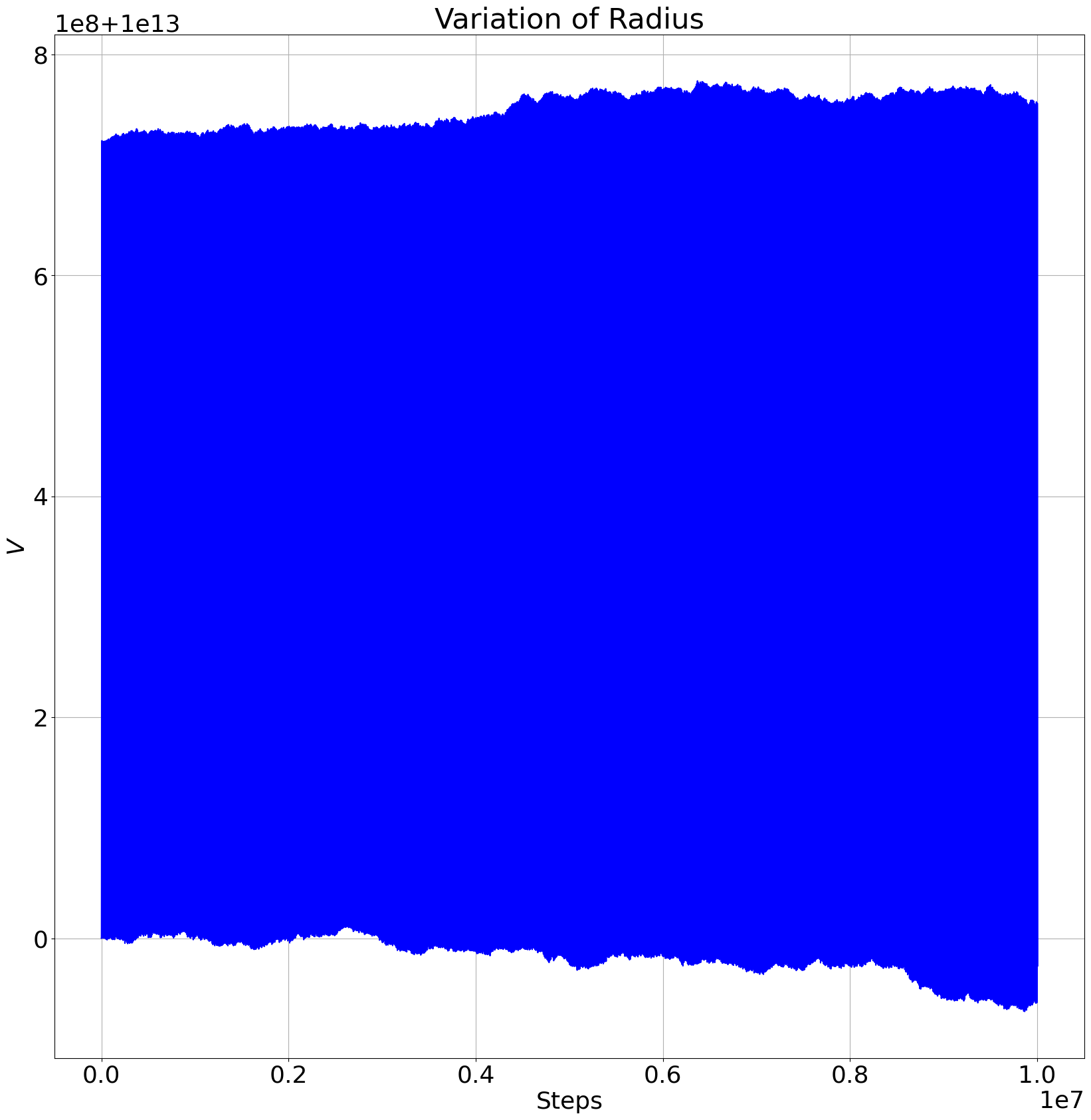}} 
    \subfigure[0 PN-with-Prop]{\includegraphics[width=0.24\textwidth]{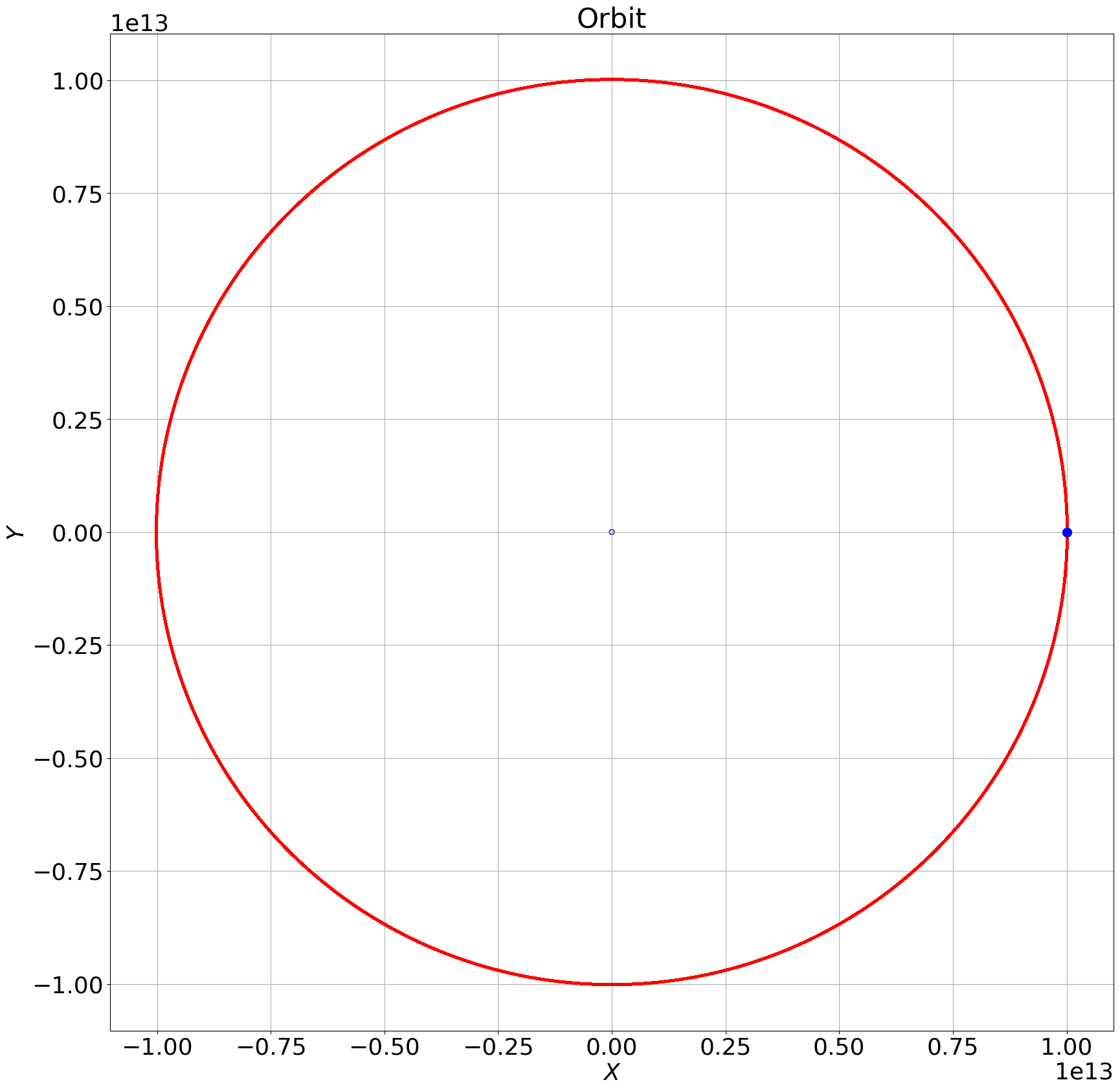}}
    \subfigure[0 PN-with-Prop]{\includegraphics[width=0.24\textwidth]{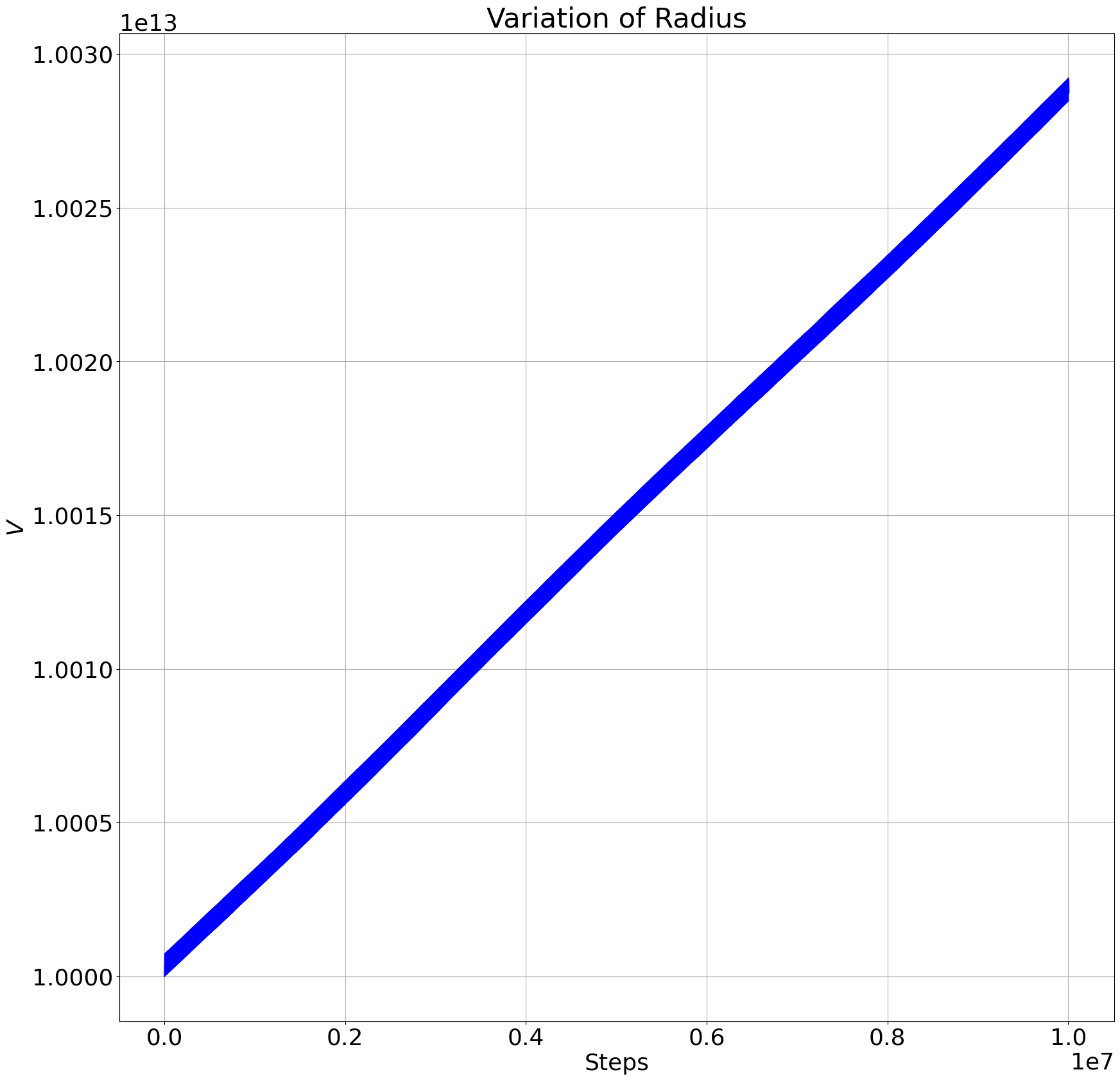}} \\
    
    \subfigure[0.5 PN-no-Prop]{\includegraphics[width=0.24\textwidth]{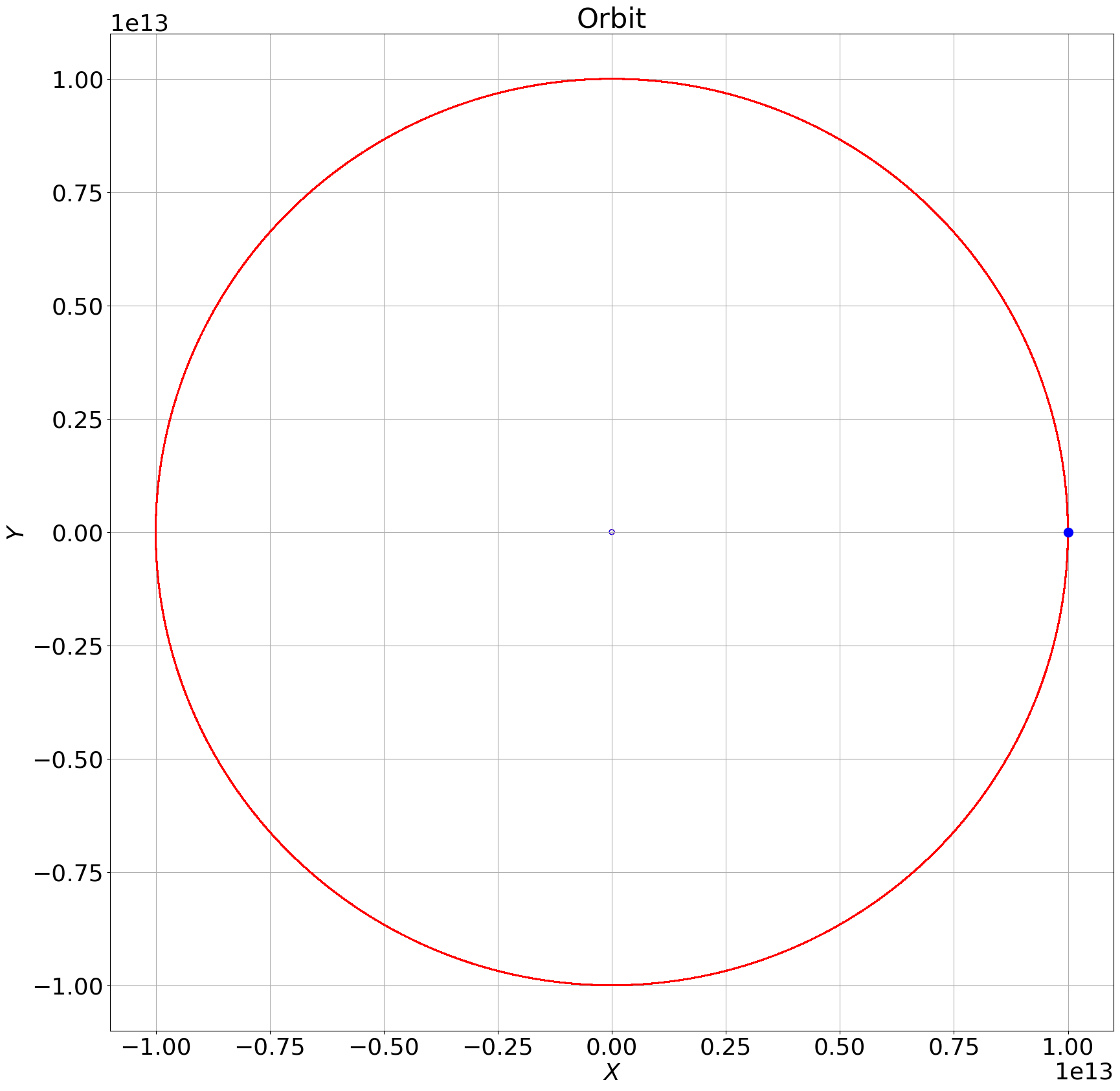}}
    \subfigure[0.5 PN-no-Prop]{\includegraphics[width=0.24\textwidth]{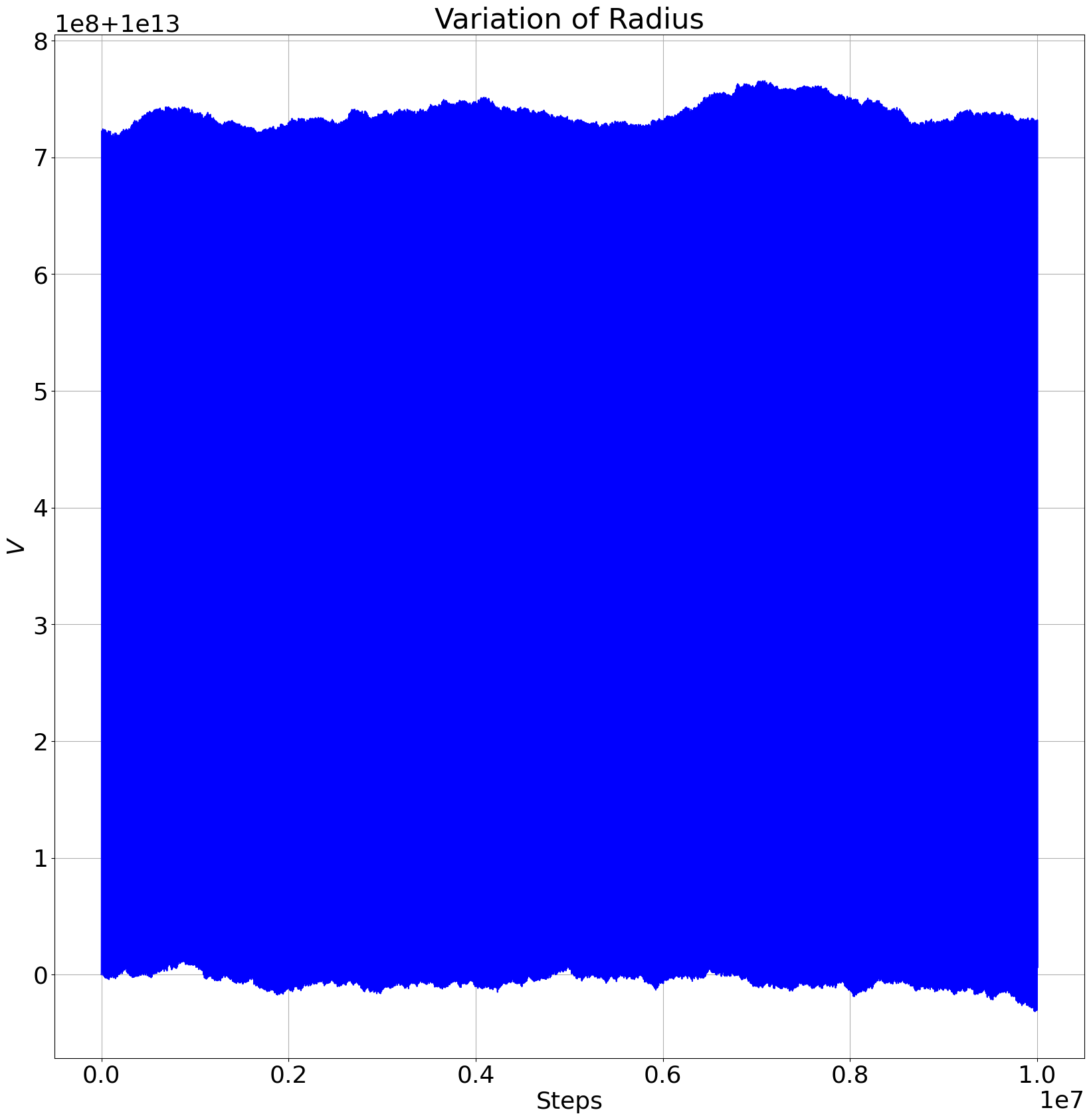}} 
    \subfigure[0.5 PN-with-Prop]{\includegraphics[width=0.24\textwidth]{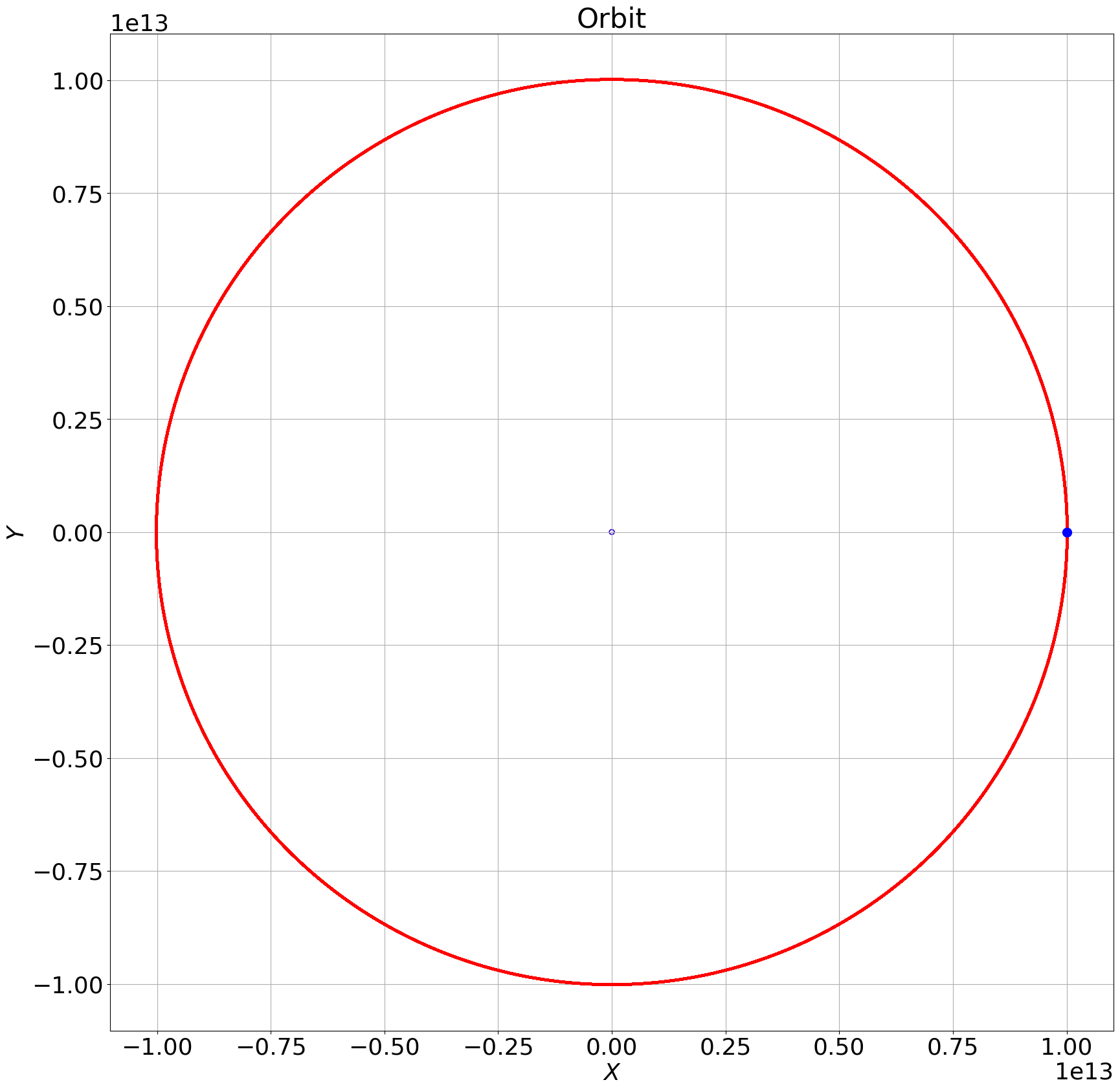}}
    \subfigure[0.5 PN-with-Prop]{\includegraphics[width=0.24\textwidth]{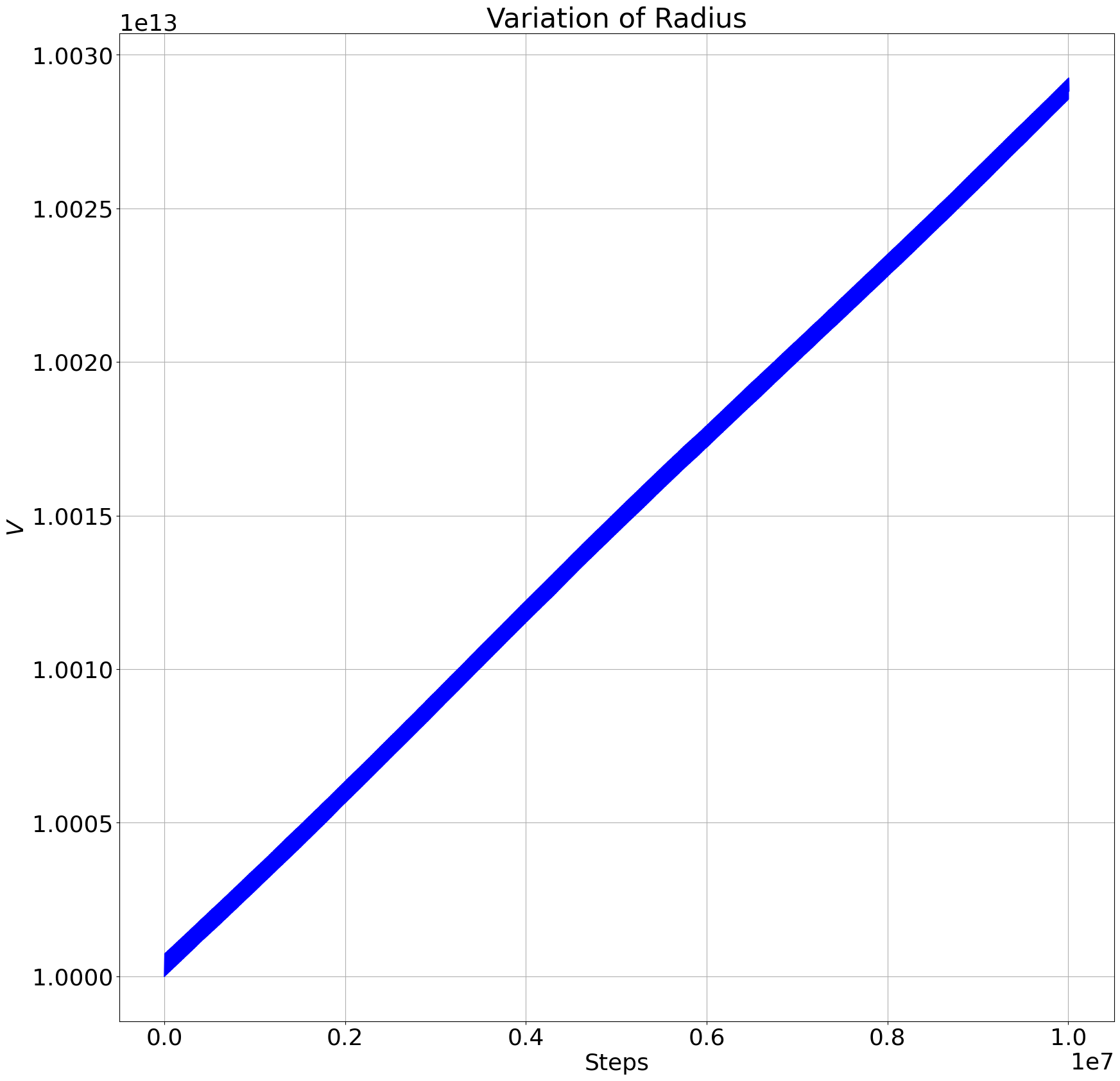}} \\
    
    \subfigure[1 PN-no-Prop]{\includegraphics[width=0.24\textwidth]{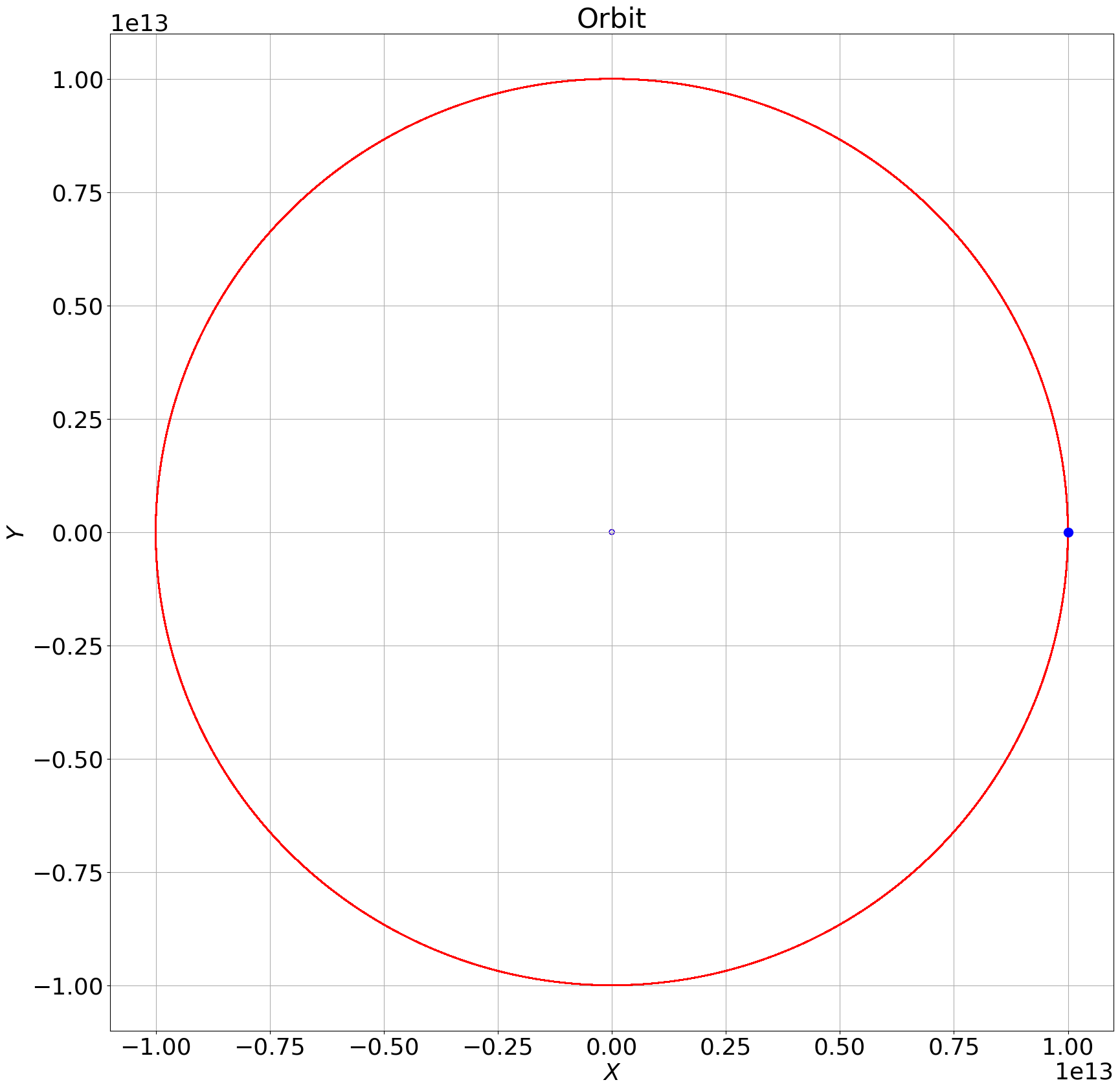}}
    \subfigure[1 PN-no-Prop]{\includegraphics[width=0.24\textwidth]{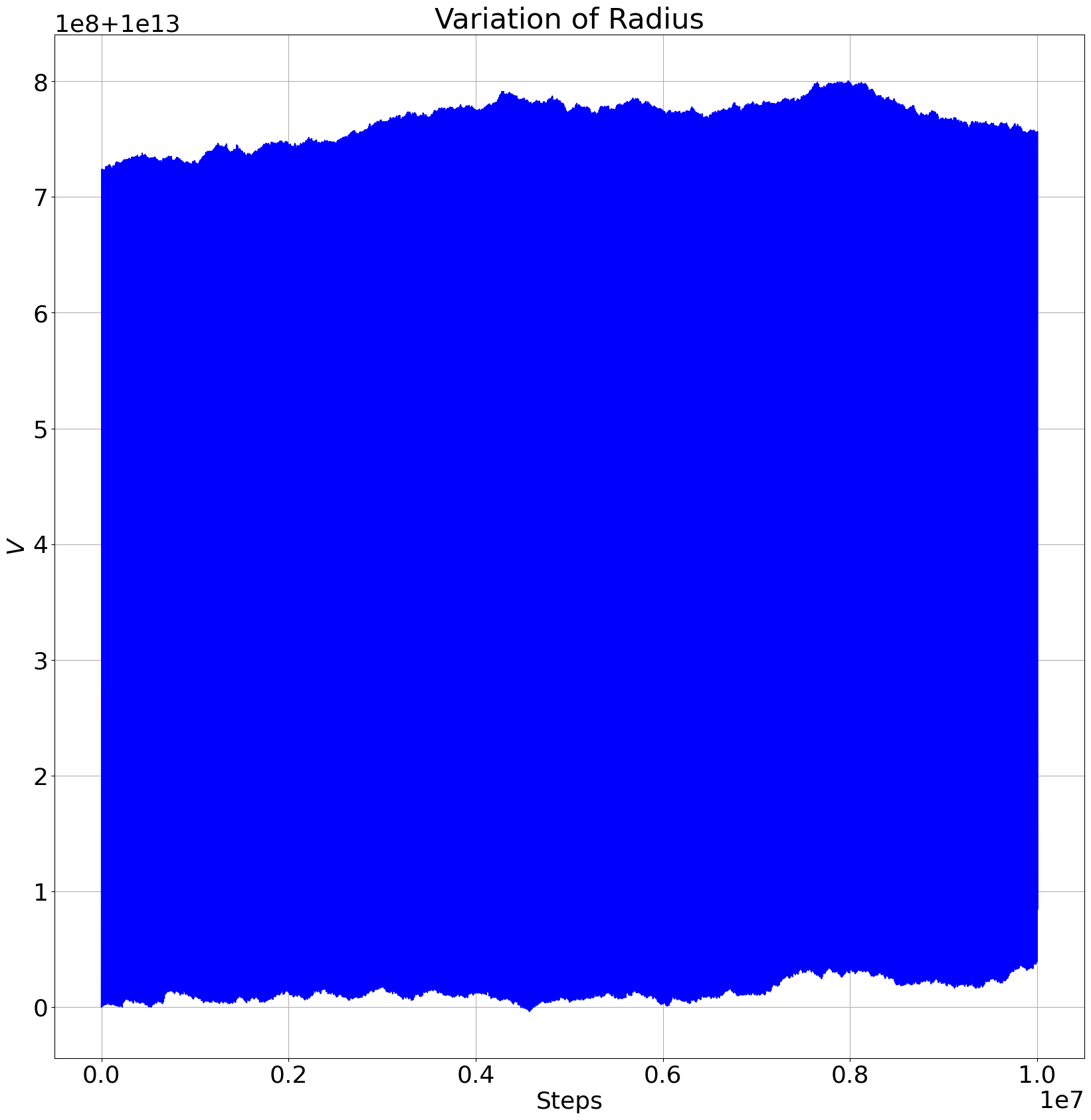}} 
    \subfigure[1 PN-with-Prop]{\includegraphics[width=0.24\textwidth]{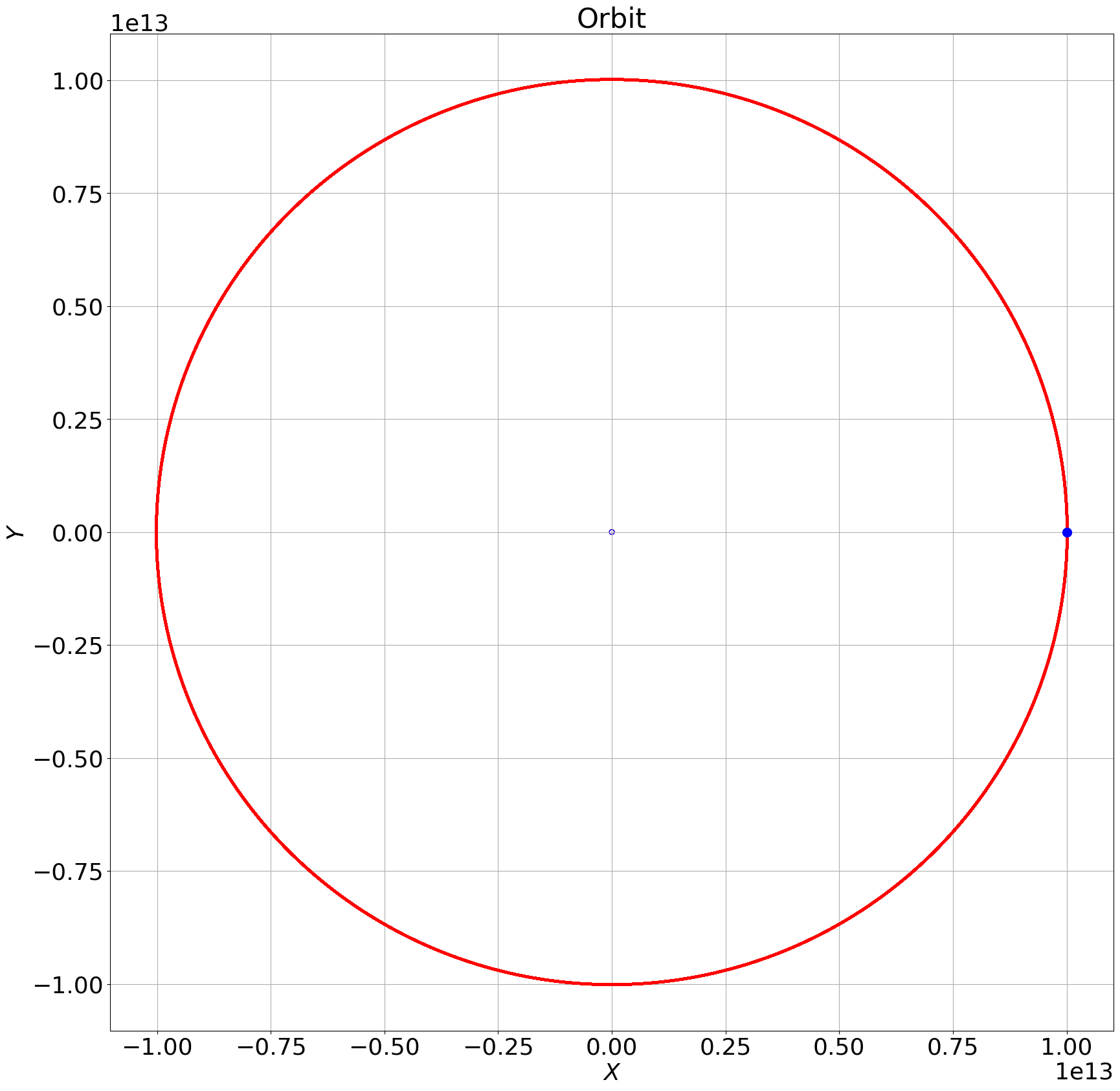}}
    \subfigure[1 PN-with-Prop]{\includegraphics[width=0.24\textwidth]{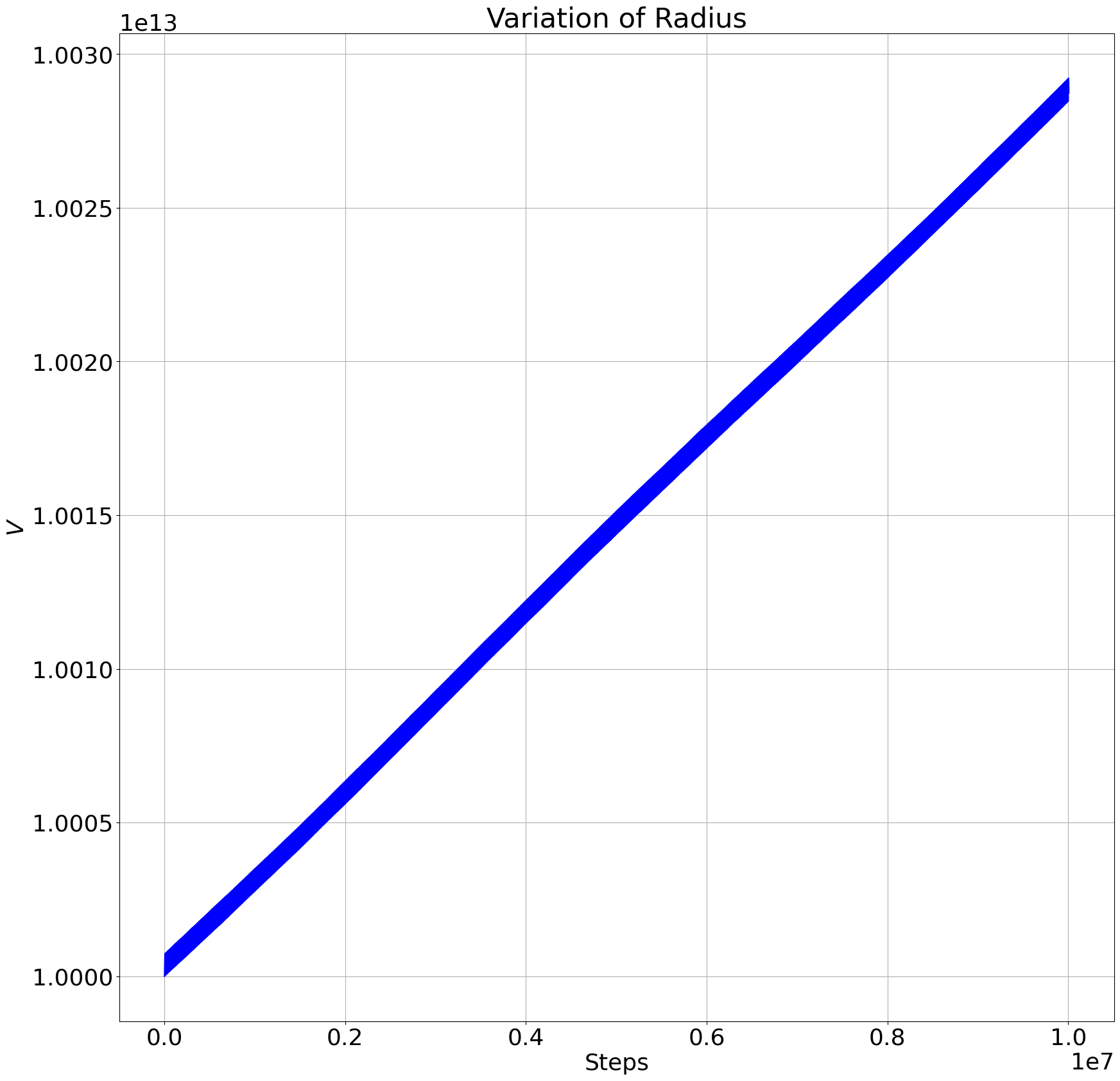}} \\
    
    \subfigure[2 PN-no-Prop]{\includegraphics[width=0.24\textwidth]{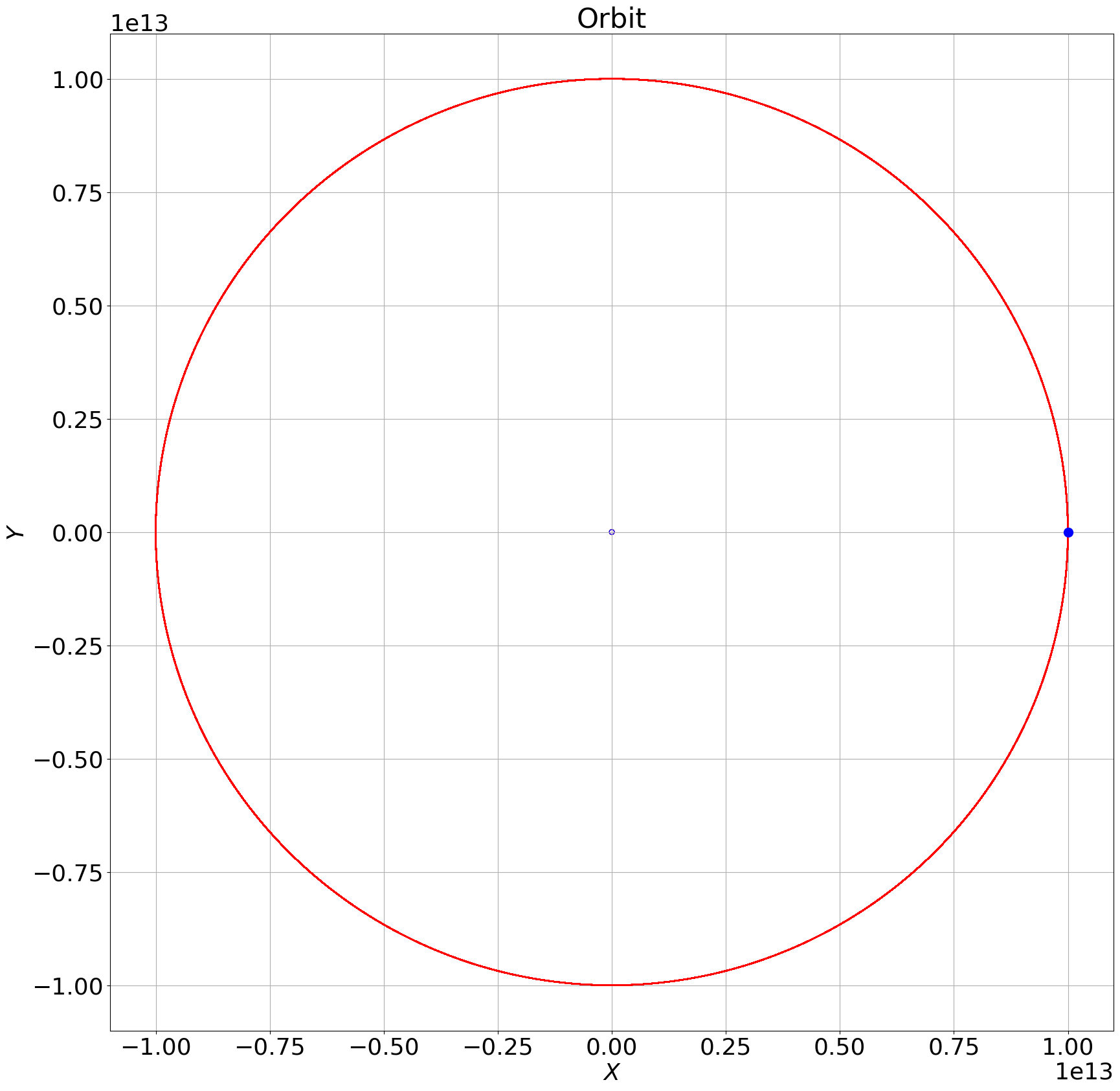}}
    \subfigure[2 PN-no-Prop]{\includegraphics[width=0.24\textwidth]{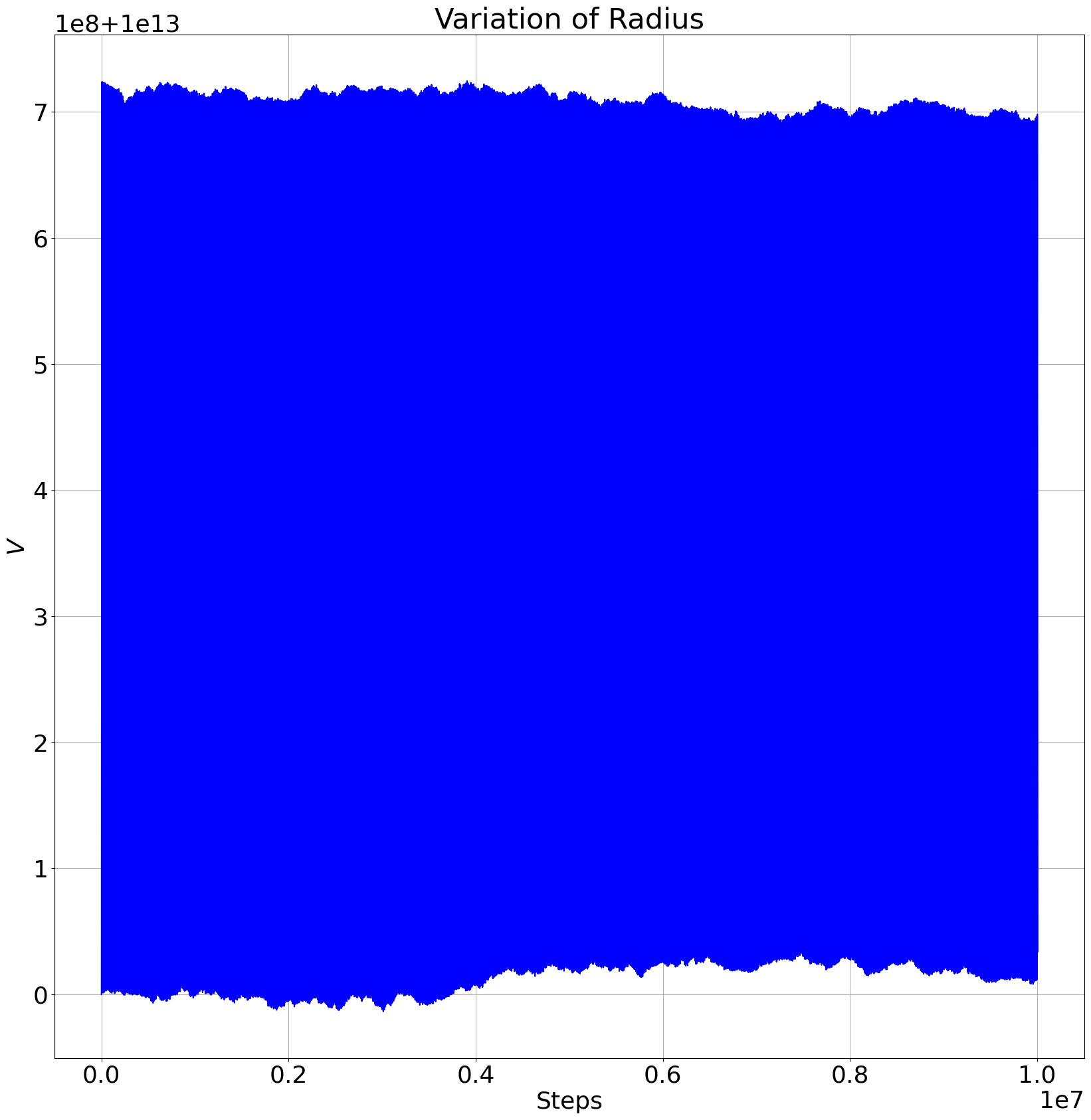}} 
    \subfigure[2 PN-with-Prop]{\includegraphics[width=0.24\textwidth]{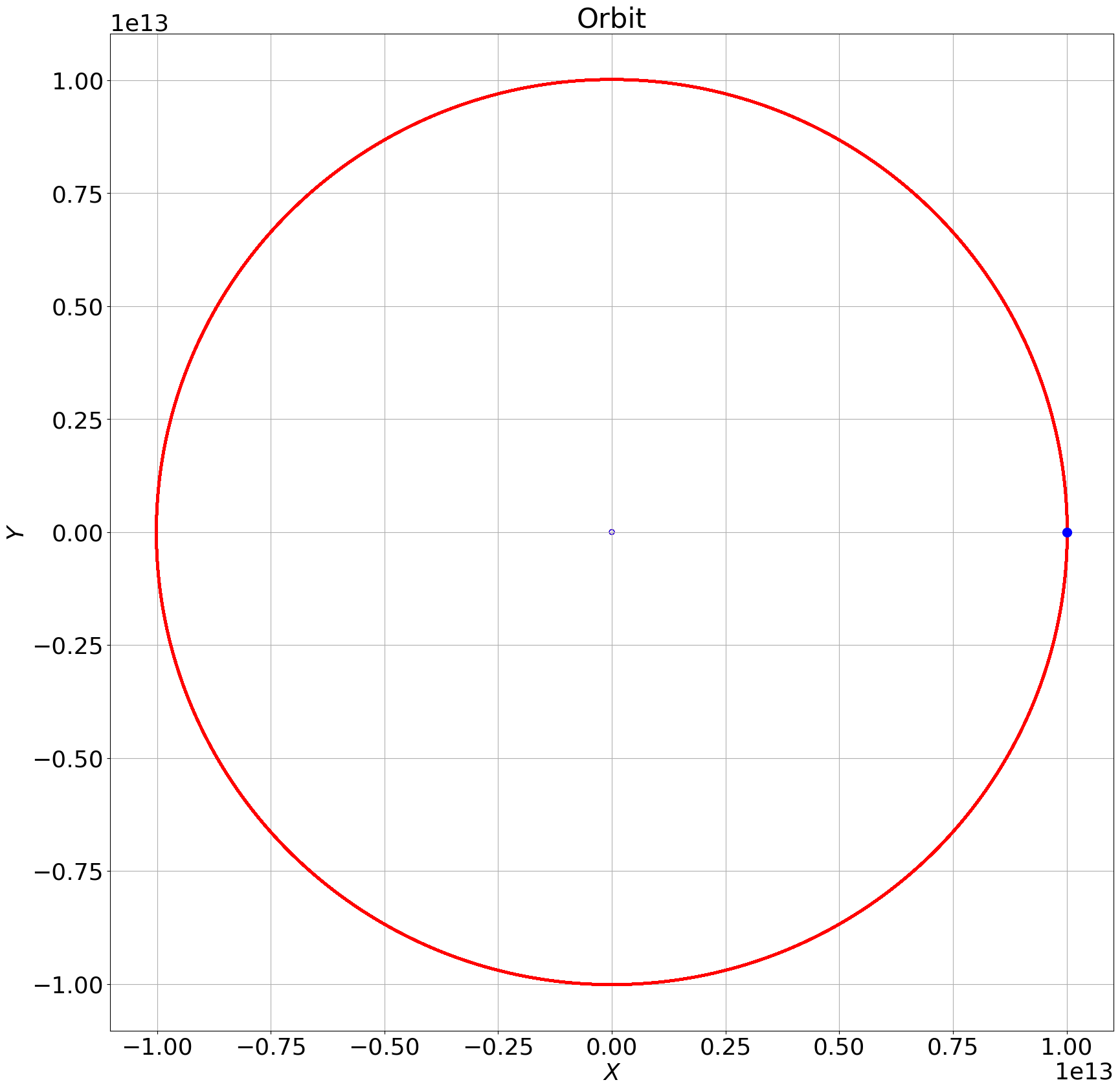}}
    \subfigure[2 PN-with-Prop]{\includegraphics[width=0.24\textwidth]{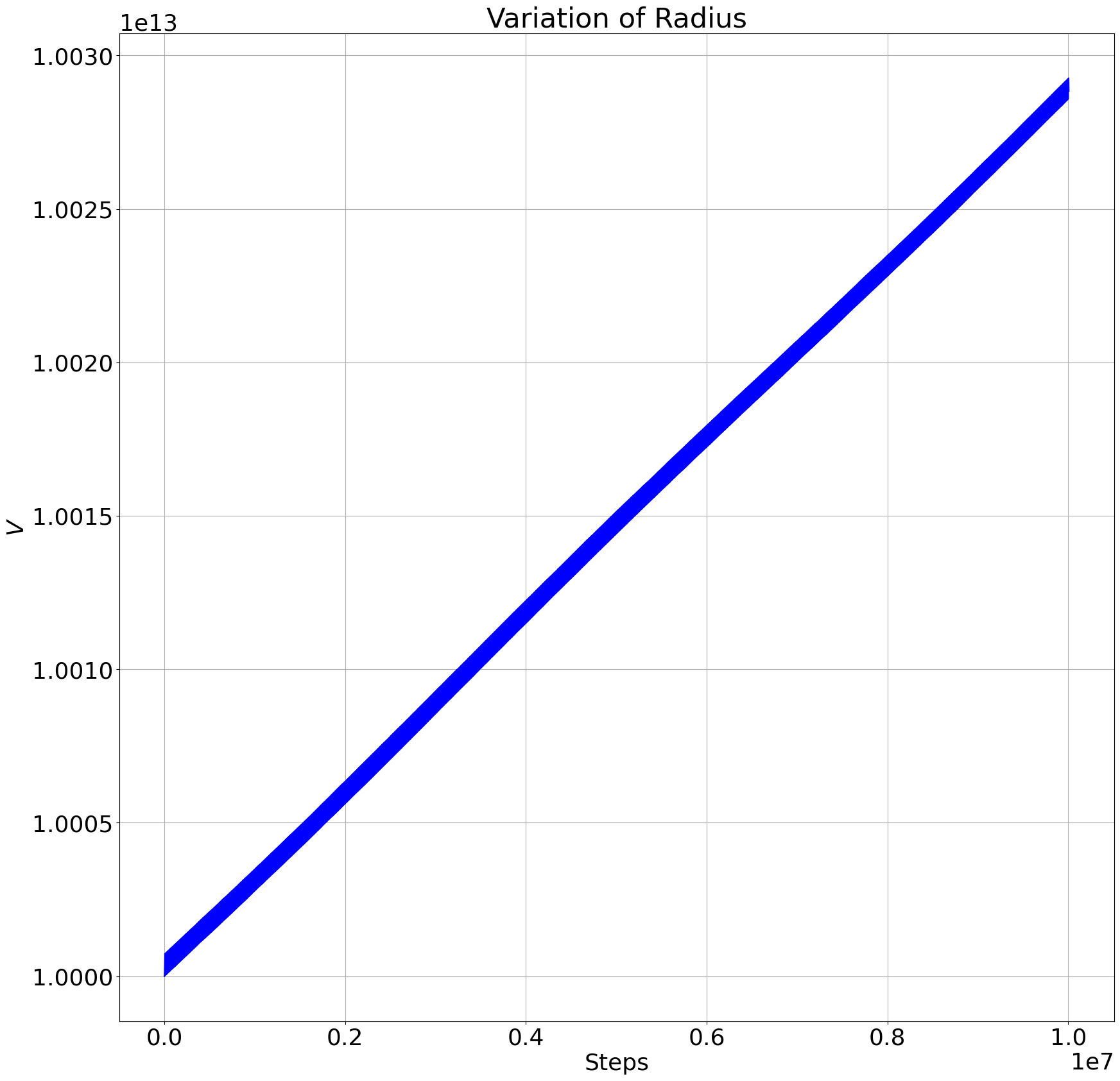}} \\
    
    \subfigure[2.5 PN-no-Prop]{\includegraphics[width=0.24\textwidth]{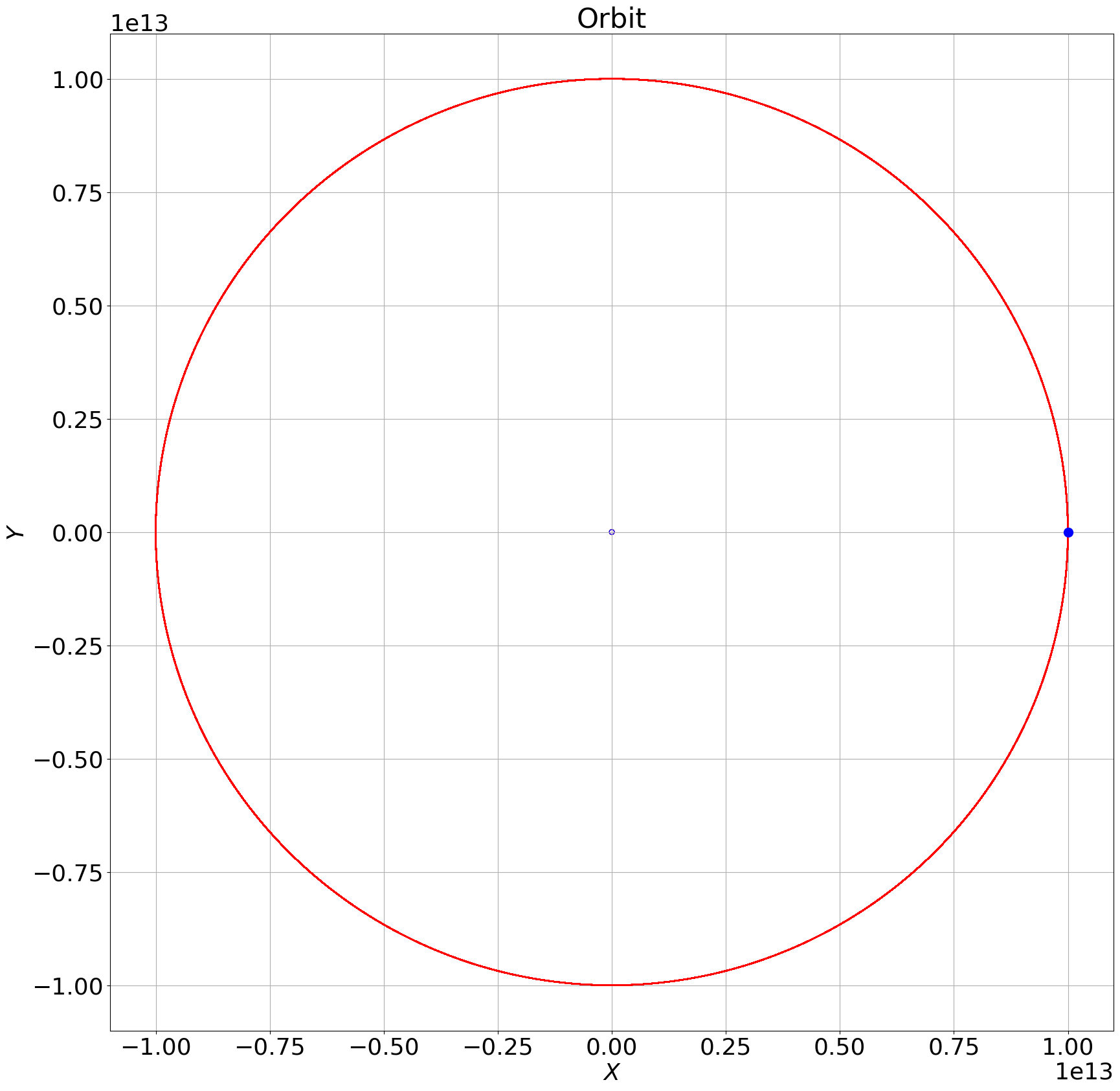}}
    \subfigure[2.5 PN-no-Prop]{\includegraphics[width=0.24\textwidth]{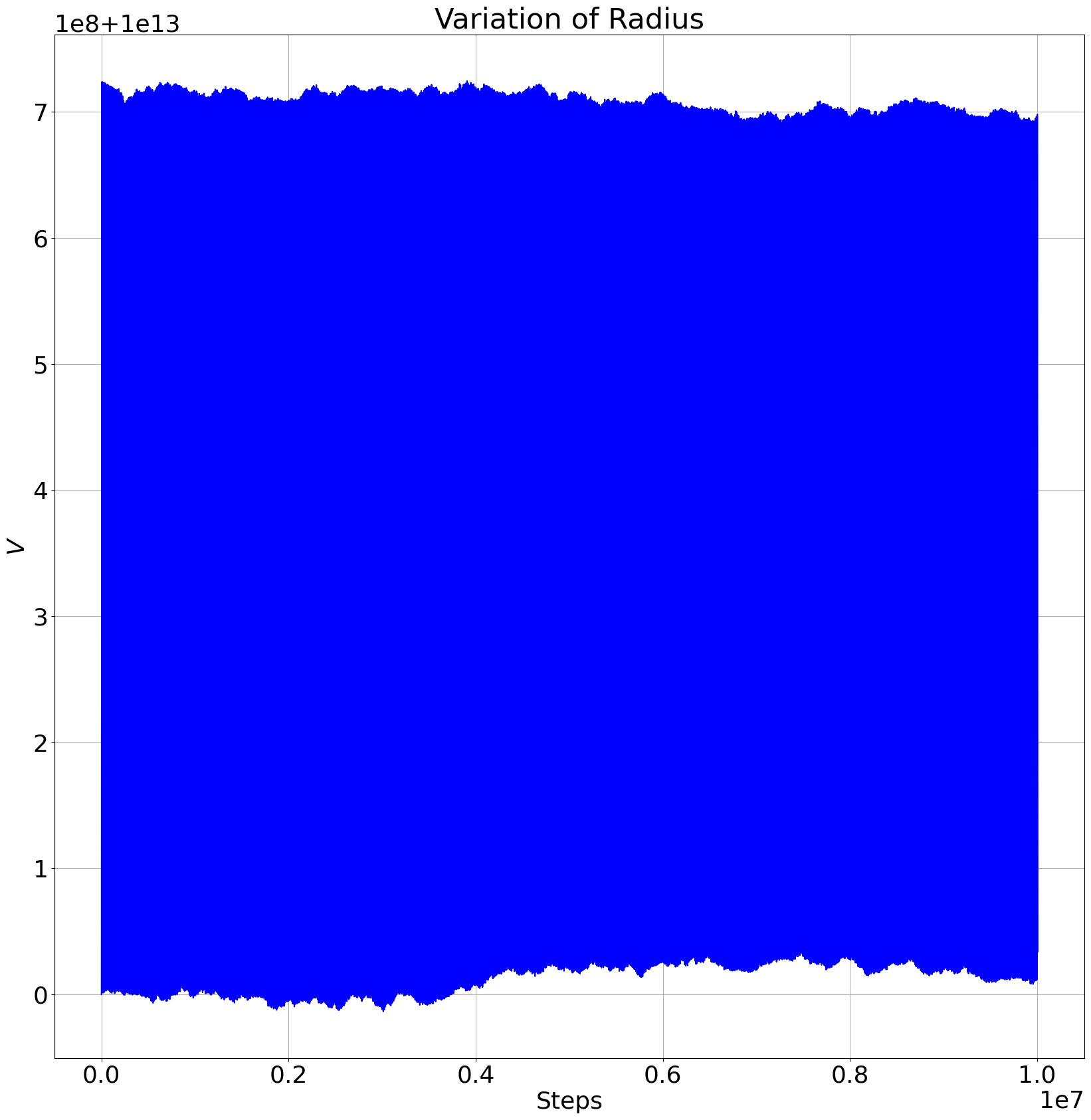}} 
    \subfigure[2.5 PN-with-Prop]{\includegraphics[width=0.24\textwidth]{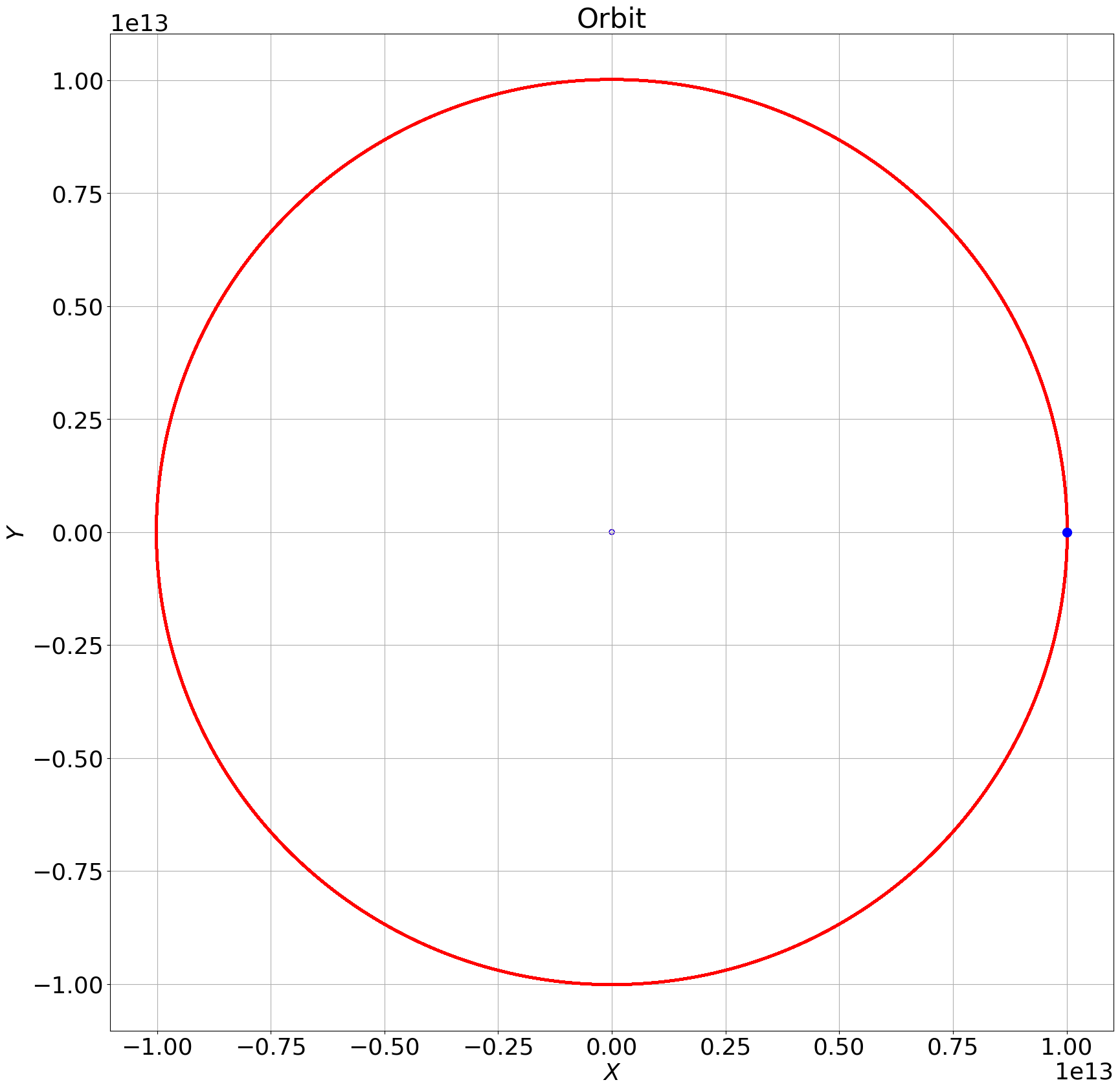}}
    \subfigure[2.5 PN-with-Prop]{\includegraphics[width=0.24\textwidth]{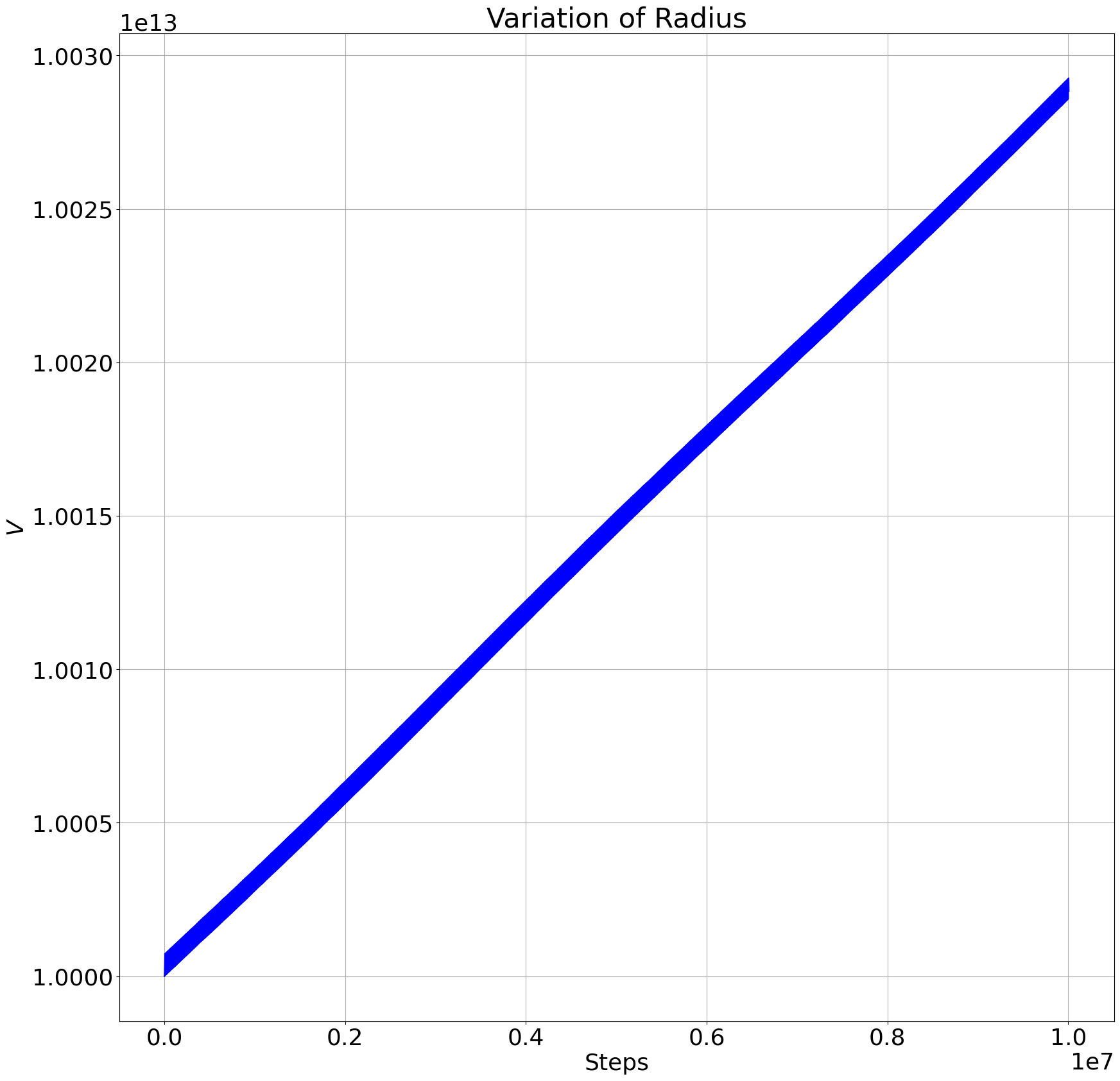}} \\
    \caption{1,000 Orbits for 5 Gravity Models With and Without Field Propagation Time Effects}
    \label{long-run}
\end{figure}

\clearpage
\section*{Aknowledgements}
I offer my sincerest thanks and gratitude to Dr Betti Hartmann and Professor Christian Böhmer both of University College London for Supervising this project.\\ \\ 
\printbibliography
\addcontentsline{toc}{chapter}{Bibliography}
\listoffigures
\listoftables
\end{document}